\documentclass[letterpaper,11pt]{yalephd}

\usepackage{geometry} 
\usepackage{brubaker} 
\usepackage[square,numbers,sort&compress,merge,elide]{natbib}
\usepackage[hidelinks]{hyperref}
\usepackage{adjustbox}

\begin{document}

\title{First results from the HAYSTAC axion search}
\author{Benjamin M. Brubaker}
\advisor{Steve K. Lamoreaux}
\date{December 2017} 

\frontmatter 

\begin{abstract}
The axion is a well-motivated cold dark matter (CDM) candidate first postulated to explain the absence of $CP$ violation in the strong interactions. CDM axions may be detected via their resonant conversion into photons in a ``haloscope'' detector: a tunable high-$Q$ microwave cavity maintained at cryogenic temperature, immersed a strong magnetic field, and coupled to a low-noise receiver. 

This dissertation reports on the design, commissioning, and first operation of the Haloscope at Yale Sensitive to Axion CDM (HAYSTAC), a new detector designed to search for CDM axions with masses above 20~$\mu\text{eV}$. I also describe the analysis procedure developed to derive limits on axion CDM from the first HAYSTAC data run, which excluded axion models with two-photon coupling $g_{a\gamma\gamma} \gtrsim 2\times10^{-14}~\text{GeV}^{-1}$, a factor of 2.3 above the benchmark KSVZ model, over the mass range $23.55 < m_a < 24.0~\mu\text{eV}$. 

This result represents two important achievements. First, it demonstrates cosmologically relevant sensitivity an order of magnitude higher in mass than any existing direct limits. Second, by incorporating a dilution refrigerator and Josephson parametric amplifier, HAYSTAC has demonstrated total noise approaching the standard quantum limit for the first time in a haloscope axion search.
\end{abstract}

\maketitle
\makecopyright{2017}
\tableofcontents
\listoffigures 
\listoftables 

\pagebreak
\vspace*{\fill}
\begin{center}\textit{For Emily}\end{center}
\vspace*{\fill}

\chapter{Acknowledgments} 
First and foremost I would like to thank my advisor, Steve Lamoreaux, for entrusting me with such a central role in the design and operation of HAYSTAC. The depth of Steve's intuition for and competence with all varieties of laboratory apparatus never ceases to amaze me, and his dedication to crossing the boundaries of subfields and learning new things in an ever more specialized world is inspiring. I learned far more than I could have ever imagined working alongside him. I am also extremely grateful for his ability to maintain his sense of humor and perspective in the face of experimental catastrophe. 

I would like to thank the members of my dissertation committee, Dave DeMille, Keith Baker, and Walter Goldberger, for many enlightening conversations and lectures over the years, and for their patience and flexibility as my schedule has slipped and my chapters have ballooned over the past few months.

To all my collaborators on HAYSTAC: thank you for all you have done to make this dissertation possible. To the grad students and postdocs in particular: thank you for making ``HAYSTAC'' possible. I learned pretty much everything I know about software from Yulia Gurevich and Ling Zhong. I'm grateful to Yulia for teaching me what was what when I was just starting out in the lab and for introducing me to Evernote. I'm very grateful to Ling for an extremely productive partnership working out the details of the HAYSTAC analysis these past two years and for her meticulous work making publication-quality figures. 

In recounting his early scientific influences, Heisenberg apparently said ``from Sommerfeld I learned optimism.'' But he never had the chance to meet Karl van Bibber, whose ability to remain upbeat through a slow trawl over parameter space (not to mention the inevitable vicissitudes of experimental physics) is incomparable; I am grateful for his pep talks and his tireless advocacy on my behalf. 

Konrad Lehnert's enthusiasm for physics is simply infectious, and appears to interact in some kind of resonant way with my own to coherently enhance the rate at which my brain is able to absorb new information. I am very much looking forward to exploring the parameter space accessible with this technique in the years to come.

I am grateful to Aaron Chou for many years of advice and inspiration, and for agreeing to be an external reader for my dissertation. I first learned to think like a physicist working with Aaron on the Fermilab Holometer, which remains near and dear to my heart for showing me how much fun experimental physics could be.

Over the past year I have had occasion to seek advice (often with vague and rambling questions) about the still somewhat surreal notion of life after grad school. I am grateful to Dave Moore, Alex Sushkov, Laura Newburgh, Reina Maruyama, and Jack Harris for their thorough and thoughtful answers.

I am grateful to everybody who has made the Yale physics department and the Wright lab such a great environment for learning and doing physics. I am grateful to Ana Malagon for many hours grappling with the axion theory literature before any of this made any sense. I would also like to thank Sid Cahn for being ubiquitous, Daphne Klemme, Sandy Tranquilli, and Paula Farnsworth for keeping logistics at bay throughout my time as a grad student, Jeff Ashenfelter and Frank Lopez for being accessible whenever I was freaking out about scheduled maintenance, and Andrew Currie for helping me through more than my share of poorly timed computer failures. 

Hands down, the best place to write in New Haven is Koffee?\ on Audubon. I would like to thank the people of Koffee?, and Nate Blair in particular for his friendship and for offering me a place to live during this last hectic month.

I very much doubt I would be where I am today without the love and support of my parents, Zsuzsa Berend and Rogers Brubaker, who have consistently encouraged me to cultivate curiosity and pursue my intellectual interests. As a token of my appreciation I have hidden the word ``sociological'' somewhere in this thesis. No cheating! I'm also very grateful that my brother Daniel was so close by during the bulk of my time in grad school: in times of exhaustion or crisis I could always count on his generosity and sense of humor.

Finally, during these past five years I have been fortunate to get to know many more wonderful people than I could possibly name here. Even so, my experience as a human in New Haven has been primarily defined by four relationships.

To Dylan Mattingly: Thank you for demonstrating to me the value of joy! I cannot wait for your fourteen-billion year opera about the slow expansion of a universe filled with an oscillating axion field.

To Zack Lasner: Thank you for the thoughtfulness and nuance you bring to every conversation: to be otherwise would betray your nature. This dissertation would have been epigraphless and thus utterly unreadable if not for you.

To Will Sweeney: Thank you for being my constant companion, from Brahms to Brago, from the blackboard to cheese boards and everything in between. As a friend and a physicist you are a perennial inspiration to me. 

To Emily Brown: I don't know how many different puddles on the floor I would be at this point if not for you but probably a lot. You keep me well and balanced and smiling a big smile. Thank you for being my partner and my friend and supporting me through the ups and downs of this past year especially. 

\mainmatter 

\chapter{Introduction}\label{chap:intro}
\setlength\epigraphwidth{0.42\textwidth}\epigraph{\itshape Afternoon, Aurbis, the reports are true, \\there is a type of zero still to be discovered, \\all [critics--?] agree.}{Traditional Dwemeri children's rhyme}

\noindent The past half century has been witness to extraordinary progress in our understanding of the universe on both the smallest and the largest scales. One especially prominent theme has been the realization that questions about the very small and questions about the very large are intimately related, often in counterintuitive ways.

\begin{figure}[h]
\centering\includegraphics[width=0.8\textwidth]{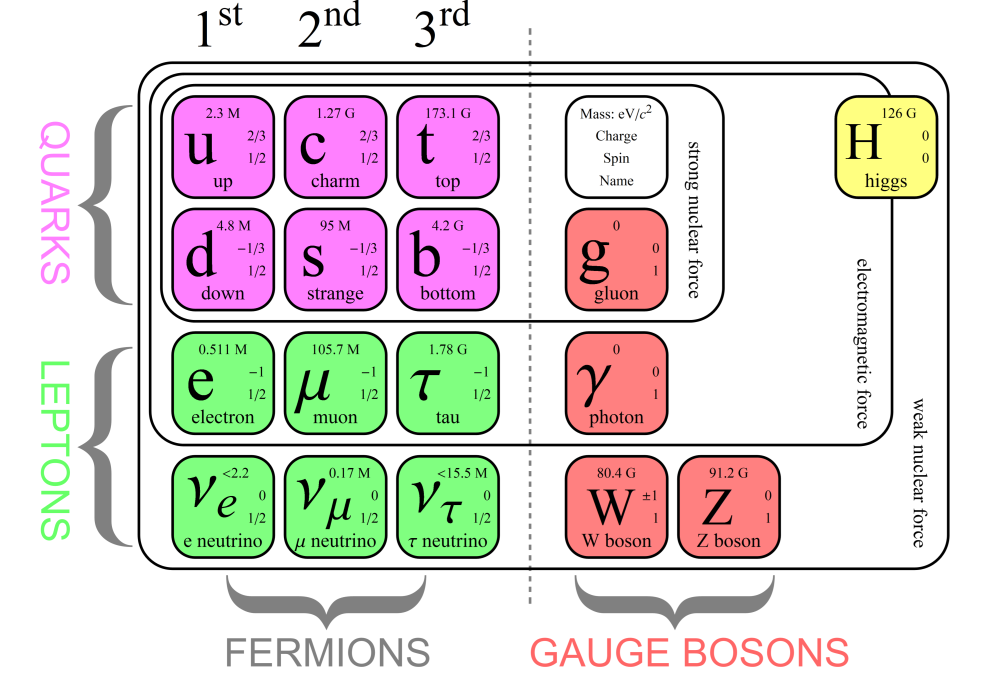}
\caption[The Standard Model of particle physics]{\label{fig:sm} The Standard Model of particle physics. Diagram by Matic Lubej~\citep{SMfig}.}
\end{figure}

The \textbf{Standard Model (SM)} of particle physics, a set of interrelated quantum field theories developed over the course of the 1960s and 1970s, explained the hundreds of ``elementary'' particles known at the time in terms of an underlying scheme so simple that all the truly elementary particles fit in Fig.~\ref{fig:sm}. Since then the SM has been the subject of intense experimental scrutiny, yet has passed every test with flying colors: with the 2012 discovery of the Higgs boson, every particle predicted by the SM has been observed, and certain predictions of the theory of quantum electrodynamics (QED) in particular have been tested with a precision of better than ten parts per billion.

Developments in cosmology (the study of the behavior of the universe on the largest spatial and temporal scales) have been if anything even more dramatic. ``Precision cosmology'' would have seemed a contradiction in terms to most practicing physicists in the first half of the twentieth century, yet now not only do we have overwhelming observational evidence for the proposition that the universe had a beginning, we also have a ``standard cosmological model'' supported by a wealth of independent observations which describes the composition of the universe with sub-percent uncertainty. This model is usually called $\gv{\Lambda}$\textbf{CDM}, where $\Lambda$ stands for \textbf{dark energy} and CDM for \textbf{cold dark matter}.

\begin{figure}[h]
\centering\includegraphics[width=0.6\textwidth]{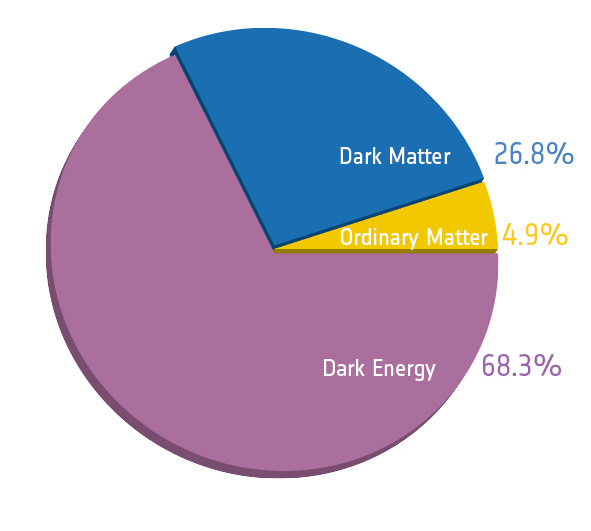}
\caption[Composition of the universe]{\label{fig:lambda} The composition of the universe. Figure from ESA/Planck~\citep{LCDMfig}.}
\end{figure}

Despite these successes, both the SM and $\Lambda$CDM face profound challenges from within. In particle physics, one particularly prominent issue is that the numerical values of many parameters appear to be balanced on the proverbial pinhead, and the SM by itself provides no mechanism to explain this seemingly implausible state of affairs. We will come to a specific example of such a ``fine-tuning problem'' shortly; for now, suffice it to say that new theoretical mechanisms to ``fix'' these problems invariably imply the existence of new particles. In cosmology, theorists deserve some credit for foregrounding the gaping holes in our understanding of the universe: the main problem with the $\Lambda$CDM model is that we understand neither $\Lambda$ nor CDM. If you thought Fig.~\ref{fig:sm} was too complicated, you'll love Fig.~\ref{fig:lambda}: everything described by the SM (quarks, leptons, atoms, stars, dust, Ph.D. theses, etc.) is relegated to the tiny sliver labeled ``ordinary matter.'' We still do not know what the other two components are.

By this I mean we do not know what either dark energy or dark matter is made of. We do know quite a lot (and we are continually learning more) about \textit{where} they are and how they behave on galactic and extragalactic scales. Dark energy appears to be everywhere, and causes the accelerated expansion of the vast regions of empty space between galaxy clusters. It is possible that dark energy is simply a manifestation of ``vacuum energy'' associated with space-time itself, in which case the question of its microscopic constituents is not meaningful. Further discussion of dark energy would take us outside the scope of this thesis.

Dark matter, on the other hand, is concentrated within galaxies and galaxy clusters. It interacts with gravity the same way that normal matter does, but no non-gravitational interactions of dark matter have been observed to date. It is invisible (hence ``dark;'' indeed it neither emits nor absorbs radiation in \textit{any} part of the electromagnetic spectrum) and appears to be spread throughout galaxies rather uniformly (at least compared to the clumpiness of normal matter). Observations indicate that as our solar system orbits the center of the galaxy at about 200 km/s, we are flying into a ``headwind'' of dark matter, which passes through us all the time without leaving a trace.

In short, although dark matter has a profound influence on the formation and dynamics of galaxies, it doesn't seem to do very much on smaller scales. It is a subject of immense theoretical and experimental interest in large part because there is strong evidence that it is non-baryonic (i.e., not made of atoms; this is of course implicit in the presentation of Fig.~\ref{fig:lambda}). Dark matter may thus be a window into ``new physics'' beyond the standard model. As noted above, theoretical extensions to the SM typically predict the existence of new particles. In some cases, hypothetical particles initially conceived in connection with completely unrelated problems in particle physics turn out to have all the right properties to explain dark matter: they would interact extremely weakly with everything in the SM and would be produced copiously in the early universe. If we could prove the existence of such particles, this would constitute a microscopic explanation for the observed astrophysical phenomena we attribute to dark matter. 

This thesis will describe an experiment designed to detect the \textbf{axion}, a hypothetical particle widely regarded as one of the best-motivated dark matter candidates. Having already (hopefully) convinced you that dark matter is worth studying, I will next briefly discuss reasons to suppose that the axion exists, and its rather improbable history as a dark matter candidate.

\section{Motivation for the axion}\label{sec:intro_theory}
The axion holds the dubious distinction of being the only particle named after a line of consumer products~\citep{soap}. It was originally postulated as part of a solution to the \textbf{strong \textit{CP} problem} proposed by theorists Roberto Peccei and Helen Quinn in 1977. The strong $CP$ problem is one the most bizarre fine-tuning problems plaguing the SM; here ``strong'' refers to Quantum Chromodynamics (QCD), the theory within the SM which describes the strong nuclear force, and $CP$ (charge-parity) symmetry is a formal symmetry of any theory in which the laws of physics do not distinguish between matter and antimatter.\footnote{Antimatter, incidentally, is \textit{not} the same thing as dark matter; it will only play a minor role in our discussion here.} 

The \textit{violation} of $CP$ symmetry is a subject of great theoretical interest because there is substantially more matter than antimatter in the observable universe today, yet the known laws of physics are mostly $CP$-symmetric. Rather awkwardly, in the theory of QCD we have precisely the opposite problem. The mathematical form of the theory generally violates $CP$ quite badly: the degree of $CP$ violation is proportional to the sum $\bar{\theta}$ of two completely independent parameters, and the ``natural'' value of this sum is order unity.\footnote{That is, $\bar{\theta}$ much smaller than 1 would seem to require an explanation. I discuss this ``naturalness criterion'' further in Sec.~\ref{sub:naturalness}.} Empirically, $\bar{\theta}<10^{-10}$, i.e., QCD is perfectly $CP$-conserving within the limits of the best (extremely precise) measurements. Within the SM, it appears that strong $CP$ violation is ``accidentally'' suppressed because the additive contributions to $\bar{\theta}$ happen to be equal and opposite to better than one part in \textit{ten billion}. 

\begin{figure}[h]
\centering\includegraphics[width=0.35\textwidth]{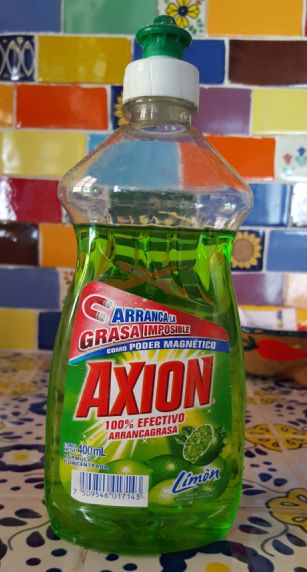}
\caption[Namesake of the axion]{\label{fig:soap} The namesake of the axion. Photo by Z.~D.~Lasner.}
\end{figure}

One of the weirdest things about the strong $CP$ problem is that it is really quite benign. Often fine-tuning problems are amenable in principle to so-called ``anthropic'' solutions, whose essence is the controversial claim that no \textit{mechanism} is required to explain fine-tuning without which sentient observers could not have evolved. For example, it remains a mystery why the parameter $\Lambda$ which controls the strength of dark energy should have the value it does, when the simplest theoretical explanation predicts that it should be larger by more than 100 orders of magnitude.\footnote{It is hard to imagine this discrepancy being displaced as the worst agreement between theory and observation in the history of science.} Some would argue that this is a meaningless question, as galaxies would not even be able to form if $\Lambda$ were much larger. The strong $CP$ problem avoids such thorny philosophical questions altogether: simply put, there is no anthropic reason to favor such a small value of $\bar{\theta}$ \citep{dd2015}. 

In the Peccei-Quinn (PQ) solution to the strong $CP$ problem, the axion works essentially like a cosmic feedback loop, which turns on in the early universe and dynamically cancels out whatever initial value $\bar{\theta}$ happens to have. This elegant theoretical mechanism was not initially thought to have any connection to the dark matter problem. However, the axion mass $m_a$ is a free parameter -- that is, the PQ mechanism does not require $m_a$ to have any particular value. Within a few years theorists realized that if $m_a$ were much \textit{smaller} than the initial formulation of the PQ mechanism assumed, axions would interact very weakly with SM particles, and moreover a large cosmic abundance of axions would be generated as a side effect of solving the strong $CP$ problem: \textit{light axions can constitute dark matter.}

A wide range of possible axion masses was quickly shown to be incompatible with experimental results in particle physics and observations in astrophysics. Thus we are left with the intriguing conclusion that if axions exist at all, they almost certainly account for at least part of the dark matter. The axion would be the lightest of the fundamental particles: the upper bound on its mass is comparable to the \textit{lower bound} on the neutrino masses, and axions may be many orders of magnitude lighter still. But if the strong $CP$ problem is solved by the PQ mechanism, the cosmic density of axions is so enormous that their collective gravitational influence dominates the motion of the largest structures in the universe! 

\section{Detecting dark matter axions}\label{sec:intro_exp}
Through astronomical observations we have detected the effects of dark matter on the motion of distant galaxies and the motion of distant stars and gas clouds within our own galaxy. We have reason to believe that dark matter is all around us all the time -- can we detect its effects more directly in a laboratory experiment? 

Detecting the \textit{gravitational} interactions of dark matter in the lab is a hopeless endeavor, because gravity is an extremely weak force whose effects only become significant on very large scales.\footnote{If this seems surprising to you, consider the fact that you can lift a paper clip with a refrigerator magnet even though the gravitational force of an entire planet is pulling down on the paper clip.} Fortunately, dark matter can have non-gravitational interactions, provided they are sufficiently weak to avoid conflict with observation -- such interactions would have no effect on galactic dynamics, but crucially might render dark matter detectable in the lab. 

Specific theories of particle dark matter candidates predict specific forms for these weak interactions.\footnote{I am using ``weak'' here in a generic descriptive sense, not in reference to the weak nuclear force specifically. Confusingly, another prominent dark matter candidates is called the WIMP (for weakly interacting massive particle), and there ``weakly interacting'' does have the more specific meaning.} Most laboratory searches for dark matter axions specifically seek to detect the vestigial interaction of the axion with a pair of photons.\footnote{Photons are particles of light. The astute reader may object that I already said dark matter does not absorb or emit light. This is true, but the axion's interaction with light is qualitatively different. Roughly speaking, regular matter can get rid of extra energy by shedding photons, whereas an axion must bump into a photon to turn into a photon.} One of the more attractive features of the axion as a dark matter candidate is that this interaction \textit{must exist} if the axion solves the strong $CP$ problem, and moreover the range of allowed values for the ``coupling constant'' quantifying the strength of this interaction is limited. 

In practice, the design of any realistic experiment must be optimized for some small slice of the allowed axion mass range. Provided we can build a sufficiently sensitive detector, we should be able to see a clear signature of axion interactions if $m_a$ happens to fall in the appropriate range; conversely, in the absence of a detection, we can rule out the existence of such axions. The most sensitive CDM axion detectors developed to date have been variations on a basic model called the \textbf{axion haloscope},\footnote{The dark matter in galaxies is sometimes referred to as a ``halo'' because it extends out past the bulk of the luminous matter in a diffuse blob.} whose essential elements are an extremely cold microwave cavity, a high-field magnet, and a low-noise microwave-frequency amplifier. With some injustice to the details, you can think of a haloscope as an exquisitely sensitive radio receiver inside an MRI magnet.

\begin{figure}[h]
\centering\includegraphics[width=0.45\textwidth]{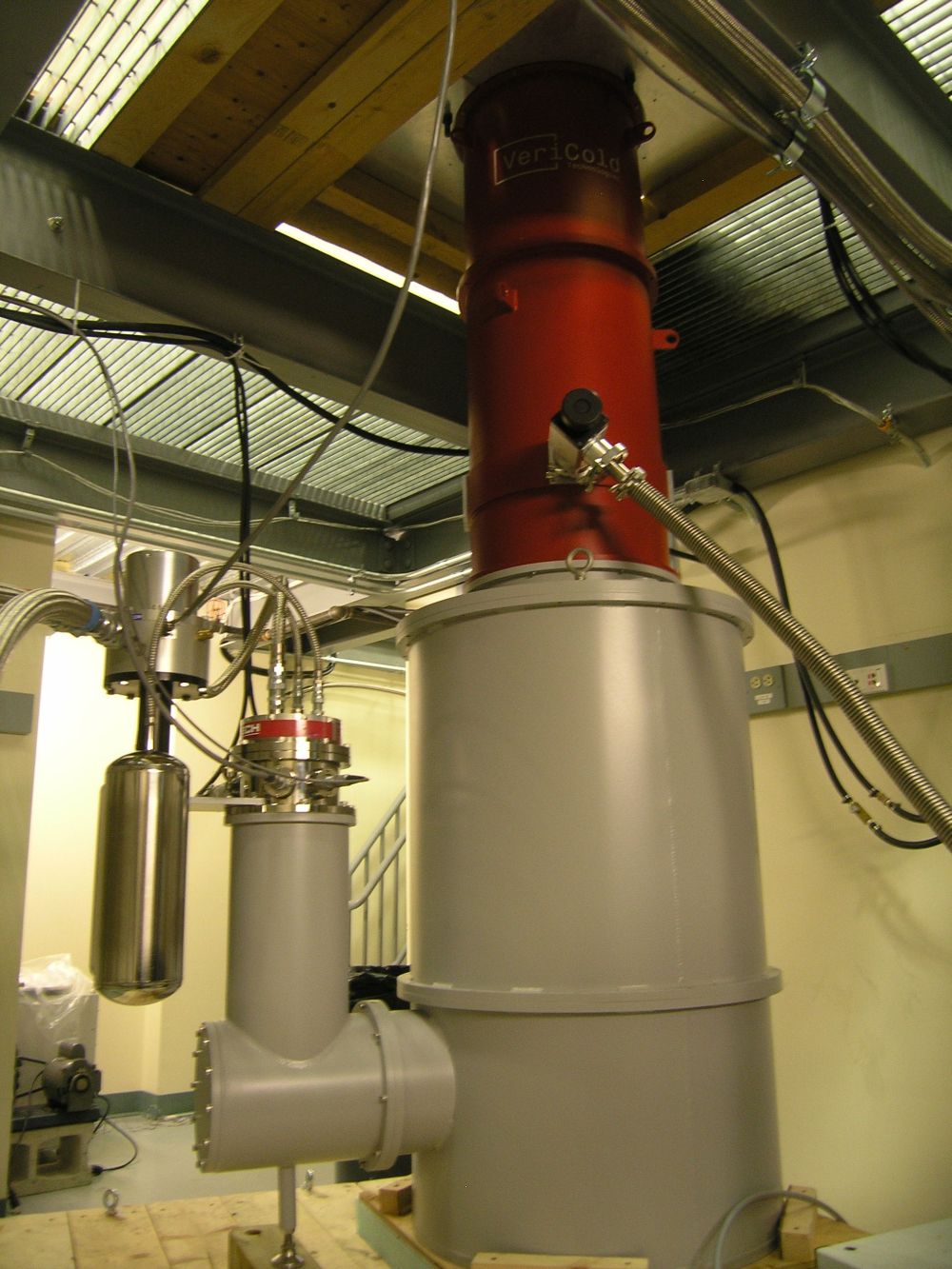}
\caption[The HAYSTAC detector]{\label{fig:haystac} The HAYSTAC detector fully assembled.}
\end{figure}

The axion haloscope is a very atypical particle detector, as the axion is anything but a typical particle. Particle physics is sometimes called ``high-energy physics,'' and indeed the axion owes its existence to new physics at extremely high energies, far beyond the reach of any conceivable collider. Nonetheless the interactions of CDM axions in the present-day universe occur at very \textit{low} energies, so haloscopes must rely on techniques and technology usually associated with fields far removed from particle physics. Similarly, particle physics is famous for enormous detectors and collaborations of hundreds of scientists, but a typical axion haloscope (see Fig.~\ref{fig:haystac}) is a laboratory-scale device which can be operated by a handful of people. 

Even the term ``particle'' is something of a misnomer -- the axion is actually so light that it behaves more like a wave, and the haloscope technique exploits this unusual behavior. Essentially, the magnet mediates a coherent transfer of energy from the axion field (which oscillates at a characteristic frequency proportional to $m_a$) to electromagnetic waves at the same frequency. Since we do not know the exact value of $m_a$, we do not know the frequency of this extremely faint electromagnetic signal, but it will fall in the microwave range for typical values of $m_a$. Thus a haloscope must be tunable -- the analogy to an ordinary car radio is actually pretty good!

Over the past five years I have had the great fortune to play a central role in the design, construction, commissioning, and first operation of the \textbf{H}aloscope \textbf{A}t \textbf{Y}ale \textbf{S}ensitive \textbf{T}o \textbf{A}xion \textbf{C}DM (HAYSTAC). Several critical components of the HAYSTAC detector were designed and fabricated by our collaborators at UC Berkeley, the University of Colorado, and Lawrence Livermore National Lab. My task as the first graduate student at the host institution was basically to put the whole thing together and make it work.

Prior to this work, only a single experiment had achieved sensitivity to realistic dark matter axion models, with masses around a few millionths of an electron-volt. HAYSTAC, designed to target axions heavier by about a factor of ten, is the second experiment to set direct laboratory limits on viable models of axion CDM. Although we did not discover the axion in the first HAYSTAC data run,\footnote{If we had, you would not be reading about it here for the first time.} we did open a new portion of the allowed axion mass range to experimental investigation.  Our most significant technical achievement was the successful integration of a Josephson parametric amplifier (JPA) into the unusual environment of an operational haloscope. This remarkable device, initially developed to facilitate research in quantum information science, can measure microwave-frequency electromagnetic fields with a precision approaching the fundamental limits imposed by the laws of quantum mechanics. 

\section{The structure of this thesis}\label{sec:outline}
In this introduction, I have presented an overview of my thesis research on the search for dark matter axions with HAYSTAC. I have done my best to keep the discussion broadly accessible to readers without a background in physics. The chapters that follow will necessarily assume familiarity with more specialized topics and methods in physics research, but I will do my best to point the reader to relevant pedagogical references wherever appropriate. My intention is that the whole thesis should be comprehensible to a first-year graduate student willing to do the requisite background reading.

The remainder of my thesis is organized as follows. In chapter~\ref{chap:theory}, I review the strong $CP$ problem, the Peccei-Quinn solution that gives rise to the axion, and features of different axion models. In chapter~\ref{chap:cosmo}, I first review the evidence for and properties of dark matter, then discuss the cosmological implications of light axions. If you are only interested in how the experiment works, you could start with chapter~\ref{chap:search}, in which I discuss the parameter space available to axions and the principles of haloscope detection. In chapter~\ref{chap:detector}, I introduce the main components of the HAYSTAC detector. In chapter~\ref{chap:data}, I discuss the measurements used to calibrate the sensitivity of HAYSTAC and the data acquisition procedure. In chapter~\ref{chap:analysis}, I explain how we derive an axion exclusion limit from the raw data written to disk during a HAYSTAC data run. Finally, I conclude in chapter~\ref{chap:conclusion} with a summary of our results and their significance, and a brief discussion of the next steps for HAYSTAC specifically and the broader outlook for the field. I have tried to adopt a pedagogical approach wherever a suitably thorough discussion aimed at non-specialists does not exist in the literature. The research described in this work has been published in Refs.~\citep{PRL2017,NIM2017,PRD2017}. All citations and section, figure, and equation references in the PDF version of this document are hyperlinked, even though they are not surrounded by ugly red boxes.

From chapter~\ref{chap:theory} through the first half of chapter~\ref{chap:search} I will use natural units with the Heaviside-Lorentz convention except where otherwise stated. In Heaviside-Lorentz units the fundamental constants have the values $c=\hbar=k_B=\varepsilon_0=1$, implying that energy, mass, temperature, and frequency can all be expressed in energy units, which we take to be electron volts (eV); length then has dimensions of eV$^{-1}$, and all electromagnetic quantities have dimensions of eV to some power. This unit system facilitates back-of-the-envelope estimates of many disparate phenomena: you can use it to calculate the approximate temperature of the surface of the sun from the fact that you can see the light it emits and understand why the same property that makes x-rays useful for imaging also makes exposure dangerous, to name just two examples. I would encourage all aspiring physicists (and armchair physicists) to familiarize themselves with its uses and limitations.\footnote{Beware of pesky dimensionless factors which are not $\mathcal{O}(1)$!} The statement that $k_B=\hbar=c=1$ in natural units is equivalent to the handy relation
\begin{equation}\label{eq:hl_units}
1~\text{eV}~\approx~1.2\times10^4~\text{K}~\approx~240~\text{THz}~\approx\ \left(1.2~\mu\text{m}\right)^{-1}.
\end{equation}
For rough estimates, just retaining the first digit in each equality is often sufficient.


\chapter{The strong $\boldsymbol{CP}$ problem and axions}\label{chap:theory}
\setlength\epigraphwidth{0.4\textwidth}\epigraph{\itshape Everything not forbidden is compulsory.\vspace{-15pt}}{Murray Gell-Mann}

\noindent In this chapter I will tell the axion's origin story with its many twists and turns. I will begin with an overview of the symmetries of the SM (Sec.~\ref{sec:sm}) and discuss how they are concealed by emergent effects at low energies. In doing so I will introduce concepts like chirality and spontaneous symmetry breaking, which will play important roles in our discussion later on. This introduction will also lead us naturally to a discussion of how the approximate symmetries exhibited by light quarks at high energies are reflected in the spectrum of hadrons at low energies. We will find that one such symmetry appears to be badly violated -- this is the so-called $U(1)_A$ problem of the strong interactions (Sec.~\ref{sec:u1a}). 

The resolution of the $U(1)_A$ problem follows from a deeper appreciation of the vacuum structure of QCD. Yet this solution begets a problem of its own -- the true vacuum state of QCD is characterized by a parameter $\bar{\theta}$ which leads to observable $CP$-violating effects for any value other than 0: specifically, the neutron will develop an electric dipole moment (EDM) proportional to $\bar{\theta}$. The severe constraints on $\bar{\theta}$ from the nonobservation of a neutron EDM give rise to the strong $CP$ problem (Sec.~\ref{sec:strong_cp}), which seems to defy all our aesthetic criteria for how a theory should behave. 

I will then discuss the proposed theoretical mechanisms for solving the strong $CP$ problem, the most compelling of which is the PQ mechanism (Sec.~\ref{sec:peccei_quinn}). One way to understand the PQ mechanism conceptually is that it amounts to adding another $U(1)_A$ symmetry to the SM -- this is how Peccei and Quinn originally conceived of what they were doing, and appreciating this perspective is the motivation for my historical detour through the $U(1)_A$ problem. Equivalently, the PQ mechanism may be regarded as a way to ``promote'' $\bar{\theta}$ from a fixed parameter of the theory to a dynamical \textit{axion} field -- the validity of this latter approach was first noted by Weinberg and Wilczek. I will explain how these two descriptions of the PQ mechanism are related.

Finally, I will discuss the generic properties of axions, and the relevant features of the most prominent axion models. In particular, I will discuss how the original PQWW axion thought to arise from electroweak symmetry breaking was ruled out by experiment. In attempting to salvage the PQ solution, theorists proposed models in which the axion emerges from new physics far above the electroweak scale. These ``invisible'' axion models were initially thought to have no observable consequences, but were later shown to have profound implications for cosmology. Discussion of axion cosmology is deferred to chapter~\ref{chap:cosmo}.

As the above outline no doubt already indicates, an account of axion theory is simply not possible without assuming some familiarity with quantum field theory (QFT) and the SM specifically. Nonetheless, I will try to keep the discussion heuristic rather than overly formal. Most papers on axion theory were written by particle theorists for particle theorists; my aim here is to present the theory in a way that is comprehensible to mere experimentalists like myself. For an introduction to the SM, see Griffiths~\citep{griffiths2008} or the very succinct summary in appendix B of the cosmology text by Kolb and Turner~\citep{KT1994}. For a more formal introduction to the techniques and concepts of QFT, see Peskin \& Schroeder~\citep{PS1995}, Schwartz~\citep{schwartz2014}, or Srednicki~\citep{srednicki2007}. For reviews of axion theory specifically, see Refs.~\citep{cheng1988,peccei1996,peccei2008}.

\section{The Standard Model}\label{sec:sm}
At energies above a few TeV, the SM is very simple and symmetric. In this limit it is a theory of massless spin-1/2 fermions subject to the gauge symmetries $SU(3)_C\times SU(2)_W\times U(1)_Y$.\footnote{The subscripts here just serve to distinguish these \textbf{gauge} (space-time-dependent) symmetries from approximate \textbf{global} symmetries based on the same symmetry groups; they denote color, weak isospin, and hypercharge, respectively. $U(1)$ refers to invariance under phase rotation; $SU(2)$ and $SU(3)$ are \textbf{non-Abelian} groups corresponding to higher-dimensional abstract rotations.} There are also spin-0 bosons (scalar fields) which are \textit{not} massless; we will ignore these scalars for now and come back to them in Sec.~\ref{sub:ckm}. 

The existence of the gauge symmetries implies that the fermions can interact with each other by exchanging massless spin-1 gauge bosons. The number of gauge bosons resulting from each gauge symmetry is the number of generators of the corresponding symmetry group. Thus interactions in QCD [for which the gauge group is $SU(3)_C$] are mediated by 8 gluons, and interactions in the unified electroweak theory [with gauge group $SU(2)_W\times U(1)_Y$] are mediated by four gauge bosons called $W^+$, $W^-$, $W^0$, and $B^0$.

Next we can enumerate the various fermions. It turns out that they come in three ``generations,'' where the fermions in the $2^\text{nd}$ and $3^\text{rd}$ generations are identical to the fermions in the $1^\text{st}$ generation in almost every respect. We can describe most of SM physics by restricting our focus to the 15 fermions in the $1^\text{st}$ generation (we shall return to the other two in Sec.~\ref{sub:ckm}). 12 of these are quarks, which come in each possible combination of three \textit{colors} (which I will call ``red,'' ``green,'' and ``blue''), two \textit{weak isospin} varieties (``up'' and ``down''), and two \textit{chiralities} (right- and left-handed). The remaining 3 fermions are the leptons, which also come in two weak isospin varieties. The ``down'' leptons (electrons) exist in both chiralities, but only left-handed versions of the ``up'' leptons (neutrinos) exist. 

Unlike the other distinctions between the fermions I introduced above, chirality is not a quantum number of the SM gauge groups. Its formal definition is rather abstract, but for massless fermions chirality is equivalent to a more physically intuitive concept called \textit{helicity}. The helicity of a particle simply indicates whether its spin is aligned (right-handed) or anti-aligned (left-handed) with its momentum.\footnote{There is great potential for confusion in the fact that fermion fields have both particle and antiparticle excitations. For a left-chiral field, the particles have left-handed helicity, but the antiparticles have right-handed helicity (and vice versa).} It is easy to see that helicity is not well-defined for a massive particle: we can always boost to a reference frame moving faster than the particle, in which case its momentum will appear to change sign, whereas its spin will not. Massless particles always travel at the speed of light so their helicity is conserved. This intuitive picture suggests that fermion mass implies an interaction between spin-1/2 fields of opposite chirality with otherwise equivalent quantum numbers.

Within the SM, the left-handed and right-handed quarks and leptons do \textit{not} have identical quantum numbers, which explains why mass terms are forbidden for SM fermions.\footnote{Fermions can also gain mass through so-called Majorana mass terms which do not mix chirality. But Majorana masses are not compatible with $U(1)$ symmetries, so this does not work for any of the SM particles with the possible exception of neutrinos. Neutrino mass is outside the scope of our discussion here.} In particular, all of the right-handed fields are neutral under $SU(2)_W$, and right-handed and left-handed versions of each field also have different \textit{hypercharges} (their phases rotate differently under a $U(1)_Y$ transformation). The leptons are distinguished from the quarks by the fact that they are neutral under $SU(3)_C$ (which is precisely why they do not come in different colors). The interactions of $SU(2)_W$ gauge bosons with leptons are exactly analogous to their interactions with quarks.

At this point, you may be inclined to contest my description of the SM as very simple and symmetric. There are certainly many aspects of the theory which remain mysterious -- we do not know \textit{why} nature chose these specific gauge groups, or why only left-handed fields transform under $SU(2)_W$. But setting aside these seemingly arbitrary features, there is a clear formal symmetry between the strong and electroweak interactions in the high-energy limit we have been considering. For example, a right-handed red up quark can turn into a right-handed red down quark by emitting a $W^+$ boson, or it can turn into a right-handed green up quark by emitting either of the two distinct gluons that couple the red and green quark states. There are also $SU(2)_W$ [$SU(3)_C$] interactions in which weak isospin [color] does not change, involving the emission or absorption of the $W^0$ boson [either of the two ``neutral'' gluons]. 

We do not observe this symmetry between $SU(2)_W$ and $SU(3)_C$ in the low-energy world of our daily experience due to two distinct emergent effects. The first is \textbf{electroweak symmetry breaking} (EWSB) via the Higgs mechanism, which gives both the weak gauge bosons and the fermions masses, and in particular gives the up-type and down-type quarks within each generation \textit{different} masses. The second is \textbf{confinement} in QCD: at low energies the strong force gets so strong that only color-neutral bound states of quarks and gluons are ever observed in nature. Together, these two effects explain why up quarks and down quarks are treated as different particles in typical descriptions of the SM (see e.g., Fig.~\ref{fig:sm}), but the differently colored states of quarks and gluons are not. 

\subsection{Spontaneous symmetry breaking}\label{sub:ssb}
Both EWSB and confinement will turn out to play important roles in axion theory; they are discussed respectively in Sec.~\ref{sub:ckm} and Sec.~\ref{sub:hadrons}. Both of these discussions will require familiarity with the phenomenon of \textbf{spontaneous symmetry breaking (SSB)}, which is also a key ingredient in the PQ solution to the strong $CP$ problem. Thus I will take this opportunity to introduce a simple model of SSB, which will suffice to describe the PQ mechanism. For a more detailed pedagogical discussion, see Refs.~\citep{PS1995,schwartz2014,aitchinson1984}.

\begin{figure}[h]
\centering\includegraphics[width=0.8\textwidth]{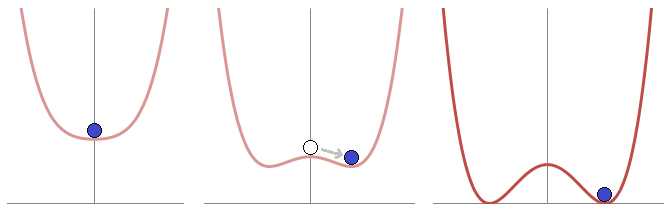}
\caption[Spontaneous symmetry breaking]{\label{fig:ssb} Radial slices of $V(\abs{\phi})$ for $\mu^2>0$ (left), $\mu^2<0$ (middle), and $\mu^2\ll0$ (right). If $\mu^2=\mu^2(T)$ decreases with decreasing temperature $T$, then at sufficiently low temperature the theory spontaneously ``chooses'' one of a continuum of equivalent minima with $\abs{\phi}=v$ characterized by different azimuthal directions in field space. Figure from wikipedia~\citep{ssbfig}.}
\end{figure}

Let us consider the Lagrangian
\begin{equation}\label{eq:ssb_lagr}
\lagr = \frac{1}{2}(\partial_\nu\phi^*)(\partial^\nu\phi) - V(\abs{\phi}),
\end{equation}
where $\phi(x)$ is a complex scalar field subject to the potential
\begin{equation}\label{eq:ssb_v}
V(\abs{\phi}) = \frac{1}{2}\mu^2\abs{\phi}^2 + \frac{\lambda}{4}\abs{\phi}^4.
\end{equation}
The Lagrangian is clearly independent of the phase of $\phi$; i.e., the theory has a $U(1)$ symmetry. If $\mu^2$ is positive, this is just the usual $\phi^4$ theory for a scalar field with mass $\mu$. If $\mu^2=0$, it is the theory of a massless interacting scalar field. The interesting case is $\mu^2<0$: then the potential $V(\abs{\phi})$ is minimized not at $\abs{\phi}=0$ but rather at $\abs{\phi}=v=\mu/\sqrt{\lambda}$. It should be emphasized that the theory has a $U(1)$ symmetry regardless of the value of $\mu^2$; thus $V(\abs{\phi})$ will look the same along any radial 1D slice of the 2D field space. Such 1D slices of $V(\abs{\phi})$ are plotted in Fig.~\ref{fig:ssb} with various values of $\mu^2$. 

We can understand the origin of the term ``spontaneous symmetry breaking'' if we suppose that $\mu^2=\mu^2(T)$, with $\mu^2(0)<0$ and $\mu^2(T) >0$ for sufficiently high temperature $T$. In general, we will be interested in small fluctuations of the field about the minimum-energy field configuration. Below the critical temperature $T_c$ for which $\mu^2(T_c)=0$, $\abs{\phi}=0$ becomes a local maximum and the field must evolve towards one of the new minima at $\abs{\phi}=v$. All such minima characterized by different values of the $U(1)$ phase are energetically equivalent, but fluctuations will randomly single out one of them -- hence ``spontaneous.''

Phase transitions that spontaneously break continuous symmetries are common in our low-energy world, with a prototypical example being a ferromagnet: all spatial directions are totally equivalent but in the low-temperature phase the magnetization must end up singling out one of them. Historically, Kirzhnits~\citep{kirzhnits1972} was the first to suggest that spontaneous breaking of symmetries in the SM could be understood as a cosmological process as the early universe cooled in the aftermath of the Big Bang; Weinberg~\citep{weinberg1974} subsequently worked out the temperature dependence of $\mu^2(T)$ in QFTs with spontaneously broken symmetry: typically we find $T_c\sim v$ in natural units. 

For our present purposes, it is sufficient to consider the properties of the theory in the low-temperature phase. We can do this by expanding $\phi(x)$ around the new minimum:
\begin{equation}\label{eq:sigma_pi}
\phi(x) = \big[v+\sigma(x)\big]e^{i\pi(x)/f_\pi}
\end{equation}
where $f_\pi$ is a parameter with mass dimension 1 whose value we will determine shortly, and $\sigma(x)$ and $\pi(x)$ are the two real scalar fields corresponding to the complex field $\phi(x)$.\footnote{There are two real degrees of freedom for \textit{any} value of $\mu^2$, but $\mu^2<0$ singles out the polar parameterization of field space whereas for $\mu^2\geq0$ polar and Cartesian coordinates are equally valid.} Plugging Eq.~\eqref{eq:sigma_pi} into Eq.~\eqref{eq:ssb_lagr}, we obtain
\begin{equation}\label{eq:broken_lagr}
\lagr = \frac{1}{2}\left(\partial_\nu\sigma\right)^2 + \frac{v^2}{2f_\pi^2}\left(\partial_\nu\pi\right)^2 + \left(\frac{1}{2}\sigma^2+ v\sigma\right)\frac{1}{f_\pi^2}\left(\partial_\nu\pi\right)^2 + V\big(v+\sigma(x)\big)
\end{equation}
The first two terms are just the kinetic energy of the $\sigma$ and $\pi$ fields; canonical normalization of these terms implies that $f_\pi=v$. The next set of terms represents interactions between $\sigma$ and $\partial_\nu\pi$. Finally, if we were to write out the potential explicitly, we would see that there is a $\sigma$ mass term as well as $\sigma$ self-interactions, which will not concern us here.

It should be emphasized that there is no $\pi$ mass term, and indeed only the \textit{derivative} of the $\pi$ field appears in the Lagrangian. Such massless fields (called \textbf{Goldstone bosons} or Nambu-Goldstone bosons) are generic features of theories with spontaneously broken continuous global symmetries. Goldstone bosons are only permitted to have derivative interactions -- even if we had added interactions between $\phi$ and other fields in the original Lagrangian [Eq.~\eqref{eq:ssb_lagr}] the only way to get a factor of $\pi$ to show up in the new Lagrangian [Eq.~\eqref{eq:broken_lagr}] is to pull it out of the exponential with a derivative. For the same reason each factor of $\partial_\nu\pi$ in the Lagrangian will be accompanied by a factor of $1/f_\pi$. Of course I have assumed that the interactions we have added respect the original $U(1)$ symmetry.

\begin{figure}[h]
\centering\includegraphics[width=0.8\textwidth]{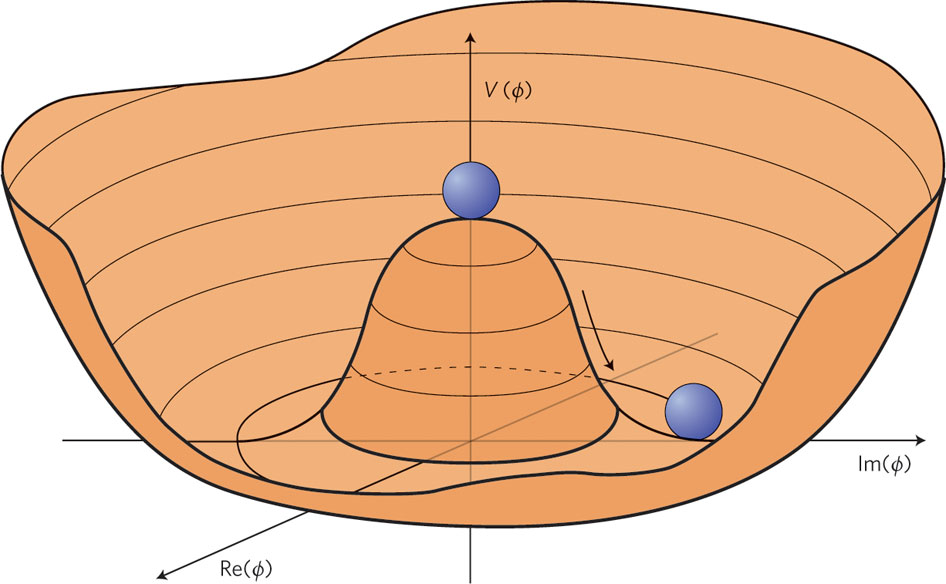}
\caption[The ``wine bottle'' potential]{\label{fig:goldstone} The ``wine bottle'' potential with spontaneously broken $U(1)$ symmetry. The Goldstone boson $\pi(x)$ is the degree of freedom corresponding to azimuthal motion in field space. Figure from Ref.~\citep{goldstonefig}.}
\end{figure}

The key qualitative features of Eq.~\eqref{eq:broken_lagr} can also be seen in Fig.~\ref{fig:goldstone}, where the ``wine bottle'' potential $V(\abs{\phi})$ is plotted in the full 2D field space. In QFT the mass of a field is the frequency of small oscillations about the minimum of its potential. It is thus clear that the radial degree of freedom $\sigma$ is massive while the azimuthal (Goldstone) degree of freedom $\pi$ is massless -- if we push a ball along the trough it will just stay wherever we put it. Under a $U(1)$ transformation with parameter $\alpha$, the $\pi$ field transforms like
\begin{equation}\label{eq:shift_sym}
\pi(x) \rightarrow \pi(x) + \alpha f_\pi.
\end{equation}
This expression is essentially why the word broken appears ``broken'' in ``spontaneously broken.'' Clearly, the value of the Goldstone field is \textit{not} invariant under $U(1)$ transformations. Instead it moves along the symmetry direction in field space, and basically looks like a dynamical version of the angle $\alpha$ parameterizing the original $U(1)$ symmetry. More formally, whenever a continuous global symmetry is spontaneously broken we will obtain $N$ massless Goldstone bosons, where $N$ is the number of generators of the broken symmetry group. The Goldstone bosons have the same quantum numbers as the symmetry generators. 

All of our discussion has been essentially classical; in the corresponding quantum theory fields are operators on a Fock space, and instead of the minimum of the classical potential we should speak of the \textbf{vacuum expectation value} (VEV) $\langle\phi\rangle = v$ (i.e., the expectation value is evaluated in the minimum-energy state with no particle excitations). However, the qualitative picture is basically unchanged.\footnote{The fact that SSB is  a classical phenomenon implies that we ought to be able to observe something like a Goldstone boson in our everyday low-energy world. A massless particle is one for which $E\rightarrow0$ as momentum $p\rightarrow0$; in a condensed matter system this implies that a ``Goldstone mode'' is a fluctuation in the order parameter which costs 0 energy as the wavelength $\lambda\rightarrow\infty$. For a ferromagnet, the Goldstone mode is a long-wavelength spin wave. Here's an experiment you can try at home: pick up a refrigerator magnet and turn it upside down -- you are rotating all the spins together, so this is the $\lambda\rightarrow\infty$ limit. Observe that the magnet does not have a preferred orientation. You have demonstrated the existence of the Goldstone mode!}

\subsection{Electroweak symmetry breaking and the CKM matrix}\label{sub:ckm}
Armed with an understanding of SSB, let us return to the scalar fields we neglected in our discussion of the standard model gauge symmetries at the beginning of Sec.~\ref{sec:sm}. The standard model contains a single \textit{Higgs doublet} of complex scalars, which transform under $SU(2)_W$ just like the quark and lepton doublets; because the Higgs fields are complex they are also charged under $U(1)_Y$. The Lagrangian also contains a SSB potential for these scalars, which looks something like a higher-dimensional generalization of Eq.~\eqref{eq:ssb_v} but may be a little more complicated. The details of the potential need not concern us: the important point is that below the \textbf{electroweak scale} $v=246$ GeV, the gauge symmetry group $SU(2)_W\times U(1)_Y$ is spontaneously broken down to $U(1)_\text{EM}$ (the gauge symmetry of QED).

Based on the discussion in Sec.~\ref{sub:ssb} above, we might expect three Goldstone bosons corresponding to the generators of the spontaneously broken gauge symmetry, and indeed the absence of such Goldstone bosons confused a lot of smart people for a long time. It turns out that when a \textit{gauge} symmetry is spontaneously broken, there are no Goldstone bosons, and instead the gauge bosons acquire masses of order $v$: this is called the \textbf{Higgs mechanism}.\footnote{Roughly speaking the Higgs mechanism works because the Goldstone bosons have the same quantum numbers as the generators of the gauge group, but so do the gauge bosons; thus the would-be Goldstone bosons have all the right properties to supply each gauge boson with an extra degree of freedom corresponding to longitudinal polarization.} In the case of EWSB, the $W^\pm$ bosons acquire mass along with one linear combination of $W^0$ and $B^0$, which we call the $Z^0$ boson. The orthogonal linear combination $A^0$ (the photon) remains massless as it is the gauge boson of the unbroken symmetry $U(1)_\text{EM}$.

Historically, the Higgs mechanism was developed to give mass to the $W^\pm$ and $Z^0$ bosons (which empirically were clearly not massless) without totally wrecking the renormalizability of the standard model. As a side effect, electroweak symmetry breaking also produces fermion masses. The basic reason for this is Murray Gell-Mann's pithy observation that in QFT ``everything not forbidden is compulsory'' (sometimes called the ``totalitarian principle''). More formally, we should expect any renormalizable interaction which does not violate the symmetries of the theory in question to exist with some nonzero coupling constant. 

There are in fact Yukawa-type interactions between Higgs fields and quarks which are renormalizable and invariant under the full SM gauge group. These terms generally have the form
\begin{equation}\label{eq:yukawa_1}
\lagr_\text{Yukawa} = -y_d\big(\bar{Q}_L\cdot \phi\big)d_R - iy_u\big(\bar{Q}_L\cdot\sigma_2\phi^*\big)u_R + \text{h.c.},
\end{equation}
where $\phi$ is the Higgs doublet, $Q_L=\big(u_L\ d_L\big)^\intercal$ is the left-handed quark doublet, $u_R$ and $d_R$ are right-handed up and down quarks, $y_u$ and $y_d$ are Yukawa coupling constants, and $\sigma_2$ is the second Pauli matrix.\footnote{The Pauli matrices are equal to the $SU(2)_W$ generators up to a normalization factor. Note also that an asterisk denotes complex conjugation and the ``bar'' denotes the combination of complex conjugation and transposition of both the $Q_L$ doublet itself and the spinors in each component of the doublet.} The inner product of 2-vectors in both terms are required to make $\lagr_\text{Yukawa}$ an $SU(2)_W$ singlet, and the right-handed quark fields get rid of color charge. The interested reader can consult Refs.~\citep{PS1995,schwartz2014} for more detailed discussion and to see that all hypercharges also cancel. 

After EWSB, we can replace the Higgs doublet with its VEV $\langle\phi\rangle=\frac{1}{\sqrt{2}}\big(0\ v\big)^\intercal$, and Eq.~\eqref{eq:yukawa_1} becomes
\begin{equation}\label{eq:yukawa_2}
\lagr_\text{Yukawa} = -\frac{1}{\sqrt{2}}y_dv\bar{d}_Ld_R - \frac{1}{\sqrt{2}}y_uv\bar{u}_Lu_R + \text{h.c.}
\end{equation}
This is precisely the form of fermion mass terms anticipated in the introduction to Sec.~\ref{sec:sm}. Thus below the electroweak scale the previously independent left- and right-chiral down quark fields are mixed together and this mixture (which we call the ``down quark'' without specifying chirality) acquires mass $m_d=y_dv/\sqrt{2}$; likewise the up quark acquires a mass $m_u=y_uv/\sqrt{2}$.\footnote{There is also a Yukawa interaction between the Higgs doublet, left-handed lepton doublet, and right-handed electron with the same form as the two terms I have included in Eq.~\eqref{eq:yukawa_1}, and this term gives the electron a mass; EWSB cannot generate neutrino mass without right-handed neutrinos. Eq.~\eqref{eq:yukawa_1} also implies Yukawa interactions between the quarks and the massive radial degree of freedom that remains after EWSB, which is analogous to $\sigma$ in the $U(1)$ example of Sec.~\ref{sub:ssb}. This is the famous Higgs boson $h$; its interactions have the same form as Eq.~\eqref{eq:yukawa_2} with $v$ replaced by $h$.}

It may bother you that no symmetry of the original Lagrangian requires the Yukawa couplings $y_u, y_d$ (and thus the quark masses) to be positive or even real. In fact these terms present another problem. Recall from Sec.~\ref{sec:sm} that the SM actually contains three generations of fermions: the strange quark $s$ and bottom quark $b$ are basically copies of the down quark in that they have all the same quantum numbers under the SM gauge groups (the charm quark $c$ and top quark $t$ are copies of the up quark in the same sense; see Fig.~\ref{fig:sm}). Therefore, no symmetry forbids interactions of the form Eq.~\eqref{eq:yukawa_1} with e.g., $d_R$ replaced with $s_R$, and $y_u$ and $y_d$ should actually be replaced with $3\times3$ matrices $Y_u$ and $Y_d$ that sit between the quark generation 3-vectors -- worse still, there is no reason to expect these matrices to be Hermitian! 

Essentially, we have seen that the ``up'' (``down'') quark given mass by EWSB is an arbitrary linear combination of the $u,c,t$ ($d,s,b$) quarks. We can recast the coupling matrix $Y_u$ into a positive diagonal form by introducing $3\times3$ unitary matrices $U_u$ and $W_u$, which can be absorbed into the definition of the left- and right-handed up-type quark fields respectively. We can diagonalize $Y_d$ in the same way with another two unitary matrices $U_d,W_d$ (see Refs.~\citep{PS1995,schwartz2014} for details). In this \textbf{mass basis}, Eq.~\eqref{eq:yukawa_2} looks like a proper mass term. The couplings of quarks to gluons, photons, and $Z^0$ bosons will not be affected by this change of variables, because such terms do not mix chiralities or different weak isospin states, so factors of e.g., $U_u$ and its adjoint $U_u^\dag$ appear in pairs and cancel out. However, $W^\pm$ bosons turn left-handed up quarks into left-handed down quarks in the original basis, so in the new basis we are left with factors of $V=U^\dag_uU_d$ and $V^\dag$ in the $W^\pm$ interactions.

$V$ is called the \textbf{Cabibo-Kobayashi-Maskawa (CKM) matrix}. By construction it is unitary but not necessarily diagonal. This implies that in the mass basis, $W^\pm$ bosons can mediate generation-changing interactions. This is more typically called ``quark flavor mixing'' and the original basis (in which the generation-changing processes were confined to the Higgs Yukawa couplings) is called the \textbf{flavor basis}.\footnote{I have thus far managed to avoid the term ``flavor,'' which is an artifact of the historical development of particle physics, and is moreover usually applied somewhat inconsistently: in the quark sector there are said to be six flavors $u,d,s,c,b,t$, whereas in the lepton sector ``flavor'' is synonymous with ``generation''). In my view the phrase ``quark flavor mixing'' obscures the fundamental difference between weak isospin mixing (which the $W^\pm$ bosons mediate in any basis) and generation mixing. Having said all of this, I will occasionally speak of ``flavor'' in the pages that follow -- it is convenient for discussing the strong interactions, which do not care that e.g., $s$ quarks are more similar to $d$ quarks than they are to $u$ quarks.} 

The elements of the CKM matrix are free parameters of the theory whose values must be specified by experiment. This is rather annoying from an aesthetic standpoint -- we started out with a beautiful theory characterized only by 3 gauge coupling constants and the 2 parameters of the Higgs potential and now we find ourselves stuck with 6 quarks masses and 3 lepton masses in addition to the elements of the CKM matrix. Nonetheless we must press on and evaluate how many independent real parameters there actually are in the CKM matrix. It is instructive to consider the general case of $n$ quark generations. The most general $n\times n$ unitary matrix has $n^2$ independent real parameters, and an $n\times n$ orthogonal (unitary and real) matrix has $n(n-1)/2$ independent real parameters. This implies that the general unitary matrix can be cast into a form where the remaining $n^2-n(n-1)/2=n(n+1)/2$ elements are phases which make the matrix complex. Thus it initially appears that the CKM matrix for $n$ quark generations contains $n(n-1)/2$ ``mixing angles'' and $n(n+1)/2$ phases. 

The situation is not quite so bad, as the theory still has the global $U(1)_V$ symmetries
\begin{eqnarray}
\big(q_L\big)_k &\rightarrow e^{i\alpha_k}\big(q_L\big)_k \label{eq:u1v}\\
\big(q_R\big)_k &\rightarrow e^{i\alpha_k}\big(q_R\big)_k, \nonumber 
\end{eqnarray}
where $k$ indexes the $2n$ quark flavors and $\alpha_k$ is the same for each $k$ in both lines. The subscript $V$ in $U(1)_V$ stands for \textbf{vector} and means that the left- and right-handed fermion fields transform in the same way under the symmetry operation. By contrast, under an \textbf{axial} or \textbf{chiral} symmetry [e.g., $U(1)_A$] the left- and right-handed fermion fields with the same quantum numbers rotate in the opposite sense. Independent transformations on left- and right-handed fermions can always be equivalently expressed as a combination of vector and axial transformations. In particular, the fact that we had to introduce \textit{two} independent unitary matrices $U$ and $W$ for each weak isospin variety to transform from the flavor basis to the mass basis implies that we used both vector and axial transformations to do so. We will return to the significance of this observation in Sec.~\ref{sub:quark_phase}.

Returning to the issue at hand, we see that for $n$ generations we can make $2n$ independent $U(1)_V$ transformations. One of these (in which $\alpha_k$ is the same for each quark flavor $k$) has no effect on the CKM matrix. The other $2n-1$ can be used to cancel out phases in the CKM matrix, implying that these phases do not actually have observable consequences. Thus, we have seen that the CKM matrix has $n(n-1)/2$ mixing angles and $n(n+1)/2-(2n-1)=(n-1)(n-2)/2$ \textit{observable} phases. We can conclude that for $n=1$ or $n=2$, the CKM matrix is real, but for $n\geq3$ it will be complex.

In general, any complex phase in the Lagrangian which cannot be absorbed by field redefinitions like Eqs.~\eqref{eq:u1v} implies that theory violates time-reversal symmetry $T$. Roughly speaking this is related to the fact that the time evolution operator in quantum mechanics is $e^{-iHt}$; see Refs.~\citep{PS1995,schwartz2014} for details. It can be proved that any reasonable QFT is invariant under the product of the discrete symmetries $C$, $P$, and $T$,\footnote{This is called the $CPT$ theorem; see Ref.~\citep{kl1997} for a good intuitive explanation.} so $T$ violation also implies $CP$ violation, whose significance was noted in Sec.~\ref{sec:intro_theory}. In the SM, there are $n=3$ fermion generations, so the CKM matrix contains a single $CP$-violating phase $\delta_{CKM}$; this is called the \textbf{KM model} of electroweak $CP$ violation.\footnote{If the SM is extended to include neutrino mass, there is also $CP$ violation in the lepton sector, which will not concern us here. Historically, $CP$ violation was observed in the decays of neutral kaons way back in 1964 -- nine years later, Kobayashi and Maskawa \citep{KM1973} postulated the existence of a third generation to explain $CP$ violation before all the particles in the \textit{second} generation were known!} I have presented this basic overview of the KM model because as we will see in Sec.~\ref{sub:quark_phase}, the nature of electroweak $CP$ violation is closely related to the strong $CP$ problem.

\subsection{Confinement and the symmetries of hadrons}\label{sub:hadrons}
Let us now turn from the electroweak interactions to QCD, which also behaves differently at low energies, albeit for a different reason. All of the SM gauge couplings are actually functions of the energy scale. A formal description of this behavior would require us to discuss the renormalization group, which would take us much farther afield from the subject of this thesis than we already are. For a simple theory like QED, one can get a qualitative sense of what is going on by supposing that the \textit{true} value of the coupling constant $e$ is the one we would observe at some very high energy scale, and at these energies $e$ is large enough to partially polarize the vacuum. The resulting plasma of virtual charged particles and antiparticles partially \textit{screens} the electric charge at larger distances [or lower energies; see Eq.~\eqref{eq:hl_units}], making it appear smaller.

This simple conceptual picture can account for gauge couplings that decrease with decreasing energy, and indeed this is the behavior exhibited by both parts of the unified electroweak gauge group. But the coupling constant in QCD actually has the opposite behavior: it \textit{increases} with decreasing energy, implying a kind of ``anti-screening.'' As a result, the theory of QCD is nice and perturbative at high energies (it exhibits ``asymptotic freedom''), but perturbation theory becomes less and less reliable with decreasing energy. In fact, if we were to compute the evolution of the QCD coupling constant with energy, we would find that it formally \textit{diverges} at the \textbf{QCD scale} $\Lambda_\text{QCD}\sim 200$ MeV. This implies the existence of confinement at energies below $\Lambda_\text{QCD}$: the theory is so strongly coupled that quarks and gluons will only appear in color-neutral bound states called \textbf{hadrons} which cannot be separated into components with nonzero color charge!

Hadrons come in two varieties: \textbf{mesons} are bound states of a quark and an anti-quark of the same color, and \textbf{baryons} are bound states of three quarks of different colors.\footnote{This description will suffice for our purposes, but it is best not to get too attached to it -- a real nuclear physicist would emphasize that many of the properties of the hadrons arise from the gluons (and virtual quark/anti-quark pairs) doing the binding.} But despite this simple classification, it is impossible to calculate low-energy hadron interactions explicitly from the more fundamental theory of quark-gluon interactions. Instead, it is fruitful to consider the global symmetries of the quark theory and see how they manifest in the spectrum of hadrons (See also Refs.~\citep{aitchinson1984,peccei1996,srednicki2007}). Historically, this perspective was very important for making sense of the plethora of hadrons discovered in particle physics experiments from the 1940s onwards, and for demonstrating the utility of the quark model.

We saw in Sec.~\ref{sub:ckm} that the full SM Lagrangian has an exact $U(1)_V$ global symmetry for each quark flavor. When we neglect the leptons and the electroweak interactions of quarks, several other \textit{approximate} symmetries become apparent. Empirically, the lightest three quarks have masses $m_u=2.2$~MeV, $m_d=4.7$~MeV, and $m_s=96$~MeV \citep{pdg2016}. Note that $m_d-m_u$, $m_u$, and $m_d$ are all $\ll\Lambda_\text{QCD}$, which is the characteristic energy scale of effects related to confinement. In this sense the $u$ and $d$ quarks may be regarded as approximately massless. $m_s\approx0$ is clearly a worse approximation, but it will also turn out to be good enough; we will ignore the $c,b,t$ quarks whose masses are $\gg\Lambda_\text{QCD}$.

We will begin by considering only the $u$ and $d$ quarks and then extend our analysis to include the $s$ quark. If the mass difference between $u$ and $d$ quarks is ignored, QCD has a global $SU(2)_V$ symmetry -- that is, the theory is invariant under abstract ``rotations'' of up quarks into down quarks and vice versa.\footnote{$SU(2)_V$ [also called isospin, short for isotopic spin] should \textit{not} be confused with the $SU(2)_W$ gauge symmetry exhibited by the weak interactions of the left-handed components of these same quarks! The fact that both theories have the same underlying symmetry group is just a coincidence, but a happy one for the historical development of particle physics: the fact that isospin so nicely explained properties of nuclei led to further investigation of $SU(2)$ symmetries and eventually to the development of electroweak theory.} In the massless limit, the theory also has an $SU(2)_A$ symmetry, in which the left- and right-handed components of the quarks rotate into each other in opposite directions, and a $U(1)_A$ symmetry, under which the quark phases rotate axially. Very generally, axial symmetries only exist in the massless limit, because quark mass terms mix the two chiralities. For this reason, the massless limit is also called the \textit{chiral} limit. 

All told, the theory of massless 2-flavor QCD is invariant under global $SU(2)_V \times SU(2)_A \times U(1)_V \times U(1)_A$ transformations, where the $U(1)_V$ piece is exact even for $m_u\neq m_d\neq0$. Now let's consider the hadrons. The $U(1)_V$ symmetry implies that we should observe anti-baryons with exactly the same mass as the corresponding baryons but opposite charge, and indeed we do.\footnote{Mesons are their own anti-particles.} The $SU(2)_V$ symmetry implies that we should observe doublets of hadrons with almost the same mass (because $m_u$ and $m_d$ are not quite equal in the real world) whose strong force interactions are identical. The most obvious such doublet comprises the proton ($m_p=938.3$~MeV) and neutron ($m_n=939.6$~MeV).

This is very encouraging for the quark model, but what about the axial symmetries? $SU(2)_A$ implies the existence of \textit{another} doublet of particles nearly degenerate in mass with the proton and neutron but with opposite parity, and we do not observe such particles. We saw in Sec.~\ref{sub:ssb} that a global symmetry which is manifest at high energies can be hidden at low energies if some operator in the theory develops a symmetry-breaking VEV. Clearly $\langle\bar{u}u\rangle\neq0$ or $\langle\bar{d}d\rangle\neq0$ would spontaneously break both axial symmetries, as these quark bilinear operators are formally similar to mass terms. There is precedent in the theory of superconductivity for the spontaneous formation of such a \textit{fermionic condensate}, and physically we expect the quarks to be tightly bound together below the confinement scale. If the axial symmetries are spontaneously broken, we should expect three Goldstone bosons corresponding to the generators of $SU(2)_A$ along with a fourth Goldstone boson from $U(1)_A$. Because these symmetries are chiral, all four Goldstone bosons will be \textbf{pseudoscalar} fields which are odd with respect to parity.

Empirically, we do not observe any massless hadrons, but we do observe a triplet of light mesons $\pi^0,\pi^+,\pi^-$ (collectively called \textit{pions}) with $m_{\pi^0}=135$~MeV, $m_{\pi^\pm}=140$~MeV which have all the expected properties of $SU(2)_A$ Goldstone bosons except nonzero mass. Of course, the $u$ and $d$ quarks are not in fact massless, so the axial symmetries were only approximate symmetries to begin with: the pions are the \textbf{pseudo-Goldstone bosons} of the approximate $SU(2)_A$ symmetry.  It may bother you that $m_{\pi^0}$ and $m_{\pi^\pm}$ are not \textit{that} small compared to $\Lambda_\text{QCD}$; similarly if we were to explicitly construct an effective Lagrangian for 2-flavor QCD in the hadronic phase analogous to Eq.~\eqref{eq:broken_lagr}, we would find that the \textit{pion decay constant} $f_\pi=93$~MeV.\footnote{As an aside we note that for pseudo-Goldstone bosons like the pions, powers of $f_\pi$ appear in the denominator in the symmetry-breaking mass term as well as the (derivative) interaction terms. The quantity $f_\pi$ is called a ``decay constant'' for largely historical reasons; of course it does enter into calculations of pion decays along with all other pion interactions.} The resolution of this apparent puzzle is just that the nucleon mass ($\approx940$~MeV; see above) is actually a better measure of the energy scale of confinement in this case; it is related to $f_\pi$ and $\Lambda_\text{QCD}$ by relatively small dimensionless factors. The important point at the end of the day is that the pions are much lighter than all the other hadrons. With a more formal analysis that we will not pursue here, it can be shown that the measured pion masses are consistent with the explicit breaking of $SU(2)_A$ by $m_u,m_d\neq0$ (see in particular Ref.~\citep{srednicki2007}). 

It will turn out to be useful to extend the preceding discussion to 3-flavor massless QCD (taking $m_s\approx0$) before we consider the $U(1)_A$ symmetry. In this case we expect a global $SU(3)_V \times SU(3)_A \times U(1)_V \times U(1)_A$ symmetry. Without going into the details, we do indeed see the expected triplets in the spectrum of baryons, as well as more doublets corresponding to different $SU(2)$ subgroups of $SU(3)_V$. If the axial symmetries are spontaneously broken in the hadronic phase, we should expect eight pseudo-Goldstone bosons from $SU(3)_A$ (c.f.\ the eight gluons) and one from $U(1)_A$. Indeed, there is a \textit{pseudoscalar meson octet} with the right quantum numbers comprising the three pions, four \textit{kaons} ($K^0,\bar{K}^0,K^+,K^-$), and one $\eta$ meson. The kaons and $\eta$ are heavier than the pions (their masses are around 500 and 550 MeV, respectively), and this should be no surprise, since $m_s\approx0$ is not a great approximation. But in a quantifiable sense, they are still light enough to be understood as pseudo-Goldstone bosons of $SU(3)_A$.

In short, there is strong evidence that the approximate $SU(2)_A$ and $SU(3)_A$ symmetries of the strong interactions of light quarks are spontaneously broken in the hadronic phase.\footnote{The precise relationship between confinement and chiral symmetry breaking is actually quite complicated, and remains the subject of current research.} This implies that expectation values of the form $\langle\bar{q}q\rangle$ are nonzero, so $U(1)_A$ should be spontaneously broken as well. In 3-flavor QCD, we should expect 3 pseudoscalar mesons with 0 values for the quark flavor quantum numbers,\footnote{Under a consistent naming scheme these would be called ``upness,'' ``downness,'' and ``strangeness.'' For historical reasons the first two are instead called positive and negative \textit{isospin projection}.} and indeed there are three such mesons: the $\pi^0$, the $\eta$, and the $\eta'$. If the $s$ quark were much much heavier we could consider only the first two in isolation, and we would then expect the $\eta$ meson to be an isospin-singlet ($I=0,I_3=0$) mixture of $\bar{u}u$ and $\bar{d}d$; the $\pi^0$ is built out of the same quark fields but it is the member of the isospin triplet with no net isospin projection ($I=1,I_3=0$).\footnote{This relationship is exactly analogous to the triplet and singlet states resulting from angular momentum addition with $s_1=s_2=1/2$ in non-relativistic quantum mechanics; hence isotopic ``spin.''} Because the $s$ quark is also pretty light, the $\eta$ meson also contains an admixture of $\bar{s}s$. Thus the $\eta$ is weirdly heavy for a $U(1)_A$ pseudo-Goldstone boson in 2-flavor QCD, but seems less out of place in the broader context of 3-flavor QCD.

Alas, this is where our luck runs out. The $\eta'$ meson is indeed a singlet under all the approximate symmetries of 3-flavor QCD, as we expect for a $U(1)_A$ Goldstone boson, but its mass is $m_{\eta'}=958$~MeV, larger than that of the proton! You may at this point be suspicious of me throwing out increasingly large numbers and declaring them sufficiently small, only to change my mind here. But you should probably trust Steven Weinberg, who worked out all the details and concluded that the pseudo-Goldstone boson of $U(1)_A$ in 3-flavor QCD should have $m_\eta'<\sqrt{3}m_\pi\approx240$~MeV \citep{weinberg1975}. Why is the $\eta'$ meson actually so much heavier? This is the essence of the $U(1)_A$ problem.

\section[The $U(1)_A$ problem]{The $\boldsymbol{U(1)_A}$ problem}\label{sec:u1a}
We have now discussed most of the aspects of the SM relevant to axion theory, with the exception of the \textbf{chiral anomaly}, which we will come to shortly. Our exploration of how the symmetries of the SM are hidden at low energies has led us to the $U(1)_A$ problem of the strong interactions. In exploring the relationship of the $U(1)_A$ problem to the chiral anomaly we will see that the topology of the QCD gauge vacuum is nontrivial; this nontrivial topology resolves the $U(1)_A$ problem but gives rise to the strong $CP$ problem. I have chosen to present all this background material not only for historical context, but also because the physics of the chiral anomaly and the $U(1)_A$ problem offer an intriguing hint as to how to resolve the strong $CP$ problem.

\subsection{The chiral anomaly}\label{sub:anomaly}
When Weinberg identified the $U(1)_A$ problem in 1975, it was already well-established that global $U(1)_A$ symmetries were \textit{anomalous}. An anomalous symmetry is a symmetry of the classical Lagrangian which is violated in the corresponding quantum theory. In this thesis I will only discuss the global chiral anomaly (also called the Adler-Bell-Jackiw anomaly after the authors of the seminal early papers on this subject \citep{adler1969,bell1969}). We will encounter several different instances of anomalous $U(1)_A$ symmetries in this chapter: in each case, the symmetry \textit{would} be spontaneously broken in the absence of the anomaly, and we will see that in some cases it nonetheless still makes sense to apply the formalism of Sec.~\ref{sub:ssb}. See Refs.~\citep{PS1995,schwartz2014} for more on anomalies in general,\footnote{Warning: the relevant chapters assume a lot of technical knowledge of QFT. In particular, they address the question of \textit{why} global $U(1)_A$ symmetries are anomalous, which is outside the scope of this thesis.} and Ref.~\citep{peccei1996} for a brief discussion of the chiral anomaly as it pertains to the $U(1)_A$ problem and the strong $CP$ problem.

You may be wondering what all the fuss is about: details aside, if the anomaly breaks the $U(1)_A$ symmetry, then this extra source of \textit{explicit} symmetry breaking implies that we should not expect Weinberg's condition to be valid; in particular, if the anomaly is ``bad enough,'' then $U(1)_A$ is not an approximate symmetry of 3-flavor QCD at all. This turns out to be the right idea, though there were good reasons at the time to think otherwise: the solution comes down to a distinction between the behavior of Abelian and non-Abelian gauge theories which is important for understanding the Strong $CP$ problem and the PQ solution.

In classical field theories, Noether's theorem (see e.g., Ref.~\citep{PS1995}) tells us how to construct the \textit{conserved current} $J^\mu$ corresponding to any symmetry of the Lagrangian. In general $J^\mu$ is a combination of the fields that transform under the symmetry operation, subject to the condition
\begin{equation}\label{eq:conserved_current}
\partial_\mu J^\mu = 0.
\end{equation}
In particular, Eq.~\eqref{eq:conserved_current} implies that the timelike component $J^0$ of the current is time-independent, and thus so is the \textit{conserved charge}
\begin{equation}\label{eq:conserved_charge}
Q = \int\mathrm{d}^3x J^0 .
\end{equation}

In most cases, the same procedure works just as well for the quantum theory in which the fields (and thus $J^\mu$) are operators. However, if the classical theory includes at least one fermion which transforms under a global $U(1)_A$ symmetry and is coupled to the gauge fields $A^\mu_a$, Eq.~\eqref{eq:conserved_current} does not hold in the quantum theory. Instead, the $U(1)_A$ current $J^\mu_A$ obeys
\begin{equation}\label{eq:anomalous_current}
\partial_\mu J^\mu_A = \frac{g^2}{32\pi^2}F^{\mu\nu}_a\tilde{F}_{\mu\nu a},
\end{equation}
where $g$ is the gauge coupling, $F^{\mu\nu}_a$ is the gauge field strength tensor
\begin{equation}\label{eq:field_strength}
F^{\mu\nu}_a = \partial^\mu A^\nu_a - \partial^\nu A^\mu_a + gf_{abc}A^\mu_bA^\nu_c
\end{equation}
and 
\begin{equation}\label{eq:dual_fs}
\tilde{F}_{\mu\nu a} = \frac{1}{2}\epsilon_{\mu\nu\alpha\beta}F^{\alpha\beta}_a
\end{equation}
is its dual, where $\epsilon_{\mu\nu\alpha\beta}$ is the Levi-Civita symbol antisymmetric in all indices. In Eq.~\eqref{eq:field_strength}, $f_{abc}$ are the \textit{structure constants} of the gauge symmetry group, which are related to the non-commutation of the generators of non-Abelian groups. If the gauge group in question is Abelian [$U(1)$], the third term in Eq.~\eqref{eq:field_strength} vanishes, and we can drop the subscripts $a$ which index the generators. 

Eq.~\eqref{eq:anomalous_current} is the formal statement of the chiral anomaly.\footnote{I will refer to this as the anomaly \textit{of} $U(1)_A$ \textit{with} the gauge group of $A^\mu_a$. As we will see, a given realization of a global $U(1)_A$ symmetry may or may not have an anomaly with any given gauge group. The potential for confusion is exacerbated by the fact that anomalies of chiral \textit{gauge} symmetries [e.g., $SU(2)_W$] with other gauge symmetries are also important in the SM. I will not discuss gauge anomalies in this thesis.} If the fermion that transforms under $U(1)_A$ is coupled to more than one gauge group, there will be one term on the RHS of Eq.~\eqref{eq:anomalous_current} for each gauge symmetry. If there are $N$ fermions with $U(1)_A$ symmetries coupled to a particular gauge group, the corresponding term should be multiplied by $N$.\footnote{I will ignore these factors of $N$ except in the one case where they actually make a qualitative difference (see Sec.~\ref{sub:defects}).} Qualitatively, Eq.~\eqref{eq:anomalous_current} says that the conservation of the chiral current $J^\mu_A$ is violated in a very specific and peculiar way. Next I will show why this symmetry violation appears to be rather benign compared to an arbitrary symmetry-breaking term we could have added to the Lagrangian.

Eq.~\eqref{eq:anomalous_current} implies that a $U(1)_A$ transformation on a single fermion field with parameter $\alpha$ does not leave the Lagrangian invariant, but instead adds a term of the form
\begin{equation}\label{eq:alpha_term}
\delta\lagr = \frac{g^2}{16\pi^2}\alpha F^{\mu\nu}_a\tilde{F}_{\mu\nu a}.
\end{equation}
For each SM gauge group, Eq.~\eqref{eq:alpha_term} is renormalizable and consistent with all the symmetries of the SM, so this appears to be an example of Gell-Mann's totalitarian principle in action: we did not include such terms in the original Lagrangian, but chiral anomalies can cause them to appear anyway. The so-called $\gv{\theta}$~\textbf{terms} we could have included in the original Lagrangian are formally equivalent to Eq.~\eqref{eq:alpha_term} with parameters $\theta_\text{EM}$, $\theta_W$, and $\theta_\text{QCD}$ in place of $\alpha$ for the corresponding gauge symmetry.\footnote{I have introduced this new notation to emphasize that $\theta_\text{EM}$, $\theta_W$, and $\theta_\text{QCD}$ are in principle parameters of the original Lagrangian, whereas $\alpha$ is an arbitrary $U(1)_A$ rotation angle which we can choose to be whatever we want.} The presence of the antisymmetric Levi-Civita symbol in Eq.~\eqref{eq:dual_fs} implies that $\theta$ terms violate $P$ and $T$ symmetries (thus they also violate $CP$ by the $CPT$ theorem). 

The reason we did not include $\theta$ terms in the Lagrangian in the first place is that each $\theta$ term is a \textit{total derivative}. We will show this explicitly for the simple Abelian case of $\theta_\text{EM}$:
\begin{align}
\frac{1}{2}\epsilon_{\mu\nu\alpha\beta}F^{\mu\nu}F^{\alpha\beta} &= \frac{1}{2}\epsilon_{\mu\nu\alpha\beta}\big(\partial^\mu A^\nu - \partial^\nu A^\mu\big)F^{\alpha\beta} \nonumber \\
&= \epsilon_{\mu\nu\alpha\beta}\partial^\mu A^\nu F^{\alpha\beta} \nonumber \\
&= \partial^\mu \big(\epsilon_{\mu\nu\alpha\beta}A^\nu F^{\alpha\beta} \big) - \epsilon_{\mu\nu\alpha\beta}A^\nu \partial^\mu \big(\partial^\alpha A^\beta - \partial^\beta A^\alpha\big) \nonumber \\
&= \partial^\mu \big(\epsilon_{\mu\nu\alpha\beta}A^\nu F^{\alpha\beta} \big).\label{eq:total_deriv}
\end{align}
The second term in the penultimate line vanishes because each of the double derivatives is totally symmetric in two Lorentz indices but the Levi-Civita tensor is antisymmetric. An analogous result may be derived for the non-Abelian case, provided one remembers that $f_{abc}$ are always totally antisymmetric in the gauge group adjoint indices. 

In QFT observable effects ultimately depend on the \textit{action} $S=\int\mathrm{d}^4x\lagr$: total derivatives in the Lagrangian thus correspond to \textit{surface terms} which can only contribute to the action at the boundaries of the (infinite) spacetime volume, where any reasonable field ought to vanish. More formally, one can show that for a $U(1)_A$ current which is not conserved due to an anomaly with QED, it is still possible to construct a perfectly well-behaved conserved \textit{charge} $Q$ \citep{adler1969,weinberg1975}. Quantum effects appear to violate the classical $U(1)_A$ symmetry in the least invasive conceivable way, provided there are no observable effects due to surface terms.

However, it turns out that surface terms \textit{can} have observable effects in non-Abelian gauge theories! This surprising result implies that chiral anomalies with non-Abelian gauge groups can result in violation of charge conservation as well as current conservation, and thus can be much less benign than chiral anomalies with QED. In particular the chiral anomaly with QCD provides an extra source of explicit symmetry breaking which resolves the $U(1)_A$ problem of 3-flavor QCD.\footnote{Although $SU(2)_W$ is also a non-Abelian gauge theory, anomalies involving $SU(2)_W$ are not relevant to the $U(1)_A$ problem or the strong $CP$ problem, and I will not discuss them here.} I will discuss the origin and interpretation of the nonvanishing surface terms responsible for all this in Sec.~\ref{sub:instantons}. But first, let us consider a simpler application of the chiral anomaly to the electromagnetic decay of the neutral pion. This example will demonstrate how the chiral anomaly can give rise to (a restricted class of) observable effects even \textit{without} nonvanishing surface terms. It will also help us understand the interactions of axions, discussed in Sec.~\ref{sub:axion_mass}.

In Sec.~\ref{sub:hadrons}, we introduced the three pions, and identified them with the pseudo-Goldstone bosons of an approximate global $SU(2)_A$ symmetry of QCD. The current corresponding to the $\pi^0$ in particular is the $z$-component of isospin $J^\mu_{A3}$, which is not exactly conserved due to nonzero quark masses. Neglecting the small quark masses, we would have $\partial_\mu J^\mu_{A3}=0$ classically. To make contact with the formalism we have developed in this section, note that $J^\mu_{A3}$ is formally equivalent to the conserved current for a $U(1)_A$ symmetry under which $u_L$ rotates by $\alpha$ and $d_L$ rotates by $-\alpha$.\footnote{This correspondence is discussed in Ref.~\citep{schwartz2014}. Note that the opposite rotation of the two quark flavors is a feature of this specific implementation of the $U(1)_A$ symmetry, and is distinct from the fact that the two chiralities rotate oppositely for each flavor under \textit{any} $U(1)_A$ transformation. In particular, the $U(1)_A$ symmetry associated with the $\pi^0$ is different from the missing $U(1)_A$ corresponding to the $\eta'$.} It can be shown that $J^\mu_{A3}$ has an anomaly with $U(1)_\text{EM}$ but not $SU(3)_C$.\footnote{The two additive contributions to the anomaly with $SU(3)_C$ associated with the $u$ and $d$ quarks are equal and opposite, so they cancel.}

The fact that the global axial symmetries of massless QCD are spontaneously broken is key to understanding how the electromagnetic anomaly of $J^\mu_{A3}$ leads to observable effects. In general, Noether currents have the same quantum numbers as the generators of the corresponding symmetry, and we saw in Sec.~\ref{sub:ssb} that Goldstone bosons also have these same quantum numbers when the symmetry is spontaneously broken. That is, there are three $SU(2)_A$ conserved currents $J^\mu_{Ai}$ in exactly the same sense as there are three pions. Thus $J^\mu_{A3}$ acts like an effective $\pi^0$ annihilation operator for the same reason that fundamental field operators in QFT are annihilation operators on the corresponding Fock space. 

Roughly speaking, this correspondence implies that $\pi^0$ acts like a dynamical version of $\theta_\text{EM}$ in the same sense as the Goldstone boson in the simple model of Sec.~\ref{sub:ssb} transformed like a dynamical version of the original symmetry parameter. In particular, we can just replace $\theta_\text{EM}$ in the Lagrangian with $\pi^0(x)/f_\pi$, because QED is an Abelian theory and thus surface terms produced by the \textit{static} $\theta_\text{EM}$ have no observable effects. Thus the existence of the chiral anomaly with QED implies an interaction of the form
\begin{equation}\label{eq:pigammagamma}
\lagr_{\pi\gamma\gamma}=\frac{e^2}{16\pi^2f_\pi}\pi^0F^{\mu\nu}\tilde{F}_{\mu\nu}
\end{equation}
where $e$ is the QED coupling (electron charge). 

This interaction has observable effects: it will contribute to the pion's decay into two photons $\pi^0\rightarrow\gamma\gamma$.\footnote{For historical reasons, photons are typically denoted by $\gamma$ in such expressions, although the photon field in the Lagrangian is always called $A^\mu$. Note also that the pion decays \textit{primarily} to two photons. Because QCD is so strong below the confinement scale, most hadrons decay into other hadrons before they get a chance to interact through QED or the weak force; the $\pi^0$ cannot do so because it is the lightest hadron.} Even without the anomaly there is another channel through which the $\pi^0\rightarrow\gamma\gamma$ decay can proceed, but it is highly suppressed by the approximate chiral symmetry; the same theoretical framework which correctly predicted the pion masses cannot account for the observed decay rate. The factor of $10^3$ discrepancy between theory and observation in the study of the $\pi^0$ lifetime was a source of great confusion for almost 20 years after the discovery of the $\pi^0$ until it was explained by Adler \citep{adler1969} in his initial derivation of the chiral anomaly. The essential takeaway point is that the chiral anomaly contributes to the $\pi^0$ decay rate without contributing to its mass. Thus it is unsurprising that theorists did not initially realize the chiral anomaly could also resolve the $U(1)_A$ problem of the $\eta'$ meson.

\begin{figure}[h]
\centering\includegraphics[width=0.8\textwidth]{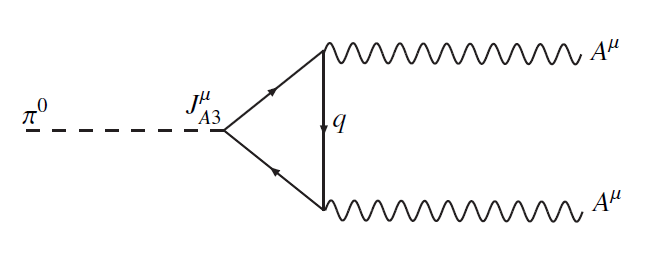}
\caption[Triangle diagram for anomalous $\pi^0\rightarrow\gamma\gamma$ decay]{\label{fig:triangle_pi} The triangle diagram which dominates the decay of the neutral pion, where $\pi^0$ is the pion field, $J^\mu_{A3}$ is the anomalous chiral symmetry, $q$ is an up or down quark, and $A^\mu$ is the photon field.}
\end{figure}

Finally, it is often helpful to represent the effects of the chiral anomaly diagrammatically. The chiral anomaly of $J^\mu_{A3}$ with $U(1)_\text{EM}$ is represented by the \textit{triangle diagram} in Fig.~\ref{fig:triangle_pi}. In such diagrams, the vertex at the left side of the triangle represents the anomalous current, and the other two vertices represent the gauge field couplings formally described by $F\tilde{F}$. Here $A^\mu$ is the photon and the anomaly is really the sum of two such diagrams with $u$ and $d$ quarks running in the loop. I emphasized above that $J^\mu_{A3}$ has the same quantum numbers as the $\pi^0$ field; this is represented by the dashed line connecting to the $J^\mu_{A3}$ vertex.

The formal structure of QFT requires that Feynman diagrams can be interpreted with time going in any direction. Thus, Fig.~\ref{fig:triangle_pi} describes not only $\pi^0$ decay but also the processes $\gamma + \gamma \rightarrow \pi^0$ and $\pi^0 + \gamma \rightarrow \gamma$. In the limit where one of the photons on the LHS is more properly thought of as a classical electromagnetic field, these two processes are the \textbf{Primakoff effect}~\citep{primakoff1951} and the inverse Primakoff effect, respectively. The axion haloscope discussed in Sec.~\ref{sec:haloscope} exploits a process analogous to the inverse Primakoff effect for axion detection.

\begin{figure}[h]
\centering\includegraphics[width=0.57\textwidth]{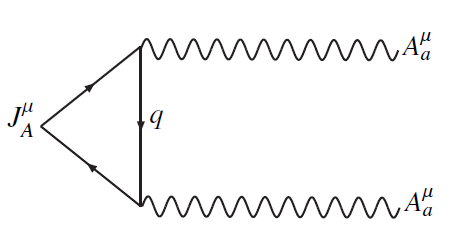}
\caption[Triangle diagram which resolves the $U(1)_A$ problem]{\label{fig:triangle_nopi} The triangle diagram for the anomalous $U(1)_A$ symmetry missing from the hadronic spectrum, where $J^\mu_{A}$ is the corresponding current, $q$ is an up, down, or strange quark, and $A^\mu_a$ is a gluon field.}
\end{figure}

The chiral anomaly of the missing $U(1)_A$ symmetry with QCD is shown in Fig.~\ref{fig:triangle_nopi}. Here $J^\mu_{A}$ is the current associated with this symmetry, for which the $\eta'$ meson \textit{would be} the pseduo-Goldstone boson, and $A^\mu_a$ is a gluon field. Because of nonvanishing surface terms, the anomaly acts like any other form of explicit symmetry breaking, and contributes to the $\eta'$ mass. In fact, the explicit symmetry breaking due to the anomaly is much larger than the explicit symmetry breaking due to the $u,d,s$ quark masses and thus the $\eta'$ mass is so large that it does not make sense to describe it as a pseudo-Goldstone boson. Roughly speaking, we can meaningfully speak of spontaneous breaking of an approximate symmetry provided $\mu \ll f$, where $\mu$ and $f$ are the characteristic energy scales of explicit and spontaneous symmetry breaking, respectively.\footnote{$\mu$ is not related to the mass parameter $\mu$ of the wine bottle potential in Sec.~\ref{sub:ssb}.} In the case of the $\pi^0$, $\mu\sim m_d$ and $f=f_\pi\sim\Lambda_\text{QCD}$, so the $\pi^0$ may be treated as a pseudo-Goldstone boson. For the $\eta'$, we still have $f\sim\Lambda_\text{QCD}$, and we will see that $\mu\sim\Lambda_\text{QCD}$ as well, so the $\eta'$ does not behave like a pseudo-Goldstone boson. We will encounter more examples of both the $\mu\ll f$ and $\mu \sim f$ cases when we come to solutions of the strong $CP$ problem in Sec.~\ref{sub:pq_mechanism}.

\subsection[Instantons and the $\theta$ vacuum]{Instantons and the $\gv{\theta}$ vacuum}\label{sub:instantons}
For non-Abelian gauge theories, Eq.~\eqref{eq:total_deriv} generalizes to $\frac{1}{2}\epsilon_{\mu\nu\alpha\beta}F^{\mu\nu}_aF^{\alpha\beta}_a=\partial^\mu K_\mu$, where 
\begin{equation}\label{eq:total_deriv_nab}
K_\mu = \epsilon_{\mu\nu\alpha\beta}\Big(A^\nu_a F^{\alpha\beta}_a - \frac{g}{3}f_{abc}A^\nu_aA^\alpha_bA^\beta_c \Big).
\end{equation}
This is still a total derivative, but there exist \textit{topologically nontrivial} gauge field configurations $A^\mu_a$ for which the corresponding surface terms do not vanish; the simplest examples of such field configurations are called \textbf{instantons}. Non-Abelian gauge theory gets very complicated, so I will only discuss the essential features; see Ref.~\citep{srednicki2007} for a good pedagogical discussion of instantons and the topological features of non-Abelian gauge groups. 

We can simplify matters by restricting our focus to \textit{vacuum field configurations}, which are solutions to the classical equations of motion for the free gauge fields described by the Lagrangian $\lagr=-\frac{1}{4}F^{\mu\nu}_aF_{\mu\nu a}$. A trivial example of a vacuum field configuration is $A^\mu_a=0$, but gauge invariance implies that there are infinitely many others related to the trivial case by gauge transformations (whether the gauge group is Abelian or non-Abelian). 

Belavin et al.~\citep{belavin1975} first noted that non-Abelian gauge theory permits vacuum field configurations for which the surface integral of Eq.~\eqref{eq:total_deriv_nab} is nonvanishing. Specifically, they showed that for $SU(2)$ gauge theory in 4D Euclidean space, there exist vacuum field configurations with
\begin{equation}\label{eq:winding}
\frac{g^2}{32\pi^2}\int\mathrm{d}^4x\,\partial^\mu K_\mu = \nu
\end{equation}
for any integer $\nu$. They also derived the explicit form of the vacuum field configuration with $\nu=1$, called the pseudoparticle or instanton configuration (the trivial vacuum of course has $\nu=0$). All we will really need to know about the explicit form of the instanton field configuration is that it contains two arbitrary parameters: the size $\rho$ and position $x_0$ of the instanton. The field is concentrated within a spherically symmetric region of Euclidean space with radius $\rho$ around $x_0$ -- hence ``pseudoparticle.'' The name ``instanton'' reflects the fact that the corresponding field configuration in (3+1)D Minkowski space is localized in time as well as space -- I will return to the question of how we should interpret this strange behavior shortly.

Belavin et al.\ were interested in the very complicated question of why certain gauge theories exhibit confinement, and thus in the large-$\rho$ limit where the mutual interactions of long-range instanton fields are essential to the low-energy behavior of the theory. 't~Hooft~\citep{thooft1976} was the first to point out that the LHS of Eq.~\eqref{eq:winding} is just the volume integral of Eq.~\eqref{eq:anomalous_current}, and a nonzero value implies that the charge corresponding to $J^\mu_A$ is not conserved, in marked contrast to the case of the chiral anomaly with QED. He then showed explicitly that the anomaly with QCD will contribute to the mass of the $\eta'$ meson, resolving the $U(1)_A$ problem.

The claim that instantons and other field configurations with $\nu\neq0$ resolve the $U(1)_A$ problem might bother you given that Belavin et al.\ showed that these field configurations exist in $SU(2)$ gauge theory specifically. But there is actually no contradiction here -- there are $SU(2)$ subgroups of the $SU(3)$ gauge theory of QCD in the same sense as 2D spatial rotations form a subgroup of 3D spatial rotations; 't~Hooft showed that the vacuum structure of these $SU(2)$ subgroups is sufficiently rich to resolve the $U(1)_A$ problem without the need for any features specific to $SU(3)$.

But there are still other reasons we might find the ``instanton solution'' to the $U(1)_A$ problem troubling. Perhaps most obvious is the fact that we do not live in 4D Euclidean space. There is a well-defined mathematical procedure in QFT called \textit{Wick rotation} for relating quantities obtained in Euclidean space to the corresponding quantities in Minkowski spacetime, but it is not \textit{a priori} clear how we should interpret the Minkowski-space field configurations corresponding to instantons (let alone the more complicated field configurations of higher $\nu$ for which we do not even have explicit expressions). We would like to understand how to interpret instantons in the real world in order to convince ourselves that they are not just artifacts of the formalism without real observable effects. 

A second reason to be suspicious of instantons is that we started this discussion by restricting ourselves to vacuum field configurations, but all such vacuum configurations are related by gauge transformations, and the principle of gauge invariance would seem to suggest that they should all be completely equivalent. Indeed, gauge invariance is sometimes called ``gauge redundancy,'' with the implication that field configurations related to each other by gauge transformations just amount to redundancies in our description of nature. Yet the instanton field configuration leads to a nonvanishing surface integral whereas the trivial gauge vacuum with $\nu=0$ does not. Clearly we are still missing a piece of this puzzle.

We can gain some insight into the proper interpretation of instantons by considering the meaning of the integer $\nu$ arising in Eq.~\eqref{eq:winding}. Belavin et al.\ noted that the restriction to vacuum field configurations can be formalized as a statement about the behavior of the gauge fields at infinity (i.e., on the surface of an arbitrarily large 3-sphere embedded in 4D Euclidean space), and that there also happens to be an isomorphism between the set of all $SU(2)$ matrices and the surface of a 3-sphere. Any given $SU(2)$ gauge field can thus be regarded as a mapping $S_3\rightarrow S_3$, and there is a discrete infinity of topologically distinct classes of such mappings distinguished by different \textbf{winding numbers}. The integer $\nu$ defined by Eq.~\eqref{eq:winding} is just the winding number of the gauge field configuration.

Around the same time as 't~Hooft demonstrated that instantons solve the $U(1)_A$ problem, a pair of papers by Callan, Dashen, and Gross~\citep{callan1976} and Jackiw and Rebbi~\citep{jackiw1976} took the topological characteristics of non-Abelian gauge theories as a starting point to clarify the physical interpretation of instantons. Both papers pointed out that non-Abelian gauge theories generically have a discrete infinity of topologically distinct vacuum states $\ket{n}$ characterized by winding number $n$. These states are related by \textit{nonlocal} gauge transformations to the trivial vacuum $\ket{0}$: states characterized by different $n$ cannot be \textit{continuously} connected by a gauge transformation without passing through gauge field configurations which are not vacuum states. This is the sense in which the different $n$-vacua are topologically distinct: they are really distinct minima separated by energy barriers.\footnote{Of course, there are also infinitely many field configurations with the same winding number which \textit{can} be related by continuous gauge transformations, as in Abelian gauge theory. These topologically equivalent gauge field configurations are physically indistinguishable in every way.} In this picture, the Euclidean gauge field configurations identified by Belavin et al.\ correspond to classically forbidden \textit{tunneling transitions} between different vacua $\ket{n_1},\ket{n_2}$ with $\nu=|n_1-n_2|$. In particular, the instanton field configuration in Euclidean space corresponds to tunneling between adjacent $n$-vacua in Minkowski space.\footnote{This is the sense in which instantons in Minkowski space are ``localized in time.'' There is generally a deep connection between localization in Euclidean space and tunneling in Minkowski space; see Refs.~\citep{thooft1976,jackiw1976} for more discussion on this point.}

Thinking of the Euclidean gauge field configurations which yield nonzero $\nu$ in Eq.~\eqref{eq:winding} as tunneling events raises the question of when and where tunneling will be significant. We might generally expect the effects of higher-$\nu$ field configurations to be suppressed relative to the effects of instantons, which correspond to ``minimal'' tunneling between adjacent vacua. More formally, it can be shown that the contribution of a single instanton to the path integral is $e^{-8\pi/g^2}$ and the contributions of higher-$\nu$ field configurations are always less important provided the gauge coupling $g$ is sufficiently small. This implies that for a confining theory like QCD, the effects of instantons are suppressed and the effects of higher-$\nu$ configurations are totally negligible at high temperatures $T\gg\Lambda_\text{QCD}$. Conversely, at lower temperatures near the confinement scale $T\sim\Lambda_\text{QCD}$, tunneling is unsuppressed and all sort of complicated field configurations contribute.\footnote{Finite-temperature QCD is not for the faint of heart; readers more intrepid than myself are referred to Ref.~\citep{gross1981}. We shall encounter this subject again in our study of axion cosmology in Sec.~\ref{sec:axion_cosmo}, and I will do my best to avoid the messy details.} For our present purposes the takeaway point is that the characteristic energy scale associated with the explicit breaking of $U(1)_A$ by topologically nontrivial gauge field configurations is $\mu\sim\Lambda_\text{QCD}$, as anticipated at the end of Sec.~\ref{sub:anomaly}.

We have seen that in QCD below the confinement scale the amplitude for tunneling between different $n$-vacua is generally quite large. Therefore it doesn't make sense to regard any of the $n$-vacua as the true vacuum state of the theory. The true vacuum state must be invariant under \textit{all} gauge transformations, including the nonlocal gauge transformations that change the vacuum field topology. The authors of Refs.~\citep{callan1976,jackiw1976} approached the problem of finding the true vacuum by working in Euclidean space, and simultaneously diagonalizing the Hamiltonian and the operator $\Omega$ implementing a gauge transform that changes the vacuum winding number by 1: $\Omega\ket{n} = \ket{n+1}$. Since $\Omega$ is a unitary operator, its eigenvalues generally have the form $e^{i\theta}$ with $0 \leq\theta<2\pi$; you can easily verify that
\begin{equation}\label{eq:theta_vac}
\ket{\theta}=\sum_n e^{-in\theta}\ket{n}
\end{equation}
are the corresponding eigenstates. By construction there is no tunneling \textit{between} $\theta$-vacua: thus different values of $\theta$ characterize different theories, perhaps with inequivalent observables.

So which $\theta$-vacuum is actually realized in the real world? The implications of this question become clearer if we follow Ref.~\citep{callan1976} and express the Euclidean path integral for vacuum-to-vacuum tunneling in the $\theta$-basis:
\begin{equation}\label{eq:theta_lagr}
\matrixel{\theta'}{e^{-S}}{\theta} = \delta(\theta-\theta')\,e^{-\left(S+i\nu\theta\right)},
\end{equation}
where $S=\int\mathrm{d}^4x\lagr$ is the action and the extra factor of $e^{-i\nu\theta}$ arises from the expression Eq.~\eqref{eq:theta_vac} for $\ket{\theta}$ in terms of $\ket{n}$; the Dirac delta function indicates that there is no tunneling between different $\theta$-vacua, as I emphasized above. The interesting part is the factor of $e^{-i\nu\theta}$: from Eq.~\eqref{eq:winding} and the definition of $\partial^\mu K_\mu$ above Eq.~\eqref{eq:total_deriv_nab}, it follows that in the theory in which the true QCD vacuum state is $\ket{\theta_\text{QCD}}$, there is an effective contribution to the Lagrangian of the form
\begin{equation}\label{eq:theta_term}
\lagr_\theta= \frac{g^2}{32\pi^2}\theta_\text{QCD}F^{\mu\nu}_a\tilde{F}_{\mu\nu a}.
\end{equation}
This is precisely the QCD $\theta$-term introduced in the discussion following Eq.~\eqref{eq:alpha_term} -- the totalitarian principle strikes again!\footnote{You may be wondering what happened to the factor of $i$ in $i\nu\theta$: it is absorbed in the Wick rotation that takes us back from the Euclidean theory of Eq.~\eqref{eq:theta_lagr} to the real world.} Thus we see that the solution to the $U(1)_A$ problem implies the existence of a QCD $\theta$-term. As we will see in the next section, Eq.~\eqref{eq:theta_term} gives rise to the strong $CP$ problem.

\section[The strong $CP$ problem]{The strong $\boldsymbol{CP}$ problem}\label{sec:strong_cp}
At this point, we have seen a term of the form Eq.~\eqref{eq:theta_term} show up in several different contexts, so a quick review is in order. We first noted that the SM can accommodate a $\theta$-term for each gauge group in Sec.~\ref{sub:anomaly}, and also noted that such terms violate the discrete symmetries $P$, $T$, and $CP$ for any nonzero value of the coefficient $\theta$. At the time, we had good reason to hope that $\theta$-terms should not generate any observable effects, because they correspond to surface terms in the action. These hopes were dashed in Sec.~\ref{sub:instantons}, where we first showed that non-Abelian gauge theories permit field configurations with nonvanishing surface terms, and then went on to show that in such theories $\theta$ has a physical interpretation as the parameter describing the true gauge vacuum state.

The resolution of the $U(1)_A$ problem required us to recognize the existence of non-Abelian gauge vacua other than $\ket{n=0}$, but does not give us any clue as to the value of $\theta_\text{QCD}$. While we have not yet explicitly demonstrated the existence of any $CP$-violating observables which depend on $\theta_\text{QCD}$, at this point it seems clear enough that such observables \textit{can} exist. In Sec.~\ref{sub:nedm} we will see that any nonzero $\theta_\text{QCD}$ produces an electric dipole moment for the neutron, and the nonobservation of the neutron EDM implies that $\theta_\text{QCD}$ must be extremely small. This is the essence of the strong $CP$ problem, though it turns out to be somewhat more subtle.

The additional subtlety has to do with the effects of quarks on the vacuum angle $\theta_\text{QCD}$. Although we originally motivated this whole discussion with the $U(1)_A$ problem arising from the approximate symmetries of the quark sector, we ignored the quarks entirely in our discussion in Sec.~\ref{sub:instantons}. When we take into account the effects of quarks in Sec.~\ref{sub:quark_phase}, we will see that $CP$ violation in QCD is in fact inseparable from electroweak $CP$ violation: the true observable parameter $\bar{\theta}$ actually depends on both $\theta_\text{QCD}$ and the Higgs Yukawa matrices discussed in Sec.~\ref{sub:ckm}. After this discussion, we will be equipped to describe somewhat more precisely in Sec.~\ref{sub:naturalness} the sense in which a small value of $\bar{\theta}$ is unnatural. Finally, I will discuss some proposed solutions to the strong $CP$ problem in Sec.~\ref{sub:solutions}.

\subsection{The electric dipole moment of the neutron}\label{sub:nedm}
My goal in the present section is to make contact with experiment by demonstrating that the neutron EDM $\mathbf{d}\propto\theta_\text{QCD}$. Since we have already seen that $\theta$-terms violate parity symmetry $P$ and time-reversal symmetry $T$, I will first show that a fundamental particle EDM would violate the same discrete symmetries. From here it is only a small conceptual step to suppose that the $CP$-violating gluon interactions arising from Eq.~\eqref{eq:theta_term} correspond in the hadronic phase to $CP$-violating nucleon-pion interactions which mediate spin-dependent interactions with external electromagnetic fields.

\begin{figure}[h]
\centering\includegraphics[width=0.5\textwidth]{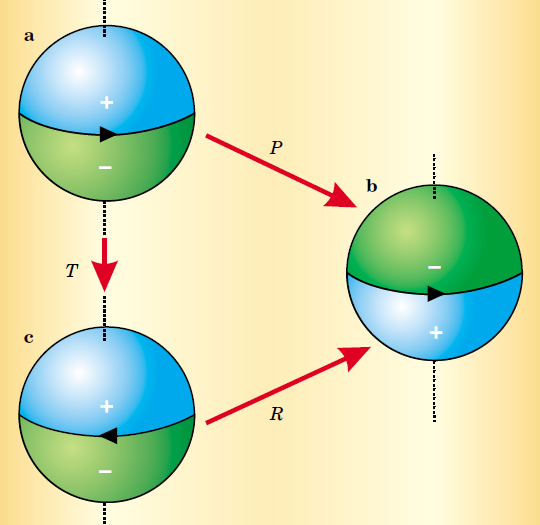}
\caption[Transformation properties of dipole moments under $C$, $P$, and $T$]{\label{fig:edm} \textbf{(a)} A particle with parallel electric and magnetic dipole moments. \textbf{(b)} A parity transformation $P$ reverses all three spatial directions, and thus reverses the electric dipole moment but not the magnetic dipole moment. \textbf{(c)} Time reversal changes the direction of the magnetic dipole moment relative to \textbf{a} but not the electric dipole moment. Thus a particle with both electric and magnetic dipole moments violates $P$ and $T$. A 180$^\circ$ rotation $R$ around the $x$ or $y$ axes demonstrates that the particles illustrated in \textbf{b} and \textbf{c} are equivalent, and so the existence of such a particle does not violate $PT$ (and therefore does not violate $C$). Figure from Ref.~\citep{edmfig}.}
\end{figure}

We can begin by noting that the neutron is a spin-1/2 particle with a nonzero \textit{magnetic} dipole moment $\gv{\mu}$, and the Wigner-Eckart theorem in quantum mechanics requires that the expectation value of any vector operator (such as an EDM) point along the spin quantization direction: thus $\mathbf{d}$ (if it exists) is either parallel or antiparallel to $\gv{\mu}$.\footnote{This result from the Wigner-Eckart theorem is often explained qualitatively by noting that a spin-1/2 particle only has enough internal degrees of freedom to specify a single preferred axis. Formally, the $x$ and $y$ components of any vector operator (i.e., those orthogonal to the spin quantization axis) may be represented as linear combinations of raising and lowering operators, so their expectation values must vanish.} Panel \textbf{a} of Fig.~\ref{fig:edm} shows a particle with $\mathbf{d}\parallel\gv{\mu}$, and illustrates that either a $P$ or $T$ transformation turns it into a particle with $\mathbf{d}$ and $\gv{\mu}$ antiparallel: this difference has observational consequences, so the EDM clearly violates $P$ and $T$ symmetries. Thus we can generically expect $CP$-violating effects in the strong interactions to give rise to baryon EDMs. We can restrict our attention to the  neutron EDM, which can be measured more precisely than other baryonic EDMs, because neutrons are both long-lived and electrically neutral. For further discussion on the experimental and theoretical study of EDMs see Ref.~\citep{kl1997}.

The first calculations of the dependence of the neutron EDM on $\theta_\text{QCD}$ were carried out in Refs.~\citep{baluni1979,crewther1979}. These calculations are quite involved (as is the relation between fundamental QCD parameters and low-energy observables more generally); for our purposes, an order-of-magnitude estimate will suffice. The magnitude of the neutron EDM is
\begin{equation}\label{eq:nedm}
d_N \sim \theta_\text{QCD}\frac{e}{m_N^2}\frac{m_um_d}{m_u+m_d} \sim 10^{-16}\cdot\theta_\text{QCD}\ e\text{-cm},
\end{equation}
where $m_N$, $m_u$, and $m_d$ are the neutron, up quark, and down quark masses, respectively, and I have neglected a small contribution from the strange quark. Realistic calculations change this naive dimensional estimate (from Ref.~\citep{cheng1988}) by an order-unity factor. To date, no neutron EDM has been observed, despite a number of extremely sensitive measurements. The best existing limit in the late 1970s was $d_N < 3\times10^{-24}\ e\text{-cm}$ \citep{dress1977}, and today it is $d_N < 3\times 10^{-26}\ e\text{-cm}$ \citep{pendlebury2015}: this implies $\theta_\text{QCD} \lesssim 10^{-10}$!

\subsection{Including quarks and electroweak interactions}\label{sub:quark_phase}
The stringent upper bound on $\theta_\text{QCD}$ obtained from the neutron EDM is rather surprising, since it implies $\theta_\text{QCD}$ is much smaller than all other dimensionless SM parameters. Nonetheless, if this were the whole story, we would be justified in treating the strong $CP$ problem with some skepticism. Perhaps we just happen to live in a world where the parameter $\theta_\text{QCD}$ is very small, or better yet, perhaps there is some yet-undiscovered feature within the theory of QCD that singles out $\ket{\theta=0}$ as the true vacuum state.

It turns out that in the SM the strong $CP$ problem actually \textit{cannot} be resolved by QCD alone: the culprits are the quarks, which are charged under both QCD and the electroweak gauge group. We ignored quarks entirely in Sec.~\ref{sub:instantons}. Of course without quarks there is no $U(1)_A$ problem to solve, and no neutron whose EDM may or may not exist. So it is more accurate to say we have implicitly been considering the theory of QCD coupled to quarks in the absence of electroweak interactions, with the additional requirement that all of the quark masses are nonzero. Recall from Sec.~\ref{sec:sm} that left- and right-handed quarks are charged differently under $SU(2)_W$ and $U(1)_Y$ but not under $SU(3)_C$ -- thus if there were no electroweak interactions, we could just include quark mass terms in the original Lagrangian with no need for the Higgs mechanism.

Before adding the electroweak interactions, it is instructive to consider the case where at least one of the quarks is exactly massless. Then Eq.~\eqref{eq:nedm} indicates that $d_N=0$ regardless of the value of $\theta_\text{QCD}$, and indeed it can be shown\,\citep{callan1976,jackiw1976} that $\theta_\text{QCD}$ does not have \textit{any} observable effects! Why does adding a single massless quark to the theory have such a dramatic effect? Qualitatively, we can understand this behavior as a consequence of the close relation of the QCD $\theta$-term to the chiral anomaly. Recall that in the presence of at least one massless quark the theory has an anomalous $U(1)_A$ symmetry: due to the chiral anomaly the effect of a $U(1)_A$ transformation on the Lagrangian is to add a term of the form Eq.~\eqref{eq:alpha_term} and thus to take $\theta_\text{QCD} \rightarrow \theta_\text{QCD} + 2\alpha$. In other words, if one of the quarks is exactly massless, we can use the chiral anomaly to shift $\theta_\text{QCD}$ to any value we like!\footnote{In Sec.~\ref{sub:hadrons} we got a lot of mileage out of approximate chiral symmetries, but here the classical chiral symmetry must be exact, such that the quantum symmetry is broken only by the chiral anomaly. Otherwise, a $U(1)_A$ transformation will also affect the Lagrangian in other ways, giving us a means to distinguish between different values of $\theta_\text{QCD}$.} This indicates that the initial value of $\theta_\text{QCD}$ cannot be of any observable consequence; see Ref.~\citep{srednicki2007} for a more formal pedagogical argument.

Having appreciated that the chiral anomaly is intimately connected to the strong $CP$ problem, we are ready to consider the effects of electroweak interactions on the strong $CP$ problem. The only thing that changes is that bare quark mass terms are now forbidden as they violate the $SU(2)_W\times U(1)_Y$ gauge symmetry, so fermion masses must be generated by EWSB as discussed in Sec.~\ref{sub:ckm}. Recall that the Higgs Yukawa coupling matrices $Y_u,Y_d$ responsible for quark mass generation are in general not diagonal, and we had to use both vector and axial phase rotations to transform from the flavor basis to the mass basis. But these axial phase rotations will contribute to $\theta_\text{QCD}$ through the chiral anomaly! More precisely, in the mass basis, the chiral anomaly will generate an additive contribution to $\theta_\text{QCD}$ of the form \citep{schwartz2014}
\begin{equation}\label{eq:quark_phase}
\theta_\text{Yukawa} = \text{arg}\big(\text{det}\big[Y_dY_u\big]\big).
\end{equation}

Though we have once again performed chiral phase rotations and observed that the value of $\theta_\text{QCD}$ changes, the situation here is very different from the case of the massless quark considered above, because the anomalous phase rotations change the quark mass terms as well as $\theta_\text{QCD}$. It is conceptually simplest to work in the mass basis where all quark mass matrices are real and diagonal -- then the coefficient of Eq.~\eqref{eq:theta_term} becomes
\begin{equation}\label{eq:theta_bar}
\bar{\theta} = \theta_\text{QCD} + \theta_\text{Yukawa}.
\end{equation}
We could instead use chiral field redefinitions to put the Lagrangian into a form in which the coefficient of Eq.~\eqref{eq:theta_term} is 0; in general the basis for the Yukawa matrices defined by this condition will coincide with neither the mass basis nor the flavor basis. But the chiral phase rotations required to accomplish this will just add the original vacuum angle $\theta_\text{QCD}$ to the original phase $\theta_\text{Yukawa}$ of the quark mass determinant: in the new basis the quark mass matrix contains complex and off-diagonal elements, with an overall $CP$-violating phase $\bar{\theta}$.

Thus we can conclude that $\bar{\theta}$ is a \textit{basis-independent}\footnote{It can be shown that the $CP$-violating phase $\theta_W$ corresponding to $SU(2)_W$ is not basis-independent and thus unphysical; see Ref.~\citep{schwartz2014} for further discussion. This is why there is no ``weak $CP$ problem'' even though the gauge theory of the weak interactions permits the existence of instantons. That $\theta_W$ is unphysical is more important in principle than in practice, since $SU(2)_W$ is not confining -- the effects of instantons would be highly suppressed at low energies in any event.} measure of $CP$ violation which receives contributions from both the strong and electroweak sectors of the SM. With this refined understanding, we can see that $\theta_\text{QCD}$ should be replaced with $\bar{\theta}$ in Eq.~\eqref{eq:nedm} for the neutron EDM, and indeed all observable effects depend only on the combination $\bar{\theta}$ rather than on $\theta_\text{QCD}$ and $\theta_\text{Yukawa}$ separately. In the absence of massless quarks, the strong $CP$ problem is real, and it has turned out to be a problem of the full SM rather than just the strong force! 

\subsection{Fine-tuning and naturalness}\label{sub:naturalness}
Far from shedding any light on the strong $CP$ problem, taking into account the effects of electroweak interactions has only made it all the more perplexing: the problem of the smallness of $\bar{\theta}$ is conceptually much thornier than the problem of the smallness of $\theta_\text{QCD}$. In particular, we clearly must abandon the hope of finding a solution to the strong $CP$ problem \textit{within} QCD: even if we could find some deep reason for $\theta_\text{QCD}=0$, $\bar{\theta}$ would still receive a contribution from $\theta_\text{Yukawa}$. Within the SM, the strong and electroweak interactions are independent, except for the fact that the same quarks are charged under both gauge groups. This suggests that there is no way for $\theta_\text{QCD}$ and $\theta_\text{Yukawa}$ to ``know'' about each other, and we need to look beyond the SM for a solution to the strong $CP$ problem.

It might seem that there is a loophole here: what if we could find some deep reason within QCD that $\theta_\text{QCD}=0$, and independently find a deep reason within the electroweak theory that $\theta_\text{Yukawa}=0$? This would solve the strong $CP$ problem without requiring the two contributions to $\bar{\theta}$ to communicate. Unfortunately, this approach does not seem very viable: as noted in Sec.~\ref{sub:ckm}, the standard model with its 3 fermion generations already has electroweak $CP$ violation in the phase of the CKM matrix $\delta_\text{CKM}$, for which the best measured value is $1.20\pm0.08$~\citep{pdg2016} -- this is hardly a small number! Essentially what we have shown is that the electroweak interactions of quarks can lead to $CP$-violating effects in two distinct ways: through $\delta_\text{CKM}$ and $\theta_\text{Yukawa}$.\footnote{It should be empahsized that $\theta_\text{Yukawa}$ and $\delta_\text{CKM}$ are not different names for the same quantity. $\delta_\text{CKM}$ governs several observable effects in $K$ and $B$ meson decays, whereas $\theta_\text{Yukawa}$ is only observable as a contribution to $\bar{\theta}$. Essentially, $\theta_\text{Yukawa}$ corresponds to the sum of phases in the Higgs Yukawa coupling matrices which would be unobservable but for the chiral anomaly.} There is no reason within the SM why one of these effects should be forbidden while the other is order-unity. In a more fundamental theory, we might expect there to be only a single parameter responsible for $CP$ violation in the weak interactions, so the most obvious ``educated guess'' for the electroweak contribution is $\theta_\text{Yukawa}\sim 1$. Then to satisfy the empirical limits on $\bar{\theta}$ without introducing any new physical mechanism, $\theta_\text{QCD}$ must also be an $\mathcal{O}(1)$ number which is equal and opposite to $\theta_\text{Yukawa}$ to better than 1 part in $10^{10}$! This is the sense in which the strong $CP$ problem is a fine-tuning problem.\footnote{I have glossed over some details here in the interest of simplicity. The CKM matrix can be expressed in more than one basis, and the value of $\delta_\text{CKM}$ I have cited is not basis-independent. A basis-independent way to parameterize $CP$ violation in the CKM matrix is through a quantity called the \textbf{Jarlskog Invariant} $J=(3.0\pm0.2)\times10^{-5}$~\citep{pdg2016}. This quantity is so much smaller than $\delta_\text{CKM}$ basically because in the standard parameterization, $\delta_\text{CKM}$ appears in a phase factor multiplying a small real number (and this is precisely why basis-specific quantities can be misleading). You may be tempted to conclude that the fine-tuning in the strong $CP$ problem is less severe than it seems, but in fact we will soon see that the ``natural'' value of $\bar{\theta}$ is $\mathcal{O}(1)$ anyway.}

As if this weren't already puzzling enough, it turns out that in the SM electroweak interactions lead to \textit{infinite renormalization} of $\theta_\text{Yukawa}$, though not before 14$^\text{th}$ order (!) in perturbation theory~\citep{ellis1979,khriplovich1994}. That is to say, for any nonzero value of $\theta_\text{Yukawa}$ we choose in the classical Lagrangian, loop corrections cause $\theta_\text{Yukawa}$ to diverge in the quantum theory. This is not quite as disastrous as it sounds -- such divergences may be thought of as artifacts of the extrapolation of the theory to very high energies where it is no longer reliable. Nonetheless, infinite renormalization implies that within the SM $\bar{\theta}$ is not a calculable quantity even in principle -- it must be regarded as a free parameter (i.e., an input to the theory). `t~Hooft~\citep{thooft1980} was the first to formalize the intuitive notion of ``naturalness'' which has long guided physicists' intuition. Roughly speaking, the principle of naturalness states that we should expect dimensionless free parameters to be $\mathcal{O}(1)$ unless a nonzero value would violate some symmetry of the theory. $\bar{\theta}<10^{-10}$ in the SM is unnatural because the theory would not be $CP$-conserving even if we somehow arranged for $\bar{\theta}$ to vanish exactly.

It should be emphasized that naturalness is ultimately an aesthetic criterion. It would not violate any law of physics if we \textit{just happened} to live in a world in which $\bar{\theta} < 10^{-10}$, and moreover the SM has other problems with naturalness, most notably the ``hierarchy problem'' of the Higgs boson mass. Maybe nature is just unnatural and that's all there is to it. Wading further into the ongoing debate about the value of naturalness in theoretical physics would take us too far afield from the subject of this thesis; see Ref.~\citep{dine2015} for a good overview of naturalness and Ref.~\citep{cheng1988} for more on the strong $CP$ problem as a naturalness problem. For the remainder of this thesis I will adopt the point of view that the strong $CP$ problem is a real problem to be solved, and its solution lies beyond the SM.

\subsection[Solving the strong $CP$ problem]{Solving the strong $\boldsymbol{CP}$ problem}\label{sub:solutions}
We already saw in Sec.~\ref{sub:quark_phase} how the strong $CP$ problem can be solved \textit{within} the SM: if one of the quarks is exactly massless, then no observable effects depend on $\bar{\theta}$, and there is no strong $CP$ problem to solve.\footnote{Technically speaking, we only showed that the neutron EDM is independent of $\theta_\text{QCD}$, but the conclusion does not change when we take into account the electroweak contribution.} This is usually called the \textbf{massless up quark} solution, because the up quark is the lightest of the SM quarks. However, the best current measurements yield $m_u=2.2\pm0.5~\text{MeV}$~\citep{pdg2016}, so the massless up quark solution appears to be ruled out.

The massless up quark solution is nonetheless of substantial historical importance, because it demonstrated an important and much more general lesson: \textit{the existence of a $U(1)_A$ symmetry broken only by the chiral anomaly with QCD solves the strong $CP$ problem}. The $U(1)_A$ symmetry associated with the $\eta'$ meson discussed in Sec.~\ref{sub:hadrons} and Sec.~\ref{sub:anomaly} is only approximate -- it is broken explicitly by the quark masses as well as the anomaly. In the presence of at least one massless quark we would have an exact $U(1)_A$ symmetry, but we will see in Sec.~\ref{sec:peccei_quinn} that there is another way to add a $U(1)_A$ symmetry to the SM: this is the \textbf{Peccei-Quinn mechanism}.

Before turning to the PQ mechanism, I will briefly mention the only other known class of solutions to the strong $CP$ problem, which do not invoke the existence of an additional $U(1)_A$ symmetry. The key ingredient in this class of solutions is \textbf{spontaneous $\boldsymbol{CP}$ violation}: essentially, these solutions posit that $CP$ is actually an exact symmetry of a more fundamental theory supplanting the SM: then by construction we must have $\theta_\text{QCD}=\theta_\text{Yukawa}=0$ in the Lagrangian. Of course, this raises the question of how to accommodate the observed $CP$-violating processes in kaon and $B$-meson systems. Within the SM, such processes are explained by the KM model of \textit{explicit} electroweak $CP$ violation (see Sec.~\ref{sub:ckm}), but if we want to solve the strong $CP$ problem by demanding that $CP$ be a symmetry of the underlying theory, we will need an alternative to the KM model in which electroweak $CP$ violation is actually an emergent effect due to spontaneous breaking of $CP$ symmetry. 

Even assuming we can come up with some other mechanism to explain $CP$-violating meson phenomenology, we are back to the problem identified above: there is an inherent tension between a small value of the spontaneously induced $\theta_\text{Yukawa}$ (which is now the only contribution to $\bar{\theta}$) and a much larger value of $\delta_\text{CKM}$. The situation is somewhat improved, in that if $CP$ symmetry is broken spontaneously rather than explicitly, $\theta_\text{Yukawa}$ only receives finite renormalization~\citep{cheng1988}. Thus in principle it is a calculable quantity which may be derived from more fundamental parameters of the theory.

The specific predictions of the KM model are in excellent agreement with experiment~\citep{pdg2016}, and most models of spontaneous $CP$ violation which were proposed to solve the strong $CP$ problem in the late 70s and early 80s have since been excluded for this reason.\footnote{In fairness to these theories, it should be noted that at the time they were proposed, the KM model was only one of several plausible ways to accommodate $CP$ violation in the theory of the electroweak interactions; it has since become a widely accepted element of the SM.} A more viable model of spontaneous $CP$ violation was proposed in 1984 by Ann Nelson~\citep{nelson1984} and generalized by Stephen Barr~\citep{barr1984a,barr1984b}. The \textbf{Nelson-Barr mechanism} invokes the existence of rather elaborate theoretical structure at energies far above the electroweak scale to ensure simultaneously that a nonzero $\theta_\text{Yukawa}$ is not generated at tree level by spontaneous $CP$ violation and that the low-energy limit of the theory is practically indistinguishable from the KM model. There are still (finite) loop corrections to $\theta_\text{Yukawa}$ which must be calculated, and ensuring that they remain sufficiently small to satisfy the experimental bound $\bar{\theta}<10^{-10}$ typically requires additional fine-tuning.

Unlike the massless up quark solution, the Nelson-Barr mechanism is not ruled out (indeed, it seeks to solve the strong $CP$ problem without observable consequences at low energies~\citep{dd2015}, so it is difficult to imagine how it could be ruled out in the near future). However, it is hardly a minimal extension to the standard model -- lots of complicated new physics is required to make it work, and new fine-tuning problems are introduced along the way. I will not consider the Nelson-Barr mechanism further in this thesis; interested readers are referred to Refs.~\citep{dd2015,cheng1988}.

\section{Peccei-Quinn theory and the axion}\label{sec:peccei_quinn}
Compared to the Nelson-Barr mechanism discussed briefly above, the massless up quark solution to the strong $CP$ problem has an appealing simplicity. In particular, it can be shown that solving the strong $CP$ problem via an extra $U(1)_A$ symmetry automatically eliminates all observable effects of $\bar{\theta}$ \textit{to all orders in perturbation theory}~\citep{cheng1988}; we don't have to worry that loop corrections will simply recreate the strong $CP$ problem we have worked so hard to eliminate. Unfortunately, nature did not see fit to supply us with a massless quark. Is there a way to embed a $U(1)_A$ symmetry in the SM without generating a massless quark?

In a pair of seminal papers in 1977~\citep{pq1977a,pq1977b}, Roberto Peccei and Helen Quinn showed that indeed there is another way to implement the required $U(1)_A$ symmetry which solves the strong $CP$ problem. Within six months Steven Weinberg~\citep{weinberg1978} and Frank Wilczek~\citep{wilczek1978} pointed out that the PQ solution implies the existence of a pseudo-Goldstone boson $a(x)$ which Wilczek dubbed the \textbf{axion}, and that in an equivalent formulation of the PQ mechanism, the solution of the strong $CP$ problem can be attributed to the dynamics of the axion field. In Sec.~\ref{sub:pq_mechanism} I will show how these two formulations of the PQ mechanism are related, and in Sec.~\ref{sub:axion_mass} I will discuss some generic features of axion models without assuming any specific implementation of the PQ mechanism.

Weinberg and Wilczek both emphasized that not only is the axion a necessary feature of the PQ solution, it also enables experimental tests of specific models in which the strong $CP$ problem is solved via the PQ mechanism. In fact, Peccei, Quinn, Weinberg, and Wilczek all assumed a particular implementation of the PQ mechanism which has come to be known as the \textbf{PQWW model}. In Sec.~\ref{sub:pqww} I will discuss the specific features of the PQWW model, which was quickly excluded by experiments. Finally, in Sec.~\ref{sub:invisible} I will discuss the generalization of the PQ mechanism to so-called \textbf{``invisible'' axion} models which remain viable to this day.

\subsection{Two perspectives on the Peccei-Quinn mechanism}\label{sub:pq_mechanism}
Let us imagine we have a pair of chiral quark fields $q_L$, $q_R$, which we would like to use to implement a global $U(1)_A$ symmetry:
\begin{eqnarray}
q_L &\rightarrow e^{i\alpha}q_L \label{eq:u1a}\\
q_R &\rightarrow e^{-i\alpha}q_R \nonumber 
\end{eqnarray}
(c.f.\ Eq.~\eqref{eq:u1v} in Sec.~\ref{sub:ckm}). I presented a qualitative argument in Sec.~\ref{sub:hadrons} that bare quark mass terms in the Lagrangian are incompatible with the existence of axial symmetries: formally, this is because mass terms have the form $m_q\big(\bar{q}_Lq_R + \text{h.c.}\big)$ -- the ``bar'' entails complex conjugation, so although $q_L$ and $q_R$ rotate oppositely under $U(1)_A$, $\bar{q}_L$ and $q_R$ rotate the same way, and the mass term is not invariant. We cannot simply add an additional massless quark to the standard model, because doing so would profoundly alter the hadronic spectrum. So we must find a way to implement a $U(1)_A$ symmetry on a massive quark.

Fortunately, we have already seen that there is another way of generating fermion masses which is moreover known to be realized in nature. Let us assume that there is no $q$ mass term in the Lagrangian, but there is a Yukawa interaction between $q$ and a complex scalar field $\sigma$ whose potential\footnote{This potential is of the form Eq.~\eqref{eq:ssb_v}, but note that here I am using the symbol $\sigma$ for the original complex field rather than its radial component, and reserving $\phi$ for the SM Higgs doublet.} has its minimum at $\abs{\sigma} = f_a$. If we demand that $\sigma$ also transforms under the symmetry operation Eq.~\eqref{eq:u1a} as $\sigma \rightarrow e^{2i\alpha}\sigma$, we see that the Yukawa term
\begin{equation}\label{eq:phiqq}
\lagr_{\sigma\bar{q}q} = \sigma\bar{q}_Lq_R + \sigma^*\bar{q}_Rq_L
\end{equation}
is invariant: the $U(1)_A$ symmetry has been restored! The implications of this simple observation are quite profound. At temperatures below $f_a$, the quark $q$ acquires a mass $m_q\sim f_a$, and yet the underlying theory still has a hidden symmetry which we can call $U(1)_\text{PQ}$ to distinguish it from the case in which the $U(1)_A$ symmetry is realized directly via a massless quark. Peccei and Quinn explicitly showed (in a more realistic model than the one presented here) that in the presence of this spontaneously broken $U(1)_\text{PQ}$ symmetry, as in the massless quark case, $\bar{\theta}$ generates no observable effects.

I have not yet specified whether $q$ and $\sigma$ are to be interpreted as an SM quark and Higgs field or entirely new fields we have postulated for the sole purpose of solving the strong $CP$ problem (likewise, I have not specified the energy scale $f_a$). In general both cases are possible, though as we will see in Sec.~\ref{sub:pqww} that the PQ mechanism is slightly more complicated when we try to implement it in such a way as to minimally augment the SM. In the case where neither $q$ nor $\sigma$ are SM fields, the simple toy model I have presented here is basically a bare-bones version of the KSVZ model to be discussed in Sec.~\ref{sub:invisible}. The important point I want to emphasize is that any realization of the PQ mechanism must at a bare minimum involve a quark\footnote{``quark'' here means a fermion which is charged under $SU(3)_C$, which may or may not have the same electroweak interactions as an SM quark. Nonzero color charge is required to give $U(1)_\text{PQ}$ a chiral anomaly with QCD.} and a scalar field which transform under $U(1)_\text{PQ}$. 

We have thus far only considered the behavior of our toy model in the high-temperature ``unbroken'' phase. At temperatures below $f_a$, $U(1)_\text{PQ}$ is spontaneously broken, or more precisely it would be spontaneously broken if it were not explicitly broken by the chiral anomaly with QCD. The anomalous nature of $U(1)_\text{PQ}$ is key to the PQ mechanism: recall from Sec.~\ref{sub:quark_phase} that it is precisely the anomaly which allows us to ``rotate away'' $\bar{\theta}$, guaranteeing that it produces no observable effects. This raises the question of whether we can still apply the formalism of Sec.~\ref{sub:ssb} and treat $U(1)_\text{PQ}$ as an approximate spontaneously broken symmetry: if so, there must exist a corresponding pseudo-Goldstone boson. 

In Sec.~\ref{sub:hadrons} we saw both an example of an approximate symmetry to which the SSB formalism can be fruitfully applied and an example of an approximate symmetry for which this formalism fails: the former is the approximate $SU(2)_A$ symmetry of 2-flavor QCD (which is explicitly broken by $m_u,m_d\neq0$ and results in three pseudo-Goldstone bosons called the pions), and the latter is the approximate $U(1)_A$ symmetry of 3-flavor QCD (which is explicitly broken by both the quark masses and the chiral anomaly with QCD). Peccei and Quinn implicitly assumed that the latter case was a better template for $U(1)_\text{PQ}$, as the approximate symmetry is broken by the chiral anomaly in both cases: in doing so, they were taking the lesson of the $U(1)_A$ problem to heart! 

However, as I emphasized at the end of Sec.~\ref{sub:anomaly}, explicit symmetry breaking by the chiral anomaly with QCD is not really fundamentally different from any other form of explicit symmetry breaking. The presence or absence of a pseudo-Goldstone boson depends only on the relative strength of the energy scales $\mu$ and $f_a$ of spontaneous and explicit breaking, respectively. I invoked the temperature-dependence of instanton effects in Sec.~\ref{sub:instantons} to argue that $\mu \sim \Lambda_\text{QCD}$. To specify the value of $f_a$ we would need to choose a particular implementation of the PQ symmetry, but very generally $f_a\gg\Lambda_\text{QCD}$. Thus, the SSB formalism can be applied to $U(1)_\text{PQ}$, and there should be a pseudoscalar pseudo-Goldstone boson, which we will call the axion. This was the essential insight that Weinberg and Wilczek contributed to the theory of the PQ mechanism.\footnote{Note that in the massless up quark solution to the strong $CP$ problem, $\mu\sim f\sim\Lambda_\text{QCD}$ (because the chiral symmetries of quarks are always spontaneously broken at the confinement scale as noted in Sec.~\ref{sub:hadrons}), so instead of the axion we would have another particle like the $\eta'$. In this sense the massless up quark solution may be regarded as the $f_a\rightarrow\Lambda_\text{QCD}$ limit of the PQ solution.}

If $U(1)_\text{PQ}$ were an exact spontaneously broken symmetry, the axion $a(x)$ would be exactly analogous to the generic Goldstone boson $\pi(x)$ introduced in Sec.~\ref{sub:ssb}. That is, the axion would be massless, with only derivative interactions suppressed by factors of $f_a^{-1}$, and qualitatively speaking $a/f_a$ would behave like a dynamical version of the $U(1)_\text{PQ}$ symmetry parameter $\alpha$ (c.f.\ Fig.~\ref{fig:goldstone}). But the chiral anomaly generates an explicit $U(1)_\text{PQ}$-breaking term
\begin{equation}\label{eq:agg}
\lagr_{agg}= \frac{g^2}{32\pi^2f_a}aF^{\mu\nu}_a\tilde{F}_{\mu\nu a}.
\end{equation}
This is not a derivative interaction, and it implies that all values of $a$ are no longer equivalent. The effects of this symmetry breaking can be represented visually as a ``tipping'' of the wine bottle potential illustrated in Fig.~\ref{fig:goldstone} in a direction specified by Eqs.~\eqref{eq:agg} and \eqref{eq:theta_term} [with $\bar{\theta}$ in place of $\theta_\text{QCD}$ in the latter]. Let us assume that we can meaningfully think of the tipping of the potential as a process that happens at a certain point in cosmological time (this assumption will be justified in Sec.~\ref{sec:axion_cosmo}). When the axion feels the tilt of its potential, it will ``roll'' from its arbitrary initial position to the new minimum. It is precisely this dynamical process that solves the strong $CP$ problem!\footnote{You can make your very own wine bottle potential at home with any punted wine bottle. First you will have to drink the wine, preferably with at least one friend. Then hold the bottle upright, and when the last few drops of wine at the bottom least expect it, tilt the bottle and watch the drops flow to the new minimum. Congratulate yourself on solving the strong $CP$ problem and remember to drink some water.}

For the dynamical evolution of the axion field to solve the strong $CP$ problem as proposed above, the axion potential must be minimized at $\bar{\theta}+a(x)/f_a=0$, i.e., the axion must roll to a point where it cancels out whatever initial value $\bar{\theta}$ happens to have. Actually, since the anomaly guarantees that Eq.~\eqref{eq:agg} has precisely the same form as Eq.~\eqref{eq:theta_term}, this is equivalent to the condition that the energy of the QCD vacuum state be minimized at $\theta_\text{QCD}=0$. There is an elegant and very concise proof of this claim by Vafa and Witten~\citep{vafa1984}. I will not summarize their argument here, except to emphasize that the statement that $\ket{\theta_\text{QCD}=0}$ has lower energy than other $\theta$-vacua does not \textit{by itself} solve the strong $CP$ problem, since in the SM alone $\theta_\text{QCD}$ is a fixed parameter. However, it does indicate that if we could ``promote'' $\theta_\text{QCD}$ from a parameter to a field, the strong $CP$ problem would solve itself dynamically. This is probably the simplest statement of the essential physics of the PQ mechanism.

\subsection{Generic axion properties}\label{sub:axion_mass}
We saw above that in one formulation of the PQ mechanism, the chiral anomaly with QCD tips the wine bottle potential, resulting in a nonvanishing \textit{axion potential} $V_a(a/f_a)=\lagr_{agg}$, and the axion's dynamical response to $V_a$ solves the strong $CP$ problem. A nonvanishing potential implies that the axion acquires a nonzero mass, which is formally given by
\begin{equation}\label{eq:axion_mass_formal}
m_a^2 = \left.\pdd{V_a}{a}\right|_{a=-f_a\bar{\theta}}.
\end{equation}
Since low-temperature QCD is strongly coupled, evaluating Eq.~\eqref{eq:axion_mass_formal} explicitly is quite nontrivial. There are a number of approaches to deriving good approximations to Eq.~\eqref{eq:axion_mass_formal}; see e.g., Refs.~\citep{cheng1988,peccei1996,peccei2008,bardeen1978}. However, we can practically guess the correct scaling of $m_a$ by noting that the Lagrangian depends only on the ratio $a(x)/f_a$, and thus the general form $\frac{1}{2}m_a^2a^2$ of an axion mass term implies that $m_a^2\propto f_a^{-2}$. By dimensional analysis $m_a^2$ must also be proportional to a quantity with mass dimension 4, which should be a measure of the energy density associated with the explicit breaking of $U(1)_\text{PQ}$ by the chiral anomaly with QCD. Thus we can guess
\begin{equation}\label{eq:axion_mass_sim}
m_a^2 \sim \frac{\Lambda_\text{QCD}^4}{f_a^2}.
\end{equation}
In the visualization of the wine bottle potential [Fig.~\ref{fig:goldstone}], $f_a$ is the radius of the trough, and $\Lambda_\text{QCD}^4/f_a=\sin(\psi)$, where $\psi$ is angle by which the potential is tilted out of the $xy$ plane. Eq.~\eqref{eq:axion_mass_sim} is just what you get if you work out the frequency of small oscillations about the minimum.\footnote{I have been somewhat cavalier with the units in making this correspondence: the $x$ and $y$ axes in field space have mass dimension 1, but the $z$ axis represents energy density and thus has mass dimension 4. To make the correspondence to the classical case of a bead on a hoop more properly, we really should have units of length for all axes and include a gravitational potential $V_g=-R\cdot g\sin(\psi)\cos(\theta)\Leftrightarrow -f_a\cdot\big(\Lambda_\text{QCD}^4/f_a\big)\cos(a/f_a)$.}

To qualitatively motivate a more precise expression for the axion mass, let us consider the interactions of axions with SM fields. Of course, the interaction Eq.~\eqref{eq:agg} between an axion and a pair of gluons must exist to solve the strong $CP$ problem, and the coefficient of this term is fixed by the chiral anomaly of $U(1)_\text{PQ}$ with QCD in the same sense as the coefficient of the $\pi^0\gamma\gamma$ interaction [Eq.~\eqref{eq:pigammagamma}] was fixed by the chiral anomaly of $J^\mu_{A3}$ with QED. In this sense Eq.~\eqref{eq:agg} is \textbf{model-independent}.\footnote{The coefficient can differ by dimensionless factors related to the particular implementation of the $U(1)_\text{PQ}$ symmetry, but these factors can be absorbed into the definition of $f_a$ in any invisible axion model. See Sec.~\ref{sub:invisible} for further discussion.} In the low-energy hadronic phase the axion-gluon-gluon interaction manifests in axion-nucleon interactions and axion-meson interactions. The interactions of axions with neutral pseudoscalar mesons such as the $\pi^0$, $\eta$ and $\eta'$ are especially important, as these fields have all the same quantum numbers as the axion. Thus, in an effective low-energy Lagrangian \citep{peccei1996,peccei2008}, mixing terms like $a\pi^0$ are permitted, and these terms are responsible for generating the axion mass. In particular, the pion mixing term dominates, and thus
\begin{equation}\label{eq:axion_mass_pion}
m_a^2 = \frac{m_\pi^2f_\pi^2}{f_a^2}\frac{m_um_d}{(m_u+m_d)^2}\approx \frac{\big(77.6~\text{MeV}\big)^4}{f_a^2}.
\end{equation}
Eq.~\eqref{eq:axion_mass_pion} is model-independent in the same sense as Eq.~\eqref{eq:agg}. Note also that Eq.~\eqref{eq:axion_mass_pion} vanishes if $m_u$ or $m_d=0$, as in this case the $U(1)_\text{PQ}$ symmetry is not necessary to solve the strong $CP$ problem.\footnote{The specific numerical value in the numerator of Eq.~\eqref{eq:axion_mass_pion} assumes the mass ratio of the lightest quarks is $z=m_u/m_d=0.56$. I will assume this value of $z$ throughout this thesis; see discussion in Sec.~\ref{sec:parameter}.}

The presence of axion-pion mixing also implies that the axion inherits the neutral pion's two-photon interaction Eq.~\eqref{eq:pigammagamma}, and this interaction will turn out to be extremely important for axion detection. The general form of the axion-photon-photon interaction is
\begin{equation}\label{eq:agammagamma_general}
\lagr_{a\gamma\gamma}=g_\gamma \frac{e^2}{16\pi^2f_a}aF^{\mu\nu}\tilde{F}_{\mu\nu},
\end{equation}
where the dimensionless coefficient $g_\gamma$ depends on the particular axion model. $\lagr_{a\gamma\gamma}$ is more model-dependent than $\lagr_{agg}$ because in addition to the fixed contribution from pion mixing, there can be another contribution if the quark or quarks charged under $U(1)_\text{PQ}$ are also charged under QED. However, the two-photon interaction is always present at some level, and except in pathological cases is relatively model-independent. We will discuss the additive contributions to $g_\gamma$ for viable axion models in Sec.~\ref{sec:parameter}.

\subsection{The PQWW model}\label{sub:pqww}
We have seen that the PQ mechanism generally requires at least one quark $q$ which is given mass by a scalar field $\sigma$ which spontaneously breaks $U(1)_\text{PQ}$. The SM of course contains many quarks which acquire mass via the spontaneous breaking of electroweak symmetry by the Higgs fields (see Sec.~\ref{sub:ckm}). Identifying $\sigma$ with the SM Higgs doublet $\phi$ would clearly be the most theoretically economical implementation of the PQ mechanism.

From Eq.~\eqref{eq:yukawa_1} we can see that there is an impediment to making this correspondence: the neutrality of the Lagrangian under the full SM gauge group requires that $\phi$ appears in the term that gives mass to down-type quarks but $\phi^*$ appears in the corresponding term for the up-type quarks. Thus the Higgs Yukawa terms will only be PQ-symmetric if the up-type and down-type quarks of the same chirality rotate oppositely under $U(1)_\text{PQ}$, but then the interactions of quarks with $W^\pm$ bosons would not respect the PQ symmetry.

Peccei and Quinn circumvented this problem by extending the Higgs sector of the SM to include two Higgs doublets, such that $\phi_1$ couples to up-type quarks and $\phi_2$ couples to down-type quarks (and leptons): this simple extension of the SM Lagrangian has a $U(1)_\text{PQ}$ symmetry under which both Higgs doublets and all SM fermions are charged. The case discussed in Sec.~\ref{sub:ckm} where the electroweak gauge symmetry is broken by the VEV of a single Higgs doublet is obviously simpler, but EWSB still works with a nonminimal Higgs sector, and when the PQWW model was first introduced, there was no experimental evidence for any particular implementation of the Higgs mechanism. The two Higgs doublets in the PQWW model have VEVs $v_1$ and $v_2$, respectively, and $f_a=\sqrt{v_1^2+v_2^2}$ is identified with the electroweak scale $v=246$~GeV. The PQWW axion may be identified with the degree of freedom corresponding to the relative phase between the neutral components of the two Higgs doublets. Its mass is given by~\citep{peccei1996,peccei2008}
\begin{equation}\label{eq:axion_mass_pqww}
m_a^\text{PQWW} = N_g\left(x+\frac{1}{x}\right)\frac{m_\pi f_\pi}{v}\frac{\sqrt{m_um_d}}{m_u+m_d},
\end{equation}
where $N_g=3$ is the number of fermion generations and $x=v_2/v_1$ is the only free parameter of the PQWW model. Plugging in numbers, we see that $m_a^\text{PQWW}\gtrsim150$ keV, where the minimum mass is obtained for $x=1$. 

Weinberg and Wilczek's original papers led to a great flurry of activity because it was immediately clear that the PQWW axion would be detectable if it existed: the relevant energies were easily accessible with the accelerator and particle detector technology of the 1970s and early 1980s. Historically, the processes most important to the experimental exclusion of the PQWW axion were radiative decays of the heavy mesons $\Upsilon$ and $J/\Psi$: within the PQWW model, the branching ratios for the processes $\Upsilon \rightarrow \gamma a$ and $J/\Psi \rightarrow \gamma a$ are proportional to $x^{-2}$ and $x^2$, respectively: their product is thus a parameter-free prediction of the PQWW model which was quickly shown to be inconsistent with experimental results \citep{edwards1982,sivertz1982,alam1983}. Many other results in particle and nuclear physics also set stringent limits on the original PQWW model and variants thereof; see Ref.~\citep{cheng1988} for more detailed discussion.

\subsection{``Invisible'' axion models}\label{sub:invisible}
By the early 1980s it was clear that the PQWW axion did not exist. Though experiments would continue through the decade, variant models with $f_a$ at the electroweak scale also looked increasingly unlikely. It did not take long for theorists to construct models in which the PQ mechanism was implemented with $f_a\gg v$. Because the axion mass and all couplings are inversely proportional to $f_a$, such models generically lead to very light ``invisible'' axions capable of evading all existing experimental bounds. 

The first specific example of an invisible axion model was the \textbf{KSVZ model} proposed by Kim~\citep{kim1979} and independently by Shifman, Vainshtein, and Zakharov~\citep{SVZ1980}. The KSVZ model is essentially just the simple toy model I presented at the beginning of Sec.~\ref{sub:pq_mechanism}, where neither the quark $q$ nor the complex scalar $\sigma$ are SM fields. Both $q$ and $\sigma$ must be $SU(2)_W$ singlets; the quark may or may not interact electromagnetically, but of course it must be charged under $SU(3)_C$. We have already seen that these fields allow us to implement a $U(1)_\text{PQ}$ symmetry whose only connection to SM fields is through the chiral anomaly with QCD. The KSVZ axion is just the phase degree of freedom of the $\sigma$ field (like the Goldstone boson in the toy model of SSB in Sec.~\ref{sub:ssb}). 

A second invisible axion model more akin to the original PQWW model was proposed by Zhitnitsky~\citep{zhitnitsky1980} and independently by Dine, Fischler, and Srednicki~\citep{DFS1981}: it has become known as the \textbf{DFSZ model}. In fact, the DFSZ model is simply the PQWW model supplemented with an additional complex scalar field $\sigma$ which is charged under $U(1)_\text{PQ}$ and has a VEV $\sim f_a\gg v$. As in the KSVZ model, $\sigma$ must be an electroweak singlet, so it cannot have direct couplings to the SM quarks, but interactions between $\sigma$ and the Higgs fields are permitted with appropriate assignments of PQ charges. The DFSZ model thus provides a way to implement a $U(1)_\text{PQ}$ symmetry under which all SM fields transform, without an additional electroweak-scale pseudo-Goldstone boson. Like the KSVZ axion, the DFSZ axion resides in the phase of the $\sigma$ field. 

In addition to the axion, we see that both prototypical invisible axion models predict the existence of very heavy fields with mass $\sim f_a$; we will ignore these fields as the relevant energies are not even remotely experimentally accessible. The axion mass may be written in the form Eq.~\eqref{eq:axion_mass_pion} for both the KSVZ and DFSZ models.\footnote{Strictly speaking, the DFSZ axion mass should be multiplied by a factor of $N_g(x+1/x)$ as in the PQWW model, but this coefficient can be absorbed by the redefinition $f_a\rightarrow f_a/[N_g(x+1/x)]$. The coefficient is related to the fact that more than one quark flavor transforms under $U(1)_\text{PQ}$, so it also appears in all anomaly-mediated interactions. Redefining $f_a$ in this way simplifies the notation for quantities most relevant to physics at the low energies of interest. However, in Sec.~\ref{sub:defects}, we will encounter one nontrivial consequence of implementing the PQ mechanism with more than one quark flavor.} In Sec.~\ref{sec:parameter} we will see that both KSVZ and DFSZ are actually families of models, distinguished by different values for the electric charge of the heavy quark in the KSVZ case or different couplings of the Higgs doublets to SM fermions in the DFSZ case.

It may seem to more cynical readers that the intent of theorists responsible for constructing invisible axion models was to render the PQ mechanism invulnerable to inconvenient experimental results. In fairness to these theorists, the late 70s and early 80s were an exciting time in particle physics: what we now call the standard model was just coming together, and a number of very well-motivated theoretical arguments pointed to new physics associated with the unification of the forces at a \textbf{Grand Unified Theory (GUT) scale} of $M_\text{GUT}\sim10^{16}$ GeV, for which incontrovertible indirect evidence would be just around the corner. In particular it was shown that the DFSZ model with $f_a\sim M_\text{GUT}$ could be easily embedded in promising theories of physics at the GUT scale (see e.g., \citep{wise1981}).

In the years to follow, while the outlook for the simplest GUT theories grew increasingly bleak, theorists realized that implementations of the PQ mechanism with high symmetry-breaking scale $f_a$ could have dramatic and sometimes disastrous effects on cosmology, and relatedly, that invisible axions might not be so invisible after all. I will address these subjects in chapters~\ref{chap:cosmo} and~\ref{chap:search}.


\chapter{Dark matter and axion cosmology}\label{chap:cosmo}
\setlength\epigraphwidth{0.61\textwidth}\epigraph{\itshape Disturbance from cosmic background radiation is something we have all experienced. Tune your television to any channel it doesn't receive, and about 1 percent of the dancing static you see is accounted for by this ancient remnant of the Big Bang. The next time you complain that there is nothing on, remember that you can always watch the birth of the universe.}{Bill Bryson}

\noindent That cosmology should have anything at all to do with particle physics is fundamentally a consequence of the \textbf{Big Bang model}, according to which the universe began in a very hot and dense state some 14 billion years ago and has been expanding and cooling since. The extremely high energies that prevailed in the very early universe suggest a twist on Gell-Mann's totalitarian principle: ``everything not forbidden already happened in the first 0.01 seconds.''\footnote{Admittedly not as pithy as the original.}

Though we do not have direct observational access to the very early universe, the basic picture outlined above has been widely accepted since the early 1980s, and tested with remarkable precision over the past two decades: the $\Lambda$CDM model introduced in chapter~\ref{chap:intro} is the particular parameterization of Big Bang cosmology which has emerged from these stringent observational tests. The question I hope to answer in the present chapter is why invisible axions (if they exist) play a particularly outsized role in cosmology.

I begin by briefly reviewing the most relevant features of Big Bang cosmology and our present understanding of the history of the universe in Sec.~\ref{sec:big_bang}. I then review the observational basis for supposing that the universe contains a large amount of non-baryonic dark matter in Sec.~\ref{sec:evidence}, and discuss some of the more prominent particle dark matter candidates in Sec.~\ref{sec:dm_candidates}. In particular, I will show that invisible axions have all the right properties to explain dark matter, provided there exist non-thermal production mechanisms in the early universe.

The remainder of the chapter is devoted to axion cosmology. We will see in Sec.~\ref{sec:axion_cosmo} that provided the axion is sufficiently ``invisible,'' the PQ mechanism itself produces dark matter axions as a side effect of solving the strong $CP$ problem! A closer look at the cosmological role of axions complicates this very economical picture considerably, precluding a definitive prediction for the cosmic abundance of dark matter axions as a function of $f_a$. Nonetheless, the basic takeaway point is that the existence of axions can profoundly affect cosmology, and axions remain among the best-motivated dark matter candidates. 

This chapter assumes that you have read the previous chapter, and is generally pitched at a similar level. My discussion of dark matter is not as thorough as my discussion of the strong $CP$ problem and axion theory in chapter~\ref{chap:theory}, and this is intentional: many more books and review papers aimed at non-specialists have been written about this subject than about axions. For the same reason, I will not be as thorough in citing the original literature. \textit{The Early Universe} by Kolb and Turner \citep{KT1994} is an extremely readable text for early universe cosmology, though there have been major developments in the field over the past two decades. Ref.~\citep{bartelmann2010} is an excellent pedagogical review paper covering many subfields of contemporary cosmological research. See Ref.~\citep{bertone2016} for a good non-technical discussion of the history of the study of dark matter. Finally, for reviews of various aspects of axion cosmology, see Refs.~\citep{cheng1988,windows1990,sikivie2008,wantz2010}.

\section{Big Bang cosmology}\label{sec:big_bang}
Here I review the elements of Big Bang cosmology which will be important for the subsequent discussion of dark matter and axions. In Sec.~\ref{sub:expansion} I will define the relevant concepts that arise as we extrapolate from the observed expansion of the universe to early times when everything was relativistic. Then in Sec.~\ref{sub:history} I will outline the major milestones in the history of the universe from the first moments we can meaningfully speak of to the present day. For a more detailed treatment, see the first three chapters of Ref.~\citep{KT1994}.

\subsection{Implications of an expanding universe}\label{sub:expansion}
The earliest observational evidence for a Big Bang cosmological model came from \textbf{Hubble's law}: empirically, distant galaxies appear to be receding from us in all directions at a rate proportional to their distance from Earth. Contemporary measurements of the proportionality factor yield $H_0=70\pm3$ km/s/Mpc.\footnote{The uncertainty I have cited reflects a persistent discrepancy between the best direct measurements~\citep{riess2016} and the value inferred from high-precision measurements of the cosmic microwave background~\citep{planck2016}; the statistical errors on both measurements are smaller by about a factor of 4. For reference, a parsec (pc) is about $3\times10^{16}$ meters.} $H_0$ is called the Hubble constant, but it is actually a function of time. More precisely, in the context of the Big Bang model, $H_0$ happens to be the present value $(t=t_0$) of the \textbf{Hubble parameter} $H(t)$. The Hubble parameter is in turn defined as $H(t)=\dot{R}/R$, where $R(t)$ is the dimensionless \textbf{Robertson-Walker scale factor} parameterizing the relative distance between any two otherwise stationary objects in the expanding universe. Conventionally, the scale factor is normalized so that the present value is $R_0=1$.

Within the Big Bang model, the expansion of the universe is driven by its total energy density. This is formalized in the \textbf{Friedmann equation}, which (with some empirically motivated assumptions about the contents of the universe) may be expressed in the form
\begin{equation}\label{eq:friedmann}
H^2 = \left(\frac{\dot{R}}{R}\right)^2 = H_0^2\left[\frac{\Omega_r}{R^4} + \frac{\Omega_m}{R^3} + \Omega_\Lambda + \frac{1-\big(\Omega_r+\Omega_m+\Omega_\Lambda\big)}{R^2} \right],
\end{equation}
where $\Omega_r$, $\Omega_m$, and $\Omega_\Lambda$ are normalized expressions of the \textit{present-day}\footnote{I have omitted the additional subscript 0 indicates that quantities are evaluated at $t=t_0$ to reduce notational clutter.} energy density associated with relativistic particles (``radiation''), nonrelativistic particles (``matter''), and dark energy, respectively. The best current measurements yield $\Omega_r=9.10\times10^{-5}$, $\Omega_m=0.309$, and $\Omega_\Lambda=0.691$, all with uncertainty at the percent level or better~\citep{planck2016}. Note that these values imply the total density $\Omega_\text{tot}=\Omega_r+\Omega_m+\Omega_\Lambda=1$, so we can ignore the last term in Eq.~\eqref{eq:friedmann}, corresponding to (apparently nonexistent) spatial curvature. There is also a strong theoretical bias in favor of a \textit{flat} universe with $\Omega_\text{tot}=1$.

The physical density $\rho_i$ of each component ($i=r,m,\Lambda$) can be obtained from the corresponding normalized density $\Omega_i$ by multiplying by the (present-day) \textit{critical density} $\rho_c=3H_0^2/(8\pi G)$, where $G$ is the Newtonian gravitational constant. In natural units $G=M_p^{-2}$, where $M_p=1.22\times10^{19}$ is the \textbf{Planck mass}.\footnote{Roughly speaking, the Planck mass is the energy scale at which both quantum mechanical and general relativistic phenomena are important. At present we do not know how to reconcile these two theories, so it is difficult to speak meaningfully of hypothetical processes at the Planck scale.} $\Omega_\text{tot}=1$ of course implies that in physical units the total energy density of the universe is (on average) equal to the critical density. It is often convenient to express cosmic densities $\rho$ in units of GeV/cm$^3$, bearing in mind that the mass energy of a proton is about 1 GeV: with the value of $H_0$ cited above, $\rho_c\approx5\times10^{-6}$ GeV/cm$^3$. Evidently we live in a substantial overdensity!

Although I have expressed Eq.~\eqref{eq:friedmann} in terms of the present-density parameters $\Omega_i$, the physical densities $\rho_i$ are in general time-dependent:\footnote{Actually, $\Omega_i$ are also time-dependent, because $\rho_c$ is a function of the Hubble parameter. To avoid confusion, I will reserve the symbol $\Omega$ for present-day density parameters, and reserve $\rho_c$ for the present-day critical density.} indeed, if we were to distribute the factor of $H_0^2$ through the RHS of Eq.~\eqref{eq:friedmann} and multiply the whole equation by $3/(8\pi G)$, each additive term on the RHS would just be the corresponding physical density $\rho_i(t)$ evaluated at the same time as the scale factor $R(t)$. All of the factors on the RHS of Eq.~\eqref{eq:friedmann} are positive, so within the framework of standard cosmology $\dot{R}(t)>0$ always. Extrapolating the present expansion of the universe backwards in time we would find that we reach $R=0$ in a finite time! It is conventional to define $t=0$ at to be the point at which $R=0$.

If we try to extrapolate all the way back to $R=0$ we will eventually run into unknown Planck-scale physics. But we do not have to extrapolate nearly that far back to arrive at a time when $R(t)$ was sufficiently small that $\rho_r(t)\gg\rho_m(t)\gg\rho_\Lambda(t)$. Thus very early in its history, the universe was \textbf{radiation-dominated} and later it became \textbf{matter-dominated}. The effects of dark energy only become relevant late in the history of the universe; thus we will ignore the $\Omega_\Lambda$ term for the remainder of this thesis.

It is easy to understand the scaling of the $\Omega_r$ and $\Omega_m$ terms in Eq.~\eqref{eq:friedmann}. Let us consider a gas of particles with mass $m$ and momentum $p$: in general, each particle has energy $E^2=m^2+p^2$. In the nonrelativistic limit $p\ll m$ the total energy density is $\rho_m =mn_m$, where $n_m$ is the number density: $\rho_m \propto R^{-3}$ just implies that while the mass remains constant, the number density is diluted by the $R^3$ expansion of the volume occupied by the gas. In the relativistic limit ($p\gg m$), the number density $n_r$ is diluted in exactly the same way, but the energy density $\rho_r = pn_r \propto R^{-4}$ since the momentum is \textbf{redshifted}\footnote{In quantum mechanics we can always associate any momentum with a de~Broglie wavelength $\lambda_p = 2\pi/p$. The cosmic redshift $p\propto R^{-1}$ is often explained as a consequence of the stretching of the wavelength by the expansion of the universe. As an intuitive explanation I find this somewhat lacking -- it makes the whole process seem more mysterious than it actually is, and fails to explain why for a massive particle the de~Broglie wavelength is stretched by the expansion but the Compton wavelength $\lambda_c = 2\pi/m$ is not. Indeed, working in the \textbf{comoving frame} singled out by the expansion of the universe, we can see that $p\propto R^{-1}$ is just a geometrical effect. A thorough discussion of coordinate systems in cosmology is beyond the scope of this thesis, but most of the essential features can be seen in the simple example of a 2D space with positive curvature (e.g., the surface of a balloon being inflated), which we can embed in a fictitious third dimension to facilitate visualization. In this example (due to Ref.~\citep{KT1994}), the scale factor is just the (appropriately normalized) time-dependent radius of the balloon, and the comoving coordinates are the angles specifying the positions on the surface. If there are two ants at fixed positions on the surface of the balloon, it is clear that the physical separation between them will increase $\propto R$ as the balloon expands (in agreement with our intuitive definition of the scale factor). If one of the ants is moving towards the other, it is equally clear that the expansion of the balloon reduces the speed at which the ant progresses across the surface.} by an extra factor of $R^{-1}$.

Fancy notation aside, Eq.~\eqref{eq:friedmann} is just a differential equation for $R(t)$, and the solution is easily obtained in two limiting regimes:
\begin{align}
R(t) &\propto t^{1/2} \qquad (\text{radiation-dominated})\label{eq:rad_dom}\\
R(t) &\propto t^{2/3} \qquad (\text{matter-dominated}).\label{eq:matter_dom}
\end{align}
The precise values of the prefactors are not very useful since these solutions are approximate anyway. In both cases, it is clear that $H=\dot{R}/R\propto1/t$. Thus we see that at any time $t$ the Hubble time $t_H=H(t)^{-1}$ is related to the current age of the universe (i.e., $t$) by an order-unity dimensionless factor: qualitatively, the universe was expanding so much faster at early times that most of the history of the universe elapsed during the most recent Hubble time, and this statement applies just as well to the early universe as to the present-day.\footnote{If the \textit{very} early universe was neither matter-dominated nor radiation-dominated, the same basic picture holds if we reinterpret $t$ as the time since beginning of the radiation-dominated era. In general physical quantities of interest will not depend on the absolute age of the universe.} Since nothing can travel faster than the speed of light, the Hubble length $r_H = ct_H$ is a measure of the size of the observable universe: at time $t$ objects separated by $D\gtrsim r_H$ are not in causal contact. In natural units the present-day Hubble scale is $H_0\sim10^{-32}$~eV.\footnote{Plugging this into Eq.~\eqref{eq:hl_units} to get nice large numbers for $t_H$ and $r_H$ is left as an exercise for the reader.}

We have gotten quite a lot of mileage out of Eq.~\eqref{eq:friedmann} alone. For our subsequent discussion we will also need to know that the early universe was very nearly in a state of \textbf{local thermal equilibrium}. Roughly speaking, we can meaningfully speak of thermal equilibrium in the early universe if $\Gamma\gg H$, where $\Gamma$ is some characteristic rate for interactions between the various particle species. For our present purposes, thermal equilibrium may be simply taken as a postulate for which there is plenty of empirical evidence; see Ref.~\citep{KT1994} for a more detailed discussion. 

\begin{figure}[h]
\centering\includegraphics[width=0.8\textwidth]{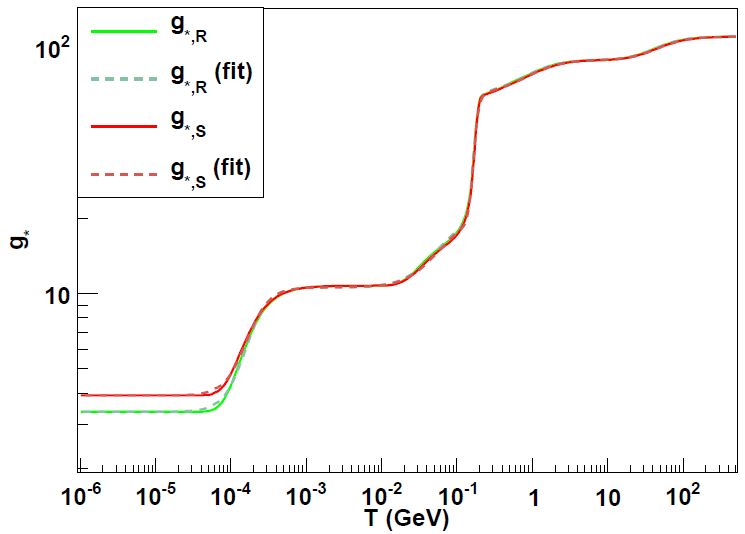}
\caption[Effective relativistic degrees of freedom vs.\ temperature]{\label{fig:dof} The effective number of relativistic degrees of freedom $g_{*}(T)$ [labeled $g_{*R}$ on the plot] that appears in the energy density of the universe, and the corresponding quantity $g_{*s}(T)$ for the entropy density. The history of the universe proceeds along the curve from the upper right to the lower left. Changes in $g_*$ correspond to temperatures at which different SM particles become non-relativistic and stop contributing significantly to the energy density (the universe is still radiation-dominated even at the left edge of this plot). The sharp drop around $T\sim 200$ MeV is the QCD phase transition. Figure from Ref.~\citep{wantz2010}.}
\end{figure}

When studying the thermodynamics of a radiation-dominated universe, we can safely neglect the matter component entirely, for two main reasons. First, as I noted above, ``radiation'' and ``matter'' are not absolute categories, but rather correspond to the limiting cases of ultrarelativistic and nonrelativistic particle behavior. Sufficiently early in the history of the universe, the temperature $T$ was so high that everything behaved like radiation. Second, provided thermal equilibrium is maintained, massive particles do not stick around once $T$ drops below $m$: their number density is suppressed by $e^{-m/T}$ due to the same physics that ensures states with $E\gg T$ are unoccupied in the statistical mechanics of more mundane quantum systems. Most of the time we are thus justified in treating the contents of the early universe using the known expressions for the energy density $\rho_r$ and the entropy density $s$ of a relativistic ideal gas:
\begin{align}
\rho_r &= \frac{\pi^2}{30}g_*T^4\label{eq:energy_rel}\\
s &= \frac{2\pi^2}{45}g_{*s}T^3 \label{eq:entropy_rel}.
\end{align}
In Eq.~\eqref{eq:energy_rel} $g_*=g_*(T)$ is the effective number of relativistic degrees of freedom at temperature $T$; it is plotted in Fig.~\ref{fig:dof}.\footnote{$g_*$ is just a generalization of the polarization degeneracy factor $g=2$ which appears in the more familiar expression for the energy density of a blackbody photon gas. The corresponding factor $g_{*s}$ in Eq.~\eqref{eq:entropy_rel} deviates slightly from $g_*$ at low temperatures because neutrinos decouple from the thermal bath before $e^+/e^-$ annihilation. This discrepancy is not important for the subjects treated in this thesis.} 

Using Eq.~\eqref{eq:energy_rel} in Eq.~\eqref{eq:friedmann} we can derive a relation between $H$ and $T$ in the radiation-dominated era; this is often more useful in practice than knowing the scale factor as a function of time. Evaluating the numerical factors, we obtain
\begin{equation}\label{eq:H_T_rad}
H = 1.66\sqrt{g_*(T)}\frac{T^2}{M_p}.
\end{equation}

Finally, thermal equilibrium implies not only that we can speak meaningfully about the temperature of the universe at any given time, but also that the expansion of the universe may be regarded as adiabatic -- thus the entropy per comoving volume $S\propto g_{*s}(T)T^3R^3$ is conserved. In fact, photons so overwhelmingly dominate the entropy of the universe that the expansion remains essentially adiabatic well after all the matter in the universe has fallen out of thermal equilibrium. 

\subsection{A short history of nearly everything}\label{sub:history}
At this point we have introduced several different ways to characterize the state of the universe as a function of time $t$: the Hubble factor $H\sim 1/t$, the temperature $T$ [given by Eq.~\eqref{eq:H_T_rad} during the radiation-dominated epoch], and the scale factor $R$ [formally given by integrating Eq.~\eqref{eq:friedmann}; see also Eqs.~\eqref{eq:rad_dom} and~\eqref{eq:matter_dom}]. In practice, $T$ and $H$ are most useful at early times when subatomic physics is relevant, and $R$ is more useful at later times when we are interested in questions of structure formation.\footnote{I have tried to simplify matters by not expressing quantities in terms of the redshift $z=R^{-1}-1$, which is sometimes used in place of or in addition to $R$.}

\begin{figure}[h]
\centering\includegraphics[width=1.0\textwidth]{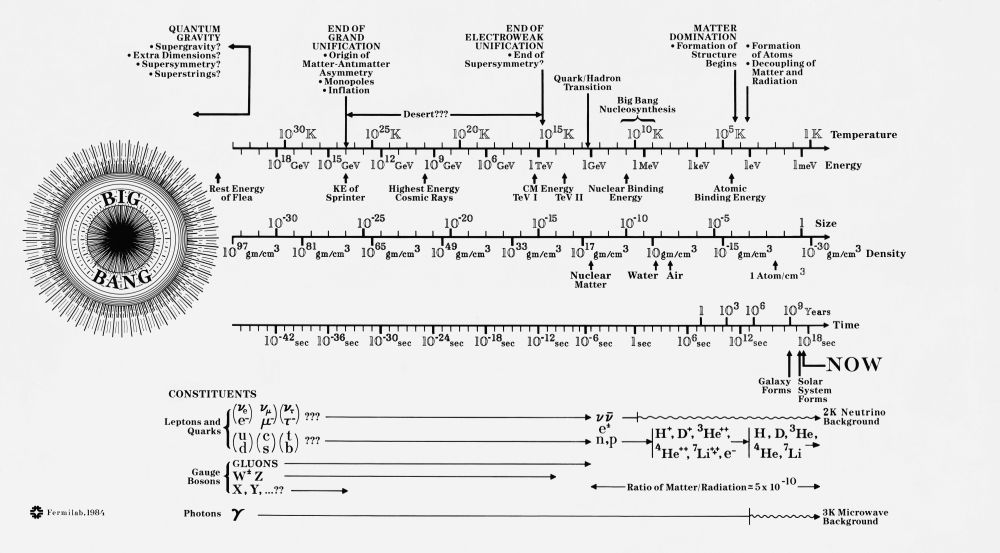}
\caption[The history of the universe]{\label{fig:history} The history of the universe. Figure from Fermilab~\citep{timeline}, with colors inverted. A version also appears in Ref.~\citep{KT1994}.}
\end{figure}

Fig.~\ref{fig:history} illustrates very concisely the major events in the history of the universe. I will discuss the most salient points below.
\begin{itemize}
\item \textbf{The very early universe}: Given our present lack of understanding of how to unify quantum mechanics with general relativity, it does not even make sense to speak of what happens at temperatures above (times before) the Planck scale ($T\sim M_p \sim 10^{19}$~GeV, $t\sim10^{-44}$~s). This is not to say we understand everything that happens below the Planck scale. I remarked at the end of chapter~\ref{chap:theory} that there is a strong prejudice among particle physicists for new physics at the GUT scale $M_\text{GUT} \sim 10^{16}$ GeV, and a radiation-dominated universe reaches this temperature at $t\sim10^{-38}$ s. But there is no clear indication of exactly what form GUT-scale physics should take, and most of the simplest models have been ruled out.
\item \textbf{Inflation} is a conjectured period of superluminal expansion in the very early universe. The theory of inflation is an extension to the standard $\Lambda$CDM model of Big Bang cosmology rather than an integral element of it. Though inflation is associated with the GUT scale in Fig.~\ref{fig:history}, the inflationary energy scale $E_I$ is quite model-dependent, and the only experimental constraint on the Hubble parameter during inflation $H_I \sim E_I^2/M_p$ is $H_I \lesssim 10^{14}$ GeV from the non-observation of inflationary gravitational waves. $E_I$ (and thus $H_I$) can be reduced by orders of magnitude without running afoul of observational bounds \citep{wantz2010}.
\par Inflation was proposed to resolve a number of issues not explained by standard Big Bang cosmology, including the observed smoothness of the universe on superhorizon scales $D>H_0^{-1}$; it also offers a mechanism for generating the \textbf{primordial density fluctuations} which seed the process of structure formation. Inflationary theory has been the subject of a spirited debate in the cosmology community over the past few years especially: it remains the dominant paradigm, but a number of prominent critics argue that it actually creates many more problems than it solves. As a non-specialist I will adopt an agnostic view of the merits of inflation, and merely summarize the generic features which are relevant for axion cosmology. 
\par Briefly, inflation posits the existence of a scalar ``inflaton'' field $\phi$ which is initially displaced from the minimum of its potential, by SSB or some other mechanism. The potential is constructed in such a way that the energy associated with this displacement dominates the energy density of the universe during the time it takes the field to return to the minimum. This potential energy density drives a period of exponential expansion, during which the Hubble parameter stays constant, and an initially causally connected volume becomes much larger than the horizon, resolving the problem of large-scale smoothness; quantum fluctuations in the $\phi$ field are also inflated to superhorizon scales and remain ``frozen'' until the normal expansion of the universe catches up with them later on. The existing radiation is dramatically diluted by inflation, so the pre-inflationary thermal state of the universe is rendered irrelevant. Eventually the $\phi$ field decays into relativistic SM particles and fields, which quickly thermalize at the \textbf{reheating temperature} $T_R\leq E_I$. At this point, the universe is once again radiation-dominated and the standard behavior discussed in Sec.~\ref{sub:expansion} applies.
\item \textbf{Electroweak symmetry breaking} ($T \sim v \sim 250$ GeV; $t \sim 10^{-11}$ s): Here we are back in familiar territory: this is precisely the mechanism discussed in Sec.~\ref{sub:ckm} for endowing the $SU(2)_W$ gauge bosons and SM fermions with mass. Before EWSB all the SM fields contribute to the radiation density; not long afterwards, $T$ falls below the mass threshold for the top quark, the Higgs, and the $W^\pm/Z_0$ bosons, which rapidly vanish from the primordial plasma. Note that while EWSB (and confinement, discussed below) are well-established elements of SM physics, the properties of the corresponding phase transitions remain the subject of ongoing research.
\item \textbf{The QCD phase transition} ($1~\text{GeV} \gtrsim T \gtrsim 100$ MeV; $ 10^{-6}~\text{s} \lesssim t \lesssim 10^{-4}$ s): I have associated this era with a rather broad temperature range around the QCD scale $\Lambda_\text{QCD}\sim200$ MeV. Below about 1 GeV the QCD coupling constant gets strong, and the topologically nontrivial field configurations discussed in Sec.~\ref{sub:instantons} begin to contribute meaningfully to the path integral, so the theory becomes difficult to treat analytically. As we will see in Sec.~\ref{sub:misalignment}, the details of how the topologically nontrivial contributions ``turn on'' with decreasing temperature affect the efficiency of dark matter axion production. \par Around $T\sim\Lambda_\text{QCD}$, the universe undergoes a phase transition associated with color confinement and possibly another phase transition associated with chiral symmetry breaking (see Sec.~\ref{sub:hadrons}). There is a sharp drop in the number of relativistic degrees of freedom (Fig.~\ref{fig:dof}), as the gluons and light quarks no longer contribute, and almost all the hadrons are too heavy to exist in significant numbers by $T\sim 100$ MeV. In a perfectly $CP$-symmetric universe, protons would be among the particles whose abundance drops rapidly to zero via particle-antiparticle annihilations, resulting in a universe without matter! Evidently, the universe has a net excess of baryons over antibaryons, resulting in a very small concentration of protons and neutrons (relative to the photon number density) which escape annihilation. This baryon asymmetry is not explained by the SM; it is generally attributed to new physics between the electroweak scale and the GUT scale, and it is beyond the scope of our discussion here.
\item \textbf{Big Bang nucleosynthesis} ($1~\text{MeV} \gtrsim T \gtrsim 100$ keV; $1~\text{s} \lesssim t \lesssim 3$ minutes): We are now entering an epoch from which direct observational evidence survives to the present day. During this three-minute window, the temperature and density are just right to enable the production of light nuclei via the fusion of protons and neutrons produced at $t\sim10^{-5}$ s. By the end of the first three minutes, essentially all the surviving neutrons are ``cooked'' into helium nuclei, though ``raw'' protons remain the most abundant baryons. Several other processes happen around this time, including the decoupling of neutrinos from the thermal bath and electron/positron annihilation.\footnote{By the overall charge neutrality of the universe, a number density of electrons equal to that of the protons must survive this process.} See Ref.~\citep{KT1994} for a lucid discussion of this extraordinarily rich subject.
\item \textbf{Matter-radiation equality} ($R\sim3\times10^{-4}$, $T \sim 1$ eV; $t \approx 5\times 10^4$ years): Not a lot happens over the next 50,000 years, except that the universe keeps expanding and the ratio $\rho_r/\rho_m$ decreases as $R^{-1}$. Eventually, at a time $t_\text{eq}$ defined implicitly by $R_\text{eq}(t_\text{eq})=\Omega_r/\Omega_m$, $\rho_m=\rho_r$, and thereafter the universe is matter-dominated.\footnote{The temperature may be obtained from the radiation density at $t_\text{eq}$. While Eq.~\eqref{eq:H_T_rad} still yields an estimate of $t_\text{eq}$ with the right order of magnitude, explicit integration of Eq.~\eqref{eq:friedmann} from 0 to $R_\text{eq}$ results in a more accurate value.} The matter-dominated universe expands more slowly [c.f.\ Eqs.~\eqref{eq:rad_dom} and \eqref{eq:matter_dom}], so gravitational structure formation becomes possible. This is not quite the whole story, as we will see in Sec.~\ref{sub:cmb}.
\item \textbf{Photon decoupling} ($R\sim10^{-3}$, $T \sim 0.3~\text{eV} \sim 3000$ K; $t \approx 3.8\times 10^5$ years): Between $t=3$ minutes and $t=380,000$ years, the protons, electrons, and photons remaining in thermal equilibrium constitute a tightly coupled plasma which cools as the universe expands. Eventually the plasma cools to the point at which electrons and protons can form stable neutral atoms, and the matter decouples from the radiation.\footnote{You may be wondering why this doesn't happen at the hydrogen reionization energy $T\sim13.6$ eV. The reason is that the equation which determines the cosmic ionization fraction contains a dimensionless prefactor which depends on the (very small) baryon-to-photon ratio of the universe. Sometimes natural units can lead us astray!} This process is usually called \textbf{recombination}, although the electrons and protons are actually forming bound states for the first time. The last photons emitted by the plasma encode a remarkable amount of information about the state of the universe at recombination; we observe them today as the \textbf{cosmic microwave background (CMB)}. Measurements of the CMB (discussed in Sec.~\ref{sub:cmb}) have resulted in the most precise constraints on many cosmological parameters, including the cosmic dark matter density.
\item \textbf{Present day} ($R_0=1$, $T_0 = 2.725$ K; $t_0 \approx 13.82\times 10^{9}$ years): I have obviously elided a lot in jumping directly to the present, including the gravitational collapse and ignition of the first stars and the formation of galaxies and galaxy clusters. The seeds of structure formation, however, are already in place well before recombination. The universe after recombination is no longer in thermal equilibrium (as our very existence demonstrates). Nonetheless we can still meaningfully define $T_0$ to be the temperature of the CMB, since CMB photons are by far the most abundant particles in the universe, and they have retained a thermal spectrum, merely cooling as $R^{-1}$ as the universe expands.\footnote{In fact that CMB is so pervasive that it was first discovered accidentally, as a persistent source of noise in a radio receiver designed for a different purpose. The advent of digital television has unfortunately made my epigraph somewhat obsolete, which just serves to illustrate that they cancel all the best shows.}
\end{itemize}

\section{Evidence for dark matter}\label{sec:evidence}
There is an abundance of observational evidence supporting the later stages of the story presented above, from $t\sim1$ s to the present. These observations indicate that the matter which begins to dominate the universe around $t_\text{eq}\sim50,000$ years is mostly non-luminous ``dark matter'' whose microscopic constituents are not known SM particles and appear to interact primarily through gravitation. In this section I will discuss a few key pieces of evidence for dark matter, related to the dynamics of galaxies (Sec.~\ref{sub:rotation}) and fluctuations in the CMB (Sec.~\ref{sub:cmb}). I chose these particular topics because they are pedagogically instructive and allow me to introduce concepts which will be relevant later in the thesis. I have omitted many other compelling pieces of evidence, from galaxy cluster dynamics, Big Bang nucleosynthesis, gravitational lensing, and baryon acoustic oscillations in the distribution of galaxies. See Ref.~\citep{bartelmann2010} for a more comprehensive overview.

\subsection{Galactic dynamics}\label{sub:rotation}
Some of the earliest and most direct evidence for dark matter comes from the observations of the motion of ordinary matter in galactic gravitational potentials. This story has been told many times (see in particular Ref.~\citep{bertone2016}); I have elected to include it primarily because we are ourselves bound to a clump of ordinary matter moving in a galactic gravitational potential, and it will be important to know how dark matter behaves on these scales if we hope to detect its interactions in the lab. Here I just summarize the general properties of galactic dark matter halos; the specific case of the Milky Way is discussed in Sec.~\ref{sec:local_dm}.

Briefly, given any spherically symmetric mass distribution, a test particle (such as a star or a cloud of hydrogen gas) in a circular orbit of radius $r$ has velocity
\begin{equation}\label{eq:rot_curve}
v_c(r) = \left(\frac{GM(r)}{r}\right)^{1/2}
\end{equation}
where $M(r)$ is the total mass inwards of radius $r$. Naively, we might expect that $M(r)$ asymptotes to a constant $M$ (essentially the total mass of all the stars in the galaxy), and thus $v(r)\propto r^{-1/2}$, for sufficiently large $r$. We can test this hypothesis by measuring the \textbf{rotation curves} of galaxies whose distance we can reliably calibrate.\footnote{See Ref.~\citep{bartelmann2010} for a lucid discussion of how astrophysical distances are calibrated.} The rotational velocity is relatively easy to measure via Doppler shifts of characteristic spectral lines. In practice, stellar spectra are most useful at small $r$ and the 21 cm emission line of neutral hydrogen is more useful at large $r$ (where there are not many stars). 
\begin{figure}[h]
\centering\includegraphics[width=0.65\textwidth]{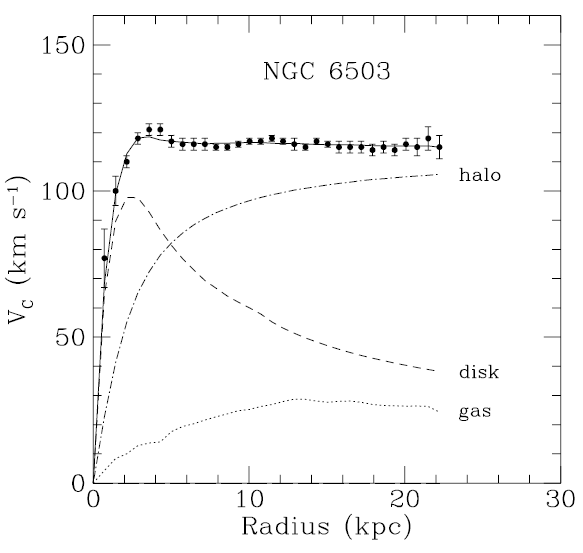}
\caption[A typical galactic rotation curve]{\label{fig:rotation} A typical galactic rotation curve measured using the 21 cm line of neutral hydrogen. The points with error bars constitute the measured rotation curve, and the dotted and dashed lines represent respectively the contributions to the rotation curve from independent measurements of the mass density in gas and stars. The dot-dashed line labeled ``halo'' represents the contribution from additional unseen mass that must be present to explain the observed rotation curve. Figure from Ref.~\citep{bertone2005}.}
\end{figure}

A typical galactic rotation curve is shown in Fig.~\ref{fig:rotation}. Apparently our intuition was totally wrong: the rotation curve is flat as far out as we can measure. Constant $v_c$ implies that $M(r)\propto r$, so we can conclude that there must be another contribution to the mass density which is relatively dilute but extends out far beyond the visible extent of the galaxy. This is what we call dark matter -- evidently it emits electromagnetic radiation at neither optical nor microwave frequencies. Studies of galactic dynamics more generally invariably indicate the presence of a substantial dark matter component. A recent study of the ultra-diffuse galaxy Dragonfly 44 is a good case in point \citep{vandokkum2016}: measurements of the velocity dispersion indicate that its mass is comparable to that of the Milky Way, with only 1\% the luminosity!

\subsection{The cosmic microwave background}\label{sub:cmb}
The dynamical measurements discussed above just tell us that there is mass we cannot see. In and of itself this is not so surprising: no law of physics requires that most of the protons in the universe had to find their way into stars; indeed, measurements on the scale of galaxy clusters tell us that there is ten times as much mass in hot ionized gas \textit{between} galaxies as there is stellar mass in galaxies \citep{bartelmann2010}. Of course, hot ionized hydrogen emits x-rays, and neutral hydrogen clouds emit 21 cm radio waves. The fact we cannot ``see'' dark matter in any part of the electromagnetic spectrum may be disturbing, but we might still hope to explain it by positing the existence of cold clumps of normal matter whose thermal emissions are masked by background radiation.

Remarkably, it turns out that the standard Big Bang cosmology gives us several observational handles on the \textbf{baryon density} $\Omega_b$ of the present-day universe as well as the net matter density $\Omega_m$: comparing $\Omega_b$ to $\Omega_m$ reveals that the universe must contain matter which is not composed of protons and neutrons!\footnote{In principle, $\Omega_b$ also includes a contribution from electrons, despite the fact that they are not baryons in the particle physics sense. However, since the total number of electrons must be the same as the total number of protons and $m_e \approx m_p/2000$, this contribution is negligible. Note that ``baryonic matter'' is \textit{not} a synonym for ``matter described by the 	SM.'' In addition to the proton, electron, and neutron, the SM contains three varieties of apparently stable neutrinos. The arguments presented here do not rule out neutrino dark matter, but we will see in Sec.~\ref{sub:neutrinos} that this possibility is excluded for other reasons.} The most precise measurements of both $\Omega_b$ and $\Omega_m$ come from \textit{anisotropies} in the CMB: slight variations in $T_0$ as a function of direction on the sky. To qualitatively understand the relevance of CMB anisotropies, we should first emphasize just how isotropic the CMB actually is: across the whole sky the fractional deviations from $T_0$ are $\mathcal{O}(10^{-5})$. This uniformity (together with the near-perfect blackbody spectrum of the CMB) is incontrovertible evidence for the claim that the universe was once in a state of thermal equilibrium, as we assumed in Sec.~\ref{sub:expansion}. 

But the fact that the CMB is so very uniform is itself puzzling. The question of how $T_0$ can be so uniform on superhorizon scales motivated the development of inflationary theory (see Sec.~\ref{sub:expansion}). Here I will focus on the uniformity on smaller scales which were in causal contact in the early universe: I will present an argument originally due to Peebles~\citep{peebles1982} (see also Ref.~\citep{bartelmann2010}) that the absence of larger fluctuations is itself evidence of non-baryonic dark matter. Quite simply, the extreme uniformity on small scales is puzzling because the present-day universe is inhomogeneous. Galaxies and galaxy clusters must have grown via gravitational interactions from much smaller density fluctuations. In full generality structure growth is a complicated nonlinear process, but (like so many problems in physics) it can be made tractable by linearization around the mean density, provided the fractional density contrast $\delta = \delta\rho/\bar{\rho}<1$. In the linear regime, fractional overdensities just grow linearly with the scale factor $R(t)$. For simplicity, we can restrict our focus to structures on large scales that are just starting to ``go nonlinear'' now -- the problem we will encounter is even more severe for smaller structures (e.g., galaxies) which have already experienced significant nonlinear evolution.

Thus let us consider a comoving scale $k$ for which $\delta_0(k)\approx1$; in practice $k$ corresponds to a physical length $\sim20$ Mpc in the present-day universe. At the time of recombination, $\delta(k)$ was smaller by a factor of $R_\text{rec}\sim10^{-3}$. If the mass contributing to the overdensity $\delta$ comes entirely from baryons, we should expect CMB temperature anisotropies $\delta T/T_0\sim \delta$ on the angular scale corresponding to $k$, because the photons and baryons are tightly coupled prior to recombination.\footnote{There are no \textit{stable} neutral baryons in the SM. Free neutrons have a $\beta$-decay lifetime of about 15 minutes, and in fact all neutrons produced in the big bang are bound up in positively charged nuclei.} The non-observation of such anisotropies suggests that the majority of the matter in the universe does not interact electromagnetically!

We can also approach this argument from the other direction. At early times, structure growth is inhibited due to the rapid expansion of the radiation-dominated universe. We saw in Sec.~\ref{sub:history} that the universe becomes matter-dominated around $t_\text{eq}\sim50,000$ years, but \textit{baryonic} structure growth is still inhibited by the pressure of the primordial plasma. In a universe containing only baryonic matter, structure growth can only proceed freely after decoupling, and there is simply not enough time for the structure we observe today to evolve. But density perturbations in dark matter can start growing as soon as the universe becomes matter-dominated; after decoupling, baryonic density perturbations can easily ``catch up'' by falling into the gravitational potential wells created by dark matter.

\begin{figure}[h]
\centering\includegraphics[width=0.9\textwidth]{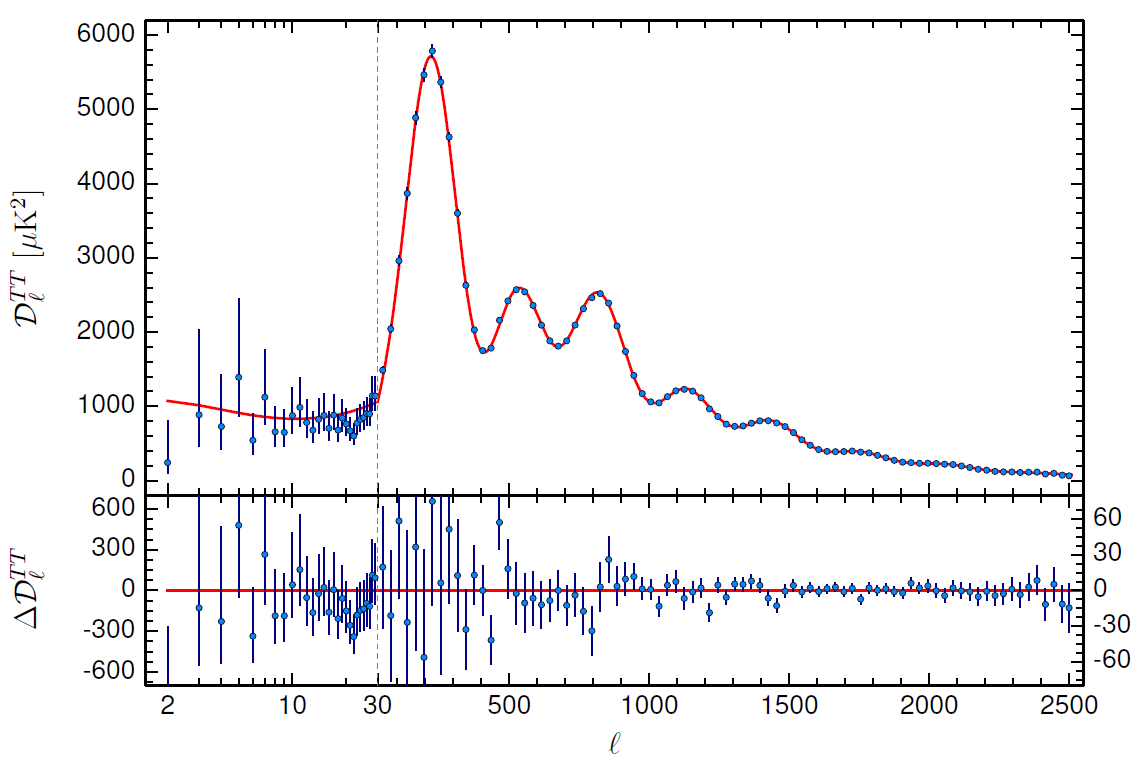}
\caption[The power spectrum of CMB temperature anisotropies]{\label{fig:planck} In the upper panel, the angular power spectrum of CMB temperature anisotropies measured by the Planck satellite, along with the best fit to a $\Lambda$CDM cosmology. The lower panel shows the residuals. Figure from Ref.~\citep{planck2016}.}
\end{figure}

The effects of dark matter on CMB anisotropies can be quantified much more precisely. In practice, $\Omega_b$ and $\Omega_m$ (and other cosmological parameters) are measured by modeling the angular power spectrum of CMB temperature anisotropies, and varying the parameters of the $\Lambda$CDM model to obtain the best fit to data. The agreement (illustrated in Fig.~\ref{fig:planck}) is extremely good, and the best contemporary measurements yield $\Omega_b~=~0.0486~\pm~0.0010$ and $\Omega_m~=~0.3089 \pm 0.0062$~\citep{planck2016} -- this is $40\sigma$ quantitative evidence for non-baryonic dark matter! Wayne Hu's cosmological parameter animations~\citep{waynehu} are an excellent resource for developing an intuitive sense of how dramatically variation in the cosmological parameters would change the observed power spectrum. For further discussion of the rich physics of the CMB, see Ref.~\citep{samtleben2007}.

\section{Dark matter particle candidates}\label{sec:dm_candidates}
We have seen compelling evidence that the universe contains a substantial amount of non-baryonic dark matter which does not interact electromagnetically. Of course we have not proved that the same stuff responsible for $\Omega_\text{DM} = \Omega_m-\Omega_b=0.26$ also explains the rotation curves of galaxies, but this is a natural assumption: lack of electromagnetic interactions would explain why galactic dark matter is non-luminous, and all that mass has to be somewhere! Studies of the distribution of dark matter suggest that not only does it not interact electromagnetically, it does not have any strong non-gravitational self-interactions. Roughly speaking, this is why dark matter appears to be so much more dilute than normal matter -- it lacks any efficient channel through which it can dissipate energy in order to condense further.

So what is dark matter made of? To account for the phenomena described thus far, any dark matter candidate must have nonzero mass, be stable on cosmological timescales, and interact very weakly. \textit{A priori}, dark matter doesn't have to be composed of just one kind of particle, but Occam's razor dictates that we consider the simplest possibility first, and fortunately, there are plausible scenarios resulting from simple extensions of the SM in which the energy density of a single new field contributes substantially to the critical density $\rho_c$. 

In this section I will consider a few of the most prominent particle dark matter candidates; a number of others are discussed in Ref.~\citep{bertone2005}. I will begin by explaining in Sec.~\ref{sub:neutrinos} how we know that dark matter is not made of neutrinos; this example will help us refine the list of qualities that a successful dark matter candidate must possess. In Sec.~\ref{sub:wimps} I briefly discuss \textbf{weakly interacting massive particles (WIMPs)}, which have been the focus of most experimental searches for dark matter particles to date. Finally, in Sec.~\ref{sub:wisps} I will note that invisible axions have all the right properties to explain dark matter, and briefly mention several other similar dark matter candidates.

\subsection{Neutrinos and hot dark matter}\label{sub:neutrinos}
Neutrinos are stable, neutral, weakly-interacting particles within the SM, and the observation of oscillations among the three neutrino species indicates that they must have nonzero mass. Thus they may seem like ideal dark matter candidates, especially given that they are already known to exist. Nonetheless, neutrino dark matter is problematic for several reasons. Here I will discuss two reasons neutrinos cannot account for the observed dark matter, both of which depend in a critical way on the smallness of the neutrino masses.

To begin, we may note that dark matter candidates may be divided into two broad categories, based on whether they were relativistic or non-relativistic at the earliest times relevant for structure formation. Though we saw in Sec.~\ref{sub:cmb} that structure formation really picks up after the universe becomes matter-dominated at $t_\text{eq}\sim5\times10^4$ years, it turns out that the behavior at earlier times is nonetheless important \citep{blumenthal1984}. Using Eq.~\eqref{eq:friedmann} with the present value of $\Omega_r$ it is easy to show that the scale factor $R(t_g)\sim 2\times10^{-6}$ when the universe is $t_g\sim3$ years old. The total mass associated with the present-day matter density contained within the horizon volume at this time is
\begin{equation}\label{eq:m_gal}
M \sim \frac{4}{3}\pi t_g^3\Omega_m\rho_cR(t_g)^3 \sim 10^{12}\ M_\odot,
\end{equation}
which is a typical galactic mass scale ($M_\odot$ being the mass of the sun). From Eq.~\eqref{eq:H_T_rad}, the temperature of the universe at time $t_g$ was $T \sim 500$ eV. Although neutrinos decouple from the rest of the matter and radiation around $t\sim1$ s, they retain their thermal distribution thereafter with $T_\nu\sim R^{-1}$ always slightly below the photon temperature $T$.\footnote{Because $e^+/e^-$ annihilation at $T\sim500$ keV heats the photons but not the neutrinos; see discussion around Fig.~\ref{fig:dof}.} Since $m_\nu \ll 500$ eV,\footnote{There are many subtleties involved in defining neutrino masses which are not important here. I will use $m_\nu$ to denote a ``typical'' neutrino mass.} neutrinos were still relativistic when galactic masses first came within the horizon at $t_g$. For this reason they constitute \textbf{hot dark matter (HDM)}. If the dark matter is already non-relativistic at time $t_g$, it is instead called \textbf{cold dark matter (CDM)}.\footnote{Thermally produced particles with $\sim\text{keV}$ mass fall between these limiting regimes, and thus can constitute \textbf{warm dark matter}, which I will not discuss. Note that dark matter neutrinos would not be relativistic today; ``hot'' only refers to their behavior in the early stages of structure formation.}

HDM and CDM lead to strikingly different patterns of structure formation. Roughly speaking, this is because HDM is flying all over the place and thus tends to smooth out primordial density fluctuations on galactic mass scales shortly after they enter the horizon: this is called ``collisionless damping.'' In the CDM case, these density fluctuations can't grow appreciably until the universe becomes matter-dominated, but at least they are not erased. All told, HDM leads to ``top-down'' structure formation (in which galaxy clusters are the first structures to gravitationally collapse, and subsequently fragment into galaxies), and CDM leads to ``bottom-up'' structure formation (galaxies form first and later merge to form clusters). Numerical simulations reveal that the top-down scenario produces a universe quite unlike the one we find ourselves living in.\footnote{The top-down scenario associated with HDM might alternatively be called ``trickle down structure formation'' after another theory of growth that doesn't work in the real world.} This suggests that dark matter is cold, and cannot be made of neutrinos.

Of course, neutrinos do exist, so they must contribute to the total density of the universe in some way. In fact, since the cosmic neutrino number density $n_\nu$ is deterministically related to $T_\nu$ while the neutrinos remain relativistic and thereafter scales as $R^{-3}$, we can turn the above argument around: to avoid conflict with observation, $\Omega_\nu\propto n_\nu\sum_\nu m_\nu$ must be small. These days, the best constraints on the sum of the neutrino masses come from the CMB power spectrum and other cosmological observables related to the distribution of structure: the resulting limits are $\sum_\nu m_\nu < 0.23$ eV and $\Omega_\nu\lesssim0.005$~\citep{planck2016}.

Independent of the physics of early-universe neutrino production, there is a simple reason neutrinos cannot constitute the dark matter halos of galaxies specifically. We will see in Sec.~\ref{sec:local_dm} that the \textit{local} dark matter density in the Milky is roughly $\rho_\text{DM}\approx 0.45~\text{GeV/cm}^3$. Using a more conservative value $m_\nu<2$ eV from purely laboratory measurements \citep{pdg2016} to avoid implicit dependence on cosmological arguments,\footnote{Since the cosmic density $n_\nu$ is known, these laboratory limits on $m_\nu$ by themselves imply that neutrinos cannot constitute all of the dark matter.} we see that if neutrinos constitute the dark matter halo of the Milky way, their local number density must be $n_\text{DM}\gtrsim 2\times10^9~\text{cm}^{-3}$.\footnote{The fact that $n_\text{DM} \gg n_\nu$ is not necessarily a problem, since we already know that the local density $\rho_\text{DM} \gg \rho_c$: dark matter is not uniformly distributed throughout the universe.} Can we cram this many neutrinos into a galaxy? In fact we cannot, because neutrinos are fermions, and this would imply the fastest neutrinos are traveling with a Fermi velocity $v_F = \frac{\hbar}{m_\nu}(3\pi^2n_\text{DM})^{1/3}\approx 10^4$ km/s, greatly exceeding typical galactic rotational velocities ($v_c\sim100~\text{km/s}$; see Fig.~\ref{fig:rotation}) and indeed exceeding the escape velocity of the Milky Way! Often this argument is turned around, and presented as a constraint on $m_\nu$ derived from the assumption that neutrinos form galactic halos with the observed values of $\rho_\text{DM}$ and $v_c$; the resulting lower bound on $m_\nu$ is inconsistent with the known upper bound $m_\nu >2$ eV. Either way, it is clear that we need to look outside the SM to find a new dark matter candidate.

\subsection{Weakly interacting massive particles (WIMPs)}\label{sub:wimps}
I have devoted so much discussion to particles that definitely do not constitute dark matter because they provide a helpful template for understanding the properties of particles that might. It is easy to see that both of the problems identified above would disappear if neutrinos were sufficiently heavy. Since the neutrinos stubbornly insist on remaining light, we can instead hypothesize a new particle $\chi$ (generically called a WIMP) that behaves just like a neutrino but is much heavier: WIMPs would constitute CDM, and galactic halos made of WIMPs have sufficiently small number density that there is no longer a problem with Fermi degeneracy. 

In fact, for reasons I will not get into, once we are out of the HDM regime, we need to keep increasing the mass until $m_\chi\gtrsim 2$ GeV to avoid \textit{overproducing} dark matter: this is called the Lee-Weinberg bound~\citep{KT1994}. As a result, WIMPs are non-relativistic not only at $t_g$ but also when they thermally decouple. If WIMPs exist, and we assume that like neutrinos they have only weak\footnote{As in ``electroweak,'' not just generically feeble.} interactions, this decoupling happens at $T\gg 1$ MeV, because the interaction rates that keep WIMPs in thermal equilibrium depend on $m_\chi$. In fact, if we depart minimally from the neutrino template and assume that the only WIMPs that are still around in the universe today are those that escaped $\chi\bar{\chi}$ annihilation at decoupling, we can relate $m_\chi$ to the present cosmic density $\Omega_\chi$: we find that $\Omega_\chi\sim\Omega_\text{DM}$ for $m_\chi\sim100$ GeV, which is indeed a sensible mass for something that interacts via the weak force! This result is often called the ``WIMP miracle.''

Thus far, we have only considered cosmological reasons that the existence of WIMPs would be nice. It turns out there are also particle physics reasons: a popular framework for extending the SM called ``supersymmetry'' generally predicts the existence of stable particles called ``neutralinos'' that behave an awful lot like WIMPs \citep{bertone2005}. The development of supersymmetry was motivated by problems seemingly unrelated to dark matter like the unification of the forces and the hierarchy problem. The possibility of solving so many big problems in one fell swoop led to a great deal of interest in WIMPs, and there has been amazing progress in improving the sensitivity of WIMP searches over the past 30 years. There are also sociological reasons for the dominance of the ``WIMP paradigm:'' both conceptual frameworks and detector technology from the existing field of neutrino physics could be adapted to the task of WIMP detection \citep{bertone2016}.

\begin{figure}[h]
\centering\includegraphics[width=0.9\textwidth]{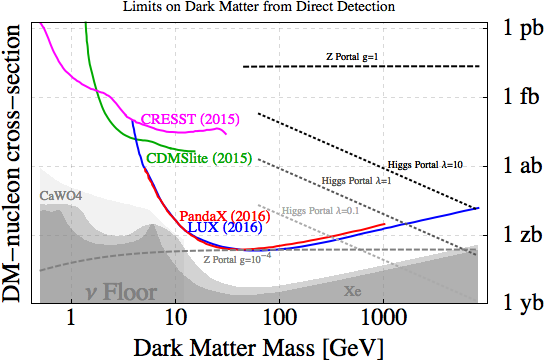}
\caption[Parameter space excluded by WIMP searches to date]{\label{fig:wimps} Parameter space for the direct detection of WIMP dark matter, with $m_\chi$ on the horizontal axis and the WIMP scattering cross section on the vertical axis. The colored curves are experimental limits; the region in the upper right corresponding to the simplest benchmark models (dashed lines) is excluded. The gray shaded region is an irreducible background due to coherent scattering of astrophysical neutrinos, which will start to affect the performance of the next generation of detectors. Figure by Adam Falkowski~\citep{resonaances}.}
\end{figure}

The only problem with WIMPs is that they do not appear to exist, at least as initially envisioned. Fig.~\ref{fig:wimps} shows the present state of WIMP parameter space. Everything above the colored curves is ruled out; in particular, the simplest supersymmetric models with $\mathcal{O}(1)$ couplings would yield WIMP scattering cross sections about seven orders of magnitude above the present experimental limits. At the same time, none of the other particles predicted by supersymmetry have been observed at the LHC, with uncomfortable implications for the whole framework. Theorists have been hard at work constructing more complicated supersymmetric models which are still viable, and of course a detection could put the matter to rest at any time. But detectors cannot get much bigger before coherent neutrino scattering (the gray shaded region in Fig.~\ref{fig:wimps}) becomes an irreducible background, and the remaining fine-tuned WIMP models become extremely difficult to test experimentally. See Ref.~\citep{bertone2010} for a relatively recent review of WIMP dark matter, which summarized the state of the field thus: ``the moment of truth has come for WIMPs: either we will discover them in the next five to ten years, or we will witness their inevitable decline.'' That was in 2010, and few would argue that the outlook for WIMPs has improved since.

\subsection{Axions, ALPs, and hidden photons}\label{sub:wisps}
At risk of oversimplifying, we have seen that light fermions cannot explain dark matter for a number of reasons, and the existence of new heavy fermions is severely constrained by null results from sensitive experiments. What about bosons? We saw in Sec.~\ref{sub:invisible} that invisible axions that arise in high-scale implementations of the PQ mechanism are electrically neutral\footnote{To restate more precisely the response to an objection anticipated in Chapter~\ref{chap:intro}, having an anomaly with $U(1)_\text{EM}$ is not the same thing as being charged under $U(1)_\text{EM}$; in particular, the axion cannot emit or absorb a single photon.} and all of their interactions are suppressed by the large energy scale $f_a$. Invisible axions are not strictly speaking stable, but they are so light that decays to almost all SM particles are kinematically forbidden, regardless of the details of the model. The lifetime of the invisible axion is always dominated by its decay to two photons, given by~\citep{moroi1998}
\begin{equation}\label{eq:axion_life}
\tau_{a\rightarrow\gamma\gamma} = \frac{64\pi^3f_a^2}{\alpha^2g_\gamma^2m_a^3} = g_\gamma^{-2}H_0^{-1}\left(\frac{26~\text{eV}}{m_a}\right)^5
\end{equation}
where $g_\gamma$ is the model-dependent coupling constant introduced in Eq.~\eqref{eq:agammagamma_general},\footnote{My convention for what is included in the coefficient I call $g_\gamma$ differs by a factor of 2 from that of Ref.~\citep{moroi1998}.} and $\alpha=e^2/4\pi$ is the fine-structure constant which we will use in place of $e$ going forward to reduce the proliferation of factors of 2 and $\pi$. In the second equality I have used Eq.~\eqref{eq:axion_mass_pion}, multiplied by $H_0/H_0$, and plugged in numbers for all factors in the numerator. We conclude that if $g_\gamma\sim1$, axions with $m_a \lesssim 26$ eV are stable on cosmological timescales. The details hardly matter, since we will see that $m_a\ll\text{eV}$ in the interesting part of parameter space.

The above arguments suggest that we should not dismiss the possibility of axion dark matter out of hand. The fact that $m_a \sim f_a^{-1}$ is very small for invisible axions presents the most obvious challenge, seeing as the demise of neutrinos as a viable dark matter candidate was ultimately a result of their low masses. The argument against neutrinos based on the available phase space in galactic halos does not apply to bosons like the axion. But the structure formation argument seems to imply that light particles like axions invariably constitute HDM, and are thus not viable.

Fortunately, this argument admits a loophole. The dark matter candidates we have considered thus far have all been \textbf{thermal relics} produced in the Big Bang which were once in equilibrium with everything else. Thermal production is a tried and true way to obtain a population of particles that survives to the present day, but it is not the only possible dark matter production mechanism.\footnote{One seeming advantage of thermal relics is that their present-day number density is in principle completely determined by their mass and the temperature at which they decouple. In practice, the example of WIMPs illustrates that this may not meaningfully constrain the parameter space, if the model does not fix the strength of the interactions that keep dark matter in thermal equilibrium in the early universe.} In Sec.~\ref{sec:axion_cosmo}, we will see that athermal axion production mechanisms necessarily arise in invisible axion models, and that the resulting axions constitute CDM even though $m_a \ll m_\nu$! 

As an aside, note the axion is not the only athermally produced light boson which could constitute CDM. An \textbf{axion-like particle (ALP)} is any hypothetical pseudo-Goldstone boson of an approximate global symmetry broken by something other than the chiral anomaly with QCD. Thus ALPs are phenomenologically similar to axions and can constitute CDM for all the same reasons, but do not solve the strong $CP$ problem. It has also been argued~\citep{nelson2011} that the ``misalignment mechanism'' for axion CDM production [to be discussed in Sec.~\ref{sub:misalignment}] can be generalized to the case of \textbf{hidden photons}, which are light vector bosons that interact with the SM only through a highly suppressed ``kinetic mixing'' with the SM photon. ALPs and hidden photons are not well-motivated in the same sense as axions and WIMPs -- they are not necessary elements of the solution to any other major extant problem in particle physics. Nonetheless, there is no good argument for the \textit{nonexistence} of such fields, and this may be reason enough to see if any parameter space can easily be probed. For further discussion of ALPs and hidden photons, see Ref.~\citep{arias2012}. \textbf{Ultra-light axions}\footnote{I ought to call such particles ``Ultra-light ALPs'' for consistency with my usage. Some authors call all light pseudoscalars ``axions,'' and call the axion which solves the strong $CP$ problem the ``QCD axion.''} have also been recently proposed as dark matter candidates: they are postulated to have masses $\sim10^{-22}$ eV, and thus de~Broglie wavelengths on kpc scales which are supposed to help with some issues with standard CDM structure formation~\citep{hui2017}. In this thesis I will restrict my focus to the hypothesis that the QCD axion constitutes dark matter.

\section{Axion Cosmology}\label{sec:axion_cosmo}
In Sec.~\ref{sub:wisps}, we saw that the invisible axion with $m_a \ll 20$ eV ($f_a \gg 3\times 10^5$ GeV) would make a good dark matter candidate if there exists a suitable non-thermal production mechanism in the early universe. In Sec.~\ref{sub:misalignment}, I will show that the PQ mechanism (often called the \textbf{misalignment mechanism} when considered as a cosmological process) is itself an axion production mechanism, and that unlike neutrinos these axions are ``born cold.''  This was first pointed out in 1983 by Preskill, Wise, and Wilczek~\citep{pww1983}, Abbott and Sikivie~\citep{as1983}, and Dine and Fischler~\citep{df1983} in back-to-back papers in the same journal. All of these papers showed that invisible axions cannot be \textit{too} invisible or the misalignment mechanism will overproduce dark matter! Of course, this also implies that if the PQ mechanism is implemented with $f_a$ near this \textbf{overclosure bound} ($f_a \sim 3\times10^{11}$~GeV, corresponding to $m_a \sim 20~\mu$eV), we can obtain $\Omega_a\sim\Omega_\text{DM}$ with the misalignment mechanism alone.

Unfortunately, it is difficult to turn the ``$\sim$'' in the above expressions into a less slippery equality. In Sec.~\ref{sub:overclosure}, I will plug in numbers to estimate the overclosure bound, and briefly discuss uncertainty in $\Omega_a$ due to non-perturbative QCD effects. Another complication will arise when we consider spatial variation in the axion field in Sec.~\ref{sub:defects}: we will see that \textbf{topological defects} are produced in the axion field during PQ symmetry breaking, and there remains considerable uncertainty in quantitative estimates of their effects on $\Omega_a$. In Sec.~\ref{sub:anthropic} we consider inflationary axion cosmology. Inflation can potentially get rid of troublesome topological defects, but introduces problems of its own. 

\subsection{Misalignment and axion CDM}\label{sub:misalignment}
Let us begin by considering the equation of motion for the axion field in the early universe:
\begin{equation}\label{eq:axion_eom}
\ddot{a} + 3H(t)\,\dot{a} + m_a^2(t)\,a = 0,
\end{equation}
where the second term is a result of applying the usual Euler-Lagrange procedure in an expanding spacetime.\footnote{The time derivative will act on $R(t)$ in the metric as well as the axion field, and the 3 just comes from the number of spatial dimensions.} Recall from Eq.~\eqref{eq:axion_mass_formal} that $m_a^2$ is formally defined as the coefficient of the leading-order term in a Taylor expansion of the potential $V_a(a/f_a)$ established by the chiral anomaly with QCD. For temperatures $T>\Lambda_\text{QCD}$, $V_a$ (and thus $m_a$) is also a function of $T$ (see discussion in Sec.~\ref{sub:instantons}), and thus a function of time.

I will use Eq.~\eqref{eq:axion_eom} to describe the evolution of the axion field over many orders of magnitude of cosmic time, from above the electroweak phase transition to the present day. Thus, it is worth taking a moment to explicitly spell out what approximations we have made. First, I have retained only the leading-order behavior of the axion potential $V_a$ (and thus implicitly assumed that $\abs{a/f_a}\lesssim1$ over the whole region of interest). This is not such an extreme constraint because $V(a/f_a)$ must be a periodic function of $a/f_a$ which is minimized at $a/f_a=0$, so values $\abs{a/f_a}\geq\pi$ are not meaningful anyway. Second, I have neglected all other axion interactions, which will be highly suppressed at all temperatures relevant to our discussion. Finally, I have assumed that the axion field is spatially homogeneous, or rather restricted my focus to the \textbf{zero mode} of the axion field (the DC term in the spatial Fourier transform of $a$). Eq.~\eqref{eq:axion_eom} is \textit{linear} as a consequence of the first approximation discussed above, so different Fourier components evolve independently. Thus, we are not actually making an independent approximation by considering only the zero mode, but merely anticipating that its behavior will be most interesting.

Lest we get bogged down in the details, I will emphasize that we have not introduced any new physics here: this is precisely the dynamical interpretation of the PQ mechanism discussed in Sec.~\ref{sub:pq_mechanism} wherein the axion field rolls down to the minimum of a potential established by the chiral anomaly with QCD. If we simplify matters for the moment by taking $H$ and $m_a$ to be constant, Eq.~\eqref{eq:axion_eom} just describes our old friend the harmonic oscillator, with a damping term due to the Hubble expansion. If $H\lesssim m_a$, the axion doesn't stop when it reaches the minimum of its potential! In retrospect this is quite obvious: the PQ mechanism solves the strong $CP$ problem by providing a way for the QCD vacuum to shed the energy associated with $\bar{\theta}\neq0$, but this energy has to go somewhere, and we have seen that it is dumped into axion field oscillations. The crucial question then becomes whether the expansion of the universe can safely dissipate the energy density associated with this oscillation, or whether it survives to the present day.

The answer to this question will of course depend on the time-dependence of $H$ and $m_a$ which we neglected above. Assuming the standard $\Lambda$CDM cosmology, $H(t)\propto1/t$ from the earliest times of interest until quite recently.\footnote{The current era of dark energy domination does not matter for our present purposes.} The time-dependence of the axion mass is much more uncertain, but for the present qualitative discussion, all we need to know is that $m_a$ increases with decreasing temperature until it attains its zero-temperature value [given by Eq.~\eqref{eq:axion_mass_pion}] for $T\lesssim\Lambda_\text{QCD}$.\footnote{This is because topological effects in QCD are suppressed at high temperatures where the gauge coupling becomes small, as noted in Sec.~\ref{sub:instantons}. Note that above the electroweak scale $v$, $m_a=0$ exactly, because the quarks are still massless, so there is no strong $CP$ problem to solve! This is an interesting bit of trivia, but it does not really matter in practice, because $v\gg\Lambda_\text{QCD}$, so $m_a$ is highly suppressed near the electroweak scale anyway.} Equivalently, $H(t)$ decreases and $m_a(t)$ increases with increasing time $t$, and there must be a time $t^*$ for which 
\begin{equation}\label{eq:t_star}
H(t^*)=m_a(t^*).
\end{equation}

The dynamics of the axion field are simple for both $t\ll t^*$ and $t\gg t^*$. At early times, the third term in Eq.~\eqref{eq:axion_eom} is negligible compared to the first two, and $a$ is frozen at its initial value. This is easy to understand conceptually if we note that the instantaneous period of axion field oscillations is $m_a^{-1}$ and the age of the universe is $t\sim H^{-1}$. $t\ll t^*$ is equivalent to $t \ll m_a^{-1}$: we are well below the characteristic timescale on which the axion feels the effects of its potential. $t^*$ may thus be interpreted as the time when the axion field starts oscillating, and the details for $t\sim t^*$ depend strongly on the behavior of $m_a(t)$. However, it is clear that at late times $t\gg t^*$, the Hubble ``friction'' becomes quite inefficient.

We can actually learn a lot by treating the $t \gg t^*$ regime more carefully. We know that $m_a(t)$ asymptotes to a constant for $T\sim\Lambda_\text{QCD}$ ($t_\text{QCD}\sim10^{-5}$ s; see Sec.~\ref{sub:history}); when precisely this happens in relation to $t^*$ of course depends on the value of $m_a$. In any event, at sufficiently late times, $m_a$ changes very slowly compared to the oscillation timescale.\footnote{Since the oscillation timescale is itself $m_a$, this condition may be formalized as $\dot{m}_a/m_a < m_a$.} In this \textit{adiabatic} regime, the solution to Eq.~\eqref{eq:axion_eom} just describes the motion of a simple harmonic oscillator whose amplitude and frequency vary slowly with time: $a(t)=a_0(t)\cos\big(m_a(t)\,t\big)$. Plugging this approximate expression for $a(t)$ back into Eq.~\eqref{eq:axion_eom}, we obtain~\citep{pww1983}
\begin{equation}\label{eq:axion_eom_adiabatic}
\frac{\mathrm{d}\big(m_aa_0^2\big)}{\mathrm{d}t}=-3H\big(m_aa_0^2\big)
\end{equation}
from the coefficient of the sine terms. Let us define $n_a=m_aa_0^2/2$ for reasons that will become clear presently. Integrating Eq.~\eqref{eq:axion_eom_adiabatic} reveals that $n_a$ scales as $R^{-3}$ as the universe expands. From the Lagrangian, the axion's potential energy density is $\frac{1}{2}m_a^2a^2$, so the total energy density\footnote{As in any harmonic oscillator, the energy sloshes back and forth between a kinetic piece and a potential piece, which are equal on average. Since we are dealing with an oscillator whose coordinate is the amplitude of a field pervading all space, the corresponding quantities have units of energy density. We will soon see that we can associate the field oscillations with a population of axion particles. The kinetic energy density of the field oscillations should \textit{not} be confused with the particle kinetic energy, which is related to spatial variation in the field.} in axion field oscillations is just $\rho_a = m_an_a$. For $t\gg t^*$ and $t\gtrsim t_\text{QCD}$, $m_a$ is constant, and this looks just like the energy of a gas of \textit{non-relativistic} particles, provided $n_a$ can be interpreted as a number density. Then Eq.~\eqref{eq:axion_eom_adiabatic} tells us that the number of axions per comoving volume is conserved, which is precisely what we expect for any decoupled particle species. Once we are out of the $t\sim t^*$ regime, $\rho_a \propto R^{-3}$: axions produced by misalignment contribute to $\rho_m$ rather than $\rho_r$ even though $T\gg m_a$! These axions are thermally decoupled, and therefore remain cold at $t\sim t_g$: they constitute CDM.

Since we managed to coax the key result $\rho_a\propto R^{-3}$ out of Eq.~\eqref{eq:axion_eom} only after a long discussion, it may be pedagogically helpful to consider another less formal perspective on why precisely axions from misalignment are so cold. Axion CDM is most simply understood as a consequence of the \textit{nonzero amplitude} of the zero mode (the average value of the axion field in the universe). Roughly speaking, particles may be regarded as ``ripples'' in fields with energy given by the oscillation frequency and momentum given by the spatial wavenumber. From the moment we wrote down Eq.~\eqref{eq:axion_eom} we have only been considering the zero mode, so it is not surprising that the resulting behavior is that of zero-momentum particles!\footnote{This is oversimplifying things slightly too much. If the particles are massless, then of course they must remain relativistic even as the wavelength $\rightarrow\infty$. Ref.~\citep{turner1983} treats the case of spatially coherent scalar field oscillations very with a general framework which can accommodate both axion CDM and inflation, and finds that the resulting particles are indeed relativistic if the leading term in the potential is $\propto a^4$.} For most SM fields $\psi$, the zero mode has zero amplitude because $\psi=0$ at the minimum of the classical potential.\footnote{Radial degrees of freedom like the SM Higgs boson are an exception, but the energy stored in the mass $m$ of such particles can be safely transfered to other particles in the thermal bath since $T$ drops below $m$ before they decouple.} The crucial feature that distinguishes the axion is that its initial value in the interval $[-\pi f_a,\pi f_a]$ is chosen randomly by \textit{spontaneous} symmetry breaking at a very early time, and only much later, after axions have totally decoupled from the thermal bath, does the \textit{explicit} symmetry breaking responsible for axion mass become effective. Thus there is no reason the initial amplitude of the zero mode should happen to be $a=0$. 

\subsection{The overclosure bound}\label{sub:overclosure}
We have seen that the misalignment mechanism always produces CDM axions, but we have yet to actually calculate the present-day CDM axion density parameter $\Omega_a$ as a function of $m_a$ (equivalently, as a function of $f_a$). We will find that $\Omega_a$ increases with \textit{decreasing} $m_a$, implying that sufficiently large $f_a$ is cosmologically untenable.\footnote{In addition to the misalignment mechanism, there will generally also be thermal production of axions in the early universe unless $f_a$ is \textit{very} large~\citep{sikivie2008}. These thermal relic axions (which may be thought of as fluctuations around the average value of the axion field) decouple in the very early universe and thereafter behave cosmologically more or less like neutrinos; in particular, the thermal axion density increases with \textit{increasing} $m_a$. Thermal axions constitute HDM, but in the viable regions of parameter space, the axion mass is so small that they contribute negligibly to the critical density~\citep{davis1986}. As in the neutrino case, the non-observation of signatures of HDM in the CMB implies a bound $m_a\lesssim0.5$~eV~\citep{divalentino2016}. However, this cosmological bound is not competitive with the astrophysical bounds to be discussed in Sec.~\ref{sec:parameter}.} The present-day density in cold axions may be calculated as
\begin{align}
\Omega_a &= \frac{1}{\rho_c}m_an_a \nonumber \\
&\approx \frac{1}{\rho_c}m_a\zeta n_a^*\big(R(t^*)\big)^3 \nonumber \\
\Rightarrow \Omega_a &\approx \frac{1}{\rho_c}\Big(\theta_i^2/2\Big)\Big(\zeta\sqrt{\chi(0)\chi(T^*)}\Big)\big(R(t^*)\big)^3. \label{eq:omega_a}
\end{align}
In the first line $m_a$ and $n_a$ are the present-day axion mass and number density. In the second line I have used the conservation of axion number per comoving volume (with $R_0=1$). Since we only expect $n_a\sim R^{-3}$ for $t\gg t^*$, I have introduced a dimensionless factor $\zeta$ with which to absorb all the complicated dynamics I ignored by extrapolating this scaling back to $t\sim t^*$. In the third line I have used $n_a^*=m_a(t^*)a^2_0(t^*)/2$ and introduced some new notation: $\chi(T)= m_a^2(T)f_a^2$ is the \textbf{topological susceptibility}\footnote{Formally, $\chi(T)$ is the second derivative of the QCD free energy with respect to $\theta_\text{QCD}$, evaluated at the minimum. This is just another way of expressing Eq.~\eqref{eq:axion_mass_formal}; the RHS of Eq.~\eqref{eq:axion_mass_pion} is thus $\chi(0)/f_a^2$.} and $\theta_i=a_0(t\ll t^*)/f_a$ is the initial misalignment angle. 

In short, the three sets of parentheses in Eq.~\eqref{eq:omega_a} encode respectively the effects of initial conditions, QCD dynamics, and the expansion of the universe on $\Omega_a$. For the moment we will just take $\theta_0\sim1$, and return to interrogate this assumption in Sec.~\ref{sub:defects} and~\ref{sub:anthropic}. Then it is clear that there are two distinct effects through which $\Omega_a$ can depend on $m_a$: complicated dynamics governed by precisely how the axion mass turns on around the QCD phase transition, and dilution since the axion field started oscillating at $t\sim t^*$. The latter effect is simple if we recall that $R(t)$ always increases with time and $t^*\sim m_a^{-1}(t^*)$ [Eq.~\eqref{eq:t_star}]. Thus lighter axions start oscillating later, and the net Hubble dilution from $t^*$ to $t_0$ is reduced. Without any specific assumptions about non-perturbative QCD dynamics, we have arrived at the somewhat counterintuitive result that $\Omega_a$ increases with decreasing $m_a$.
 
To obtain a specific estimate of the value of $f_a$ for which $\Omega_a\sim\Omega_\text{DM}$, we must confront the non-perturbative QCD dynamics we have thus far managed to ignore. The original papers to treat this subject \citep{pww1983,as1983,df1983} used the \textbf{dilute instanton gas approximation (DIGA)}, which ignores higher-$\nu$ field configurations and also neglects interactions between instantons: DIGA predicts $\chi(T)\sim T^{-b}$, with $b\approx 8$~\citep{gross1981}.\footnote{DIGA also predicts a particular form for the $\theta$-dependence of the QCD free energy, and thus $V_a\approx \chi(T)\big[1-\cos(a/f_a)\big]$. Since we already knew that $V_a$ was a periodic function of $a/f_a$, this is basically the simplest behavior it could have. Note that there is a slight tension between our assumption that $\theta_i\sim1$ and our decision to retain only the leading-order behavior of $V_a$. Corrections for anharmonic behavior can be absorbed into $\zeta$.} This approximation yields
\begin{equation}\label{eq:overclosure}
\Omega_a \sim \left(\frac{f_a}{10^{12}~\text{GeV}}\right)^{7/6},
\end{equation}
implying that for an axion with $f_a\sim10^{12}$ GeV ($m_a\sim6\ \mu$eV), the misalignment mechanism provides the closure density all by itself. These days, we know that there is stuff in the universe besides dark matter, but in 1983 $\Omega_m$ and $\Omega_\text{tot}$ were not yet precisely measured and $\Omega_\Lambda$ had not been measured at all: it was common to assume that $\Omega_\text{DM}\approx\Omega_\text{tot}\approx1$, because it was already difficult to reconcile estimates of $\Omega_b$ from big bang nucleosynthesis with measurements of $\Omega_m$. From a contemporary point of view, Eq.~\eqref{eq:overclosure} tells us that $\Omega_a\sim\Omega_\text{DM}$ for $f_a\sim3\times10^{11}$ GeV ($m_a\sim20\ \mu$eV). For more detailed pedagogical summaries of this derivation, see Refs.~\citep{turner1986,cheng1988}.

In any event, the bound implied by Eq.~\eqref{eq:overclosure} is several orders of magnitude below $f_a\sim M_\text{GUT}$, which was widely regarded as the natural scale for new physics beyond the SM.\footnote{It is interesting to note that while all three original papers on the misalignment mechanism arrived at similar quantitative results, they interpreted their results differently. Refs.~\citep{pww1983} and~\citep{as1983} entertained the possibility that $f_a\sim10^{12}$ GeV and axions play a significant role in cosmology, whereas Ref.~\citep{df1983} emphasized the cosmological disaster that would result from a GUT-scale PQ mechanism, implying that this strongly disfavored the PQ solution to the strong $CP$ problem.} Thus we should ask how robust this overclosure bound is to changes in the assumptions that went into deriving it. Our treatment of non-perturbative QCD clearly deserves more scrutiny. DIGA is expected to be valid at high temperatures where vacuum-to-vacuum tunneling is suppressed, but must break down as we approach $T\sim\Lambda_\text{QCD}$ and large-$\rho$ instantons contribute significantly to the path integral: a gas of arbitrarily large particles clearly cannot be dilute! Modern approaches have used more complex analytic approximations~\citep{wantz2010} and attempted to evaluate $\chi(T)$ numerically using the techniques of lattice QCD, but these calculations are computationally demanding and great care must be taken to avoid numerical artifacts. As of this writing, the most recent lattice calculations obtain results for the misalignment contribution to $\Omega_a$ that differ by a factor of 5~\citep{borsanyi2016,bonati2016}.

Moreover, the overclosure bound can be dramatically changed by relaxing assumptions of the standard $\Lambda$CDM cosmology. While this may seem rather unpalatable given how well $\Lambda$CDM seems to work, it remains a viable possibility because our earliest direct evidence for a radiation-dominated universe comes from the era of big bang nucleosynthesis. In particular, the out-of-equilibrium decay of a new heavy particle between the QCD phase transition and nucleosynthesis would cause the temperature of the universe to decrease much more slowly during this time, thus increasing the dilution of the initial misalignment energy density \citep{steinhardt1983}. Equivalently, this hypothetical particle decay increases the entropy of the universe, and $R(t^*)$ in Eq.~\eqref{eq:omega_a} is calculated by assuming adiabatic expansion from $t^*$ to the present.

\subsection{Topological defects in the axion field}\label{sub:defects}
To summarize, we have seen that the misalignment contribution to the CDM axion density $\Omega_a$ can only be reliably estimated up to a factor of 5 or so at best. Here we will look more closely at the assumption that the initial misalignment angle $\theta_i\sim1$: we will see that this assumption can indeed be justified, but compels us to consider contributions to $\Omega_a$ from topological defects associated with the spontaneous breaking of PQ symmetry. In particular, we must consider the effects of both \textbf{cosmic strings} and \textbf{domain walls}. The string contribution $\Omega_a$ turns out to be even harder to compute than the contribution from misalignment, and the domain wall contribution must be avoided at all costs to avert cosmological disaster. For a good introduction to topological defects in cosmology, see Ref.~\citep{vilenkin1985}.

The initial misalignment angle $\theta_i$ may be regarded as the sum of $\bar{\theta}$ and the random azimuthal position chosen by spontaneous PQ symmetry breaking.\footnote{Physically, these contributions to $\theta_i$ are totally indistinguishable.} Thus to treat the initial conditions more carefully, we should ask consider the correlation length for fluctuations in $\theta_i$ as a function of cosmic time. For $t\ll t^*$, domains corresponding to different values of $\theta_i$ will grow to become horizon-sized, because even though all values of $a$ are energetically equivalent sufficiently far above the QCD phase transition, there is still energy associated with gradients in the $a$ field. The axion field in each domain begins to oscillate at time $t^*$, so the present-day cosmic density $\Omega_a$ should correspond to the average over a very large number [$\sim(t_0/t^*)^3$] of causally distinct domains. Thus we can take $\theta_i=\theta_\text{rms}\approx\pi/\sqrt{3}$; this is the value used to derive Eq.~\eqref{eq:overclosure}.\footnote{A careful treatment of the misalignment mechanism would also take into account the contribution of non-zero modes with horizon-scale wavelengths, corresponding to axions with finite momentum $p\sim H\sim (T^*)^2/M_p\ll T^*$. For any reasonable value of $T^*$, $p$ is much smaller than the asymptotic value of the axion mass, so these axions are still non-relativistic. The non-zero modes contribute an $\mathcal{O}(1)$ factor to Eq.~\eqref{eq:overclosure} but do not change the qualitative result obtained using the zero mode only~\citep{sikivie2008}.}

However, if the axion field takes on different values in different causally distinct regions of space, there is nothing preventing the local value of $a$ from varying continuously from $-\pi f_a$ to the energetically equivalent value $\pi f_a$ on a sufficiently long closed loop through space: any such loop must enclose a point at which $a$ is not well-defined. This is the cross-section of a cosmic string defect, and it is easy to show that such strings must either be infinite or form closed loops.\footnote{For $a$ to be undefined we must have $\langle\sigma\rangle=0$ (not the energetically favored value $\langle\sigma\rangle=f_a$) at this point. In practice the string core has a finite radius of order $f_a^{-1}$ determined by the tradeoff between potential energy and gradient energy. There is a finite energy per unit length (tension) $\sim f_a^2$ associated with the field configuration of the string core.} Generally, strings form whenever a continuous symmetry is spontaneously broken, with a prototypical example in low-energy physics being vortices in superconductors.\footnote{Unfortunately, the literature contains references to both ``string axions'' (string theory) and ``axion strings'' (topological defects), so it is worth emphasizing that cosmic strings have nothing to do with string theory.} That cosmic strings are more important than mere ripples in the axion field (i.e., thermal axions) was first noted by Ref.~\citep{davis1986}. This has everything to do with their topological character: strings can move around and oscillate but in the absence of interactions cannot evolve into smooth field configurations. 

In fact, in the absence of interactions, a simple dimensional argument indicates that the string energy density scales like $\rho_s\sim R^{-2}$: expansion can dilute the string density but can't do anything to get rid of the string tension. Thus we might naively expect strings to dominate the energy of the universe if they exist at all, leading to a major conflict with $\Lambda$CDM cosmology. However, simulations of cosmic string dynamics indicate that interactions between strings enable them to break apart and eventually decay via radiation of field quanta: string decay constitutes an independent athermal axion production mechanism! Unfortunately, string dynamics are very complicated, and estimates of the contribution to $\Omega_a$ vary by more than a factor of 100: string decay may be anywhere from subdominant to the misalignment mechanism to significantly more important in determining $\Omega_a$. See Refs.~\citep{sikivie2008,wantz2010,windows1990} for overviews of axion string dynamics and Ref.~\citep{fleury2016} for a recent study.

We must also consider the cosmological role of domain walls, which are two-dimensional topological defects produced in the spontaneous breaking of \textit{discrete} symmetries. Generically, axion models have a $Z(2N)$ discrete symmetry, where $N$ is the number of quark flavors charged under $U(1)_\text{PQ}$. Thus far, we have been able to get away with absorbing this factor of $N$ into the definition of $f_a$ so that the formal expression for the axion mass is the same for the KSVZ and DFSZ models. But we have ignored the fact that this redefinition implies $a$ is periodic with period $2\pi f_a/N$, and thus there are $N$ energetically equivalent (but topologically inequivalent!) $CP$-conserving minima in the range $[-\pi f_a,\pi f_a]$.\footnote{Roughly speaking, this is because the effects of chiral rotations on $\bar{\theta}$ are enhanced by a factor of $N$, so physics is invariant under a phase rotation by an angle $\alpha=2\pi m/N$ for any integer $m$.} Causally disconnected domains generally start oscillating around different minima, which are separated by domain walls \citep{sikivie1982}.

The problem with domain walls is basically the problem we narrowly avoided with strings, made even worse by the fact that they are 2D: the surface energy (which like the string tension is determined by the field configuration which minimizes the sum of potential and gradient terms) is unaffected by dilution, so $\rho_w\sim R^{-1}$. Moreover there is no efficient mechanism for the decay of topologically stable walls. Thus the production of stable domain walls is simply incompatible with cosmological observations, and any model which predicts them is not viable. There are a number of ways to avoid this domain wall problem, most of which involve embedding the PQ mechanism in a more elaborate theoretical framework to either eliminate the $Z(2N)$ symmetry or introduce a continuous symmetry and thereby render the different minima topologically equivalent; see Ref.~\citep{cheng1988} for a summary.\footnote{KSVZ models (in which $N=1$) do not have a domain wall problem. There is still a discrete $Z(2)$ (reflection) symmetry which is spontaneously broken, so domain walls are generated by the PQ mechanism. However, there is only one minimum, so these walls are not topologically stable; their decay contributes negligibly to $\Omega_a$ \citep{sikivie2008}.}

We have seen that both misalignment and string decay can produce axion CDM. What happens to these cold axions between $t^*$ and the present day? Very generally structure must grow gravitationally from some initial inhomogeneity. Both primordial density fluctuations produced by inflation \citep{twz1983,axenides1983} and topological defects \citep{ipser1983,stecker1983} have been studied as possible seeds of axionic structure formation; the latter scenario is now disfavored by the absence of characteristic patterns it would imprint on the CMB.\footnote{See discussion in Sec.~\ref{sub:anthropic}.} Structure formation is a complex topic which is entirely outside the scope of this thesis, but the general consensus appears to be that axions can account for the observed patterns of structure in the universe as well as any other kind of CDM.\footnote{Sikivie~\citep{sikivie2009} has proposed that axion CDM undergoes Bose-Einstein condensation on a cosmic scale, leading to observable differences in the structures formed by axions and other dark matter particles such as WIMPs, but this conclusion has been disputed~\citep{guth2015}.}

\subsection{Inflation and the anthropic axion window}\label{sub:anthropic}
How does axion cosmology change if we add inflation to the $\Lambda$CDM model? Recall from Sec.~\ref{sub:history} that inflation is a period of exponential expansion in the early universe governed by physics at an energy scale $E_I$. During inflation the Hubble parameter is $H_I\sim E_I^2/M_P$, and inflation ends with  the universe radiation-dominated at temperature $T_R\leq E_I$. For axion physics, the relevant question is the value of the reheating temperature $T_R$ relative to $f_a$. If $T_R>f_a$ (PQ symmetry breaking happens after inflation), axion cosmology proceeds as described in Sec.~\ref{sub:misalignment} -- Sec.~\ref{sub:defects}, and we can remain completely agnostic about whether inflation happened at all.

However, if inflation occurs with $T_R<f_a$ (PQ symmetry breaking happens before or during inflation), then a patch of the pre-inflationary universe containing a single value of $\theta_i$ becomes larger than the observable universe today. This dilutes away all the topological defects in the axion field, solving the domain problem if there was one to solve and implying that the only CDM axions in this scenario are produced by misalignment. The catch is that $\theta_i$ is now an additional free parameter. For $\theta_i\lesssim1$, the overclosure bound becomes \citep{sikivie2008}
\begin{equation}\label{eq:overclosure_inflation}
\Omega_a \sim 0.15\theta_i^2\left(\frac{f_a}{10^{12}~\text{GeV}}\right)^{7/6}.
\end{equation}
Evidently we can evade the overclosure bound by supposing that we happen to live in a part of the universe with very small $\theta_i$! This may seem very dubious -- after all, we introduced the axion in order to explain the apparent fine-tuning of the QCD $\theta$ angle, and now we find ourselves once again postulating an initially fine-tuned $\theta$ angle! In fairness to the proponents of the inflationary scenario, the GUT-scale axion favored by many theorists appears to require only $\sim 1\%$ fine-tuning to be viable, making it in some sense ``less unnatural'' than the value $\bar{\theta}<10^{-10}$ that we started with. Moreover, unlike the strong $CP$ problem, this inflationary fine-tuning \textit{does} admit an anthropic explanation:\footnote{Whether \textit{you} want to admit an anthropic explanation is another matter entirely.} in a world with $f_a\sim M_\text{GUT} > T_R$, patches of the universe with $\theta_i\sim1$ would have $\Omega_\text{tot}\gg1$, and thus would lead brief and tragic lives unsuitable for the evolution of conscious observers. For this reason, the region of parameter space with $f_a\gtrsim10^{12}$ GeV is often called the \textbf{anthropic axion window}.

Inflation is most often invoked to motivate large values of $f_a$ (small $m_a$), but $\Omega_a \sim \Omega_\text{DM}$ with large $m_a$ is also possible in the inflationary scenario if $\theta_i\approx\pm\pi$: then the small-angle approximation used in Eq.~\eqref{eq:overclosure_inflation} is not valid. Taking $V_a\propto[1-\cos(a/f_a)]$ as a specific example of a periodic potential, it is easy to see that this reverse fine-tuning \textit{increases} $\Omega_a$ relative to the value obtained from the harmonic potential with a given $m_a$: the axion spends some time perversely perched on top of its potential before eventually rolling down in one direction or the other, so oscillations start later than $t^*\sim m_a^{-1}$; see Ref.~\citep{wantz2010} for details. Allowing for some QCD uncertainty and fine-tuning of $\theta_i$, inflation evidently makes it possible to obtain $\Omega_a \sim \Omega_\text{DM}$ for any value of $f_a$ between $M_p$ and $6\times10^9$~GeV ($5\times 10^{-13}~\text{eV} < m_a < 1$~meV). 

Fortunately, it is possible to constrain the anthropic axion parameter space. $f_a>T_R$ implies that the axion exists as a massless field during inflation, so quantum fluctuations in the axion field are inflated to superhorizon scales, just like the quantum fluctuations of the inflaton. A key distinction is that the inflaton has decayed into all the particles of the SM and possibly others by the end of reheating, whereas the axion field does not have any interesting dynamics until QCD time. As a result, at late times relevant to structure formation, quantum fluctuations in the inflaton field produce \textbf{adiabatic perturbations} (fluctuations in $\rho_m$ \textit{correlated} with fluctuations in $\rho_r$) and quantum fluctuations in the axion field produce \textbf{isocurvature perturbations} (fluctuations in $\rho_m$ \textit{equal and opposite} to fluctuations in $\rho_r$). Basically, the fluctuations in the inflaton field get translated into radiation overdensities at a time when all SM fields are radiation, whereas fluctuations in the axion field redistribute energy from relativistic gluons to cold axions. The amplitude of the isocurvature axion fluctuations is
\begin{equation}\label{eq:isocurvature}
\delta\rho_a \sim H_I/(f_a\theta_i),
\end{equation}
and $\theta_i$ is itself deterministically related to $f_a$ if we assume the axions constitute all of dark matter. 

The reason all of this is relevant is that CMB observations favor a purely adiabatic spectrum of initial perturbations, with isocurvature perturbation amplitudes constrained to the few percent level~\citep{planck2016}.\footnote{Generally, causal microphysical processes after inflation can only ever redistribute energy, and thus correspond to isocurvature density perturbations. Structure formation seeded by topological defects (mentioned briefly in Sec.~\ref{sub:defects}) is looking increasingly unlikely for this reason.} The nonobservation of isocurvature perturbations at this level, together with Eq.~\eqref{eq:isocurvature}, allows us to cut a swath through the $f_a$ vs.\ $H_I$ parameter space; see e.g., Ref.~\citep{wantz2010}. Qualitatively, the conclusion is that the theoretically favored region $E_I\sim M_\text{GUT}$ is ruled out by isocurvature constraints if PQ symmetry breaking happens before or during inflation. Low-scale inflation models evade this bound, as do models in which PQ symmetry breaking happens after inflation or does not happen at all. As noted in Sec.~\ref{sub:history}, observations of the CMB are generally only able to put upper bounds on $E_I$; thus the region of the anthropic axion window with low $E_I$ seems less vulnerable to exclusion in the near future. Conversely, a \textit{detection} of the imprint of inflationary gravitational waves in the CMB which survives community scrutiny would measure $E_I$ and thereby single-handedly exclude the pre-inflationary PQ breaking scenario.\footnote{Several papers to this effect \citep{visinelli2014,divalentino2014} were written in response to a detection claimed by the BICEP2 experiment, which turned out to be foreground contamination.}


\chapter{Searching for axions}\label{chap:search}
\setlength\epigraphwidth{0.53\textwidth}\epigraph{\itshape This is going to be like finding a needle in the world's biggest haystack\dots Fortunately, I brought a magnet!}{Tony Stark}

\noindent In chapter~\ref{chap:cosmo} we saw that despite their exceptionally weak interactions with SM fields, invisible axions play an outsized role in cosmology.\footnote{Indeed, it is precisely because axion couplings are so weak that the energy stored in oscillations of the axion field cannot be dissipated except by Hubble dilution.} In this chapter, I will address another way the invisible axion fails to live up to its name: if axions constitute dark matter, their electromagnetic interactions may be observable in a sufficiently sensitive laboratory-scale detector.

I will begin in Sec.~\ref{sec:parameter} by reviewing the remaining viable parameter space for axions, with an emphasis on their interactions with photons. In Sec.~\ref{sec:local_dm}, I will review what is known about the local dark matter mass and velocity distributions and their implications for experimental probes of axion CDM: we will see that throughout most of the allowed mass range the axion field is coherent on laboratory scales. In Sec.~\ref{sec:haloscope} I will discuss the conceptual design for a detector sensitive to coherent effects of the axion field: this is the axion haloscope, which was introduced briefly in chapter~\ref{chap:intro}. I will go on to derive the figures of merit for a haloscope search. Finally, in Sec.~\ref{sec:searches} I will review past efforts in haloscope detection and motivate the HAYSTAC experiment whose first operation has been the focus of my graduate research. 

The discussion in the first half of this chapter will draw on material established in chapters~\ref{chap:theory} and \ref{chap:cosmo}. From Sec.~\ref{sec:haloscope} onwards, the subjects discussed in this thesis will decouple more or less completely from the details of the particle theory and cosmology. If you are willing to accept the form of the axion-photon interaction [Eq.~\eqref{eq:agammagamma}] and the proposition that axion CDM is described by a spatially homogeneous field oscillating at a frequency $\nu_a=m_a$ (Sec.~\ref{sec:local_dm}), you could skip all the fancy theory stuff and starting reading here. That said, I encourage you to give the fancy theory stuff a chance!

\section{Axion Parameter Space}\label{sec:parameter}
I will begin by reviewing the properties of invisible axions most relevant to experiment, and establishing the notation to be used in the remainder of this thesis, which will differ slightly from that of Sec.~\ref{sub:axion_mass}. The most important fundamental parameter in any invisible axion model is the PQ symmetry breaking scale $f_a$. The axion mass $m_a$ is given in terms of $f_a$ by the model-independent expression 
\begin{equation}\label{eq:axion_mass}
m_a = \frac{m_\pi f_\pi}{f_a}\frac{\sqrt{z}}{1+z} = \frac{\sqrt{\chi}}{f_a},
\end{equation}
where $z$ is the light quark mass ratio $z=m_u/m_d$ and $\chi$ is the zero-temperature QCD topological susceptibility (see Sec.~\ref{sub:overclosure}). Confinement precludes a direct measurement of $m_u$ or $m_d$ and thus $z$ must be extracted from hadronic observables with the aid of theory input; the best current data constrains $z$ to the range $0.38 < z < 0.58$ \citep{pdg2016}. Historically, axion papers have tended to use $z=0.56$ obtained from chiral perturbation theory, and I will assume this value throughout the thesis for consistency with numerical values that appear in the literature. With $z=0.56$, $\chi = [77.6~\text{MeV}]^4$.\footnote{$\chi$ may also be calculated using the methods of lattice QCD. A recent zero-temperature lattice calculation~\citep{borsanyi2016} obtained the value $\chi = [75.6~\text{MeV}]^4$, which agrees with the chiral perturbation theory expression Eq.~\eqref{eq:axion_mass} for $z=0.415$. We will see shortly that the axion-photon coupling increases with decreasing $\chi$, so in this sense $z=0.56$ is also a conservative choice.}

In any invisible axion model, the axion has a coupling to two photons of the form
\begin{equation}\label{eq:agammagamma}
\lagr_{a\gamma\gamma}=\frac{1}{4}g_{a\gamma\gamma}aF^{\mu\nu}\tilde{F}_{\mu\nu} = -g_{a\gamma\gamma}a\mathbf{E}\cdot\mathbf{B},
\end{equation}
where I have introduced the physical coupling
\begin{equation}\label{eq:photon_coupling_1}
g_{a\gamma\gamma} = \frac{\alpha g_\gamma}{\pi f_a} = \frac{\alpha g_\gamma}{\pi\sqrt{\chi}}m_a,
\end{equation}
and $\alpha$ is the fine-structure constant. Note that while the model-dependent coefficient $g_\gamma$ introduced in Eq.~\eqref{eq:agammagamma_general} is dimensionless, the physical coupling $g_{a\gamma\gamma}$ has dimension $\text{eV}^{-1}$. See appendix~\ref{app:maxwell} for the derivation of the second equality in Eq.~\eqref{eq:agammagamma}, which will be our starting point for the conceptual design of a haloscope detector.\footnote{This derivation really has nothing to do with axion physics specifically, but may be helpful for readers not used to working with the covariant formulation of electrodynamics.}

In the second equality in Eq.~\eqref{eq:photon_coupling_1} I have exchanged the more fundamental parameter $f_a$ for the axion mass $m_a$. Until now I have worked sometimes in terms of $f_a$ and sometimes in terms of $m_a$, depending on what seemed more appropriate for the problem at hand. For the purposes of axion phenomenology, the most relevant parameters are $m_a$ and $g_{a\gamma\gamma}$.

\begin{figure}[h]
\centering\includegraphics[width=1.0\textwidth]{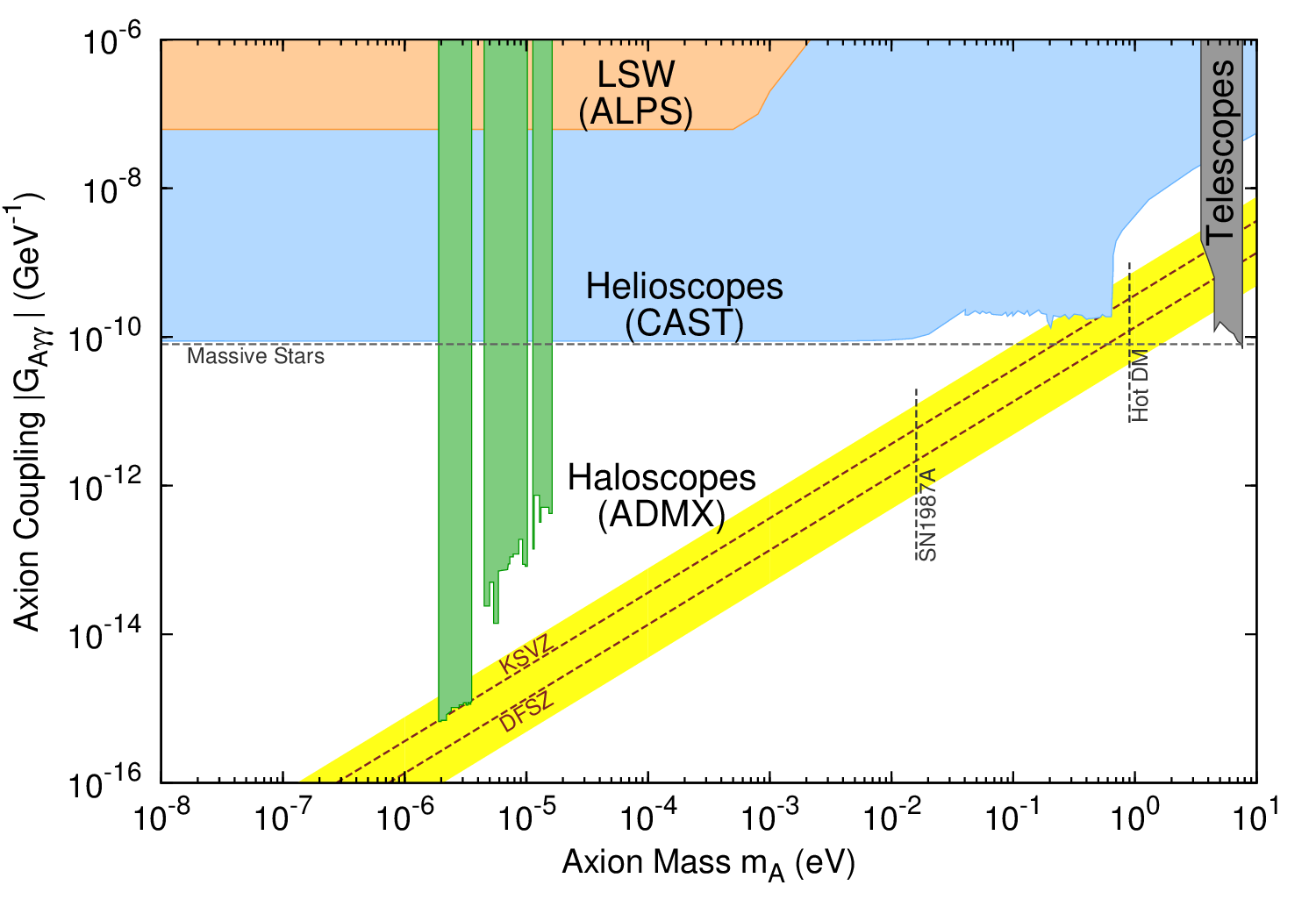}
\caption[Axion parameter space circa 2014]{\label{fig:paramspace_big} Parameter space for generic axion-like particles (ALPs); axions that solve the strong $CP$ problem are constrained to the yellow diagonal band with $g_{a\gamma\gamma}\propto m_a$. Other solid colored regions represent direct experimental exclusion limits, and dashed horizontal and vertical lines represent indirect astrophysical bounds. Note that some bounds have changed slightly since this plot was published (in Ref.~\citep{pdg2014}) in 2014.}
\end{figure}

More specifically, Eq.~\eqref{eq:photon_coupling_1} tells us that the requirement that axions solve the strong $CP$ problem implies $g_{a\gamma\gamma}\propto m_a$, where the precise value of the proportionality constant depends on the model-dependent coefficient $g_\gamma$. We can see this in the plot of axion parameter space shown in Fig.~\ref{fig:paramspace_big}: the range of possible values for $g_\gamma$ defines a \textbf{model band}, which is shown in yellow and encompasses the two benchmark invisible axion models discussed in Sec.~\ref{sub:invisible}.\footnote{Fig.~\ref{fig:paramspace_big} is a log-log plot, so any linear relationship will appear as a line of slope 1, and the intercept encodes the constant of proportionality} In Sec.~\ref{sub:model_band} I will discuss the considerations that go into defining the model band.

The solid colored regions in Fig.~\ref{fig:paramspace_big} which are not the model band represent direct limits from laboratory experiments, and the dashed horizontal and vertical lines represent indirect limits from astrophysical and cosmological observations. In Sec.~\ref{sub:overclosure} we were able to set an upper bound on the axion mass by noting that $m_a\gtrsim 0.5$~eV would imply a cosmic density of axion HDM incompatible with CMB observations: this is represented by the vertical line marked ``Hot DM'' in Fig.~\ref{fig:paramspace_big}. In Sec.~\ref{sub:astro} we will see that astrophysical observations allow us to set more restrictive constraints on the high-$m_a$ end of the parameter space. 

Even taking into account astrophysical bounds, we see that a wide range of possible values for $m_a$ is still viable.\footnote{The most sensitive experimental probes of axion parameter space rely on the assumption that axions constitute the Milky Way's dark matter halo. In principle, in the standard cosmological scenario in which inflation precedes the spontaneous breaking of PQ symmetry or does not happen at all, the assumption that $\Omega_a=\Omega_\text{DM}$ fixes $m_a$. In practice, the systematic uncertainties associated with axion CDM production within the standard cosmology and the possibility of an inflationary loophole together preclude a definitive calculation of the relationship between $m_a$ and $\Omega_a$. In this thesis I will simply treat $m_a$ as a free parameter.} Within the viable parameter space, only haloscope searches have set limits (shown in green) that come anywhere near the model band.\footnote{The other limits shown in Fig.~\ref{fig:paramspace_big} are from experiments that do not assume axions constitute CDM. Such experiments are intrinsically less sensitive because they must rely on processes in the present-day universe to produce axions, and broadband because these axions will generally be relativistic. Searches which do not assume axions constitute CDM are outside the scope of this thesis; see Ref.~\citep{annurev2015} for a recent review.} We will see in Sec.~\ref{sec:searches} that the mass range accessible to haloscopes is $1 \lesssim m_a \lesssim 50~\mu$eV, give or take a factor of 2 on either end. Throughout this thesis, whenever a specific value of $m_a$ is useful for illustrating some point, my default example will be $m_a\sim 20~\mu$eV. 

We can imagine two distinct reactions to the openness of axion parameter space (relative to the WIMP parameter space depicted in Fig.~\ref{fig:wimps}): a pessimist might bemoan the slow pace of experimental progress in the 34 years since axion CDM was first proposed, whereas an optimist would note that the naive parameter values that the authors of Refs.~\citep{pww1983} and \citep{as1983} had in mind are still viable -- the same certainly cannot be said for the simplest models of WIMP dark matter! 

Lest we become too jaded by the perspective of our hypothetical pessimist, let us briefly consider why axion searches are so hard. Fig.~\ref{fig:paramspace_big} indicates that for $m_a\sim20~\mu$eV, $g_{a\gamma\gamma}\sim10^{-15}~\text{GeV}^{-1}$ typically. It seems reasonable that to suppose that observable effects involving the electromagnetic interaction of a single axion scale as a positive power of the dimensionless quantity $g_{a\gamma\gamma}\Delta E$, where $\Delta E$ is some measure of the energy transfer. Dark matter axions are very non-relativistic, and thus the maximum energy which can be transfered from any given halo axion to a pair of photons is $\Delta E=m_a$: then $g_{a\gamma\gamma}\Delta E\lesssim2\times10^{-29}$! We shall see in Sec.~\ref{sec:haloscope} that this isn't actually how a practical axion detector works. The point of this crude qualitative argument is just to emphasize that axions with $f_a$ far above the electroweak scale were dubbed ``invisible'' for a reason -- what is really remarkable is that they are detectable at all!

Having whetted our appetite by discussing axion parameter space at the broadest level, we can now return to examine in somewhat greater detail two issues glossed over in the above discussion. Readers who skipped chapters~\ref{chap:theory} and \ref{chap:cosmo} may wish to skip ahead to Sec.~\ref{sub:axion_cdm}.

\subsection{The axion model band}\label{sub:model_band}
It would be nice to have some sense of how much the coefficient $g_\gamma$ varies across different invisible axion models. We can begin by writing $g_\gamma$ in the form~\citep{cheng1995}
\begin{align}
g_\gamma&=\frac{1}{2}\left[\frac{E}{N} - \frac{2}{3}\frac{4+z}{1+z}\right] \label{eq:photon_coupling_2}\\
&=\frac{1}{2}\left[\frac{E}{N} - 1.95\right]\nonumber
\end{align}
where $E$ and $N$ are the electromagnetic and color anomaly coefficients of $U(1)_\text{PQ}$, respectively, and I have used $z=0.56$ in the second line. The ratio $E/N$ can generally be positive, negative, or zero.

On the RHS of Eq.~\eqref{eq:photon_coupling_2}, the first term corresponds to a triangle diagram like Fig.~\ref{fig:triangle_pi} with the axion in place of the pion field. For this term to contribute there must exist a quark which is charged under both $U(1)_\text{PQ}$ (so that it can couple to the axion field at the left vertex) and $U(1)_\text{EM}$ (so that it can couple to photons at the other two vertices): this is precisely the statement that $U(1)_\text{PQ}$ have an anomaly with QED ($E\neq0$). The second term in Eq.~\eqref{eq:photon_coupling_2} corresponds to a diagram in which an axion mixes with a virtual pion via the same process that generates axion mass, and this virtual pion couples to a pair of photons through the triangle diagram of Fig.~\ref{fig:triangle_pi}.

The coefficient of the pion's electromagnetic anomaly is fixed by SM physics [see discussion surrounding Eq.~\eqref{eq:pigammagamma}], so the contribution of axion-pion mixing only depends on known hadronic physics and on $N$, the coefficient of the anomaly of $U(1)_\text{PQ}$ with QCD. I have followed the usual convention in absorbing a factor of $N$ into the definition of the PQ scale $f_a$ (see discussion in Sec.~\ref{sub:invisible}): with this convention the second term in Eq.~\eqref{eq:photon_coupling_2} is model-independent, and a factor of $N^{-1}$ appears in the first term. 

Evidently, if we are unlucky we can get destructive interference between the two additive terms in $g_\gamma$~\citep{kaplan1985}. So it is worth taking a closer look at how easy it is to get such a cancellation by generalizing the prototypical invisible axion models discussed in Sec.~\ref{sub:invisible}. I will elide many details which may be found in Refs.~\citep{cheng1995,kaplan1985,kim1998}.

Recall that in the KSVZ model, none of the SM fermions are charged under $U(1)_\text{PQ}$, and we instead introduced a new heavy quark $q$ which must be a singlet under $SU(2)_W$ to avoid messing up the symmetries of the SM. In the original KSVZ model $q$ was also assumed to be electrically neutral, in which case $E=0$. Variants of the KSVZ model may be generated by assigning a different electric charge $e_q$ to the quark $q$.\footnote{$e_q=0$ may seem to be the natural choice given the close relation between the weak and electromagnetic interactions, but the SM already contains right-handed quarks and charged leptons which have electric charge but are $SU(2)_W$ singlets, and we may as well keep an open mind.} For these models, $E/N=6e^2_q$.

In the DFSZ model the situation is precisely reversed. We do not posit the existence of new quarks, but we must assign PQ charges to the existing SM quarks and also to the charged leptons. PQ charges for the leptons are necessary because the two Higgs doublets in the DFSZ model must have PQ charges deterministically tied to those of the quarks for the entire Lagrangian to be invariant under $U(1)_\text{PQ}$, and one of these doublets must also interact with the charged leptons to give them mass.\footnote{Incidentally, you may be wondering how the DFSZ model with its two Higgs doublets is faring now that one (and only one) Higgs field has been discovered at the LHC. According to Ref.~\citep{espriu2015}, there is still plenty of Higgs-sector parameter space available for DFSZ models.} 

The assignment of PQ charges in the DFSZ model with two Higgs doublets is highly constrained by the symmetries of the SM. We must first specify which Higgs doublet gives mass to the charged leptons: the case where the leptons couple to $\phi_2$ along with down-type quarks is sometimes called the type-I DFSZ model, and the case where the leptons couple to $\phi_1$ along with up-type quarks is called type-II DFSZ. In both cases, a minor miracle happens: both $E$ and $N$ turn out to be proportional to the sum $X_u+X_d$ of PQ charges for the up-type and down-type quarks \citep{cheng1995}. Thus $g_\gamma$ (which depends only on $E/N$) is independent of the actual values of $X_u$ and $X_d$.\footnote{$E/N$ is also independent of the PQ charge of $q$ in the KSVZ-type models considered above, but this is perhaps less surprising given that there is only one PQ-charged quark in this case.}

Let's review what we have learned thus far. In the original ($e_q=0$) KSVZ model, we have $E/N=0$ and thus $g_\gamma=-0.97$. We might also consider KSVZ-type models with $e_q=(\pm1/3,\pm2/3,\pm1)$, given that these electric charges are known to exist in nature already. We obtain respectively $E/N=(2/3,8/3,6)$ and $g_\gamma=(-0.64,0.36,2.03)$. The type-I (type-II) DFSZ model has $E/N=8/3$ ($E/N=2/3$) and $g_\gamma=0.36$ ($g_\gamma=-0.64$). In all of these cases, $g_\gamma$ has turned out to be quite robust. We do find significant model dependence if we extend our consideration to type-III DFSZ models, which introduce a \textit{third} Higgs doublet to give mass to the charged leptons~\citep{cheng1995,kim1998}. In type-III DFSZ models the PQ charge assignments do matter: it is easy to obtain both large suppression ($E/N=2$) and large enhancement ($E/N=-22/3$) even if we only permit $X_\text{PQ}=\pm1/3,\pm2/3,\pm1$. 

I will adopt a conservative definition of the model band spanned by these latter two examples, which are due to Ref.~\citep{cheng1995}: since only the magnitude of $g_\gamma$ will matter in practice, we can then write $0.03 < \abs{g_\gamma} < 4.64$.\footnote{See Ref.~\citep{diluzio2017} for a recent alternative approach.} Note that the lower bound here is highly sensitive to the value of the quark mass ratio $z$, and if it so happens that $z=0.5$, $g_\gamma$ vanishes identically! The point of defining this model band is then not to be totally exhaustive, but rather to define a range of values in which we might reasonably expect $g_\gamma$ to lie, given our ignorance of the details of the PQ mechanism.\footnote{Neither of the ``standard'' invisible axion models seems entirely satisfactory -- the singlet quark in KSVZ-type models seems rather out of place in the context of the SM, and while the DFSZ model fares better in this respect it has a domain wall problem (see Sec.~\ref{sub:defects}) in its simplest form. Thus it is good to be open-minded about how exactly the PQ mechanism is implemented.} Conversely, we may find it reassuring that it is easy to obtain $g_\gamma\sim\mathcal{O}(1)$ with relatively simple models.

\subsection{Astrophysical limits}\label{sub:astro}
From the discussion in chapters~\ref{chap:theory} and \ref{chap:cosmo} alone, we might be tempted to conclude that there is no reason the PQ mechanism cannot be realized with $m_a \sim 100$ meV -- such an axion would not be produced in large numbers either as HDM or CDM. However, while cosmologically harmless, such an axion would not be \textit{astrophysically} harmless. The basic mechanism responsible for astrophysical bounds on axions is very general: in any given astrophysical body whose core temperature is $T_\text{core}$, particles with mass $m\lesssim T_\text{core}$ can be efficiently produced by thermal processes inside the core. If the particles in question interact sufficiently weakly that they are likely to be radiated out into space without interacting again, they can provide a very efficient energy loss channel compared to photons. 

The relevant temperatures for our purposes are $1~\text{keV} \lesssim T_\text{core} \lesssim 10~\text{MeV}$, where the lower end of the range describes typical stars like our sun and the upper end is characteristic of supernovae; clearly there is no kinematic barrier to the production of axions. Of course, if axions interact too weakly, they are not copiously produced in astrophysical bodies at all, and no interesting bounds can be set. Clearly, the ``optimal'' coupling for the purposes of obtaining dramatic effects on astrophysics is whatever results in a mean free path comparable to the size of the body in question. To appreciate just how dramatic these effects can be, it is sufficient to note that the mean free path for a photon produced in the center of the sun is only $\sim 1$~cm \citep{windows1990}, and the longevity of stars is basically governed by how long it takes energy produced by nuclear fusion in the interior to diffuse out to the surface and escape.

My treatment of this subject has been rather ahistorical: astrophysical bounds on axions and other light particles have been studied since the mid 1970s, well before the cosmological implications of axions were appreciated. A wide variety of astrophysical objects have been considered: the basic procedure is to identify some observable effect which is a sensitive probe of excess energy loss and then work out all the messy details to set limits on particle masses or couplings. See Refs.~\citep{windows1990,raffelt2008} for detailed reviews; here I will just briefly summarize the best current limits. 

The cross sections for most axion production mechanisms turn out to have strong temperature dependence, so stronger constraints are obtained by considering hotter astrophysical bodies. If we consider only production by the $\lagr_{a\gamma\gamma}$ interaction, the best bounds come from the abundance of hot horizontal branch stars relative to red giants in globular clusters in which all the stars were known to be produced at the same time: this is indicated by the horizontal line marked ``massive stars'' on Fig.~\ref{fig:paramspace_big}. For most of the 1980s this was the best model-independent astrophysical bound. Axion-electron interactions provide a more efficient source of stellar axion production, and thus more stringent limits could be placed on DFSZ-type models in which axion-lepton interactions are necessarily present than on KSVZ models in which such interactions are absent at tree level.\footnote{For this reason, the KSVZ axion is sometimes called a ``hadronic'' axion~\citep{kaplan1985}.}

This distinction has been less important since the detection of neutrinos from SN1987a. The importance of neutrinos as an energy loss mechanism in supernovae is itself an excellent illustration of the general principle used to set astrophysical bounds on axions. Sufficiently light axions would stream out of the core without interacting at all, and thus ``steal'' much of the energy that would otherwise be carried away by neutrinos: the main observable effect would be a reduction in the duration of the neutrino burst. Axion production in supernovae turns out to mainly depend on the axion-nucleon coupling, which is quite model-independent due to its origin in the QCD anomaly of $U(1)_\text{PQ}$. The result is a limit on the axion mass $m_a\lesssim 16$~meV applicable to both KSVZ and DFSZ axions~\citep{raffelt2008}; this is indicated on Fig.~\ref{fig:paramspace_big} by the vertical line marked ``SN1987A.''\footnote{The supernova bound is sometimes cited in the literature as $m_a\lesssim1$~meV. I have quoted the more conservative value. The discrepancy likely has something to do with the numerical modeling required to translate the very robust limit on the rate of excess energy loss (which nobody disputes) into constraints on axion parameters.}

In short, by considering the astrophysical effects of axions, we find that the entire axion mass range $m_a\gtrsim16$~meV is excluded for one reason or another. Achieving $\Omega_a\sim\Omega_\text{DM}$ with $m_a\sim 10$~meV requires some fine-tuning, but this scenario has been has been discussed by at least one group~\citep{kawasaki2015}; by $m_a\sim1$~meV we are back in the region where both pre-inflationary and post-inflationary PQ breaking can produce $\Omega_a\sim\Omega_\text{DM}$. There does not appear to be much room in the parameter space for axions which do not play a large role in cosmology! 

\section{Axion CDM in the Milky Way}\label{sec:local_dm}
For the remainder of this thesis, I will assume that axions constitute all of dark matter.\footnote{This is standard practice for quoting the sensitivity of any experiment aiming to directly detect dark matter: limits can always be trivially rescaled if the hypothesis that the candidate in question accounts for any given fraction of the local density is of interest.} Then
\begin{equation}\label{eq:rho_a_dm}
\rho_a=\rho_\text{DM},
\end{equation}
where $\rho$ denotes a \textit{local} mass density. I emphasized in Sec.~\ref{sub:expansion} that the density in any galactic halo is many orders of magnitude larger than the critical density. From this point onwards we will no longer need to speak of the average cosmic density in axion CDM, but clearly we must have some rough idea of the local density of axions in our corner of the Milky Way if we want to detect them in a terrestrial lab.

In Sec.~\ref{sub:halo} I review what is known about the distribution and kinematics of dark matter in the solar neighborhood, and what sorts of approximations we can make given our incomplete knowledge of the galactic halo. In Sec.~\ref{sub:axion_cdm} I will discuss the implications of these distributions for axion CDM specifically. This discussion will also help motivate the conceptual design of the haloscope detectors to be discussed in Sec.~\ref{sec:haloscope}.

\subsection{Halo models and the local density}\label{sub:halo}
Experiments aiming to directly detect non-gravitational interactions of dark matter must invariably make assumptions about the mass and velocity distributions of the dark matter in the Milky Way. These two distributions are related by the assumption that the galactic halo has \textbf{virialized}, i.e., settled into an equilibrium state in which its time-averaged kinetic and potential energy are related by the virial theorem. Virialization is a rather weak assumption which does not require thermodynamic equilibrium at the microscopic level, and numerical simulations of CDM structure formation usually lead to virialized halos.\footnote{You may be wondering \textit{how} the halo reaches virial equilibrium, given that the self-interactions of axions (and other CDM particle candidates) are so weak that they are effectively collisionless. This is a very complicated subject, but the zeroth-order answer is that a time-dependent gravitational potential (e.g., during the initial gravitational collapse of an overdensity) can cause even collisionless particles to equilibrate. This process goes by the delightful name of ``violent relaxation.''} Throughout this thesis I will assume a fully virialized halo; searches for a non-virialized axionic component are briefly discussed in Sec.~\ref{sub:admx}.

For a virialized dark matter halo, all we need to do is specify the mass distribution (and the values of a few parameters). The topics we have covered thus far already give us a rough sense of what this mass distribution should look like. Recall from Sec.~\ref{sub:rotation} that the flatness of observed galactic rotation curves implies [via Eq.~\eqref{eq:rot_curve}] that the total mass $M(r)$ contained within a sphere of radius $r$ scales as $M(r)\sim r$. This in turn implies that the mass density $\rho(r)\sim r^{-2}$. Of course, this scaling is only approximate insofar as the rotation curve is not perfectly flat. Nonetheless, this crude argument suggests that the spherically symmetric density profile $\rho(r)\propto r^{-2}$ is not a bad place to start: this halo model is usually called the \textbf{isothermal sphere}.\footnote{``Isothermal'' is somewhat misleading in this context -- it merely implies that, with $\rho(r)\propto r^{-2}$, the typical particle kinetic energy ($K\propto v_c^2$) is independent of $r$, which is just how we derived the isothermal sphere in the first place. If the galaxy were composed of only a single particle species of a given mass, this would imply a common temperature $T$. For a galaxy containing multiple particle species (e.g., axions and baryons), all particles will move at the same rotational velocity but not share a common temperature.}

There are two very obvious problems with the isothermal sphere which indicate that it cannot possibly be the whole story: it has an unphysical cusp as $r\rightarrow0$, and if extrapolated to $r\rightarrow\infty$ it implies that the total galactic mass is infinite! A more reasonable density profile would have milder $r$-dependence at small $r$ and fall off faster for large $r$: such behavior is realized for example in the popular Navarro-Frenk-White (NFW) profile. Here I will instead consider an ad hoc patch to fix the small-$r$ behavior of the isothermal sphere, and not worry about what happens at large $r$, which will not be relevant for a terrestrial experiment anyway. Specifically, we will assume that the halo approaches a constant density ``core'' at sufficiently small radius:
\begin{equation}\label{eq:pits}
\rho(r)=\frac{\rho_c}{1+(r/r_c)^2}.
\end{equation}
This is usually called a \textbf{pseudo-isothermal sphere}. The rotation curve of a pseudo-isothermal sphere model has a simple expression in terms of the two parameters $r_c$ and $\rho_c$~\citep{jimenez2003}. In practice it is often more convenient to parameterize the halo model using the measured values of the local circular velocity $v_c$ and the local density $\rho_\text{DM}$ evaluated at our position relative to the center of the Milky Way, $r=r_0\approx8$~kpc.\footnote{It should also be noted that the measured value of $v_c$ includes the effects of baryonic contributions to the Milky Way's rotation curve, which account for about half the mass within a sphere of radius $r_0$~\citep{read2014}.} Applying the pseudo-isothermal sphere model to the Milky Way we obtain $r_c \approx 3$~kpc; $r_0$ is then sufficiently far outside the core that we can safely ignore the fact that the halo is not truly isothermal.\footnote{Some authors even go so far as to call Eq.~\eqref{eq:pits} an isothermal density profile, to add insult to terminological injury.}

A truly isothermal sphere would imply a Maxwell-Boltzmann distribution of dark matter particle velocities about the modal velocity $v_c$ in the galactic rest frame. In the pseudo-isothermal sphere, the velocity distribution should be roughly Maxwellian for $r\sim r_0$, and for simplicity we will assume it to be Maxwellian. In this case the dark matter velocity distribution is completely specified by the value of the local circular velocity $v_c=220$~km/s,\footnote{In principle, the velocity distribution should also be truncated at the galactic escape velocity, $v_\text{esc}\approx530$~km/s. In practice, this cutoff will not have any significant impact on the haloscope search, so I will ignore it. The existence of the galactic escape velocity implies that even if a sizable fraction of the local dark matter is not virialized, it certainly cannot be moving much \textit{faster} than in the virialized case.} and the corresponding distribution for the kinetic energy $K$ is a $\chi^2$ distribution of degree 3:
\begin{equation}\label{eq:f_dist_E}
f(K) = \frac{2}{\sqrt{\pi}}\sqrt{K}\left(\frac{3}{m_a\langle v^2\rangle}\right)^{3/2}e^{-\frac{3K}{m_a\langle v^2\rangle}},
\end{equation}
where $\langle v^2\rangle = 3v_c^2/2$ is the second moment of the Maxwell-Boltzmann velocity distribution. Eq.~\eqref{eq:f_dist_E} indicates that $\langle v^2\rangle$ is an approximate measure of the width of the kinetic energy distribution, which is what we are ultimately interested in for a haloscope search. For this reason, axion papers often quote the value of the RMS velocity $\langle v^2\rangle^{1/2}=270$~km/s instead of $v_c$; $\langle v^2\rangle^{1/2}$ is usually referred to as the \textbf{virial velocity}.

\begin{figure}[h]
\centering\includegraphics[width=0.7\textwidth]{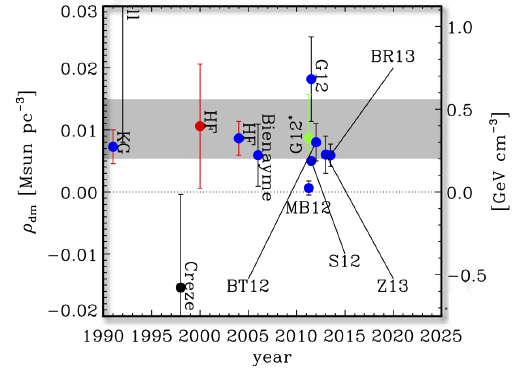}
\caption[Recent measurements of the local dark matter density]{\label{fig:read_dm} Recent measurements of $\rho_\text{DM}$. Points with error bars represent \textit{local} measurements derived from the vertical velocities of tracers near the sun. The gray band indicates the range of values obtained from \textit{global} measurements that use the whole Milky Way rotation curve. Figure from Ref.~\citep{read2014}.}
\end{figure}

Finally, we must specify the value of the local dark matter density $\rho_\text{DM}$. In practice this is the hardest part: the required dynamical measurements are complicated by our position in the middle of the galaxy of interest and by the importance of baryonic effects for $r\approx r_0$. The first paper~\citep{iocco2015} to claim model-independent constraints excluding $\rho_\text{DM}=0$ at $>3\sigma$ was quite recent! See Ref.~\citep{read2014} for a thorough review of measurements of the local density. For our present purposes, everything we need to know is summarized in Fig.~\ref{fig:read_dm}. Searches for axion CDM have typically assumed $\rho_\text{DM}=0.45~\text{GeV/cm}^3$, while WIMP searches typically cite $\rho_\text{DM}=0.3~\text{Gev/cm}^3$ instead; both values clearly fall within the range of recent measurements. I will adopt the former value for consistency with the axion search literature.

In the preceding paragraphs I motivated a simple model of the distribution and kinematics of dark matter in the galactic rest frame. Of course, we do not actually live in the galactic rest frame: the motion of any terrestrial laboratory relative to the galactic halo is dominated by the orbital velocity of the sun about the center of the galaxy $v_s\approx v_c$.\footnote{If we wanted to get really fancy, we could also consider the earth's orbital velocity about the sun ($v_o\approx 30$~km/s) and its rotational velocity about its own axis ($v_r\approx 0.5$~km/s) \citep{turner1990}. In practice the rotational velocity is negligible and the earth's orbital velocity is also usually irrelevant.} A reasonable approximation to the lab frame kinetic energy distribution may be obtained by taking $\langle v^2\rangle\rightarrow1.7\langle v^2\rangle$ in Eq.~\eqref{eq:f_dist_E}~\citep{turner1990}. I will discuss the impact of this broadening on the haloscope search further in Sec.~\ref{sec:rebin}. For the present qualitative discussion, I will continue to use $\langle v^2\rangle^{1/2}$ as an estimate of the typical spread in dark matter particle velocities, ignoring the factor of $\sqrt{1.7}$ along with other order-unity coefficients.

\subsection{Properties of axion CDM}\label{sub:axion_cdm}
Let us consider the behavior of axion CDM in a fully virialized pseudo-isothermal halo with local density $\rho_a=\rho_\text{DM}=0.45~\text{GeV/cm}^3$ and rest-frame virial velocity $\langle v^2\rangle^{1/2}=270$~km/s, bearing in mind that the chief virtues of this halo model are its simplicity and the absence of strong evidence for any particular alternative.

We can immediately make several simple observations with profound implications for axion detection. First, the axion number density in apparently ``empty'' space within our galaxy is quite large: $n_a=\rho_a/m_a \sim 2\times10^{13}~\text{cm}^{-3}$ for $m_a\sim20~\mu$eV. Of course this is still many orders of magnitude smaller than the number of atoms in an average cubic centimeter of anything in our daily experience. The large value of $n_a$ will seem more remarkable in light of our second observation: the axion has an enormous de~Broglie wavelength
\begin{equation}\label{eq:lambda_a}
\lambda_a = \frac{2\pi}{m_av} \sim 3\times10^{8}~\text{eV}^{-1} \sim 70~\text{m},
\end{equation}
for $m_a\sim20~\mu$eV, where I have introduced $v=\langle v^2\rangle^{1/2}/c\sim10^{-3}$ for notational convenience.\footnote{There is a factor of $2\pi$ in the second equality in Eq.~\eqref{eq:lambda_a} because the de~Broglie wavelength is $\lambda=h/p$, but my convention in Eq.~\eqref{eq:hl_units} was to set $\hbar=1$.} Normally we are used to thinking about the de~Broglie wavelength as a microscopic quantity; CDM axions are different because they are both very light and very non-relativistic. We can now see that the \textbf{phase space occupancy} of axion CDM in the galactic halo is $\mathcal{N}_a\sim n_a\lambda_a^3\sim 10^{25}$, which is a more meaningfully impressive number than $n_a$.

Large values of $\lambda_a$ and $\mathcal{N}_a$ together indicate that axion CDM will behave more like a classical field than a sea of particles. Let's work out the implications of this claim: $\lambda_a$ is related by an $\mathcal{O}(1)$ factor to the \textbf{coherence length} of axion CDM, which we can loosely define as the region over which the axion field is approximately constant. The precise value is not very relevant, since the coherence length will be much larger than the linear scale of a laboratory experiment in any event: I will simply use $\lambda_a$ for the coherence length to avoid introducing too much notation. 

In one sense we are just encountering again what we already saw in our foray into early-universe cosmology in Sec.~\ref{sub:misalignment}: axion CDM can be usefully regarded as coherent oscillations of a spatially homogeneous field. More precisely, the value of the field acts like the coordinate of a harmonic oscillator with angular frequency $m_a$. In the case of present-day galactic halos, nonzero axion momenta imply a finite \textbf{coherence time}
\begin{equation}\label{eq:tau_a}
\tau_a=\frac{\lambda_a}{v} \sim 200~\mu\text{s}
\end{equation}
for $m_a\sim20~\mu$eV. Basically, $\tau_a$ quantifies how long it takes for a local detector monitoring the oscillations in the axion field to move into a patch of the field where those oscillations are appreciably out of phase. It is usually more useful to express this as a number of oscillation periods than as a time. Thus we define the ``axion quality factor''
\begin{equation}\label{eq:q_a}
Q_a = \frac{1}{2\pi}m_a\tau_a = \frac{1}{v^2} \sim 10^{6},
\end{equation}
independent of the axion mass. Of course $v^2$ was introduced as a measure of the fractional width of the CDM axion energy distribution -- thus $Q_a$ also measures how sharply peaked axion field oscillations are in the frequency domain, like the $Q$ factor of any other oscillator.

To formally express axion CDM as an oscillating field let us focus on a region of space which is small compared to $\lambda_a$: then we can write $a(\mathbf{x},t)\approx a(t)$, where the time-dependence includes both rapid oscillations on a timescale $(m_a/2\pi)^{-1}$ and slow frequency modulation with a characteristic timescale~$\tau_a$. We can then write the local axion density in the form\footnote{A factor of $1/2$ which appeared in our earlier discussion following Eq.~\eqref{eq:axion_eom_adiabatic} is absent here because he have expressed the energy density in terms of the mean-squared field amplitude instead of the peak amplitude.}
\begin{equation}\label{eq:rho_a_field}
\rho_a = m_a^2\avg{a^2},
\end{equation}
where $\avg{a^2}$ is defined by
\[
\avg{a^2} = \frac{1}{2t_0}\int_{-t_0}^{t_0}\mathrm{d}t\, a^2(t)
\]
with $t_0$ a sufficiently long reference time. Following Ref.~\citep{krauss1985}, we define the Fourier transform convention
\begin{align*}
a(\omega) &= (2t_0)^{-1/2}\int_{-t_0}^{t_0}\mathrm{d}t\, a(t)e^{i\omega t} \\
a(t) &= (2t_0)^{1/2}\int_{-\infty}^{\infty}\frac{\mathrm{d}\omega}{2\pi}\, a(\omega)e^{-i\omega t}
\end{align*}
from which
\begin{equation}\label{eq:Parseval}
\avg{a^2} = \int_{-\infty}^{\infty}\frac{\mathrm{d}\omega}{2\pi}\abs{a(\omega)}^2
\end{equation}
follows by Parseval's Theorem. 

In the limit $Q_a\rightarrow\infty$, none of this formalism would be necessary, as we could trivially evaluate $\avg{a^2}$. In principle, given any halo model we can still evaluate it: the Fourier component $a(\omega)$ is just proportional to Eq.~\eqref{eq:f_dist_E} evaluated at $K=\omega-m_a$, with normalization set by Eq.~\eqref{eq:rho_a_field}. In practice, expressions which are agnostic as to the precise functional form of $a(\omega)$ are useful for deriving the signal power in an axion haloscope.

\section{The axion haloscope}\label{sec:haloscope}
We have seen that a search for axion CDM has at least one and perhaps two distinct advantages over a more generic axion search: the large local axion number density $n_a$ and the coherence of axion field oscillations on sufficiently small spatial and temporal scales. An ideal axion CDM detector should be designed to maximally exploit this unique behavior. In 1983, Pierre Sikivie~\citep{sikivie1983} first proposed an idea for such a detector, which he called the axion haloscope. The conceptual design of the haloscope was refined in a pair of follow-up publications by Sikivie~\citep{sikivie1985} and Krauss, Moody, Wilczek, and Morris~\citep{krauss1985} in 1985. 

\begin{figure}[h]
\centering\includegraphics[width=0.7\textwidth]{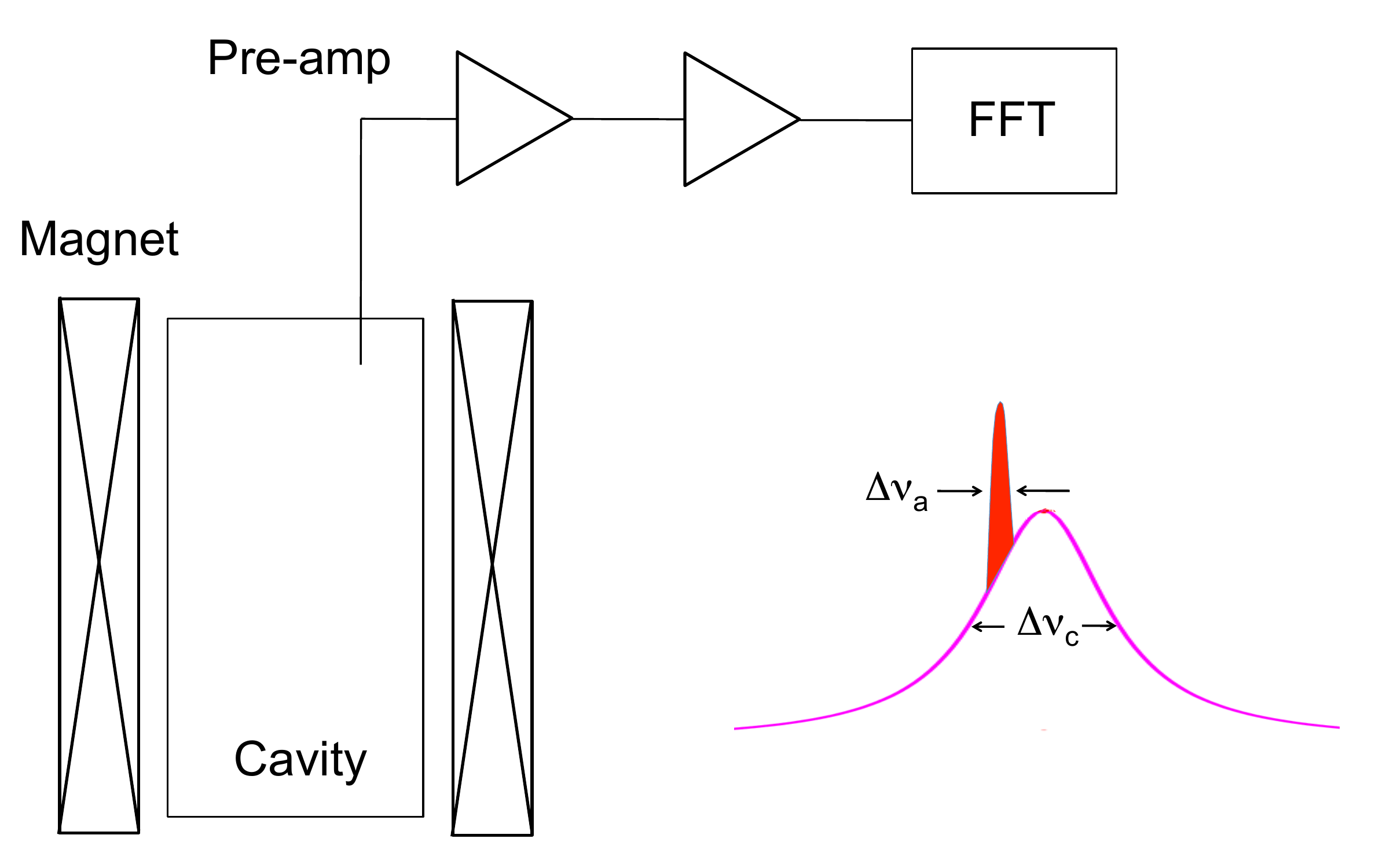}
\caption[The essential elements of an axion haloscope]{\label{fig:haloscope_simple} Left: the essential elements of an axion haloscope. Right: illustration of the usual case (not to scale) where the axion linewidth $\Delta\nu_a$ is much smaller than the cavity linewidth $\Delta\nu_c$.}
\end{figure}

An axion haloscope relies on the interaction Eq.~\eqref{eq:agammagamma} to locally convert the axion CDM energy density into an observable electromagnetic signal, which is enhanced by the large local density, by the coherence of the axion field oscillations, and finally by the application of a large external magnetic field: the latter is an example of the inverse Primakoff effect, mentioned in connection with the pion's two-photon coupling in Sec.~\ref{sub:anomaly}. These three enhancement factors collectively explain how the haloscope signal evades the extreme suppression suggested by the naive dimensional argument in Sec.~\ref{sec:parameter}. 

After elaborating on this qualitative picture of the haloscope in Sec.~\ref{sub:haloscope_qual}, I will present a detailed derivation of haloscope \textbf{signal power} in Sec.~\ref{sub:signal}, following Ref.~\citep{krauss1985}. Then I will consider how this conceptual design holds up in the presence of noise in Sec.~\ref{sub:radiometer} and Sec.~\ref{sub:sql}, and finally derive in Sec.~\ref{sub:scan} an expression for the \textbf{scan rate}, which is the best figure of merit for the haloscope search given that the axion mass $m_a$ is unknown. My discussion in this section will be quite thorough, and it may be useful to refer to the first few pages of Ref.~\citep{NIM2017} for a much more concise overview of the big picture. Relative to the analogous material in past dissertations describing haloscope detectors, Sec.~\ref{sub:sql} is likely to be of particular interest, as it discusses quantum noise limits which were not relevant for the operation of haloscopes before HAYSTAC.

\subsection{Essential physics of the haloscope}\label{sub:haloscope_qual}
\begin{figure}[h]
\centering\includegraphics[width=0.5\textwidth]{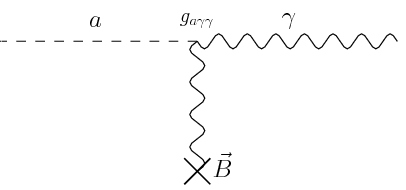}
\caption[The inverse Primakoff effect]{\label{fig:primakoff} The inverse Primakoff effect: an axion $a$ comes in from the left and absorbs a photon from the magnetic field $\mathbf{B}$ in the process of transforming into a photon $\gamma$. Figure adapted from Ref.~\citep{cadamuro2012}.}
\end{figure}

An axion haloscope is a \textit{cryogenic, tunable high-$Q$ microwave cavity immersed in a strong magnetic field and coupled to a low-noise receiver} (see Fig.~\ref{fig:haloscope_simple}). In the next few sections, we will see how all of these elements work together to enable the direct detection of axion CDM. I will begin by considering the role of the magnetic field. Fig.~\ref{fig:primakoff} shows a Feynman diagram representing the inverse Primakoff effect, in which the interaction Eq.~\eqref{eq:agammagamma} is realized with one of the external photon lines provided by a classical electromagnetic field.

The Feynman diagram representation implicitly prompts us to think about the Primakoff effect as a scattering process in particle physics. In Sec.~\ref{sub:wisps} we saw that the invisible axion's lifetime is many orders magnitude longer than the age of the universe as a direct consequence of the weakness of the coupling $g_{\gamma\gamma}$. The inverse of Eq.~\eqref{eq:axion_life} is of course the decay rate for a single axion, and a observable axion CDM decay signal would be enhanced by the large number density of axions $n_a$ associated with the external axion line in Fig.~\ref{fig:primakoff}.\footnote{At least one radio telescope search for extragalactic axion decay lines has been conducted~\citep{blout2001}, but the resulting limit is an order of magnitude weaker than the bound from massive stars discussed in Sec.~\ref{sub:astro}.} By employing the inverse Primakoff effect we are taking this logic one step further: taking one of the photon lines to be an input instead of an output enables us to crank up the density of incoming photons to compensate for the small coupling. 

A classical electromagnetic field corresponds to a large density of virtual photons: in particular, a large static field may be regarded as a superposition of many photons with zero net momentum.\footnote{The probability of finding a photon with any given momentum $\mathbf{p}$ is proportional to the square of the Fourier component of the field at wavenumber $\mathbf{k}=\mathbf{p}$. ``Virtual'' in this context just means that these photons don't obey the propagating photon dispersion relation.} Removing a single photon does not appreciably change the energy of this field configuration, and in this sense a classical field is kinematically ``heavy:'' thus in a Primakoff process the incoming axion scatters elastically and emerges as a photon with the same total energy, and polarization parallel to the applied magnetic field [c.f.~Eq.~\eqref{eq:agammagamma}].\footnote{In principle the interaction Eq.~\eqref{eq:agammagamma} would also work with a large classical \textit{electric} field, but the presence of electric charge in the universe renders this approach distinctly less attractive. It is relatively easy to arrange for large magnetic fields to coexist peacefully with other detector components.}

Neglecting the small kinetic energy of CDM axions, elastic scattering implies that the outgoing photon has frequency
\begin{equation}\label{eq:nu_a}
\nu_a = m_a \sim 5~\text{GHz}
\end{equation}
for our benchmark value $m_a\sim20~\mu$eV; the mass range $1 \lesssim m_a \lesssim 50~\mu$eV corresponds to microwave frequencies $250~\text{MHz} \lesssim \nu_a \lesssim 12~\text{GHz}$.\footnote{In keeping with the $\hbar=1$ convention of Heaviside-Lorentz units, Eq.~\eqref{eq:nu_a} should more properly read $m_a=\omega_a\sim2\pi\times5~\text{GHz}$, but the linear frequency $\nu$ is more relevant for experimental design. In Sec.~\ref{sub:signal} I will be more careful with factors of $2\pi$.} Of course, as we saw in Sec.~\ref{sec:local_dm}, CDM axions don't have identically zero kinetic energy. We will see that the spectral distribution of the haloscope signal is proportional to $\abs{a(\omega)}^2$ and is thus given by Eq.~\eqref{eq:f_dist_E} for a pseudo-isothermal halo. I will always reserve $\nu_a$ for the frequency corresponding to the axion mass.

We have had some success understanding the phenomenological features of the Primakoff effect (and the role of the magnetic field in a haloscope detector) from a particle viewpoint. To more fully appreciate how crucial the large axion density $n_a$ is to making axion interactions observable, and to understand the role of the microwave cavity in the axion haloscope, it is most useful to regard Eq.~\eqref{eq:agammagamma} as generating an additional source term in Maxwell's equations. The equations of axion electrodynamics are derived from the Lagrangian in appendix~\ref{app:maxwell}, and solved with boundary conditions appropriate to a haloscope detector in Sec.~\ref{sub:signal}. For the present qualitative discussion, the important feature of these equations is that in the presence of large static magnetic field $B_0$, a homogeneous oscillating axion field induces an electric field oscillating at the same frequency $\nu_a$, with amplitude
\begin{equation}\label{eq:E_field_axion}
E_0\sim g_{a\gamma\gamma}B_0a_0.
\end{equation}
Note that the statement that the $a$ and $E$ fields oscillate at the same frequency is completely equivalent to the result we obtained by considering the kinematics of the Primakoff effect. Neglecting numerical factors and using Eqs.~\eqref{eq:photon_coupling_1} and \eqref{eq:rho_a_field}, Eq.~\eqref{eq:E_field_axion} may be rewritten as 
\begin{align}
E_0&\sim \frac{\alpha g_\gamma m_a}{\pi\sqrt{\chi}}B_0\frac{\sqrt{\rho_a}}{m_a} \nonumber\\
&\sim \frac{\alpha g_\gamma }{\pi}\sqrt{\frac{\rho_a}{\chi}}B_0 \label{eq:E_field_no_f},
\end{align}
which is independent of $m_a$. More precisely, we see that under the assumption that axions constitute the local dark matter density, observable effects in an axion haloscope are suppressed only by powers of the dimensionless quantity $\rho_\text{DM}/\chi\sim\rho_\text{DM}/\Lambda_\text{QCD}^4$: the much larger energy scale $f_a$ has dropped out of the problem entirely! This observation goes a long way towards explaining why haloscopes have sensitivity to $g_{a\gamma\gamma}$ so much better than experiments that do not assume axions constitute dark matter (see Fig.~\ref{fig:paramspace_big}).

Unforuntately, Eq.~\eqref{eq:E_field_no_f} is still too small to be detectable in a reasonable experiment. Assuming $g_\gamma\sim\mathcal{O}(1)$ and a large laboratory field $B_0=10$~T, Eq.~\eqref{eq:E_field_no_f} indicates that the induced electric field amplitude is only $E_0\sim 10^{-12}$~V/m. The final effect we have yet to consider is the enhancement of the haloscope signal by the coherence of axion field oscillations. To understand this effect, let us first imagine that $m_a$ is known. Then we can enclose the high field region in a \textbf{microwave cavity} with resonant frequency $\nu_a$. Microwave resonators are complicated three-dimensional structures with many different resonant modes, but we can understand the essential physics by abstracting the situation further, and treating the axion field and cavity mode as a pair of coupled oscillators.\footnote{I thank Aaron Chou for this simple yet extremely illuminating analogy. If you have not seen a demonstration of energy transfer between coupled oscillators, I encourage you to check out the video on his website~\citep{aaron}.} By assumption the axion field and cavity mode have the same frequency $\nu_a$: the only other parameters of this simplified system are the oscillator linewidths $\Delta\nu_a$ and $\Delta\nu_c$ and their mutual coupling $\kappa$, all with units of frequency. At $t=0$, the axion field is oscillating and the cavity field is stationary.

It is instructive to first consider the interactions of coupled oscillators in the strong coupling regime ($\kappa\gg\Delta\nu_a,\Delta\nu_c$), although this is not the limit in which the haloscope operates. For strong coupling, we can neglect damping entirely and the equations of motion are easy to solve: the normal modes of the system are symmetric and antisymmetric linear combinations of the two oscillations.\footnote{If you prefer quantum mechanics, the physics is exactly analogous to Rabi flopping in a two-level system not initialized in one of its eigenstates.} With the initial conditions we have specified, the oscillators exchange energy at the beat frequency $\kappa$.

Let us assume the coupled axion-cavity system is characterized by the nominal parameters $m_a=20~\mu$eV and $B_0=10~\text{T}$. Then $\Delta\nu_a = \tau_a^{-1}=1/(200~\mu\text{s})$ [Eq.~\eqref{eq:tau_a}], and we will see in Sec.~\ref{sec:searches} that in practice we will always have $\Delta\nu_c\gg\Delta\nu_a$, as illustrated in Fig.~\ref{fig:haloscope_simple}; equivalently, the lifetime of an excitation in the cavity $\tau_c=\Delta\nu_c^{-1}$ will always be shorter than $\tau_a$. With these parameters, Eq.~\eqref{eq:E_field_axion} indicates that $\kappa=(g_{a\gamma\gamma}B_0)/2\pi\sim1/(3.5~\text{days})$!\footnote{Magnetic field strength has dimension $\text{eV}^2$ in natural units, so you need to remember the value of $\mu_0$ in addition to the relations in Eq.~\eqref{eq:hl_units} for a back-of-the-envelope estimate of $\kappa$. Alternatively, you can remember that $1~\text{T} \approx 200~\text{eV}^2$.} The axion and cavity mode are \textit{exceptionally} weakly coupled. Physically, the amplitude of the cavity field oscillations will grow coherently until time $t=\tau_c$, after which the power dissipated in the cavity will balance the incoming axion conversion power: thus we can approximate the behavior in this regime by just cutting off the slow dynamics of the strongly coupled solution at $t=\tau_c$. To set the normalization, we can note that the energy density of the axion field oscillations is $\rho_a$, and the energy density in the cavity field oscillations is $\sim E_0^2$. Thus
\begin{align}
E(t) &\sim \sqrt{\rho_a}\cos(m_at)\sin(2\pi\kappa \tau_c) \nonumber \\
&\sim m_aa_0\cos(m_at)g_{a\gamma\gamma}B_0\tau_c\nonumber\\
\Rightarrow E_0 &\sim g_{a\gamma\gamma}B_0a_0Q_L, \label{eq:E_field_axion_q}
\end{align}
which is just Eq.~\eqref{eq:E_field_axion} enhanced by the cavity quality factor $Q_L=\tau_c\nu_a$.\footnote{I use $Q_L$ instead of $Q_c$ to emphasize that the cavity is loaded by some external circuitry which couples the axion conversion power out to the receiver we use to measure it.} Evidently the optimal value is $Q_L=Q_a$; improving the cavity quality factor further beyond this point would not help because the haloscope would then be limited by the finite coherence time of the axion field itself. In practice, as noted above, experiments have only operated in the $Q_L\ll Q_a$ regime. 

The presence of $Q_L$ in Eq.~\eqref{eq:E_field_axion_q} is often attributed to the cavity ``deforming the density of states'' for the outgoing photons, which in my view obfuscates the essentially classical nature of this resonant enhancement. In particular, it should be emphasized that the cavity does not \textit{amplify} the axion-photon interaction, as it is just a hunk of metal with no external power source. Large $Q_L$ just allows the coherent oscillations to grow for longer, and the larger the amplitude of the cavity field oscillations, the easier it becomes for the axion to ``push'' the cavity mode.\footnote{Another enlightening approach I will not discuss in detail is to treat the cavity using the input-output formalism of quantum optics~\citep{zheng2016}. In this picture, the coupling to the axion field is a port like any other, and the cavity acts like an impedance matching network between the axion port and the output port.} 

We have now worked out qualitatively most of the essential physics of the signal in an axion haloscope. The three critical parameters are the local density $\rho_a$ (which we get for free if axions constitute dark matter), the magnetic field strength $B_0$ (which enhances the interaction through the Primakoff effect), and the cavity quality factor $Q_L$ (which enhances the signal due to the coherence of the interaction on short timescales). The coupled oscillator model does not however capture the effects of spatial coherence, which will emerge naturally from the more formal derivation of the haloscope signal power in Sec.~\ref{sub:signal}. In particular, we will find that the observable signal scales with the cavity volume, but the spatial profile of the cavity mode electric field must be well-matched to that of external magnetic field.

\subsection{Signal power derivation}\label{sub:signal}
A field theoretic derivation of Maxwell's equations [Eqs.~\eqref{eq:div_E}, \eqref{eq:curl_B}, \eqref{eq:curl_E}, and \eqref{eq:div_B}] for the axion-coupled electromagnetic field is included in appendix~\ref{app:maxwell}. We can begin our study of the axion haloscope by invoking the largeness of $\lambda_a$ [Eq.~\eqref{eq:lambda_a}] to discard the terms involving $\gv{\del}a$. We can safely assume that the axion-sourced $B$ field of the cavity mode is negligible compared to the static external $B$ field. I will also assume that the external field is spatially constant throughout the cavity volume: then we can replace $\mathbf{B}$ on the RHS of Eq.~\eqref{eq:curl_B} with $B_0\uv{z}$.\footnote{Typically the haloscope geometry is cylindrical and $B_0$ is supplied by a solenoid, so a constant field is a reasonable approximation; in HAYSTAC it is an excellent approximation. Taking into account spatial variation in the external field requires more notation, but the physics is not appreciably different.} Taking the time derivative of Eq.~\eqref{eq:curl_B}, we obtain
\begin{align*}
\gv{\del}\times\partial_t\mathbf{B} - \partial^2_t\mathbf{E} &= -g_{a\gamma\gamma}B_0\partial^2_ta\,\uv{z} \\
\Rightarrow -\gv{\del}\times\gv{\del}\times\mathbf{E} - \partial^2_t\mathbf{E} &= -g_{a\gamma\gamma}B_0\partial^2_ta\,\uv{z}
\end{align*}
using Eq.~\eqref{eq:curl_E}. Note that
\[
\gv{\del}\times\gv{\del}\times\mathbf{E} = \gv{\del}\left(\gv{\del}\cdot\mathbf{E}\right) - \del^2\mathbf{E} = - \del^2\mathbf{E}
\]
by Eq.~\eqref{eq:div_E}. Thus together Maxwell's equations yield the inhomogeneous wave equation
\begin{equation}\label{eq:wave}
\del^2\mathbf{E} - \partial^2_t\mathbf{E} = -g_{a\gamma\gamma}B_0\partial^2_ta\,\uv{z}.
\end{equation}

The cavity walls define the boundary conditions for Eq.~\eqref{eq:wave}. I will initially assume the cavity is lossless, which implies the existence of a complete set of orthogonal modes $\mathbf{e}_m$ satisfying
\begin{equation}\label{eq:Helmholtz}
\left(\omega_m^2 + \del^2\right)\mathbf{e}_m(\mathbf{x}) = 0.
\end{equation}
We can write the orthogonality condition for the modes as
\begin{equation}\label{eq:ortho}
\int\mathrm{d}^3\mathbf{x}\,\mathbf{e}_m(\mathbf{x})\cdot\mathbf{e}^*_n(\mathbf{x}) = \lambda_n\delta_{nm},
\end{equation}
where the integration is over the cavity volume, and the normalization $\lambda_n$ is arbitrary and may be different for each mode. We may express $\mathbf{E}$ in the basis of cavity modes:
\begin{equation}\label{eq:decomposition}
\mathbf{E}\left(\mathbf{x},t\right) = \sum_m E_m(t)\mathbf{e}_m(\mathbf{x}).
\end{equation}
Note that there's also another complete set of modes $\mathbf{b}_m$ satisfying Eq.~\eqref{eq:Helmholtz} and Eq.~\eqref{eq:ortho}, where $\mathbf{e}_m$ and $\mathbf{b}_m$ are orthogonal for each $m$, the cavity magnetic field may be expressed as a sum over $\mathbf{b}_m$, and the electric and magnetic field amplitudes for a given mode will be equal. Using Eq.~\eqref{eq:decomposition} and Eq.~\eqref{eq:Helmholtz} in Eq.~\eqref{eq:wave} and taking the inner product with $\mathbf{e}^*_n(\mathbf{x})$, we obtain
\[
-\sum_m \left(\omega_m^2 + \partial^2_t \right)E_m(t)\,\mathbf{e}_m(\mathbf{x})\cdot\mathbf{e}^*_n(\mathbf{x}) = -g_{a\gamma\gamma}B_0\partial^2_ta(t)\,\uv{z}\cdot\mathbf{e}^*_n(\mathbf{x}).
\]
Integrating over the cavity volume and using Eq.~\eqref{eq:ortho}, this becomes
\begin{equation}\label{eq:eom_undamped}
\left(\omega_n^2 + \partial^2_t \right)E_n(t) = g_{a\gamma\gamma}B_0\frac{\kappa_n}{\lambda_n}\partial^2_ta(t),
\end{equation}
where I have defined 
\[
\kappa_n = \int\mathrm{d}^3\mathbf{x}\,\uv{z}\cdot\mathbf{e}^*_n(\mathbf{x}).
\]

Eq.~\eqref{eq:eom_undamped} is now the equation of motion for a driven, undamped harmonic oscillator. The damping coefficient for a mode with resonant frequency $\omega_n$ and quality factor $Q_n$ is its linewidth $\omega_n/Q_n$, so we take loss into account by adding a term of the form $\left(\omega_n/Q_n\right)\partial_tE_n$ to the LHS of Eq.~\eqref{eq:eom_undamped}. In the Fourier domain,
\[
\left(\omega_n^2 - \omega^2 - i\frac{\omega\omega_n}{Q_n}\right)E_n(\omega) = g_{a\gamma\gamma}B_0\frac{\kappa_n}{\lambda_n}\omega^2a(\omega).
\]
So the amplitude of the Fourier component at $\omega$ is just
\begin{equation}\label{eq:E_field_axion_formal}
E_n(\omega) = g_{a\gamma\gamma}B_0\frac{\kappa_n}{\lambda_n}\frac{\omega^2a(\omega)}{\omega_n^2 - \omega^2 - i\omega\omega_n/Q_n},
\end{equation}
which may be compared to Eq.~\eqref{eq:E_field_axion_q}: we have obtained the expected scaling with $g_{a\gamma\gamma}$, $a$, and $B_0$. To simplify the dependence on parameters of the cavity mode, it is easiest to proceed directly to evaluating the total steady-state electromagnetic energy stored in the cavity mode:
\begin{align}
U_n &= \frac{1}{2}\avg{E^2_n(t)}\int\mathrm{d}^3\mathbf{x}\left[\abs{\mathbf{e}_n(\mathbf{x})}^2 + \abs{\mathbf{b}_n(\mathbf{x})}^2\right] \nonumber\\
&= \lambda_n\int_{-\infty}^{\infty}\frac{\mathrm{d}\omega}{2\pi}\abs{E_n(\omega)}^2 \nonumber\\
&= g^2_{a\gamma\gamma}B_0^2\frac{\kappa_n^2}{\lambda_n}\int_{-\infty}^{\infty}\frac{\mathrm{d}\omega}{2\pi}\frac{\omega^4\abs{a(\omega)}^2}{\left(\omega_n^2 - \omega^2\right)^2 + \omega^2\omega_n^2/Q_n^2} \label{eq:axion_mode_energy}.
\end{align}
using the same Fourier transform convention defined for the axion field [Eq.~\eqref{eq:Parseval}].

Now we invoke the assumption that $\abs{a(\omega)}^2$ is sharply peaked near the cavity resonance. Specifically, we are assuming $\omega_\text{max} \approx m_a = \omega_n -\delta\omega_a$, where the detuning $\delta\omega_a$ is $\sim\mathcal{O}\left(\omega_n/Q_n\right)$, and $Q_a \gg Q_n \gg 1$ (see discussion in Sec.~\ref{sub:haloscope_qual}). Then the coefficient of $\abs{a(\omega)}^2$ is approximately constant over the range of frequencies that contributes appreciably to the integral, and we can set $\omega=m_a$ everywhere outside of $\abs{a(\omega)}^2$. For a high-$Q$ cavity, we can expand to lowest nonvanishing order in the small quantity $\delta\omega_a/\omega_n$. The factors we pulled out of the integral using $Q_a\gg Q_n$ then become
\begin{align*}
\frac{1}{\left[\big(\omega_n/m_a\big)^2-1\right]^2 + \big(\omega_n/m_a\big)^2/Q_n^2}  &\approx \frac{Q_n^2}{Q_n^2\left[\big(1 + 2\delta\omega_a/\omega_n\big)-1\right]^2 + 1} \\
&= \frac{Q_n^2}{\left(2Q_n\delta\omega_a/\omega_n\right)^2 + 1}.
\end{align*}
In the second additive term in the denominator I dropped terms which appear to be higher order in $\delta\omega_a/\omega_n$ but are relatively suppressed by a factor of $Q_n^2$. This is just the well-known result that the transfer function of a high-$Q$ harmonic oscillator is approximately Lorentzian. Applying Eqs.~\eqref{eq:Parseval} and \eqref{eq:rho_a_field} to the $\abs{a(\omega)}^2$ term remaining inside the integral, Eq.~\eqref{eq:axion_mode_energy} becomes
\begin{equation}\label{eq:axion_mode_energy_2}
U_n = g^2_{a\gamma\gamma}B_0^2\frac{\kappa_n^2}{\lambda_n}\frac{Q_n^2}{1 + \left(2Q_n\delta\omega_a/\omega_n\right)^2}\frac{\rho_a}{m_a^2}.
\end{equation}

I have thus far used $n$ to index the cavity modes; for a cylindrical cavity we should really have the three indices $m,n,\ell$. It is conventional to define the \textbf{form factor}
\begin{equation}\label{eq:form_factor}
C_{mn\ell} = \frac{\kappa_{mn\ell}^2}{V\lambda_{mn\ell}} = \frac{\left(\int\mathrm{d}^3\mathbf{x}\,\uv{z}\cdot\mathbf{e}^*_{mn\ell}(\mathbf{x})\right)^2}{V\int\mathrm{d}^3\mathbf{x}\,\varepsilon(\mathbf{x})\abs{\mathbf{e}_{mn\ell}(\mathbf{x})}^2},
\end{equation}
where $V$ is the cavity volume and $\varepsilon(\mathbf{x})$ is the permittivity.\footnote{I have assumed that the vacuum permittivity ($\varepsilon_0=1$ in Heaviside-Lorentz units) is a good approximation throughout this derivation, but a realistic calculation of the form factor should account for any dielectrics inside the cavity.} $C_{mn\ell}$ is a dimensionless quantity independent of both the cavity volume and the normalization of the eigenmodes $\mathbf{e}_{mn\ell}$. It is easy to see that $C_{mn\ell}\leq1$, where the bound is only saturated with $\mathbf{e}_{mn\ell}(\mathbf{x})$ constant and pointing along $\mathbf{z}$ everywhere in the cavity volume. Such a field configuration is of course unphysical, since $\mathbf{e}_{mn\ell}(\mathbf{x})$ must go smoothly to zero at the conducting walls of the cavity: thus very generally $C_{mn\ell}<1$. Physically, the form factor parameterizes the overlap between the cavity mode and the external magnetic field -- this dependence comes directly from the interaction Lagrangian Eq.~\eqref{eq:agammagamma}.

An empty cylindrical cavity generally has both \textbf{transverse electric (TE)} modes with $\mathbf{e}_{mn\ell}\perp\uv{z}$ and \textbf{transverse magnetic (TM)} modes with $\mathbf{b}_{mn\ell}\perp\uv{z}$ (and thus a nonvanishing projection of $\mathbf{e}_{mn\ell}$ onto $\uv{z}$); a cavity with internal conducting rods can also have transverse electromagnetic (TEM) modes. Clearly, $C_{mn\ell}$ will vanish identically for any TE or TEM mode, so we can restrict our attention to TM modes. Moreover, it is easy to see that for the homogeneous external field $B_0\uv{z}$ we have assumed, spatial oscillations in the electric field profile lead to cancellations in $C_{mn\ell}$. For an empty cylindrical cavity, $C_{mn\ell}\neq0$ only for $\text{TM}_{0n0}$ modes, and the form factor for these modes falls rapidly with increasing mode index $n$, which counts the number of radial antinodes in the field profile. For deviations from this ideal cavity geometry, the mode structure can be significantly more complicated, but the above qualitative conclusions still hold; there is generally still a ``$\text{TM}_{010}$-like'' mode which is used in practice by most haloscope detectors. Restricting our focus to the $\text{TM}_{010}$ mode implies that the radius of the cavity will be comparable to the photon wavelength $\nu^{-1}$. We will return to the significance of this point in Sec.~\ref{sub:highfreq}.

We can also take this opportunity to clean up some more notation in Eq.~\eqref{eq:axion_mode_energy_2}. From here on we will only be concerned with a single mode, so I will drop the mode subscript on $U$ and take $\omega_n\rightarrow\omega_c$, $Q_n\rightarrow Q_L$. Our final result for the steady-state energy transferred to the cavity mode by axion field oscillations is
\begin{equation}\label{eq:mode_energy}
U = g^2_{a\gamma\gamma}\frac{\rho_a}{m_a^2}B_0^2VC_{mn\ell}Q_L^2\frac{1}{1 + \left(2Q_L\delta\omega_a/\omega_c\right)^2}.
\end{equation}
Note that $U\propto Q_L^2$ is precisely what we would have obtained from Eq.~\eqref{eq:E_field_axion_q}. The only qualitatively new feature here is the form factor $C_{mn\ell}$; we could have also used an educated guess to obtain the Lorentzian dependence on the detuning $\delta\omega_a$. 

Ultimately, what we care about is not the energy deposited in the mode, but how much power we are able to couple into our receiver chain. For any resonator with multiple loss channels, the total linewidth is just the sum of partial linewidths, which implies the $Q$ factors associated with internal and external losses obey
\begin{equation}\label{eq:q_partial_sum}
\frac{1}{Q_L} = \frac{1}{Q_0} + \frac{1}{Q_r}.
\end{equation}
Here $Q_0$ represents the unloaded quality factor, limited by the nonzero resistance of the cavity walls, and $Q_r$ parameterizes the coupling to the receiver. We can define the dimensionless \textbf{cavity coupling parameter}
\begin{equation}\label{eq:beta_def}
\beta = \frac{Q_0}{Q_r},
\end{equation}
in terms of which
\begin{equation}\label{eq:Q_L}
Q_L = \frac{Q_0}{1+\beta}.
\end{equation}
Referring to $\beta$ as a ``coupling'' is potentially confusing for two distinct reasons. First, we are already using ``coupling'' for both $g_{a\gamma\gamma}$ and $g_\gamma$. Second, in discussions of electromagnetic resonators more generally, ``coupling'' is usually reserved for quantities with dimensions of frequency (such as $\kappa$ in our coupled oscillator example of Sec.~\ref{sub:haloscope_qual}). Nonetheless, this usage is standard in the haloscope literature, so I will stick with it. In Sec.~\ref{sec:cavity} we will see that $\beta$ is more properly regarded as an \textit{impedance ratio}.

The total power dissipated in any electromagnetic resonator is given by
\begin{equation}\label{eq:total_power_res}
P_\text{tot} = \omega_c\frac{U}{Q_L},
\end{equation}
and likewise the signal power coupled to the receiver is
\begin{align*}
P_\text{sig} &= \omega_c\frac{U}{Q_r} \\
&= \omega_c\frac{U}{Q_L}\left(1 - \frac{1}{1+\beta}\right) \\
&=\frac{\beta}{1+\beta}\omega_c\frac{U}{Q_L} \\
&= g^2_{a\gamma\gamma}\frac{\rho_a}{m_a^2}\frac{\beta}{1+\beta}\omega_cB_0^2VC_{mn\ell}Q_L\frac{1}{1 + \left(2Q_L\delta\omega_a/\omega_c\right)^2}.
\end{align*}
As we are now making contact with experiment, we must bid farewell to natural units. Using Eq.~\eqref{eq:photon_coupling_1} and inserting the appropriate fundamental constants, we obtain
\begin{equation}\label{eq:signal_power}
P_{\text{sig}} = \left(g_\gamma^2\frac{\alpha^2}{\pi^2} \frac{\hbar^3c^3\,\rho_a}{\chi}\right)\left(\frac{\beta}{1+\beta}\omega_c\frac{1}{\mu_0}B_0^2VC_{mn\ell}Q_L\frac{1}{1+\left(2\delta\nu_a/\Delta\nu_c\right)^2}\right),
\end{equation}
where the cavity mode linewidth is $\Delta\nu_c = \nu_c/Q_L$ and $\nu_c=\omega_c/2\pi$. Eq.~\eqref{eq:signal_power} is our final result for the observable signal in a haloscope detector. It is valid in any self-consistent set of units, and moreover units may be chosen independently in the two sets of parentheses, which contain theory and detector parameters, respectively. 

We are now in a position to better appreciate the role of the cavity coupling $\beta$. I have written Eq.~\eqref{eq:signal_power} in terms of the loaded cavity quality factor $Q_L$, which is more directly related to measured quantities than $Q_0$ (see Sec.~\ref{sec:cavity}). However, $Q_0$ is the more ``fundamental'' quantity in that it is fully determined by the geometry and material composition of the cavity, whereas $\beta$ (hence $Q_L$) is generally adjustable. Using Eq.~\eqref{eq:Q_L} we can see that $P_\text{sig}\propto\beta/(1+\beta)^2$ is the true dependence of the haloscope signal power on the cavity coupling: then $P_\text{sig}$ is maximized at $\beta=1$. When $\beta=1$ the cavity is said to be \textbf{critically coupled} to the receiver chain: half the signal power is dissipated by the resistance of the cavity walls, and the other half is coupled into the receiver. When the cavity is \textbf{undercoupled} ($\beta<1$), $Q_L$ increases towards $Q_0$, and thus the total axion conversion power $P_\text{tot}$ increases, but most of this power is absorbed internally. When the cavity is \textbf{overcoupled} ($\beta>1$), most of the conversion power is measurable, but the resonant enhancement of the signal is reduced. This is a simple and pleasing picture, but we will see in Sec.~\ref{sub:scan} that critical coupling is nonetheless not the optimal haloscope operating point.

As anticipated in Sec.~\ref{sub:haloscope_qual}, $P_\text{sig}$ is not suppressed by powers of the PQ scale $f_a$ as a consequence of the assumption that axions constitute all of dark matter. Reading out the cavity signal has cost as a factor of $Q_L$ relative to Eq.~\eqref{eq:mode_energy}, but the signal is nonetheless enhanced by the Primakoff effect and temporal coherence as anticipated in Sec.~\ref{sub:haloscope_qual}. We can also appreciate the role of \textit{spatial coherence}. $P_\text{sig}\propto V$ should not be surprising given that the linear scale of the cavity is $\nu_c^{-1}\ll\lambda_a$, but it is quite unlike the usual case in particle physics, where typical figures of merit have the same sublinear scaling with total detector mass ($\propto V$) as with integration time, and thus exposures are quoted in $\text{kg}\cdot\text{yrs}$.

Unfortunately, despite all these enhancement factors, the haloscope signal power is still very small. Plugging in some benchmark parameter values for a detector operating at $\nu_c\sim 5~\text{GHz}$ (see Tab.~\ref{tab:params} in Sec.~\ref{sub:scan}), and assuming the coupling coefficient $g_\gamma^\text{KSVZ}=-0.97$,\footnote{$g_\gamma^\text{KSVZ}$ is the coupling for the original KSVZ model with $e_q=0$ (see Sec.~\ref{sub:model_band}), which will be my standard point of reference.} we obtain $P_\text{sig}\sim5\times10^{-24}~\text{W}$.\footnote{That's 5~\textit{yoctowatts}, where yocto is the smallest SI prefix anybody has bothered to name.} We will get a better sense of just how small $P_\text{sig}$ is when we compare it to the noise power that will inevitably accompany the signal in any real haloscope in Sec.~\ref{sub:radiometer}. The smallness of the signal is challenge number one for any haloscope detector. Challenge number two is that $m_a$ is unknown, so the cavity resonance $\nu_c$ must be tunable: we will return to this latter point in Sec.~\ref{sub:scan}.

\subsection{Coherent detection and the radiometer equation}\label{sub:radiometer}
To understand the noise limiting the sensitivity of a haloscope search, it is important to appreciate that the haloscope signal is itself a \textit{noise signal} in any measurement of duration $\tau>\tau_a\sim 200~\mu\text{s}$. That is, it does not have a fixed relationship to any laboratory reference, and the conversion power will be distributed over a finite bandwidth
\begin{equation}\label{eq:delta_nu_a}
\Delta\nu_a=\frac{1}{\tau_a} = \frac{\nu_a}{Q_a} \sim 5~\text{kHz}
\end{equation}
for $m_a\sim20~\mu\text{eV}$.\footnote{It is important to emphasize that $\Delta\nu_a$ is only an order-of-magnitude estimate and we will need to be more careful when we actually analyze real data in chapter~\ref{chap:analysis}. For our present purposes, Eq.~\eqref{eq:delta_nu_a} will suffice.} Even if we happen to be looking around the right frequency, the presence of axions will manifest in incoherent (if narrowband) fluctuations of the cavity's electric field, which are \textit{in principle} indistinguishable from other contributions to the total noise in the same bandwidth. The situation is not completely hopeless: in this section we will see that we can indeed use statistical methods to constrain the presence of excess noise due to axion conversion, and we can always study the scaling of any putative axion signal with the quantities on the RHS of Eq.~\eqref{eq:signal_power}.\footnote{The scaling with $B_0$ is a particularly powerful probe, in that it is very difficult to imagine how any detector systematic could result in a single nearly monochromatic microwave tone whose amplitude depends on the strength of an applied DC magnetic field. In this respect the haloscope is atypical of detectors used for precision measurements: it is more or less completely free of systematics, and reducing the level of statistical noise is the main challenge.} But the general point merits some emphasis: no matter how hard we look, we will not find a convenient label which tells us whether any individual photon belongs to the signal or the noise.

So what are these other contributions to the total noise that we must contend with? One contribution that will always be present is \textbf{Johnson (thermal) noise}~\citep{nyquist1928} associated with the resistance of the cavity walls. At any finite temperature, power is dissipated by the thermal motion of electrons in an imperfect conductor, and these fluctuating currents imply the presence of fluctuating electromagnetic fields in the interior of the cavity.\footnote{More formally, Johnson noise is a specific manifestation of the fluctuation-dissipation theorem~\citep{cw1951}.} In 1946 Robert Dicke showed using an exceptionally clear and simple thought experiment that the voltage noise on an antenna receiving blackbody radiation is completely equivalent to the Johnson noise of a matched load at the same temperature \citep{dicke1946}. Thus we can also think about Johnson noise as arising from a blackbody photon gas inside the cavity, in thermal equilibrium with the walls. 

In fact, many different physical processes can give rise to contributions to the total noise which are indistinguishable from Johnson noise at some effective temperature. For the present we will simply assume we can write the total noise power in any bandwidth $\Delta\nu$ in the form
\begin{equation}\label{eq:Johnson}
P_N = k_BT_\text{sys}\Delta\nu,
\end{equation}
where the \textbf{system noise temperature} $T_\text{sys}$ is not necessarily equal to the physical temperature $T$ of the haloscope, and I have put in the factor of $k_B$ explicitly because we will henceforth measure temperature in kelvins instead of eV.\footnote{In using the term ``system noise temperature" to refer to the \textit{total} noise power per unit bandwidth, I am following the convention in the haloscope literature. In microwave engineering parlance this term is reserved for noise added by the receiver, and excludes thermal noise that enters the receiver along with the signal of interest.}

For a sense of scale, let's consider the room-temperature thermal noise ($T_\text{sys}\sim300~\text{K}$) in a bandwidth $\Delta\nu_a$: we obtain $P_N\sim2\times10^{-17}~\text{W}$, which is larger than $P_\text{sig}$ in the KSVZ model by a factor of $4\times10^6$! Clearly, cryogenic temperatures are necessary for the haloscope search, but at this point it is by no means obvious that they will be sufficient: a \textbf{dilution refrigerator} can be used to reach a physical temperature of $T\sim30~\text{mK}$, but even if we could just plug this number into Eq.~\eqref{eq:Johnson} to obtain the total noise,\footnote{As we shall see in Sec.~\ref{sub:sql}, this naive estimate would significantly \textit{underestimate} the true noise.} we would still have $P_\text{sig}/P_N\sim2\times10^{-3}$.

The key insight which explains how axion detection is possible in the presence of thermal noise is the following: the figure of merit for a haloscope search should be
\begin{equation}\label{eq:snr_simplest}
R=\frac{P_\text{sig}}{\delta P_N},
\end{equation}
where $\delta P_N$ is the \textit{uncertainty} in our estimate of the total noise power, and $R$ is usually called the \textbf{signal-to-noise ratio (SNR)}. Eq.~\eqref{eq:snr_simplest} is just the statement that if we understood what the total noise power ought to be with zero uncertainty, we could attribute any arbitrarily small excess power to axions or other new physics! To make further progress, we will need to write down a specific expression for $\delta P_N$ in an axion haloscope, which in turn requires that we specify whether we intend to couple our microwave cavity to a \textbf{coherent} or \textbf{incoherent} receiver. 

The defining feature of a coherent or \textbf{linear} receiver is that its output is a linear function of the electric field at the receiver input. This definition implies that a coherent receiver has \textit{intrinsic spectral resolution}: the Fourier component of the input signal at frequency $\omega$ will show up at the same frequency in the output signal, and the receiver itself does not limit how precisely we can measure the frequency $\omega$.\footnote{As a corollary, any coherent phase relationship between Fourier components at the input is preserved at the output, and this feature gives the coherent receiver its name.} In an incoherent or \textbf{bolometric} receiver, the output is proportional to the average \textit{intensity} (i.e., the square of the input), and thus does not preserve the phase information in the input signal. The spectral resolution of an incoherent receiver is limited by the receiver bandwidth, which may be controlled by the actual nonlinear detector element or by an input filter. The fact that we can always place linear components such as filters at the input of an incoherent receiver or measure intensity at the output of a coherent receiver suggests that coherent and incoherent receivers are not fundamentally different. Treating both regimes in a unified formalism \citep{zmuidzinas2003} reveals that the distinction really hinges on whether the first element to \textit{amplify} the input signal and protect the SNR from further degradation is linear or nonlinear.

In principle, the discussion in the above paragraph applies to detectors of electromagnetic radiation at any frequency. In practice, technological limitations conspire to favor coherent detection at low frequencies $\nu\ll1$~THz and bolometric detection at high frequencies $\nu\gg1$~THz. Roughly speaking, this is because the electronic response of common materials is too slow to implement a linear amplifier at IR and optical frequencies, while at microwave frequencies the small energies of individual photons make high-efficiency bolometric detection (\textbf{single-photon detection}) challenging. For this reason, all haloscope searches to date including HAYSTAC have used coherent receivers, and I will restrict my focus to the coherent case in this thesis. It should be noted however that the past few years have seen tremendous progress towards microwave single-photon detection; we will encounter one respect in which a single-photon detector would have a significant advantage over a coherent receiver in Sec.~\ref{sub:sql}.

In practice, we couple the microwave cavity to a coherent receiver by inserting a small antenna through a port in one of the endcaps as illustrated in Fig.~\ref{fig:haloscope_simple}: the motion of electrons in the antenna then transduces the cavity's fluctuating electric field $E_z$ into a fluctuating voltage at the receiver input. This signal is amplified by the receiver chain, digitized at room temperature, and ultimately processed using a discrete Fourier transform and analyzed in the frequency domain as a power spectrum. To understand the sensitivity of the haloscope search, it is sufficient to consider the statistics of the noise voltage at the receiver input, where we evaluated the signal power in Eq.~\eqref{eq:signal_power}.

As noted above, I will assume the total noise may be modeled as Johnson noise at some effective temperature $T_\text{sys}$. This essentially amounts to the assumption that the amplitude distribution of the noise voltage within a bandwidth $\Delta\nu$ is Gaussian. The noise power Eq.~\eqref{eq:Johnson} is proportional to the variance of this voltage distribution. Invoking the formula for the standard error of the variance of a Gaussian distribution, we obtain
\begin{equation}\label{eq:delta_power_initial}
\delta P_N = \sqrt{\frac{2}{n-1}}k_BT_\text{sys}\Delta\nu,
\end{equation}
where $n$ is the number of independent samples drawn from the Gaussian distribution, invariably sufficiently large that we take $n-1\approx n$. Finally, the Nyquist sampling theorem guarantees that noise limited to a bandwidth $\Delta\nu$ and measured for a time $\tau$ is completely represented by $n=2\Delta\nu\tau$ independent samples. Using this result in Eq.~\eqref{eq:delta_power_initial}, we obtain
\begin{equation}\label{eq:delta_power}
\delta P_N = \frac{k_BT_\text{sys}\Delta\nu}{\sqrt{\Delta\nu\tau}}.
\end{equation}

Eq.~\eqref{eq:delta_power} describes the uncertainty in the (effective) Johnson noise power in any bandwidth $\Delta\nu$. It is easy to see that the optimal bandwidth for a haloscope search is $\Delta\nu\approx\Delta\nu_a$: for $\Delta\nu<\Delta\nu_a$ we are not capturing the whole signal, and for $\Delta\nu > \Delta\nu_a$ we are adding more noise without appreciably increasing the signal power.\footnote{Note that up to small corrections due to the details of the axion signal lineshape, the total signal power and total noise power both scale linearly with $\Delta\nu$ for $\Delta\nu<\Delta\nu_a$: in this regime the sensitivity only improves with increasing $\Delta\nu$ because a larger bandwidth implies that we obtain more independent samples of the noise in any integration time $\tau$. This is a standard result in radiometry~\citep{pozar2012}, and is a good example of why it is useful to think of axion conversion as a noise signal.} Taking $\Delta\nu=\Delta\nu_a$ in Eq.~\eqref{eq:delta_power} and using Eq.~\eqref{eq:snr_simplest}, we obtain the \textbf{Dicke radiometer equation}
\begin{equation}\label{eq:dicke}
R = \frac{P_\text{sig}}{k_BT_\text{sys}}\sqrt{\frac{\tau}{\Delta\nu_a}}.
\end{equation}
Eq.~\eqref{eq:dicke} tells us the time $\tau$ required to detect or exclude axion conversion power $P_\text{sig}$ assuming the system noise temperature $T_\text{sys}$ is known. We will consider the SNR required for a haloscope search more carefully when we work out the details of the analysis in Sec.~\ref{sec:candidates}: for now I will assume we demand $R=5$ to estimate the sensitivity of a search with a given set of experimental parameters.

To appreciate the consequences of Eq.~\eqref{eq:dicke} for experimental design, recall from Eq.~\eqref{eq:signal_power} that the signal power is appreciable within a bandwidth of order $\Delta\nu_c\gg\Delta\nu_a$ centered on $\nu_c$: the intrinsic spectral resolution of a coherent receiver then implies that a single measurement simultaneously probes $\sim\Delta\nu_c/\Delta\nu_a$ independent possible values that of $m_a$.\footnote{Another consequence of this observation is that a haloscope employing a coherent receiver is more sensitive by a factor of $\sqrt{\Delta\nu_c/\Delta\nu_a}$ than one that uses a bolometric receiver with the same value of $k_BT_\text{sys}$ (i.e., the same total noise power per unit bandwidth). This enhancement factor can be obtained in either of two ways. If the bandwidth in the bolometric case is set by the cavity, then clearly we must take $\Delta\nu_a\rightarrow\Delta\nu_c$ in Eq.~\eqref{eq:dicke}. We can recover the optimal bandwidth by placing a narrowband filter at the input of the bolometric receiver, but then we would have to make $\Delta\nu_c/\Delta\nu_a$ independent measurements of duration $\tau$ with different filter settings.} At most one of these values will be realized in nature, and this implies that if $T_\text{sys}$ were independent of frequency, axion conversion would manifest as a small ``bump'' at frequency $\nu_a$ in an otherwise flat power spectrum. In practice, $T_\text{sys}$ \textit{will} exhibit frequency-dependence, but only on spectral scales $\gtrsim\Delta\nu_c$, which is the smallest frequency scale in the detector design.

The separation of spectral scales $\Delta\nu_a\ll\Delta\nu_c$, which is ultimately a consequence of the difficulty of achieving high $Q$ factors at microwave frequencies with normal metals (see Sec.~\ref{sub:highfreq}), thus has important consequences for how haloscope searches are conducted in practice. A spectrally localized power excess is a much less ambiguous signature of axion conversion than a discrepancy between the measured value of $T_\text{sys}$ and the value it \textit{ought} to have across the whole spectrum; the latter could be caused by any number of more mundane detector effects. Specifically, $\Delta\nu_a\ll\Delta\nu_c$ implies that we do not need to measure $T_\text{sys}$ itself with high precision, and thus we do not require exquisite control over receiver gain fluctuations as in many applications of sensitive microwave radiometry (e.g., measurements of the CMB).

\subsection{Quantum noise limits}\label{sub:sql}
Having clarified the nature of the receivers used in real haloscope detectors and derived the radiometer equation, we can now consider the various contributions to $T_{\text{sys}}$. For any haloscope at physical temperature $T$ coupled to a coherent receiver, the most general expression for the system noise temperature is
\begin{equation}\label{eq:system_noise}
k_BT_{\text{sys}} = h\nu N_{\text{sys}} = h\nu\left(\frac{1}{e^{h\nu/k_BT} -1} + \frac{1}{2} + N_A\right).
\end{equation}
The first additive term on the RHS represents the actual thermal noise, which we see is proportional to the average number of blackbody photons inside the cavity at physical temperature $T$: this is Dicke's correspondence between Johnson noise and blackbody radiation in action! The second term is \textbf{quantum noise} associated with the zero-point fluctuations of the blackbody gas. The \textbf{added noise} $N_A$ encompasses the total \textit{input-referred} noise originating in the receiver itself.\footnote{Eq.~\eqref{eq:system_noise} indicates that $T_\text{sys}$ is explicitly frequency-dependent in principle. In practice, the small fractional bandwidth of a high-$Q$ cavity implies that this frequency-dependence is negligible in any given power spectrum centered on the cavity resonance $\nu_c$: $\nu\approx\nu_c$ is a good approximation. The \textit{implicit} frequency dependence of $N_A$ (discussed in Sec.~\ref{sec:noise}) is a much larger effect.}

Prior to the work described in this thesis, all haloscope detectors had operated in the Rayleigh-Jeans limit $k_BT \gg h\nu_c$, where Eq.~\eqref{eq:system_noise} reduces to 
\begin{equation}\label{eq:system_noise_rj}
T_\text{sys}\approx T + T_A,
\end{equation}
with $T_A=(h\nu_c/k_B)N_A$. This expression is often cited as an equality in the haloscope literature, but it is clearly invalid at sufficiently low temperatures. In the Wien limit $k_BT \ll h\nu_c$, the thermal contribution is exponentially suppressed and quantum effects dominate: the total noise is then independent of the physical temperature $T$. The behavior in the Wien limit implies that units of quanta, used throughout this thesis, are more appropriate than temperature units for sufficiently low-noise receivers.

The HAYSTAC detector described in this thesis has reached the regime where the approximate expression Eq.~\eqref{eq:system_noise_rj} is no longer accurate, so we ought to consider more carefully the limits imposed by quantum mechanics on the haloscope signal measurement. We can begin with the observation that any classical narrowband ($\Delta\nu_c \ll \nu_c$) voltage signal may be written in the two equivalent forms
\begin{align}
V(t) &= V_0\Big[X_1(t)\cos(2\pi\nu_c t) + X_2(t)\sin(2\pi\nu_c t)\Big] \label{eq:E_quadratures}\\
 &= \frac{V_0}{2}\Big[a(t)e^{-2\pi i\nu_ct} + a^*(t)e^{2\pi i\nu_ct} \Big] \label{eq:raising_lowering}
\end{align}
where $X_1$ and $X_2$ are called the signal \textbf{quadratures}.\footnote{We will not need to consider the precise time-dependence of the quadrature amplitudes except to note that it occurs on long timescales $\sim1/\Delta\nu_c$. For the purpose of developing intuition, it may be useful to note that if we peel off the DC term ($X_1(t)\rightarrow A_0+X_1(t)$), then $X_1$ and $X_2$ map onto amplitude and phase modulation respectively in the limit of small modulation depth $X_1/A_0,X_2/A_0\ll 1$.}  In the case of the haloscope, $V(t)$ is the voltage on the antenna at the receiver input, which is proportional to the electric field of the cavity mode of interest.

The standard procedure for quantizing field amplitudes is to promote $a$ and $a^*$ to operators $\hat{a}$ and $\hat{a}^\dag$ with the commutation relation $[\hat{a},\hat{a}^\dag]=1$. Then the Hamiltonian for the cavity mode is just that of the quantum harmonic oscillator: $\hat{H} = h\nu_c\big(\hat{N} + 1/2\big)$ with $\hat{N}=\hat{a}^\dag\hat{a}$. The equivalence of Eqs.~\eqref{eq:raising_lowering} and \eqref{eq:E_quadratures} then implies that the Hamiltonian may be written in the form
\begin{equation}\label{eq:cavity_hamiltonian}
\hat{H} = \frac{h\nu_c}{2}\big(\hat{X}_1^2 + \hat{X}_2^2\big),
\end{equation}
with
\begin{equation}\label{eq:quadrature_commutation}
\big[\hat{X}_1,\hat{X}_2\big]=\frac{i}{2}.
\end{equation}
That is, the quadrature amplitudes are just like the position and momentum in the prototypical quantum harmonic oscillator, which of course also oscillate $90^\circ$ out of phase.

We can now understand the origin of the quantum noise term in Eq.~\eqref{eq:system_noise} as a consequence of the basis in which we choose to measure the cavity mode. A coherent receiver as defined above measures the quadrature amplitudes $\hat{X}_1$ and/or $\hat{X}_2$, whereas a bolometric receiver measures $\hat{N}$. Quantum noise arises because neither $\hat{X}_1$ nor $\hat{X}_2$ commutes \textit{with the Hamiltonian}: this is the hefty price we pay for the intrinsic spectral resolution of a coherent receiver.

Unfortunately, the second term in Eq.~\eqref{eq:system_noise} is not the only limit imposed by quantum mechanics on the sensitivity of a haloscope detector. We have thus far totally ignored the third term representing the input-referred added noise $N_A$. Before we consider the constraints that quantum mechanics places on $N_A$, it is useful to consider in very general terms where this contribution comes from. As noted in Sec.~\ref{sub:radiometer}, the crucial element which defines a coherent receiver is a \textbf{linear amplifier} whose output is just the input signal multiplied by some gain $G\gg1$.\footnote{In practice, $G$ may be frequency-dependent, but we assume this frequency-dependence is known or measurable.} The energy required to achieve this amplification has to come from somewhere: totally independent of operational details, any device capable of amplifying a signal can also couple noise into the output signal through whatever channel serves as the ``power line.'' 

By definition, the added noise of any amplifier is not already physically present in the input signal. However, it is conventional to define the input-referred added noise $N_A$ as the physical added noise divided by the gain $G$. The input-referred added noise is just the additional noise that would have to be present at the input of an ideal noiseless amplifier with gain $G$ to reproduce the noise we actually observe at the real amplifier's output. This definition is convenient because it allows us to define the input-referred added noise of the entire receiver chain in an analogous way. It is easy to see that if the receiver chain includes several linear amplifiers with gain $G_i$ and added noise $N_{Ai}$ in series, the net input-referred added noise of the receiver chain will be
\begin{equation}\label{eq:amp_chain}
N_A = N_{A1} + \frac{1}{G_1}N_{A2} + \frac{1}{G_1G_2}N_{A3} + \cdots
\end{equation}
The takeaway point is that the noise performance of the first amplifier (or \textbf{preamplifier}) determines the added noise of the whole receiver chain, provided it has sufficiently high gain $G_1$.\footnote{Going forward, I will always quote contributions to the total noise referred to the input of the receiver, and not explicitly indicate ``input-referred'' except where omitting it would lead to confusion.} A low-noise preamplifier is thus just as important to the feasibility of the haloscope search as low physical temperature $T$.\footnote{In the preceding discussion I have focused on contributions to $N_A$ from amplifiers, but we must also account for other contributions in the design of a practical low-noise receiver (see discussion in Sec.~\ref{sub:receiver_noise}). In particular, lossy elements in the signal chain will degrade the SNR, and the preamplifier should be placed as close to the input of the receiver chain as possible in order to minimize this loss contribution.}

Naively, it seems that we should be able to reduce $N_A$ as far as we want by being sufficiently careful in the design of a high-gain preamplifier. However, it can be proved that any \textbf{phase-insensitive linear amplifier} must contribute
\begin{equation}\label{eq:haus_caves}
N_A\geq1/2
\end{equation}
noise quanta to the output signal: this is called the \textbf{Haus-Caves theorem}~\citep{haus1962,caves1982}. A phase-insensitive linear amplifier is just a linear amplifier which applies the same gain to both quadratures $X_1$ and $X_2$: since our choice of reference phase in Eq.~\eqref{eq:E_quadratures} was arbitrary, this definition implies that the gain is independent of the phase of the input signal.\footnote{Beware! A phase-insensitive linear amplifier should not be confused with a (nonlinear) bolometric amplifier, which is occasionally called ``phase-insensitive'' for a different reason: it throws away all information about the phase of the input signal.} The simple quantum optics picture of the cavity mode developed above suggests an explanation for this result: a phase-insensitive linear amplifier is trying to measure both quadrature amplitudes, which do not commute \textit{with each other} [c.f.\ Eq.~\eqref{eq:quadrature_commutation}].

The same formalism may be used to treat the more general case of phase-sensitive linear amplifiers which apply different gains to the two quadratures. Of particular interest is the case of a \textbf{single-quadrature linear amplifier}, which amplifies the signal in the $X_1$ quadrature and \textit{deamplifies} $X_2$, such that the product of the quadrature gains is unity. There is no lower bound on the added noise of single-quadrature amplifiers~\citep{caves1982}: this is consistent with our qualitative picture in which the Haus-Caves theorem is a consequence of the non-commutation of the two quadratures, since the single-quadrature amplifier does not ``measure'' $\hat{X}_2$ in any meaningful way. See Ref.~\citep{caves1982} for a much more thorough and very lucid account of quantum limits on the added noise of linear amplifiers.

In summary, we have learned that ``quantum mechanics extracts its due twice''~\citep{caves1982}. The Haus-Caves theorem and the input quantum noise together imply the \textbf{Standard Quantum Limit (SQL)}
\begin{equation}\label{eq:sql}
N_\text{sys} \geq 1
\end{equation} 
on the total noise in any haloscope coupled to a phase-insensitive linear amplifier.\footnote{Sometimes Eq.~\eqref{eq:haus_caves} is referred to as the SQL on the noise performance of a linear amplifier. I will reserve the term SQL for the statement about the total noise $N_\text{sys}$, which is what ultimately limits the sensitivity of any haloscope search.} The SQL may also be parameterized in temperature units as
\begin{equation}\label{eq:t_sql}
T_\text{SQL} = (h/k_B)\nu_c\sim 240~\text{mK}
\end{equation}
for $\nu_c\sim5~\text{GHz}$. Eq.~\eqref{eq:t_sql} is a useful approximate relation to bear in mind when considering the design of a haloscope using a coherent receiver. At any given operating frequency, the SNR will stop improving with decreasing physical temperature $T$ for $T\lesssim T_\text{SQL}$. A hypothetical microwave single-photon detector compatible with other requirements of haloscope design would not be subject to the SQL, and the realization of such a detector would have profound implications for future high-frequency haloscope searches (or searches at sufficiently low temperatures; see Ref.~\citep{lamoreaux2013}). I will discuss the prospects for single-photon detection in the haloscope search briefly in Sec.~\ref{sub:haystac}.

But I am getting ahead of myself in discussing possible schemes for circumventing the SQL when \textit{reaching} the SQL is still quite challenging with present amplifier technology. All of my discussion thus far has been based on general principles, with no indication of how we might actually realize an amplifier whose noise performance is consistent with Eq.~\eqref{eq:haus_caves}. Two such amplifiers are introduced in Sec.~\ref{sub:admx} and Sec.~\ref{sub:haystac}. For the moment, I will assume the existence of such an amplifier, and take $N_\text{sys}=2$~quanta as a benchmark value for a haloscope search at $\nu_c\sim5~\text{GHz}$.

\subsection{Haloscope scan rate}\label{sub:scan}
Eqs.~\eqref{eq:signal_power}, \eqref{eq:dicke}, and \eqref{eq:system_noise} may be combined to estimate the sensitivity of a haloscope operating at a fixed microwave frequency $\nu_c$, and obtain a constraint on $\abs{g_\gamma}$ in the mass range $h(\nu_c-\Delta\nu_c)/c^2\lesssim m_a\lesssim h(\nu_c+\Delta\nu_c)/c^2$. In practice, haloscopes are designed to be tunable to probe a wider range of possible values of $m_a$. Roughly speaking, we want to sit at one frequency $\nu_c$ and measure the cavity noise for whatever integration time $\tau$ is required to attain the desired sensitivity, then increment $\nu_c$ by $\delta\nu_c$ and repeat this process many times.

The axion-sensitive resonant mode of any microwave cavity can be tuned by using an automated mechanical system to adjust the position of internal conductors or dielectrics.\footnote{The physical implementation of haloscope tuning systems is discussed in Sec.~\ref{sub:admx} and Sec.~\ref{sec:cavity}. Generally speaking a given microwave resonator can be tuned by at most a factor of 2, and often the useful contiguous tuning range will be much smaller. The tuning range of any given haloscope can of course be extended by swapping out the cavity, but in practice the range of accessible frequencies will also be limited by the bandwidth of receiver components.} In practice other cavity parameters such as $Q_L$, $\beta$, and $C_{mn\ell}$ will vary throughout the tuning range, and the total noise $N_\text{sys}$ can exhibit both fine frequency dependence (i.e., dependence on the detuning from $\nu_c$ within any given spectrum) and coarse frequency dependence (i.e, variation with $\nu_c$ across spectra). In analyzing real data, we must of course take this frequency-dependence into account, and I will explain how we do so in chapter~\ref{chap:analysis}. For now I will assume all haloscope parameters except $\nu_c$ remain constant throughout the tuning range, and moreover that the tuning step size is some constant faction $1/F$ of the cavity linewidth (i.e., $\delta\nu_c=\Delta\nu_c/F$). These assumptions will enable us to determine an appropriate value for $\delta\nu_c$ and derive a useful approximate expression relating the haloscope scan rate $\mathrm{d}\nu/\mathrm{d}t\approx \delta\nu_c/\tau$ to the coupling sensitivity $\abs{g_\gamma}$. 

We can begin with the radiometer equation Eq.~\eqref{eq:dicke} for a single measurement in which the cavity resonance was at frequency $\nu_c=\nu_0$, and take $\tau$ to be fixed. Then the SNR $R(\nu_a)$ will depend on the detuning $\delta\nu_a$ between the putative axion mass and the mode frequency; specifically, it will inherit the Lorentzian dependence of Eq.~\eqref{eq:signal_power} if $N_\text{sys}$ is frequency-independent. At $\nu_a=\nu_0$, this measurement alone yields
\begin{equation}\label{eq:radiometer_nu0}
R(\nu_0) = \frac{1}{h\nu_0N_\text{sys}}\sqrt{\frac{\tau}{\Delta\nu_a}}P_\text{sig}(\delta\nu_a=0).
\end{equation}
If $\delta\nu_c\lesssim\Delta\nu_c$ (equivalently, $F\gtrsim1$), measurements at several consecutive tuning steps will contribute to the SNR at each frequency $\nu_a$ in the tuning range. Let us restrict our focus to $\nu_a=\nu_0$ for definiteness: clearly Eq.~\eqref{eq:radiometer_nu0} represents the largest single contribution to $R(\nu_0)$, but the symmetric dependence of Eq.~\eqref{eq:signal_power} on the detuning indicates that in general we should also consider the contributions from $K$ consecutive tuning steps in either direction from $\nu_c=\nu_0$, where the appropriate value of $K$ will be determined shortly.

Contributions to the SNR at any frequency from different integrations add in quadrature,\footnote{We can motivate this scaling by considering a single measurement of duration $\tau_0$ with peak SNR $R_0$, and noting that merely dividing the data into two segments of duration $\tau_1$ and $\tau_2=\tau_0-\tau_1$ should not change the SNR. From Eq.~\eqref{eq:dicke} we have $R_i=R_0\sqrt{\tau_i/\tau_0}$ for $i=1,2$, from which the desired result follows readily. In Sec.~\ref{sec:rescale_combine} we will obtain this result from a different perspective, and see explicitly that it still holds if quantities other than the integration time differ between measurements.} so we can include the contributions from the surrounding tuning steps by defining the total squared SNR at $\nu_a=\nu_0$ to be
\begin{equation}\label{eq:radiometer_nu0_total}
R(\nu_0)^2 = \frac{1}{(h\nu_0N_\text{sys})^2}\frac{\tau}{\Delta\nu_a}\sum_{k=-K}^{K}P_\text{sig}^2(\delta\nu_a = k\delta\nu_c).
\end{equation}
Next we multiply both sides of Eq.~\eqref{eq:radiometer_nu0_total} by $(\delta\nu_c/\tau)R(\nu_0)^{-2}$ to obtain an expression for the scan rate
\begin{equation}\label{eq:scan_rate_0}
\frac{\mathrm{d}\nu}{\mathrm{d}t} = \frac{\delta\nu_c}{\tau} = \frac{1}{R(\nu_0)^2}\frac{P^2_\text{sig}(\delta\nu_a=0)}{(h\nu_0N_\text{sys})^2}\frac{1}{\Delta\nu_a}\frac{\Delta\nu_c}{F}\sum_{k=-K}^{K}\left(\frac{1}{1 + (2k/F)^2}\right)^2.
\end{equation}
The scan rate is proportional to the dimensionless expression
\begin{equation}\label{eq:Z_F_K}
Z(F,K) = \frac{1}{F}\sum_{k=-K}^{K}\left(\frac{1}{1 + (2k/F)^2}\right)^2 \approx \frac{4}{5}
\end{equation}
for any $F \geq 2$ and any $K\gtrsim F$.\footnote{Some examples: $Z(2,\infty)=0.81$, $Z(2,2)=0.79$, $Z(20,\infty)=0.79$, $Z(20,20)=0.76$.} Qualitatively, a smaller step size implies that the same sensitivity can be achieved with smaller integration time $\tau$ at each step, and the two effects cancel out in the scan rate. Note also that taking $K=F$ corresponds to designating an \textbf{analysis band} of full width $2\Delta\nu_c$ centered on $\nu_c$ in the power spectrum derived from each measurement and ignoring all data at larger detunings, which won't contribute appreciably to the SNR anyway.\footnote{This is of course an arbitrary cutoff: the scan rate required to achieve a given SNR would change by by $<10\%$ if we were to take the analysis band width to be $1.5\Delta\nu_c$ instead. Some conservatism in parameter choices is warranted, however, since in a real experiment neither the cavity linewidth $\Delta\nu_c$ nor the fractional step size $F$ will be constant over the tuning range. I will discuss how we set the analysis band in practice in Sec.~\ref{sub:if_band}.}

Before we use Eq.~\eqref{eq:Z_F_K} to write out the expression for the scan rate in full, we can note that the ratio $\Delta\nu_c/\Delta\nu_a$ that appears in Eq.~\eqref{eq:scan_rate_0} is equivalent to $Q_a/Q_L$, and the factor of $Q_L$ in the denominator will cancel one power of $Q_L$ in $P_\text{sig}^2$. Also, the factor of $\nu_0$ in Eq.~\eqref{eq:signal_power} will cancel the factor multiplying $N_\text{sys}$ in the denominator. Our final expression for the scan rate is
\begin{equation}\label{eq:scan_rate}
\frac{\mathrm{d}\nu}{\mathrm{d}t} \approx \zeta\frac{4}{5}\frac{Q_LQ_a}{R^2}\left(g_\gamma^2\frac{\alpha^2}{\pi^2}\frac{\hbar^3c^3\rho_a}{\chi} \right)^2\left(\frac{1}{\hbar\mu_0}\frac{\beta}{1+\beta}B_0^2VC_{mn\ell}\frac{1}{N_{\text{sys}}}\right)^2.
\end{equation}
where I have introduced an efficiency factor $\zeta$ to account for dead time during a data run. We see that provided we take frequency steps $\delta\nu_c\lesssim\Delta\nu_c/2$, we can obtain essentially uniform parameter space coverage, up to variation in the tuning step size and frequency-dependence of detector parameters; thus I have taken $R(\nu_0)=R$.

Eq.~\eqref{eq:scan_rate} is the most useful figure of merit for the design of a haloscope detector. Most of the scaling with detector parameters follows directly from Eqs.~\eqref{eq:signal_power} and Eq.~\eqref{eq:dicke}, but the scaling with $Q_L$ (or equivalently with $Q_0$) is \textit{weaker} than that of Eq.~\eqref{eq:signal_power}: thus cavity design studies should seek to maximize $C_{mn\ell}^2Q_0$ rather than $C_{mn\ell}^2Q_0^2$. This makes sense: when we increase $Q_L$, we enhance the signal power but also reduce the cavity linewidth, slowing the scan rate. Once the cavity design (hence $Q_0$) is fixed, we can similarly re-evaluate the optimal value of $\beta$. We see that $\frac{\mathrm{d}\nu}{\mathrm{d}t}\propto\beta^2/(1+\beta)^3$, which is maximized when the cavity is overcoupled with $\beta=2$. This is a consequence of exactly the same physics responsible for changing the scaling with $Q$.

In practice, we choose a target sensitivity $\abs{g_\gamma}$ for which the scan rate is reasonable assuming some nominal values for the detector parameters, and then determine the required measurement time $\tau$ at each step using
\begin{equation}\label{eq:tau_scan}
\tau = \frac{\zeta}{M}\delta\nu_c\left(\frac{\mathrm{d}\nu}{\mathrm{d}t}\right)^{-1},
\end{equation}
where we may opt to split the total integration time required at each frequency across $M$ different scans over the tuning range: $M>1$ helps ensure that transient phenomena during a data run do not completely eliminate sensitivity at any frequency.

Benchmark parameter values for the HAYSTAC detector are shown in Tab.~\ref{tab:params}. The quoted values of the cavity volume $V$ and external magnetic field strength $B_0$ represent constants of the HAYSTAC detector design; other parameter values represent reasonable estimates of what is achievable with a detector operating in the $\nu_c\sim5~\text{GHz}$ range.\footnote{The value of $V$ quoted here is the clear volume of the cavity, excluding the space occupied by a large tuning rod. The same convention is used in the normalization of $C_{mn\ell}$.} In the first HAYSTAC data run over the frequency range $5.7<\nu_c <5.8~\text{GHz}$, the measured values of most parameters were typically quite close to those listed in Tab.~\ref{tab:params}; $N_\text{sys}$ was roughly 50\% larger due to a poor thermal link between the tuning rod and the rest of the cryogenic system. See Sec.~\ref{sec:cavity} for measured values of the cavity parameters, Sec.~\ref{sec:noise} for measurements of $N_\text{sys}$, and Sec.~\ref{sec:daq} for the values of $\zeta$ and $\delta\nu_c$ in the first HAYSTAC data run. 
\begin{table}[htbp]
\centering
\begin{tabular}{|c|c|c|c|c|c|c|c|c|c|}
\hline
\rule{0pt}{2.8ex}$\boldsymbol{\nu_c}$& $\boldsymbol{Q_L}$ & $\boldsymbol{\beta}$ & $\boldsymbol{C_{010}}$ & $\boldsymbol{V}$ &  $\boldsymbol{B_0}$ &  $\boldsymbol{N_\text{sys}}$ & $\boldsymbol{\zeta}$ & $\boldsymbol{\delta\nu_c}$ \\ \hline
\rule{0pt}{2.8ex}5~GHz & $10^4$ &2 & 0.5 & 1.5~L& 9 & 2 & 0.7 & 75~kHz \\ \hline
\end{tabular}
\caption[Benchmark parameters for the HAYSTAC detector]{\label{tab:params} Benchmark detector parameters for estimating HAYSTAC sensitivity. See Sec.~\ref{sec:cavity}, Sec.~\ref{sec:daq}, and Sec.~\ref{sec:noise} for discussion of the values actually obtained during the first HAYSTAC data run.}
\end{table}

Using the parameters in Tab.~\ref{tab:params} in Eq.~\eqref{eq:scan_rate} along with $R=5$, $Q_a=10^6$, $\rho_a=0.45~\text{GeV/cm}^3$, and $\chi=[77.6~\text{MeV}]^4$, we find that the scan rate required for KSVZ sensitivity is $\frac{\mathrm{d}\nu}{\mathrm{d}t} \sim 40$~MHz/year. Backing off to $\abs{g_\gamma}=2\times\abs{g_\gamma^\text{KSVZ}}$ we obtain $\frac{\mathrm{d}\nu}{\mathrm{d}t} \sim 600$~MHz/year, which is much more palatable. With $\delta\nu_c\approx75~\text{kHz}$ ($F\approx7$), the latter scan rate implies 45~minutes of integration are required at each frequency. In practice we can split this integration time across $M=3$ full scans over the range, to obtain $\tau=15$~minutes of integration time per step.

\section{Haloscopes in practice}\label{sec:searches}
A number of experiments seeking to directly detect the interactions of CDM axions have been conducted since the publication of Ref.~\citep{sikivie1983}. In Sec.~\ref{sub:admx}, I will briefly review the design and results of haloscope detectors from the late 1980s to 2012, to put my work on the HAYSTAC detector in context. I will discuss the challenges facing haloscopes at high frequencies in Sec.~\ref{sub:highfreq}, and the motivation for HAYSTAC specifically in Sec.~\ref{sub:haystac}. 

\subsection{The pilot experiments and ADMX}\label{sub:admx}
The first haloscope detector was designed and constructed at Brookhaven National Lab within a few years of Sikivie's proposal, operated by a Rochester-Brookhaven-Fermilab collaboration (hereafter \textbf{RBF}). An independent group at the University of Florida (\textbf{UF}) created the second operational haloscope shortly thereafter. The first results from the RBF~\citep{RBF1987} and UF~\citep{UF1990} experiments were in many ways quite similar: both demonstrated sensitivity within a factor of 10 of the axion model band within distinct $100~\text{MHz}$ windows near $\nu_a\sim1.2~\text{GHz}$ ($m_a\sim5~\mu\text{eV}$). The RBF [UF] experiment used the TM$_{010}$ mode of a $V=10~\text{L}$ [$7~\text{L}$] cylindrical copper cavity with $Q_0\approx1.9\times10^5$ [$1.5\times10^5$] and a solenoidal magnet with central field $B_0=6~\text{T}$ [$7.5~\text{T}$]; in both experiments, cavity mode tuning was accomplished by the axial insertion of a dielectric rod. 

Both experiments operated with the cavity immersed in liquid helium and coupled to a cryogenic transistor-based preamplifier, resulting in total noise well into the Rayleigh-Jeans regime [Eq.~\eqref{eq:system_noise_rj}]. The helium bath in the RBF experiment was at atmospheric pressure ($T\approx4.2~\text{K}$), and the amplifier contributed $T_a\sim12~\text{K}$; thus the total noise was $N_\text{sys}\sim280$ in units of quanta. The UF experiment achieved better noise performance using pumped liquid helium ($T\approx2.2~\text{K}$) and a state-of-the art \textbf{high-electron mobility transistor (HEMT)} amplifier with $T_a\sim3\text{K}$. The RBF group took more data after this initial result, ultimately expanding their coverage up to almost $4~\text{GHz}$ ($16~\mu\text{eV}$) with somewhat reduced sensitivity~\citep{RBF1989}. These first-generation haloscopes achieved impressive sensitivity for what were essentially proof-of principle experiments, but nonetheless it was clear that reaching the axion model band above $\nu_a\sim1~\text{GHz}$ was simply not feasible with the technology of the late 1980s.

A decade later the first operation of the \textbf{Axion Dark Matter eXperiment (ADMX)} set the first limits in the model band, excluding axion models with $\abs{g_\gamma}\geq\abs{g_\gamma^\text{KSVZ}}$ and $700~\text{MHz} \leq \nu_a \leq 800~\text{MHz}$~\citep{ADMX1998}. The initial incarnation of the ADMX detector, sited at Lawrence Livermore National Lab, comprised a 7.6~T magnet with a 60~cm bore diameter, a cylindrical copper cavity with $V=200~\text{L}$ and $Q_0\approx1.5\times10^5$ for the $\text{TM}_{010}$ mode, and a HEMT preamplifier with $T_a\approx2$~K; the cavity and HEMT were maintained in equilibrium with a pumped helium bath with $T\approx 1-2$~K~\citep{ADMX2000}. 

Most of these parameters are comparable to those of the RBF and UF detectors: the improved sensitivity was mainly due to the much larger cavity volume $V$, which was in turn a consequence of lower-frequency operation. The $\text{TM}_{010}$ mode frequency for an empty cylindrical cavity with radius $r$ is $\nu_{c}\sim c/r$ [the exact expression will be given in Eq.~\eqref{eq:nu0n0}], which implies $V\propto \nu_{c}^{-2}$ if the cavity length $L$ is fixed, or $V\propto \nu_{c}^{-3}$ if the aspect ratio is preserved. The resonant frequency of the empty ADMX cavity is actually 460~MHz: higher frequencies are accessible through the use of internal metal rods whose off-axis rotation tunes the mode (by varying the radial boundary condition). It is significant that metal rods allow microwave cavities to be tuned \textit{up} in frequency, and thus enable high-frequency $\text{TM}_{010}$-like modes to be realized in relatively large-volume resonators. In contrast, the dielectric tuning scheme employed by RBF and UF works by increasing the permittivity rather than changing the boundary conditions: thus dielectric insertion can only \textit{decrease} $\nu_c$ from the value determined by the dimensions of the empty cavity.

The ADMX collaboration continued to operate their detector after this initial result, swapping out tuning rods and amplifiers whenever necessary, and extended their exclusion at KSVZ sensitivity down to the minimum cavity frequency $\nu_c\sim460~\text{MHz}$~\citep{ADMX2002,ADMX2004}. Also in the 1990s, a Japanese group developed an ambitious proposal to use a beam of Rydberg atoms as microwave single-photon detectors for the haloscope search and thereby circumvent the practical $\sim 1~\text{K}$ lower bound on the added noise of HEMT amplifiers, then the best available linear amplifiers at microwave frequencies. A prototype experiment called CARRACK I was constructed~\citep{matsuki1996}, and a full-scale version was planned~\citep{tada1999}, but ultimately operational complexity proved too great, and the concept was not pursued.

While hopes for a breakthrough in single RF photon detection in the late 1990s proved premature, the prospects for coherent detection improved with the development of low-noise RF amplifiers based on \textbf{DC SQUIDs (Superconducting QUantum Interference Devices)} by the Clarke group at UC Berkeley~\citep{msa1998,msa1999}. A DC SQUID (when appropriately biased) is essentially an extremely sensitive magnetic flux-to-voltage transducer based on the DC Josephson effect:\footnote{There are also so-called ``RF SQUIDs'' which exploit the AC Josephson effect. Despite its name, the RF SQUID is not useful for axion detection. In the rest of this thesis I will drop the ``DC'' from ``DC SQUID'' to simplify the terminology.} thus it can also serve as a linear voltage amplifier if an input coil is used to convert the input voltage into a flux. As this description indicates, the natural application of a SQUID amplifier is to low-frequency signals ($\nu<100~\text{MHz}$) for which the impedance of the input coil is low compared to that of the inevitable parasitic capacitance in the system. The innovation of the Clarke group was basically replacing the input coil with a tunable microstrip resonator, enabling operation up to $\sim1~\text{GHz}$.

SQUID amplifiers behave in many respects very much like ``normal'' transistor amplifiers such as HEMTs: a DC voltage or current is used to bias the amplifier in the middle of some desired operating range in which a small fluctuation in the input signal produces a much larger fluctuation in the output: both the power used for amplification and the added noise come from the bias line; the SQUID is only slightly more complicated in that it also requires a DC flux bias. Nonetheless, the physical nature of the amplification process in a SQUID is completely different than in a transistor amplifier, and empirically the noise performance of SQUID amplifiers continues to improve roughly linearly with decreasing physical temperature well below $1~\text{K}$.

Since 2004, the efforts of the ADMX collaboration have been directed towards a major detector upgrade to realize near-quantum limited operation with SQUID amplifiers. The chief technical challenges are the rather large ($\sim1~\text{m}$) scale of the detector, which necessitates lots of custom cryogenic hardware, and the extreme sensitivity of SQUID amplifiers to magnetic flux: this flux sensitivity is of course precisely what makes the SQUID such a sensitive low-noise amplifier, but it implies that great care must be taken with magnetic shielding to incorporate a SQUID into the high-field environment of a haloscope detector. A 2008--2010 data run using a SQUID amplifier at $T\approx2~\text{K}$ served as a proof of principle~\citep{ADMX2011} and extended the ADMX exclusion marginally: the full ADMX exclusion at KSVZ sensitivity is now $460 < \nu_a < 892~\text{MHz}$ ($1.9 < m_a < 3.69~\mu\text{eV}$)~\citep{ADMX2010,ADMX2016}.\footnote{ADMX has also conducted several searches for non-virialized CDM in the form of cold, high-density streams of axions that fell into the galaxy relatively recently: such streams would manifest as sharp features in the spectrum of a putative haloscope signal~\citep{duffy2006,hoskins2011,ADMX2016}. For fixed experimental parameters, a haloscope search specifically targeting non-virialized axions will generally be sensitive to smaller couplings $|g_\gamma|$ because $\Delta\nu_a$ is smaller, but the converse is not true: the sensitivity of a search that assumes virialization is not appreciably degraded if there is non-virialized structure in the true signal. In this sense virialization is a conservative assumption. Searches for non-virialized axion signals are also complicated by time-dependent Doppler shifts arising from the orbital motion of the earth, which can eliminate the gain in sensitivity unless signal candidates are rescanned promptly or additional assumptions are made about the direction of the axion stream.} However, lower physical temperature is required to reap the full benefit of the SQUID's improved noise performance. Installation of a dilution refrigerator at the new ADMX site at the University of Washington was completed in 2016. The proximate goal of the ADMX collaboration is to rescan the same $\mathcal{O}\left(\mu\text{eV}\right)$ mass range at DFSZ sensitivity or better; it remains to be seen what value of $N_\text{sys}$ will be obtained in practice with this new configuration.

\begin{figure}[h]
\centering\includegraphics[width=1.0\textwidth]{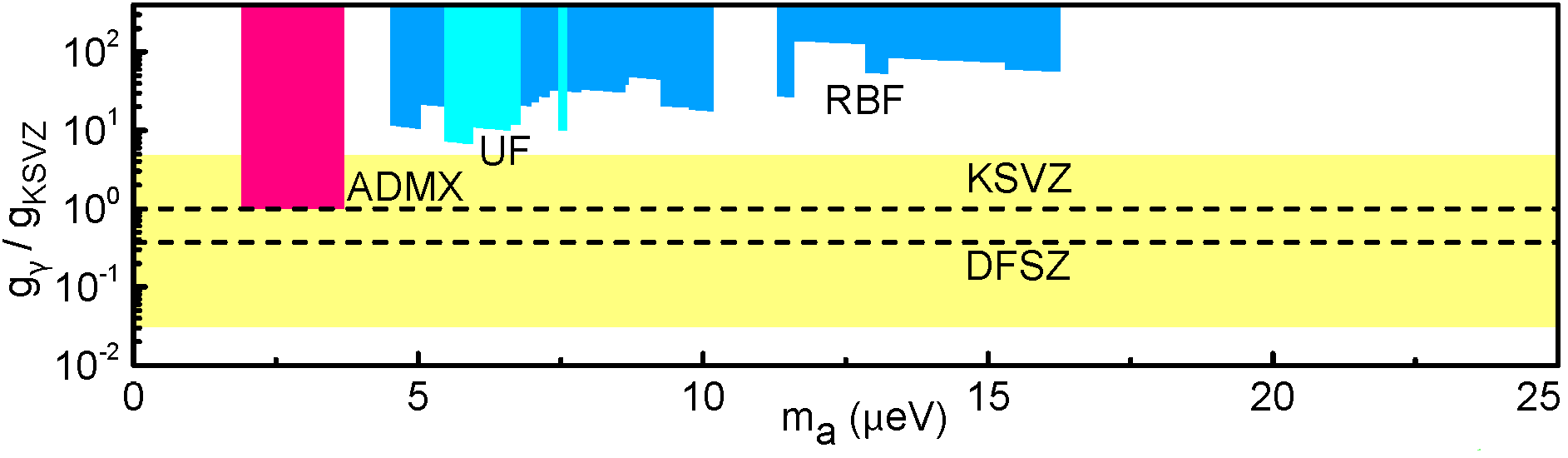}
\caption[Haloscope limits before HAYSTAC]{\label{fig:paramspace_lin} Haloscope parameter space as of summer 2016, with limits from the RBF (blue, Refs.~\citep{RBF1987,RBF1989}), UF (cyan, Refs.~\citep{UF1990,hagmann1990}), and ADMX (magenta, Refs.~\citep{ADMX1998,ADMX2002,ADMX2004,ADMX2010,ADMX2016}) experiments. For uniformity of presentation, both RBF and UF limits have been rescaled to $\rho_a = 0.45$~GeV/cm$^3$ from their original published values, where $\rho_a = 0.3$~GeV/cm$^3$ was used. The axion model band (Ref.~\citep{cheng1995}) is shown in yellow.}
\end{figure}

\subsection{High-frequency challenges}\label{sub:highfreq}
The results of past haloscope searches are summarized in Fig.~\ref{fig:paramspace_lin}, where I have restricted the axes to the region of parameter space most accessible to the haloscope search. More than thirty years after haloscope detection was first proposed, experimental sensitivity to pessimistically coupled axion models with $m_a\lesssim5~\mu\text{eV}$ ($\nu_a \lesssim 1~\text{GHz}$) finally appears possible. But a great deal of unexplored parameter space remains, especially at high frequencies. 

The absence of limits in the model band at higher frequencies is ultimately a consequence of the poor scaling of Eq.~\eqref{eq:scan_rate} with increasing frequency. The dominant effect is the frequency-dependence of the product $C_{mn\ell}V$, which may be thought of as the effective volume occupied by the cavity mode. For the $\text{TM}_{0n0}$ modes of an empty cylindrical cavity, the frequency and form factor are \citep{sikivie1985}
\begin{align}
\nu_n &= \frac{x_{n}}{2\pi}\frac{c}{r}, \label{eq:nu0n0}\\ 
C_{0n0} & = \frac{4}{x^2_{n}}, \label{eq:C0n0}
\end{align}
respectively, where $x_n$ is the $n$th root of the zeroth-order Bessel function $J_0(x)$; $x_1=2.405$, from which it follows that $C_{010}=0.69$. But Eqs.~\eqref{eq:nu0n0} and \eqref{eq:C0n0} also imply that the product
\begin{equation}\label{eq:cv_scaling}
C_{0n0}V = \frac{c^2L}{\pi \nu_n^2}
\end{equation}
whether we use the fundamental mode of a small cavity or a higher-order mode of a larger cavity! There are some tricks we can play with judicious placement of tuning rods in a realistic cavity, but operational complexity and sensitivity to machining and alignment imperfections increase rather quickly if we go down this route. ADMX has attempted to circumvent this poor scaling by coherently combining the signals from multiple smaller cavities occupying the magnet bore volume~\citep{kinion2001}; work in this direction is ongoing, but synchronizing the cavity frequencies and couplings has proven to be quite difficult.

Moreover, we have thus far considered only a single mode in isolation, but any microwave cavity will also support TE modes which do not couple to the axion and do not tune together with the TM modes:\footnote{In particular, TE modes do not tune at all as we change the radial boundary conditions by rotating internal conducting rods.} it can be shown that the asymptotic spectral density of TE modes scales as $\mathrm{d}N_\text{TE}/\mathrm{d}\nu \sim (V/c^3)\nu^2$~\citep{cavity1990}. At frequencies where the TM mode of interest becomes degenerate with a TE mode, mode-mixing occurs and the $Q$ and form factor are significantly degraded. Ideally we would like to keep $\mathrm{d}N_\text{TE}/\mathrm{d}\nu$ relatively low to minimize the number of such \textbf{mode crossings}, and this implies that we cannot easily evade the poor scaling implied by Eq.~\eqref{eq:cv_scaling} with an extreme cavity aspect ratio: $L/r\approx 5$ is usually taken as a practical upper bound.

On top of the poor scaling with $C_{mn\ell}V$, the other detector parameters in Eq.~\eqref{eq:scan_rate} are difficult to improve significantly with the constraints of present technology. The haloscopes described in Sec.~\ref{sub:admx} all used copper cavities, because the low-temperature surface resistance of copper is smaller than that of other normal metals. The surface impedance of superconductors is much smaller still, and superconducting microwave cavities which achieve $Q\sim10^6$ are commonplace in quantum information research, but such cavities would not remain superconducting in the presence of the strong magnetic fields required for haloscope detection. 

In fact, the $Q$ factors of normal metal cavities are bounded from above by the \textbf{anomalous skin effect}~\citep{pippard1947}, and the values cited in Sec.~\ref{sub:admx} are relatively close to this theoretical maximum. To understand the anomalous skin effect, it is convenient to parameterize the unloaded quality factor of a microwave cavity in terms of its skin depth $\delta$:
\begin{equation}\label{eq:q_delta}
Q_0 = \varsigma\frac{V}{A\delta},
\end{equation}
where $A$ is the surface area and $\varsigma$ is a mode-dependent numerical coefficient which is typically $\mathcal{O}(1)$ for the low-order modes of interest; e.g., $\varsigma=2$ for the $\text{TM}_{010}$ mode of an empty cylindrical cavity. Ignoring the factor of $\varsigma$, an intuitive interpretation of Eq.~\eqref{eq:q_delta} is that electromagnetic energy is stored in the cavity volume whereas dissipation occurs at the surface within a shell of thickness $\delta$. This is precisely what we expect from the classical theory of the skin depth, which tells us that at any finite frequency $\nu$, electromagnetic fields only penetrate a distance 
\begin{equation}\label{eq:delta_classical}
\delta = \big(\pi\nu\mu\sigma\big)^{-1/2}
\end{equation}
into a metal with permeability $\mu\approx\mu_0$ and DC conductivity $\sigma$. According to classical electrodynamics, $\delta$ should decrease (and thus $Q$ will increase) as we decrease the temperature $T$, because $\sigma$ increases with decreasing temperature. In particular, we should expect the $Q$ to stop improving with decreasing $T$ only when $\sigma(T)$ flattens out due to material impurities. 

Empirically, this is not what happens: instead, at microwave frequencies and cryogenic temperatures, $\delta$ asymptotes to a constant value $\delta_a$ while $\sigma$ continues to improve.\footnote{A useful rule of thumb for good-quality OFHC (oxygen-free high-conductivity) copper is that the $Q$ of a 1~GHz cavity will improve by about a factor of 4 in cooling from room temperature to 4~K, while the DC conductivity improves by a factor of $\gtrsim100$~\citep{gp}.} In a semiclassical picture~\citep{pippard1947,finger2008}, this ``anomalous'' behavior arises because the classical expression Eq.~\eqref{eq:delta_classical} becomes invalid when $\delta$ drops below the electrons' mean free path $\ell = (\sigma m_ev_F/n_ee^2)$, where $v_F$ is the metal's Fermi velocity and $n_e$ is its conduction electron density.\footnote{For copper, $v_F=1.57\times10^8~\text{cm/s}$ and $n_e=8.5\times10^{22}~\text{cm}^{-3}$.} In the anomalous regime, only a small fraction of the electrons can contribute to conduction at frequency $\nu$, resulting in a suppression of the effective microwave conductivity (relative to $\sigma$) by a factor of $\mathcal{O}(\delta/\ell)$. We thus take $\sigma\rightarrow\gamma(\delta/\ell)\sigma$ in Eq.~\eqref{eq:delta_classical}, where $\gamma$ is a numerical factor whose value we could determine with a more formal derivation, and solve for $\delta$. We see that the $\sigma$-dependence drops out, and we obtain a material-dependent constant
\begin{equation}\label{eq:delta_anom}
\delta_a = \left(\frac{m_ev_F}{\gamma\pi\nu\mu n_ee^2}\right)^{1/3}.
\end{equation}
Ref.~\citep{hagmann1990} quotes $\delta_a=2.8\times10^{-5}~\text{cm}$ for copper at $\nu=1~\text{GHz}$, which corresponds to $\gamma=7.6$ in Eq.~\eqref{eq:delta_anom}. Using Eq.~\eqref{eq:delta_anom} in Eq.~\eqref{eq:q_delta} with the dimensions of the empty ADMX cavity, we find that the value $Q_0\sim1.5\times10^5$ obtained by ADMX is a factor of 3.7 from the upper bound from the anomalous skin effect, which is about as close as one can reasonably hope to get with real materials and tuning rods inside the cavity.\footnote{The RBF and UF experiments obtained comparable or higher $Q$ factors despite higher operating frequencies because they used sapphire rods with very low loss for tuning. Metal tuning rods necessarily increase the cavity's surface area $A$ and thus reduce $Q$ via Eq.~\eqref{eq:q_delta}.} Moreover, Eqs.~\eqref{eq:q_delta} and \eqref{eq:delta_anom} together imply that $Q_0\propto\nu^{-2/3}$ assuming we maintain the cavity aspect ratio as we scale up to higher frequencies.

We have nearly exhausted the parameters with which we can hope to improve the haloscope scan rate even in principle. Having already devoted Sec.~\ref{sub:sql} to a discussion of quantum noise limits, I will not say much about $N_\text{sys}$, except to emphasize a minor point that is often not made explicit: the form of Eq.~\eqref{eq:scan_rate} indicates that the scan rate (and the SNR) are the same for a given number of noise quanta $N_\text{sys}$ at any frequency (up to changes in cavity parameters). As long as we remain stuck with phase-sensitive linear amplifiers, the SQL bounds how much we can improve the SNR by reducing $N_\text{sys}$.

Finally, we can consider the magnetic field strength $B_0$. Fortunately for the haloscope community, superconducting magnets are useful for a wide variety of scientific applications from nuclear magnetic resonance to accelerator physics. As a result, NbTi solenoids with $B_0 \lesssim 9~\text{T}$ are relatively affordable, but a single such magnet with a sufficiently large bore still costs upwards of $\$200,000$, and is typically the single most expensive component of any haloscope detector. To achieve $B_0 \gtrsim 9~\text{T}$, we would need to switch from NbTi to Nb$_3$Sn, which has a higher critical field at a standard $T\approx4~\text{K}$ operating temperature. The market for Nb$_3$Sn magnets is small and specialized, and for this reason among others sufficiently large-bore designs are more expensive than NbTi magnets by at least an order of magnitude. Finally, magnets wound from high-$T_c$ superconductors offer the only route to $B_0\gtrsim20~\text{T}$, but this technology is still in its infancy, and in any event, prohibitively expensive for a typical haloscope experiment. If magnet prices drop and haloscope searches attract more attention (due to e.g., the continued non-discovery of WIMP dark matter), this ``brute force'' approach to increasing sensitivity may become attractive: after all, Eq.~\eqref{eq:scan_rate} has stronger dependence on $B_0$ than on any other parameter.

\subsection{Motivation for HAYSTAC}\label{sub:haystac}
Despite the unfavorable scaling of Eq.~\eqref{eq:scan_rate}, there are several compelling reasons to operate a haloscope detector a decade higher in frequency than the current ADMX limits. We saw in chapter~\ref{chap:cosmo} that predictions of the axion mass from the cosmic CDM abundance are rife with uncertainty even without inflation; in an inflationary universe such predictions are not possible even in principle. Nonetheless, within the standard cosmology, most calculations seem to agree that $\Omega_a > \Omega_\text{DM}$ in the $m_a\lesssim 4~\mu\text{eV}$ range probed by ADMX. While we should always be open to surprises (and perhaps even to inflation), the physics case for developing experiments capable of probing higher masses is very strong. Despite the emergence of a number of promising new ideas in recent years (see Sec.~\ref{sec:other_exp}), haloscope detection remains the only technique with proven sensitivity to the CDM axion model band, and thus should not be discounted.

There are also technical advantages to operating a haloscope at $\nu_c\gtrsim 5~\text{GHz}$. First, although the smaller scale of a high-frequency experiment implies a loss of signal power, it also simplifies both magnet design and cryogenic design substantially, thus enabling the rapid deployment of a relatively cheap detector by a small team of experimentalists.\footnote{Conversely, a haloscope detector much bigger than ADMX would present quite a significant engineering challenge, and thus $1~\mu\text{eV}$ may be taken as an approximate lower bound on the region of parameter space accessible with the haloscope technique.} Second, progress in the measurement of microwave-frequency electromagnetic fields at the single-quantum level in the past decade has been truly phenomenal. Much of this progress has been driven by the prospect of realizing a quantum computer using superconducting microwave circuits in the 4--12~GHz range as ``qubits'' (quantum bits). Superconducting qubits coupled to high-$Q$ superconducting cavities are actually excellent microwave single-photon detectors. Rendering such a qubit detector tunable and incorporating it into a operational haloscope will be extremely challenging, but may prove transformative for the haloscope search~\citep{zheng2016}. The long-term goal of microwave single-photon detection would be reason enough to develop the haloscope technique at GHz frequencies.

More importantly for near-term haloscope searches, progress in microwave quantum measurement has also led to renewed interested in \textbf{Josephson Parametric Amplifiers (JPAs)}, exquisitely sensitive linear amplifiers whose operation relies on the physics of the Josephson effect. For this reason JPAs (like the SQUID amplifiers considered in Sec.~\ref{sub:admx}) must be shielded extremely carefully from external magnetic fields to be used in haloscope detectors. Unlike SQUID amplifiers, JPAs do not necessarily exploit this flux sensitivity itself for amplification. I will discuss the design and operating principle of JPAs in greater detail in Sec.~\ref{sec:jpa}; for now, let me just distinguish the JPA by emphasizing that it is naturally a resonant device designed to operate away from DC; in practice, JPAs are most conveniently realized at microwave frequencies in the 2--12~GHz range.\footnote{The upper end of this range is set by the requirement that $\nu$ remain well below the Nb plasma frequency, and corresponds to $m_a=50~\mu\text{eV}$. This range may be extended by perhaps a factor of 2 with unconventional JPA designs~\citep{konrad}, but at such high frequencies single-photon detection will likely be necessary to reach the model band in any event, and radically new approaches to cavity design will also be necessary ($\nu_c=12~\text{GHz} \rightarrow r=1~\text{cm}$ for the $\text{TM}_{010}$ mode of a cylindrical cavity). Thus $m_a\approx50~\mu\text{eV}$ may be regarded as an approximate upper bound on haloscope parameter space. One intrepid group has explored haloscope design at $\nu_c=34$~\text{GHz}~\citep{YMCE2015}, but the sensitivity achieved in their initial search~\citep{malagon2014} was four orders of magnitude away from the QCD axion model band.} 

The defining feature of any parametric amplifier is that the power for the amplification process comes from an AC ``pump'' (near the resonant frequency for the designs I will consider) rather than a DC bias. This fact has several profound consequences. First, it implies that (at least in principle) a parametric amplifier can be realized entirely without resistive elements, enabling extremely good control over sources of noise. Second, the pump tone provides a natural phase reference, and thus parametric amplifiers are \textit{phase-sensitive} devices. The role of the ``J'' in the JPA is twofold: it enables a nearly ideal realization of the (much more general) physics of parametric amplification, and allows these naturally narrowband devices to be made tunable, as was first demonstrated by Ref.~\citep{castellanos2007}. 

Though natively phase-sensitive, JPAs may also be operated as phase-insensitive linear amplifiers, and real JPAs operated in this way achieve added noise consistent with Eq.~\eqref{eq:haus_caves}. In quantum measurement applications, single-quadrature JPA operation is more common, and in this mode JPAs routinely achieve sub-quantum-limited noise performance~\citep{castellanos2008,mallet2011}. We will see in Sec.~\ref{sec:jpa} that phase-insensitive operation is simpler and more readily applicable to the haloscope search, and I will assume phase-insensitive JPA operation throughout this thesis. Nonetheless, single-quadrature operation with an appropriately configured receiver may offer an alternative to single-photon detection for enhancing the haloscope scan rate~\citep{zheng2016}.

The challenges and opportunities discussed above serve as excellent motivation for the \textbf{Haloscope at Yale Sensitive To Axion CDM (HAYSTAC)}, which was conceived to target the region of parameter space with $\nu_a\gtrsim5~\text{GHz}$. The initial HAYSTAC design incorporated a JPA and a dilution refrigerator to compensate for the loss of cavity volume by reducing $N_\text{sys}$ as close to the SQL as possible (c.f.\ Tab.~\ref{tab:params}). In the rest of this thesis I will describe the design and operation of the HAYSTAC detector and the analysis of data from the first axion search with HAYSTAC. This first data run achieved $N_\text{sys}\approx 3$ and set limits in the axion model band for $5.7 < \nu_a < 5.8~\text{GHz}$ ($23.55 < m_a < 24.0~\mu\text{eV}$)~\citep{PRL2017}. This makes HAYSTAC the second experiment to probe viable models of axion CDM, about an order of magnitude higher in mass than those probed by ADMX.

In addition to taking data in this well-motivated and unexplored region of parameter space, HAYSTAC was designed to serve as a testbed for R\&D which will be critical to the success of next-generation haloscopes. In particular, the next incarnation of the HAYSTAC detector will incorporate a \textbf{squeezed-state receiver} to explore the application of single-quadrature JPA operation to the haloscope search. I will mention some of these R\&D efforts briefly in Sec.~\ref{sec:haystac_future}.


\chapter{The HAYSTAC detector}\label{chap:detector}
\setlength\epigraphwidth{0.55\textwidth}\epigraph{\itshape If we start making a list of things that aren't here,\\ we could be here all night. You know, pens for instance. Let's stick with things we can see.}{Wheatley}

\noindent In chapter~\ref{chap:search} I motivated the conceptual design of an axion haloscope detector, comprising a cryogenic microwave cavity immersed in a strong magnetic field and coupled to a low-noise receiver. I also discussed the challenges facing haloscope operation at frequencies $\nu_c\sim5~\text{GHz}$, as well as the unique opportunities afforded by operation in this range. In the present chapter I will describe how the haloscope concept is realized in the design of HAYSTAC. 

I begin in Sec.~\ref{sec:detector_overview} with an overview of the main elements of HAYSTAC and how they interface with each other. In Sec.~\ref{sec:infra} I describe the cryogenic and magnet systems and the many layers of magnetic shielding required to null out the external field near the JPA. In Sec.~\ref{sec:cavity} I describe the design and operation of the first HAYSTAC microwave cavity, and the mechanical systems used to tune the cavity and adjust its coupling to the receiver. In Sec.~\ref{sec:jpa} I discuss the essential physics responsible for JPA operation and the design of the JPA serving as the HAYSTAC preamplifier. Finally, in Sec.~\ref{sec:receiver} I discuss the design of the receiver chain used to read out noise around the cavity resonance. Parts of this chapter were adapted from Ref.~\citep{NIM2017}.

Up to this point, my approach in this dissertation has been essentially pedagogical: I have tried to motivate how and why one might design an experiment to search for cosmic axions. From this point onwards, I will adopt a more straightforwardly descriptive approach. I assume that if you are still reading at this point you are interested in operating such an experiment yourself,\footnote{Or else you are on my committee and have no choice in the matter.} and that you have a general familiarity with cryogenics, microwave engineering, and frequency-domain signal processing. I will make exceptions and return to a more pedagogical approach when I discuss the design (Sec.~\ref{sec:jpa}) and operation (Sec.~\ref{sec:jpa_op}) of JPAs, and our approach to haloscope noise calibration (Sec.~\ref{sec:noise}), as these topics are not treated in detail elsewhere.

For an introduction to the design of cryogenic experiments and dilution refrigerators in particular, see Ref.~\citep{pobell1996}. A great introduction to microwave engineering may be found in Ref.~\citep{pozar2012}. Finally, I found Ref.~\citep{brigham1988} invaluable when I was first developing intuition about the discrete Fourier transform and digital signal processing.

\section{Detector overview}\label{sec:detector_overview}
\begin{figure}[h]
\centering{\includegraphics[width=.6\textwidth]{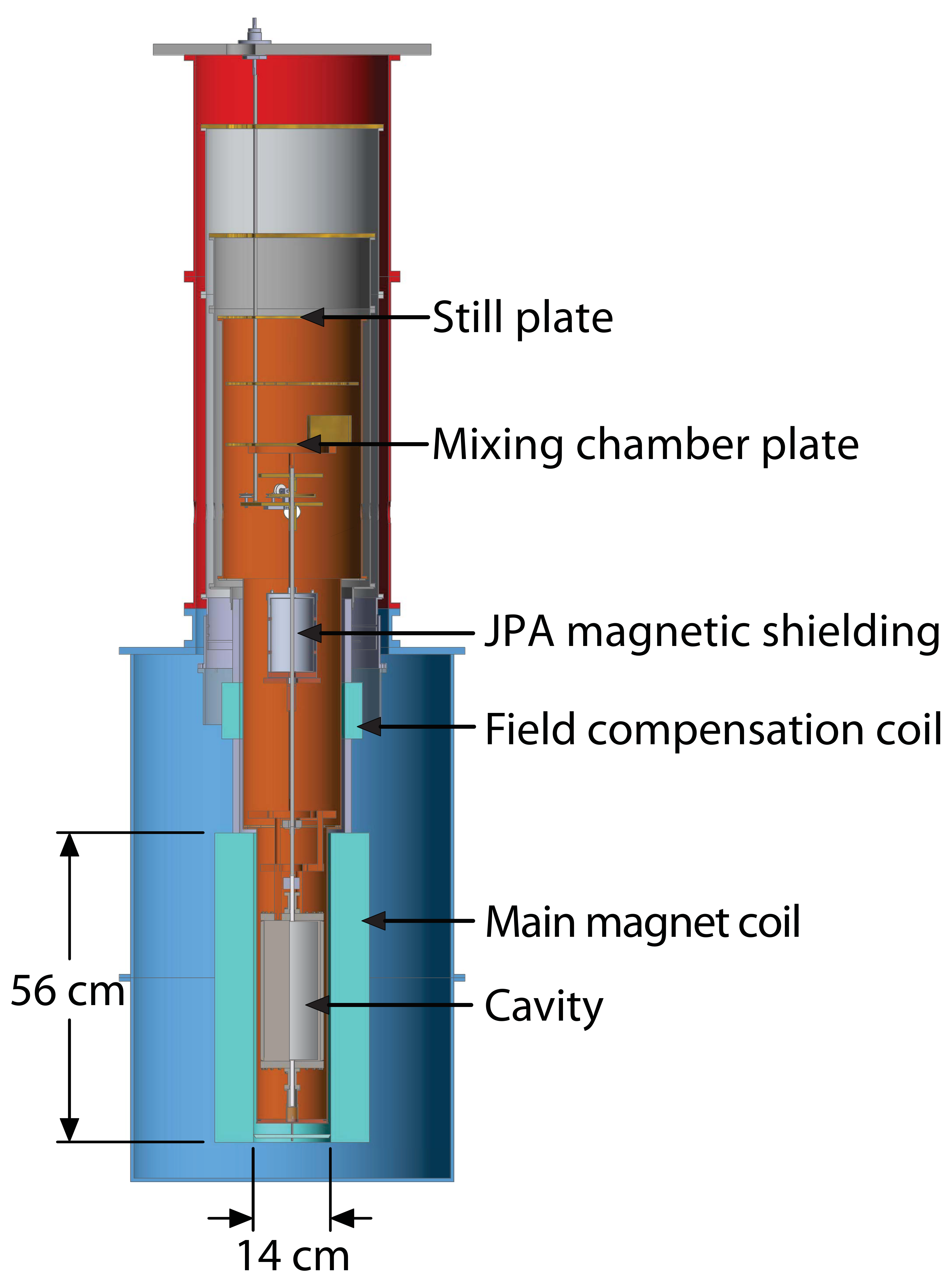}}
\caption[HAYSTAC detector schematic]{\label{fig:solidworks} HAYSTAC detector schematic: the red volume is the DR vacuum shield, the blue volume is the magnet body, and the orange shaded region is the volume bounded by the still shield and its extensions. The magnet's 70 K and 4 K radiation shields are not shown below their interface with the DR shields. The persistent superconducting coils surrounding the JPA shielding can are also not shown.}
\end{figure}

The main elements of the HAYSTAC detector are depicted schematically in Fig.~\ref{fig:solidworks}. The footprint of the experiment is set by a 9~T superconducting solenoid integrated with a cryogen-free dilution refrigerator (DR). The cavity hangs in the center of the magnet bore, supported by a gold-plated copper ``gantry'' thermally anchored to the DR's mixing chamber plate, which is maintained at $T_\text{mc}=127~\text{mK}$ during operation of the detector (see discussion in Sec.~\ref{sub:cryo}).  The long G10 fiberglass shafts which enable the cavity to be tuned via actuation of a room-temperature stepper motor are also shown.

\begin{figure}[h]
\centering{\includegraphics[width=.3\textwidth]{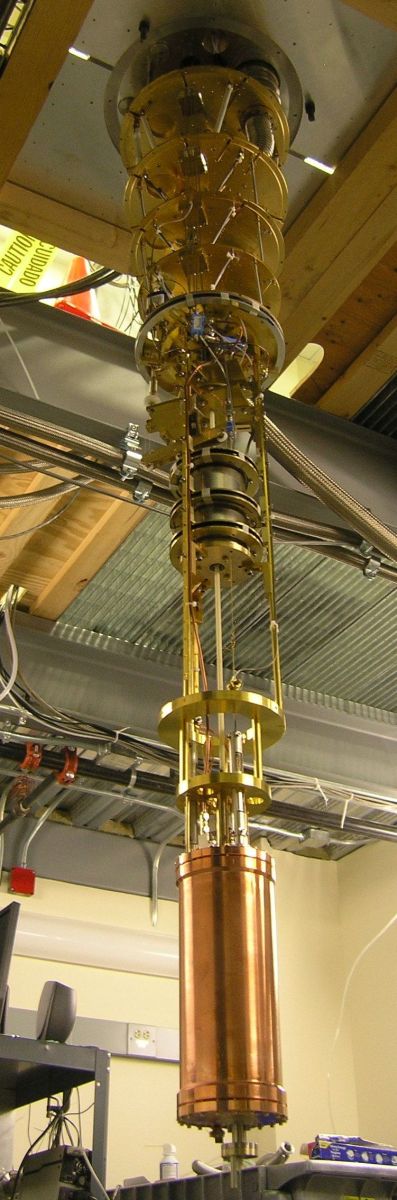}}
\caption[HAYSTAC partially assembled]{\label{fig:haystac_open} The HAYSTAC detector before the installation of radiation shields and insertion into the magnet bore.}
\end{figure}

The 14~cm magnet bore diameter sets the scale for the available axion-sensitive volume, and it is important to use this volume as efficiently as possible. However, the magnet coils are maintained by a cryocooler at an operating temperature $T_\text{mag}\approx3.6~\text{K}$, and the much colder cavity needs to be thermally shielded from this relatively hot environment. Thus the cavity and gantry are enclosed in radiation shields anchored to the DR's still plate at $T_\text{still}= 775~\text{mK}$. The volume enclosed by these still shields is shaded orange in Fig.~\ref{fig:solidworks}. The lowest segment of the still shield has 13~cm outer diameter (OD) and 12.7~cm inner diameter (ID). The cavity endcaps themselves have an OD of 11.4 cm.\footnote{The barrel is slightly thinner (see Fig.~\ref{fig:haystac_open}) to reduce the total mass that needs to be cooled to $T_\text{mc}$.} Thermal shorts between these systems in close physical proximity would seriously compromise operation of the experiment: thus a thin stainless steel pin attached to the bottom flange of the cavity is used to center the cavity inside the still shields, and a G10 disk mounted to the bottom of the still shield assembly keeps the assembly centered in the magnet bore.

The JPA used in the HAYSTAC detector is essentially a superconducting (Nb) $LC$ circuit whose inductance comes from an array of SQUIDs. As noted in Sec.~\ref{sub:haystac}, the JPA is exceptionally sensitive to magnetic flux: careful magnetic shielding is required to achieve stable JPA operation even in experiments which do not intentionally introduce large magnetic fields. In HAYSTAC, we enclosed the JPA in a canister comprising both ferromagnetic and superconducting shields, surrounded the shields with persistent superconducting coils, and positioned this assembly in the center of the gantry, in a region where the applied field is suppressed by a compensation coil built in to the main magnet. In this way we reduced the field near the JPA to a very uniform $B\sim 10^{-3}~\text{G}$, which is acceptably small.\footnote{$1~\text{G} = 10^{-4}~\text{T}$.} 

\begin{figure}[h]
\centering{\includegraphics[width=.7\textwidth]{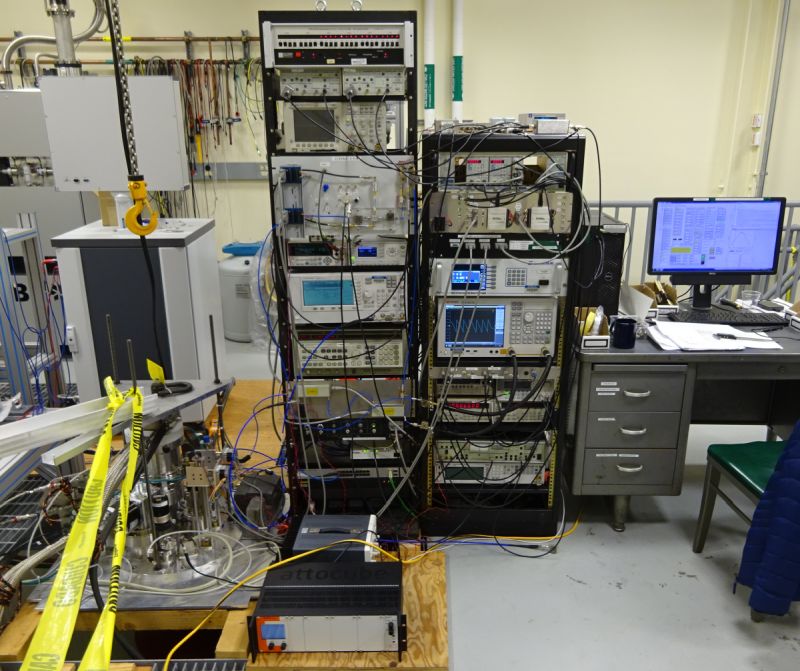}}
\caption[The upper level of the HAYSTAC lab]{\label{fig:upper_lab} The upper level of the HAYSTAC lab, showing the top level of the DR, the room-temperature electronics, and the DAQ computer. The DR computer and gashandling system are not pictured.}
\end{figure}

HAYSTAC is sited at the Wright Laboratory of Yale University in a two-level lab equipped with cooling water and 3-phase electrical power, which are required to run the DR and cryocoolers for both the DR and magnet. The top flange of the DR is bolted to a stainless steel plate; when the experiment is fully assembled, this plate rests on vibration isolation pads flush with the floor of the upper level (see Fig.~\ref{fig:haystac}). When the DR/gantry assembly is not inserted into the bore of the magnet, it is supported from above, allowing convenient access to the detector components from the lower level; Fig.~\ref{fig:haystac_open} is a photograph of the detector at this stage, before the installation of thermal shields. Feedthroughs for thermometry and other DC wiring, microwave signal lines, and motion control systems are accessible from the upper level, shown in Fig.~\ref{fig:upper_lab}, which also hosts electronics racks, the data acquisition (DAQ) computer, and an independent computer which controls the DR. A small room on the lower level houses helium compressors for both the DR and magnet cryocoolers.

\begin{figure}[h]
\centering{\includegraphics[width=.55\textwidth]{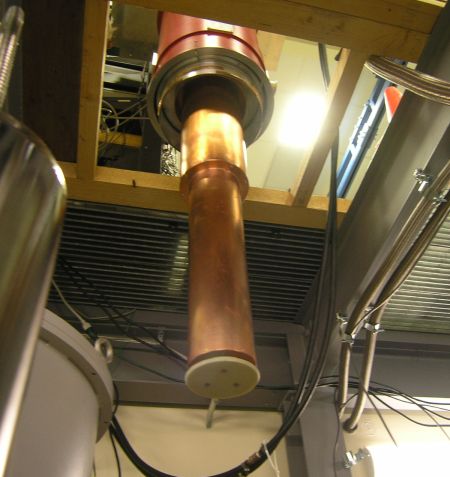}}
\caption[Insertion of HAYSTAC into the magnet bore]{\label{fig:insertion} Insertion of HAYSTAC into the magnet bore. Note the copper fingerstock on the aluminum 70~K shield and the G10 centering disk.}
\end{figure}

The DR and magnet share a single common insulating vacuum space. Extensions of the 4~K and 70~K DR thermal shields overlap the corresponding stages at the top of the magnet, and thermal contact is provided both by fingerstock that presses between the extensions and the corresponding magnet thermal stages and by blackbody radiation between the overlapping surfaces. This design allows the DR/gantry assembly to be easily inserted into and removed from the magnet by use of an overhead hoist (Fig.~\ref{fig:insertion}). The magnet is on a cart equipped with casters and jack screws. Insertion is accomplished by lifting the DR, rolling the magnet cart into place, leveling the cart and raising it to the appropriate height using the jack screws, and finally lowering the DR/gantry assembly into the magnet until the DR and magnet's vacuum flanges contact, at which point the flanges are bolted together. The gas flow and electronic control lines are long enough that only a few cables must be disconnected to insert or remove the cryostat.  

\begin{figure}[h]
\includegraphics[width=1.0\textwidth]{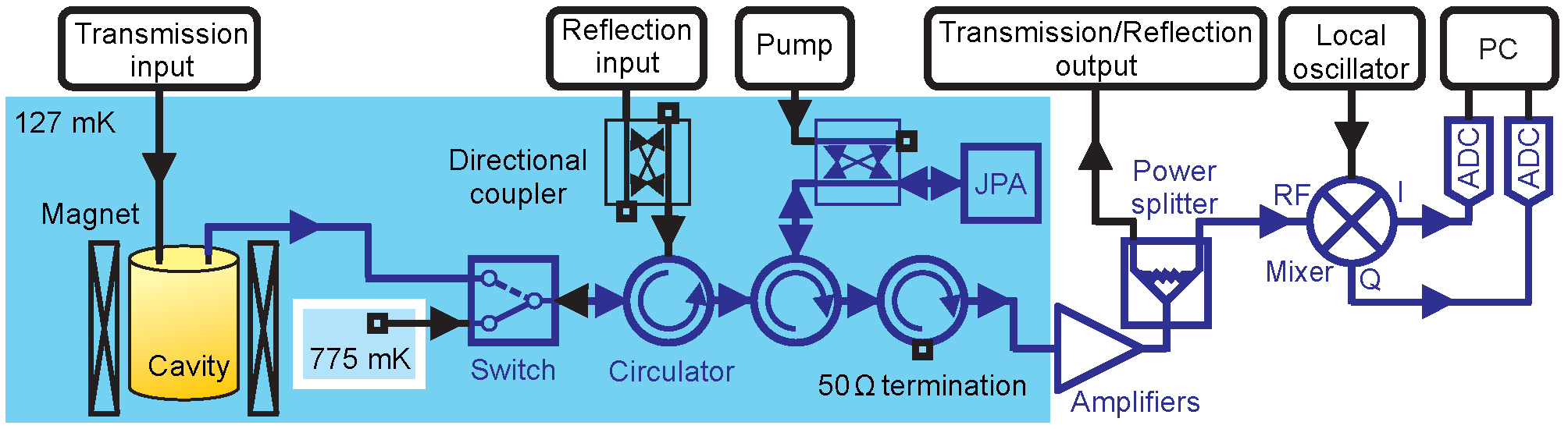}
\caption[Simplified HAYSTAC receiver diagram]{\label{fig:receiver_simple} Simplified receiver diagram: blue arrows indicate the path that a putative axion signal would take through the system, and black arrows indicate other paths.}
\end{figure}

The final crucial component of HAYSTAC is a low-noise receiver chain. A simplified version of the receiver chain is schematically illustrated in Fig.~\ref{fig:receiver_simple}. The first element in the receiver signal path is a microwave switch that may be actuated to toggle the receiver between the cavity and a $50~\Omega$ termination thermally anchored to the DR still plate (i.e., a Johnson noise source at the known temperature $T_\text{still}$). This switch allows us to calibrate the receiver's added noise $N_A$ via a procedure described in Sec.~\ref{sec:noise}. The HAYSTAC preamplifier comprises the magnetically shielded JPA circuit itself together with a directional coupler to deliver the JPA's pump tone and a circulator to separate input and output signals. Signals exiting the JPA are amplified further at 4~K and room-temperature, and downconverted to an \textbf{intermediate frequency (IF)} band using an IQ mixer whose local oscillator (LO) is normally set 780~kHz above the cavity resonance.\footnote{We will also occasionally find it useful to study the detector's noise performance at frequencies far detuned from any cavity mode (see Sec.~\ref{sub:noise_offres}). In chapters~\ref{chap:data} and \ref{chap:analysis} especially, we will need to keep track of both the IF and RF frequencies associated with Fourier components in spectra obtained from the HAYSTAC receiver. I will use ``RF'' and ``microwave'' synonymously throughout the rest of this thesis.} After further amplification and filtering the IF signals are digitized and processed by the DAQ computer.

The HAYSTAC microwave signal path features three input lines as well as the receiver output line. The ``pump line'' bypasses the cavity altogether; in normal operation it is used for the CW pump tone which powers the JPA. Two other lines may be used to inject signals into the cavity in transmission and reflection. These lines are not needed for the acquisition of axion-sensitive data,\footnote{If axions constitute CDM, the cavity's ``axion port'' is always powered!} but they are still useful for cavity characterization. We use a vector network analyzer (VNA) to measure the cavity's frequency response in both transmission and reflection. 

\section{Experimental infrastructure}\label{sec:infra}
Here I will discuss in more detail the systems responsible for maintaining the thermal and magnetic environments critical to succesful operation of the HAYSTAC detector. The DR and magnet cryogenic systems are discussed in Sec.~\ref{sub:cryo}, the magnet itself is discussed in Sec.~\ref{sub:magnet}, and the JPA's magnetic shielding is discussed in Sec.~\ref{sub:shielding}.

\subsection{Cryogenic systems}\label{sub:cryo}
As noted above, both the DR and the magnet are ``dry systems:'' their 70~K and 4~K stages are cooled to and maintained at their operating temperatures by closed-cycle helium cryocoolers rather than by external liquid cryogens. This choice resulted in a substantial reduction of both operational complexity and operating cost, both important practical considerations for any haloscope experiment given the long periods of continuous operation required to achieve significant parameter space coverage. However, a cryogen-free system relies on an uninterrupted supply of electrical power: a power outage while the experiment is operating will result in a magnet quench, whose consequences can be quite severe. See appendix~\ref{app:quench} for discussion of the quench experienced during the first operation of HAYSTAC.

The DR was manufactured in 2007 by VeriCold Technologies (subsequently acquired by Oxford Instruments). The only major modification was the replacement of the VeriCold's pulse tube cooler with a Sumitomo RDK-415D Gifford-McMahon cryocooler which achieved better cooling power at the 70~K and 4~K stages.\footnote{In retrospect, this was a poor choice because the GM cryocooler produced larger vibrations; see discussion in Sec.~\ref{sub:jpa_fluct}.} Calibrated ruthenium oxide temperature sensors are used to monitor $T_\text{still}$, $T_\text{mc}$, and the cavity temperature $T_\text{cav}$. These RuO sensors (along with Cernox and platinum sensors used at the 4~K and 70~K stages) are periodically measured by a Lakeshore 370 AC resistance bridge. The temperatures at the magnet's 4~K and 70~K stages are independently measured by a Cryomagnetics TM-600 temperature monitor.

The Lakeshore scanner also has temperature control functionality which is integrated into the DR software. We did not use this functionality, and instead incorporated a custom analog temperature controller comprising an SRS SIM960 PID controller, a homemade AC bridge circuit, and a homemade differential current source, along with a heater and a thermometer installed on the DR's mixing chamber plate. This system measures $T_\text{mc}$ independent of the measurement of other sensors, resulting in reduced delay and improved stability; it was used to maintain the system at $T_\text{mc}=127\pm1~\text{mK}$ during normal operations.

The mixing chamber plate of the DR can reach temperatures as low as 10~mK when the DR is operated on its own, and tends to flatten out around 25~mK when integrated with the gantry/cavity assembly. The elevated operating temperature was chosen mostly to mitigate the temperature-dependent modulation of the JPA gain by vibrational fluctuations: this effect was first observed during system commissioning, and is discussed further in Sec.~\ref{sub:jpa_fluct}. Operating at $T_\text{mc}=127~\text{mK}$ does not have a large impact on the sensitivity of the experiment: the thermal contribution to Eq.~\eqref{eq:system_noise} yields an additional 0.13 quanta relative to operation at 0~K in the $5.7 < \nu_c < 5.8$~GHz range scanned during the first HAYSTAC data run. Moreover, at $T_\text{mc}=127$~mK the DR has a larger cooling power (about $150~\mu\text{W}$), and the heat capacity of the mixing chamber is also larger than at lower temperatures: the latter effect results in reduced temperature excursions from mechanical motions (Sec.~\ref{sub:cav_tuning}) and actuation of the microwave switch (Sec.~\ref{sub:receiver_cryo}).\footnote{Repeated switching and prolonged actuation of the motion control systems produced temperature changes of at most a few mK, and the small tuning steps characteristic of normal data run operation produced no observable effect on the DR temperature.}

For the purposes of the haloscope search, the mixing chamber temperature is really a proxy for the cavity temperature and the temperature at the JPA. The gantry is a tripod with gold-plated copper alloy legs and gold-plated copper rings at both ends, clamped to the DR's mixing chamber plate on one side and to the cavity's upper endcap on the other. Empirically, changes in $T_\text{cav}$ (measured on the cavity's lower endcap) track changes in $T_\text{mc}$ after a delay of only a few minutes, implying a thermal conductivity which is very good for the $T\sim100$~mK range.\footnote{The response to changes is a better metric than equality in the measured values of $T_\text{cav}$ and $T_\text{mc}$, since low-temperature thermometer calibrations are very finicky and in particular can exhibit hysteretic dependence on both temperature and magnetic fields. There was a 20~mK offset between the measured values of $T_\text{cav}$ and $T_\text{mc}$ at $B_0=9~\text{T}$. The calibration of the mixing chamber thermometer (which was never exposed to the full 9~T field) is likely more reliable. Uncertainty in thermometer calibration is unlikely to be the dominant contribution to the uncertainty in our exclusion limit (see discussion in appendix~\ref{app:error}).} In addition to validating that the gantry provides a good thermal link to the mixing chamber, this observation indicates that the 0.125~mm copper plating on the cavity (which is machined out of stainless steel; see Sec.~\ref{sub:cav_design}) is also sufficient for a good thermal link. The JPA is thermally anchored to the gantry through a thick copper braid, and all lossy receiver components are thermalized to the DR in a similar way. 

As noted in Sec.~\ref{sec:detector_overview}, the cavity and gantry are thermally shielded with an extension of the DR still shield. This extension comprises two demountable sections with dimensions $17.8~\text{cm}~\text{OD} \times 45~\text{cm}$ and $13.0~\text{cm}~\text{OD} \times 53.3~\text{cm}$ for the upper and lower sections, respectively. The original still shield extension was constructed from $0.16$ cm sheet copper with welded seams; it was deformed during the magnet quench (appendix~\ref{app:quench}). Subsequently, we constructed a replacement shield using heavily plated stainless steel ($0.125 - 0.250~\text{mm}$), prompted by our experience with the thermal link to the copper-plated cavity.

\begin{figure}[h]
\centering{\includegraphics[width=1.0\textwidth]{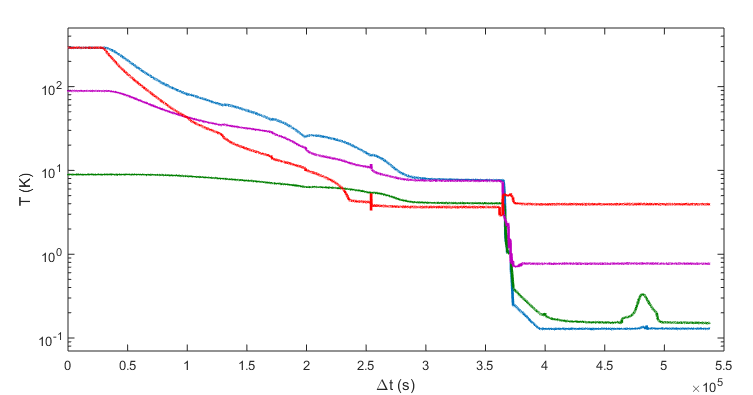}}
\caption[Sample DR cooldown temperature record.]{\label{fig:cooldown} A typical cooldown to steady-state operation: $T_\text{mc}$, $T_\text{cav}$, $T_\text{still}$ and $T_\text{4\,K}$ are plotted in blue, green, purple, and red, respectively. The fact that $T_\text{cav}$ and $T_\text{still}$ flatten out below 293~K at the left edge of the plot just reflects the limited operating ranges of these thermometers; the plot of $T_\text{mc}$ is stitched together from two thermometers which operate above and below $T=10~\text{K}$, respectively. From $\Delta t\approx0.3\times10^5~\text{s}$ to $\Delta t\approx2.8\times10^5~\text{s}$, the cryocooler cools the 4~K and 70~K stages, and the cooling power is transmitted to the other plates by circulating the DR's $^3$He/$^4$He mixture through a precooling circuit. At $\Delta t\approx3.7\times10^5~\text{s}$, the mixture is evacuated from the precooling line, pressurized by the room-temperature gashandling system, and forced through a large flow impedance, cooling the lower stages of the DR through the Joule-Thomson effect. Once $T_\text{mc}\lesssim1$~K, the mixture condenses into the mixing chamber, and cooling by dilution proceeds until the system equilibrates at a value of $T_\text{mc}$ determined by the still temperature and the setpoint of the PID control loop. The feature in $T_\text{cav}$ between $\Delta t\approx4.6\times10^5~\text{s}$ and $\Delta t\approx5.0\times10^5~\text{s}$ is due to ramping up the external magnetic field.} 
\end{figure}

The DR operating alone has a cooldown time of about 14~hours. When integrated with the gantry, cavity, and magnet, the cooldown time to $T_\text{mc}=127~\text{mK}$ is slightly over 3~days. The temperature record for a typical cooldown is plotted in Fig.~\ref{fig:cooldown}.

\subsection{Magnet}\label{sub:magnet}
The magnet was designed and manufactured to custom HAYSTAC specifications by Cryomagnetics, Inc. The inductance of the magnet is 189~H, and at the peak field of $B_0=9~\text{T}$ the current in the coils is 72~A. The magnet operates in persistent mode where the current flows through a superconducting loop instead of a room-temperature supply: the field is thus very stable and requires no external power to maintain. During steady-state operation the temperature of the magnet coils equilibrates at $T_\text{mag}\approx3.6~\text{K}$.\footnote{The precise value depends slightly on the ambient temperature of the lab, which affects the helium gas used in the refrigeration cycle.}

To ramp the field up or down, a heater is used to raise a localized segment of the NbTi loop (the ``persistent switch'') above its superconducting transition, forcing the current to flow through through a room-temperature supply (Cryomagnetics 4~G); $T_\text{mag}$ also increases to about 3.8~K when the heater is on. The supply current can then be changed very slowly to ramp the field produced by the main solenoid to the desired value. We set a ramp rate of $\frac{\mathrm{d}B}{\mathrm{d}t}=2.5$~mA/s (implying 8 hours to ramp from 0 to 9~T). The ramp rate is limited by thermal considerations: varying the field heats the magnet itself and also induces eddy currents in the still shields which heat the lower stages of the DR. 

In practice, the limit from the heating of the magnet coils is more restrictive. The critical temperature of the NbTi coils is a decreasing function of the circulating current (equivalently, the field), and at the full field it is $T_c(9~\text{T})\approx 4.15$~K. Thus it is critical that $T_\text{mag}$ not exceed this value near the end of the field ramping procedure. Empirically, $T_\text{mag}$ rises quickly to $\approx4.3$~K when the field is ramped from 0 to 0.5~T, due to the polarization of electronic spins in the magnet's aluminum support structure, and subsequently falls, asymptoting to $\approx4.0$~K by the end of the ramp procedure.\footnote{Of course $T_\text{mag}$ falls back to 3.6~K once the field stops changing and the persistent switch is turned off.} Our present ramp rate is thus acceptable, but it should be emphasized that there is not a lot of overhead if the magnet coils happen to be running hotter than usual while field ramping is underway.

The effects of ramping the field at $\frac{\mathrm{d}B}{\mathrm{d}t}=2.5$~mA/s are also visible in the DR temperature record (see Fig.~\ref{fig:cooldown}). The effect on $T_\text{cav}$ is the largest, since the temperature controller compensates for the heating of the mixing chamber and the effects of this low ramp rate on $T_\text{still}$ are negligible. The eddy current heating of the cavity is largest when the field is between about 4 and 7 T; we attribute this effect to electrons being polarized in the stainless steel cavity body.

\begin{figure}[h]
\centering{\includegraphics[width=1.0\textwidth]{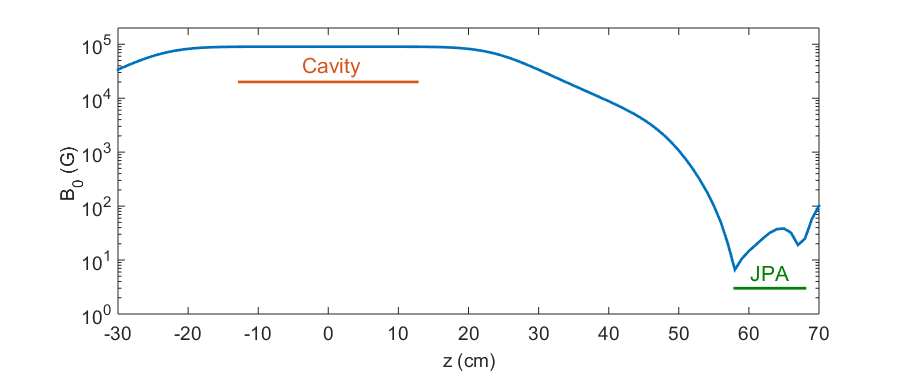}}
\caption[Magnetic field profile along the $\uv{z}$ axis]{\label{fig:bfield} $B_0$ vs.\ $z$ data from Cryomagnetics, radially averaged within 2~cm of the center of the magnet bore ($z=0$ indicates the axial center of the bore). The axial positions of the cavity and the JPA canister are indicated. The data suggests that the field at the JPA would be $\approx 600$~G in the absence of the field compensation coil.}
\end{figure}

The field is extremely homogeneous within 6~cm radially and $\pm13$~cm axially of the center of the magnet bore, where the cavity is supported: in particular the radial component of the field is $<50$~G within this region.\footnote{The specification on the radial field component is relevant to an R\&D proposal in which the inner surface of the cavity barrel would be coated with a superconducting thin film. With careful material selection and deposition procedures, superconductivity can  be maintained in the presence of a large parallel $B$ field, but still cannot tolerate transverse fields.} As noted in Sec.~\ref{sec:detector_overview}, the magnet design includes a second set of oppositely oriented superconducting coils in series with the main coil, which cancel the field to less than 50~G in the vicinity of the JPA shielding canister (see Fig.~\ref{fig:solidworks}). The magnetic field profile is plotted in Fig.~\ref{fig:bfield}.

\subsection{Magnetic shielding}\label{sub:shielding}
As noted in Sec.~\ref{sub:magnet}, the compensation coil built into the magnet produces a region where the field is always less than 50~G. The freestanding superconducting coils and multilayer ferromagnetic/superconducting shielding introduced in Sec.~\ref{sec:detector_overview} reduce the field by another factor of $\sim5\times10^{5}$. The residual field inside the shielded volume must also be uniform at a similar level on $\sim100~\mu\text{m}$ scales, both because the flux through the JPA's SQUID array (see Sec.~\ref{sub:jpa_tuning}) must be uniform, and because mechanical vibrations will couple field gradients into time-varying magnetic flux (see Sec.~\ref{sub:jpa_fluct}).

\begin{figure}[h]
\centering{\includegraphics[width=.6\textwidth]{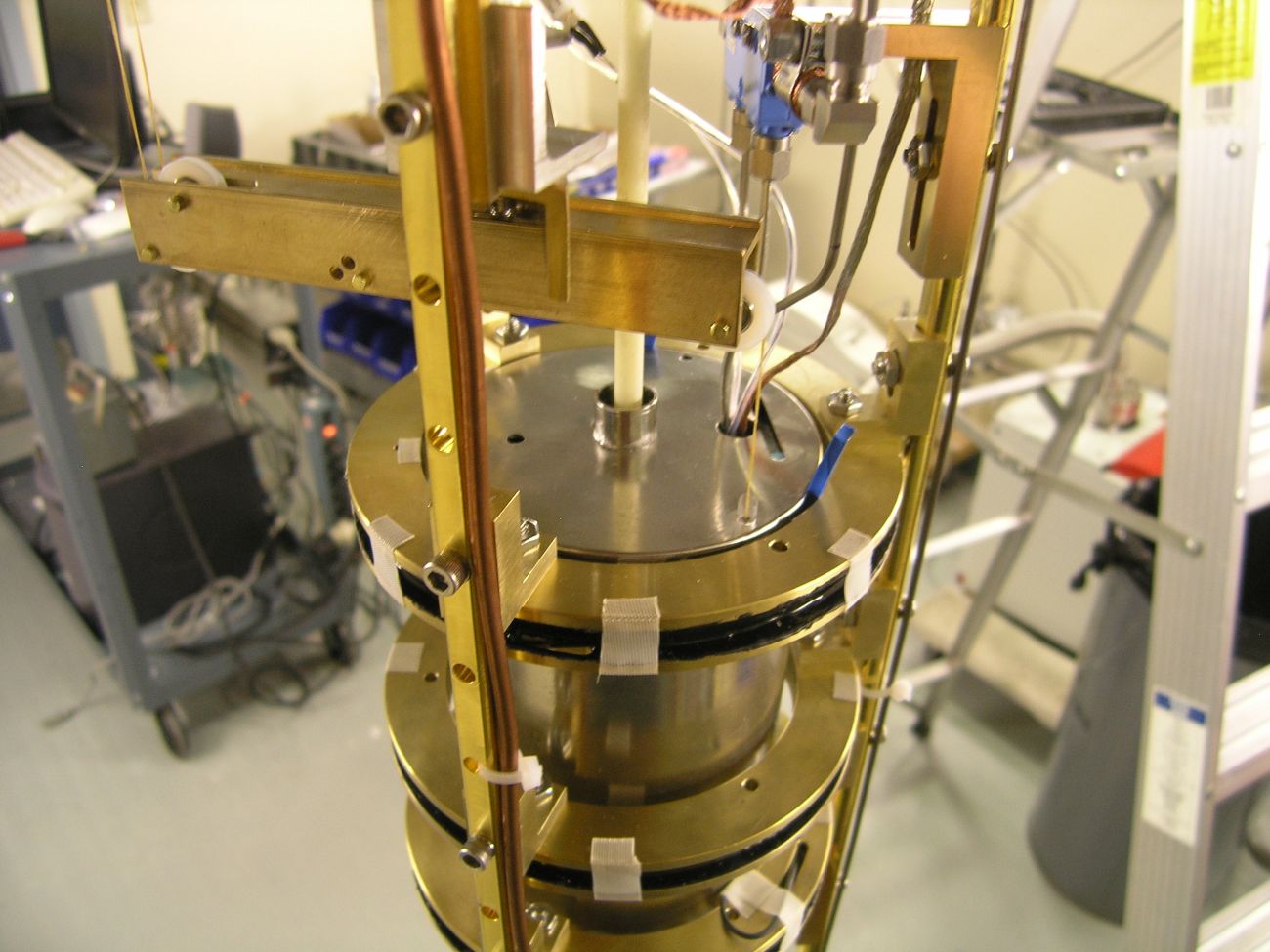}}
\caption[Magnetic shielding for the JPA]{\label{fig:mag_shields} Magnetic shielding of the JPA, including three persistent superconducting coils.}
\end{figure}

The outer elements of the JPA shielding can be seen in Fig.~\ref{fig:mag_shields}. The three persistent superconducting coils provide a quasi-active shield against flux changes: because they become superconducting before the 9~T field is ramped up, the flux inside will be maintained at the original ambient level. The superconducting shield coils each comprise 100 turns of 0.7~mm copper-clad NbTi wire (Supercon Inc.\ SC-T48B-M-0.7mm) wound around 9.3~cm ID brass hoop forms; the wire ends were cold welded to make a persistent superconducting connection (see Ref.~\citep{NIM2017} for details). A similarly wound coil, with 24~cm OD, is clamped to the bottom of the DR mixing chamber plate (see Fig.~\ref{fig:haystac_open}). Without the coil the fringe field of the magnet would be about 300~G in this region, which is large enough to affect the operation of the shielded cryogenic circulators discussed in Sec.~\ref{sec:receiver}.   

The ferromagnetic shield comprises a nested pair of cylinders made of 1.5~mm thick Amumetal 4~K. They are nearly hermetically sealed, with welded-on bottom plates and tight-fitting lids, and only a few small penetrations necessary to permit the passage of the G10 shaft and Kevlar lines used by the motion control systems (Sec.~\ref{sub:cav_tuning}) as well as the JPA's thermal link and microwave and DC signal lines. The outer shield (FS1) has length 15.2~cm and OD 8.9~cm, while the inner shield (FS2) has length 13.3~cm and OD 7.6~cm.

Between the two ferromagnetic shields is a 1.6~mm Pb superconducting shield with geometry similar to the ferromagnetic shields; it fits closely over the outside of FS2. The inside of the Pb can is coated with clear acrylic paint to avoid electrical contact and thermal currents; it is glued to FS2 using Loctite~680. The final passive shield component is a 0.13~mm thick Nb sheet that lines the inner surface of FS2. This sheet forms an ``open cylinder'' with no endcaps and an axial slit to allow flux trapped inside to escape. The sheet is coated with acrylic paint, again to prevent the formation of electrical contacts, and glued to the inner surface. A G-10 disk and nylon standoffs support the enclosure housing the JPA (Sec.~\ref{sub:jpa_design}) at the axial center of the shield assembly, with a radial offset to allow the tuning shaft to pass. The shields are thermally anchored to a Cu base plate affixed to the gantry. 

The Nb sheet enforces a boundary condition that the magnetic field must be axially homogeneous along its surface. The ferromagnetic boundary condition at the FS2 endcaps requires that the magnetic field be perpendicular to the surface, which matches perfectly the boundary condition imposed by the Nb sheet. Together the superconducting and ferromagnetic boundary conditions ensure that the residual field at the JPA is both very small and very homogeneous. A final active layer of magnetic shielding is provided by feedback to the JPA flux bias coil (discussed in Sec.~\ref{sub:feedback}), which ensures that the residual field is also temporally stable.

In addition to ensuring an acceptable flux environment during operations at $T_\text{mc}=127$~mK, the persistent coils and shields ensure that flux is not trapped in the Nb JPA circuit when it cools through its superconducting transition at $T_\text{Nb}\approx 9~\text{K}$: even trapped flux from the earth's field can compromise the stability of the JPA (see Sec.~\ref{sub:jpa_fluct}). The permeability of the ferromagnetic shields is maximized at 4~K, so the net shielding is even larger around the transition than in steady-state operation.

\section{Microwave cavity}\label{sec:cavity}
HAYSTAC collaborators at Berkeley and Livermore designed the cavity currently installed in the experiment, and the Berkeley group continues to develop cavity designs to probe different parts of the $\nu_c\gtrsim5~\text{GHz}$ parameter space. I discuss the design and fabrication of the current HAYSTAC cavity in Sec.~\ref{sub:cav_design}, the \textit{in situ} characterization of its performance in the $5.7 - 5.8$~GHz range scanned for the first HAYSTAC data run in Sec.~\ref{sub:cav_meas}, and the mechanical systems used for tuning and adjusting the receiver coupling in Sec.~\ref{sub:cav_tuning}.

\subsection{Cavity design}\label{sub:cav_design}
The initial HAYSTAC cavity is a right circular cylinder with barrel length 25.4~cm and 10.2~cm~ID (and thus 2~L total volume). It was machined from stainless steel, electroplated with oxygen-free high-conductivity (OFHC) copper via the UBAC process, and finally annealed; this last step increases the grain size of the copper crystals and thus decreases the surface resistance.\footnote{Stainless steel is used because copper deforms easily and is not conducive to precision machining. The $125\mu\text{m}$ copper plating is very thick compared to the skin depth (Sec.~\ref{sub:highfreq}) and thus is sufficient for excellent RF conductivity.} Knife edges on both ends of the barrel ensure good electrical contact with the endcaps, each of which is bolted to the barrel with 20 equally spaced 4-40 screws. The $\text{TM}_{010}$ mode of the empty cavity has frequency $\nu_c=2.25$~GHz [c.f.\ Eq.~\eqref{eq:nu0n0}]. From Eq.~\eqref{eq:q_delta} with $\varsigma=2$ we obtain the theoretical upper bound
\begin{equation}\label{eq:q010}
Q_0 \leq \frac{L/r}{1+L/r}\frac{r}{\delta_a} \approx 2\times10^5,
\end{equation}
where I have used $\nu_c$ for the empty cavity in Eq.~\eqref{eq:delta_anom}. The empty HAYSTAC cavity achieved $Q_0$ within a factor of 2 of this limit in initial tests at 4~K.

\begin{figure}[h]
\centering{\includegraphics[width=.7\textwidth]{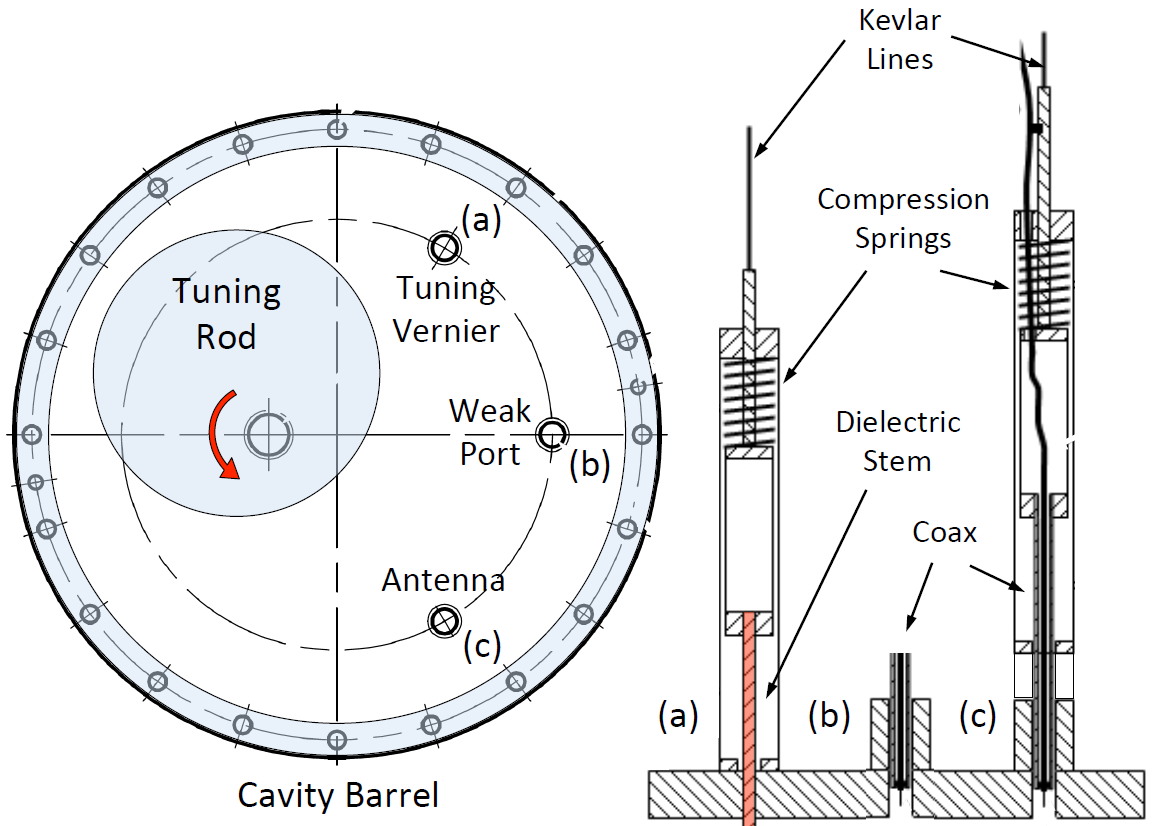}}
\caption[Cavity ports and mechanical controls]{\label{fig:cav_top}Top and side view of the cavity upper endcap, detailing the positions and design of the tuning rod, tuning vernier, signal injection port, and variable-coupling antenna. The arrow indicates the direction of increasing $\theta$.}
\end{figure}

\begin{figure}[t]
\centering{\includegraphics[width=1.0\textwidth]{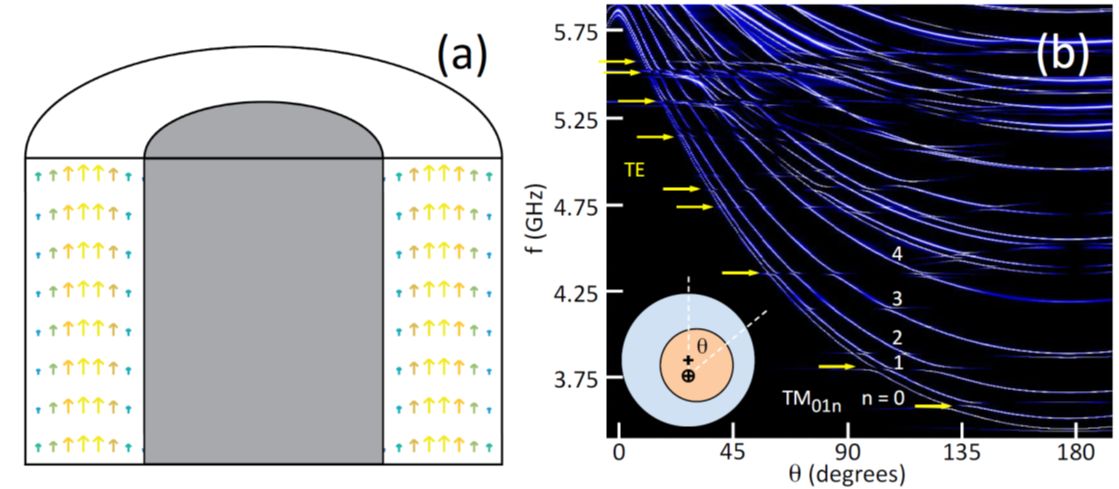}}
\caption[$\text{TM}_{010}$ field profile and mode map]{\label{fig:cav_mode} (a) CST Microwave Studio simulation of the TM$_{010}$-like mode with the tuning rod centered in the cavity ($\theta=0^\circ$). (b) Measured cavity mode map. For an aspect ratio $L/r\approx5$, the lowest-frequency TM modes are the $\text{TM}_{01n}$ modes, where $n$ is the number of nodes along $\uv{z}$. The frequencies of these modes decrease steeply with increasing radial distance of the tuning rod from cavity center (increasing $\theta$). TE modes (indicated by yellow arrows) do not couple to the antenna, and thus only become visible in the mode map near mode crossings where they hybridize with TM modes. The TE mode frequencies are largely insensitive to the position of the tuning rod.}
\end{figure}

A single copper-plated stainless rod with $5.1~\text{cm}$ diameter is used for coarse tuning of the cavity (see Fig.~\ref{fig:cav_top}). The rod pivots around an off-axis dielectric axle, nearly touching the wall at one end of its tuning range and centered within the cavity at the other end. I will label the position with the rod centered as $\theta=0^\circ$: the direction of increasing $\theta$ is given by the right-hand rule looking down into the cavity from above. The full range of motion $0<\theta<180^\circ$ resulted in a tuning range $5.84 > \nu_c > 3.55$~GHz for the $\text{TM}_{010}$-like mode.\footnote{The cavity's other TM modes also tune as the rod is rotated, maintaining a constant frequency spacing which is useful for mode identification.} The tuning rod occupies 25\% of the cavity volume, and also increases the surface-area-to-volume ratio relative to that of the empty cavity (reducing $Q_0$). However, throughout the tuning range of the $\text{TM}_{010}$-like mode, the cavity volume is larger than that of an empty cylinder whose $\text{TM}_{010}$ resonance is at the same frequency (see discussion in Sec.~\ref{sub:admx}). In fact, at $\theta=0^\circ$, the cavity is essentially an annular resonator, which has the largest volume for a TM$_{010}$-like mode at any given frequency. The electric field profile of the $\text{TM}_{010}$ mode at $\theta=0^\circ$ is shown in Fig.~\ref{fig:cav_mode}~\textbf{(a)}, and a ``mode map'' illustrating the cavity spectrum as a function of the rod angle $\theta$ is shown in Fig.~\ref{fig:cav_mode}~\textbf{(b)}.

Tuning rod design presents several challenges. Unless the gaps between the tuning rod and the cavity endcaps are very small relative to the cavity dimensions, the field tends to get concentrated in these gaps (which act essentially like parallel plate capacitors), leading to significant degradation of the form factor $C_\text{010}$ over large swathes of the tuning range. Simulations indicated that gaps $\lesssim 250~\mu\text{m}$ were sufficiently small to achieve good performance given the HAYSTAC cavity dimensions. It might seem that this issue could be circumvented with DC electrical contact between the tuning rod and the cavity endcaps, but in practice it is difficult to make contact at DC without compromising the RF performance of the cavity; in particular a conducting axle would act like an additional antenna that would siphon power out of the cavity mode and substantially reduce the $Q$. The initial HAYSTAC cavity was designed with $250~\mu\text{m}$ gaps; Teflon ``washers'' are used to ensure that the gaps on either side remain equal during cavity assembly without compromising the free rotation of the rod.

The considerations discussed above imply that the tuning rod axle must be dielectric, which in turn makes it difficult to thermalize the rod to the cavity walls (and the mixing chamber). The axle used in HAYSTAC was a polycrystalline alumina tube with 0.25''~OD and 0.125''~ID, chosen for its high thermal conductivity relative to other dielectrics.\footnote{In retrospect, single-crystal sapphire would have been a better choice.} However, this design did not provide a sufficiently good thermal link to the tuning rod; the consequences of this poor thermal link for the first HAYSTAC data run are discussed in Sec.~\ref{sub:hotrod}. After the first HAYSTAC data run, we improved the tuning rod thermalization by inserting a solid copper ``finger'' partially into the alumina shaft, as discussed in Ref.~\citep{zhong2017}. I will not discuss upgrades to HAYSTAC since the first data run in detail in this thesis.

We will see in Sec.~\ref{sub:cav_tuning} that tuning rod angular steps $\delta\theta$ corresponding to frequency steps $\delta\nu_c\approx100$~kHz can be readily achieved with our initial rotary motion control system design, though not without some complications. The HAYSTAC cavity design also featured an independent fine-tuning system based on varying the axial insertion of a 3.2~mm diameter alumina rod ($\varepsilon_r=10$), which I will refer to as the ``tuning vernier'' to distinguish it from the copper ``tuning rod.'' The upper cavity endcap has ports for two antennas (see Fig.~\ref{fig:cav_top}): the transmission line antenna has weak coupling (nominally $-50$~dB attenuation at $T\leq4~\text{K}$; thus it does not affect $Q_L$) and the coupling of the receiver antenna can be varied by adjusting its insertion depth. The tuning vernier and antenna are adjusted using Kevlar-based systems described in Sec.~\ref{sub:cav_tuning}.

Each antenna is an electric field probe constructed by simply stripping away a section of the outer conductor from a piece of 0.141'' coax connectorized on one end with an SMA plug. Electrical contact between the antenna outer conductors and the endcap is maintained by RF fingerstock; the receiver antenna was rhodium-plated and the fingerstock was gold-plated to ensure good contact is maintained after repeated antenna adjustments. The E-field probe measures $E_z$ and thus is insensitive to TE modes (c.f.\ Fig.~\ref{fig:cav_mode}). Increasing the insertion depth of the antenna increases the cavity-receiver coupling $\beta$ provided that the antenna is not so far in that it appreciably distorts the cavity's mode structure.

\begin{figure}[h]
\centering{\includegraphics[width=.85\textwidth]{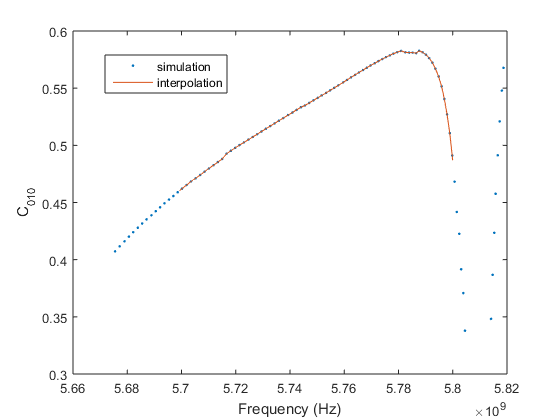}}
\caption[$C_{010}$ vs.\ frequency in the initial HAYSTAC scan range]{\label{fig:c010} $C_{010}$ vs.\ $\nu_c$ obtained from a simulation in CST microwave studio. The spacing between blue dots corresponds to the angular resolution $\delta\theta$ of the simulation, and the red curve is an interpolation. Higher resolution is evidently needed near the mode crossing at 5.81~GHz, but this region was not scanned in the initial HAYSTAC data run}
\end{figure}

Simulations of the spatial profile of the $\text{TM}_{010}$ electric field as a function of the rod angle $\theta$ were conducted in CST Microwave Studio at Berkeley and Livermore, and validated by bead perturbation measurements (see Ref.~\citep{NIM2017} for further discussion). These simulations were used to compute the form factor $C_{010}$ as a function of frequency. $C_{010}$ is plotted in Fig.~\ref{fig:c010} over the 5.7 -- 5.8~GHz range covered in the initial HAYSTAC data run; it is close to our nominal target value of 0.5 throughout this range. In general, the cavity characterization campaign at Berkeley has demonstrated that for cavities with $\nu_c\gtrsim5$~GHz, extremely good machining and alignment tolerances are required to avoid issues with mode localization~\citep{cavity1990}: $\mathcal{O}(50~\mu\text{m})$ misalignments have significant effects on the mode's field profile.\footnote{The stringent tolerances led us to abandon an earlier cavity design based on studies of multi-rod resonators described in Ref.~\citep{stern2015}.} The alumina spindles protruding from both ends of the tuning rod were affixed to a stainless steel ``backbone'' running through the rod to ensure good alignment.

\subsection{In-situ characterization}\label{sub:cav_meas}
Although the cavity described in Sec.~\ref{sub:cav_design} achieved a $\sim50\%$ fractional tuning range, the demands of a practical haloscope search constrained our initial data run to a small slice of this range. Optimizing haloscope performance over a wide frequency range is a nontrivial and not very glamorous aspect of detector design: it is necessary to simultaneously maximize $Q_0$ and $C_{010}$, and avoid both mode crossings and regions of the tuning range where the mode is concentrated far from the antenna, such that only very weak coupling $\beta \ll 1$ is possible. In practice, the initial HAYSTAC cavity achieved optimal performance in several distinct windows which were each largely free of mode crossings. We elected to focus on the $5.6-5.8~$GHz range, and ended up restricting the first HAYSTAC data run to the upper half of this range, due to issues with the tuning system described in Sec.~\ref{sub:cav_tuning}.

After installing the cavity in HAYSTAC, we characterized it by using the VNA to inject swept-frequency signals through the cavity in transmission or reflection (see Fig.~\ref{fig:receiver_simple}). Wide sweeps of the cavity's amplitude response in transmission look like vertical slices of Fig.~\ref{fig:cav_mode}~\textbf{(b)}: the signal gets through the cavity only on resonance, and so each mode shows up as a narrow peak (or a bright spot in an intensity plot). In reflection modes appear as sharp notches relative to a background with ripples on large spectral scales due to slight impedance mismatches in the microwave signal path.\footnote{Impedance mismatches in microwave systems are usually parameterized by the voltage standing wave ratio (VSWR). Typical VSWR observed in HAYSTAC is $\approx1.05$ on scales $\gtrsim50$~MHz.}

The symmetry of the cavity design naively seems to suggest that positive and negative rod angles $\theta$ are equivalent. In practice, this symmetry is not observed due to fabrication imperfections: studies of the electric field distribution were consistent with form factor simulations in the 5.6--5.8~GHz range only on the $\theta>0$ side of the cavity (i.e., with the rod in the upper half of Fig.~\ref{fig:cav_top}). The symmetry is also broken by the position of the receiver antenna, which can be interchanged with the tuning vernier without requiring any other design changes. We put the antenna at $\theta = -45^\circ$ as illustrated in Fig.~\ref{fig:cav_top} after early tests indicated that it was not possible to critically couple to the mode with the rod on the same side of the cavity as the antenna; this effect is independent of the superior performance with the rod on the $\theta>0$ side of the cavity.

Fig.~\ref{fig:cav_mode}~\textbf{(b)} indicates that the antenna couples to many TM modes which all tune together with increasing $\theta$. However, most of these modes do not have appreciable form factor (i.e., they are not efficiently coupled to the axion field): in particular, the prominent $\text{TM}_{01n}$ modes with $n>0$ have $C_{01n}=0$ exactly. This makes it very important to have a robust procedure for correctly identifying the $\text{TM}_{010}$ mode. In the first HAYSTAC data run mode identification was greatly facilitated by our operating point near the top of the tuning range: we can tune the rod clockwise past $\theta=0$ and verify that the mode we believe to be $\text{TM}_\text{010}$ turns around at 5.84~GHz and that no other $\theta$-dependent modes appear at lower frequencies. We repeat this foolproof but time-consuming test once per cooldown; after tuning the $\text{TM}_{010}$ mode back down into the desired operating range we keep track of it using the characteristic spacing between the lowest three modes (30~MHz between $\text{TM}_{010}$ and $\text{TM}_{011}$; 90~MHz between $\text{TM}_{011}$ and $\text{TM}_{012}$).

\begin{figure}[h]
\centering{\includegraphics[width=.85\textwidth]{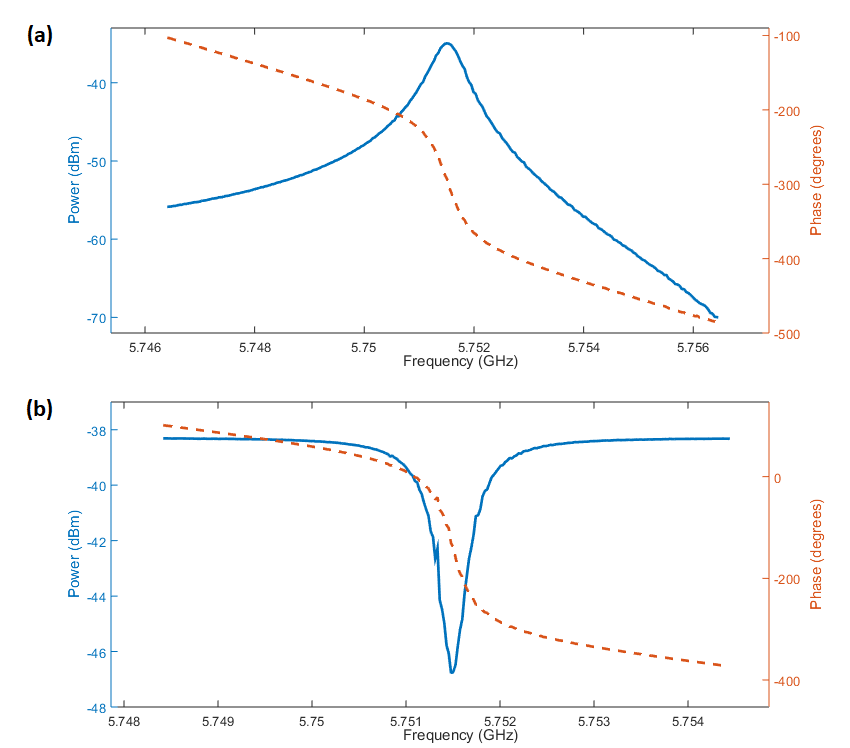}}
\caption[Cavity transmission and reflection measurements]{\label{fig:tx_rfl} \textbf{(a)} Power measured at the VNA input (blue solid curve) and cavity phase response (red dashed curve) vs.\ frequency for a swept signal injected through the transmission line near the $\text{TM}_{010}$ resonance. The sweep comprised 251 points over a total span of 10~MHz approximately centered on $\nu_c$; the power at the VNA output was +10~dBm and the measurement bandwidth at each point was 100~Hz. Fitting this measurement yields $Q_L=9800$. \textbf{(b)} A swept signal injected through the reflection line near the $\text{TM}_{010}$ resonance. The color coding and sweep parameters are the same as in \textbf{(a)}, except the span is 6~MHz. From this measurement we obtain $\beta=2.2$.}
\end{figure}

Having identified the $\text{TM}_\text{010}$ mode, we can ``zoom in'' on the mode in transmission or reflection to characterize its behavior in the frequency range of interest. Fig.~\ref{fig:tx_rfl} shows the magnitude and phase response for typical sweeps over the $\text{TM}_{010}$ resonance in both transmission and reflection. This data was taken with the cavity cold, and the JPA off but all subsequent amplifiers on. I have normalized the magnitude response to the total power measured at the VNA input, to give a sense of the typical power levels of interest for cryogenic measurements of the cavity mode.\footnote{Power transmission in microwave systems is usually characterized in logarithmic units. A decibel (dB) is a logarithmic measure of power gain $G$: $G_\text{log}~[\text{dB}] = 10\log_{10}(G_\text{lin})$, where $G_\text{log}<0$ implies attenuation ($G_\text{lin}<1$). Absolute power levels are often quoted in dBm (decibels relative to a milliwatt): $P_\text{log}~[\text{dBm}] = 10\log_{10}\big(P_\text{lin}/(1~\text{mW})\big)$. Convenient correspondences to remember are $G_\text{log}=3~\text{dB} \leftrightarrow G_\text{lin} \approx 2$, $G_\text{log}=5~\text{dB} \leftrightarrow G_\text{lin} \approx 3$, and $G_\text{log}=10~\text{dB} \leftrightarrow G_\text{lin} = 10$.} But the absolute power level is not needed to extract cavity parameters, and for transmission measurements we also do not need the phase response. The transmission magnitude response is just the usual Lorentzian characteristic of a high-$Q$ harmonic oscillator. We thus fit the measured power (in arbitrarily normalized linear units) to 
\begin{equation}\label{eq:tm010_tx}
P(\nu) = \frac{P_0}{1 + \big(2Q_L(\nu-\nu_c)/\nu_c\big)^2}
\end{equation}
to obtain best-fit values for $\nu_c$ and $Q_L$.\footnote{$Q_L$ could also be obtained from a reflection measurement, and indeed the transmission line is not strictly necessary for the haloscope search at all. However, it is a little more intuitive to look at the cavity response in transmission, and we can also use this line to inject synthetic axion-like signals into the experiment to validate the analysis (see appendix~\ref{app:fake_axions}).}

To extract the coupling parameter $\beta$ we use both the magnitude and phase data from reflection measurements of the $\text{TM}_{010}$ mode. For developing intuition, it is useful to model the cavity as series $RLC$ circuit, whose quality factor is inversely proportional to the resistance $R$, which is just the value of the resonator impedance $Z_r$ exactly on resonance where the reactive components cancel. In the equivalent circuit model, the impedance $Z_c$ of the transmission line as seen from the cavity is given by its actual $50~\Omega$ characteristic impedance in parallel with a small coupling capacitor representing the antenna~\citep{pozar2012}. It is straightforward to show that Eq.~\eqref{eq:beta_def} is equivalent to
\begin{equation}\label{eq:beta_imp}
\beta = \frac{\abs{Z_c}}{R},
\end{equation}
and critical coupling is just the condition that the cavity be impedance-matched to the transmission line. The impedance mismatch associated with any arbitrary interface may be parameterized by the amplitude reflection coefficient
\begin{equation}\label{eq:gamma_refl}
\Gamma = \frac{Z_r-Z_c}{Z_r+Z_c},
\end{equation}
where $Z_r$ and $Z_c$ are both generally complex. I have defined $\Gamma$ as the reflection coefficient for a signal incident on the cavity from the transition line. Restricting our focus to the behavior exactly on resonance, we see that the signal is completely absorbed by the load at critical coupling ($\Gamma=0$). When the cavity is far undercoupled ($\beta\ll1$), the cavity looks like an open circuit to an incoming wave, and $\Gamma\rightarrow+1$. When the cavity is far overcoupled ($\beta\gg1$), the cavity looks like a short circuit, so $\Gamma\rightarrow-1$. Also making use of the fact that the resonant phase shift $\Delta\phi<\pi$ ($\Delta\phi>\pi)$ for an undercoupled (overcoupled) cavity, we can write
\begin{equation}\label{eq:beta_gamma}
\beta = \frac{1+\text{sgn}(\Delta\phi-\pi)\abs{\Gamma}}{1-\text{sgn}(\Delta\phi-\pi)\abs{\Gamma}}.
\end{equation}

Eq.~\eqref{eq:beta_gamma} is not the most compact expression but it is convenient for measurement, since $\abs{\Gamma}$ is just the value of the amplitude response on resonance normalized to the off-resonance level, and $\Delta\phi$ is also easily obtained from the data.\footnote{In practice, to obtain $\Delta\phi$ we must subtract off the electrical delay, which is just the phase shift associated with the total cable length in the VNA signal path. The electrical delay is thus simply proportional to frequency, as the off-resonance behavior in both panels of Fig.~\ref{fig:tx_rfl} indicates.}

\begin{figure}[h]
\centering{\includegraphics[width=.85\textwidth]{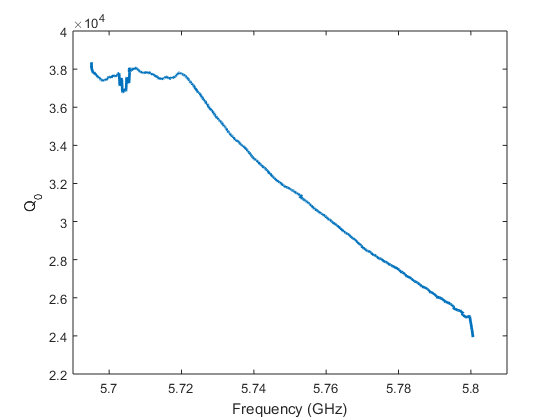}}
\caption[$Q_0$ vs.\ frequency in the initial HAYSTAC scan range]{\label{fig:Q0} $Q_0$ vs.\ $\nu_c$ obtained from measurements during the initial HAYSTAC data run. The feature around 5.704~GHz is associated with a narrow mode crossing discussed in Sec.~\ref{sub:badscans}.}
\end{figure}

By measuring the cavity in both transmission and reflection throughout the scan range, we can obtain the frequency-dependence of $Q_0$, depicted in Fig.~\ref{fig:Q0}: we see that $Q_0\sim3\times10^4$ is a typical value throughout this range, implying that the effect of the extra surface area of the tuning rod and other deviations from ideal behavior associated with the more complicated geometry is to reduce the $Q$ of the empty cavity by a factor of $\sim3-4$. We can choose $Q_L$ [Eq.~\eqref{eq:Q_L}] to be any value $\leq Q_0$ by adjusting the antenna insertion to set $\beta$ to the desired value. As noted in Sec.~\ref{sub:scan}, $\beta=2$ optimizes the scan rate of the haloscope search, which implies a target value $Q_L\sim10^4$ in HAYSTAC. The relationship between $\beta$ and the antenna insertion depth is frequency-dependent; empirically, at fixed insertion depth $\beta$ remains roughly constant as we tune down from 5.8~GHz to 5.75~GHz, and then begins to increase with decreasing frequency.\footnote{The frequency dependence of $\beta$ at constant insertion depth does not merely track that of $Q_0$, because the efficiency of coupling to the cavity mode depends on the frequency-dependent spatial profile of the mode's $E$ field.}

\subsection{Motion control systems}\label{sub:cav_tuning}
The HAYSTAC cavity relies on three motion control systems, which implement tuning rod rotations and axial adjustment of both the antenna and the tuning vernier. The main requirement for these systems is that they reliably deliver sufficiently small angular or linear motions without any appreciable heating of the DR: step size uniformity and minimal backlash are desirable but not strictly necessary. We opted to avoid the cryogenic gear-based system used by ADMX~\citep{ADMX2000}, as such designs run into problems with excessive heat loads and the freezing of residual gas at low temperatures; instead we transmitted motion from room-temperature stepper motors to the cavity using a system primarily based on springs, Kevlar lines, and G10-CR (cryogenic G10) shafts.

 In their initial incarnations, all three systems used double-shaft stepper motors (Applied Motion Products HT-23-595D NEMA 23) driven by STR2 microstepping controllers from the same manufacturer. Each motor's shaft is coupled on one side to a ten-turn potentiometer that encodes the net rotation angle. The vernier and antenna stepping motors are coupled to the experiment through 10:1 worm gear reductions, while the tuning rod motor is coupled directly. The rotary motions are coupled into the cryostat using Lesker FMH-25A Dynamic O-Ring Shaft Seal feedthroughs on top of the DR. The microstepping resolution of the tuning rod stepper was set at 5000 steps/revolution, and the other two steppers were set to 200 steps/revolution.

Within the cryostat, the antenna and vernier are controlled by 0.36~mm Kevlar thread lines that are wound directly onto the 0.25'' rotary feedthrough shaft that extends into the vacuum. At each stage of the DR, the lines pass through 1/8''~ID, 1'' long tubes that are thermally linked to the DR stage and serve as radiation shields. The lines are routed to the cavity by use of nylon pulleys; part of this system is visible in Fig.~\ref{fig:mag_shields}.\footnote{The ball bearings in these pulleys (McMaster-Carr 3434T13) were ultrasonically cleaned to remove all lubricants that would freeze at low temperatures.} The antenna and vernier are supported by fixtures that allow only axial motion (see Fig.~\ref{fig:cav_top}). A spring is compressed when the antenna or vernier is pulled out of the cavity by the Kevlar thread, ensuring smooth and reversible motion.\footnote{It should be noted that Kevlar \textit{expands} on cooling (like a rubber band), and the spring tensioning of the lines must absorb the length change lest the lines fall out of the pulley grooves.}

Rotary motion for the tuning rod system is delivered to the mixing chamber level by a G10-CR tube (0.25'' OD, 1/32'' thick). The ends of the tube were fitted with glued-in brass extensions which enable the use of set screws to couple to both the feedthrough at the top and the mechanics at the mixing chamber end.  While G10-CR was chosen for its good thermal insulation, careful heat sinking of this solid link from $T_\text{mc}$ to 300~K is nonetheless critical. At each DR stage, a 1'' section of brass tubing (0.25'' ID, 1/64'' wall) is glued to the G10-CR, and a loose-fitting brass tube (0.75'' long, 5/32'' ID) is slipped over the 1'' section and coupled to the stage using copper braid.  At the 4 K stage, a brass blackbody radiation block was inserted into the G10-CR tube.

Just below the mixing chamber, a pulley and torsion spring system is used to transfer rotary motion from the upper shaft to a second 0.25'' G10-CR tube collinear with the magnet axis. A Kevlar line connects the brass extension of the upper shaft to the 3.22"~diameter pulley, and the torsion spring ensures that the line is always pulled taut. The lower G10-CR shaft runs through the JPA magnetic shield and down to the cavity, where a 1.7:1 anti-backlash gear reduction provides the final radial displacement required to couple to the tuning rod axle. The system inside the DR for transmitting rotary motion to the cavity thus provides a net reduction ratio of 22:1.

With the microstepping settings and reduction ratios cited above, a single step with the tuning rod system resulted in a typical rod angle change of $\delta\theta\approx3\times10^{-3}\,^\circ$ and a typical frequency step of $\delta\nu_c\approx100$~kHz, and a single step on either the vernier or antenna systems resulted in $\delta z\approx10~\mu\text{m}$ of linear travel. The full range of travel on the linear systems is about 2.9~cm. Backlash for the linear systems (rotary system) was typically about 50 (200) steps. 
All three mechanical systems produced negligible heat loads at the operating temperature with typical actuation rates of 5~Hz.

The Kevlar-based linear systems were very robust and provided the required levels of control over the antenna and vernier. The rotary system was problematic in one significant respect: due to stiction in the mechanical system, the rod angle $\theta$ drifted slowly to its final equilibrium position after actuation. The resulting mode frequency drifts were typically $\sim 100$~kHz over $\approx$15~minutes, with no consistent temporal profile; rarely after large rod motions the mode frequency drifted a total of several hundred kHz on timescales of hours. Mode frequency drifts during a long cavity noise measurement are problematic because any haloscope analysis must weight the resulting power spectra by the Lorentzian mode lineshape in order to combine them (see Sec.~\ref{sec:rescale_combine}). We could always simply try to wait out the drift, but this would significantly reduce the live-time efficiency $\zeta$ introduced in Sec.~\ref{sub:scan}.

In the first HAYSTAC data run we circumvented this mode drift by using a ``hybrid'' tuning scheme: we took fine frequency steps using the tuning vernier and actuated the rod only intermittently. The frequency shift resulting from vernier insertion is highly nonlinear (increasing deeper into the cavity), so we restricted ourselves to a 1600 step total range of vernier travel which we divided across 16 iterations. Every 17 iterations, the DAQ code calculates the number of steps on the rotary system corresponding to the frequency range covered in the most recent full vernier cycle, and actuates the tuning rod by that many steps. Then the code slowly (1.67~Hz step rate) resets the vernier to its starting position over 16 minutes while waiting out the mode frequency drift. Though this may sound like a long time, it was not actually the largest factor limiting the overall data run efficiency $\zeta=0.72$ (see Sec.~\ref{sub:daq_procedure}) given our integration time per iteration of $\tau=15$~minutes.

The only issue with the hybrid tuning scheme was that the frequency shift resulting from vernier insertion was not uniform as a function of frequency: this is to be expected, given that the vernier couples to the mode's changing $E$ field profile. At 5.75 GHz the frequency shift due to a single 100-step ``iteration'' was about 50~kHz; at 5.8~GHz it was larger by a factor of 3, and at 5.7~GHz it was smaller by about a factor of 2. This resulted in very nonuniform tuning, which is not a problem per se, but makes the automation of the experiment more difficult. Already by 5.7~GHz the scan rate with vernier tuning was quite slow, and the ``hybrid'' tuning scheme quickly peters out at lower frequencies. For this reason among others, we decided to scan only 5.7 -- 5.8~GHz for the initial HAYSTAC data run. Since the first data run, we have replaced the rotary system entirely with a cryogenic piezoelectric rotator (Attocube ANR240) mounted on the gantry. This system enables much more direct coupling to the tuning rod, and provided very uniform small frequency steps with negligible drift. Details of the piezo tuning system are discussed in Ref.~\citep{zhong2017}.

\section{Josephson parametric amplifier}\label{sec:jpa}
HAYSTAC collaborators at CU Boulder/JILA fabricated the JPA currently installed in the experiment, and conduct R\&D towards optimizing JPA designs for the haloscope search. I discuss the basic physical principles of parametric amplification and JPAs specifically in Sec.~\ref{sub:jpa_intro}, describe the design and fabrication of the present HAYSTAC JPA in Sec.~\ref{sub:jpa_design}, and describe the JPA's flux tuning system in Sec.~\ref{sub:jpa_tuning}. Discussion of JPA commissioning and operation is deferred to Sec.~\ref{sec:jpa_op}.

\subsection{Principles of JPA operation}\label{sub:jpa_intro}
The phenomenon of \textbf{parametric gain} may be realized by taking any simple harmonic oscillator and modulating one of the parameters that governs its resonant frequency $\omega_0$ at a modulation frequency $\omega_m=2\omega_0$. Whatever provides the energy of this modulation is called the \textbf{pump}, and the oscillator when operating in this regime is called a parametric amplifier. The basic physics of parametric gain is energy transfer between the pump and Fourier components of the oscillator's motion with fixed phase relationships to the modulation. In particular, oscillations initially in phase with the modulation are linearly amplified and oscillations $90^\circ$ out of phase are deamplified. This implies that a parametric amplifier is inherently a single-quadrature linear amplifier, whose performance is not constrained by the Haus-Caves theorem (see discussion in Sec.~\ref{sub:sql}).

\begin{figure}[h]
\centering{\includegraphics[width=.8\textwidth]{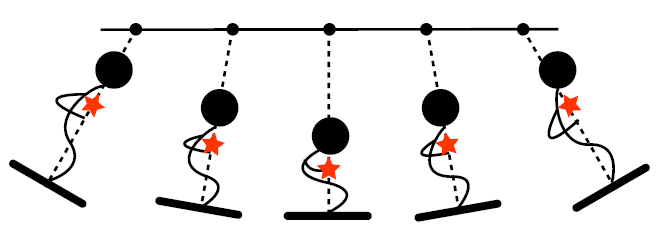}}
\caption[A mechanical parametric amplifier]{\label{fig:swing} A mechanical parametric amplifier in the form of a child on a swing. Figure from Ref.~\citep{swing}.}
\end{figure}

A simple classical example might serve to make the above discussion more clear. The swing in Fig.~\ref{fig:swing} is a parametric amplifier, whose pump is provided by a child with impeccable coordination alternately standing and crouching\footnote{It should be emphasized that the more common swing-driving practice of extending and retracting one's legs at frequency $\omega=\omega_0$ is not related to parametric amplification.} to modulate the center of mass position $\ell$ and thus the swing's frequency $\omega_0=\sqrt{g/\ell}$. If the child has studied the physics of parametric amplifiers, she will know that the right thing to do is stand when approaching the turning points (to boost her potential energy) and crouch when passing through the origin (to flatten the potential and make her kinetic energy go further). This corresponds to a specific choice of phase at the modulation frequency $\omega_m=2\omega_0$. Were the child to try this at any other frequency, the effect would quickly become incoherent, while standing at the origin and crouching at either end would deamplify the initial oscillations.

To realize a parametric amplifier in an electrical circuit, we need to modulate either the inductance $L$ or capacitance $C$ in a Hamiltonian of the form
\begin{equation}\label{eq:lc_hamilton}
H = \frac{1}{2}CV^2 + \frac{1}{2}LI^2,
\end{equation}
where $\omega_0=1/\sqrt{LC}$. A convenient trick is to use \textit{nonlinear} electrical elements. For example, if we can find a component whose inductance is 
\begin{equation}\label{eq:quad_nonlinearity}
L = L_0 + \Delta L\big(I/I_0\big)^2,
\end{equation}
then driving the circuit with a pump tone at $\omega_p=\omega_0$ will create the desired $2\omega_0$ modulation of the inductance, provided the pump is strong enough to excite the nonlinearity.

It should be emphasized that this ``quadratic nonlinearity'' is not essential to the physics of parametric amplifiers or JPAs specifically: JPA designs exist where $2\omega_0$ oscillations in the inductance are obtained in a different way.\footnote{There is a terminological pitfall to be aware of here: ``linear'' in this context would mean the current $I$ obeys a linear differential equation, and thus the parameters $C$ and $L$ are independent of $I$. It so happens that in this case the nonlinear ($I$-dependent) term is itself a nonlinear function of $I$.} Nonetheless, a quadratic nonlinearity like that of Eq.~\eqref{eq:quad_nonlinearity} is the basis for a conceptually simple and common JPA design; in particular, the JPA installed in HAYSTAC operates in this way. For definiteness, I will restrict my focus in the remainder of this thesis to the case where parametric gain is obtained using Eq.~\eqref{eq:quad_nonlinearity} and a pump tone at $\omega_p=\omega_0$.
 
\begin{figure}[!h]
\centering{\includegraphics[width=.7\textwidth]{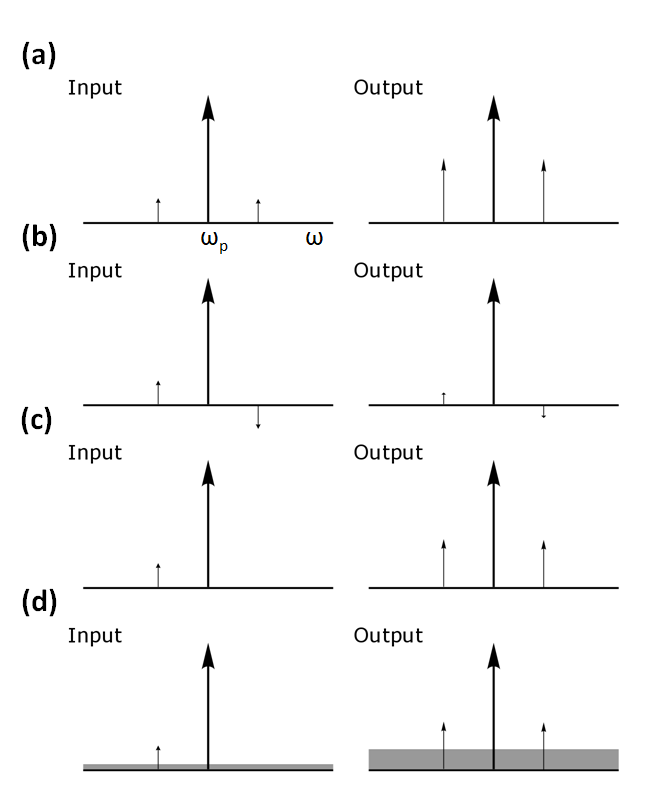}}
\caption[Schematic of JPA response to different inputs]{\label{fig:jpa_in_out} Response of a parametric amplifier to various different input signals (not to scale). \textbf{(a)} A signal in phase with the pump is amplified. \textbf{(b)} A signal $90^\circ$ out of phase with the pump is deamplified. \textbf{(c)} A single-sideband signal is a superposition of the in-phase and quadrature-phase inputs, and thus experiences both direct gain and intermodulation gain. \textbf{(d)} Intermodulation gain also couples broadband input noise into the output signal: this is the origin of the added noise $N_A$ for an ideal JPA operating in phase-insensitive mode.}
\end{figure}

For this kind of parametric amplifier, the simple qualitative picture above indicates that input signals in phase with the pump tone will be amplified and signals $90^\circ$ out of phase will be deamplified. A realistic parametric amplifier will have some small finite bandwidth (see Sec.~\ref{sub:jpa_design}), so we can apply the narrowband signal formalism introduced in Eqs.~\eqref{eq:E_quadratures} and Eq.~\eqref{eq:raising_lowering}. The in-phase signals (i.e., those whose Fourier spectra are symmetric about the pump) are said to reside in the JPA's \textbf{amplified quadrature} $X_1$, and quadrature-phase signals (with Fourier spectra antisymmetric about the pump) constitute the \textbf{squeezed quadrature} $X_2$. This situation is illustrated schematically for the simplest in-phase and quadrature-phase signals (cosine and sine) in Fig.~\ref{fig:jpa_in_out}~\textbf{(a)} and \textbf{(b)}, respectively. We see that signals in $X_1$ are amplified and signals in $X_2$ are deamplified as expected.

The quadratures $X_1$ and $X_2$ form a natural basis for discussing the operation of a parametric amplifier. But of course we can just as well consider input signals confined to the upper or lower sideband (relative to the pump), which are just superpositions of signals in $X_1$ and $X_2$. The linearity of the parametric amplification process\footnote{That is, any input $I_s$ regard as a ``signal'' must be sufficiently small that the nonlinear term in Eq.~\eqref{eq:quad_nonlinearity} is negligible. Essentially, we are using the strong pump tone to deform the Hamiltonian in some desired way, and then considering the linear response to small perturbations around this operating point. See Ref.~\citep{yurke2006} for a detailed discussion.} implies that the output follows directly from the superposition principle, as illustrated in Fig.~\ref{fig:jpa_in_out}~\textbf{(c)}. We see that when applied to a signal detuned to one side of the pump, a parametric amplifier exhibits both \textbf{direct gain} (an input signal at $\omega_s$ produces an output at $\omega_s$) and \textbf{intermodulation gain} (an input at $\omega_s$ produces an output at the \textbf{image frequency} $2\omega_p-\omega_s$), with $G_I \rightarrow G_D$ in the high-gain limit. In this limit, the output spectrum of a parametric amplifier is always symmetric about the pump, whether or not the input spectrum was symmetric.

Note that the operation shown in Fig.~\ref{fig:jpa_in_out}~\textbf{(c)} is that of a phase-insensitive linear amplifier: an arbitrarily complicated spectrum confined to the lower sideband (which by definition has no fixed phase relationship to the pump) is reproduced faithfully at the amplifier output insofar as there is nothing in the upper sideband. Then the Haus-Caves theorem [Eq.~\eqref{eq:haus_caves}] says that a parametric amplifier must add noise to any single-sideband signal, and it is instructive to consider where this noise comes from. We have thus far neglected the noise that must be physically present at the input of the parametric amplifier: typically the amplifier will ``see'' some thermal environment at temperature $T$ in addition to the signal of interest, and the input noise will be given by the first two terms in Eq.~\eqref{eq:system_noise} with $\nu=\omega_0/2\pi$. 

The case in which this broadband thermal noise (which is spectrally flat within the amplifier's small fractional bandwidth) and a single monochromatic signal are the only features in the input spectrum is illustrated in  Fig.~\ref{fig:jpa_in_out}~\textbf{(d)}. We see that the input noise at the image frequency is coupled into the output at the signal frequency by intermodulation gain. Insofar as the noise spectrum was flat, the total noise at $\omega_s$ in the output spectrum is just double the input noise multiplied by the amplifier gain. If we make the input noise as small as possible by taking $T\rightarrow0$, the input-referred added noise is just equal to the input quantum noise. Then $N_\text{sys}=1$: the amplifier precisely saturates the standard quantum limit! 

Our discussion motivating the SQL in Sec.~\ref{sub:sql} showed only that any phase-insensitive linear amplifier must add at least the minimum noise given by Eq.~\eqref{eq:haus_caves} to be consistent with the laws of quantum mechanics; it gave no hint as to where this added noise originates in any given system. In the qualitative discussion above, I have presented a heuristic picture of the physical origin of $N_A$ in an ideal parametric amplifier operated in the phase-insensitive mode.\footnote{A parametric amplifier operating in single-quadrature mode is often called a degenerate parametric amplifier; the same device operating in the phase-insensitive mode is said to be nondegenerate.} A more formal treatment would be required to show that an ideal parametric amplifier \textit{noiselessly} amplifies $X_1$ when operated in the single-quadrature mode: roughly speaking, the amplifier correlates noise Fourier components above and below the pump, such that they destructively interfere in the noise added to the amplified quadrature. Moreover, the $X_2$ quadrature is deamplified even if only quantum noise was present at the input, and thus a parametric amplifier may be used to produce a \textbf{squeezed state}, with noise variance $<1/2$~quantum in one quadrature. These two features -- noiseless single-quadrature amplification and squeezing of vacuum noise -- are essentially why parametric amplifiers are interesting to physicists working in quantum measurements and quantum information. 

In the haloscope search, we expect the axion signal to appear as a localized power excess at some unknown frequency $\nu_a$ which is either on one side of the pump or the other: thus phase-insensitive operation (Fig.~\ref{fig:jpa_in_out}~\textbf{(d)}) is most natural for the haloscope search, and we have thus far operated our JPA in phase-insensitive mode during HAYSTAC data runs. Of course, the physics of parametric amplification is the same in any basis: the only thing that changes is which Fourier components we include in the definition of the ``signal.'' We can always choose to read out a JPA in such a way that we only measure $X_1$, but it can be shown that doing so offers no improvement in the axion search sensitivity without a second JPA and significantly greater operational complexity.\footnote{Operating in single-quadrature mode by itself does not eliminate the quantum limit on the input noise [i.e., the second additive term in Eq.~\eqref{eq:system_noise}]. It may still seem that single-quadrature operation improves the SNR by a factor of 2 due to the absence of the $N_A$ term, but this improvement is exactly canceled by the loss of information about the $X_2$ component of the axion signal. Thus, single-quadrature operation offers no improvement in axion search sensitivity unless we can also eliminate the zero-point motion of the input noise by initializing the cavity in a squeezed state~\citep{zheng2016}. Realization of such a squeezed state receiver is a major R\&D project within the HAYSTAC collaboration (see Sec.~\ref{sec:haystac_future}).}

This is all I'm going to say about the essential physics of parametric amplifiers; for a more detailed pedagogical discussion, I recommend Refs.~\citep{yurke2006} and \citep{castellanos2010}. Next we can consider the specific case of the Josephson parametric amplifier. In a JPA, nonlinear inductance of the form Eq.~\eqref{eq:quad_nonlinearity} is provided by a \textbf{Josephson junction}, comprising a pair of superconductors separated by a very thin insulating layer or other ``weak link.'' The superconducting state on either side of the junction is described in Landau-Ginzburg theory by a complex order parameter $\psi=\abs{\psi}e^{i\delta}$. The Josephson effect is the tunneling of Cooper pairs from superconductor $A$ to superconductor $B$ in response to the phase difference $\delta=\delta_A-\delta_B$.\footnote{Qualitatively, $\psi$ is sort of like a macroscopic wavefunction for the correlated dynamics of all the particles in each superconductor, and in this picture the Josephson effect is something like a macroscopic manifestation of the ``probability current'' you may have seen in an elementary quantum mechanics course.} Because these Cooper pairs carry charge, the tunneling manifests at DC as a dissipationless ``supercurrent'' in the absence of any applied voltage! 

Extending our consideration to time-dependent $\delta$, the Josephson effect is described by the equations
\begin{align}
V(t) &= \frac{\Phi_0}{2\pi}\frac{\partial\delta}{\partial t}, \label{eq:jos_volt}\\
I(t) &= I_0\sin\big(\delta(t)\big), \label{eq:jos_curr}
\end{align}
where $\Phi_0=h/2e=2\times10^{-15}~\text{T\,m}^2$ is the \textbf{flux quantum}, a fundamental constant with dimensions of magnetic flux, and $I_0$ is the \textbf{critical current} of the Josephson junction, which depends on material properties, geometry, and temperature. Comparing Eq.~\eqref{eq:jos_volt} to the integral form of Faraday's law reveals that the appropriately rescaled Josephson phase $(\Phi_0/2\pi)\delta(t)$ behaves formally like a magnetic flux. Motivated by this correspondence, we can define the Josephson inductance 
\begin{align}
L_J &= \frac{\Phi_0}{2\pi}\frac{\delta(t)}{I(t)} \nonumber \\
&= \frac{\Phi_0}{2\pi I_0}\frac{\text{arcsin}\big(I/I_0\big)}{I/I_0}, \label{eq:jos_induct}
\end{align}
where I have used Eq.~\eqref{eq:jos_curr}. Expanding Eq.~\eqref{eq:jos_induct} to second order in $I/I_0$, we obtain precisely the form of Eq.~\eqref{eq:quad_nonlinearity}, with $L_0=\Phi_0/(2\pi I_0)$ and $\Delta L = L_0/6$. Thus, a superconducting $LC$ circuit whose inductance is due in part to one or more Josephson junctions behaves like a parametric amplifier when driven with a sufficiently intense pump tone near resonance. We will see in Sec.~\ref{sub:jpa_design} that with a well-designed JPA circuit the effects of higher-order nonlinearities in the expansion of Eq.~\eqref{eq:jos_induct} can be made negligible. The Josephson equations themselves do not specify a preferred frequency scale, but in practice JPA designs are most conveniently realized at microwave frequencies, where cooling to temperatures $T< T_\text{SQL}$ is feasible, and the required circuit dimensions are well within the capabilities of modern microfabrication.

The Josephson effect enables the realization of a parametric amplifier in the form of a microwave resonant circuit. In practice, the JPA has several other important advantages. One is the low internal loss which we get ``for free'' by fabricating the whole circuit out of a superconductor: this makes it easy to realize a design in which the \textit{only} noise added by a JPA operating in phase-insensitive mode is the input noise at the image frequency. A second advantage derives from the intimate relationship between the Josephson phase and magnetic flux already noted above. The quantization of the net magnetic flux through a superconducting loop implies that a \textbf{DC SQUID} (a pair of Josephson junctions connected in parallel with superconducting leads) acts like a single Josephson junction with an effective critical current
\begin{equation}\label{eq:I_squid}
I_0^S \approx 2I_0\abs{\cos\left(\frac{\pi\Phi}{\Phi_0}\right)},
\end{equation}
where $\Phi$ is the magnetic flux through the SQUID loop~\citep{castellanos2010}. 

We already encountered the SQUID in our discussion of earlier haloscope searches in Sec.~\ref{sub:admx}; there the modulation of a DC current by a time-varying flux $\Phi(t)$ was used for amplification. Using SQUIDs instead of Josephson junctions in the design of a JPA allows us to exploit Eq.~\eqref{eq:I_squid} in a different way. Application of a DC flux $\Phi_0$ adjusts $L_J$ and thus the resonant frequency: thus JPAs can be made \textbf{flux-tunable}. I will discuss the considerations that go into determining the JPA bandwidth in Sec.~\ref{sub:jpa_design}, but the intrinsically resonant nature of a parametric amplifier indicates that we should not expect such devices to be very broadband. Thus flux-tunability greatly facilitates the integration of JPAs into various applications; it is especially critical for the haloscope search.

\subsection{JPA design and fabrication}\label{sub:jpa_design}
\begin{figure}[h]
\centering\includegraphics[width=0.6\textwidth]{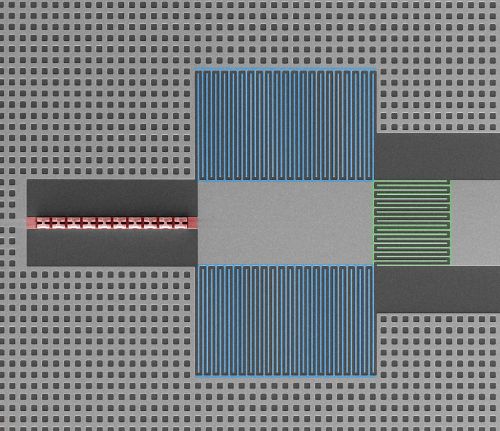}
\caption[Microphotograph of JPA circuit]{\label{fig:jpa_layout} Microphotograph of JPA circuit. The SQUID array (approximately 150 $\mu$m long) is highlighted in red on the left; the circuit's resonance is determined by the SQUID inductance and the geometric capacitance (blue). The circuit is coupled to a $50~\Omega$ transmission line through a smaller capacitance (green). The surrounding superconducting ground plane is waffled in order to pin magnetic flux vortices in place and keep them from the SQUID array.}
\end{figure}

In the previous section I described how a tunable JPA could be realized in the form of a superconducting $LC$ circuit with one or more SQUIDs comprising the nonlinear inductor. The HAYSTAC JPA is shown in Fig.~\ref{fig:jpa_layout}: it is a lumped element single-port resonator comprising an interdigitated capacitor $C$, a series array of $N_S=20$~SQUIDs, and a much smaller capacitor $C_c$ used to couple to a $ 50~\Omega$ transmission line.

The single-port design indicates that the JPA functions as an amplifier in reflection. Indeed, up to the nonlinear behavior, it is described by the same equivalent circuit model applied to the cavity's response in reflection in Sec.~\ref{sub:cav_meas}, with very small $R$ because the entire circuit is fabricated from superconducting niobium and operated at $T\ll T_\text{Nb}$.\footnote{The circuit was fabricated using a Nb/AlO/Nb trilayer process, where the aluminum oxide forms the tunnel barrier in the Josephson junctions. The high critical temperature $T_\text{Nb}\approx 9~\text{K}$ of niobium (compared to $T_\text{Al}\approx 1~\text{K}$ for aluminum, which is the other superconductor commonly used in JPA design) is desirable given our high operating temperature relative to most applications using JPAs.} Thus the JPA is always overcoupled to the transmission line, and the $Q$-factor of the circuit is determined by the coupling capacitor $C_c$. 

The parameters that can be adjusted to optimize the JPA design are $I_0$, $N_S$, $C$, and $C_c$.\footnote{Each SQUID brings with it a fixed ratio of linear to nonlinear inductance [see discussion below Eq.~\eqref{eq:jos_induct}]. In general the (linear) geometric inductance of the circuit will also contribute to the total $L$, but for most applications it is not beneficial to further dilute the nonlinearity, so designs attempt to minimize geometrical inductance. I will ignore geometrical inductance for the most part in this discussion, but practical designs must of course account for it.} For the JPA designs used in HAYSTAC, $I_0=6~\mu\text{A}$ is fixed by the material choice and the SQUID geometry. Then $N_S$ and $C$ together determine both the 0-field operating frequency $\nu_{J}(\Phi=0)\approx1/\big(2\pi\sqrt{N_SL_JC}\big)$ and the characteristic impedance $Z_r\approx\sqrt{N_SL_J/C}$. Typically designs fix $Z_r\approx 50~\Omega$: then the desired operating frequency determines $N_S$.

We have yet to discuss the role of the JPA circuit's $Q$ factor. With a more formal treatment~\citep{castellanos2010}, it may be shown that the JPA's \textbf{critical power} scales as 
\begin{equation}\label{eq:jpa_crit_pwr}
P_c \propto I_0^2/Q^3.
\end{equation}
Roughly speaking, the critical power sets the scale for the pump power $P_p$ required to obtain parametric gain at a useful level; we will motivate a more precise definition in Sec.~\ref{sub:jpa_bias}. Eq.~\eqref{eq:jpa_crit_pwr} indicates that by increasing the $Q$, we can get useful gain out of the JPA with pump amplitude $I_p\ll I_0$, such that we can safely neglect higher-order nonlinear terms in the expansion of Eq.~\eqref{eq:jos_induct}. Higher $Q$ also helps isolate the JPA circuit from external perturbations. However, there is a tradeoff, as the power of any signal to be amplified must be $\ll P_p$ to avoid saturating the JPA: increasing the $Q$ thus reduces the JPA's dynamic range as well as its bandwidth. The HAYSTAC JPA is a $Q\approx250$ design.

The parameters described above result in a JPA with a 0-field frequency $\nu_J(\Phi=0)=6.5~\text{GHz}$, a bare linewidth $\gamma_J = 26$~MHz, and a 0-field critical power $P_c =-90~\text{dBm}$. When the JPA is operated as an amplifier, it has an approximately constant (amplitude) gain-bandwidth product $\sqrt{G_J}\Delta\nu_J\approx\gamma_J$, where $G_J$ is its peak power gain and $\Delta\nu_J$ is the FWHM of the gain profile. For the purposes of the haloscope search, we want $G_J$ to be sufficiently large that the added noise of the second-stage amplifier is negligible (see Sec.~\ref{sub:receiver_noise}) and $\Delta\nu_J$ to be large compared to $2\Delta\nu_c\approx1~\text{MHz}$, so low noise is maintained through the analysis band. Both of these conditions are realized at our typical operating point $G_J\approx21$~dB, $\Delta\nu_J\approx2.3$~MHz (see also discussion in Sec.~\ref{sub:jpa_bias}).

The circuit described above is enclosed in a light-tight copper box with linear dimension $\approx2.5$~cm, whose only connection to the outside world is a short length of 0.085'' NbTi/NbTi coaxial cable connectorized outside with an SMA plug. The end of the cable inside the JPA enclosure is connected to a right-angle SMP surface mount connector mounted on a printed circuit board. The SMP connector couples signals from the coax to a coplanar waveguide transmission line, which is finally coupled to the JPA chip itself through wirebonds. The efficient coupling of a microfabricated circuit with characteristic dimensions of order $100~\mu\text{m}$ to a 2~mm coaxial cable is a nontrivial technical task which should not be overlooked. More recent JPA designs proposed for HAYSTAC attempt to streamline this coupling to reduce the loss as much as possible (see Sec.~\ref{sub:noise_offres}).

\subsection{Flux tuning}\label{sub:jpa_tuning}
\begin{figure}[h]
\centering{\includegraphics[width=.7\textwidth]{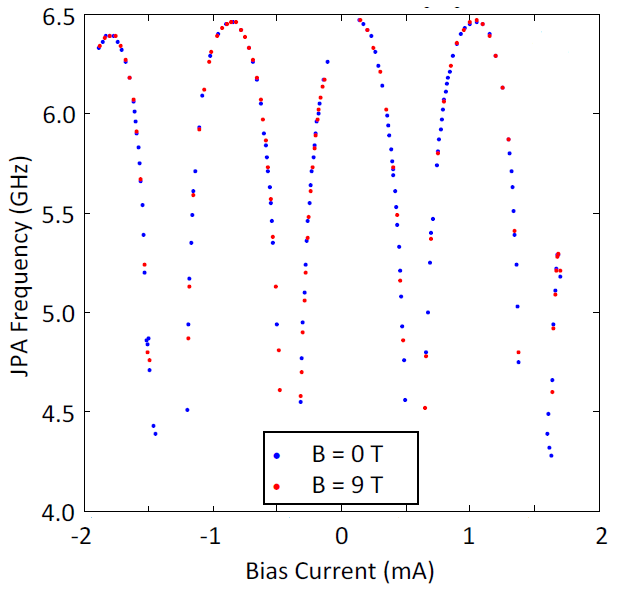}}
\caption[JPA resonant frequency vs.\ flux bias]{\label{fig:flux_quanta} $\nu_J$ vs.\ bias current $I_\Phi\propto\Phi$. There is essentially no change in frequency as the main magnetic field is ramped from 0 to 9~T. This data was acquired using a smaller bias coil that was later replaced with the large-area coil described in the text.}
\end{figure}

The final feature of the JPA design we have yet to discuss is the flux tuning. The formalism developed in the preceding sections indicates that adjusting the flux $\Phi$ threading each SQUID loop away from an integer number of flux quanta reduces the SQUID critical current $I_0^S$ [Eq.~\eqref{eq:I_squid}], and thus increases $L_J$ [Eq.~\eqref{eq:jos_induct}], reduces $P_c$ [Eq.~\eqref{eq:jpa_crit_pwr}], and tunes $\nu_J$ down in frequency. Naive application of Eq.~\eqref{eq:I_squid} would seem to indicate that $I_0^S$ can be nulled out completely, and the JPA can be tuned all the way down to DC, but Eq.~\eqref{eq:I_squid} neglects the inevitable geometrical inductance associated with the SQUID loop itself, which limits the tuning range in practice. The total tuning range of the HAYSTAC JPA is about 2 GHz, with the minimum frequency achieved for $\Phi\approx\Phi_0/2$ (see Fig.~\ref{fig:flux_quanta}).

The area of each SQUID loop in the HAYSTAC JPA is $A\sim10~\mu\text{m}^2$, implying that a flux quantum corresponds to an incident magnetic field $B\sim\Phi_0/A\sim2~\text{G}$. While Eq.~\eqref{eq:I_squid} is formally periodic for arbitrarily large $\Phi$, in practice the performance is degraded for $\Phi\gtrsim\text{few}\times\Phi_0$ due to imperfection in SQUID array fabrication. For flux tuning to work as described above, the flux through each SQUID in the array must be equal. These considerations motivate the requirements for the magnetic shielding described in Sec.~\ref{sub:shielding}: we want to ensure the flux through the SQUID array remains stable and uniform at a level $\Phi\ll\Phi_0$ when the magnet is ramped to the full 9~T field. 

Fig.~\ref{fig:flux_quanta} indicates that our present system achieves the requisite level of shielding required to operate a JPA in a haloscope search. Measurements of the shift of the JPA resonance between 0 and 9~T indicate the residual flux is $\sim10^{-3}~\Phi_0$, from which we obtain the value $B\sim10^{-3}~\text{G}$ cited in Sec.~\ref{sec:detector_overview}. Fig.~\ref{fig:flux_quanta} also indicates that there is a ``dead zone'' around half-integer multiples of $\Phi_0$,  where the JPA's critical power becomes so small that it is in practice inoperable. At the practical minimum of $\nu_J\approx4.5$~GHz, the critical power is $P_c\approx-120~\text{dBm}$. The enhanced flux sensitivity towards the bottom of the JPA's tuning range also complicates operation in this region. The $5.7 - 5.8$~GHz region scanned during the first HAYSTAC data run falls in the middle of the JPA's tuning range, where performance is more reliable. In this region, a typical cavity frequency step $\delta\nu_c\sim100~\text{kHz}$ corresponds to $\delta\Phi\sim2\times10^{-5}~\Phi_0$, so fine control over the flux bias is required for JPA tuning.

The JPA flux bias is delivered by passing a current through a superconducting coil wound around the copper JPA enclosure: the large cross-sectional area $\sim6~\text{cm}^2$ helps ensure the flux through the SQUID array is very uniform. The bias current required to produce a flux $\Phi_0$ through each SQUID loop with this system is $I_{\Phi_0}\approx14~\text{mA}$; this rather large value is also a consequence of the large coil dimension. The large value of $I_{\Phi_0}$ implies that only a single flux quantum is accessible with this system, but the flux is both more uniform and more stable than in early tests with a small coil for which $I_{\Phi_0}$ was larger.

The bias current is delivered into the DR by a homemade balanced and temperature-compensated current source with 1~mA/V transconductance, which is in turn driven by a homemade 20-bit DAC with $10~\mu\text{V}$ resolution controlled by the DAQ computer. This system provided an extremely stable bias current to the JPA, with $I_\Phi\sim4$~mA being a typical value during the first HAYSTAC data run.

\section{Receiver and electronics}\label{sec:receiver}
The cryogenic microwave layout and room-temperature microwave/IF layout are depicted schematically in Fig.~\ref{fig:cryo_setup} and Fig.~\ref{fig:RT_setup} in appendix~\ref{app:diagrams}; the receiver signal path is shown in blue in both diagrams.  I will make frequent reference to the component labels in these diagrams (associated with manufacturers and part numbers in Tab.~\ref{tab:parts}) in my description of the HAYSTAC receiver and electronics in this section. 

In Sec.~\ref{sub:receiver_noise} I discuss the amplifier chain and the cabling used to approximate the ideal case in which the preamplifier (JPA) is the only contribution to the receiver's added noise $N_A$. In Sec.~\ref{sub:receiver_cryo} I discuss the required cryogenic microwave components and the microwave and DC lines inside the DR. In Sec.~\ref{sub:receiver_rt} I discuss the room-temperature microwave and IF electronics, excluding elements used only in the path for feedback to the JPA's flux bias, which are discussed in Sec.~\ref{sub:feedback}. Finally, in Sec.~\ref{sub:grounding} I discuss efforts to mitigate excess electronic noise coupling into the room-temperature part of the receiver chain.

The microwave and IF portions of the receiver chain together constitute an impedance-matched network with $50~\Omega$ characteristic impedance. Components in the microwave portion of the chain are connected with 0.085'' semi-rigid coaxial cables terminated in SMA connectors.\footnote{Cable stock was purchased in meter-length segments, and cut to length and connectorized in the lab with an AMP 1055835-1 crimping tool.} Components in the IF chain are connected with double-shielded RG223/U BNC cables (see Sec.~\ref{sub:grounding}).

\subsection{Added noise of the receiver chain}\label{sub:receiver_noise}
In Eq.~\eqref{eq:amp_chain} and the surrounding discussion we saw that the added noise $N_A$ of an lossless receiver chain is dominated by the contribution of the preamplifier in the high-gain limit. In a real experiment with finite preamplifier gain $G$, component losses and subsequent amplifiers will necessarily contribute at some level. It is intuitively obvious that imperfect power transmission efficiency $\eta$ between the receiver antenna and the JPA input should reduce the SNR of the haloscope search. Formally, this is because any inefficiency ($\eta<1$) in the receiver chain at temperature $T$ acts like a source of Johnson noise
\begin{equation}\label{eq:noise_eta}
N_\eta = \left(1-\eta\right)\left(\frac{1}{e^{h\nu/k_BT} - 1} + \frac{1}{2}\right).
\end{equation}
Eq.~\eqref{eq:noise_eta} indicates that a lossless element ($\eta=1$) produces no noise; for a termination ($\eta=0$) it reduces to the usual result for Johnson noise.\footnote{Eq.~\eqref{eq:noise_eta} may be derived by noting that if our receiver is connected to a passive network in thermal equilibrium, the distribution of loss among the elements of the network cannot make any difference: the \textit{total} noise per unit bandwidth is always given by Eq.~\eqref{eq:noise_eta} with $\eta=0$. Thus any component which is capable of absorbing noise power internally must also generate noise; this is the essence of the fluctuation-dissipation theorem.} We will see in Sec.~\ref{sec:noise} that the effects of loss can be modeled as another contribution to $N_A$, and that the technique we use to measure $N_A$ is sensitive to this contribution.

To minimize this loss contribution, all receiver components between the cavity and the 4~K stage of the DR are connected with NbTi/NbTi coaxial cables (marked SC in Fig.~\ref{fig:cryo_setup}) which remain superconducting in the 9~T field due to flux pinning.\footnote{Measurements confirmed that the net insertion loss of the receiver chain is not affected at any measurable level by ramping up the field.} The loss per unit length of these cables is negligible compared to connector and component losses. The elements contributing to the net loss before the JPA are the microwave switch S1, two circulators C, a directional coupler D, as well as SMA connectors and the elements used to inject a signal into the JPA chip (described in Sec.~\ref{sub:jpa_design}). The discrete elements S1, C, D (discussed in Sec.~\ref{sub:receiver_cryo}) are all required for the proper operation of HAYSTAC.

Before we turn our attention to the amplifier contributions to $N_A$, we can briefly consider the input lines. All three input lines provide paths for room-temperature thermal noise to couple into the DR, and we need to ensure that this noise does not contribute substantially to $N_\text{sys}$. 20~dB cryogenic attenuators\footnote{NiCr resistors are used in all cryogenic attenuators and terminations.} interrupt all three input lines at the 4~K stage, so the total thermal noise incident on the base-temperature stage through these lines (neglecting cable losses) is $\sim 4~\text{K} + 300~\text{K}/100\approx 7~\text{K}$. In the pump and reflection lines this noise is further attenuated by 30~dB at base temperature, so the thermal noise from each line appears in the receiver as a $\sim7~\text{mK}$ contribution we can safely neglect.\footnote{We can afford to use only 20~dB of attenuation at base temperature in the transmission line because its coupling to the cavity mode is so weak.} 

We must also consider the phase noise of the JPA pump generator, which always accompanies the pump tone and is thus invariably incident on the JPA during operations. Phase noise per unit bandwidth is specified in dB relative to the carrier (dBc/Hz), and decreases with detuning from the carrier frequency. In HAYSTAC the pump tone is provided by an Agilent E8257D microwave signal generator with ultra-low phase noise (option UNY), which achieves about $-130~\text{dBc/Hz}$ at 100~kHz detuning for operating frequencies $\sim5~\text{GHz}$. The JPA critical power at any frequency is $P_c\leq-90~\text{dBm}$ (see Sec.~\ref{sub:jpa_design} and Sec.~\ref{sub:jpa_tuning}; at 5.75~GHz $P_c\approx-100~\text{dBm}$). Thus the incident phase noise at 100~kHz detuning is smaller than $-220~\text{dBm/Hz}\sim 7~\text{mK}$.

As noted in Sec.~\ref{sub:jpa_design}, we operate the JPA with peak gain $G_J=21~\text{dB}$ ($G_J\approx125$ in linear units). The amplified signal at the JPA output is routed by circulators to the second-stage amplifier, a commercial cryogenic HEMT (Low Noise Factory LNF-LNC4\_8A) installed at the 4~K stage. The HEMT datasheet quotes a noise temperature $T_\text{HEMT}\approx3~\text{K}$ and gain $G_\text{HEMT}\approx40~\text{dB}$ over a $4-8$~GHz bandwidth at any physical temperature $T<10$~K.\footnote{At the 4~K stage the DR still has enough cooling power to dissipate the heat generated by the HEMT bias. At any lower temperature this would present an unacceptable heat load, and at any higher temperature the HEMT's added noise would be higher.} Even if we assume the input-referred system noise $N_\text{sys}$ saturates the SQL, the amplified noise at the HEMT input in temperature units [Eq.~\eqref{eq:t_sql}] is $G_J\,T_\text{SQL} \approx 12\,T_\text{HEMT}$: the HEMT is indeed a small contribution. In this estimate I have neglected losses between the JPA and the HEMT (which reduce the effective JPA gain), but allowing for loss (and for the possibility that $T_\text{HEMT}$ differs from the nominal value specified by the manufacturer) clearly does not change the qualitative result, especially given that $N_\text{sys}>1$ in practice.

Neglecting the additional contribution from the HEMT itself and again assuming the SQL at the receiver input, the effective temperature of the cavity noise at the HEMT output is $G_\text{HEMT}\,G_J\,T_\text{SQL}\approx3.5\times10^5$~K, next to which room-temperature thermal noise is completely negligible: clearly we no longer need to worry about a few dB of losses from cables or components. Before downconversion the signal is amplified again by a room-temperature transistor amplifier (A1 in Fig.~\ref{fig:RT_setup}), which has 32~dB gain (and $120$~K noise temperature which we can safely ignore). At low frequencies it is conventional to quote voltage gain rather than power gain and to quote signal noise levels in $\text{V}/\sqrt{\text{Hz}}$ instead of temperature units; the two parameterizations are just related by the $50~\Omega$ characteristic impedance of the receiver. Accounting for losses, a typical cavity noise level at the IF amp (A2) inputs is about $200~\text{nV}/\sqrt{\text{Hz}}$, which is again large compared to the $\sim5~\text{nV}/\sqrt{\text{Hz}}$ noise characteristic of low-noise room-temperature electronics at $\text{kHz} - \text{MHz}$ frequencies; the IF amps add about this much noise and provide a factor of 200 voltage gain. Finally, at the input of the analog-to-digital converters (ADCs), the RMS noise is about 60~mV within the $2.5~\text{MHz}$ bandwidth of the IF filters F2, compared to which noise added by the ADCs themselves is negligible.\footnote{In practice, the ADCs are operated with a 500~mV input range, because the dynamic range of the receiver is limited not by the noise power level but by the power in the JPA pump tone, which follows the same path as the signal after exiting the JPA. Both the ADCs and constraints on the IF configuration due to the presence of the pump tone are discussed in Sec.~\ref{sec:daq}. Because the noise power level is always less than the pump power, which is always less than $-90~\text{dBm}$ at the JPA output, saturation of microwave and IF components is never a significant concern.}

In summary, we have seen that we should expect nontrivial contributions to the receiver added noise $N_\text{A}$ in HAYSTAC to come from the JPA's added noise, the HEMT added noise referred to the receiver input, and component losses between the receiver antenna and the JPA; contributions from the rest of the receiver chain will be at the few \% level or smaller. Measurements of $N_A$ are discussed in Sec.~\ref{sec:noise}.

\subsection{Cryogenic components and cabling}\label{sub:receiver_cryo}
The cryogenic microwave switch labeled S1 in Fig.~\ref{fig:cryo_setup} is required to calibrate the receiver's added noise (Sec.~\ref{sec:noise}); it is a magnetically actuated latching transfer switch physically mounted on the mixing chamber plate of the DR.\footnote{Mounting the switch directly on the mixing chamber enables the temperature controller to respond rapidly to the switch actuation heat load. Moreover, operation of the switch in a higher ambient field is not necessarily guaranteed, so we did not want to mount it further down in the magnet bore. Thus in the present HAYSTAC design, signals are routed from the cavity to the mixing chamber plate, then back down to the JPA, and then up to the HEMT via the circulators on the mixing chamber plate. Although this configuration is somewhat contrary to the reasoning which led us to position the field-free region lower down in the magnet bore in the initial design, the extra superconducting cable length required does not contribute significantly to the loss.} A rather large 4.3~V pulse is required to actuate the switch at base temperature, where its resistance is about $50~\Omega$. The instantaneous power is thus very large, but the pulse is only applied for a very short time controlled by a variable capacitor on a custom switch driver. We estimate the total pulse energy associated with a single switch actuation to be $\sim 10~\text{mJ}$. We verified that the insertion loss of S1 was extremely stable over repeated actuations. The effect of switch actuation on the base temperature $T_\text{mc}$ was only $\sim1$~mK; its effect on JPA performance was larger, as we will see in Sec.~\ref{sub:jpa_fluct}. The unused microwave port of the switch is terminated in a $50~\Omega$ load. The function of the bias tee BT1 on the still terminator is to eliminate any systematics which might result from a DC path to ground in this switch configuration.

Two commercial ferrite circulators (labeled C in Fig.~\ref{fig:cryo_setup}) are required to isolate the JPA from both the backaction of the HEMT and its own backaction in reflection from the cavity, while a third is required to turn the JPA into a proper amplifier with separate input and output ports. Circulators are nonreciprocal three-port devices which rely on the magnetic properties of ferrites to single out a preferred direction. Terminating one of the ports of a circulator with a matched load turns it into an isolator, which transmits microwave signals unidirectionally. Ferrite circulators must be specifically designed for cryogenic operation; the ones installed in HAYSTAC operate in the $4 - 8$~GHz band. These circulators were individually magnetically shielded by the supplier, but we found their directivity was significantly degraded by the fringing field prior to the installation of the shielding coil described in Sec.~\ref{sub:shielding}. The circulators are anchored to the mixing chamber plate along the DR axis to maximize the shielding factor.

Lastly, the directional couplers (labeled D in Fig.~\ref{fig:cryo_setup}) are four-port reciprocal devices containing two transmission lines in close proximity: incident signals at any port are transmitted to the opposite side of the transmission line with low loss; 1\% of the incident power is also coupled out to one of the ports on the other transmission line (indicated by the diagonal arrows in Fig.~\ref{fig:cryo_setup}), while the fourth port is highly isolated. The directional coupler on the pump line is used to couple the pump tone into the JPA; the other two essentially serve as cryogenic 20~dB attenuators without DC connectivity.

To operate HAYSTAC we need to deliver several DC electrical signals (to tune the JPA, power the 4~K HEMT, and actuate S1) to the base plate of the DR, in addition to the DC wiring required for thermometry. The DC signals pass through API 56F715-005 EMI filters mounted directly on the DB15 feedthroughs on top of the DR; 36~AWG phosphor-bronze wire is used to carry these signals down to the still plate, below which superconducting wire is used. The microwave input and output signals are carried by 0.085'' coaxial cables as noted above. The outer conductors of all four coax lines are thermalized at each stage of the DR with gold-plated copper clamps which also serve to block stray light at the higher-temperature stages. The inner conductors in the receiver output line are thermalized using the bias tee BT2 at the HEMT input, with its RF+DC port oriented towards the low-temperature stages and its DC input shorted to ground.\footnote{The attenuators discussed in Sec.~\ref{sub:receiver_noise} also serve to thermalize the inner conductors in the input lines.} Stainless $0.085''$~coax is used between room temperature and 4~K in all four lines and down to base temperature in the input lines; as noted above, the lower-temperature segments of the receiver line are superconducting.

\subsection{Room-temperature electronics and the IF chain}\label{sub:receiver_rt}
Signals exiting the DR are directed via a commercial power splitter/combiner to both a commercial Agilent E5071C vector network analyzer and a commercial IQ mixer (labeled M2 in Fig.~\ref{fig:RT_setup}) serving as the input of what is essentially a homemade spectrum analyzer; we tracked the cavity noise level through the spectrum analyzer system in Sec.~\ref{sub:receiver_noise}.

Signals produced by the VNA may be routed via software-controlled solid-state switches S2 to the desired DR input line: the use of the transmission and reflection lines to characterize the cavity mode was discussed in Sec.~\ref{sub:cav_meas}, and JPA measurements using the pump line will be discussed in Sec.~\ref{sec:jpa_op}. The fixed room-temperature attenuators shown in Fig.~\ref{fig:RT_setup} were chosen to equalize the net attenuation experienced by a signal propagating from the VNA to the JPA input through each path,\footnote{The net attenuation is about $-115~\text{dB}$ with the variable attenuators AT~1 and AT~2 both set at 0~dB. These variable attenuators are relics of an earlier design; during operation, the VNA output power is always adjusted internally.} to minimize the effects of interference between the desired signal and leakage through the S2 switches.

The room-temperature microwave chain includes three microwave signal generators in addition to the VNA. An HP~8340B provides the local oscillator for both M2 and the flux feedback system discussed in Sec.~\ref{sub:feedback}. An Agilent~E8257D (with an ultra-low phase noise option discussed in Sec.~\ref{sub:receiver_noise}) provides the pump tone for the JPA. Finally, a Keysight~N5183B is used to inject synthetic axion-like signals into the cavity transmission line, as described in appendix~\ref{app:fake_axions}. All three generators and the VNA are phase-locked to a common 10~MHz reference provided by a SRS FS725m rubidium source (not pictured); the two analog-to-digital converters (ADCs) shown in Fig.~\ref{fig:RT_setup} are components of a PCIe digitizer board (GaGe Oscar CSE4344) installed in the DAQ computer, which also locks to the same reference. Isolators are placed at all generator outputs, along with the VNA and IQ mixer inputs and the input and output of the microwave amplifier A1.

Our spectrum analyzer comprises all the discrete components between M2 and the ADCs in Fig.~\ref{fig:RT_setup}, along with LabVIEW code that computes FFTs, implements image rejection, and averages power spectra in parallel with timestream data acquisition (to be discussed in Sec.~\ref{sub:power_spectra}). A mixer is a circuit element which downconverts incident RF signals to an IF band (DC to a few MHz in our case), while maintaining the phase relations between Fourier components in the input signal; it is thus a linear element as far as the signal is concerned, even though frequency conversion is an inherently nonlinear process. Downconversion is realized by superimposing the signal of interest with a much larger local oscillator (LO) tone at the input of a nonlinear element, whose Taylor expansion contains a quadratic term which multiplies the signal and the LO. Then a familiar trigonometric identity yields
\begin{align}
V_\text{out} &\propto V_\text{LO}V_\text{in}\cos(\omega_\text{LO}t)\cos(\omega_\text{RF} t) + \cdots \nonumber \\
&\propto \frac{V_\text{LO}V_\text{in}}{2}\Big[\cos\big((\omega_\text{LO}-\omega_\text{RF})t\big) + \cos\big((\omega_\text{LO}+\omega_\text{RF})t\big)\Big] + \cdots \label{eq:trig}
\end{align}
where $\omega_\text{RF}$ is any Fourier component in the input signal and $\omega_\text{IF}=\omega_\text{LO}-\omega_\text{RF}$ is the corresponding Fourier component in the IF output. The components at the sum frequency are removed by filtering; mixers are designed for operation at a particular LO drive level $V_\text{LO}$ for which the other terms I have elided are negligible. 

For definiteness, let us assume that the LO tone is set at a higher RF frequency than all Fourier components of interest in the input signal, as is the case in HAYSTAC. Then it is clear that the Fourier component of the input at the ``image'' frequency $\omega_\text{IM}= 2\omega_\text{LO} - \omega_\text{RF}$ corresponds to an IF output at $-\omega_\text{IF}$. Eq.~\eqref{eq:trig} is manifestly symmetric in the sign of the LO-signal frequency difference: the mixer thus couples image noise from the opposite side of the LO into the signal Fourier components of interest. 

The similarity of the problem of mixer image noise to the physics of parametric amplification we already discussed in Sec.~\ref{sub:jpa_intro} suggests that we can disambiguate signal and image frequencies if we combine both quadrature phases of the IF signal in the appropriate way. The simple mixer we have discussed thus far singles out the IF component in phase with the LO. By splitting the input signal and sending it to two identical mixers driven by the same LO but with a $90^\circ$ relative phase shift, we realize an IQ mixer, which has two quadrature outputs $I$ (for in-phase) and $Q$ (quadrature-phase). Image rejection in HAYSTAC is implemented in software and will be discussed in Sec.~\ref{sub:power_spectra}; here I will just note that the signal paths from the $I$ and $Q$ mixer outputs should ideally be totally equivalent. 

The function of the F1 low-pass filters in Fig.~\ref{fig:RT_setup} is to attenuate RF leakage into the IF chain, since components designed to operate in the IF range will not necessarily present $50~\Omega$ impedance at microwave frequencies. The F2 filters limit the usable IF bandwidth to 2.5~MHz as noted in Sec.~\ref{sub:receiver_noise}; after filtering, the cavity noise signal is amplified by the IF amps A2 and sampled at 25~MS/s by the ADCs on the GaGe digitizer board. Significant oversampling relative to the F2 filter bandwidth (which is in turn comfortably larger than the width of the analysis band in each spectrum; see Sec.~\ref{sub:if_band}) is required because the A2 output filters are single-pole low-pass filters with relatively slow rolloff.

\subsection{IF interference and grounding}\label{sub:grounding}
The procedures used to bias the JPA to high gain and construct power spectra from the noise timestream sampled at the receiver output are discussed in Sec.~\ref{sec:jpa_op} and Sec.~\ref{sec:daq}, respectively. During detector commissioning we observed many narrow peaks above the noise floor $N_\text{sys}$ in power spectra obtained from measurements of duration $\tau\approx15$~minutes, which was our target value for the data run. These peaks always appeared in the same IF bins, independent of the LO frequency $\nu_\text{LO}$, and indeed independent of whether $\nu_\text{LO}$ was set 780~kHz above the mode frequency $\nu_c$ as usual, or far from $\nu_c$, such that the receiver measures noise generated by the terminator on the reflection line directional coupler after reflection from the cavity (see Fig.~\ref{fig:cryo_setup} and discussion in Sec.~\ref{sub:yfactor}). The above observations suggests that these spikes are due to IF interference coupling in to the receiver through room-temperature electronics, rather than RF interference coupling into the DR.\footnote{However, it should be noted that small narrowband RF signals generated at fixed detuning from the LO or pump tones can mimic the behavior of IF interference, as $\nu_\text{LO}$ defines the relationship between IF and RF frequencies, and the pump is always maintained at a fixed 1.6~MHz detuning from the LO (see Sec.~\ref{sub:if_band}).} This is fortunate, because narrowband RF interference is more insidious for the haloscope search: the axion itself looks like a small narrowband RF power excess!\footnote{The absence of any observed RF interference implies that the cryostat is an effective Faraday cage, and may also reflect the fact that the $5.7-5.8$~GHz range scanned during the initial HAYSTAC data run is occupied primarily by broadband WLAN channels (``5~GHz WiFi'') which are each 20~MHz wide. WLAN channels leaking into our detector would manifest as broadband contributions to $N_\text{sys}$ appearing in many adjacent spectra.}

The various IF features observed in the HAYSTAC power spectra had no single common origin. One particularly prominent comb of peaks with 16~kHz periodicity originated in switching power supplies in the STR2 stepper motor drivers, and coupled into the grounded DR chassis capacitively within the motors themselves; we eliminated this feature by galvanically isolating the motor support structure from the DR. Other features whose origin was never determined were nonetheless observed to exhibit strong dependence on the grounding of the DR and the room-temperature receiver chain; we were able to eliminate many of them by adding the DC blocks and baluns shown in Fig.~\ref{fig:RT_setup}, and removing the third prong of the AC-DC converter for the HEMT power supply (part number LNF-PS\_2). In general, our strategy was to separately ground the DR and the receiver chain, and eliminate most paths from the receiver outer conductors to the chassis of the electronics racks (Fig.~\ref{fig:upper_lab}).\footnote{Manufacturers of commercial microwave generators typically do not bother to isolate the SMA outer connectors on their devices from the chassis, since the long inductive paths characteristic of ground loops have high impedance at microwave frequencies. But nothing stops interference at lower frequencies from following this same path from a noisy ground; thus DC blocks on generator outputs are usually a good idea for sensitive applications.} 

We were able to significantly reduce the contamination of the HAYSTAC power spectra by careful attention to the grounding configuration; replacing all RG58C/U BNC cables in the IF chain with double shielded RG223U cables also proved helpful. However, certain features only appeared when the system was cold, making it difficult to determine their origin. The narrowband interference we were unable to eliminate with hardware changes is flagged and removed in the analysis procedure (Sec.~\ref{sub:badbins}).


\chapter{Measurements with HAYSTAC}\label{chap:data}
\setlength\epigraphwidth{0.32\textwidth}\epigraph{\itshape A billion is a thousand million? Why wasn't I informed of this?}{Robert Mankoff}

\noindent I have now discussed all the essential elements of the HAYSTAC detector, but said relatively little about its operation. It seemed most straightforward to discuss the cavity measurement and tuning procedures along with the design of the cavity and motion control systems in Sec.~\ref{sec:cavity}. In the present chapter, I will discuss the characterization and operation of the JPA and the overall noise performance of HAYSTAC in greater detail. I will also describe the automated HAYSTAC data acquisition procedure, to set the stage for the discussion of offline data analysis in chapter~\ref{chap:analysis}. I have chosen to treat these subjects separately because the use of a JPA is what most differentiates HAYSTAC from the haloscopes that preceded it, and the difference between $N_\text{sys}\approx2$ and $N_\text{sys}\approx20$ is quite significant!

In Sec.~\ref{sec:jpa_op} I describe the procedures used to determine JPA operating parameters and automate JPA tuning, as well as the measures we took to ensure sufficiently stable JPA operation during the long integrations required for a sensitive haloscope search. In Sec.~\ref{sec:daq}, I describe the HAYSTAC power spectra and the data acquisition procedure. Finally, in Sec.~\ref{sec:noise} I discuss the principles of \textbf{$\boldsymbol{Y}$-factor measurements}, which we use to calibrate the added noise of the HAYSTAC receiver, and the results of \textit{in situ} noise calibrations. Parts of this chapter were adapted from Refs.~\citep{NIM2017} and \citep{PRD2017}.

\section{JPA operations}\label{sec:jpa_op}
JPAs are widely used in quantum information research, but we cannot simply copy well-established protocols in HAYSTAC because the requirements of the haloscope search differ in several key respects. First, as noted in Sec.~\ref{sub:jpa_intro}, most quantum measurement applications of JPAs use them as noiseless single-quadrature amplifiers and/or squeezers, whereas we want to operate the JPA like a more conventional phase-insensitive amplifier which just happens to have exceptionally low noise. I will discuss the implications of this distinction for the HAYSTAC downconversion architecture in Sec.~\ref{sub:if_band}, and the implications for noise calibration in Sec.~\ref{sub:yfactor}. 

Second, the haloscope search requires continuous stable operation over periods of months, and automated procedures to maintain the JPA in a high-gain state as its resonant frequency is tuned over many linewidths. For quantum measurement applications, the tunability of JPAs just makes them more flexible: the same device can interface with microwave circuit elements in different experiments with different fixed resonant frequencies. I am not aware of any application prior to HAYSTAC which has required a JPA to be tuned continuously over any significant fraction of its total operating frequency range.

In Sec.~\ref{sub:jpa_bias}, I will describe how we automate the adjustment of the bias parameters to maintain high gain as the JPA is tuned, after providing the necessary context with a pedagogical introduction to the practical measurement of JPA parameters. In Sec.~\ref{sub:jpa_fluct}, I will describe several issues encountered in JPA commissioning and what we did to mitigate them. Finally, in Sec.~\ref{sub:feedback}, I will describe the feedback system we designed to control the JPA's flux bias and thereby enable extended stable operation.

\subsection{JPA characterization and biasing}\label{sub:jpa_bias}
When only a single tone with power $P\ll P_c$ is present at the JPA input, the JPA is just a linear (albeit tunable) $LC$ circuit excited in reflection, and we can apply the model developed in Sec.~\ref{sub:cav_meas} to describe cavity reflection measurements. In Sec.~\ref{sub:jpa_design} I emphasized that the very low dissipation of niobium at $T\sim100~\text{mK}$ implies that the circuit is in the extreme-overcoupled limit. From the discussion around Eq.~\eqref{eq:beta_gamma}, we see that we should expect no resonant dip in the amplitude reflection coefficient $\abs{\Gamma}$, but we can still identify the resonance $\nu_J$ through the phase shift $\Delta\phi$ [c.f.\ Fig.~\ref{fig:tx_rfl}].

Thus our first task is to measure the linear response of the JPA by exciting it with a weak probe tone (setting the probe power at least $30$~dB below our initial estimate of the critical power at the frequency of interest). In HAYSTAC we use the VNA to inject the probe tone (whose frequency is swept over a rather broad range) through the pump line, to avoid any filtering by the cavity. We can read off $\nu_J$ from the center of the JPA phase response, and verify that we are indeed looking at the JPA resonance by adjusting the flux bias $\Phi$ and checking that $\nu_J(\Phi)$ exhibits the expected behavior. In this way we can map out the JPA tuning range, as shown in Fig.~\ref{fig:flux_quanta}; note that very low probe power is required for accurate linear-response measurements at low frequencies.

\begin{figure}[h]
\centering\includegraphics[width=0.7\textwidth]{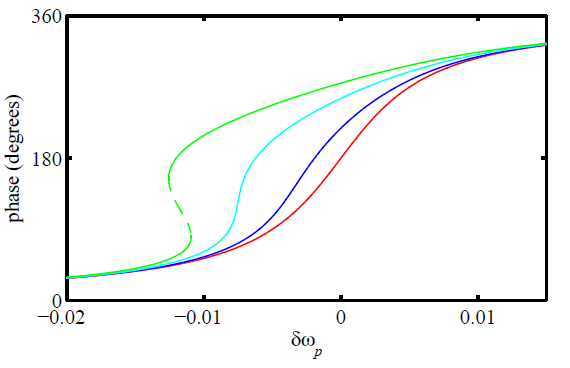}
\caption[JPA phase response, illustrating bistability at high power]{\label{fig:jpa_phase} Calculations of a JPA's phase response at fixed $\Phi$ and several different values of the probe power $P$: red, blue, cyan, and green curves represent $P\ll P_c$, $P\approx 0.25\,P_c$, $P\approx0.8\,P_c$, and $P\approx1.5\,P_c$, respectively. We see that the JPA's phase response moves towards lower frequencies and gets sharper as we increase $P$, and becomes bistable for $P>P_c$. $\delta\omega_p$ on the horizontal axis represents detuning from the $P=0$ resonant frequency. Figure from Ref.~\citep{castellanos2010}.}
\end{figure}

The next step is to begin to engage the JPA's nonlinearity, by fixing $\Phi$ and increasing the power $P$ of the swept tone (which is still the only tone incident on the circuit). Repeating the sweep at several values of $P$ we would see the behavior illustrated in Fig.~\ref{fig:jpa_phase}: the phase response moves towards lower frequencies and also gets sharper with increasing $P$. The fact that $\nu_J$ decreases is simply a consequence of $\nu_J\propto1/\sqrt{L_J}$, where $L_J$ is given by Eq.~\eqref{eq:quad_nonlinearity} with positive $\Delta L$ (see discussion below Eq.~\eqref{eq:jos_induct}). To understand the sharpening of the JPA's phase response would require a more formal approach (see Refs.~\citep{yurke2006,castellanos2010}).\footnote{It should be emphasized that the behavior exhibited in Fig.~\ref{fig:jpa_phase} has nothing to do with JPAs or the Josephson effect specifically -- it is just the classical dynamics of a \textbf{Duffing oscillator}, which is a simple harmonic oscillator with an extra cubic term in its equation of motion or equivalently a quartic term in its Hamiltonian [c.f.\ Eq.~\eqref{eq:lc_hamilton} with $L$ given by Eq.~\eqref{eq:quad_nonlinearity}].} Working out the details, we would find that the slope of the phase response diverges at the critical power $P=P_c$, and at this point the resonance has shifted down by $\nu_J(P=0) - \nu_J(P_c)=\sqrt{3}\gamma_J$, where $\gamma_J$ is the bare linewidth introduced in Sec.~\ref{sub:jpa_design}. For $P>P_c$, the JPA phase response becomes bistable as illustrated in Fig.~\ref{fig:jpa_phase}. We can repeat this measurement at various values of $\Phi$ to map out the frequency-dependence of the critical power $P_c(\nu_J)$.

\begin{figure}[h]
\centering\includegraphics[width=0.65\textwidth]{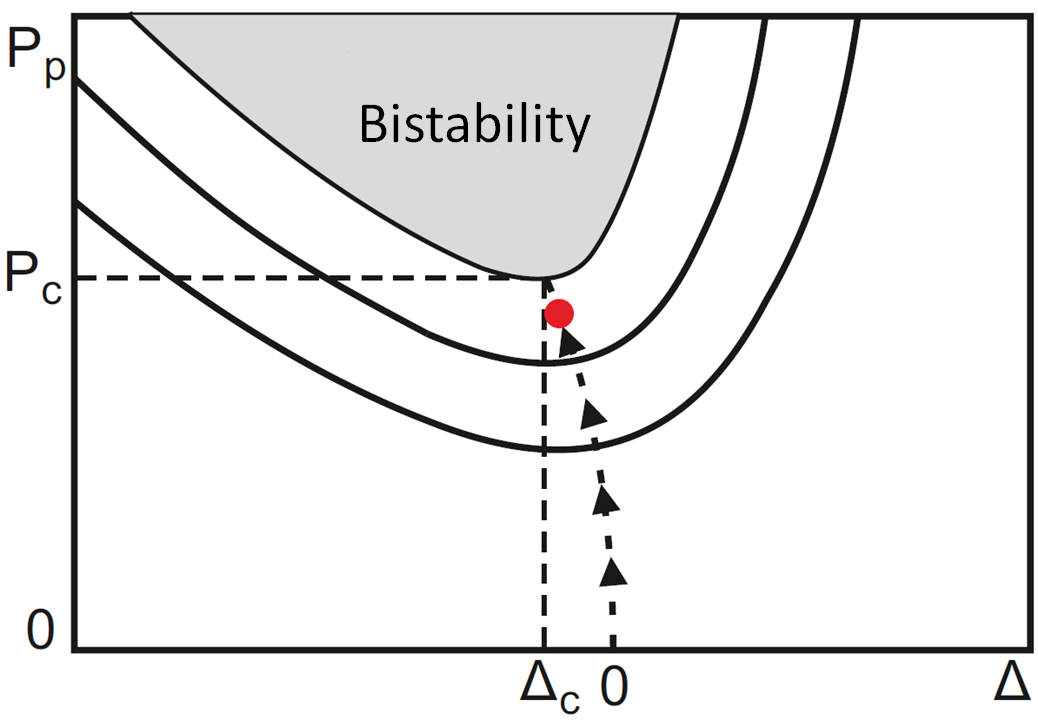}
\caption[JPA biasing parameter space]{\label{fig:jpa_biasing} Schematic representation of the parameter space for JPA biasing. All features are intended for illustrative purposes only; the shape of the bifurcation region was based on a theory plot in Ref.~\citep{krantz2013}. The pump power is on the vertical axis and the detuning $\Delta$ between the pump frequency and the 0-power $LC$ resonance of the JPA circuit is on the horizontal axis. The solid curves are contours of constant gain, and the arrows represent the path taken by the bias procedure outlined in the text, intersecting each gain curve at the minimum-power point. Beyond the critical point $(\Delta_c,P_c)$, the system is bistable. Our operating point is indicated by the red dot.}
\end{figure}

Having obtained a rough estimate of $P_c$ at some desired operating frequency, we can begin to operate the JPA as an amplifier, by turning down the probe tone power and introducing a CW pump tone with power $P_p$ and frequency $\nu_p$. The fact that we can independently set $\nu_p$ and $P_p$ suggests that JPA biasing parameter space is actually two-dimensional. A schematic representation of this parameter space is shown in Fig.~\ref{fig:jpa_biasing}, where $\Delta=\nu_p-\nu_J(\Phi,P_p=0)$ is the detuning of the pump from the JPA's zero-power resonance. We see that $\Delta$ may be adjusted either by varying the flux bias $\Phi$ or the pump frequency $\nu_p$: thus we can fix one of these parameters and use the other (together with $P_p$) to adjust the JPA's operating point). 

The solid contour lines in Fig.~\ref{fig:jpa_biasing} represent lines of constant gain -- if we were to plot such lines at any fixed (logarithmic) gain interval, we would see that they get closer and closer together in ($\Delta,P_p$) space, and the gain formally diverges at the boundary of the shaded region (of course this divergence is not observed in practice). The minimum pump power for which it is possible to reach this bistable region is the critical power $P_c$ defined above. We see that divergent gain at $P_p=P_c$ is achieved only when $\Delta=\Delta_c=-\sqrt{3}\gamma_J$, which provides some intuition as to the physical role of the detuning: at this critical point, we have just positioned the pump tone exactly at the JPA's resonant frequency, which has moved down as a result of the increase in inductance.

Fig.~\ref{fig:jpa_in_out} suggests (and a more formal derivation would confirm) that the amplifying bandwidth of the JPA is \textit{always} centered on $\nu_p$, with peak gain $G_J$ and bandwidth $\Delta\nu_J$ determined by $\Delta$ and $P_p$.\footnote{If $\Delta$ is sufficiently large that $\nu_p$ is nowhere near the JPA's resonant frequency, you will still in principle have an amplifier operating at $\nu_p$, though you will likely not be very happy with its gain.} In an axion haloscope, we would like to keep the JPA's amplifying band fixed relative to the $\text{TM}_{010}$ resonance, so that $N_\text{sys}$ remains relatively constant as we tune the cavity mode. Thus, in each iteration of the haloscope search we will set $\nu_p$ at constant detuning from $\nu_c$ (see discussion in Sec.~\ref{sub:if_band}), and move around the JPA biasing parameter space by varying the flux bias $\Phi$ and pump power $P_p$.\footnote{This is perhaps slightly less intuitive than fixing $\Phi$ and varying $\nu_p$ if one is accustomed to thinking in the microwave domain, where frequency is literally a knob on every generator. All we are doing is tuning a resonator to dial it in to a fixed drive frequency instead of tuning the drive until it coincides with a fixed resonance.}

In practice, in the haloscope search we will want to operate on the dashed curve in Fig.~\ref{fig:jpa_biasing}, which intersects each contour of constant gain at the minimum power point: I refer to this curve as the ``spine.'' From Fig.~\ref{fig:jpa_biasing} we can see that the condition that $P_p(\Delta)$ is \textit{minimized} at constant $G_J$ is equivalent to the condition that $G_J(\Delta)$ [or $G_J(\Phi)$] is \textit{maximized} at fixed $P_p$. This in turn implies that in a noisy flux environment, the JPA will be most stable when operated on the spine, where $G_J$ is only quadratically sensitive to small perturbations $\delta\Phi$ about the nominal flux bias $\Phi$. This choice of operating point also allows us to implement a feedback loop to stabilize the flux bias, discussed in Sec.~\ref{sub:feedback}.

We now know everything we will need to know about the conceptual procedure for biasing the JPA. Before we discuss how we bias the JPA in practice, we should understand what considerations enter into determining the target gain $G_J$. As noted in Sec.~\ref{sub:jpa_design}, the JPA maintains a roughly constant gain-bandwidth product $\sqrt{G_J}\Delta\nu_J\approx\gamma_J$, and we typically operate with $G_J\approx21$~dB ($\Delta\nu_J\approx2.3$~MHz).\footnote{The gain-bandwidth product can be quite different if the JPA is operated away from the spine in Fig.~\ref{fig:jpa_biasing}.} In Sec.~\ref{sub:receiver_noise}, we motivated this operating point by requiring that the total noise at the JPA output overwhelms the added noise of the HEMT amplifier at the 4~K stage; we saw that with our typical operating parameters we should expect the HEMT to contribute at the $\sim10\%$ level. It is by no means clear that increasing the peak gain further would help, since this would decrease the gain farther from $\nu_p$, and $\Delta\nu_J$ is not \textit{that} much larger than $\Delta\nu_c$. 

But there is another important constraint on $G_J$: we must ensure that we are sufficiently far from the critical point that the JPA still behaves like a linear amplifier. This condition turns out to be most easily verified by studying the noise at the JPA's output. Specifically, we bias up the JPA to some desired operating gain $G_J$, then set $\nu_\text{LO}=\nu_{p}$ and measure noise timestreams on both the I and Q IF outputs of the IF chain described in Sec.~\ref{sub:receiver_rt}. In this configuration (which is \textit{not} how we normally operate the JPA; see Sec.~\ref{sub:if_band}) the mixer's $I$ and $Q$ quadratures are related by some phase rotation $\phi$ to the JPA quadratures $X_1,X_2$. In an $I$-$Q$ plot of the quadrature noise spectra, the output of an ideal single-quadrature amplifier would look like an ellipse whose major axis ($X_1$) is rotated by $\phi$ with respect to the $I$ axis.\footnote{This is the way single-quadrature amplification is usually presented in quantum measurement contexts. In this phase-space picture, the single-quadrature amplifier evades the requirement that amplification add noise by preserving the original phase-space volume: it merely squeezes the variance out of $X_2$ and into $X_1$.} For a real JPA operated at high gain, the ellipse will eventually get deformed into a shape that looks more like a banana.\footnote{It should be emphasized that this is not due to the JPA departing from Duffing oscillator behavior, but rather due to the Duffing oscillator being a poor approximation to an ideal single-quadrature amplifier at high gain, essentially as a consequence of energy conservation. I thank Dan Palken and Maxime Malnou for enlightening conversation on this point.} We do not actually need to look at an $I$-$Q$ scatter plot to see this effect, if we note that the noise amplitude distribution obtained by collapsing the ``ellipse'' scatter plot into any 1D slice of phase space is always Gaussian, albeit with a phase-dependent variance: this is just the statement that the output noise of the JPA is completely described by the Gaussian input noise with different gains applied to the two quadratures. 1D slices of the ``banana'' distribution will generally not be Gaussian, which implies that the amplifier is doing something nonlinear to the input noise in this regime. The upshot of all of this is that we can simply histogram the $I$ and $Q$ noise timestreams at various operating gains and look for signs of non-Gaussianity. We started to see small deviations from Gaussianity for $G_J\geq24$~dB, which were quite prominent by 26~dB. We thus conclude that $G\approx21$~dB is a reasonable operating point, but it was important to not take this for granted.

Now we can discuss biasing in practice. In HAYSTAC we use the VNA to produce a probe tone which is swept rapidly over a small frequency range centered on $\nu_p$, with power about $40$~dB below our estimate of the critical power.\footnote{We could alternatively use a CW probe tone, but in practice even a narrow sweep range gives us more confidence that we know what it is we're looking at. When the CW pump tone and the swept probe tone are both injected through the pump line, we can get interference whenever they are at the same frequency. We can avoid having to deal with such interference by simply setting the VNA to an even number of frequency steps, with the sweep centered exactly on the pump frequency. In HAYSTAC, we use 6~frequency steps, a 20~Hz IF~bandwidth, and a span range of 1~MHz to bias the JPA.} The fact that we can turn the JPA into a passive mirror by simply turning off the pump and detuning the flux bias makes absolute gain calibration much easier for the JPA than for most amplifiers. We begin (before turning on the pump) by adjusting the flux bias sufficiently far to detune the JPA by more than the bare linewidth $\gamma_J$: in the first HAYSTAC data run we set a ``JPA-off shift current'' $\Delta I_\Phi=0.14$~mA, which detunes $\nu_J$ by about 50~MHz.\footnote{In practice $\Delta I_\Phi$ cannot be too large, because the response of the ferromagnetic shields to large changes in the applied flux is hysteretic.} We take a normalization sweep in this configuration, then undo the flux bias shift and turn the pump generator on, initializing $P_p$ about 10~dB below $P_c$.\footnote{In practice this requires measuring or estimating the difference in the attenuation of the pump generator -- JPA and VNA -- JPA signal paths. $P_p=P_c-10$~dB is a conservative starting point far from the bistable region.} We then sweep the probe tone repeatedly over the range used for the normalization measurement, adjusting $\Phi$ after each sweep until we have maximized $G_J$ with respect to $\Phi$ at constant $P_p$. If $G_J<21$~dB, we increment $P_p$ by some small fixed amount and repeat the optimization of $G_J$ with respect to $\Phi$, repeating this whole process until we obtain $G_J\approx21$~dB. Schematically, this procedure corresponds to moving horizontally on Fig.~\ref{fig:jpa_biasing} until we find the spine, and then zigzagging up the spine to the desired point.\footnote{If at any point a measurement yields $G_J>21$~dB, we decrease $P_p$ continuously until $G_J<21$~dB, and then continue with the procedure described above. The purpose of this asymmetry between increasing and decreasing pump power is to quickly get us out of the supercritical regime where the JPA's behavior is less intuitive.}

\begin{figure}[h]
\centering\includegraphics[width=0.8\textwidth]{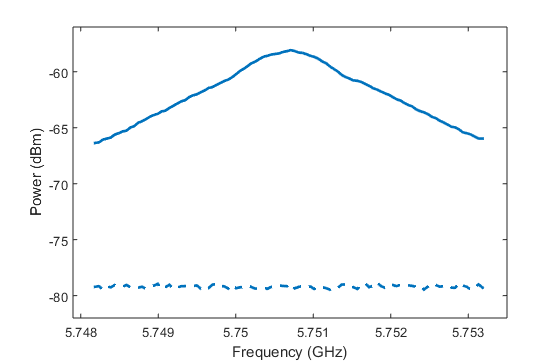}
\caption[Sample JPA gain measurement]{\label{fig:gain_sample} A sample JPA gain measurement: power measured at the VNA input with the JPA on (solid curve) and off (dashed curve) vs.\ frequency for a swept probe tone injected through the pump line. The sweep comprised 100 points over a 5~MHz span centered on $\nu_p$; the power at the VNA output was $-29$~dBm and the measurement bandwidth at each point was 20~Hz. By normalizing the JPA-on curve to the JPA-off curve and fitting to a Lorentzian, we obtain $G_J=20.8$~dB and $\Delta\nu_J=2.1~\text{MHz}$.}
\end{figure}

After biasing, we can sweep over a wider range to measure the JPA's gain profile: such a measurement is shown in Fig.~\ref{fig:gain_sample}. During normal HAYSTAC operation, we measure the gain profile over 5~MHz centered on the pump, which is wide enough to get a sense of its (Lorentzian) shape; we ultimately only care about the gain in a smaller bandwidth on the high-frequency side of the pump (see Sec.~\ref{sub:if_band}).\footnote{Note that $G_J$ measured using the amplification of a swept tone (or a CW tone, for that matter) is the gain of the JPA in phase-insensitive operation, because at any given point in the sweep, the tone is either on one side of the pump or the other. $G_J$ is thus exactly the quantity that we care about for the haloscope search; it is smaller than the single-quadrature power gain by a factor of 2.} By varying the probe power and repeating this measurement, we can measure the JPA's dynamic range and fix the probe power at a level where we are in no danger of saturation. We typically operate the JPA $5-6~\text{dB}$ below its 1~dB compression point, which is approximately given (in logarithmic units) by $P_{1\,\text{dB}}\approx P_p-G_J-13~\text{dB}$~\citep{castellanos2009}.

I have now discussed how we measure $\nu_J$ and $P_c$ throughout the tuning range, how we determine the optimal JPA operating point $G_J$, and how we bias the JPA. The only remaining topic is the automation of this biasing procedure, which is critical to the practical realization of the haloscope search. By this point it should be clear that there is no fundamental obstacle to such automation; I presented the biasing procedure above in an algorithmic manner which makes the correspondence to a practical software-controlled implementation straightforward. Our task is further simplified by the fact that we do not need to automate the process of biasing up the JPA starting from unity gain. Rather, the scenario relevant to the haloscope search is one in which we must periodically tune the pump frequency $\nu_p$ and the cavity mode resonance $\nu_c$ together by $\delta\nu_c\approx100~\text{kHz} \ll \Delta\nu_J$, and reoptimize $\Phi$ and $P_p$ at each step to make sure we remain at the maximally stable operating point. The program must also be able to accommodate somewhat larger initial offsets from the optimal parameters that arise from switching between the cavity and the still terminator used for intermittent noise calibrations (see Sec.~\ref{sub:jpa_fluct} and Sec.~\ref{sec:noise}), but we never have to automate the much more difficult decision of which direction to vary the bias parameters when there is no appreciable gain at all.

\begin{figure}[h]
\centering\includegraphics[width=1.0\textwidth]{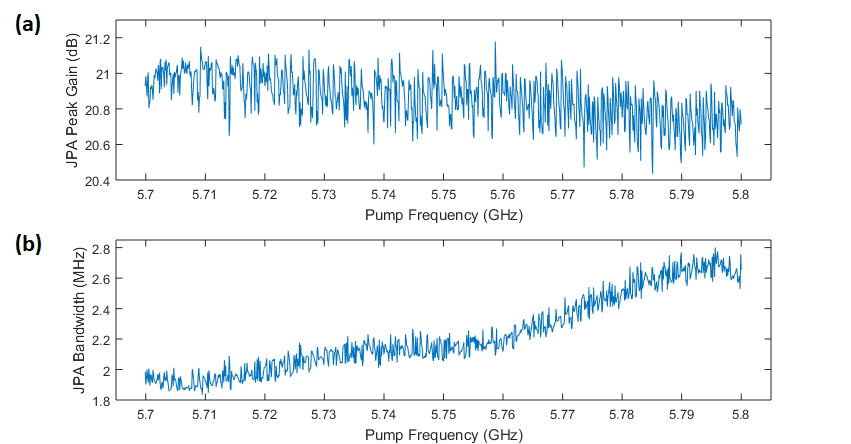}
\caption[JPA gain and bandwidth vs.\ frequency]{\label{fig:jpa_gain_bw} \textbf{(a)} $G_J$ vs.\ $\nu_p$ obtained from a test of automated JPA tuning and biasing over the $5.7-5.8$~GHz range scanned during the first HAYSTAC data run. \textbf{(b)} $\Delta\nu_J$ vs.\ $\nu_p$ obtained from the same measurement.}
\end{figure}

The HAYSTAC autobiasing procedure is incorporated into the LabVIEW code that controls the data acquisition (see Sec.~\ref{sub:daq_procedure}). We specified a minimum current step size of $\delta I_\Phi=300$~nA (corresponding to the value $\delta\Phi=2\times10^{-5}~\Phi_0$ cited in Sec.~\ref{sub:jpa_tuning}), a minimum pump power step size of $\delta P_p = 0.01$~dB, and a target value of $20.35 \leq \avg{G} \leq 20.65$~dB for the average gain $\avg{G}$ within a 1~MHz window around the pump -- the latter condition typically corresponds to $20.7 \lesssim G_J \lesssim 21.0$~dB. With these parameters, the program typically takes $\sim6$~s to adjust the bias parameters after each cavity tuning step. Fig.~\ref{fig:jpa_gain_bw} shows the results of an ``autotuning'' test, wherein the JPA is initialized in a high-gain state and the pump frequency is tuned over a wide range in small steps, with autobiasing at each iteration -- essentially a haloscope data run without the cavity or the actual axion-sensitive noise measurements. The autobiasing procedure is clearly sufficiently robust for the requirements of the haloscope search.

\subsection{JPA commissioning}\label{sub:jpa_fluct}
Having determined that we can reliably bias the JPA to the desired gain $G_J\approx21$~dB, we must consider how stably this gain is maintained on long timescales $\sim15$~minutes. A convenient way to characterize the stability of the JPA is to inject a CW probe tone at some small fixed detuning from the pump: then the probe tone power at the receiver output is essentially a proxy for the JPA gain.\footnote{The probe tone power (measured either at the VNA input or in the output spectrum) also responds to any other changes in the net receiver gain, but in practice all other elements of the receiver are very stable.} In HAYSTAC, we use the VNA to inject a CW probe tone through the pump line at 30~kHz detuning from $\nu_p$; the probe tone power is $-30$~dBm power at the VNA output, corresponding to $\approx-145$~dBm at the JPA input. This tone is always on during noise measurements in which the VNA would otherwise be idle; during commissioning we also found it useful to monitor the probe tone at the VNA input when not making any other measurements.

The most prominent JPA instability observed early in commissioning (when we were still operating the DR at a base temperature of $T_\text{mc}\sim25$~mK) was a strong modulation of the JPA gain by vibrational fluctuations. These fluctuations were initially observed the first time we assembled the full detector and ramped the magnet up to a few~T: their most striking manifestation was total (i.e., 100\% modulation depth) ``chopping'' of the gain vs.\ time, correlated with the audible $\approx1$~Hz cryocooler pulses which will be all too familiar to any graduate student who has worked in a cryogenics lab. 

The most obvious mechanism through which DR vibrations can couple into the JPA gain is motion of the JPA's SQUID array in an inhomogeneous field. After these early tests, we significantly improved the JPA's magnetic shielding (adding the Nb~cylinder and the persistent coils described in Sec.~\ref{sub:shielding}), and also redesigned the JPA mounting: in the original scheme the JPA was mounted on a relatively springy copper brace supported from above the ferromagnetic shields. We also added the persistent coil to shield the circulators at this time. The vibrational fluctuations were indeed significantly mitigated after this round of detector upgrades. But now that we knew what we were looking for, we observed a similar but smaller effect whether or not the field in the main magnet was on, with $\sim40\%$ fractional amplitude at the pulse tube frequency.

The absence of any field-dependence in this second round of measurements was clear evidence that the residual vibrational fluctuations were not due to appreciable penetration of the magnetic shielding by the external field. After additional tests demonstrated that the spectrum of vibrationally induced fluctuations extended up to a few~kHz, we realized these gain fluctuations probably did not originate in the JPA's flux-sensitivity at all. Bandwidths of order kHz are difficult to reconcile with eddy current screening in the copper JPA enclosure: empirically, this screening limited modulation of the flux bias $\Phi$ delivered through an external coil to frequencies below $\sim10$~Hz, and Faraday's law implies that all external flux fluctuations should be screened in the same way. Simply comparing the spectrum around the pump tone injected through the DR (without any JPA gain) to the spectrum obtained when the pump generator is directly connected to the room-temperature portion of the receiver chain demonstrates that vibrations of the DR imprint small AM sidebands on the pump tone: there is evidently some mechanism through which fridge vibrations can modulate the pump power.

Fortunately, we discovered that the vibrational fluctuations also exhibit strong temperature dependence, and moreover they depend on the thermal history within each cooldown: the fluctuations persist (at the 20\% level) when the fridge is heated from $T_\text{mc}\sim25$~mK to 127~mK, but completely disappear if the fridge is never allowed to cool below $T_\text{mc}\sim100$~mK. This temperature-dependence could be explained by some localized heating induced by DR vibrations, because all heat capacities drop precipitously at these very low temperatures; the hysteric behavior seems to implicate some magnetic material. Based on these observations and the noise bandwidth considerations discussed above, the prime culprit appears to be the magnetic shielding of the circulators: we can imagine that vibrations somehow modulate the magnetic properties of the shields, which in turn modulate the circulator $S$ parameters, which modulate the pump power. While this story seems plausible, I should emphasize that we never conclusively confirmed the origin of this effect, since we were ultimately able to get rid of it simply by operating at $T_\text{mc}=127$~mK.

More recently, we observed a second distinct flavor of JPA gain fluctuation, which manifested in intermittent sharp notches in the probe tone power (measured at the VNA input) vs.\ time. Looking more closely at individual notches, we saw that the gain dropped to 0 dB essentially instantaneously, and then bounced back on a timescale of seconds to its original value. This temporal profile was quite distinct from that of vibrational fluctuations. We observed a worst-case notch periodicity of $\sim5$~per minute, which was reduced at a higher temperature. The effect disappeared completely when we ramped down the field, cycled the DR above $T_\text{Nb}\approx9$~K, and then ramped the field up again at base temperature, suggesting the problem may have been the result of trapped flux inside the JPA circuit itself hopping back and forth between different pinning sites. Later the problem reappeared, and disappeared again next time we ramped down the field for unrelated reasons; we have not observed it since. 

Neither of the effects discussed above actually affected the JPA during the first HAYSTAC data run. However we did have to contend with the effects on the JPA of actuating the cryogenic microwave switch S1. The most obvious such effect is that slightly different JPA bias parameters are required to obtain $G_J\approx21$~dB depending on whether the receiver is coupled to the ``hot load'' (the still terminator) or the ``cold load'' (the cavity): typical offsets are $\sim0.2$~dB in pump power and $\sim2~\mu$A in the bias current. This steady-state effect may be due to impedance mismatches being slightly different in the two switch configurations, resulting in a very small change in the net transmission of the path from the pump generator to the JPA input.\footnote{This shifts the required value of $P_p$ as measured at the E8257D output; the shift in the required $I_\Phi$ follows from the curvature of the ``spine'' in Fig.~\ref{fig:jpa_biasing}.} This effect merely implies that we must rebias the JPA whenever we actuate the switch in order to meaningfully compare hot and cold noise spectra.

There are also transient effects associated with actuating the switch. Although switching had minimal impact on the DR temperature records, monitoring the CW probe tone at the VNA input as we actuated the switch revealed that it caused the JPA gain to drop out completely, with a recovery time of about a minute (the system equilibrated at a slightly lower net gain due to the switch state dependence described above). Based on these observations, we included a $\Delta t_\text{switch}=2$~minute wait after each switch actuation in the data acquisition procedure (Sec.~\ref{sub:daq_procedure}) for the first HAYSTAC data run. 

Noise measurements since the first HAYSTAC data run revealed that although the gain is completely stable two minutes after switching, $N_\text{sys}$ is systematically larger by about 15\% for about \textit{ten} minutes after switching: evidently the JPA does not actually equilibrate on shorter timescales! We have adapted our procedure to actuate the switch less frequently and set $\Delta t_\text{switch}=15$~minutes to wait out this effect whenever we do actuate the switch.\footnote{The overall effect on the first run sensitivity due to not catching this systematic effect earlier is quite small, and if anything it implies that we slightly \textit{overestimated} the noise prevailing in most cavity noise measurements which were far from any switch actuations.}

\subsection{Flux feedback system}\label{sub:feedback}
More persistent than all of the issues discussed in Sec.~\ref{sub:jpa_fluct} were flux drifts on long timescales comparable to the integration time per step $\tau=15$~minutes, which were likely due to very slow dynamics in the ferromagnetic shielding itself. In order to mitigate the effects of such flux drifts, we implemented feedback to the bias current $I_\Phi$ to hold the net flux bias $\Phi$ (from the bias current plus environmental contributions) constant. The feedback scheme works by modulating $I_\Phi$ at a low frequency $\nu_\text{mod}$ and using a lock-in amplifier to measure the component of the JPA gain at $\nu_\text{mod}$. As we will see presently, this system relies crucially on the fact that we operate at a point where $G_J(\Phi)$ is maximized, as discussed in Sec.~\ref{sub:jpa_bias}. 

The room-temperature signal paths used exclusively by the JPA flux feedback system are shown in pink in Fig.~\ref{fig:RT_setup}. The JPA flux bias is modulated at $\nu_\text{mod}=26$~Hz by adding an oscillating current (controlled by an Agilent 33220A function generator) to the DC bias current set in software. The modulation frequency is limited by eddy currents in the copper JPA enclosure (see Sec.~\ref{sub:jpa_fluct}); we pushed $\nu_\text{mod}$ as high as possible, such that the chosen value is above the 3~dB point of this effective low-pass filter. Thus the 33220A modulation amplitude actually must be quite large ($6~\mu$A RMS) to deliver sufficient AC flux at $\nu_\text{mod}$ to the JPA; this amplitude was set empirically to produce a modulation depth of $\sim0.1\,\Delta\nu_J$. When the net DC flux $\Phi$ is offset from the value that maximizes $G_J$, the JPA gain is modulated at $\nu_\text{mod}$ with a phase determined by the sign of the offset $\Delta\Phi$. When the bias modulation is centered on the optimal value, the Taylor expansion of $G_J(\Phi)$ has no linear term, so only higher harmonics of $\nu_\text{mod}$ are present in the gain variation. Thus the Fourier component of $G_J$ at the modulation frequency may be used as an error signal in an analog feedback system to stabilize the flux bias.\footnote{Another way to think about this is that any feedback system requires an antisymmetric error signal to tell us which way we have drifted from the operating point. Roughly speaking, modulating $I_\Phi$ rapidly (compared to the characteristic timescales of the fluctuations we would like to control) locally samples the (antisymmetric) derivative of $G_J$ with respect to $\Phi$.}

As noted in Sec.~\ref{sub:jpa_fluct}, we use the VNA to inject a weak CW probe tone through the pump line at 30~kHz detuning from $\nu_p$ during noise measurements. The function of the feedback circuitry in the upper left corner of Fig.~\ref{fig:RT_setup} is to provide an LO at the appropriate IF frequency to bring the probe tone to DC on the IF side of the mixers M3. Both quadrature outputs are amplified and squared using AD633 analog multipliers; the sum of squared voltage signals is a measure of the received power in the probe tone and thus a measure of the JPA gain. An SRS SR510 lock-in amplifier measures the variation of the probe tone power at $\nu_\text{mod}$. By adding the SR510 output to the bias current we obtain a simple proportional feedback loop; the SR510 gain, phase, and filtering are chosen to provide a stable feedback signal. During JPA biasing and sweep measurements the probe tone is not present because the VNA is otherwise occupied, so both modulation and feedback are interrupted by switching off the signal and reference outputs of the 33220A in software. 

This feedback system was quite successful, reducing the empirical frequency of cavity noise measurements compromised by flux drifts from $\sim20\%$ to $<1\%$.

\section{Data acquisition}\label{sec:daq}
The goal of haloscope design is basically to realize a precise measurement of the noise power as a function of frequency (i.e., the noise power spectrum) around the cavity resonance at each discrete tuning step in a data run. We now know enough about the JPA to understand the downconversion architecture required for proper operation of HAYSTAC. The resulting constraints on the HAYSTAC IF configuration, which governs the features we observe in the power spectra, are discussed in Sec.~\ref{sub:if_band}. 

In Sec.~\ref{sub:power_spectra} I discuss the procedure for \textit{in situ} computation of these power spectra from the noise timestream which is directly sampled by the ADCs at the receiver output. Acquisition of the power spectrum is the most critical measurement at each iteration of a haloscope search, but other operations are of course required to maintain stable operation and calibrate the axion search sensitivity. In Sec.~\ref{sub:daq_procedure} I describe our data acquisition procedure in general and the first HAYSTAC data run specifically.

\subsection{The analysis band}\label{sub:if_band}
The hardware elements of the homemade HAYSTAC spectrum analyzer are described in Sec.~\ref{sub:receiver_rt}. To measure noise around the $\text{TM}_{010}$ mode frequency $\nu_c$ we set the LO frequency to $\nu_\text{LO}=\nu_c+780$~kHz and the JPA pump frequency to $\nu_p=\nu_c-820$~kHz, then sample both IF channels at 25~MS/s with the ADCs on the PCIe digitizer board. The upper sideband of the LO (which does not contain any Fourier components of interest) is eliminated by combining the digitized signals in the frequency domain (see Sec.~\ref{sub:power_spectra}), and the result is a single power spectrum with components between DC and the 12.5~MHz Nyquist frequency, corresponding to signals detuned $0-12.5$~MHz below the pump.\footnote{Note that the (arbitrary) decision to put the signals of interest in the lower sideband of the LO implies that increasing IF frequency corresponds to decreasing RF frequency in all HAYSTAC power spectra. In retrospect it would have been conceptually simpler to position the LO such that the Fourier components of interest were in its upper sideband.} 

\begin{figure}[h]
\centering\includegraphics[width=.8\textwidth]{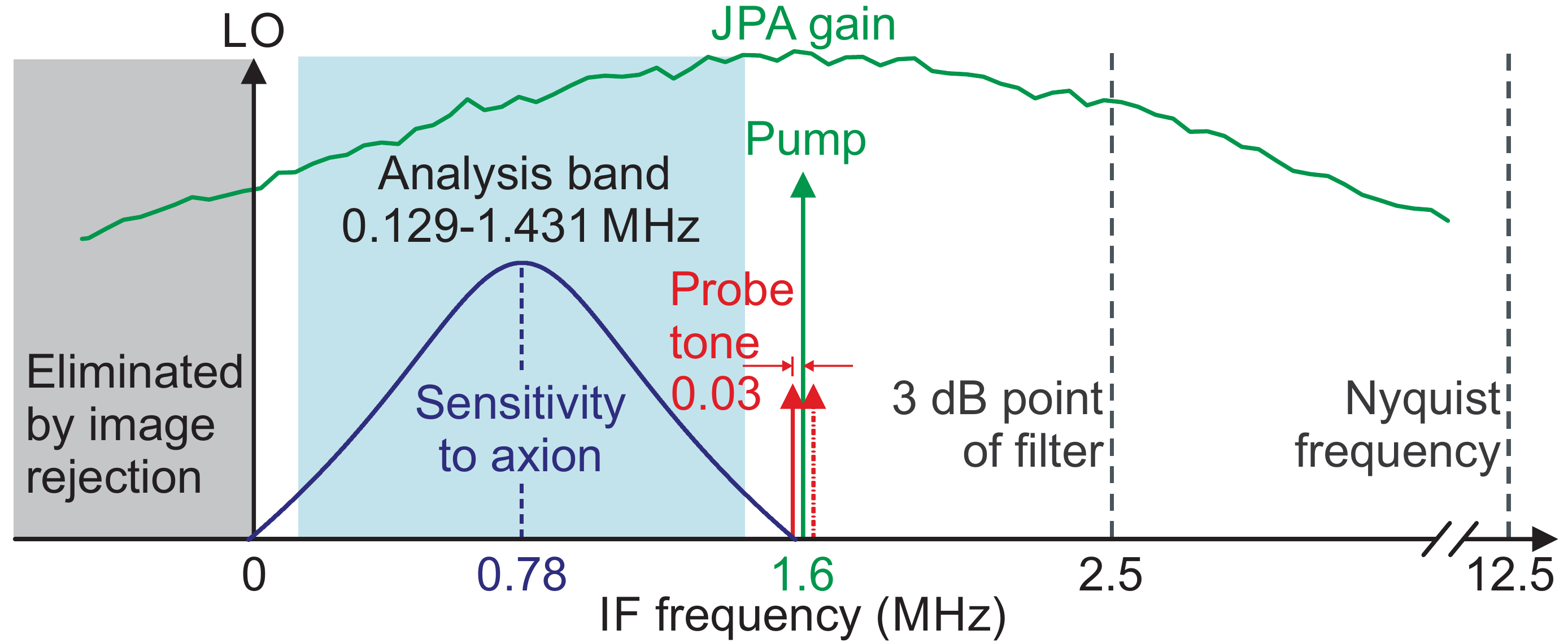}
\caption[Schematic diagram of IF setup]{\label{fig:IF_layout} Diagram illustrating the HAYSTAC IF configuration. The JPA gain profile and TM$_{010}$ Lorentzian profile are plotted using real data and a fit to real data, respectively. Both plots have logarithmic y axes; the absolute scale of the sensitivity plot is arbitrary. The solid red arrow is the probe tone introduced in Sec.~\ref{sub:jpa_fluct}; the dot-dashed red arrow represents the probe tone image created on the opposite side of the pump by the JPA's intermodulation gain. Images of the axion-sensitive Fourier components around the cavity, created on the other side of the pump by the same process, are omitted in the diagram for clarity.}
\end{figure}

Fig.~\ref{fig:IF_layout} is a simplified diagram of the HAYSTAC IF configuration, showing the features discussed above. It may be useful at this point to summarize the relations between the various frequency scales that will play a role in our subsequent discussions. When appropriately biased, the JPA has $G_J\approx21$~dB peak gain in a bandwidth $\Delta\nu_J \approx 2.3$~MHz centered on $\nu_p$. $\Delta\nu_J$ is larger than the typical cavity linewidth $\Delta\nu_c \approx 500$~kHz, which ensures that the total input-referred noise remains low over all frequencies of interest in each spectrum. The cavity linewidth $\Delta\nu_c$, which sets the width of the axion-sensitive region in each spectrum, is in turn much larger than the typical virialized axion linewidth $\Delta\nu_a$ [Eq.~\eqref{eq:delta_nu_a}]. Finally, we will see in Sec.~\ref{sub:power_spectra} that the spectra are written to disk with frequency resolution $\Delta\nu_b=100~\text{Hz}\ll\Delta\nu_a$.

Fig.~\ref{fig:IF_layout} also indicates that we limit our analysis to 1.302~MHz centered on $\nu_c$. The motivation for defining this analysis band of width $\approx2\Delta\nu_c$ was discussed in Sec.~\ref{sub:scan}: essentially we found that axion conversion at larger detunings from $\nu_c$ would contribute negligibly to the SNR.\footnote{1.3~MHz is a conservative choice that accounts for variation of $\Delta\nu_c$ (and the tuning step size $\delta\nu_c$) over the scan range. The precise 1.302~MHz width is a legacy of design choices that were not ultimately relevant in the final version of the analysis.} But we have not yet motivated the position of the analysis band in the IF spectrum (equivalently, the choice of detunings $\nu_\text{LO}-\nu_c$ and $\nu_c-\nu_p$). Image rejection requires that all Fourier components of interest be detuned in the same direction from $\nu_\text{LO}$ (see discussion in Sec.~\ref{sub:receiver_rt}), and phase-insensitive JPA operation (Sec.~\ref{sub:jpa_intro}) requires that all Fourier components of interest be detuned in the same direction from $\nu_p$. A configuration with $\nu_\text{LO}=\nu_p$ would be the natural choice for single-quadrature operation (see Sec.~\ref{sub:jpa_bias}), but is a bad choice for phase-insensitive operation: in this configuration the mixer quadratures $I$ and $Q$ are related by some arbitrary phase rotation to the JPA quadratures $X_1$ and $X_2$. Thus the quadrature variances $\text{Var}(I)$ and $\text{Var}(Q)$ are not necessarily equal, and image rejection will not work. 

These constraints collectively indicate that the center of the analysis band (i.e., $\nu_c$) should be roughly halfway between $\nu_\text{LO}$ and $\nu_p$ in each spectrum. Close to $\nu_p$, the spectrum is contaminated by both the CW probe tone used by the flux feedback system (Sec.~\ref{sub:feedback}) and by the pump tone's phase noise, discussed in Sec.~\ref{sub:receiver_noise}. Close to DC, $1/f$ noise dominates, and the relative contribution of the HEMT to $N_A$ also grows with detuning from the center of the JPA's amplifying band at $\nu_p$. The precise placement of the analysis band was tweaked to exclude bins in which IF interference was most persistent. 

\subsection{Measuring power spectra}\label{sub:power_spectra}
Now that we understand which Fourier components are relevant in each power spectrum, I will briefly discuss how these power spectra are constructed in practice. We could in principle save the full timestream data from both IF channels and construct power spectra offline. However, the 25~MS/s sampling rate and the 14-bit resolution of the ADCs on the GaGe digitizer board together imply a timestream data rate of $\approx90$~MB/s. Using the nominal parameters  with which we estimated HAYSTAC sensitivity in Sec.~\ref{sub:scan} ($M=3$, $\tau=15$~minutes, $\delta\nu_c\approx75$~kHz), we see that the total size of the timestream data set from the first HAYSTAC data run would be roughly $90~\text{MB/s}\times M\tau\times (100~\text{MHz}/\delta\nu_c)\sim300~\text{TB}$. This data set is so massive because it preserves all the information necessary to construct a power spectrum with frequency resolution $\Delta\nu_b=1/\tau\approx 1~\text{mHz}$.

A technical advantage of fine spectral resolution is that it facilitates flagging and cutting individual IF bins contaminated by narrowband interference (discussed in Sec.~\ref{sub:grounding}). A power spectrum bin width $\Delta\nu_b$ which is small compared to $\Delta\nu_a$ also enables us to optimally tailor our analysis to the expected spectral distribution of axion signal power around $\nu_a$ (see Sec.~\ref{sub:lineshape}). But $\Delta\nu_b=1~\text{mHz}\lll\Delta\nu_a$ is clearly excessive on both counts. In HAYSTAC we average power spectra down to a resolution $\Delta\nu_b=100$~Hz \textit{in situ}, thereby reducing the timestream data rate by a factor of $10^5$ while retaining both advantages of narrow bins. 

There is also another benefit of this \textit{in situ} averaging independent of the reduction of disk space required to store the data. If we tried to save the full timestream dataset, the live-time efficiency introduced in Eq.~\eqref{eq:scan_rate} would be limited to $\zeta\leq0.25$ by the time required to transfer such large data sets from the PCIe digitizer to the hard drive of the DAQ computer. We avoid this inefficiency by transferring the timestream data in small segments to a 10~GB RAM disk and performing all of the required \textit{in situ} processing in RAM.\footnote{The DAQ computer is a Dell Precision T3600 running 64-bit Windows~7, with an Intel Xeon E5-1620 processor and 32~GB of RAM.}

The required \textit{in situ} processing includes image rejection as well as FFT and power spectrum computation. As noted in Sec.~\ref{sub:receiver_rt}, each IF channel is sensitive to both the Fourier components of interest in the lower sideband of the LO and unwanted image frequencies in the upper sideband. The 90$^\circ$ relative phase shift between the two otherwise identical IF outputs of an IQ mixer may be exploited for image rejection: adding the $Q$ output to the $I$ output with a $+(-)90^\circ$ phase shift suppresses the upper (lower) LO sideband. This is the operating principle of commercial image reject mixers, which have limited IF bandwidth because they implement the required $90^\circ$ phase shift in hardware.

We implement image rejection in the frequency domain in software: this scheme works at any IF frequency, with the only possible drawback being that amplitude and phase mismatches in the discrete IF components in the two channels can limit the degree of image rejection.\footnote{We compensate for this partially by multiplying the $I$ channel timestream in software by the amplitude imbalance due to component variations in the A2 amplifiers (measured to be 1.14).} By taking FFTs of both the I and Q channels and defining $X(\omega) = \left(\text{Re}[I(\omega)] - \text{Im}[Q(\omega)]\right) + i\left(\text{Im}[I(\omega)] + \text{Re}[Q(\omega)]\right)$ we obtain rejection of the upper sideband better than 20~dB throughout the analysis band in the power spectrum $\left|X(\omega)\right|^2$.

During noise measurements, we digitize both IF channels simultaneously in 5~s segments, then transfer each segment to RAM. The total data in each segment across both IF channels is 437.5 MB, and the time required to transfer this data to RAM is 1.2~s, which caps the data acquisition efficiency at $\zeta\leq80\%$. The \textit{in situ} processing code divides the data from each channel into 500 non-overlapping records of duration $1/\Delta\nu_b=10$~ms, computes the FFT of each record with no windowing, combines the $I$ and $Q$ FFTs corresponding to the same 10~ms time slice to implement image rejection as described above, constructs a power spectrum from each sample of $X(\omega)$, and finally averages all 500 power spectra corresponding to each segment. All processing for each 5~s segment occurs in RAM in parallel with the acquisition of the next segment. At the end of data acquisition (whose total duration is $\tau/\zeta$), the power spectra from all 5~s segments are averaged together to obtain a single spectrum obtained from $\Delta\nu_b\tau=9\times10^4$ averages, which is written to disk.

\subsection{Data run procedure}\label{sub:daq_procedure}
A haloscope axion search consists of a sequence of iterations separated by discrete tuning steps, with a measurement of the noise power spectrum and various auxiliary measurements at each iteration. The HAYSTAC auxiliary data consists of VNA measurements of the cavity mode and JPA gain profile at each step and periodic $Y$-factor measurements (discussed in Sec.~\ref{sec:noise}) to calibrate the receiver's added noise. Temperatures and pressures throughout the cryogenic system are also monitored independently throughout the run as described in Sec.~\ref{sub:cryo}. The purpose of the auxiliary data is to characterize detector parameters that can vary during the run, both to calibrate the axion search sensitivity (Sec.~\ref{sub:rescale}) and to define data quality cuts (Sec.~\ref{sub:badscans}).

\begin{figure}[h]
\centering\includegraphics[width=1.0\textwidth]{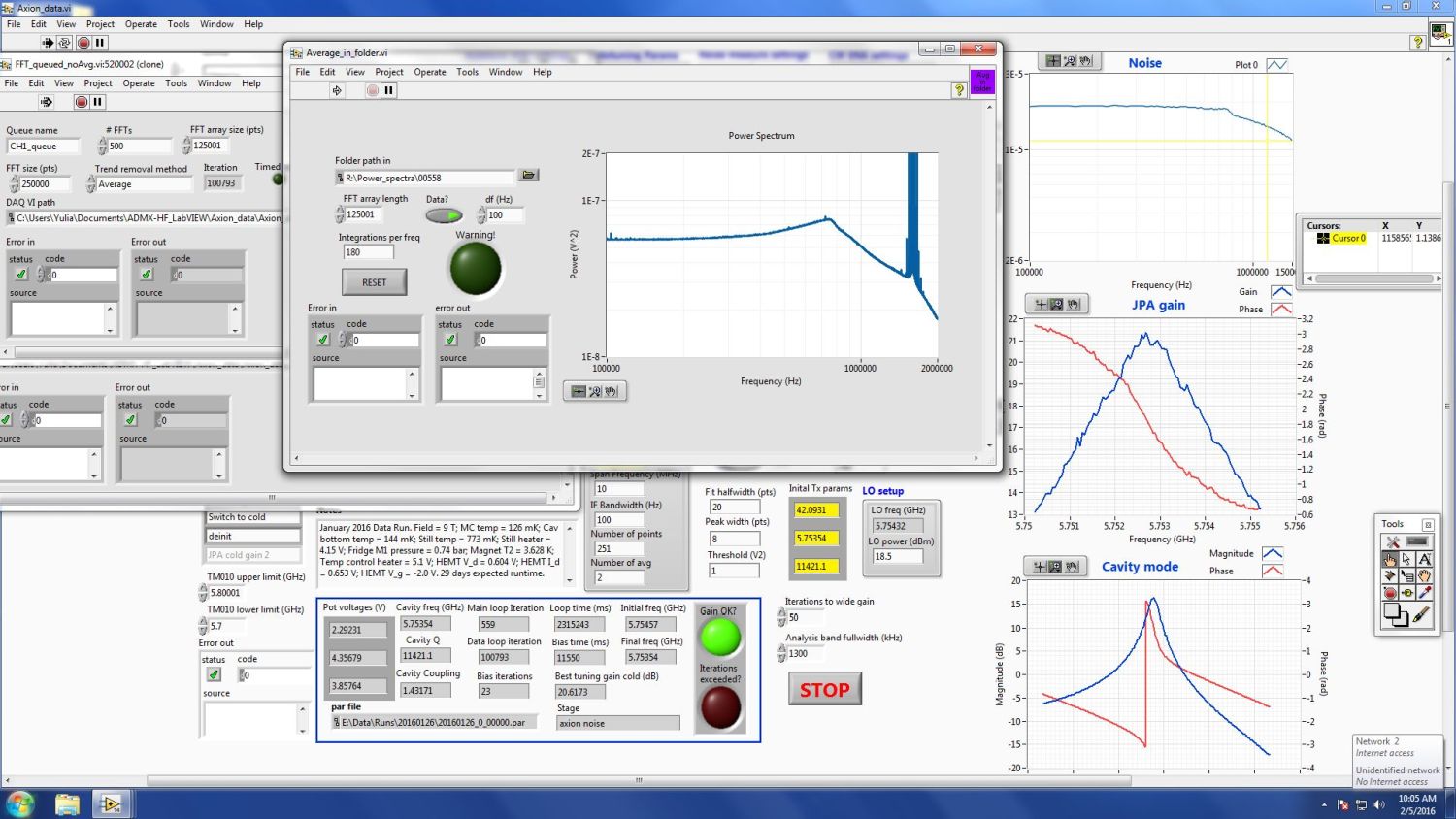}
\caption[HAYSTAC DAQ user interface]{\label{fig:daq} A screenshot of the user interface for the LabVIEW program that controls HAYSTAC data acquisition.}
\end{figure}

Data acquisition for the experiment is fully automated and controlled by a LabVIEW program (Fig.~\ref{fig:daq}). At the beginning of each iteration, the DAQ program tunes the $\text{TM}_{010}$ mode with a fixed mechanical actuation of either the tuning rod or the tuning vernier (see Sec.~\ref{sub:cav_tuning} for discussion of the tuning scheme adopted for the first HAYSTAC data run). Next it measures $\nu_c$ and $Q_L$ using a VNA sweep through the cavity transmission line (Sec.~\ref{sub:cav_meas}), and sets $\nu_\text{LO}$ by rounding the measured value of $\nu_c$ to the nearest 100~Hz and then adding 780~kHz.\footnote{Coercing $\nu_\text{LO}$ to the nearest 100~Hz ensures that the bin boundaries in different spectra are always aligned. As a result the analysis band is not exactly centered on $\nu_c$ in each spectrum, but the maximum offset is always $<\Delta\nu_b$.} Then it sets $\nu_p=\nu_\text{LO}-1.6$~MHz: setting the JPA pump frequency at a fixed offset from the LO instead of the cavity resonance ensures that the $1/\Delta\nu_b = 10$~ms integration time of each subspectrum is an integer number of periods at frequency $\nu_p$, which minimizes spreading of the pump power throughout the spectrum.\footnote{Sinusoidal signals of arbitrary frequency will generally not be confined to single bins in the spectrum because we do not apply a window function to the timestream data during \textit{in situ} processing. The ``rectangular window'' (equivalent to not windowing at all) is the correct choice for a haloscope search as it has the smallest equivalent noise bandwidth. Given the constraint of the rectangular window, another advantage of a bin width $\Delta\nu_b\ll\Delta\nu_a$ is that it ensures negligible distortion of the axion signal lineshape by the FFT.} 

After updating the value of $\nu_p$, the program optimizes the JPA gain using the autobiasing procedure described in Sec.~\ref{sub:jpa_bias}, measures the gain profile over 5~MHz centered on $\nu_p$ (see Fig.~\ref{fig:gain_sample}), and then turns on the LO and the CW tone and engages the flux feedback system described in Sec.~\ref{sub:feedback}. The system then acquires $\tau=15$~minutes of axion-sensitive timestream data, while the \textit{in situ} processing described in Sec.~\ref{sub:power_spectra} runs in parallel. At the end of the cavity noise measurement, the JPA gain and cavity transmission profiles are each measured again, to flag any unusually large mode frequency drifts (see Sec.~\ref{sub:cav_tuning}) and large bias flux drifts.\footnote{Because the feedback system described in Sec.~\ref{sub:feedback} only operates during the noise measurement, comparing the ``before'' and ``after'' JPA gain profiles can tell us that the environmental contribution to the flux drifted over the course of the measurement, but not whether the feedback system was perfectly correcting for this drift up to the moment we turned the feedback off to measure the gain profile. Thus the second JPA gain profile at each iteration is mainly useful for giving us a sense of how noisy the JPA's flux environment was during different parts of the run.} Finally, $\beta$ is measured with a reflection sweep as discussed in Sec.~\ref{sub:cav_meas}. $Y$-factor measurements (Sec.~\ref{sec:noise}) are also performed intermittently to give us a handle on any frequency-dependent contributions to $N_\text{sys}$. In the first HAYSTAC data run, we made $Y$-factor measurements every 10 iterations, resulting in an average live-time efficiency of $\zeta=0.72$.\footnote{Note that $\zeta$ is capped at $0.8$ by the inefficiency of the power spectrum data acquisition itself. The time required for stepping and auxiliary measurements at each iteration further limits it to $\zeta=0.76$. Finally $Y$-factor measurements and waiting out the delayed stepping of the tuning rod every 17 iterations together reduce the efficiency to $\zeta=0.72$.} We thus make \textit{in situ} measurements of every parameter appearing in Eqs.~\eqref{eq:signal_power} and \eqref{eq:dicke} that can change between iterations, with the exception of $C_{010}$, whose frequency dependence (Fig.~\ref{fig:c010}) is obtained from simulation. All sweep data is saved to disk along with the averaged cavity noise power spectrum and critical parameter values such as the LO frequency; this amounts to about 3~MB per iteration.  

In the procedure described above, the receiver tracks the $\text{TM}_{010}$ mode without any need for active control over the tuning step size $\delta\nu_c$.  In principle the whole sequence can be repeated as many times as desired. In practice, due to a minor memory leak somewhere deep in the code, LabVIEW starts to slow down when this program runs continuously for more than four days. We thus divided the data acquisition period into four-day segments; it typically took less than 10~minutes to restart LabVIEW and initialize the first iteration of each new segment with parameters from the last iteration of the previous segment. 

Note also that antenna insertion depth is not automatically adjusted to reoptimize $\beta$ during the run: this simplifies operations and does not appreciably affect the sensitivity, both because the value of $\beta$ at fixed insertion depth changes relatively slowly over the frequency range covered in a single 4-day subscan, and because the dependence of $g_\gamma$ on $\beta$ is relatively weak. We periodically readjusted the antenna manually between subscans. In the ideal case discussed in Sec.~\ref{sub:scan}, the scan rate is optimized at $\beta=2$. However, in the presence of excess thermal noise originating in the cavity, the scan rate is instead optimized for $\beta>2$ (see Sec.~\ref{sub:hotrod}), so we aimed to err on the high side. The average coupling in the first HAYSTAC data run was $\beta\approx2.3$.

For the first HAYSTAC data run described in this dissertation, we chose to scan the 100~MHz range between 5.7 and 5.8~GHz. As discussed in Sec.~\ref{sub:cav_meas}, this range represents about half of a window (nearly) free of mode crossings in which the $\text{TM}_{010}$ mode parameters were optimized; nonuniform frequency steps resulting from our ``hybrid'' tuning scheme (Sec.~\ref{sub:cav_tuning}) further constrained efficient operation to roughly the upper half of this window. In choosing a 100~MHz window specifically for our first publication, we were also following the precedent set by the RBF~\citep{RBF1987}, UF~\citep{UF1990}, and ADMX~\citep{ADMX1998} experiments.

\begin{figure}[h]
\centering\includegraphics[width=0.8\textwidth]{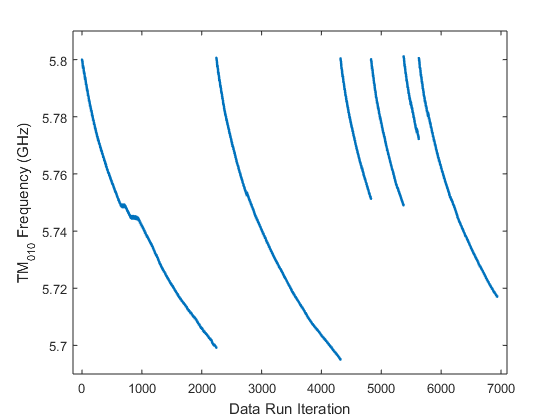}
\caption[$\nu_c$ vs.\ iteration in the first HAYSTAC data run]{\label{fig:tx_f0_data} The $\text{TM}_{010}$ frequency $\nu_c$ vs.\ iteration index in the first HAYSTAC data run; slower tuning at low frequencies is evident. The two short horizontal segments correspond to regions where the tuning system briefly got stuck.}
\end{figure}

The full axion search dataset comprised 6936 power spectra (along with the associated auxiliary data), obtained from two continuous scans over the full tuning range and several shorter scans to compensate for the nonuniformity of hybrid tuning (see Fig.~\ref{fig:tx_f0_data}). The first 2244 spectra were acquired between January~26 and March~5, 2016; the rest of the data was obtained between May~16 and August~2 after a power outage which damaged the system and disrupted operations (appendix~\ref{app:quench}). After repairing the system, we found its operating parameters were largely unchanged during the summer scan. The only significant difference between the two scans was that we injected synthetic axion signals into the experiment at several random frequencies during the winter scan, to validate our data acquisition and analysis procedures (appendix~\ref{app:fake_axions}). After confirming that our analysis procedure was sensitive to these synthetic axion signals, we elected not to inject signals during the summer scan.

Finally, it is worth reiterating in the context of haloscope data acquisition that the haloscope technique is limited by statistical fluctuations of the total noise power rather than systematics (see also discussion in Sec.~\ref{sub:radiometer}). We discriminate axion signals from fluctuations in the noise by requiring a large SNR (Eq.~\eqref{eq:dicke}) and applying an excess power threshold to the (appropriately processed) data. This procedure is discussed in detail in Sec.~\ref{sec:candidates}; here I just want to emphasize that false positives are inevitable for any reasonable value of the threshold. The only way to determine whether a given large power excess is an axion or a statistical fluctuation is to declare all frequencies exceeding the threshold to be \textbf{rescan candidates}, and acquire more data at each such frequency. Rescan data acquisition differs in a few important respects from the continuous data acquisition described in this section: see Sec.~\ref{sec:rescan} for further discussion.

\section{Noise calibration}\label{sec:noise}
I have now discussed how we quantify each parameter which affects the haloscope search SNR except the total noise $N_\text{sys}$ defined by Eq.~\eqref{eq:system_noise}: $N_\text{sys}$ deserves its own section as it is in many respects the most critical such parameter for the success of HAYSTAC specifically. I showed in Sec.~\ref{sec:jpa_op} that we were able to achieve and maintain high JPA gain $G_J$ during HAYSTAC operations. Given this success, we should expect meaningful contributions to the receiver's added noise $N_A$ to arise from the added noise of the JPA, the input-referred added noise of the HEMT, and imperfect power transmission efficiency between the input of the receiver and the JPA (see Sec.~\ref{sub:receiver_noise}); $N_\text{sys}$ will of course have contributions from quantum and thermal noise at the receiver input in addition to $N_A$.

As I emphasized at the end of Sec.~\ref{sub:radiometer}, we do not require an especially \textit{precise} calibration of $N_\text{sys}$ for the haloscope search, because the axion signal is spectrally localized. We will see that there are many subtleties associated with  noise calibration in HAYSTAC, so this is quite fortunate! Nonetheless, a robust and reliable noise calibration procedure is of paramount importance, not least because it allows us to determine how close we are able to get to the SQL [Eq.~\eqref{eq:sql}] in practice.

The presence of the cryogenic microwave switch at the HAYSTAC receiver input allows us to calibrate $N_A$ using $Y$-factor measurements (defined in Sec.~\ref{sub:yfactor}), which are a standard technique in microwave radiometry. However, naive application of the usual expression for $N_A$ obtained from a $Y$-factor measurement tends to \textit{underestimate} the true value of $N_A$, for reasons intimately tied to the essential physics of parametric amplification discussed in Sec.~\ref{sub:jpa_intro}.\footnote{This effect is particularly insidious because the systematically low value for $N_A$ tends to roughly cancel out the effects of loss, etc. Thus the incorrect result obtained from naive application of the $Y$-factor technique ends up being quite close to the SQL, and physicists (like other humans) have a tendency to scrutinize things less carefully when it appears that everything is working.} Even accounting for this effect, we should not necessarily take for granted that the ``total noise'' we measure by adding the known input noise to the value for ``$N_A$'' obtained through the $Y$-factor technique is in fact identical to the quantity $N_\text{sys}$ which matters for the sensitivity of the haloscope search.

For these reasons, I will devote Sec.~\ref{sub:yfactor} to a pedagogical derivation of the methodology applied to $Y$-factor measurement analysis in HAYSTAC. In Sec.~\ref{sub:noise_offres} I will describe our $Y$-factor measurement procedure, and the results of measurements in which the JPA is biased up far from the cavity mode, such that the input noise comes from a $50~\Omega$ termination on the mixing chamber plate of the DR. Finally, in Sec.~\ref{sub:hotrod} I discuss the results of noise measurements on resonance, in which we observe an extra contribution associated with the poor tuning rod thermal link noted in Sec.~\ref{sub:cav_design}.

\subsection[Principles of $Y$-factor calibration]{Principles of $\boldsymbol{Y}$-factor calibration}\label{sub:yfactor}
A simple way to measure the added noise of any microwave receiver is via a $Y$-factor measurement, in which we connect the receiver input to two matched (50~$\Omega$) loads at different known temperatures $T_C$ and $T_H>T_C$, and measure the noise power in each configuration: $T_H$ and $T_C$ are referred to as the ``hot'' and ``cold''  loads, respectively. It is typically assumed that if this is done carefully, the noise added by the receiver can be made independent of the input noise. We then define the $Y$ factor as the measured hot/cold noise power ratio. We see that
\begin{equation}\label{eq:yfactor_simple}
Y = \frac{P_H}{P_C} = \frac{T_H + T_A}{T_C + T_A},
\end{equation}
which is easily solved for $T_A$. 

Eq.~\eqref{eq:yfactor_simple} is the standard expression that appears in textbooks (e.g., Ref.~\citep{pozar2012}); it clearly assumes the Rayleigh-Jeans limit of the blackbody distribution. But the more insidious assumption is the one we used to derive Eq.~\eqref{eq:yfactor_simple}. We saw in Sec.~\ref{sub:jpa_intro} that for an ideal parametric amplifier operating in phase-insensitive mode, the added noise is precisely thermal noise which originates on the opposite side of the pump from the signal and is coupled into the ouptut by intermodulation gain: as a matter of principle, it cannot be made independent of the input noise!

To reformulate Eq.~\eqref{eq:yfactor_simple} in a form conducive to the haloscope search, let us first consider how $Y$-factor measurements are realized in HAYSTAC. The cold load is the cavity itself, which I will assume to be at temperature $T_\text{mc}=127$~mK based on the rapid response of the cavity thermometer to temperature changes at the mixing chamber (Sec.~\ref{sub:cryo}). The hot load is the terminator connected to the other input of the switch S1, which is physically fixed to the still plate at temperature $T_\text{still}=775$~mK. $T_\text{still}$ is hot enough that the Rayleigh-Jeans approximation is accurate, but $T_\text{mc}$ is not. Thus we will return to our usual notation $N_\text{mc} = N(T_\text{mc})$, where $N(T)$ is given by Eq.~\eqref{eq:noise_eta} with $\eta=0$; note that it includes the vacuum contribution. With the above operating temperatures, $N_\text{mc} = 0.63$ quanta and $N_\text{still}=2.8$ quanta.

Actually, we should be a little more careful in defining the cold load, since the cavity only looks like a $50~\Omega$ load exactly on resonance when it is critically coupled [see Eq.~\eqref{eq:gamma_refl}]. With any other coupling $\beta$ (or any nonzero detuning from resonance even at $\beta=1$), the cavity does not provide the full thermal noise because some of the noise generated by the surface resistance of the walls is reflected off the antenna and absorbed internally. But the difference is made up by the thermal noise of the upper terminated port on the directional coupler in the reflection input line (see Fig.~\ref{fig:cryo_setup}), which reflects off the cavity and comes out through the receiver. Insofar as they are at the same temperature, the Johnson noise generated in the cavity and the Johnson noise reflected off the cavity always sum to $N_\text{mc}$ independent of $\beta$ and the detuning: this is precisely what it means to be in thermal equilibrium!

Having established the requisite notation, we now define $N_r=N_A-N(T)$ as the added noise of the receiver \textit{excluding} the contribution of an ideal parametric amplifier which is necessarily present in HAYSTAC.\footnote{Note that we are not assuming that the JPA acts like an ideal parametric amplifier. More mundane sources of JPA added noise, due to e.g., unexpected internal loss, can be included in $N_r$.} Assuming $N_r$ does not change between switch configurations, the expression for the $Y$ factor becomes
\begin{equation}\label{eq:yfactor_jpa}
Y  = \frac{2N_\text{still} + N_r}{2N_\text{mc} + N_r},
\end{equation}
which may be inverted to obtain
\begin{equation}\label{eq:added_noise_jpa}
N_r = \frac{2(N_\text{still} - YN_\text{mc})}{Y - 1}.
\end{equation}

Eqs.~\eqref{eq:yfactor_jpa} and Eq.~\eqref{eq:added_noise_jpa} have the distinct benefit relative to the usual expressions of not being necessarily wrong, though they do still make various assumptions which we will address presently. But before we generalize Eq.~\eqref{eq:added_noise_jpa}, we should verify that the quantity in the denominator of Eq.~\eqref{eq:yfactor_jpa} is actually equivalent to $N_\text{sys}=N_\text{mc}+N_A$ which appears in the radiometer equation Eq.~\eqref{eq:dicke}. Referring the added noise of the HEMT to the receiver input [Eq.~\eqref{eq:amp_chain}] is conceptually simple enough, and for now we will continue to assume thermal equilibrium between the cavity and the DR base plate: then the question of whether we are actually measuring the right thing reduces to the question of how loss before the JPA affects the SNR of the haloscope search vs.\ how it contributes to $N_r$ as defined by Eq.~\eqref{eq:added_noise_jpa}.

Let's first consider the SNR [Eq.~\eqref{eq:dicke}] for an idealized version of HAYSTAC with no loss between the receiver input and the JPA: I will denote the total noise in the lossless case by $N'_\text{sys}$. Now let us introduce some loss before the JPA, formalized as a power transmission efficiency $\eta<1$. Clearly, the signal power is just multiplied by $\eta$, whereas the total noise power does not change [c.f.\ Eq.~\eqref{eq:noise_eta}] as long as the lossy elements are in thermal equilibrium with everything else. The experiment with loss evidently has the same SNR as a lossless experiment with total noise $N_\text{sys}=N'_\text{sys}/\eta$. Equivalently, we may model loss before the JPA as an additive contribution to $N_A$ given by 
\begin{equation}\label{eq:noise_loss}
N_\text{loss} = \frac{1-\eta}{\eta}\left(N_\text{mc} + N_A'\right),
\end{equation}
where the primed notation again indicates the case where loss is not included as an effective contribution to the noise. 

Now we turn to the effects of $\eta$ on a $Y$-factor measurement, and define $\eta=\eta_0\eta_1$, with $\eta_0$ between the cavity and S1 and $\eta_1$ between S1 and the JPA input. We will see that we can reproduce Eq.~\eqref{eq:noise_loss} if $\eta_0=1$: thus in this limit the $Y$-factor technique measures precisely the sum of contributions which we care about for the haloscope search. Essentially, the reason for this correspondence is that both the hot load noise and the axion signal are attenuated by the same factor $\eta_1$, whereas the cold load noise is unchanged. To obtain this result more rigorously, let's consider a hypothetical $Y$-factor measurement in the lossless gedanken-HAYSTAC:
\begin{equation}\label{eq:yprime_jpa}
Y' = \frac{2N_\text{still}+N_r'}{2N_\text{mc}+N_r'}.
\end{equation}
Loss will replace a fraction of the hot load noise (on both sides of the pump) with cold load noise, so what we actually measure is
\begin{equation}\label{eq:yfactor_loss}
Y = \frac{2\eta_1N_\text{still} + 2(1-\eta_1)N_\text{mc} + N_r'}{2N_\text{mc} + N_r'}.
\end{equation}
Note that $\eta_0$ does not enter this expression no matter how small it is, because the hot noise never sees it and the cold noise is at the same temperature. Relating Eqs.~\eqref{eq:yprime_jpa} and \eqref{eq:yfactor_loss} we obtain
\begin{equation}\label{eq:y_yprime}
Y - 1 = \eta_1\left(Y'-1\right).
\end{equation}
We now instead model the loss in the $Y$-factor measurement as an additional additive contribution to $N_r'$:
\[
Y = \frac{2N_\text{still}+N_r' + N_\text{loss}}{2N_\text{mc}+N_r'+N_\text{loss}}.
\]
Solving for $N_\text{loss}$ yields
\begin{align}
N_\text{loss} &= \frac{2N_\text{still}+N_r' - Y(2N_\text{mc}+N_r')}{Y-1} \nonumber \\
&= \frac{2N_\text{still}+N_r'  - \big(2\eta_1N_\text{still} + 2(1-\eta_1)N_\text{mc} + N_r'\big)}{\eta_1(Y'-1)} \nonumber \\
&= \frac{1-\eta_1}{\eta_1}\frac{2(N_\text{still} - N_\text{mc})}{\left(Y'-1\right)}, \label{eq:noise_loss_blah}
\end{align}
where I have used Eq.~\eqref{eq:yfactor_loss} in the numerator and Eq.~\eqref{eq:y_yprime} in the denominator of the second line. Finally, since Eq.~\eqref{eq:yprime_jpa} is formally identical to Eq.~\eqref{eq:yfactor_jpa}, 
\[
N_r' + 2N_\text{mc} = \frac{2(N_\text{still} - Y'N_\text{mc})}{Y' - 1} + 2N_\text{mc} = \frac{2(N_\text{still} - N_\text{mc})}{\left(Y'-1\right)},
\]
which is just the second factor on the RHS of Eq.~\eqref{eq:noise_loss_blah}. Thus, we obtain
\begin{equation}\label{eq:noise_loss_2}
N_\text{loss} = \frac{1-\eta_1}{\eta_1}\left(2N_\text{mc} + N_r'\right)
\end{equation}
which is equivalent to Eq.~\eqref{eq:noise_loss} for $\eta_0=1$ as anticipated. 

We have learned that provided the assumption of thermal equilibrium at the base temperature is satisfied, a $Y$-factor measurement in HAYSTAC is sensitive to the sum of input-referred amplifier noise and effective noise due to $\eta_1$, and these are the only contributions to  $N_\text{sys}$ in the limit $\eta_0\rightarrow1$. The effect of $\eta_0$ on the SNR cannot be measured \textit{in situ} with our present system, and thus for the purposes of the analysis we treat it as a factor in the signal power calibration (Sec.~\ref{sub:rescale}) rather than an effective contribution to the loss. We estimated $-0.6$~dB for the contributions from the long superconducting cable and connector losses: thus in linear units $\eta_0=10^{-0.6/10}=0.87$. As we will see in Sec.~\ref{sub:hotrod}, violating the assumption of thermal equilibrium messes up the correspondence between $N_\text{sys}$ and what we obtain from $Y$-factor measurements in a way that is not quite so easy to correct. The broader point to take away from all of this is that a careful analysis is required to determine whether any given general method for quantifying the noise performance of a microwave receiver actually measures the quantity of interest to the haloscope search.

Now that we understand the contributions to $N_r$, we can generalize Eq.~\eqref{eq:added_noise_jpa} to account for two practical measurement effects. First, hot and cold noise power measurements in HAYSTAC are really measurements of the noise power spectrum, so Eq.~\eqref{eq:added_noise_jpa} should in principle be a function of frequency: in particular the lower JPA gain at the low-IF-frequency end of the analysis band (see Fig.~\ref{fig:IF_layout}) implies less suppression of the input-referred HEMT noise and thus larger $N_A$. Second, we saw in Sec.~\ref{sub:jpa_fluct} that actuating the switch affects the JPA gain. In practice we always rebias the JPA after actuating the switch, but small changes in $G_J$ and $\Delta\nu_J$ between switch configurations are still possible. We account for both effects by rewriting Eq.~\eqref{eq:added_noise_jpa} as
\begin{equation}\label{eq:added_new}
N_A(\nu) = \frac{2\left[A(\nu)N_\text{still} - Y(\nu)N_\text{mc}\right]}{Y(\nu) - A(\nu)} + N_\text{mc},
\end{equation}
where $A(\nu)=G_H(\nu)/G_C(\nu)$ is obtained from the measured hot/cold JPA gain profiles. I have written this expression with $N_A$ on the LHS as it is a more standard quantity to reference than $N_r$, which is of course just the first additive term on the RHS.\footnote{Note that this expression breaks down if $A(\nu)$ deviates too far from 1: then the hot and cold HEMT contributions to $N_A$ will differ, and the assumption of a single effective added noise source at the JPA input is no longer valid. Furthermore, if $A(\nu)\sim Y(\nu)$, the estimate of $N_A$ will be extremely sensitive to measurement uncertainty.}

\subsection{Off-resonance measurements}\label{sub:noise_offres}
In practice, our $Y$-factor measurement procedure comprises the following steps, repeated for each switch configuration: we bias up the JPA to the target gain, measure the gain profile, and take 5~s of noise data, to which we apply the same \textit{in situ} processing used to construct power spectra from the axion search data. Immediately after the noise measurement we make a second measurement of the JPA gain profile, and then measure the noise spectrum with the JPA off, which we expect to be dominated by the added noise of the HEMT.\footnote{One can in principle estimate certain contributions to $N_A$ by determining the SNR improvement in the measurement of a small coherent tone with the JPA on relative to the same measurement with the JPA off, but the effect of losses on such on/off measurements is much more convoluted. Our receiver is also very suboptimally configured for a measurement of $T_\text{HEMT}$, which is significantly larger than both $T_\text{still}$ and $T_\text{mc}$; the hot load noise experiences nontrivial attenuation along the way which exacerbates the problem. For these reasons, we did not end up using the JPA-off noise data from either switch configuration in our analysis.} We do not engage the flux feedback system due to the short measurement duration; the second JPA gain profile just allows us to discard rare measurements during which the gain was clearly unstable.\footnote{Note also that we do not normalize the JPA gain measurements when computing $A(\nu)$ (i.e., we take the ratio of JPA-on sweeps rather than the gain ratio strictly speaking). This is the correct thing to do in principle in case some other contribution to the net receiver gain changes when we actuate the switch. In practice, it doesn't appear to make a difference.} We average the calibration spectra down to 10~kHz resolution, for a total of 130 data points in the analysis band. 

As noted in Sec.~\ref{sub:daq_procedure}, this whole sequence was carried out every 10~iterations during the data run, with 2~minutes of waiting after each switch actuation (see Sec.~\ref{sub:jpa_fluct}). These measurements (discussed in Sec.~\ref{sub:hotrod}) enable us to calibrate $N_\text{sys}$ around $\nu_c$, which is just what we need to know for the haloscope analysis. However, due to the issue with the tuning rod thermal link discussed in Sec.~\ref{sub:cav_design}, the resonant noise measurements exhibit signs of thermal disequilibrium which complicate the analysis significantly. Because of this it is helpful to first consider the results of off-resonant noise calibration measurements conducted during detector commissioning, for which we tuned the JPA to some frequency far from any cavity modes, and set $\nu_\text{LO}$ 1.6~MHz above $\nu_p$. Thus the IF configuration only differs from Fig.~\ref{fig:IF_layout} in the absence of the cavity mode between $\nu_p$ and $\nu_\text{LO}$. In off-resonance measurements, the cold load noise comes entirely from the terminated port on the reflection line directional coupler (see Fig.~\ref{fig:cryo_setup} and discussion in Sec.~\ref{sub:yfactor}).

From off-resonance measurements, we consistently obtained $N_A\approx1.35$~quanta on average throughout the analysis band. The noise exhibited relatively mild dependence on IF frequency within the analysis band: $N_A$ was minimized ($\approx1.3$~quanta) around $\sim1.1$~MHz, rising to $\approx1.45$~quanta towards low IF frequencies presumably due to the JPA gain rolloff. The measurements exhibited no systematic dependence on the RF frequency $\nu_p$: repeated measurements at the same frequency exhibited just as much variation as measurements at different frequencies. Empirically, this variation was correlated across the analysis band in each measurement, with $\delta N_A/N_A\sim4\%$. While this is certainly sufficient precision for the haloscope search, it is worth emphasizing that it is much larger than the statistical fluctuations in the noise in each power spectrum bin. The origin of this residual systematic variation in the noise is unknown.

Eq.~\eqref{eq:dicke} only depends on the total noise $N_\text{sys}$, not on any of the distinct contributions to $N_A$ from the JPA, the HEMT, and losses. Moreover we do not have a way to precisely measure these contributions independently. Nonetheless, it is instructive to try to obtain rough estimates of the different contributions to $N_A$, to get a sense of where we can improve. Just under half of the noise ($N_\text{mc}=0.63$~quanta) may be attributed to an ideal JPA operating in phase insensitive mode at temperature $T_\text{mc}$; this leaves us with $N_r=0.73$ quanta to account for. Once we attribute any portion $N_r'$ to amplifiers (i.e., the input referred noise of the HEMT and/or JPA internal loss), we must have $N_\text{loss}=N_r-N_r'$ in Eq.~\eqref{eq:noise_loss_2}, which fixes $\eta_1$. For definiteness, let's take $N'_r=N_\text{HEMT}=0.17$~quanta, from the manufacturer's spec for the HEMT noise and a rough estimate of the loss between the JPA and the HEMT. Plugging the relevant numbers into Eq.~\eqref{eq:noise_loss_2}, we obtain $\eta_1 = 0.72$ ($-1.4$~dB of loss in logarithmic units). Recent efforts to optimize JPA designs within the HAYSTAC collaboration suggest that coupling out of the JPA enclosure (Sec.~\ref{sub:jpa_design}) may account for as much as 0.5~dB of this. It seems plausible that the cryogenic insertion losses of the switch, two circulators, and a directional coupler could together account for the other 0.9~dB.

All of these numbers should be taken as rough estimates, but collectively they do indicate that what we are seeing is reasonable: in particular there is no need to invoke internal loss in the JPA. The \textit{total} noise inferred from these measurements is $N_\text{sys}\approx2$~quanta, which was precisely the value used to estimate HAYSTAC sensitivity in Sec.~\ref{sub:scan}: in the real world, we can unfortunately only claim it as the value we would obtain in the absence of thermal disequilibrium in the cavity, as we will see presently. Considering the various contributions to $N_A$ also illustrates how difficult it is to eliminate the last few quanta of noise in an already exceptionally quiet system. If we were able to operate at $T_\text{mc}\approx70$~mK ($N_\text{mc}\approx0.52$), this would reduce the noise by about 0.25~quanta, because we suffer the thermal contribution twice due to nature of the JPA added noise. We can also attempt loss reduction: a target value of $\eta_1=0.9$ ($-0.5$~dB in logarithimic units) seems ambitious but feasible, and would reduce $N_\text{sys}$ to 1.4~quanta. $N_\text{sys}<1.4$~quanta seems unlikely without a qualitatively different measurement scheme.

\subsection{Thermal disequilibrium and the ``hot rod'' problem}\label{sub:hotrod}
\begin{figure}[h]
\centering\includegraphics[width=0.8\textwidth]{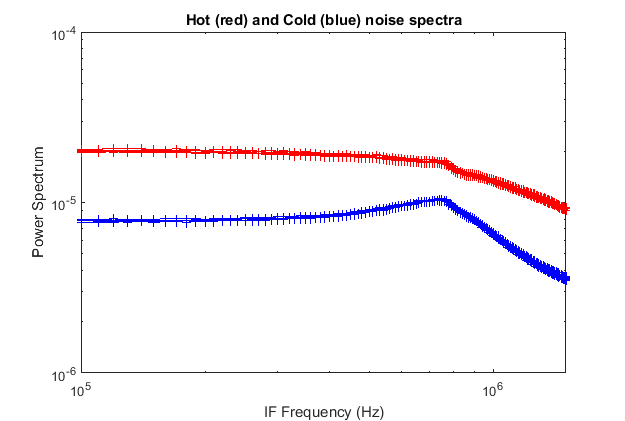}
\caption[Hot load and cold load noise spectra]{\label{fig:noise_spectra} Hot load (red) and cold load (blue) power spectra (with arbitrary units on the vertical axis) centered on the $\text{TM}_{010}$ mode at IF frequency 780~kHz. The frequency-dependence of the hot load noise is due to the frequency-dependence of components in the receiver, and would also be observed in the cold load in the absence of the resonant excess.}
\end{figure}

In noise measurements around the TM$_{010}$ resonance, we observe an additional Lorentzian excess $\Delta N_\text{cav}(\nu)$ centered on the IF frequency corresponding to $\nu_c$, which is not present with the switch pointed at the hot load (Fig.~\ref{fig:noise_spectra}). One straightforward interpretation is that this excess represents a contribution to the \textit{input} noise, rather than the receiver's added noise. From this perspective, it appears that the microwave system which the receiver probes when pointed at the cold load is not in thermal equilibrium in a suspiciously cavity-mode-shaped region of the spectrum. 

In fact, this is precisely the behavior we would expect from a cavity at temperature $T_\text{cav}>T_\text{mc}$: the noise from the reflection line directional coupler and the noise from the cavity no longer conspire to produce a flat spectrum, resulting in an apparent excess (deficit) for $T_\text{cav}>T_\text{mc}$ ($T_\text{cav}<T_\text{mc}$).\footnote{The latter case is physically implausible since the cooling power comes from the mixing chamber.} We can describe this situation formally with the expression
\begin{equation}\label{eq:yfactor_cav}
Y(\nu) = A(\nu)\frac{2N_\text{still} + N_r(\nu)}{2N_\text{mc} + \eta\Delta N_\text{cav}(\nu) + N_r(\nu)},
\end{equation}
where $\Delta N_\text{cav}$ is the \textit{excess} noise relative to $N_\text{mc}$: we expect $\Delta N_\text{cav}(\nu)\rightarrow0$ for $\nu$ far from $\nu_c$ in either direction, but it is not clear that this is a good approximation at the edges of the analysis band (which after all was chosen to be comparable to the cavity linewidth). I have also made the (probably safe) assumption that the loss $\eta=\eta_0\eta_1$ still occurs at temperature $T_\text{mc}$. The fundamental problem with Eq.~\eqref{eq:yfactor_cav} is clear enough: there are now two unknown frequency-dependent quantities, and we simply do not have enough information within each $Y$-factor measurement to uniquely determine both of them. Eq.~\eqref{eq:yfactor_cav} may be rearranged to yield
\begin{equation}\label{eq:added_cav}
N_A(\nu) + \frac{Y(\nu)}{Y(\nu)-A(\nu)}\eta\Delta N_\text{cav}(\nu) = \frac{2\left[A(\nu)N_\text{still} - Y(\nu)N_\text{mc}\right]}{Y(\nu) - A(\nu)} + N_\text{mc},
\end{equation}
where all the quantities on the RHS are measurable or known. The coefficient of the $\eta\Delta N_\text{cav}$ term on the LHS reflects the fact that it appears only in the denominator of Eq.~\eqref{eq:yfactor_cav}, whereas $N_A$ (or $N_r$) appears in both the numerator and the denominator. Since this coefficient is $>1$, we see that naively applying Eq.~\eqref{eq:added_new} to measurements exhibiting this feature would \textit{overestimate} the impact of the real physical quantity $\eta\Delta N_\text{cav}$. Hereafter I will absorb the factor of $\eta$ into the normalization of $\Delta N_\text{cav}$, since we do not have a precise independent measurement of $\eta$.

I should emphasize that thus far we have simply assumed $\Delta N_\text{cav}$ originates in thermal disequilibrium, which is plausible but by no means the only possibility: for example the existence of a feedback path producing an instability of the cavity-JPA system could produce a resonant excess in the noise spectrum. The feedback scenario seems \textit{a priori} less likely given the absence of other observable effects of such an instability, but we don't have to lean heavily on intuition when we can distinguish these two explanations experimentally. A thermal effect should be maximized at $\beta=1$, and thereafter decrease with increasing $\beta$, because reflected noise dominates in the far-overcoupled limit (c.f., the JPA).\footnote{For this reason, the value of $\beta$ that optimizes haloscope search sensitivity increases when the noise produced within the cavity is hotter than the reflected noise. This observation underlies the squeezed state receiver concept discussed briefly in Sec.~\ref{sec:haystac_future}.} With feedback the behavior depends more strongly on the details, but generically we might expect it to increase for large $\beta$, due to the large resonant phase shift. The dependence of the resonant excess on $T_\text{mc}$ however provides an unambiguous test: any kind of feedback should get worse as we increase $T_\text{mc}$, whereas a thermal excess should get relatively smaller until it eventually recedes into the background.

We measured the dependence of $\Delta N_\text{cav}$ on both coupling and temperature and in both cases observed the expected behavior for a thermal effect: in particular, the excess vanished completely for $T_\text{mc} \gtrsim 550$~mK.\footnote{This is only a very rough estimate of the degree of thermal disequilibrium, because empirically $\Delta N_\text{cav}$ has some systematic dependence on $\nu_c$, discussed below. The presence of $\eta$ multiplying the excess noise indicates that it is probably a slight underestimate.} The demonstration that $\Delta N_\text{cav}$ has a thermal origin offers some support for the ansatz that $N_A$ is the same on-resonance as off-resonance, which allows us to break the degeneracy in Eq.~\eqref{eq:added_cav} by subtracting the average of off-resonance measurements $\bar{N}_A(\nu)\approx1.35$.\footnote{Our ansatz is rather crude, but physically speaking there is no reason to imagine that this thermal effect would affect any of the contributions to $N_A$. In particular the ``ideal parametric amplifier'' contribution to the JPA's added noise will not change, because the resonant excess is confined to one side of the pump in the JPA's input spectrum.} By doing so we obtain an independent estimate of $\Delta N_\text{cav}$ from each resonant $Y$-factor measurement, though this method implies that variations in $\bar{N}_A$ among spectra (see Sec.~\ref{sub:noise_offres}) are attributed instead to variation in $\Delta N_\text{cav}$.\footnote{It is easy to conceive of a slightly different procedure in which the measured fluctuations in the LHS of Eq.~\eqref{eq:added_cav} are instead attributed mostly to $N_A$; we have explored this approach more recently in HAYSTAC.}

\begin{figure}[h]
\centering\includegraphics[width=.7\textwidth]{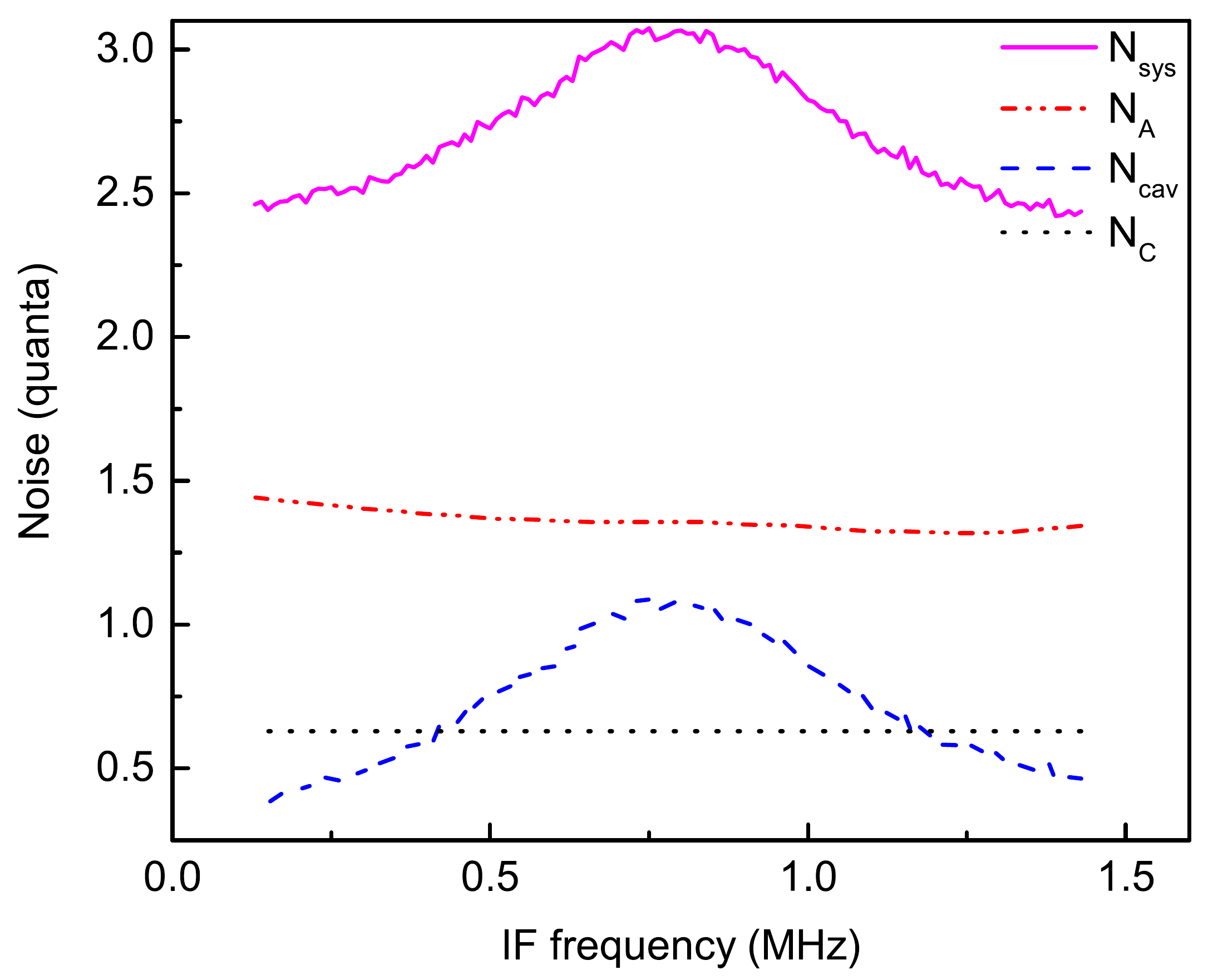}
\caption[Contributions to $N_\text{sys}$ measured \textit{in situ}]{\label{fig:Yfactor} A representative noise measurement. $N_\text{mc}$ (black dotted line; labeled $N_C$ in the legend) is obtained from thermometry, $N_A$ (red dot-dashed line) is obtained from the average of off-resonance $Y$-factor measurements, and $\Delta N_{\text{cav}}$ (blue dashed line; labeled $N_\text{cav}$ in the legend) is obtained from a single $Y$-factor measurement during the data run using the ansatz outlined in the text. $N_{\text{sys}}$ (pink solid line) is the sum of these contributions.}
\end{figure}

The procedure described above allows us to quantify the peak excess contribution $\Delta N_\text{cav}(\nu_c) \sim 1$~quantum, albeit with substantial ($\sim17\%$) variation across spectra, whose implications for the HAYSTAC analysis are discussed in Sec.~\ref{sub:rescale}. This amplitude implies some element in the cavity with an effective temperature $T\sim600$~mK, consistent with the crude result from the direct test of temperature-dependence. The likely culprit is a poor thermal link between the tuning rod and the cavity barrel (see Sec.~\ref{sub:cav_design}).\footnote{Recall from Sec.~\ref{sub:cryo} that thermometry suggests the barrel itself is well-coupled to the mixing chamber plate.} For this reason we have taken to calling this effect the ``hot rod'' problem. A breakdown of the contributions to the total noise in a representative spectrum is shown in Fig.~\ref{fig:Yfactor}. The typical total noise is $N_\text{sys}\approx3$ quanta on resonance, falling to $\approx2.2$ quanta at the edges of the analysis band. Even with a hot rod inside, HAYSTAC is an extremely quiet system!

As noted in Sec.~\ref{sub:cav_design}, we improved the tuning rod thermal link in the aftermath of the first HAYSTAC data run (see also discussion in Ref.~\citep{zhong2017}). This did not completely eliminate the hot rod effect, but did reduce the peak amplitude from $\Delta N_\text{cav}(\nu_c)\sim 1$~quantum to $\Delta N_\text{cav}(\nu_c)\sim 0.35$~quanta, confirming that our understanding of the origin of the resonant excess is correct. The fact that the excess did not disappear completely might be interpreted to indicate the existence of some mechanism which \textit{heats} the rod, such that improving the coupling to the mixing chamber merely pushes down the temperature at which the heating and cooling power balance. Alternatively, it could indicate that the thermal conductivity of the path through the rod has very sharp temperature dependence, such that the rod never truly equilibrates, but its thermal time constant becomes effectively infinite below a certain temperature.


\chapter{The axion search analysis}\label{chap:analysis}
\setlength\epigraphwidth{0.49\textwidth}\epigraph{\itshape My liege, and madam, to expostulate\\ What majesty should be, what duty is,\\ Why day is day, night night, and time is time,\\ Were nothing but to waste night, day, and time.\\ Therefore, since brevity is the soul of wit,\\ And tediousness the limbs and outward flourishes, \\I will be brief.}{Polonius}

\noindent In Sec.~\ref{sec:daq} I discussed the procedure used to acquire axion-sensitive power spectra with the HAYSTAC detector, and described the data acquired during the first HAYSTAC data run spanning several months in 2016. In this chapter I will provide a detailed pedagogical account of the analysis procedure used to generate an axion CDM exclusion limit from this data. The basic framework of our analysis owes much to the haloscope analysis procedure published by ADMX~\citep{ADMX2001}; in the course of my dissertation research I attempted to generalize and improve on it wherever possible.

This chapter was adapted in large part from Ref.~\citep{PRD2017}. It is the most technical chapter in this thesis, and likely to be the least interesting to any readers who are not engaged in haloscope data analysis. Nonetheless, I think it is important to thoroughly document the HAYSTAC analysis procedure for the benefit of future HAYSTAC students and also students and scientists working on other haloscopes and similar experiments. The chapter is outlined at the end of the overview in Sec.~\ref{sec:analysis_overview}.

\section{Analysis overview}\label{sec:analysis_overview}
\begin{figure}[h]
\centering\includegraphics[width=0.7\textwidth]{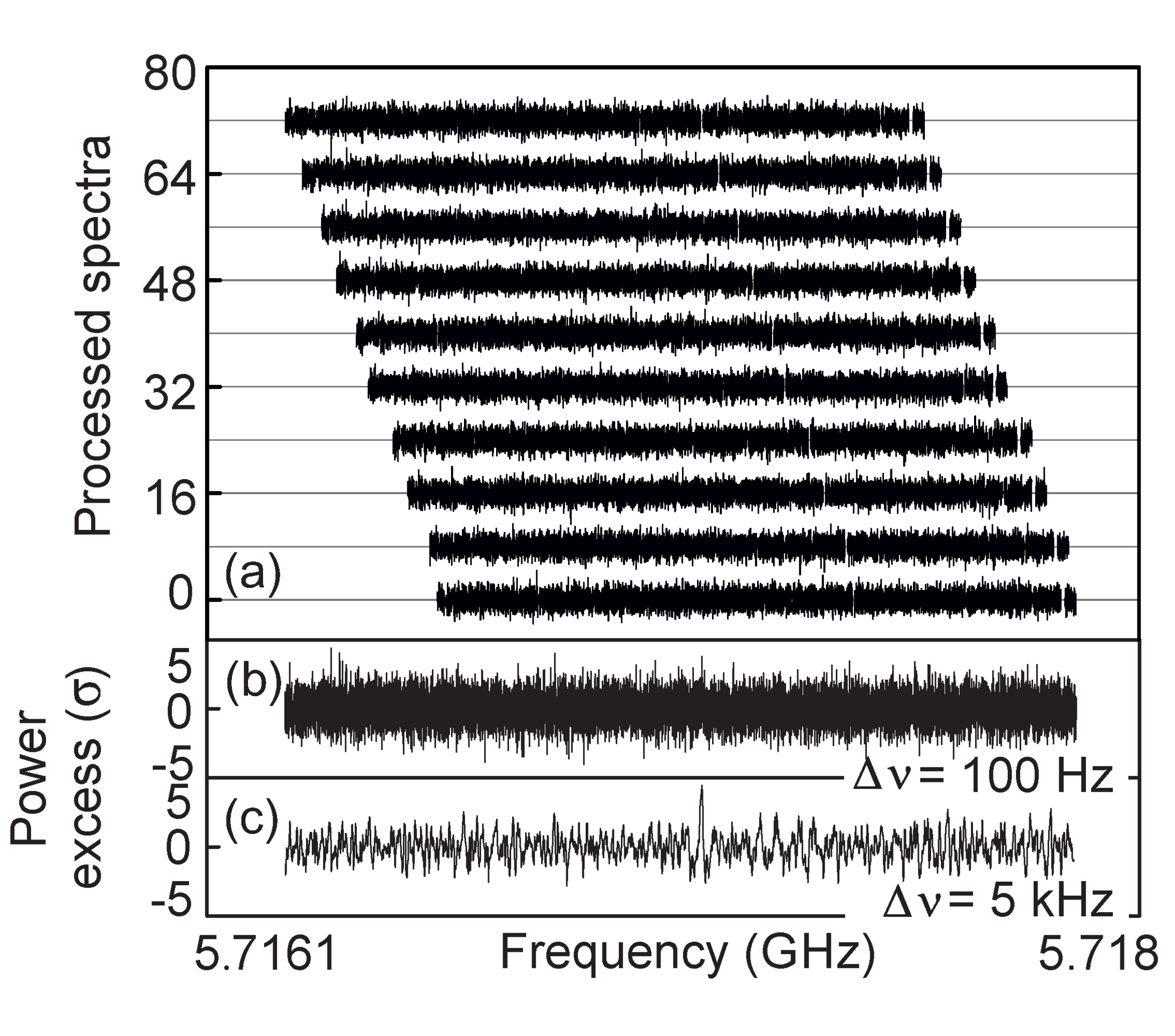}
\caption[A synthetic axion signal in the HAYSTAC data]{\label{fig:fakeaxion} Schematic illustration of how haloscope data is combined to produce a single axion-sensitive spectrum, using real HAYSTAC data around a synthetic axion signal injected into the experiment (see appendix~\ref{app:fake_axions}). \textbf{(a)} 100~Hz-resolution power spectra from individual 15-minute integrations around the injection frequency. \textbf{(b)} The optimally weighted combined spectrum, still at 100 Hz resolution. \textbf{(c)} The grand spectrum obtained by rebinning to 5~kHz with weights that take into account the axion lineshape; the synthetic axion is clearly visible.}
\end{figure}

The goal of a haloscope analysis is to combine a set of overlapping axion-sensitive power spectra to produce a single spectrum that optimizes the SNR throughout the scan range (see Fig.~\ref{fig:fakeaxion}). 

Put another way, if there exists an axion with $\nu_a$ within the scan range and photon coupling $|g_\gamma|$ sufficiently large, the conversion power should almost always result in a large excess relative to noise in the bin corresponding to $\nu_a$ in the final spectrum. The minimum coupling $|g^\text{min}_\gamma|$ for which this statement will hold is set primarily by the detector design, but we must still understand how much the analysis procedure degrades this intrinsic sensitivity. The analysis should ideally allow us to write down an explicit expression for $|g^\text{min}_\gamma|$ as a function of the desired confidence level (which quantifies the ``almost always'' in the informal description above). 

When we consider how best to combine spectra, one issue that immediately arises is that the shape and normalization of each spectrum depend both on quantities that affect the SNR (e.g., $N_\text{sys}$), and quantities that do not (e.g., the net gain of the receiver chain, including the frequency-dependent attenuation of all room-temperature components). Rather than try to tease apart the relevant and irrelevant contributions, we can remove the spectral baseline entirely using a fit or filter, then rescale the resulting spectra using parameters extracted from the auxiliary data from each iteration. In this way we can properly account for variation in sensitivity among spectra and within each spectrum. 

After baseline removal the bins in each spectrum may be regarded as samples drawn from a single Gaussian distribution.\footnote{The spectra are approximately Gaussian because each spectrum saved to disk is the average of a large number of subspectra, so the bin variance is much smaller than the mean squared bin amplitude. This point is discussed further in Sec.~\ref{sub:stats}.} This is a convenient reference point for understanding the effects of subsequent processing on the statistics of the spectra. Of course, we need to make sure that the baseline removal procedure does not fit out bumps in the spectra on frequency scales comparable to $\Delta\nu_a$, or we will significantly degrade the axion search sensitivity. This point suggests that baseline removal is more fruitfully regarded as a problem in filter design than a fitting problem, as it has been described in previous ADMX analyses~\citep{ADMX2001}. The filter perspective will turn out to be quite useful in understanding the statistics of the spectra. 

The task of removing the spectral baseline without appreciably attenuating any axion signal is made tractable by their different characteristic spectral scales, or in other words by $\Delta\nu_c \gg \Delta\nu_a$. As emphasized in Sec.~\ref{sub:radiometer}, this inequality is ultimately the result of the difficulty of achieving high cavity $Q$ factors with normal metals at microwave frequencies: a haloscope detector with sufficiently high $Q$ that $\Delta\nu_c \approx \Delta\nu_a$ would in principle be more sensitive. Since such a detector has yet to be built, we can exploit the fact that $\Delta\nu_c \gg \Delta\nu_a$ where it simplifies the analysis. 

Because our spectra have $\Delta\nu_b \ll \Delta\nu_a$ (see Sec.~\ref{sub:power_spectra}), the analysis procedure will generally involve taking appropriately weighted sums both ``vertically'' (i.e., combining IF bins from different spectra corresponding to the same RF bin) and ``horizontally'' (i.e., combining adjacent bins in the same spectrum). One of the main innovations of our analysis procedure is that we use the same maximum likelihood principle to obtain the optimal weights in both cases. Various statistical subtleties arise in the latter case because nearby bins in the same spectrum can be correlated. I will demonstrate in Sec.~\ref{sub:correlations} that we understand the origin of these correlations sufficiently well to obtain the relationship between  $|g^\text{min}_\gamma|$ and the confidence level from the statistics of the combined data, rather than from Monte Carlo as in the published ADMX analyses~\citep{ADMX2001}.
 
In the preceding paragraphs I have emphasized the main themes of this chapter, which may be helpful to keep in mind as we work through the details. For ease of reference, I have outlined the steps of our procedure below, and indicated the section in which each step is discussed more thoroughly.
\begin{enumerate}
\item Use the auxiliary data to identify spectra that appear to be compromised and cut them from further analysis (Sec.~\ref{sub:badscans}).
\item Average the remaining \textbf{raw spectra} together aligned according to IF frequency to identify compromised IF bins and cut them from further analysis (Sec.~\ref{sub:badbins}). This procedure also yields an estimate of the average shape of the spectral baseline in the analysis band.
\item Normalize the analysis band in each raw spectrum to the average baseline, then use a \textbf{Savitzky-Golay (SG) filter} to remove the remaining spectral structure in each \textbf{normalized spectrum} (Sec.~\ref{sub:sg_filter}). Then subtract 1 from each spectrum to obtain a set of dimensionless \textbf{processed spectra} described by a single Gaussian distribution (Sec.~\ref{sub:stats}).
\item Multiply each processed spectrum by the average noise power per bin and divide by the Lorentzian axion conversion power profile to obtain a set of \textbf{rescaled spectra} (Sec.~\ref{sub:rescale}). Construct a single \textbf{combined spectrum} across the whole scan range by taking an optimally weighted sum of all the rescaled spectra (Sec.~\ref{sub:combine}).
\item Rebin the combined spectrum via a straightforward extension of the optimal weighted sum from the previous step to non-overlapping sets of adjacent combined spectrum bins (Sec.~\ref{sub:rebinned_spectrum}). Then, taking into account the expected axion lineshape (Sec.~\ref{sub:lineshape}), construct the \textbf{grand spectrum} by adding an optimally weighted sum of adjacent bins to each bin in the \textbf{rebinned spectrum} (Sec.~\ref{sub:grand_spectrum}).
\item After correcting for the effects of the SG filter on both the statistics of the grand spectrum (Sec.~\ref{sub:correlations}) and the SNR (Sec.~\ref{sub:axion_atten}), set a threshold $\Theta$ for which some desired fraction of axion signals with a given SNR would result in excess power $>\Theta$. Then flag all bins with excess power larger than $\Theta$ as rescan candidates (Sec.~\ref{sub:target_confidence}).
\item Acquire sufficient data around each rescan candidate to reproduce the sensitivity at that frequency in the original grand spectrum (Sec.~\ref{sub:rescan_daq}). Follow the procedure above, with a few minor differences, to construct a grand spectrum for the rescan data, and determine if any candidate exceeds the corresponding threshold (Sec.~\ref{sub:rescan_analysis}). If no candidate exceeds the second threshold, the corrected SNR obtained in step 6 sets the exclusion limit. Any persistent candidates can be interrogated manually.
\end{enumerate}

\section{Data quality cuts}\label{sec:cuts}
\subsection{Cuts on spectra}\label{sub:badscans}
Our first task is to flag and cut any spectra whose sensitivity to axion conversion we cannot reliably calculate. One way spectra can be compromised is through anomalous drifts in the TM$_{010}$ mode frequency $\nu_c$ (see Sec.~\ref{sub:cav_tuning}) large enough to systematically distort the subsequent weighting of the spectrum by the Lorentzian profile of the cavity mode (see Sec.~\ref{sub:rescale}). Another concern is large JPA bias flux drifts that the feedback system described in Sec.~\ref{sub:feedback} is unable to correct for: the average JPA gain in such iterations is reduced and thus the input-referred noise is systematically higher than what we infer from periodic \textit{in situ} $Y$-factor measurements.

Cutting measurements compromised by mode frequency drift is straightforward, because we make VNA measurements of the cavity mode in transmission both before and after the cavity noise measurement at each iteration during the data run (see Sec.~\ref{sub:daq_procedure}). Our analysis routine fits both measurements to Lorentzians and cuts iterations exceeding the conservative threshold $|\nu_{c1}-\nu_{c2}|>60~\text{kHz}\approx\Delta\nu_c/10$ from subsequent analysis. 

\begin{figure}[h]
\centering\includegraphics[width=0.75\textwidth]{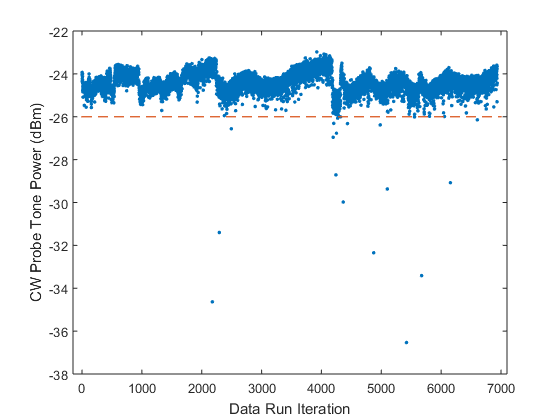}
\caption[Probe tone power vs.\ iteration in the first HAYSTAC data run]{\label{fig:cw_tone} Probe tone power vs.\ iteration $i$ in the first HAYSTAC data run: the blue dots represent the probe tone power measured at the ADCs, and the red dashed line represents the threshold used to flag outliers. Slow variation in the received power encodes frequency-dependence in the net receiver gain. The sharp drop in the average level for $4200 \lesssim i \lesssim 4300$ is due to dependence of the receiver gain exclusive of $G_J$ on the ambient temperature in the lab, which was higher during this period due to a broken ventilation unit. The lab temperature generally fluctuated more during the winter run than the summer run, leading to sharper features for $i\lesssim2200$ than for the second scan over the same range ($2200 \lesssim i \lesssim 4200$).}
\end{figure}

We flag iterations compromised by gain drifts using the spectra themselves. The weak CW probe tone introduced in Sec.~\ref{sub:jpa_fluct} provides one measure of the average JPA gain during the cavity noise measurement which is accessible in the power spectrum; the average level of each spectrum in a 100~kHz window close to $\nu_p$ is another. We set thresholds for both measures of the average JPA gain empirically to separate obvious outliers from the normal variation among spectra (see Fig.~\ref{fig:cw_tone}). In both cases, the thresholds were approximately 1~dB below the typical power averaged across all spectra.\footnote{These thresholds are consistent with independent measurements indicating that flux feedback holds the JPA gain constant to within 10\% on timescales comparable to $\tau$. Small gain fluctuations during a cavity noise measurement will cause the normalization of each 10~ms subspectrum averaged by the \textit{in situ} processing code to differ, but this variation is correlated across all the bins in each subspectrum; it affects the precision with which we can measure the mean noise power, but not the variance of the noise power within each spectrum; see also discussion at the end of Sec.~\ref{sub:radiometer}. At our operating gain, the effect of such small fluctuations on the system noise temperature is small compared to the uncertainty.}

We also scanned the rest of the auxiliary data for any other anomalies that might motivate a cut, and observed a narrow ($\approx60$~kHz) notch around 5.7046~GHz superimposed on measurements of the cavity response in transmission and reflection near this frequency. The absence of any analogous feature in the corresponding JPA gain profiles indicates that the notch originates in the cavity, most likely due to the TM$_{010}$ mode hybridizing with an ``intruder'' TE or TEM mode. The observation that the precise notch frequency depends on the tuning vernier insertion depth supports this interpretation. As a consequence of the ``hybrid'' tuning scheme used in the first HAYSTAC data run (Sec.~\ref{sub:cav_tuning}), the notch frequency appeared to wander over a range of a few hundred kHz. 

We noticed that the notch was also visible in the spectra around the same frequency, which suggests that the effective temperature of the intruder mode was lower than that of the TM$_{010}$ mode (see Sec.~\ref{sub:hotrod}). These measurements collectively indicate that our basic assumption of the axion interacting with a single cavity mode fails around the intruder mode, and neither the VNA measurements of the cavity nor noise calibrations are likely to be reliable here. To be conservative, we simply cut all spectra containing any sign of the intruder mode.

Other auxiliary data, such as JPA-off receiver gain measurements (obtained at each iteration from the JPA gain normalization sweep shown as a dashed line in Fig.~\ref{fig:gain_sample}) and the DR temperature records, did not not prompt us to define additional cuts. Overall, of the 6936 spectra obtained during our first data run, we cut 170 from the subsequent analysis, of which 128 were cut in connection with the intruder mode. Of the remaining 42 spectra, 33 were cut because of JPA gain drifts, and 9 because of mode frequency drifts.

\subsection{Cuts on IF bins}\label{sub:badbins}
Narrowband interference can contaminate individual bins in spectra that are otherwise sensitive to axion conversion: the collective effect of such contaminated bins is to distort the statistics of the spectra. As noted in Sec.~\ref{sub:grounding}, we observed narrowband IF features in HAYSTAC power spectra which we were able to mitigate but not completely eliminate during commissioning.

We flag the ``bad bins'' contaminated by IF interference using the following procedure. First, we divide the set of spectra (ordered chronologically) into three approximately equally sized groups. We truncate each spectrum to the analysis band plus $W=500$~bins (50~kHz) on either side. We then average all truncated spectra within each group aligned according to IF frequency; this averaging reveals many sharp features due to IF interference too small to be visible above the noise floor of individual spectra. We apply an SG filter with polynomial degree $d=10$ and half-width $W$ to the averaged spectrum to obtain an estimate of the spectral baseline. The SG filter is described in more detail in Sec.~\ref{sub:sg_filter}; for our present purposes it is sufficient to regard it as a low-pass filter with a very flat passband (i.e., it perfectly preserves features on large spectral scales).

Dividing the averaged spectrum by the SG filter output and subtracting 1 produces a spectrum whose statistics (in the absence of IF interference) are Gaussian, with mean 0 and standard deviation $\sigma^{\text{IF}}=(N_{\text{IF}}\Delta\nu_b\tau)^{-1/2}$, where $N_{\text{IF}}\approx2200$ is the number of spectra in the group.\footnote{The procedure used here to flag IF interference is similar to the baseline removal procedure described in Sec.~\ref{sec:baseline} with a few key differences. Here the SG filter is applied to a spectrum that is more heavily averaged by a factor of $N_{\text{IF}}$, and we do not divide out the average shape of the spectrum before applying the SG filter. Both effects imply that the polynomial degree $d$ of the SG filter must be higher here than in the main analysis.} The most obvious effect of IF interference is to produce a surplus of bins with large positive power excess. We flag all bins that exceed a threshold value of $\theta^\text{IF}=4.5\sigma^{\text{IF}}$; in the 14020 bins of the truncated spectrum, we expect on average only 0.05~bins exceeding this threshold due to statistical fluctuations. As noted in Sec.~\ref{sub:daq_procedure}, the fact that we do not apply any windowing in the construction of HAYSTAC power spectra implies that the excess power associated with narrowband IF interference will not be entirely confined to isolated bins. To be conservative, for every set of contiguous bins exceeding the threshold, we add the 3 adjacent bins on either side to the list of bad bins. Empirically, while many features due to IF interference are indeed quite sharp, others consist of $\sim30$ consecutive bins exceeding the threshold. Averaging smaller numbers of adjacent spectra reveals that these broader features are the result of narrow IF peaks that wander back and forth across a range of a few kHz over the course of the data run. 

A second, more subtle effect of IF interference is to distort the local estimate of the spectral baseline around any sufficiently large power excess. To mitigate this effect we repeat the process described above iteratively. We remove all flagged bins from the averaged spectrum and apply the SG filter again to obtain a refined baseline estimate; using this improved baseline we generally find some additional bins with values $\geq \theta^\text{IF}$; again 3 bins on either side are also flagged. In practice this procedure takes only 2 or 3 iterations to converge. The output of this iterative process is a list of bad bins within the truncated spectrum for each group of spectra; we also obtain an estimate of the average spectral baseline that we will use in the next stage of the analysis procedure.

The bad bin lists we obtain from our three distinct groups of spectra are quite similar: roughly 75\% of the bins that appear on each list also appear on the other two, and most discrepancies amount to shifting the boundaries of contiguous sets of bad bins. Because the three lists appear to describe IF interference that does not change throughout the run, we combined them into a single final list of bad bins to be cut from every spectrum. Any minimal group of 7 consecutive bins is included in the final list if it appears in two of the three lists and excluded if it appears on only one list. For all other features the final list is the union of the three lists. 11\% of the bins in the analysis band (1456 bins) appear on this final list.

Finally, we also want to flag narrowband interference that would average out in the procedure described above. Thus we set a threshold $\theta^\text{p}$ in each processed spectrum in units of the standard deviation $\sigma^\text{p}$ (Sec.~\ref{sub:stats}). We cannot afford to be as aggressive in cutting such features because Gaussian statistics dictates that roughly 300 bins will exceed 4.5$\sigma^\text{p}$ across all 6766 processed spectra. Thus we set $\theta^\text{p}=6\sigma^\text{p}$, resulting in an additional 0--30 anomalous bins cut from each processed spectrum. The distribution of these bins throughout the spectra implicates temporally intermittent IF interference rather than RF interference.

\section{Removing the spectral baseline}\label{sec:baseline}
\begin{figure}[h]
\centering\includegraphics[width=1.0\textwidth]{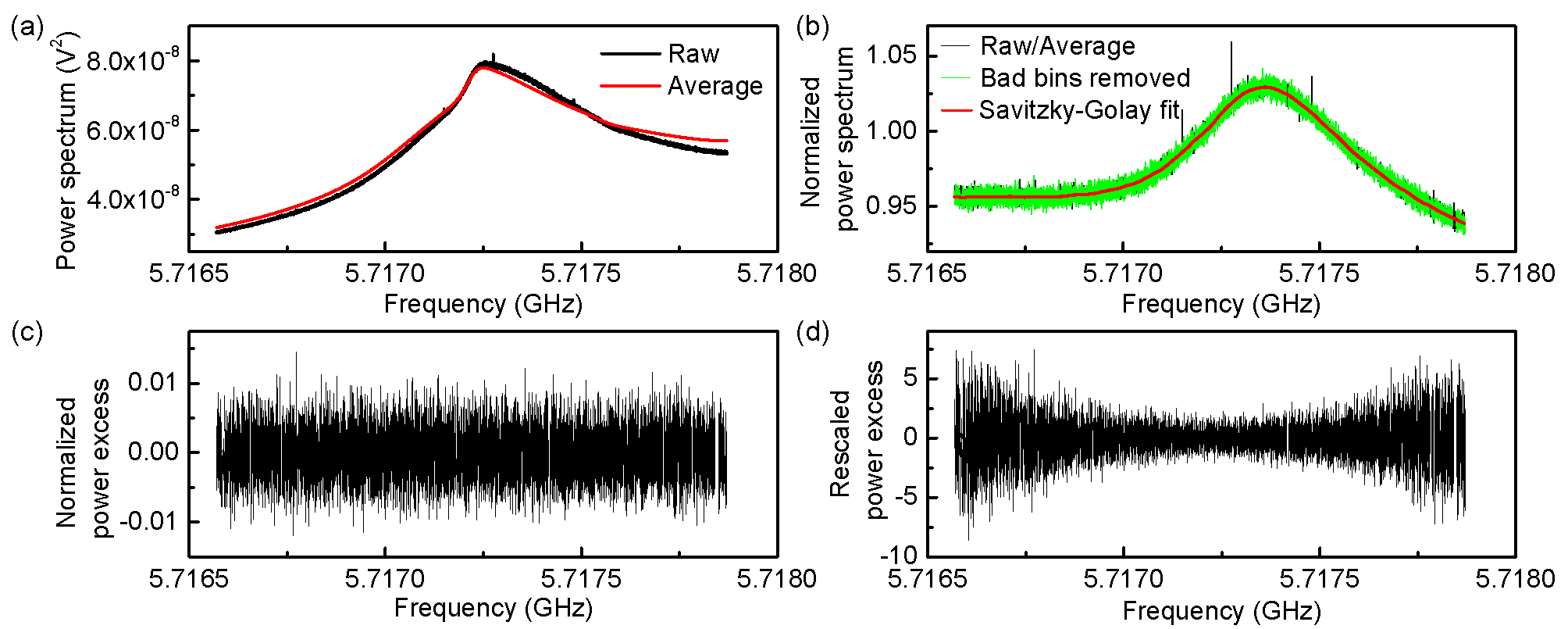}
\caption[A representative power spectrum at various stages of processing]{\label{fig:sample} The analysis band in a representative power spectrum at various stages of processing. (a) Black: the raw spectrum, whose shape is determined by the cavity noise spectrum and the net gain of the receiver. Red: the average baseline, which is the output of a Savitzky-Golay filter applied to the average of many such spectra (Sec.~\ref{sub:badbins}). (b) Black: the normalized spectrum obtained by dividing the raw spectrum by the average baseline. Green: the same spectrum after removing bins contaminated by IF interference. Red: the Savitzky-Golay fit to the residual baseline of this spectrum (Sec.~\ref{sub:sg_filter}). (c) The processed spectrum obtained by dividing the normalized spectrum by the Savitzky-Golay fit and subtracting 1 (Sec.~\ref{sub:stats}). Gaps in the spectrum are the result of removing the contaminated bins. (d) The rescaled spectrum obtained by multiplying the processed spectrum by $h\nu_{c}N_\text{sys}\Delta\nu_b/P_\text{sig}$, where the system noise $N_\text{sys}$ and signal power $P_\text{sig}$ each depend on both the mode frequency $\nu_c$ and on the IF frequency within each spectrum (Sec.~\ref{sub:rescale}). The combined spectrum is given by a maximum-likelihood weighted sum of the complete set of rescaled spectra.}
\end{figure}

A typical raw power spectrum from the HAYSTAC detector, truncated to the analysis band, is shown in black in Fig.~\ref{fig:sample}(a). As emphasized in Sec.~\ref{sec:analysis_overview}, the spectral baseline is in principle the product of $N_\text{sys}$ (which affects the sensitivity of the axion search) and the net gain of the receiver (which does not). On large spectral scales the shape of the baseline is mainly due to three effects. Rolloff at the low-RF (high-IF; see Fig.~\ref{fig:IF_layout}) end of the spectrum is due to room-temperature IF components, rolloff on the high-RF side comes from the JPA gain profile, and the intermediate region around the cavity resonance is enhanced by the heightened temperature of the tuning rod (Sec.~\ref{sub:hotrod}). We see that there can be as much as $\sim4$~dB variation in the ``gain'' within a single spectrum.

An average baseline obtained via the process described in Sec.~\ref{sub:badbins} is shown in red in Fig.~\ref{fig:sample}(a). Systematic deviations of the raw spectrum from the average baseline indicate that the spectral baseline can change from one iteration to the next. Such variation is not surprising, as the JPA is a narrowband amplifier for which gain fluctuations imply bandwidth fluctuations. The shape of the ``hot rod'' excess also depends on frequency-dependent parameters of the cavity mode (Sec.~\ref{sub:hotrod}), and there may be many other effects that can cause the spectral baseline to vary.

Nonetheless, normalizing each raw spectrum to the average baseline does reduce the typical variation across each spectrum from $\sim4$~dB to $\sim0.5$~dB; the normalized spectrum (which is now dimensionless) is shown in black in Fig.~\ref{fig:sample}(b). At this point we also remove all the bins compromised by IF interference from each spectrum. The normalized spectrum with bad bins removed is shown in green in Fig.~\ref{fig:sample}(b). Although only the analysis band is shown in Fig.~\ref{fig:sample}, we actually apply the above steps to the analysis band plus 500 bins on either side. These extra bins essentially serve as buffer regions for the SG filter that we now employ to remove the residual baseline of each spectrum.

\subsection{The Savitzky-Golay filter}\label{sub:sg_filter}
The simplest way to understand the SG filter is as a polynomial generalization of a moving average characterized by two parameters $d$ and $W$. For each point $x_0$ in the input sequence (assumed to be much longer than $W$), we fit a polynomial of degree $d$ in a $(2W+1)$-point window centered on $x_0$. The value of the SG filter output at $x_0$ is defined to be the least-squares-optimal polynomial evaluated at the center of the window, and this process is repeated for each $x_0$; thus the filter output is a smoothed version of the input sequence, with edge effects within $W$ points of either end.

Savitzky and Golay~\citep{sg1964} showed that this moving polynomial fit is equivalent to a discrete convolution of the input sequence with an impulse response that depends only on $d$ and $W$. This correspondence implies that we can fruitfully think about least-squares-smoothing from the perspective of filtering rather than fitting. The even symmetry of the SG filter impulse response implies that only even values of $d$ generate unique filters. We can gain further insight into the properties of SG filters by considering their performance in the frequency domain~\citep{schafer2011}. In the haloscope analysis considered here, we convolve the SG filter impulse response with an input sequence which is itself a power spectrum. Describing the Fourier transform of the SG impulse response as the filter's ``frequency response'' may thus be misleading; we will instead refer to this Fourier transform as a transfer function in the ``inverse bin domain.''

\begin{figure}[h]
\centering\includegraphics[width=0.7\textwidth]{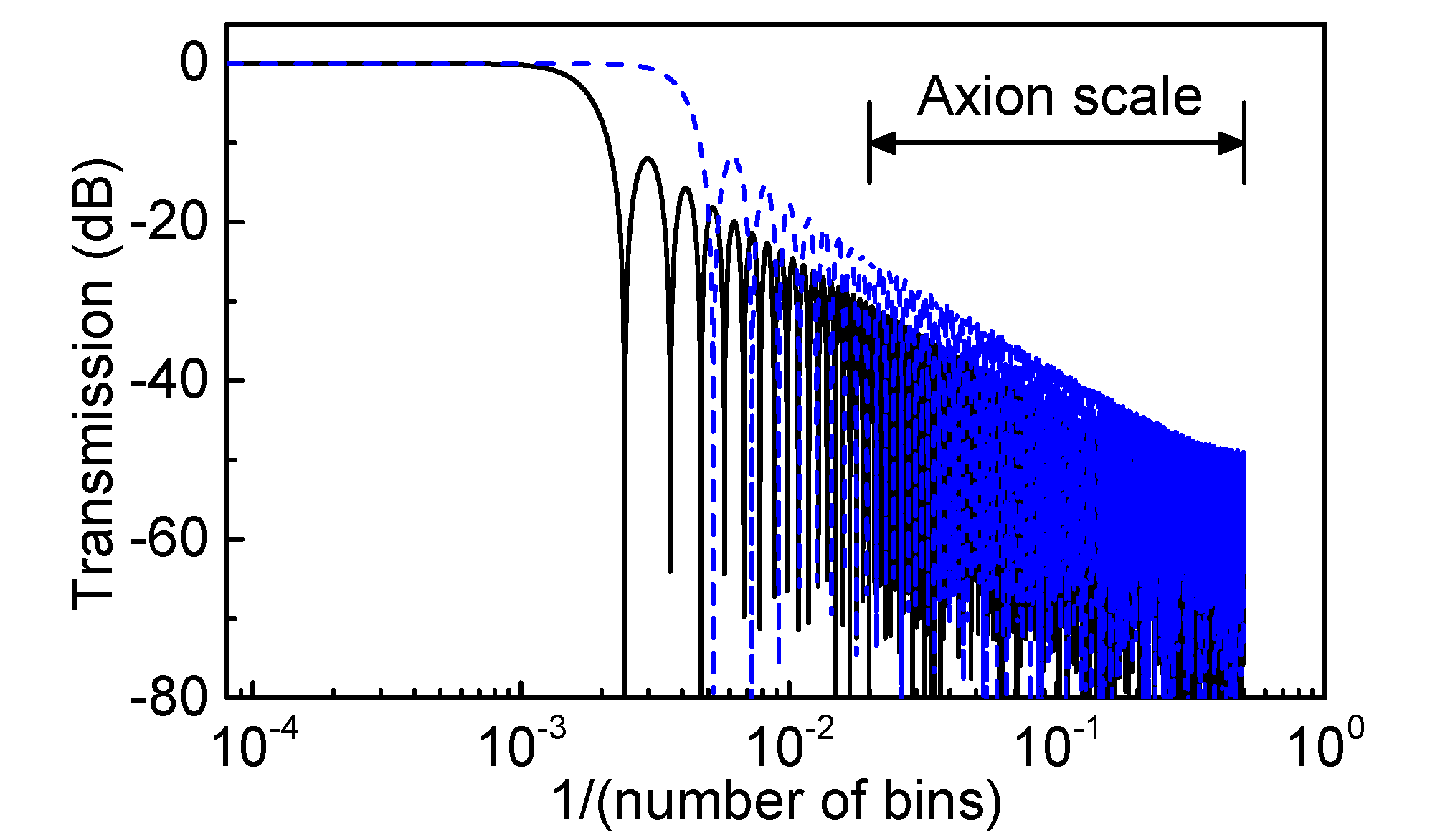}
\caption[Savitzky-Golay filter transfer functions]{\label{fig:filter} Transfer functions of the Savitzky-Golay filters used in our analysis. The solid black curve depicts the filter with $W=500$ and $d=4$ used in the initial scan analysis; the dashed blue curve depicts the filter with $W=300$ and $d=6$ used in the rescan analysis (Sec.~\ref{sec:rescan}). We exploit the very flat passband of the filter on large spectral scales for baseline removal. The behavior of the filter on small spectral scales of 2-50 bins determines its effects on the axion signal and the statistics of the grand spectrum.}
\end{figure}

Two SG filter transfer functions used in the HAYSTAC analysis are plotted in Fig.~\ref{fig:filter}. In general, SG filters are low-pass filters with extremely flat passbands and mediocre stopband attenuation. The 3~dB point that marks the transition between these two regions scales approximately linearly with $d$ and approximately inversely with $W$. In particular, the 3~dB point for an SG filter with $d=4$ and $W=500$ (black solid line in Fig.~\ref{fig:filter}) is $\approx 1/(517~\text{bins})$. Thus when this filter is applied to one of the normalized spectra discussed above, features in the residual baseline on spectral scales sufficiently large compared to 51.7~kHz will be essentially perfectly preserved in the filter output, and features on smaller spectral scales are suppressed to varying degrees. The output of the SG filter applied to the normalized spectrum in Fig.~\ref{fig:sample}(b) is shown in red on the same plot. After dividing each normalized spectrum by the corresponding SG filter output to remove the residual baseline, we can discard the 500 bins at either edge of each spectrum, whose only purpose has been to keep edge effects out of the analysis band; all subsequent processing is applied to the analysis band of each spectrum only.

The design of any digital filter involves some tradeoff between passband and stopband performance, and we have seen that SG filters generally sacrifice some stopband attenuation to optimize passband flatness. It remains to be shown that this is the correct choice for a haloscope analysis. To see this, note that appreciable passband ripple implies the presence of systematic structure on large scales in the processed spectra. Such structure in turn implies that we cannot assume all processed spectrum bins are samples drawn from the same Gaussian distribution (see Sec.~\ref{sub:stats}); thus we cannot construct a properly normalized measure of excess power in an arbitrary IF bin, which is a central assumption of the rest of the analysis. 

Imperfect stopband attenuation, on the other hand, implies that features and fluctuations on small spectral scales are slightly suppressed when we divide each normalized spectrum by the SG filter output; equivalently, the SG filter slightly attenuates axion signals and imprints small negative correlations between processed spectrum bins. We will show that we can quantify both the filter-induced signal attenuation (Sec.~\ref{sub:axion_atten}) and the effects of correlations on the statistics of the grand spectrum in which we ultimately conduct our axion search (Sec.~\ref{sub:correlations}). Computing the axion search sensitivity directly from the statistics of the spectra requires a thorough understanding of both effects.\footnote{The application of SG filters to spectral baseline removal in a haloscope search was first explored by Ref.~\citep{malagon2014}, which did not however adopt the frequency-domain approach used here or consider the effects of filter-induced correlations.}

The above discussion implies that passband flatness is a more important consideration than stopband attenuation for estimating spectral baselines in a haloscope analysis, and thus the SG filter is a good choice.\footnote{It is possible that a different filter may be able to achieve better attenuation than the SG filter in the relevant part of the stopband while retaining the requisite passband flatness. HAYSTAC is currently exploring the application of other FIR and IIR digital filters to baseline removal in the haloscope search analysis.} Acceptable values of the filter parameters $d$ and $W$ are constrained by the integration time $\tau$ at each tuning step. Increasing $\tau$ would make us sensitive to smaller-amplitude systematic structure in the baseline on smaller spectral scales, and we must push the 3~dB point of the SG filter up towards smaller scales to ensure that this structure remains confined to the passband (see appendix~\ref{app:sg_params} for a more detailed discussion). We will see in Sec.~\ref{sub:rescan_analysis} that different values of $d$ and $W$ are appropriate for the analysis of rescan data.

\subsection{Statistics of the processed spectra}\label{sub:stats}
At each data run iteration, the total noise referred to the receiver input is statistically equivalent to thermal noise at some effective (possibly frequency-dependent) temperature; thus the noise voltage distribution is Gaussian, and the fluctuations in each Nyquist-resolution subspectrum will have a $\chi^2$ distribution of degree 2. During data acquisition we average $\Delta\nu_b\tau = 9\times10^4$ such subspectra together, so the noise power fluctuations about the slowly varying baseline of each raw spectrum will be Gaussian by the central limit theorem. 

The baseline removal procedure described above should thus yield a set of flat dimensionless spectra, each with small Gaussian fluctuations about a mean value of 1. Ultimately, we are interested in \textit{excess power} (which may be positive or negative) relative to the average noise power in each bin, so we subtract 1 from each spectrum after dividing out the SG filter output. We refer to the set of spectra obtained this way as the processed spectra; a representative processed spectrum is shown in Fig~\ref{fig:sample}(c). 

\begin{figure}[h]
\centering\includegraphics[width=1.0\textwidth]{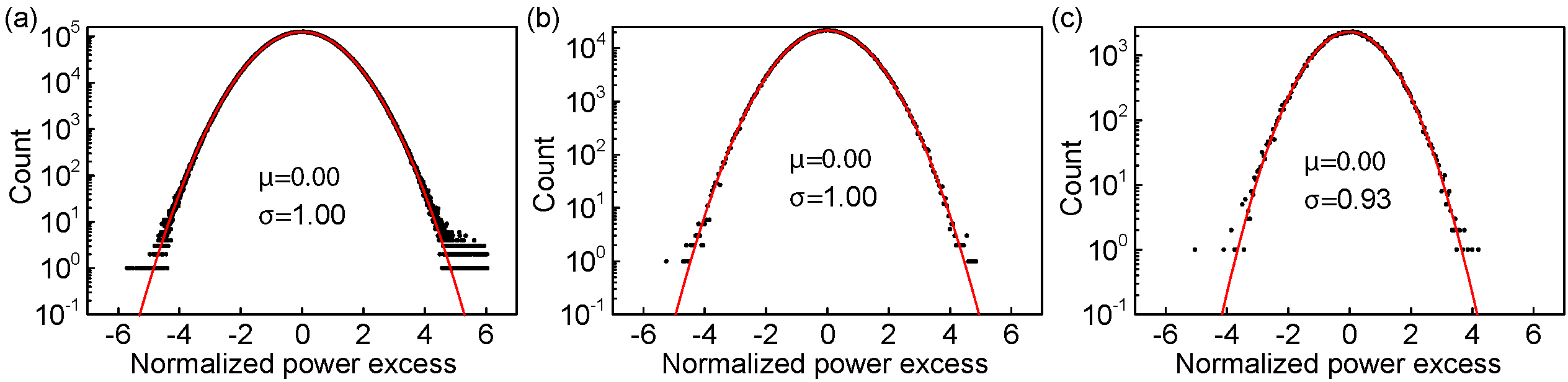}
\caption[Histograms of power spectra at various stages of processing]{\label{fig:data_hist} Histograms of HAYSTAC power spectra at various stages of the processing, with each bin in each spectrum normalized to its expected standard deviation. In each plot, the histogram (black circles) is fit with a Gaussian (red curve), and the mean $\mu$ and standard deviation $\sigma$ obtained from the fit are displayed. (a) Histogram of all bins $\delta^\text{p}_{ij}/\sigma^\text{p}$ from all processed spectra (Sec.~\ref{sub:stats}). There is a surplus of bins at large positive excess power (with a cutoff at $\theta^\text{p}=6\sigma^\text{p}$) due to narrowband IF interference (Sec.~\ref{sub:badbins}). Otherwise, the distribution of bins is Gaussian with the expected standard deviation. (b) Histogram of all combined spectrum bins $\delta^\text{c}_k/\sigma^\text{c}_k$ (Sec.~\ref{sub:combine}), demonstrating Gaussian statistics with the expected standard deviation. (c) Histogram of all grand spectrum bins $\delta^\text{g}_\ell/\sigma^\text{g}_\ell$ (Sec.~\ref{sub:grand_spectrum}). The statistics of the spectrum are still Gaussian, but the standard deviation is reduced by a factor $\xi=0.93$ due to small-scale correlations ultimately traceable to the imperfect SG filter stopband attenuation (Sec.~\ref{sub:correlations}).}
\end{figure}

In the absence of axion conversion, the bins in each processed spectrum should be samples drawn from a single Gaussian distribution with mean $\mu^\text{p}=0$ and standard deviation $\sigma^\text{p}=1/\sqrt{\Delta\nu_b\tau}=3.3\times10^{-3}$. In Fig.~\ref{fig:data_hist}(a) we have histogrammed all IF bins from all processed spectra together in units of $\sigma^\text{p}$. The excess power distribution is indeed Gaussian out to $\approx5\sigma$, and the excess above $5\sigma$ is likely due to intermittent IF interference slightly too small to exceed the threshold $\theta^\text{p}$ (Sec.~\ref{sub:badbins}). These large single-bin power excesses will be significantly diluted when we combine and rebin spectra.

Fig.~\ref{fig:data_hist}(a) indicates that each bin in each processed spectrum may be regarded as a random variable drawn from the same Gaussian distribution, and this is an important check on our baseline removal procedure. It does not follow that each spectrum is a sample of Gaussian white noise, because nearby bins in each spectrum will be correlated due to the imperfect stopband attenuation of the SG filter. 

We can observe effects of these correlations if we regard each spectrum (rather than each bin) as a sample of the same Gaussian process. Let $\delta^\text{p}_{ij}$ represent the value of the $j$th IF bin in the $i$th processed spectrum, for $i=1,\dots,N$ and $j=1,\dots,J$; $N=6766$ and $J=11564$ for the first HAYSTAC run after the cuts discussed in Sec.~\ref{sec:cuts}. The $i$th processed spectrum has sample mean 
\begin{equation}\label{eq:mu_p}
\mu^\text{p}_i = \frac{1}{J}\sum_j\delta^\text{p}_{ij}
\end{equation}
and sample variance
\begin{equation}
\left(\sigma^\text{p}_i\right)^2 = \frac{1}{J-1}\sum_j\left(\delta^\text{p}_{ij}-\mu^\text{p}_i\right)^2.
\label{eq:sigma_p}
\end{equation}
In the absence of correlations, the set of sample means should be Gaussian distributed about $\mu^\text{p}$ with standard deviation $\sigma_\mu = \sigma^\text{p}/\sqrt{J}$, and the set of sample variances should be Gaussian distributed about $(\sigma^\text{p})^2$ with standard deviation $\sigma_{\sigma^2} = \sqrt{2/(J-1)}(\sigma^\text{p})^2$, again by the central limit theorem. The presence of negative correlations on small spectral scales will reduce $\sigma_\mu$ substantially and also increase $\sigma_{\sigma^2}$ slightly, without appreciably changing the mean value of either distribution. Empirically, we find that $\sigma_\mu$ is smaller than the above estimate by an order of magnitude, and $\sigma_{\sigma^2}$ is larger by about 8\%. 

The distortions of the sample mean and variance distributions noted above do not themselves affect the axion search sensitivity. But the correlations responsible for them are still important, since the remainder of our analysis procedure will involve taking both horizontal and vertical weighted sums of processed spectrum bins. A weighted sum of any number of \textit{independent} Gaussian random variables is another Gaussian random variable, with mean given by the weighted sum of component means, and standard deviation given by the quadrature weighted sum of component standard deviations. If instead the random variables are jointly normal but \textit{correlated}, the sum is still Gaussian and has the same mean, but computing the variance of the sum requires knowledge of the full covariance matrix. We will return to this point in Sec.~\ref{sub:correlations}.

\section{Combining spectra vertically}\label{sec:rescale_combine}
The $NJ$ processed spectrum bins $\delta^\text{p}_{ij}$ correspond to $M < NJ$ unique RF bins ($M\approx1.07\times10^6$ for the first HAYSTAC data run). For notational convenience we define the symbol $\Gamma_{ijk}=1$ if the $j$th IF bin in the $i$th spectrum is one of the $m_k$ bins corresponding to the $k$th RF frequency ($\Gamma_{ijk}=0$ otherwise). Our next task is to construct a single combined spectrum by taking an optimally weighted vertical sum of all $m_k$ IF bins corresponding to each RF bin $k$. The $m_k$ bins in each sum will be statistically independent, since each processed spectrum contains at most one IF bin to corresponding to any given RF bin $k$.

To gain insight into the form of the optimally weighted sum, let us consider the simple case where all axion conversion power is confined to a single RF bin $k'$. Then each processed spectrum bin with $\Gamma_{ijk'}=1$ may be regarded as a sample from a Gaussian distribution whose mean is nonzero. We will initially assume that all of these bins have the same mean $\mu_{k'}=1$ but possibly different standard deviations; of course, all bins with $\Gamma_{ijk'}=0$ also share a mean value, namely 0.  

This assumption allows us to formulate the requirement for an optimally weighted vertical sum more precisely: for each $k$ we will choose weights that yield the \textbf{maximum likelihood (ML)} estimate of the true mean value $\mu_k$ shared by all the contributing bins. ML estimation is briefly summarized in appendix~\ref{app:mle}. In Sec.~\ref{sub:combine} we will see that ML weighting maximizes the SNR among all choices that yield unbiased estimates of the power excess.

In reality, the sensitivity of any given processed spectrum bin to axion conversion depends on both $i$ and $j$, so each of the bins with $\Gamma_{ijk'}=1$ is actually a Gaussian random variable with a \textit{different} nonzero mean. Moreover, we saw in Sec.~\ref{sub:stats} that each bin in each processed spectrum has the \textit{same} standard deviation $\sigma^\text{p}$ -- we did not consider axion signals when discussing the statistics of the processed spectra, but we should expect the fluctuations of the noise power to be independent of the presence or absence of axion conversion power. 

Evidently the assumption we used above to motivate the ML-weighted vertical sum was precisely backwards. We can cast the problem into a form amenable to ML weighting by rescaling the processed spectra so that axion conversion would yield the same mean power excess in any rescaled spectrum bin. Determining the appropriate rescaling factor is the subject of the next section. After rescaling the spectra, we can meaningfully define ML weights and thus construct the combined spectrum.

\subsection{The rescaling procedure}\label{sub:rescale}
We rescale the processed spectra by multiplying each spectrum by the mean noise power per bin and dividing by the signal power. The $j$th bin in the $i$th rescaled spectrum is then
\begin{equation}
\delta^\text{s}_{ij} = \frac{h\nu_{ci}N_{ij}\Delta\nu_b\delta^\text{p}_{ij}}{P_{ij}},
\label{eq:delta_s}
\end{equation}
where $\nu_{ci}$ is the $\text{TM}_{010}$ mode frequency in the $i$th spectrum, $N_{ij}$ is the total noise $N_\text{sys}$ with its $i$ and $j$ dependence made explicit, and $P_{ij}$ is the total conversion power $P_\text{sig}$ we would obtain from a KSVZ axion signal confined to the $j$th bin of the $i$th spectrum.\footnote{By assuming a KSVZ axion signal confined to a single 100~Hz bin here we are just choosing a simple but physically implausible normalization for $P_{ij}$. The exclusion limit which is the final product of our analysis will not depend on this arbitrary normalization.}

It may be helpful to discuss qualitatively why Eq.~\eqref{eq:delta_s} is the appropriate form for the rescaling factor. An axion signal with any given conversion power will be relatively suppressed by baseline removal if it happens to appear in a noisier spectrum or a noisier region of a given spectrum; multiplying by $N_{ij}$ undoes this suppression. Dividing by the signal power undoes the relative suppression of conversion power in spectra that are less sensitive overall due to implicit frequency dependence in Eq.~\eqref{eq:signal_power}, and undoes the Lorentzian suppression of the conversion power for axions at nonzero detuning $\delta\nu_a$ from $\nu_c$.

The net result is that in the absence of noise, the hypothetical single-bin axion signal we have considered will yield $\delta^\text{s}_{ij}=1$ in each bin with $\Gamma_{ijk'}=1$. In the presence of noise, each of these bins is a Gaussian random variable with mean $\mu^\text{s}_{ij}=1$ and every other bin is a Gaussian random variable with $\mu^\text{s}_{ij}=0$. The rescaled spectra are no longer flat: each bin has a standard deviation
\begin{equation}
\sigma^\text{s}_{ij} = \frac{h\nu_{ci}N_{ij}\Delta\nu_b\sigma^\text{p}_{i}}{P_{ij}}.
\label{eq:sigma_s}
\end{equation}
Note that $\sigma^\text{s}_{ij} = (R^\text{\,s}_{ij})^{-1}$, where $R^\text{\,s}_{ij}$ is the SNR for our hypothetical single-bin axion signal [c.f.\ Eq.~\eqref{eq:dicke}]; this is just another way of saying that an axion signal in any bin of any rescaled spectrum produces a mean power excess of 1. A representative rescaled spectrum is shown in Fig.~\ref{fig:sample}(d). Its overall shape is mainly due to the Lorentzian mode profile.

We obtain an explicit expression for $P_{ij}$ by making explicit and discretizing the frequency-dependence in Eq.~\eqref{eq:signal_power}:
\begin{equation}\label{eq:power_ij}
P_{ij} = U_0\left(\nu_{ci}\frac{\beta_i}{1+\beta_i}C_i\frac{Q_{Li}}{1+\left[2(\nu_{ij}-\nu_{ci})/\Delta\nu_{ci}\right]^2}\right),
\end{equation}
where 
\begin{equation}\label{eq:U_0}
U_0=g_\gamma^2\frac{\alpha^2}{\pi^2}\frac{\hbar^3c^3\rho_a}{\chi}\frac{2\pi}{\mu_0}\eta_0B_0^2V
\end{equation}
is a constant with dimensions of energy. 

The factors we have absorbed into the definition of $U_0$ are independent of both $i$ and $j$ and thus only affect the overall normalization of the rescaled spectra.\footnote{Of course $B_0$ can change in principle, but in practice the magnet is extremely stable in persistent mode.} All of these factors are defined in the derivation of Eq.~\eqref{eq:signal_power} in chapter~\ref{chap:search}, with the exception of $\eta_0$, which is the signal attenuation factor due to loss between the cavity and microwave switch, introduced in Sec.~\ref{sub:yfactor}. We need to temporarily fix $\abs{g_\gamma}$ to set a definite normalization for the rescaled spectrum; as noted above we took $|g_\gamma|=|g^\text{KSVZ}_\gamma| = 0.97$, corresponding to the standard KSVZ model (Sec.~\ref{sub:model_band}).

The remaining factors in Eq.~\eqref{eq:power_ij} can vary as the mode is tuned; $\nu_{ij}$ is the RF frequency of the $j^\text{th}$ bin in the $i^\text{th}$ spectrum. We average the two cavity transmission measurements from each iteration (Sec.~\ref{sub:daq_procedure}) and fit the average to a Lorentzian (Sec.~\ref{sub:cav_meas}) to obtain $\nu_{ci}$ and $Q_{Li}$ (and thus $\Delta\nu_{ci}$). From the reflection measurement at each iteration we extract the coupling $\beta_i$ using the procedure described in Sec.~\ref{sub:cav_meas}. The frequency-dependence of the form factor $C_i$ is obtained from simulation (Fig.~\ref{fig:c010}). 

Next we can consider contributions to $N_{ij}$. Using Eq.~\eqref{eq:system_noise} and the notation established in Sec.~\ref{sec:noise}, we may write
\begin{equation}
N_{ij} = N_\text{mc} + (\Delta N_\text{cav})_{ij} + (N_A)_{ij}.
\label{eq:noise_ij}
\end{equation}
As noted in Sec.~\ref{sub:hotrod}, we made $Y$-factor analysis in the presence of thermal disequilibrium ($\Delta N_\text{cav}\neq0$) tractable by assuming $N_A$ in each spectrum is given by the average of off-resonance measurements during commissioning [i.e., $(N_A)_{ij}=(\bar{N}_A)_j$ for each $i$]. We expect $\Delta N_\text{cav}$ to exhibit $i$-dependence due to variation in $Q_i$ and $\beta_i$ in addition to its roughly Lorentzian $j$-dependence.\footnote{Technically, $N_\text{mc}$ is a function of frequency evaluated at $\nu_{ci}$, but it changes negligibly over the scan range, so we suppress its $i$-dependence; $j$-dependence due to the finite analysis band width is of course much smaller still.} Moreover, the effective temperature of the cavity mode is determined by a competition between the walls, which are well coupled to the mixing chamber, and the thermally isolated rod; the relative strength of these contributions could depend on the shape of the cavity mode and thus on $\nu_{ci}$. 

Empirically, the $\Delta N_\text{cav}$ profiles obtained from nearby $Y$-factor measurements were clearly correlated, but there was no deterministic frequency dependence strong enough to justify any particular interpolation scheme. Thus, we simply set $N_{ij}$ for each spectrum at which we did not make a $Y$-factor measurement using the nearest measured value of $\Delta N_\text{cav}$. In appendix~\ref{app:error} we estimate the uncertainty in our exclusion limit resulting from possible miscalibration of the noise temperature.

\subsection{Constructing the combined spectrum}\label{sub:combine}
I have shown that the rescaled spectrum IF bins corresponding to each RF bin are independent Gaussian random variables with the same mean (1 in the presence of a single-bin KSVZ axion and 0 in the absence of a signal) and different variances. To obtain the ML estimate of this mean value (see appendix~\ref{app:mle}) we weight each bin by its inverse variance:
\begin{equation}
w_{ijk} = \frac{\Gamma_{ijk}(\sigma^\text{s}_{ij})^{-2}}{\sum_{i'}\sum_{j'}\Gamma_{i'j'k}(\sigma^\text{s}_{i'j'})^{-2}},
\label{eq:weights}
\end{equation}
where the denominator ensures that the weights are normalized.\footnote{Many of the expressions to follow contain sums over $i$ and $j$ in both the numerator and denominator. I will avoid cumbersome primes through slight abuse of notation by using the same indices $i$ and $j$ in both sums. $k$, which is not summed over, is understood to have the same value in the numerator and denominator. Sums whose upper and lower limits are elided are to be interpreted as running over all possible values of the index.} Then the ML estimate of the mean in each combined spectrum bin $k$ is given by the weighted sum of contributing bins: 
\begin{eqnarray}
\delta^\text{c}_k &&= \sum_i\sum_jw_{ijk}\delta^\text{s}_{ij} \nonumber\\
&&= \frac{\sum_i\sum_j\Gamma_{ijk}\left(P_{ij}/h\nu_{ci}N_{ij}\Delta\nu_b\sigma^\text{p}_{i}\right)^2\left(h\nu_{ci}N_{ij}\Delta\nu_b\delta^\text{p}_{ij}/P_{ij}\right)}{\sum_i\sum_j\Gamma_{ijk}\left(P_{ij}/h\nu_{ci}N_{ij}\Delta\nu_b\sigma^\text{p}_{i}\right)^2} \nonumber\\
\Rightarrow \delta^\text{c}_k &&= \frac{\sum_i\sum_j\Gamma_{ijk}\left(P_{ij}\delta^\text{p}_{ij}/h\nu_{ci}N_{ij}\Delta\nu_b(\sigma^\text{p}_{i})^2\right)}{\sum_i\sum_j\Gamma_{ijk}\left(P_{ij}/h\nu_{ci}N_{ij}\Delta\nu_b\sigma^\text{p}_{i}\right)^2}\label{eq:delta_c}.
\end{eqnarray}
The standard deviation of each bin in the combined spectrum is the quadrature weighted sum of contributing standard deviations:
\begin{eqnarray}
\sigma^\text{c}_k &&= \sqrt{\sum_i\sum_jw_{ijk}^2\left(\sigma^\text{s}_{ij}\right)^2} \nonumber\\
&&= \sqrt{\frac{\sum_i\sum_j\Gamma_{ijk}\left(\sigma^\text{s}_{ij}\right)^{-4}\left(\sigma^\text{s}_{ij}\right)^2}{\left[\sum_{i}\sum_{j}\Gamma_{ijk}(\sigma^\text{s}_{ij})^{-2}\right]^2}} \nonumber\\
&&= \frac{\sqrt{\sum_i\sum_j\Gamma_{ijk}\left(\sigma^\text{s}_{ij}\right)^{-2}}}{\sum_{i}\sum_{j}\Gamma_{ijk}(\sigma^\text{s}_{ij})^{-2}} \nonumber\\
\Rightarrow \sigma^\text{c}_k &&= \left[\sum_i\sum_j\Gamma_{ijk}\left(\frac{P_{ij}}{h\nu_{ci}N_{ij}\Delta\nu_b\sigma^\text{p}_{i}}\right)^2\right]^{-1/2}.\label{eq:sigma_c}
\end{eqnarray}

For each $k$, there are $m_k$ nonvanishing contributions to the sums in the expressions above. In the first HAYSTAC data run, typical values of $m_k$ ranged from 50 to 120 across the combined spectrum due to nonuniform tuning.\footnote{There are two $\sim$ MHz-width peaks in the distribution of $m_k$ with peak values of 150 and 200, corresponding to the flat regions in Fig.~\ref{fig:tx_f0_data}. $m_k$ also drops precipitously around the frequency of the intruder mode where we cut spectra (Sec.~\ref{sub:badscans}) and at the edges of the scan range. On spectral scales small compared to the analysis band width, $m_k$ fluctuates by $\pm2$ due to the presence of missing bins in the processed spectra.}

Two numbers are required to characterize the combined spectrum at each frequency: $\delta^\text{c}_k$ and $\sigma^\text{c}_k$ describe respectively the actual power excess in each combined spectrum bin and the power excess we expect from statistical fluctuations. Absent any axion signals, each $\delta^\text{c}_k$ should be a Gaussian random variable drawn from a distribution with mean $\mu^\text{c}_k = 0$ and standard deviation $\sigma^\text{c}_k$. Thus the distribution of normalized bins
\begin{equation}
\frac{\delta^\text{c}_k}{\sigma^\text{c}_k} = \frac{\sum_i\sum_j\Gamma_{ijk}\left(P_{ij}\delta^\text{p}_{ij}/h\nu_{ci}N_{ij}\Delta\nu_b(\sigma^\text{p}_{i})^2\right)}{\sqrt{\sum_i\sum_j\Gamma_{ijk}\left(P_{ij}/h\nu_{ci}N_{ij}\Delta\nu_b\sigma^\text{p}_{i}\right)^2}}
\label{eq:ds_c}
\end{equation}
should be Gaussian with zero mean and unit variance; we can see in Fig.~\ref{fig:data_hist}(b) that this is indeed the case.\footnote{In practice $\delta^\text{c}_k/\sigma^\text{c}_k$ will still appear to have a standard normal distribution even in the presence of axion conversion, since $\mu^\text{c}_k\neq0$ in only a few bins.}

We can equivalently describe the combined spectrum by specifying the values of $\delta^\text{c}_k/\sigma^\text{c}_k$ and $R^\text{\,c}_k = \left(\sigma^\text{c}_k\right)^{-1}$ for each $k$. The normalization of the ML weights implies that, for a single-bin KSVZ axion at frequency $k'$, $\mu^\text{c}_{k'}=1$ and thus $E\big[\delta^\text{c}_{k'}/\sigma^\text{c}_{k'}\big]=R^\text{\,c}_{k'}$. Physically, $R^\text{\,c}_k$ is the SNR that a single-bin KSVZ axion \textit{would have} in the $k$th bin of the combined spectrum (whether or not such an axion exists). In terms of the SNR, eq.~\eqref{eq:sigma_c} becomes
\begin{equation}
R^\text{\,c}_k = \sqrt{\sum_i\sum_j\Gamma_{ijk}\left(R^\text{\,s}_{ij}\right)^2}\label{eq:snr_c},
\end{equation}
which tells us that the SNR in each bin of the combined spectrum is the (unweighted) quadrature sum of the SNR across contributing bins, as anticipated in Sec.~\ref{sub:scan}.

As discussed in appendix~\ref{app:mle}, the ML estimate of the mean of a Gaussian distribution also has the smallest variance among unbiased estimates. The variance of the mean of a Gaussian distribution is simply proportional to the variance of the distribution, so equivalently ML weights yield the smallest $\sigma^\text{c}_k$ and thus the largest $R^\text{\,c}_k$ among all possible weights that do not systematically bias $\delta^\text{c}_k$. Thus, ML weighting is optimal for the haloscope analysis in a real physically intuitive sense.

\section{Combining bins horizontally}\label{sec:rebin}
The parameterization of the combined spectrum in terms of $\delta^\text{c}_k/\sigma^\text{c}_k$ and $R^\text{\,c}_k$ lends itself naturally to identifying axion candidates and setting exclusion limits, via the procedure outlined in Sec.~\ref{sec:candidates}. However, $R^\text{\,c}_k$ is the (unrealistically large) SNR for an axion signal confined to the single-bin bandwidth $\Delta\nu_b$, whereas our goal here is to construct an analysis tailored to the detection of virialized axions with $\Delta\nu_a \gg \Delta\nu_b$. Thus, our next task is to determine an explicit expression for the grand spectrum $\delta^\text{g}_\ell/\sigma^\text{g}_\ell$ as a weighted sum of adjacent combined spectrum bins. As in Sec.~\ref{sec:rescale_combine}, we take the optimal weights to be those that yield the ML estimate of the mean grand spectrum power excess, after rescaling to make the expected excess due to axion conversion uniform across all contributing bins. The discussion above indicates that ML weights in the horizontal sum will maximize $R^\text{\,g}_\ell$, the SNR for a virialized axion signal concentrated in the $\ell$th grand spectrum bin. 

In the choice of ML weights for the vertical sums that define the combined spectrum, I have followed the published ADMX analysis procedure~\citep{ADMX2001}, albeit with a somewhat different approach for pedagogical purposes.\footnote{See also Refs.~\citep{ADMX2000,daw1998,yu2004,hotz2013} for different presentations of ML weighting in the ADMX analysis; note that there are a number of errors in the expressions corresponding to Eqs.\ \eqref{eq:delta_c} and \eqref{eq:sigma_c} in Refs.~\citep{ADMX2001}, \citep{ADMX2000}, and \citep{daw1998}.} In extending ML weighting to horizontal sums of adjacent bins in the combined spectrum, I am deviating from the procedure used by ADMX. I discuss the key differences between my present approach and the ADMX procedure further in Sec.~\ref{sub:grand_spectrum}.

Though the principles of ML estimation remain valid, horizontal sums differ from the vertical sums considered in Sec.~\ref{sec:rescale_combine} in two important respects. First, we can no longer assume that the bins in each sum are independent random variables; indeed, as noted in Sec.~\ref{sec:baseline}, we have reason to expect correlations on small spectral scales in the processed spectra, and thus also in the combined spectrum. ML estimation of the mean of a multivariate Gaussian distribution with arbitrary covariance matrix is in principle straightforward (see appendix~\ref{app:mle}). In practice, it requires additional information about off-diagonal elements of the covariance matrix that are not as easily estimated as the variances. In the present analysis, I take ML weights that neglect correlations as approximations to the true ML weights, and define the horizontal sum using expressions appropriate for the uncorrelated case. I will quantify the effects of correlations in Sec.~\ref{sub:correlations}.

Second, independent of any subtleties involving correlations, we have some additional freedom in how we define the horizontal sum besides the choice of weights. The simplest approach is to define each grand spectrum bin as a ML-weighted sum of all bins within a segment of length $C\approx\Delta\nu_a/\Delta\nu_b$ in the combined spectrum, such that the segments corresponding to different grand spectrum bins do not overlap. The total number of grand spectrum bins is then $n\approx M/C$. The disadvantage of this approach is that the signal power will generally be split across multiple bins unless $\nu_a$ happens to line up with our binning. We need to introduce an attenuation factor $\eta_m$ (Sec.~\ref{sub:lineshape}) to account for the average effect of misalignment on the SNR.

We can minimize misalignment effects by allowing the segments of the combined spectrum corresponding to different grand spectrum bins to overlap: if each such segment is $K\approx\Delta\nu_a/\Delta\nu_b$ bins long, then the first grand spectrum bin will be a ML-weighted sum of the first through $K$th combined spectrum bins, the second grand spectrum bin will be a ML-weighted sum of the second through $(K+1)$th bins, and so on. But this procedure implies a total of $n\approx M$ grand spectrum bins, and thus the number of statistical rescan candidates (Sec.~\ref{sec:candidates}) will be larger at any given sensitivity than in the non-overlapping case; equivalently the total integration time required to exclude axions of a given coupling will be longer.

The two approaches considered above may be regarded as limiting cases of a more general procedure in which we split the construction of the grand spectrum into two steps. First we take ML-weighted sums of adjacent bins in non-overlapping segments of the combined spectrum to yield a rebinned spectrum with resolution $\Delta\nu_r = C\Delta\nu_b$. Then we construct the grand spectrum via ML-weighted sums of adjacent bins in overlapping segments of length $K$ in the rebinned spectrum. $C$ and $K$ should be chosen so that the product $CK\approx\Delta\nu_a/\Delta\nu_b$; it should be emphasized that I have thus far cited only a very rough estimate for $\Delta\nu_a$, and we are free to choose $C$ and $K$ independently within a reasonable range. 

In the two-step procedure described above, the rebinned spectrum weights and grand spectrum weights are each obtained from the ML principle, but of course we must specify a supposed distribution of signal power before we can define ML weights. The $\ell$th grand spectrum bin should be a sum over bins in the rebinned spectrum frequency range $[\nu_\ell,\nu_{\ell+K-1}]$ weighted so that the SNR is maximized if $\nu_a\approx\nu_\ell$. I will articulate this condition more precisely in Sec.~\ref{sub:lineshape}, but we can already see that the grand spectrum weights will depend on the axion lineshape. 

The weights used to construct the rebinned spectrum cannot themselves depend on the lineshape: the above example demonstrates that any given $\nu_\ell$ will correspond to the axion mass in one grand spectrum bin and the tail of the axion power distribution in another. We thus define weights to yield the ML estimate of the mean power excess in each bin of the rebinned spectrum assuming the axion signal distribution is uniform across contributing combined spectrum bins. As we reduce $C$, the distribution of signal power on scales smaller than $\Delta\nu_r$ becomes more uniform, and we can also use a finer approximation to the axion lineshape in the grand spectrum weights. 

For the analysis of the first HAYSTAC data run we used $C=10$ and $K=5$, informed by the tradeoffs noted above. In the next section, I  will briefly discuss the expected axion lineshape and its implications for the analysis. Then I will explain how we construct the rebinned spectrum (in Sec.~\ref{sub:rebinned_spectrum}) and the grand spectrum (in Sec.~\ref{sub:grand_spectrum}).

\subsection{The axion signal lineshape}\label{sub:lineshape}
As noted in Sec.~\ref{sub:haloscope_qual}, the spectral shape of a haloscope signal is proportional to the axion kinetic energy distribution. We saw in Sec.~\ref{sub:halo} that for a pseudo-isothermal halo, dark matter particle velocities in the galactic rest frame obey a Maxwell-Boltzmann distribution, and the corresponding kinetic energies have a $\chi^2$ distribution of degree 3 [Eq.~\eqref{eq:f_dist_E}]. For axion CDM we can rewrite Eq.~\eqref{eq:f_dist_E} in terms of the measured signal frequency $\nu\geq\nu_a$:
\begin{equation}
f(\nu) = \frac{2}{\sqrt{\pi}}\sqrt{\nu-\nu_a}\left(\frac{3}{\nu_a\left<\beta^2\right>}\right)^{3/2}e^{-\frac{3(\nu-\nu_a)}{\nu_a\left<\beta^2\right>}},
\label{eq:f_dist}
\end{equation}
where $\left<\beta^2\right>=\left<v^2\right>/c^2$ and $\left<v^2\right> = (270\text{ km/s})^2$ is the squared virial velocity defined in Sec.~\ref{sub:halo}. In a terrestrial lab frame the spectrum of the axion signal thus becomes~\citep{turner1990}
\begin{equation}\label{eq:f_dist_2}
f'(\nu) = \frac{2}{\sqrt{\pi}}\left(\sqrt{\frac{3}{2}}\frac{1}{r}\frac{1}{\nu_a\left<\beta^2\right>}\right)\sinh\left(3r\sqrt{\frac{2(\nu-\nu_a)}{\nu_a\left<\beta^2\right>}}\right)\exp\left(-\frac{3(\nu-\nu_a)}{\nu_a\left<\beta^2\right>}-\frac{3r^2}{2}\right),
\end{equation}
where $r=v_s/\sqrt{\left<v^2\right>}\approx\sqrt{2/3}$, and $v_s$ is the orbital velocity of the solar system about the center of the galaxy. Eq.~\eqref{eq:f_dist_2} is not a $\chi^2$ distribution, but is reasonably well approximated by Eq.~\eqref{eq:f_dist} with $\left<\beta^2\right>\rightarrow1.7\left<\beta^2\right>$; of course, it approaches Eq.~\eqref{eq:f_dist} in the limit $r\rightarrow0$. 

We used Eq.~\eqref{eq:f_dist} where we should have used Eq.~\eqref{eq:f_dist_2} it our initial analysis of the first HAYSTAC data run.\footnote{I thank B. R. Ko for drawing my attention to this point.} Specific parameter values cited throughout Sec.~\ref{sec:rebin} and \ref{sec:candidates} assume Eq.~\eqref{eq:f_dist}, as this was used to derive the exclusion limit published in Ref.~\citep{PRL2017}, but it should be emphasized that the formal procedure outlined in this chapter is independent of any specific assumptions about the signal shape. For an axion signal whose spectrum is given by Eq.~\eqref{eq:f_dist_2}, our exclusion limit is degraded by $\approx20\%$ (quantified more precisely in appendix~\ref{app:axion_width}) due to the combination of an irreducible effect from the wider signal bandwidth and the fact that this analysis was not optimized for this wider signal, as future HAYSTAC analyses will be.

A haloscope analysis can ultimately depend on the spectral shape of the axion signal only through the grand spectrum weights, which in turn can only depend on slices of $f(\nu)$ integrated over the resolution of the rebinned spectrum $\Delta\nu_r=C\Delta\nu_b$. Thus we define the integrated signal lineshape to be
\begin{equation}
L_q(\delta\nu_r) = K\int_{\nu_a+\delta\nu_r+(q-1)\Delta\nu_r}^{\nu_a+\delta\nu_r+q\Delta\nu_r}f(\nu)\,\mathrm{d}\nu,
\label{eq:int_lineshape}
\end{equation}
where $q=1,\dots,K$, and $\delta\nu_r$ is the misalignment of $\nu_a$ relative to the bin boundaries in the rebinned spectrum. Clearly, the effects of such misalignment will be periodic with period $\Delta\nu_r$; more precisely, we define $\delta\nu_r$ in the range $-z\Delta\nu_r < \delta\nu_r \leq (1-z)\Delta\nu_r$, with $0<z<1$. The value of $z$ should be chosen so that for any $\delta\nu_r$ in this range, $\eta_c(\delta\nu_r) = \sum_qL_q(\delta\nu_r)/K$ is larger than the value we would obtain by shifting the range over which the $q$ index is defined up or down by 1.\footnote{We might naively imagine a symmetric interval (corresponding to $z=0.5$) would be optimal in this sense. In practice, given the asymmetry of the axion lineshape, there will be more power in the $K$-bin sum if the lower bound of the integral in the $q=1$ bin is detuned below $\nu_a$ than at an equal detuning above $\nu_a$. This implies that we should consider $z>0.5$; the optimal value will depend on the choice of $C$ and $K$.} Physically, $\eta_c$ is the fraction of signal power contained within a grand spectrum bin; it approaches 1 independent of $\delta\nu_r$ for $K$ sufficiently large. At any fixed value of $K$, the sum also depends on $\delta\nu_r$ and thus on $C$.

We can gain some insight into the considerations that enter into the choice of $C$ and $K$ by imagining for the moment that we take the grand spectrum weights to be uniform, as in Ref.~\citep{ADMX2001}. Then, with $C=1$, $\eta_c\rightarrow1$ as $K$ increases, but the RMS noise power grows as $\sqrt{K}$, so the grand spectrum SNR ($\propto\eta_c/\sqrt{K}$) is maximized at a finite value of $K$. The SNR is relatively insensitive to $\delta\nu_r$ at $C=1$; as we increase $C$, keeping $CK$ fixed, $\eta_c$ remains unchanged in the best-case scenario $\delta\nu_r=0$, but larger misalignments are possible, so dependence of the SNR on $\delta\nu_r$ grows more pronounced.

In order to define ML weights for the grand spectrum (Sec.~\ref{sub:grand_spectrum}), we will need an expression for some ``typical'' lineshape $\bar{L}_q$ that is independent of misalignment. The best approach is to define $\bar{L}_q$ as the average of $L_q(\delta\nu_r)$ over the range in which $\delta\nu_r$ is defined.\footnote{$\bar{L}_q$ has no $\ell$ index because in practice we evaluated Eq.~\eqref{eq:int_lineshape} with $\nu_a = 5.75\text{ GHz}$ both in the limits of integration and within $f(\nu)$. It would be trivial to instead calculate the lineshape with $\nu_a=\nu_\ell$ in the $\ell$th grand spectrum bin, but the variation of the lineshape over the initial HAYSTAC scan range was negligible.} Then the misalignment attenuation can be defined as $\eta_m= \text{SNR}(\{\bar{L}_q\})/\text{SNR}(\{L_q(\delta\nu_r=0)\})$.\footnote{With this definition, $\eta_m$ is a useful figure of merit for comparing different values of $C$ and $K$, but we will not have to explicitly account for it in our analysis procedure, as the average effect of misalignment on the SNR is included in the definition of $\bar{L}_q$.} In the ML-weighted grand spectrum the SNR is no longer proportional to $\eta_c$ (indeed, it asymptotes to a constant value rather than degrading as we continue to increase $K$). However, the above prescription for defining $\eta_m$ still holds if we use the correct expression for the SNR [see Eq.~\eqref{eq:phi_align} in appendix~\ref{app:error}]. With $C=10$ and $K=5$, the optimal range for $\delta\nu_r$ is obtained for $z=0.7$, and the misalignment attenuation is $\eta_m=0.97$.

\subsection{Rebinning the combined spectrum}\label{sub:rebinned_spectrum}
After choosing the values of $C$ and $K$ to be used in the remainder of the analysis, we rescale the combined spectrum, taking $\delta^\text{c}_k \rightarrow (CK)\delta^\text{c}_k$ and $\sigma^\text{c}_k \rightarrow (CK)\sigma^\text{c}_k$. This rescaling leaves $\delta^\text{c}_k/\sigma^\text{c}_k$ formally unchanged and takes $R^\text{\,c}_k \rightarrow R^\text{\,c}_k/(CK)$, just what we would have obtained had we normalized Eq.~\eqref{eq:power_ij} to a more physically plausible fraction $1/(CK)$ of the expected KSVZ signal power in the first place.\footnote{The only reason I did not define $U_0$ with an additional factor of $1/(CK)$ in Eq.~\eqref{eq:U_0} is to emphasize that the construction of the combined spectrum is independent of $C$ and $K$.} After this rescaling we expect $\mu^\text{c}_{k'}=1$ if a KSVZ axion signal deposits a fraction $1/(CK)$ of its power in the combined spectrum bin $k'$.

In Sec.~\ref{sub:combine} I wrote rather verbose expressions for Eqs.~\eqref{eq:delta_c} and \eqref{eq:sigma_c} to make the dependence on physically meaningful quantities such as $P_{ij}$ explicit. The ML-weighted sum can be written more succinctly in terms of
\begin{equation}
D^\text{c}_k = \frac{\delta^\text{c}_k}{(\sigma^\text{c}_k)^2} = \frac{1}{CK}\sum_i\sum_j\Gamma_{ijk}\frac{\delta^\text{s}_{ij}}{(\sigma^\text{s}_{ij})^2},
\label{eq:d_c}
\end{equation}
which is just the sum in the numerator of Eq.~\eqref{eq:delta_c} rescaled by $1/(CK)$ as discussed above. Each $D^\text{c}_k$ is a Gaussian random variable with standard deviation $R^\text{\,c}_k$. We obtain the ML-weighted rebinned spectrum from
\begin{equation}
D^\text{r}_\ell = \sum_{k=k_i(\ell)}^{k_f(\ell)}D^\text{c}_k \label{eq:d_r}
\end{equation}
and
\begin{equation}
\left(R^\text{\,r}_\ell\right)^2 = \sum_{k=k_i(\ell)}^{k_f(\ell)}\left(R^\text{\,c}_k\right)^2,\label{eq:snr_r}
\end{equation}
where $k_i(\ell)=(\ell-1)C+1$, $k_f(\ell)=\ell C$, $\ell=1,\dots,n$, and $n\approx M/C$; $n\approx1.07\times10^5$ for the first HAYSTAC data run. 

In the absence of correlations between combined spectrum bins, each $D^\text{r}_\ell$ is a Gaussian random variable with standard deviation $R^\text{\,r}_\ell$. Defining $\sigma^\text{r}_\ell = (R^\text{\,r}_\ell)^{-1}$ and $\delta^\text{r}_\ell = D^\text{r}_\ell(\sigma^\text{r}_\ell)^2$ as in the combined spectrum, it follows that each rebinned spectrum bin $\delta^\text{r}_\ell$ is a Gaussian random variable with standard deviation $\sigma^\text{r}_\ell$ (and mean $\mu^\text{r}_\ell=0$ in the absence of axion signals). Each $\delta^\text{r}_\ell$ is the ML-weighted estimate of the mean power excess in $C$ adjacent combined spectrum bins $\delta^\text{c}_k$ if the axion power distribution is uniform on scales smaller than $\Delta\nu_r$. More precisely, $\mu^\text{r}_{\ell'}=1$ if a KSVZ axion deposits a fraction $1/(CK)$ of its power in each of the $C$ adjacent combined spectrum bins corresponding to the rebinned spectrum bin $\ell'$, and $R^\text{\,r}_{\ell'}$ is the SNR for such a signal. 

Neglecting small-scale variation in $R^\text{\,c}_k$, Eq.~\eqref{eq:snr_r} implies that the SNR in each bin of the rebinned spectrum has increased by $\sqrt{C}$. This is exactly what we should expect given that the signal power grows roughly linearly with bandwidth $\Delta\nu$ (for $\Delta\nu$ sufficiently small compared to $\Delta\nu_a$) and the RMS noise power grows as $\sqrt{\Delta\nu}$ [see discussion around Eq.~\eqref{eq:dicke}]. Empirically, the RMS variation in $\sigma^\text{c}_k$ is typically $\lesssim1\%$ on 10-bin scales (and $\approx3\%$ on 50-bin scales), so the rebinned spectrum would not change much if we used uniform weights instead of ML weights. We will see in Sec.~\ref{sub:grand_spectrum} that ML weighting of the grand spectrum leads to a larger improvement relative to an unweighted analysis.

In the absence of correlations, each $\delta^\text{r}_\ell$ has standard deviation $\sigma^\text{r}_\ell$, so $\delta^\text{r}_\ell/\sigma^\text{r}_\ell$ should have a standard normal distribution, like the analogous quantity in the combined spectrum. Empirically, in the first HAYSTAC data run, $\delta^\text{r}_\ell/\sigma^\text{r}_\ell$ was Gaussian with standard deviation $\xi^\text{r}=0.98$. $\xi^\text{r}\neq1$ is a consequence of the fact that the expression for the variance of a sum of Gaussian random variables used in Eq.~\eqref{eq:snr_r} does not hold in the presence of correlations, as noted at the end of Sec.~\ref{sub:stats}.\footnote{A similar reduction in the standard deviation following a horizontal sum was observed in Ref.~\citep{yu2004}, pg. 122, and attributed to the baseline removal procedure, but not discussed further.} An analogous effect will arise in the construction of the grand spectrum, so we will defer further discussion of this point to Sec.~\ref{sub:correlations}.

\subsection{Constructing the grand spectrum}\label{sub:grand_spectrum}
To extend the ML-weighted horizontal sum further, we must account for the fact that, for any given value of $\nu_a$, the distribution of axion signal power in the $K$ bins containing most of the signal is nonuniform. Specifically, for a KSVZ axion with $\nu_a\approx\nu_{\ell'}$, we expect $\mu^\text{r}_{\ell'+q-1}=\bar{L}_q$ for $q=1,\dots,K$. As in Sec.~\ref{sec:rescale_combine}, we must rescale the contributing bins so that they all have the same mean power excess before defining ML weights. For the $\ell$th grand spectrum bin, the appropriate rescaling is obtained by dividing both $\delta^\text{r}_{\ell+q-1}$ and $\sigma^\text{r}_{\ell+q-1}$ by $\bar{L}_q$, or equivalently by multiplying both $D^\text{r}_{\ell+q-1}$ and $R^\text{\,r}_{\ell+q-1}$ by $\bar{L}_q$. The quantities of interest in the ML-weighted grand spectrum are then given by
\begin{equation}
R^\text{\,g}_\ell = \sqrt{\sum_q(R^\text{\,r}_{\ell+q-1}\bar{L}_q)^2}\label{eq:snr_g}
\end{equation}
and
\begin{equation}
\frac{\delta^\text{g}_\ell}{\sigma^\text{g}_\ell} = \frac{D^\text{g}_{\ell}}{R^\text{\,g}_\ell} = \frac{\sum_q D^\text{r}_{\ell+q-1}\bar{L}_q}{\sqrt{\sum_q(R^\text{\,r}_{\ell+q-1}\bar{L}_q)^2}}\label{eq:delta_sigma_g},
\end{equation}
Neglecting effects of the SG filter stopband, each $\delta^\text{g}_\ell$ should be a Gaussian random variable with standard deviation $\sigma^\text{g}_\ell=(R^\text{\,g}_\ell)^{-1}$ and mean $\mu^\text{g}_\ell$. Our definition of $\bar{L}_q$ in Sec.~\ref{sub:lineshape} implies that $\mu^\text{g}_{\ell'}=1$ (equivalently, $E[\delta^\text{g}_{\ell'}/\sigma^\text{g}_{\ell'}]=R^\text{\,g}_{\ell'}$) for a KSVZ axion signal with average misalignment in bin $\ell'$.\footnote{Here and elsewhere in this chapter, ``an axion signal in the grand spectrum bin $\ell'$'' should be taken as shorthand for the condition $-0.7\Delta\nu_r < \nu_{\ell'}-\nu_a < 0.3\Delta\nu_r$, where $\nu_\ell$ refers to the frequency at the lower edge of bin $\ell$. For detunings outside this range, the SNR will be larger in a different grand spectrum bin, and we should speak of the signal ``in'' that bin instead.} The small uncertainty in $\mu^\text{g}_{\ell'}$ associated with the range of possible misalignments will contribute to the uncertainty in our exclusion limit, discussed in appendix~\ref{app:error}. Within $K$ bins of $\ell'$, $0<\mu^\text{g}_\ell<1$, because the overlapping horizontal sum correlates nearby grand spectrum bins.\footnote{It should be emphasized that these correlations are independent of, and would occur even in the absence of, the correlations between \textit{combined spectrum} bins responsible for $\xi^\text{r}\neq1$. The implications of these grand spectrum correlations for the analysis will be discussed further in Sec.~\ref{sub:target_confidence}.} Of course, $\mu^\text{g}_\ell=0$ for $|\ell-\ell'|\geq K$. 

Empirically, $\delta^\text{g}_\ell/\sigma^\text{g}_\ell$ [histogrammed in Fig.~\ref{fig:data_hist}(c)] has a Gaussian distribution with mean 0 and standard deviation $\xi=0.93$. We saw above that correlations within each bin of the rebinned spectrum already reduced the width of the histogram by a factor $\xi^\text{r}=0.98$, which implies that the reduction we can attribute specifically to correlations between \textit{different} rebinned spectrum bins is $\xi^\text{g}=\xi/\xi^\text{r}=0.95$.

Setting aside the issue of correlations, we can gain further insight into the properties of our ML-weighting horizontal sum by considering how it differs from the corresponding step in the ADMX haloscope analysis procedure. ADMX analyses tailored to the detection of virialized axions have consistently used $\Delta\nu_b/\Delta\nu_a$ approximately a factor of 10 larger than in the present analysis and $C=1$ (i.e., no rebinning after data combining). The original ADMX analysis~\citep{ADMX2001,ADMX2000,daw1998} took the grand spectrum to be an unweighted sum of $K=6$ combined spectrum bins. This is not quite the same as setting $\bar{L}_q=1$ in Eqs.~\eqref{eq:snr_g} and \eqref{eq:delta_sigma_g} because our sums are still ML-weighted by $(\sigma^\text{r}_\ell)^{-2}$ in this limit. However, as noted in Sec.~\ref{sub:rebinned_spectrum}, the variation in $\sigma^\text{c}_k$ on the relevant scales is small enough that in practice there is not much difference. 

Thus we will compare our ML analysis to the unweighted $K$-bin sum in the limit that $\sigma^\text{r}_\ell$ (equivalently $R^\text{\,r}_\ell$) is equal in all contributing bins. In this limit, the grand spectrum SNR may be written in the form
\begin{equation}
R^\text{\,g}_\ell=F\big(K,\Delta\nu_r,\{\bar{L}_q\}\big)K\sqrt{\Delta\nu_r}R^\text{\,r}_\ell,
\label{eq:uw_limit}
\end{equation}
where we have introduced a figure of merit $F$ to encode the dependence of $R^\text{\,g}_\ell$ on $K$, $C$, and $\bar{L}_q$. It becomes apparent that $R^\text{\,g}_\ell$ only depends on these quantities through $F$ when we rewrite Eq.~\eqref{eq:dicke} for the rebinned spectrum SNR in the form 
\begin{equation}
R^\text{\,r}_\ell=[(P_\ell/K)/(h\nu_\ell N_\ell)]\sqrt{\tau/\Delta\nu_r},\label{eq:snr_r_approx}
\end{equation}
where $P_\ell$ ($N_\ell$) is an appropriately weighted average of the total axion conversion power (system noise) across all contributing processed spectrum bins.

For our ML-weighted analysis, we obtain an explicit expression for $F$ by comparing Eq.~\eqref{eq:uw_limit} to Eq.~\eqref{eq:snr_g}:
\begin{equation}
F_\text{ML} = \sqrt{\frac{1}{\Delta\nu_r}\sum_q(\bar{L}_q/K)^2}\label{eq:fom_ml}.
\end{equation}
The figure of merit for an unweighted sum follows from Eq.~\eqref{eq:uw_limit} and $R^\text{\,g}_\ell=\sqrt{K}\eta_cR^\text{\,r}_\ell$ (see Sec.~\ref{sub:lineshape}):
\begin{equation}
F_\text{uw} = \frac{1}{\sqrt{K\Delta\nu_r}}\sum_q \bar{L}_q/K\label{eq:fom_rect}.
\end{equation}

For a meaningful comparison between analyses, we must assume the same underlying signal spectrum $f(\nu)$ in both cases. If we also assume that both analyses are characterized by the same values of $K$ and $\Delta\nu_r$, and thus the same $\bar{L}_q$, then $F_\text{uw}$ is just the mean of $\bar{L}_q$ multiplied by $(K\Delta\nu_r)^{-1/2}$, whereas $F_\text{ML}$ is the RMS of $\bar{L}_q$ times the same factor. Thus $F_\text{ML}\geq F_\text{uw}$ independent of any specific features of the lineshape; this is another way to understand the improvement in sensitivity from ML weighting.\footnote{Eq.~\eqref{eq:fom_ml} only quantifies the true improvement in the SNR from a ML analysis if our analysis has assumed the correct signal lineshape, but insofar as the true signal distribution is closer to the nominal lineshape than to a ``boxcar'' of width $K\Delta\nu_r$, the ML analysis will still be more sensitive than an unweighted sum.}

We can also use Eqs.~\eqref{eq:fom_ml} and \eqref{eq:fom_rect} to compare the sensitivity of analyses based on the same model $f(\nu)$ but characterized by different $\Delta\nu_r$ and/or $K$ and thus different $\bar{L}_q$; this is a convenient way to quantify the considerations discussed in Sec.~\ref{sub:lineshape}. For $f(\nu)$ given by Eq.~\eqref{eq:f_dist}, the improvement in the SNR from an optimal ML-weighted analysis relative to an optimal unweighted analysis is about 7.5\%.\footnote{Here ``optimal'' means the SNR is maximized with respect to $\Delta\nu_r$ and $K$ (or, for ML weighting, it is sufficiently close to its asymptotic value). The values of $\Delta\nu_r$ and $K$ adopted for the present analysis are not optimal in this sense, and indeed the SNR for our present analysis is only about 2\% better than the SNR in the optimal unweighted case. However, this optimization does not take into account the fact that the integration time required for rescans increases as we reduce $\Delta\nu_r$, as emphasized at the beginning of Sec.~\ref{sec:rebin}. A better comparison would consider the improvement in SNR for a ML analysis relative to an unweighted analysis that results in comparable total rescan time. Our present ML analysis has $11.5\%$ better SNR than the unweighted analysis with the same $\Delta\nu_r$ and $K$.}

In more recent ADMX analyses~\citep{ADMX2004,yu2004,hotz2013,lyapustin2015}, the grand spectrum is defined as a weighted sum of combined spectrum bins, with weights corresponding to the coefficients of a Wiener Filter (WF). In my notation, the WF weight for the bin $\delta^\text{r}_{\ell+q-1}$ is
\begin{equation}
u^\text{WF}_q = \frac{\bar{L}_q^2}{\bar{L}_q^2+(\sigma^\text{r}_{\ell+q-1})^2},
\label{eq:wf}
\end{equation}
up to a normalization factor. These weights are obtained as solutions to the least-squares minimization of the difference between the noisy observations $\delta^\text{r}_{\ell+q-1}$ and the mean power $\bar{L}_q$ independently in each bin. In the high-SNR limit $\sigma^\text{r}_{\ell+q-1}\ll \bar{L}_q$, $u^\text{WF}_q\rightarrow1$, whereas in the low-SNR limit, $u^\text{WF}_q\rightarrow (\bar{L}_q/\sigma^\text{r}_{\ell+q-1})^2$. In neither limit do they agree with the unnormalized ML weights,\footnote{Here we are comparing the coefficients of the bins $\delta^\text{r}_{\ell+q-1}$ in the ML and WF analyses. The ML weights are more properly defined as the coefficients of the rescaled bins $\delta^\text{r}_{\ell+q-1}/\bar{L}_q$. With this definition the numerator is $(\bar{L}_q/\sigma^\text{r}_{\ell+q-1})^2$, but there is no such rescaling step in the WF analysis.} $u^\text{ML}_q = \bar{L}_q/(\sigma^\text{r}_{\ell+q-1})^2$.

The origin of this discrepancy is the fact that, while the ML and WF schemes are both based on least-squares optimization, they are obtained by minimizing the mean squared error with respect to different quantities: the ML procedure yields the least-squares optimal estimate of the mean power excess in the (appropriately rescaled) contributing bins (and thus results in larger SNR than all other unbiased analyses, as noted in Sec.~\ref{sub:combine}), whereas the WF procedure yields the least-squares optimal estimates of the weights that most robustly undo the smearing of the axion lineshape due to the presence of noise. In my view, the ML scheme relates more directly to the fundamental quantities of interest in the haloscope search.

Finally, we briefly note one more practical difference between the WF and ML methods: the WF weights depend on the SNR, whereas the ML weights only depend on the shape of the axion signal independent of any overall normalization. In practice the WF should be evaluated at an estimate of the average threshold sensitivity $|g^\text{min}_\gamma|$ to be obtained from the analysis. In the high-SNR limit, the WF sum becomes unweighted, and the SNR improvement from ML weighting may be estimated from $F_\text{ML}/F_\text{uw}$ as noted above.

\subsection{Accounting for correlations}\label{sub:correlations}
In the discussion above we noted two distinct effects on the rebinned spectrum [Eqs.~\eqref{eq:d_r} and \eqref{eq:snr_r}] and grand spectrum [Eqs.~\eqref{eq:snr_g} and \eqref{eq:delta_sigma_g}] due to correlations between nearby combined spectrum bins. First, we have not used the correct expression for the variance of a weighted sum of correlated Gaussian random variables in Eqs.~\eqref{eq:snr_r} and \eqref{eq:snr_g}. Second, in the presence of correlations, the weights we have used are not actually the optimal ML weights. The former effect is responsible for $\xi^\text{r},\xi^\text{g}\neq1$; note that it is completely independent of whether or not the weights are optimal. We will consider the effect on the variance first; doing so will allow us to estimate the sum of off-diagonal elements in the relevant covariance matrices, and thus quantify the deviation from the optimal weights.

The most general expression for the variance of a weighted sum of $K$ Gaussian random variables $X_q$ is 
\begin{equation}\label{eq:var_general}
\text{Var}\Big(\sum_qw_qX_q\Big) = \sum_q w_q^2\,\text{Var}(X_q) +2\sum_{q}\sum_{q'=1}^{q-1}w_qw_{q'}\text{Cov}(X_q,X_{q'}).
\end{equation}
I will apply this expression to obtain the correct variance $(\hat{\sigma}^{\text{g}}_\ell)^2$ of the $\ell$th grand spectrum bin. With $X_q=\delta^\text{r}_{\ell+q-1}/\bar{L}_q$ and the grand spectrum weights used in Sec.~\ref{sub:grand_spectrum}, we obtain
\begin{eqnarray}
\big(\hat{\sigma}^{\text{g}}_\ell\big)^2 = \big(\sigma^\text{g}_\ell\big)^2+2\big(\sigma^\text{g}_\ell\big)^4\sum\limits_{q=1}^K\sum\limits_{q'=1}^{q-1}\frac{\bar{L}_q\bar{L}_{q'}\Sigma^\text{r}_{\ell qq'}}{(\sigma^\text{r}_{\ell+q-1}\sigma^\text{r}_{\ell+q'-1})^2},\ \ \ \ \ \ \label{eq:sigma_g_hat}
\end{eqnarray}
where $\Sigma^\text{r}_{\ell qq'} = \text{Cov}(\delta^\text{r}_{\ell+q-1},\delta^\text{r}_{\ell+q'-1})$, and the factor of $(\sigma^\text{g}_\ell)^4$ multiplying the second term comes from the normalization of the ML weights. The analogous expression for the correct variance of the $\ell$th rebinned spectrum bin is
\begin{equation}
\big(\hat{\sigma}^{\text{r}}_\ell\big)^2 = \big(\sigma^\text{r}_\ell\big)^2+2\big(\sigma^\text{r}_\ell\big)^4\sum\limits_{k=k_i(\ell)}^{k_f(\ell)}\sum\limits_{k'=k'_i(\ell)}^{k-1}\frac{\Sigma^\text{c}_{kk'}}{(\sigma^\text{c}_{k}\sigma^\text{c}_{k'})^2}, \label{eq:sigma_r_hat}
\end{equation}
with $\Sigma^\text{c}_{kk'} = \text{Cov}\big(\delta^\text{c}_{k},\delta^\text{c}_{k'}\big)$. 

Having established the requisite formalism, we can now ask whether taking correlations into account explains the observed reduction of the grand spectrum and rebinned spectrum standard deviations. We see immediately that $\hat{\sigma}^{\text{g}}_\ell$ can be smaller than $\sigma^{\text{g}}_\ell$ if the sum over off-diagonal elements of the covariance matrix is on average slightly negative. Formally the ratio $\hat{\sigma}^{\text{g}}_\ell/\sigma^{\text{g}}_\ell$ is frequency-dependent, but if nonzero $\Sigma^\text{r}_{\ell qq'}$ is a consequence of the stopband properties of the SG filter, we should expect the correlation matrix $\rho^\text{r}_{\ell qq'}=\Sigma^\text{r}_{\ell qq'}/(\sigma^\text{r}_{\ell+q-1}\sigma^\text{r}_{\ell+q'-1})$ to depend only on the bin spacing $\Delta q= q-q'$. Analogous arguments also apply to the ratio $\hat{\sigma}^{\text{r}}_\ell/\sigma^{\text{r}}_\ell$. Thus we expect
\begin{equation}\label{eq:xi_g}
\xi^\text{g}=\hat{\sigma}^{\text{g}}_\ell/\sigma^{\text{g}}_\ell 
\end{equation}
and
\begin{equation}\label{eq:xi_r}
\xi^\text{r}=\hat{\sigma}^{\text{r}}_\ell/\sigma^{\text{r}}_\ell
\end{equation}
in the case of filter-induced correlations. 

We used a simulation to show that the observed values of $\xi^\text{r}$ and $\xi^\text{g}$ are indeed fully explained by processed spectrum correlations imprinted by the SG filter. Each iteration in the $\xi^\text{g}$ simulation generates a set of $m$ 14020-bin Gaussian white noise spectra with mean 1 and standard deviation $\sigma^\text{p}=1/\sqrt{\Delta\nu_b\tau}$, multiplies each spectrum by a random sample baseline derived from data, then uses the baseline removal procedure described in Sec.~\ref{sec:baseline} to obtain a set of simulated processed spectra.\footnote{The sample baselines used here and in the simulation described in Sec.~\ref{sub:axion_atten} were each obtained by applying a high-order SG filter (as in Sec.~\ref{sub:badbins}) to the average of about 50 consecutive raw spectra after removing contaminated bins.} The $m$ processed spectra are averaged without weighting or offsets to obtain a single simulated combined spectrum, in which we average non-overlapping 10-bin segments. We calculate the product of each pair of bins with $0 \leq \Delta q \leq 5$ in the simulated rebinned spectrum. Averaging each such product over $\approx500$ iterations of the simulation, we obtain reasonably precise estimates of $(\sigma^{\text{r}}_\ell)^2$ and $\Sigma^\text{r}_{\ell qq'}$ for each bin $\ell$ in the rebinned spectrum. Then we calculate $\sigma^{\text{g}}_\ell$ and $\hat{\sigma}^{\text{g}}_\ell$ from Eqs.~\eqref{eq:snr_g} and \eqref{eq:sigma_g_hat}, and $\xi^\text{g}$ from Eq.~\eqref{eq:xi_g}. 

We find that $\xi^\text{g}=0.95$ is constant throughout the analysis band, independent of $m$ for values ranging from $m=1$ out to at least $m=400>\text{max}(m_k)$ and independent of $\tau$ out to at least $\tau=900\text{ s}$.\footnote{Our simulation and Eq.~\eqref{eq:xi_g} measure $\xi^\text{g}$ rather than $\xi=\xi^\text{g}\xi^\text{r}$ because we use $\sigma^{\text{r}}_\ell$ rather than $\hat{\sigma}^{\text{r}}_\ell$ in Eqs.~\eqref{eq:sigma_g_hat} and \eqref{eq:snr_g}. Note also that $m_k$ is itself an upper bound on the averaging in each bin, because contributing spectra are not uniformly weighted.} From an analogous simulation to quantify the effects of correlations on the rebinned spectrum we obtain a constant $\xi^\text{r}=0.98$. To verify that the implementation of the simulation was correct, we calculate the same quantities from the simulated Gaussian white noise spectra directly (bypassing the steps where we imprint and then remove the baseline); we obtain $\xi^\text{g}=\xi^\text{r}=1$ as expected for this null test.

These results demonstrate conclusively that the observed values of $\xi^\text{r}$ and $\xi^\text{g}$ depend only on the stopband properties of the SG filter. Fig.~\ref{fig:filter} indicates that the filter-induced negative correlations increase at larger bin separations, consistent with the empirical result $1-(\xi^\text{g})^2>1-(\xi^\text{r})^2$. The explicit demonstration that $\xi^\text{g}$ and $\xi^\text{r}$ are independent of $m$ is critical because in the real data $m_k$ varies throughout the combined spectrum: $m$-independence implies that nonuniform weighting and frequency offsets between processed spectra will not affect our results. We conclude that $\xi^\text{r}$ and $\xi^\text{g}$ are frequency-independent, as indeed the numerical agreement between the simulated and observed values already indicates.\footnote{The values of $\xi^\text{g}$ and $\xi^\text{r}$ obtained from the real data were also unchanged when we divided the axion search dataset in half in various ways (winter/summer, high/low RF frequency, upper/lower half of analysis band) and constructed the grand spectrum separately from each subset of the data.}

It follows that each grand spectrum bin $\delta^\text{g}_\ell$ is a Gaussian random variable with standard deviation 
\begin{equation}
\tilde{\sigma}^\text{g}_\ell \, = \, \xi\sigma^\text{g}_\ell \, = \, \xi^\text{g}\xi^\text{r}\sigma^\text{g}_\ell.
\label{eq:sigma_g_tilde}
\end{equation}
and mean $\mu^\text{g}_\ell=0$ in the absence of axion signals. Now let us suppose there exists a KSVZ axion in bin $\ell'$ of the grand spectrum. If the only effect of the imperfect SG filter stopband were to correlate the statistical fluctuations of the noise in nearby bins, we would still have $\mu^\text{g}_{\ell'}=1$, since the mean of a weighted sum of Gaussian random variables is independent of whether or not they are correlated. 

However, the imperfect SG filter stopband will also lead to slight attenuation of any locally correlated power excess (e.g., an axion signal) in the raw spectra, so we should expect $\mu^\text{g}_{\ell'}=\eta'<1$. It follows that $\delta^\text{g}_{\ell'}/\tilde{\sigma}^\text{g}_{\ell'}$ is a Gaussian random variable with standard deviation 1 and mean
\begin{equation}
\tilde{R}^\text{\,g}_{\ell'} = \eta'/\tilde{\sigma}^\text{g}_{\ell'} = \eta R^\text{\,g}_{\ell'},
\label{eq:snr_g_tilde}
\end{equation}
where $\eta=\eta'/\xi$. Thus we see that filter-induced \textit{signal} attenuation $\eta'$ actually only reduces the SNR by the smaller factor $\eta$, because the RMS fluctuations of the noise power within the axion bandwidth are also reduced. The procedure we use to quantify $\eta$ is described in detail in Sec.~\ref{sub:axion_atten}; though formally Eq.~\eqref{eq:snr_g_tilde} allows $\eta>1$, we will find that $\eta<1$, indicating that the net effect is indeed reduction of the SNR.

Finally, we can return to the second effect of correlations neglected in the construction of the grand spectrum: in the presence of correlations, neither the rebinned spectrum weights nor the grand spectrum weights are actually the true ML weights. We are now equipped to show that in practice deviations from the optimal weights are negligibly small in both cases. 

I noted in appendix~\ref{app:mle} that the true ML weights in the presence of correlations are sums over rows of the inverse covariance matrix. Applying the approximation in Eq.~\eqref{eq:weights_cor},\footnote{It can be shown using Eq.~\eqref{eq:sigma_g_hat} that the average of the off-diagonal elements of the correlation matrix is $1.5\big[\big(\xi^\text{g}\big)^2-1\big]/(K-1)\approx-0.035$, where the numerical factor is due to lineshape weighting. Thus a first-order approximation is appropriate.} we find that the (properly normalized) true ML weights for the grand spectrum are
\begin{eqnarray}
\tilde{w}_{\ell q} &&= \frac{\big(\sigma^\text{g}_\ell\big)^2}{2-\big(\xi^\text{g}\big)^2}\Bigg[\frac{\bar{L}_q^2}{\big(\sigma^\text{r}_{\ell+q-1}\big)^2} - \sum_{q'\neq q}\frac{\bar{L}_q\bar{L}_{q'}\Sigma^\text{r}_{\ell qq'}}{\big(\sigma^\text{r}_{\ell+q-1}\sigma^\text{r}_{\ell+q'-1}\big)^2}\Bigg] \nonumber \\
&&= w_{\ell q}^0 + \delta w_{\ell q}. \label{eq:w_tilde}
\end{eqnarray}
Up to a change in the normalization, $w_{\ell q}^0=w_{\ell q}$, the ML weights in the absence of correlations. The mean value of $\delta w_{\ell q}$ just compensates for this rescaling such that $\tilde{w}_{\ell q}$ remain normalized. The typical change in the relative weighting is given by the standard deviation of $\delta w_{\ell q}$, which is easy to calculate given the covariances obtained in our simulation: we find that the RMS fractional change in the weights is about 5\%. 

The resulting fractional change in $\delta^\text{g}_\ell$ will be much smaller because it is the average of $K$ 5\% deviations that are mutually negatively correlated (because the weights remain normalized). Thus, the systematic effect from neglecting correlations in the grand spectrum ML weights is small compared to the sources of error considered in appendix~\ref{app:error}; the analogous effect in the rebinned spectrum is smaller still due to the smaller value of $1-(\xi^\text{r})^2$.

\section{Candidates and exclusion}\label{sec:candidates}
Via the procedure described in the previous sections, we have condensed our axion search data into the $2n$ numbers $\delta^\text{g}_\ell/\tilde{\sigma}^\text{g}_\ell$ and $\tilde{R}^\text{\,g}_\ell$. The statistical fluctuations of the total noise power result in a standard normal distribution for the corrected grand spectrum $\delta^\text{g}_\ell/\tilde{\sigma}^\text{g}_\ell$ in the absence of axion signals, and a KSVZ axion signal in a particular bin $\ell'$ would displace the mean of $\delta^\text{g}_{\ell'}/\tilde{\sigma}^\text{g}_{\ell'}$ by $\tilde{R}^\text{\,g}_{\ell'}$. Now I will explain how we use these quantities to interrogate the presence of axion conversion power in our scan range and derive an exclusion limit if there are no persistent signals.

As noted at the end of Sec.~\ref{sub:daq_procedure}, we have no \textit{a priori} knowledge of which bin $\ell'$ (if any) corresponds to the axion mass, and the only qualitative difference between an axion signal and a positive excess power fluctuation in any given bin is that a true signal should be persistent across different scans at the same frequency. Thus the best we can do is set a threshold $\Theta$ and define any bin with $\delta^\text{g}_\ell/\tilde{\sigma}^\text{g}_\ell\geq\Theta$ as a \textit{rescan candidate}. In the absence of grand spectrum correlations, we would expect
\begin{equation}
\hat{S}=n\big[1-\Phi(\Theta)\big]
\label{eq:candidates}
\end{equation}
such rescan candidates from statistics alone, where $\Phi(x)$ is the cumulative distribution function of the standard normal distribution. We can then collect sufficient data at each rescan frequency to reproduce the sensitivity in the initial scan (Sec.~\ref{sub:rescan_daq}), and thereby distinguish any real axion signal from statistical fluctuations (Sec.~\ref{sub:rescan_analysis}). 

In light of the above discussion, our proximate task is to determine an appropriate value for $\Theta$. To simplify matters, let us first assume $\tilde{R}^\text{\,g}_\ell=R_T$ is constant throughout the scan range. Perhaps the simplest choice of threshold is $\Theta=R_T$. Taking  $n\approx1.07\times10^5$ for the first HAYSTAC data run and assuming for now that $R_T=5$, we obtain $\hat{S}=0.03$; thus any bin exceeding the threshold is very unlikely to be a statistical fluctuation. The problem with this choice of threshold becomes clear when we suppose there is an axion signal with SNR $R_T$ in some bin $\ell'$: then $\delta^\text{g}_{\ell'}/\tilde{\sigma}^\text{g}_{\ell'}$  is a Gaussian random variable with mean $R_T$ and standard deviation 1. $\Theta=R_T$ is a poor choice of threshold because the probability that $\delta^\text{g}_{\ell'}/\tilde{\sigma}^\text{g}_{\ell'}\geq\Theta$ is only 50\%. 

\begin{figure}[h]
\centering\includegraphics[width=0.8\textwidth]{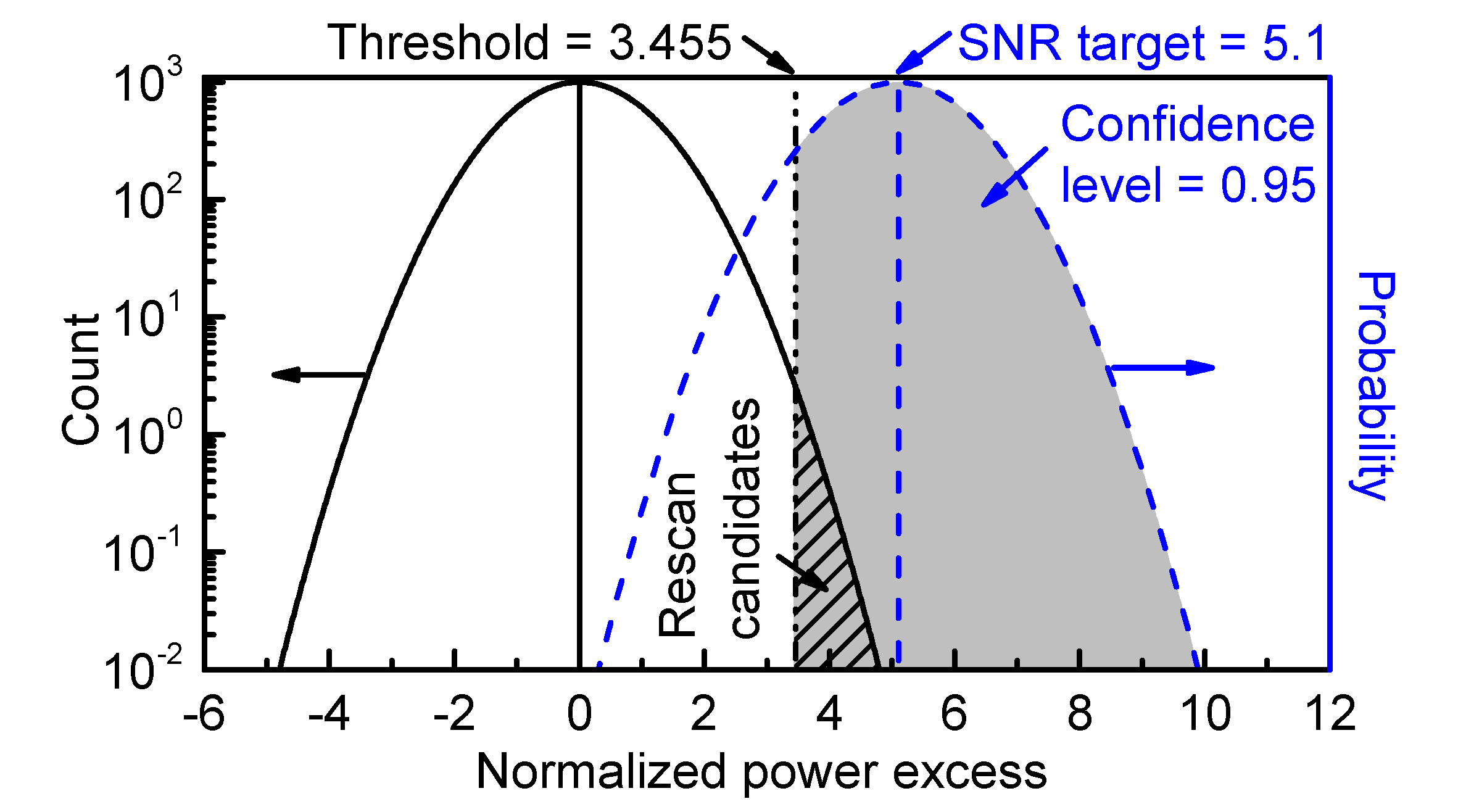}
\caption[The relationship between SNR, rescan threshold, and confidence level]{\label{fig:confidence} Schematic illustration of the relationship between the SNR target $R_T$, rescan threshold $\Theta$, rescan yield $\hat{S}$, and initial scan confidence level $c_1$. The vertical axis on the left applies to the solid black curve representing the expected standard normal distribution of grand spectrum bins $\delta^\text{g}_\ell/\tilde{\sigma}^\text{g}_\ell$; the integral is the total number of grand spectrum bins $n$. The vertical axis on the right applies to the dashed blue curve, which is a normalized Gaussian distribution with unit standard deviation and mean $R_T$, and represents the expected distribution of excess power $\delta^\text{g}_{\ell'}/\tilde{\sigma}^\text{g}_{\ell'}$ in a single grand spectrum bin $\ell'$ containing an axion signal. The threshold $\Theta$ (dot-dashed vertical line) intersects both distributions: $\hat{S}$ (hatched region) is the integral of the grand spectrum distribution above $\Theta$, and $c_1$ (gray shaded region) is the integral of the signal distribution above $\Theta$.}
\end{figure}

For arbitrary $\Theta$ (again assuming a signal with SNR $R_T$ in bin $\ell'$), the probability that $\delta^\text{g}_{\ell'}/\tilde{\sigma}^\text{g}_{\ell'}\geq\Theta$ in the presence of noise is called the axion search \textbf{confidence level}. If we require a confidence level $\geq c_1$ for the initial scan, the appropriate threshold is
\begin{equation}
\Theta=R_T - \Phi^{-1}(c_1),
\label{eq:threshold}
\end{equation}
and the expected rescan yield $\hat{S}$ follows from Eq.~\eqref{eq:candidates}. The relationship between all of these quantities is illustrated in Fig.~\ref{fig:confidence}. In Sec.~\ref{sub:target_confidence} we will see that grand spectrum correlations modify the expected rescan yield slightly, so we should actually expect $\bar{S}<\hat{S}$ candidates.

In the above discussion I assumed constant SNR throughout the scan range, when in fact $\tilde{R}^\text{\,g}_\ell$ varied significantly on scales $\gtrsim1$~MHz in the first HAYSTAC data run, with typical values between 0.7 and 1.2, due to nonuniform tuning and frequency-dependence of the cavity $Q$, form factor, etc. Recall that $\tilde{R}^\text{\,g}_\ell$ is the SNR for an axion signal with photon coupling $|g_\gamma|=|g^\text{KSVZ}_\gamma|$, and as I emphasized in Sec.~\ref{sub:rescale}, our decision to normalize Eq.~\eqref{eq:U_0} to the KSVZ coupling specifically was completely arbitrary. To obtain a frequency-independent threshold, we can simply define
\begin{equation}
G_\ell=\left(R_T/\tilde{R}^\text{\,g}_\ell\right)^{1/2},
\label{eq:g_ell}
\end{equation}
from which it follows that $R_T$ is the SNR for an axion with frequency-dependent coupling
\begin{equation}
|g^\text{min}_\gamma|_\ell=G_\ell|g^\text{KSVZ}_\gamma|.
\label{eq:g_min}
\end{equation}

Eqs.~\eqref{eq:threshold} -- \eqref{eq:g_min} completely determine the confidence level at which we can exclude axions as a function of the two-photon coupling $|g_\gamma|$ in each bin of the grand spectrum.\footnote{In contrast, in most ADMX analyses~\citep{ADMX2001,ADMX2000,daw1998,ADMX2004,yu2004} the confidence level is obtained from Monte Carlo. The Monte Carlo procedure involves constructing a mock grand spectrum containing a large number of simulated axion signals with known SNR $R_T$, setting a threshold $\Theta$, and defining $c_1$ as the fraction of simulated axions flagged as rescan candidates; the simulation may be repeated many times to determine the behavior of $c_1$ as a function of $R_T$ and/or $\Theta$. This more involved approach was originally adopted to circumvent the effects of correlations on the horizontal sum, which I have shown we can quantify.} By varying $\Theta$ we can adjust the tradeoff between $\bar{S}$ (which determines the total time we need to spend acquiring rescan data) and $|g^\text{min}_\gamma|_\ell$, the minimum coupling to which our search is sensitive.\footnote{A coupling $|g_\gamma|_\ell>|g^\text{min}_\gamma|_\ell$ corresponds to a signal with $\text{SNR} > R_T$. At any given threshold $\Theta$, a result $\delta^\text{g}_{\ell'}/\tilde{\sigma}^\text{g}_{\ell'}<\Theta$ implies that axions with mass $\nu_{\ell'}$ and coupling $|g^\text{min}_\gamma|_{\ell'}$ are excluded with confidence $c_1$, and axions with the same mass but larger coupling are excluded at higher confidence.} The validity of these expressions hinges crucially on our ability to regard each grand spectrum bin as a sample drawn from a Gaussian distribution with known mean and standard deviation. I demonstrated in Sec.~\ref{sub:correlations} that we are justified in treating any bin that does not contain an axion signal in this way. In Sec.~\ref{sub:axion_atten}, I will show that we can also quantify the mean and standard deviation for any bin containing an axion signal, thus validating the above procedure. Then I will return to the choice of threshold in Sec.~\ref{sub:target_confidence}.

\subsection{SG filter-induced attenuation}\label{sub:axion_atten}
In Sec.~\ref{sub:correlations} I claimed that with a KSVZ axion signal in the grand spectrum bin $\ell'$, $\delta^\text{g}_{\ell'}/\tilde{\sigma}^\text{g}_{\ell'}$ is a Gaussian random variable with mean given by Eq.~\eqref{eq:snr_g_tilde} and standard deviation 1. Let us consider each of the claims here more carefully. In writing Eq.~\eqref{eq:snr_g_tilde} I have implicitly assumed that $\eta$ is frequency-independent. While we could of course write a similar expression with $\eta\rightarrow\eta_\ell$, the utility of Eq.~\eqref{eq:snr_g_tilde} lies in the fact that we only need to specify a single correction factor to know the SNR in each bin. It is reasonable to expect $\eta'$ (and thus $\eta$) to be frequency-independent, as $\eta'\neq1$ is ultimately a consequence of the same imperfect SG filter stopband attenuation that led to frequency-independent $\xi\neq1$. We will see more directly that $\eta$ is constant below.

In claiming that the distribution of excess power about the mean value $\tilde{R}^\text{\,g}_{\ell'}$ is Gaussian with standard deviation 1, I am only assuming that the statistical fluctuations of the total noise power in any given bin are independent of whether or not that bin also includes excess power due to axion conversion. This is certainly a valid assumption for the raw data. We quantify $\eta$ using a simulation which will also demonstrate explicitly that this assumption still holds in the grand spectrum.

The simulation we use to quantify $\eta$ begins by defining a set of $m$ uniformly spaced simulated mode frequencies $\nu_{ci}$ and a frequency axis for a 14020-bin spectrum with resolution $\Delta\nu_b=100$~Hz centered on each mode frequency. With a tuning step size $\delta\nu_c=1.402\text{ MHz}/m$, the low-frequency end of the last spectrum lines up with the high-frequency end of the first, and $m_k$ (the number of spectra contributing to the $k$th combined spectrum bin) will vary from 1 to $m$ over the tuning range. Each spectrum is initialized to the expected signal power for an axion with coupling $|g_\gamma|$ and mass $\nu_a$ near the middle of the simulated frequency range. The signal power in the $j$th bin of the $i$th spectrum is proportional to the integral of Eq.~\eqref{eq:f_dist} over an interval $\Delta\nu_b$ around the RF frequency $\nu_k$ for which $\Gamma_{ijk}=1$, multiplied by the \textit{inverse} of the rescaling factor defined in Sec.~\ref{sub:rescale}. For simplicity we take $Q_{Li}$, $C_i$, $\beta_i$, and $N_{ij}$ to be the same for each spectrum $i$, so that variation in the rescaling factor only comes from the $j$-dependence of $N_{ij}$ and the Lorentzian mode profile.

After the initialization described above, each iteration of the simulation adds simulated Gaussian white noise with mean 1 and standard deviation $\sigma^\text{p}$ to each spectrum, and sends the full set of spectra through two analysis pipelines in parallel. The ``standard'' analysis multiplies each spectrum by a random sample baseline (see Sec.~\ref{sub:correlations}), then applies the baseline removal procedure of Sec.~\ref{sec:baseline} to obtain simulated processed spectra, and finally combines the simulated spectra both vertically and horizontally, following the procedure of Sec.~\ref{sec:rescale_combine}--\ref{sec:rebin}, to obtain a simulated grand spectrum. The ``ideal'' analysis is identical except that it bypasses the steps that imprint and then remove the baseline; thus we expect no effects associated with the SG filter in the ideal grand spectrum.

From each iteration, we record the values of the normalized power excess $\delta^\text{g}_{\ell}/\sigma^\text{g}_{\ell}$ (not $\delta^\text{g}_{\ell}/\tilde{\sigma}^\text{g}_{\ell}$) and the uncorrected SNR $R^\text{\,g}_{\ell}$ in $\approx 2K$ bins around $\nu_{\ell'}\approx\nu_{a}$ in both the standard and ideal grand spectra. We also record the value of $\delta^\text{g}_{\ell}/\sigma^\text{g}_{\ell}$ in a few other bins far from $\nu_{\ell'}$ in different parts of the standard grand spectrum. The coupling $|g_\gamma|$ is chosen to yield $R^\text{\,g}_{\ell'}\approx5$ for $m\approx200$. We let the simulation run for $\sim10^4$ iterations, after which we can histogram the distribution of any of the recorded bins across iterations. 

\begin{figure}[h]
\centering\includegraphics[width=0.7\textwidth]{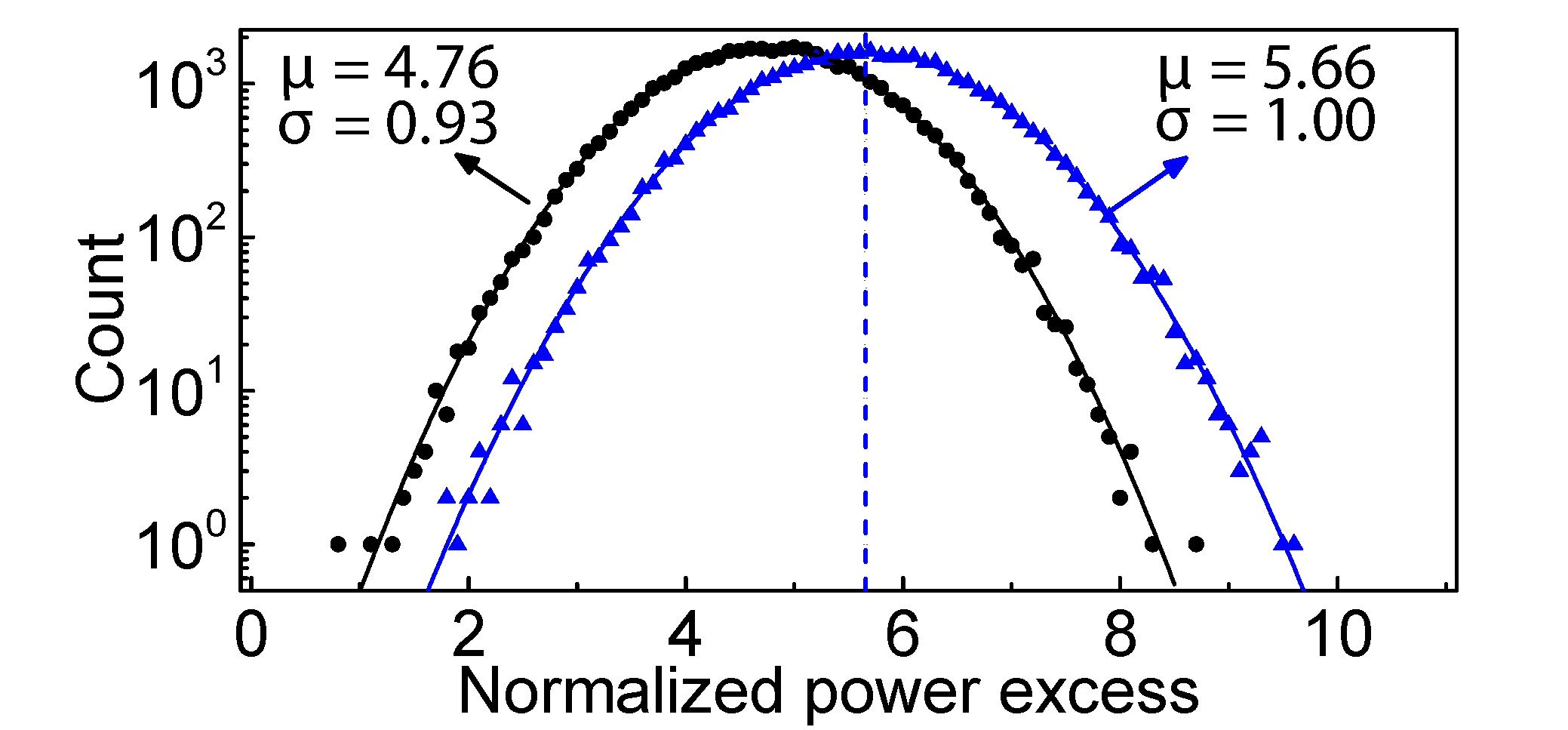}
\caption[SG filter-induced signal attenuation]{\label{fig:simu_hist} The results of a simulation to quantify the filter-induced attenuation $\eta$ of the axion search SNR for a simulated axion signal with $\nu_a=\nu_{\ell'}$ in the presence of Gaussian white noise. The grand spectrum power excess $\delta^\text{g}_{\ell'}/\sigma^\text{g}_{\ell'}$ is obtained in each iteration of the simulation using two different analysis pipelines. For each analysis, we have histogrammed the distribution of excess power across iterations of the simulation (data points) and fit the histogram with a Gaussian (solid curves); the best-fit mean and standard deviation are shown on the plot. When the contributing spectra are combined and rebinned directly (blue triangles), the distribution is Gaussian with standard deviation 1 and mean equal to the calculated SNR $R^\text{\,g}_{\ell'}$ (indicated by the dashed line). When each contributing spectrum is scaled by an empirical baseline and the standard analysis procedure is then applied (black circles), the distribution is still Gaussian but with a smaller standard deviation $\xi=0.93$, equal to the value obtained in real data. The ratio of the two mean values is $\eta'$, from which we obtain $\eta=\eta'/\xi=0.90$.}
\end{figure}

We are primarily interested in comparing the excess power distribution in bin $\ell'$ of the standard grand spectrum to the excess power distribution in the same bin of the ideal grand spectrum. This comparison is shown in Fig.~\ref{fig:simu_hist}. We see that in the ideal grand spectrum, the fluctuations of the noise power in the bin $\ell'$ containing an axion signal are Gaussian with standard deviation $\sigma^\text{g}_{\ell'}$, as they would be in any other bin; we can also see that our standard analysis procedure correctly calculates the SNR $R^\text{\,g}_{\ell}$ in the absence of SG filter effects.\footnote{That is, $E[\delta^\text{g}_{\ell'}/\sigma^\text{g}_{\ell'}]_\text{i} = (R^\text{\,g}_{\ell'})_\text{i} = (R^\text{\,g}_{\ell'})_\text{s}$, where the subscripts ``s'' and ``i'' refer to the standard and ideal analyses, respectively. The calculated values of $(R^\text{\,g}_{\ell})_\text{i}$ and $(R^\text{\,g}_{\ell})_\text{s}$ are nearly equal in each bin $\ell$ because they only depend on the measured data through the distribution of processed spectrum standard deviations (Sec.~\ref{sub:stats}), which is changed only very marginally by the presence of the SG filter.}

In the standard grand spectrum, we find that the fluctuations of the noise power in bin $\ell'$ are still Gaussian, with a reduced standard deviation $\tilde{\sigma}^\text{g}_{\ell'}=\xi\sigma^\text{g}_{\ell'}$ and $\xi=0.93$ as in real data. We also obtain Gaussian fluctuations with standard deviation $\tilde{\sigma}^\text{g}_{\ell}$ in other bins $\ell$ far from the axion mass. This provides strong evidence for the assertion that \textit{each} $\delta^\text{g}_{\ell}$ is a Gaussian random variable with standard deviation $\tilde{\sigma}^\text{g}_{\ell}$, whether or not bin $\ell$ contains an axion signal. Since we histogrammed $\delta^\text{g}_{\ell'}/\sigma^\text{g}_{\ell'}$ rather than $\delta^\text{g}_{\ell'}/\tilde{\sigma}^\text{g}_{\ell'}$ to more directly see the effects of the SG filter on $\sigma^\text{g}_{\ell'}$, the ratio of the two bin $\ell'$ excess power distributions measures $\eta'$ rather than $\eta$; formally, $E[\delta^\text{g}_{\ell'}/\sigma^\text{g}_{\ell'}]_\text{s}/E[\delta^\text{g}_{\ell'}/\sigma^\text{g}_{\ell'}]_\text{i}=(\mu^\text{g}_{\ell'})_\text{s}(R^\text{\,g}_{\ell'})_\text{s}/(R^\text{\,g}_{\ell'})_\text{i}=\eta'$. Dividing the value of $\eta'$ obtained this way by $\xi$ we find $\eta=0.90$.

This result for $\eta$ is independent of $m$ out to at least $m=400$ (c.f.\ the analogous result for $\xi$ from the simulation described in Sec.~\ref{sub:correlations}). It also does not change if we vary $|g_\gamma|^2$ by $\pm50\%$; this linearity implies that we do not have worry about the simulation reproducing the precise value of $R_T$ to be used in the analysis. Finally, $\eta$ is independent of the misalignment of $\nu_a$ relative to the grand spectrum binning: with arbitrary misalignment $E[\delta^\text{g}_{\ell'}/\sigma^\text{g}_{\ell'}]_\text{i}\neq R^\text{\,g}_{\ell'}$, but $E[\delta^\text{g}_{\ell'}/\sigma^\text{g}_{\ell'}]_\text{s}$ always changes by the same factor.\footnote{For the simulation plotted in Fig.~\ref{fig:simu_hist}, we set $\nu_a$ to coincide with a bin boundary in the rebinned spectrum, and used $L_q(\delta\nu_r=0)$ rather than $\bar{L}_q$ in the grand spectrum weights. This choice made it simpler to confirm $E[\delta^\text{g}_{\ell'}/\sigma^\text{g}_{\ell'}]_\text{i} = R^\text{\,g}_{\ell'}$ and thereby verify the correct implementation of the analysis procedure (recall that with the lineshape $\bar{L}_q$, $E[\delta^\text{g}_{\ell'}/\sigma^\text{g}_{\ell'}]_\text{i} = R^\text{\,g}_{\ell'}$ if we average over the range of possible misalignments, but is not necessarily true for any given misalignment). We confirmed that we obtain the same value of $\eta$ using $\bar{L}_q$ in the grand spectrum weights.}

Taken together, the results of the simulation are entirely consistent with the interpretation of $\eta\neq1$ as a result of the imperfect stopband attenuation of the SG filter. Thus we conclude that Eq.~\eqref{eq:snr_g_tilde} correctly describes the SNR in each bin of the grand spectrum. Although we have seen that filter-induced attenuation is a small effect, we may still ask whether we can avoid this slight SNR degradation by using different SG filter parameters. This question is explored further in appendix~\ref{app:sg_params}. 

\subsection{Setting the threshold}\label{sub:target_confidence}
We now return to the question of how we set appropriate values for $c_1$ and $\Theta$. In many subfields of particle physics it is conventional to cite parameter exclusion limits at $90\%$ or $95\%$ confidence. For the analysis of the first HAYSTAC data run we set $c_1=0.95$, for which Eq.~\eqref{eq:threshold} becomes $R_T - \Theta = 1.645$.

\begin{figure}[h]
\centering\includegraphics[width=0.8\textwidth]{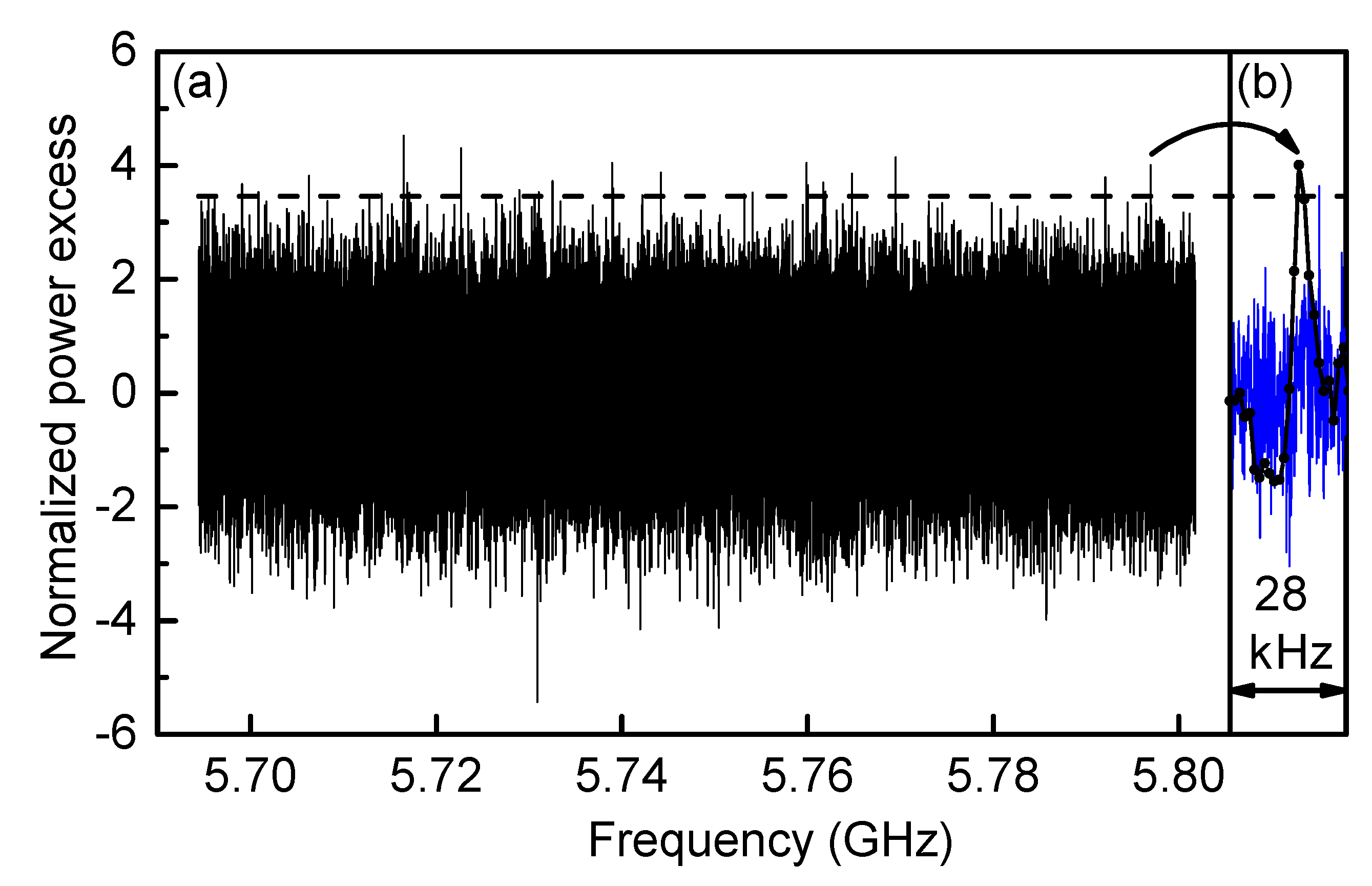}
\caption[The corrected grand spectrum and candidate threshold]{\label{fig:threshold} (a) The corrected grand spectrum $\delta^\text{g}_\ell/\tilde{\sigma}^\text{g}_\ell$ plotted as a function of frequency $\nu_\ell$ along with the (frequency-independent) threshold $\Theta=3.455$. The 28 rescan candidates are those bins for which $\delta^\text{g}_\ell/\tilde{\sigma}^\text{g}_\ell\geq\Theta$; some are hard to see because of the finite line thickness. (b) In black, the corrected grand spectrum in a small region around the highest-frequency rescan candidate. The vertical scale is the same as in (a) and the horizontal scale has been expanded by a factor of $\sim500$. In blue, the combined spectrum $\delta^\text{c}_k/\sigma^\text{c}_k$ in the same frequency range. As expected, the large power excess at the candidate frequency in the grand spectrum is due to $\sim CK$ consecutive combined spectrum bins in which the power excess is on average slightly positive, rather than a few combined spectrum bins with extremely high power excess.}
\end{figure}

Given a value for $c_1$, the considerations that enter into the choice of $\Theta$ are best illustrated with an explicit example. For the first HAYSTAC data run we chose $\Theta=3.455$, corresponding to a threshold SNR of $R_T=5.1$. $S=28$ grand spectrum bins exceeded this threshold and were flagged as rescan candidates; they are listed in Tab~\ref{tab:rescans}. The corrected grand spectrum $\delta^\text{g}_\ell/\tilde{\sigma}^\text{g}_\ell$ and threshold $\Theta$ are shown in Fig.~\ref{fig:threshold}. Visual inspection suffices to demonstrate qualitatively the important point that many of the candidates are quite marginal; more precisely 11 of the 28 candidates exceed the threshold by less than $\Delta\Theta=0.1$, implying that we could have eliminated all of these candidates at the cost of a $\Delta G_\ell/G_\ell = [(R_T+\Delta\Theta)/R_T]^{1/2}-1\approx 1\%$ degradation of our exclusion limit. Conversely, reducing $\Theta$ by 0.1 would have improved the exclusion limit by 1\% at the cost of 10 additional rescan candidates.

Of course, this strong dependence of the rescan yield on the threshold is just what we expect from Gaussian noise statistics.\footnote{One consequence of the sensitivity of $S$ to small changes in $\Theta$ is that the rescan lists for even relatively similar analyses (characterized by e.g., slightly different choices of $K$ and/or $C$, or WF instead of ML weights) typically only overlap by $\sim60-80\%$.} It is common for haloscope searches to set $R_T=5$ in estimates of the sensitivity that can be achieved with a given set of design parameters, but there is nothing special about this choice. In principle, $\Theta$ (and thus $R_T$) should be chosen to optimize the coupling sensitivity at fixed \textit{total} integration time (initial scan plus rescans). For any haloscope detector using a coherent receiver, rescans are intrinsically less efficient than the initial scan, so the time spent on rescans should be a small fraction of the time spent acquiring the initial scan data.\footnote{This is because each measurement improves the SNR in $\sim\Delta\nu_c/\Delta\nu_a$ non-overlapping grand spectrum bins simultaneously. In the initial scan each of these bins is relevant whereas in rescans we only care about the SNR within $K$ bins of each candidate. In practice the discrepancy is further exacerbated by the fact that rescans are more difficult to fully automate than the continuous initial scan, and thus have worse live-time efficiency.} By this criterion, the optimal threshold is higher still than the value $\Theta=3.455$ we adopted for the first HAYSTAC data run.

Thus far in this section I have discussed the real data rescan yield $S(\Theta)$ without reference to any theoretical model. To confirm that we have obtained a rescan yield consistent with statistics, we must take into account the fact that any two grand spectrum bins $\ell$ and $\ell'$ will be positively correlated if $|\ell-\ell'|\leq K-1$ because the segments of the rebinned spectrum contributing to the bins $\ell$ and $\ell'$ will overlap. These correlations imply that both real axion signals and statistical fluctuations in the excess power are likely to result in several adjacent bins exceeding the threshold. We should not define all such bins as rescan candidates because they are largely redundant. Thus we add bins to the list of rescan candidates in order of decreasing excess power, and remove $K-1$ bins on either side of each candidate from consideration before moving on to the next candidate. The values of $S(\Theta)$ cited above were obtained using this procedure, which was originally proposed by Ref.~\citep{daw1998}. 

Recall that $\hat{S}(\Theta)$ defined by Eq.~\eqref{eq:candidates} describes the expected rescan yield for a grand spectrum whose bins are samples drawn from a standard normal distribution: it does not depend on whether or not nearby bins are correlated provided $n$ is much larger than the correlation length. Thus, Eq.~\eqref{eq:candidates} would correctly describe the expected rescan yield if we did not exclude the correlated bins around each candidate; given that we do exclude these bins, we should actually expect a rescan yield $\bar{S}(\Theta)<\hat{S}(\Theta)$. Note also that though the presence of grand spectrum correlations affects the rescan yield, it does not affect the initial scan confidence level $c_1$.\footnote{For any value of $\nu_a$ within our scan range, there will be some grand spectrum bin $\ell'$ in which the SNR is maximized, and the best limits we can set will come from this bin. $R_T$ is the SNR in bin $\ell'$ if the axion has mass $\nu_a$ and coupling $|g_\gamma| = |g^\text{min}_\gamma|_{\ell'}$ (up to an uncertainty quantified in appendix~\ref{app:error}); it follows from Eq.~\eqref{eq:threshold} that if bin $\ell'$ does not exceed the threshold $\Theta$, we can exclude such axions with confidence $c_1$. The non-observation of excess power above the threshold in adjacent correlated bins just gives us an additional, strictly less restrictive constraint on the coupling of the axion with mass $\nu_a$.} Our procedure for cutting correlated bins from the rescan yield \textit{will} affect the rescan analysis procedure, discussed in Sec.~\ref{sub:rescan_analysis}.

We obtain the $\Theta$-dependence of the expected rescan yield $\bar{S}(\Theta)$ from a simple simulation. We generate a simulated rebinned spectrum containing Gaussian white noise, apply the ML-weighted sum of Sec.~\ref{sub:grand_spectrum} to obtain a simulated grand spectrum, and then flag rescan candidates with the same procedure used for real data, cutting $K-1$ bins on either side of each candidate. We repeat this simulation with different values of $\Theta$ between 2.3 and 4.3, and then repeat it $\approx50$ times at our chosen value of $\Theta=3.455$ to obtain a range of probable values for $\bar{S}$.

From this simulation we obtain $\bar{S}(\Theta)$ consistently smaller than $\hat{S}(\Theta)$ as expected: at $\Theta=3.455$, $\hat{S}=29.5$, $\bar{S} = 24\pm5$ and $S=28$.\footnote{The fact that $S$ is closer to $\hat{S}$ than $\bar{S}$ at $\Theta=3.455$ is just a fluke made possible by the small candidate statistics at such a large threshold. At $\Theta=2.5$, for example, we would have $S=396$, $\bar{S}=372$, and $\hat{S}=588$. Note that for any value of $\Theta$, $\hat{S} > \bar{S} > \hat{S}/K$, where the latter is the rescan yield we would obtain from $n/K$ uncorrelated bins (this was also noted by Ref.~\citep{daw1998}). The second inequality gets at the reason (anticipated in Sec.~\ref{sec:rebin}) that we did not take $C=1$ and $K=50$ in constructing the grand spectrum: the number of rescan candidates would be much larger at comparable sensitivity even after we ensure that no two candidates fall within $K$ bins of each other.} We conclude that the observed rescan yield $S$ is consistent with statistics -- by itself this result does not disfavor the hypothesis that any of our candidates could be a real axion signal, since the expected variation in $\bar{S}$ is larger than one, and we expect at most one axion in the data set. To settle the question one way or another, we now turn to the acquisition and analysis of rescan data around each candidate.

\section{Rescan data and analysis}\label{sec:rescan}
Three numbers are required to fully characterize each of the $S=28$ candidates obtained from the initial scan data set: the signal frequency $\nu_{\ell(s)}$, the threshold coupling $G_{\ell(s)}$ (relative to the KSVZ coupling), and the properly normalized power excess $\delta^\text{g}_{\ell(s)}/\tilde{\sigma}^\text{g}_{\ell(s)}$, where $\ell(s)$ is the index of the grand spectrum bin that exceeded the threshold and $s=1,\dots,S$. Only the first two quantities explicitly appear in our subsequent analysis (though of course the power excess determines whether any given bin is flagged as a rescan candidate in the first place). The rescan candidates are listed in Tab.~\ref{tab:rescans}.

\begin{table}[htbp]
\centering
\begin{tabular}{|c|c|c|c|}
\hline
\rule{0pt}{2.8ex} $\boldsymbol{s}$ & $\boldsymbol{\nu_{\ell(s)}~[\text{GHz}]}$ & $\boldsymbol{G_{\ell(s)}}$ & $\boldsymbol{\delta^\text{g}_{\ell(s)}/\tilde{\sigma}^\text{g}_{\ell(s)}}$ \\ \hline
\rule{0pt}{2.8ex} 1 & 5.79697 & 2.62 & 4.00\\ \hline
\rule{0pt}{2.8ex} 2 & 5.79207 & 2.51 & 3.79\\ \hline
\rule{0pt}{2.8ex} 3 & 5.76952 & 2.33 & 4.14\\ \hline
\rule{0pt}{2.8ex} 4 & 5.76479 & 2.33 & 3.86\\ \hline
\rule{0pt}{2.8ex} 5 & 5.76195 & 2.22 & 3.54\\ \hline
\rule{0pt}{2.8ex} 6 & 5.76168 & 2.21 & 3.69\\ \hline
\rule{0pt}{2.8ex} 7 & 5.76003 & 2.25 & 3.65\\ \hline
\rule{0pt}{2.8ex} 8 & 5.75986 & 2.25 & 4.04\\ \hline
\rule{0pt}{2.8ex} 9 & 5.75406 & 2.16 & 3.52\\ \hline
\rule{0pt}{2.8ex} 10 & 5.75316 & 2.05 & 3.50\\ \hline
\rule{0pt}{2.8ex} 11 & 5.74418 & 2.03 & 3.87\\ \hline
\rule{0pt}{2.8ex} 12 & 5.74222 & 2.25 & 3.48\\ \hline
\rule{0pt}{2.8ex} 13 & 5.73908 & 2.26 & 3.60\\ \hline
\rule{0pt}{2.8ex} 14 & 5.73898 & 2.28 & 4.04\\ \hline
\rule{0pt}{2.8ex} 15 & 5.73251 & 2.31 & 3.73\\ \hline
\rule{0pt}{2.8ex} 16 & 5.73105 & 2.31 & 3.53\\ \hline
\rule{0pt}{2.8ex} 17 & 5.73060 & 2.29 & 3.46\\ \hline
\rule{0pt}{2.8ex} 18 & 5.72897 & 2.29 & 3.56\\ \hline
\rule{0pt}{2.8ex} 19 & 5.72648 & 2.19 & 3.46\\ \hline
\rule{0pt}{2.8ex} 20 & 5.72269 & 2.25 & 4.30\\ \hline
\rule{0pt}{2.8ex} 21 & 5.71711 & 2.53 & 3.52\\ \hline
\rule{0pt}{2.8ex} 22 & 5.71691 & 2.27 & 3.69\\ \hline
\rule{0pt}{2.8ex} 23 & 5.71652 & 2.33 & 4.53\\ \hline
\rule{0pt}{2.8ex} 24 & 5.71413 & 2.33 & 3.50\\ \hline
\rule{0pt}{2.8ex} 25 & 5.70634 & 2.21 & 3.82\\ \hline
\rule{0pt}{2.8ex} 26 & 5.70087 & 2.42 & 3.53\\ \hline
\rule{0pt}{2.8ex} 27 & 5.69911 & 2.57 & 3.67\\ \hline
\rule{0pt}{2.8ex} 28 & 5.69613 & 2.72 & 3.47\\ \hline
\end{tabular}
\caption[Rescan candidates from the first HAYSTAC data run]{\label{tab:rescans} The frequency $\nu_{\ell(s)}$, threshold coupling $G_{\ell(s)}$ relative to the KSVZ coupling, and properly normalized power excess $\delta^\text{g}_{\ell(s)}/\tilde{\sigma}^\text{g}_{\ell(s)}$ for each of the $S=28$ rescan candidates from the first HAYSTAC data run. Note that $G_{\ell(s=21)}$ is unusually large because the candidate frequency coincided with reduced coverage resulting from the presence of synthetic signal injections in the winter scan. No candidates persisted in the rescan data.}
\end{table}

To establish whether any of our rescan candidates is persistent, we must first determine for each candidate the rescan time $\tau^*_s$ required to obtain SNR $R^*_T$ for an axion signal at frequency $\nu_{\ell(s)}$ with coupling $G_{\ell(s)}$.\footnote{In general, I will use the superscript $^*$ to denote previously defined quantities whose values differ in the rescan analysis.} Then we can acquire rescan data at each candidate frequency. The considerations that enter into these steps are described in Sec.~\ref{sub:rescan_daq}.

We can imagine two alternative approaches to processing the rescan data. One possibility is to process the rescan and initial scan data sets together to produce a single combined spectrum, from which we obtain a modified grand spectrum by following the procedure in Sec.~\ref{sec:rebin}. The extra integration time at each candidate frequency implies that each $\tilde{R}^\text{\,g}_{\ell(s)}$ will increase by roughly a factor of $\sqrt{2}$. Since we are interested in probing the same value of $G_{\ell(s)}$, we can impose a higher threshold $\Theta^*_{\ell(s)}$ around each candidate. We can thus ensure that a real axion signal exceeds the new threshold with some desired confidence $c_2$, while simultaneously greatly reducing the probability that a statistical fluctuation does so.

Alternatively, we can process the rescan data separately, following the procedure of Sec.~\ref{sec:baseline} -- \ref{sec:rebin} to produce a rescan grand spectrum, and leaving the initial scan grand spectrum unchanged. The rescan data set should allow us to set a \textbf{coincidence threshold} $\Theta^*_{\ell(s)}$ around each candidate frequency which a real axion signal should exceed with confidence $c_2$. If $c_2\approx c_1$, we do not expect $\Theta^*_{\ell(s)}$ to be substantially greater than $\Theta$ in this case, so the probability that a statistical fluctuation exceeds the threshold in any given bin will not change, but it is much less likely that this should happen in any of the same bins as in the initial scan.

If no changes to the analysis procedure are required for rescan data, these two approaches are completely equivalent. Here we take ``separate processing'' approach, which is conceptually cleaner in that we process spectra together whenever we want to \textit{improve} the coupling sensitivity $|g^\text{min}_\gamma|$ and separately when we want to \textit{reproduce} the coupling sensitivity of a previous scan. As we will see in Sec.~\ref{sub:rescan_analysis}, the rescan analysis differs from the initial scan analysis in a few crucial respects, such that we must use separate processing to obtain correct expressions for the coincidence thresholds $\Theta^*_{\ell(s)}$.

\subsection{Rescan data acquisition}\label{sub:rescan_daq}
The most efficient way to acquire rescan data at the candidate frequency $\nu_{\ell(s)}$ is to take one long measurement with the axion-sensitive cavity mode fixed at frequency $\nu_{cs}\approx\nu_{\ell(s)}$. We can calculate the integration time $\tau^*_s$ required to obtain SNR $R^*_T$ by starting with an expression analogous to \eqref{eq:snr_r_approx} and using Eqs.~\eqref{eq:power_ij}, \eqref{eq:uw_limit}, \eqref{eq:snr_g_tilde}, and \eqref{eq:g_ell}. The result is
\begin{equation}
\tau^*_s= \frac{1}{1-\epsilon}\Bigg[\frac{R^*_ThN_sH\big(\delta\nu_{as}\big)}{\eta^*F_\text{ML}G^2_{\ell(s)}U_0C_sQ_{Ls}\frac{\beta_s}{1+\beta_s}}\Bigg]^2,\label{eq:tau_*}
\end{equation}
where $\eta^*=0.76$ is the filter-induced attenuation for the rescan analysis (see Sec.~\ref{sub:rescan_analysis}), the system noise $N_s$ is evaluated in the middle of the analysis band, and I have lumped all dependence on the detuning $\delta\nu_{as}=\nu_{cs}-\nu_{\ell(s)}$ into the factor $H(\delta\nu_{as})$ normalized so that $H(0)=1$; I  have also assumed that only a fraction $1-\epsilon$ of the integration time at each cavity setting $\nu_{cs}$ will contribute to improving the SNR at the candidate frequency.

Eq.~\eqref{eq:tau_*} indicates that in order to know how long to integrate at each candidate frequency, we must estimate the values of the parameters $Q_{Ls}$, $\beta_{s}$, $C_s$, and $N_s$ (see Sec.~\ref{sub:rescale}) and the detuning $\delta\nu_{as}$ between the mode and candidate frequencies. If the true value any of these parameters during the rescan measurement deviates from the value we assume in the calculation of $\tau^*_s$, the true SNR $\hat{R}^*_{\ell(s)}$ calculated from the rescan data (see Sec.~\ref{sub:rescan_analysis}) will deviate from the target value $R^*_T$. 

This observation motivates the question of what nominal value to assign to $R^*_T$ in Eq.~\eqref{eq:tau_*} -- there is no \textit{a priori} reason we must set $R^*_T= R_T$. Note that $\hat{R}^*_{\ell(s)}\neq R^*_T$ for any given candidate is not a problem provided that the probability $p_s$ of a statistical fluctuation exceeding the corresponding coincidence threshold $\Theta^*_{\ell(s)}$ remains $\ll 1$. This probability may be roughly estimated as
\begin{equation}
p_s=n_K\big[1-\Phi\big(\Theta^*_{\ell(s)}\big)\big],\label{eq:p_s}
\end{equation}
where
\begin{equation}
\Theta^*_{\ell(s)} =\hat{R}^*_{\ell(s)} - \Phi^{-1}(c_2)\label{eq:threshold_*}
\end{equation}
and I have defined an effective number of independent bins $1<n_K<2K-1$ to account for the fact that we will reject the hypothesis of an axion in bin $\ell(s)$ only if $\delta^{\text{g}*}_\ell/\tilde{\sigma}^{\text{g}*}_\ell$ exceeds the appropriate coincidence threshold in neither the original bin $\ell(s)$ nor any of the $(K-1)$ correlated bins on either side (see discussion in Sec.~\ref{sub:rescan_analysis}). $n_K = 1$ $(n_K = 2K-1)$ would correspond to treating the $2K-1$ bins associated with each candidate as perfectly correlated (uncorrelated); the appropriate value is clearly somewhere in between these two extremes.

We would like to demand that $\sum_sp_s \ll 1$ in order to avoid a second round of rescans in the absence of axion signals. If we assume for now that $\Theta^*_{\ell(s)}$ will not vary too much around the nominal value obtained by taking $\hat{R}^*_{\ell(s)}\rightarrow R^*_T$ in Eq.~\eqref{eq:threshold_*}, we should set 
\begin{equation}
R^*_T = \Phi^{-1}\Big(1-\Big[\sum_sp_s/(S\times n_K)\Big]\Big)+\Phi^{-1}(c_2).\label{eq:snr_t_*}
\end{equation}
For the first HAYSTAC analysis, we estimated $n_K\approx K$ and demanded that simultaneously $\sum_sp_s\leq0.05$ and $c_2=0.95$; with these choices, Eq.~\eqref{eq:snr_t_*} yields $R^*_T=5.03$ (equivalently, $\Theta_{\ell(s)}^*\approx3.28$).

Next we need to specify how we evaluate the other parameters that enter into the calculation of $\tau^*_s$. For each candidate we set $N_s$ by averaging $N_{ij}$ [Eq.~\eqref{eq:noise_ij}] over all initial scan $Y$-factor measurements $i$, and evaluating the average in the IF bin $j$ corresponding to the cavity resonance. The form factor $C_s$ and the unloaded cavity quality factor $Q_{0s}$ depend deterministically on the cavity frequency (see Figs.~\ref{fig:c010} and \ref{fig:Q0}); $Q_{Ls}=Q_{0s}/(1+\beta_s)$ then follows from our ability to control the cavity-receiver coupling $\beta$ by adjusting the antenna insertion. We set $\beta_s=2$ (comparable to typical values of $\beta_i$ throughout the initial scan) for each candidate.\footnote{$\beta\approx2$ is optimal for a continuous data run as discussed in Sec.~\ref{sub:scan}. For a rescan measurement in which we only care about the SNR in a few bins around $\nu_{\ell(s)}$, critical coupling ($\beta=1$) is better if $\delta\nu_{as}\approx0$ and $\Delta N_\text{cav}=0$. With $\Delta N_\text{cav}\neq0$, $N_\text{sys}$ also depends on $\beta$ as noted in Sec.~\ref{sub:hotrod}. With $\beta_s=1$, we would systematically underestimate the total noise in the rescan measurement by using a value of $N_s$ obtained from initial scan measurements.}

The detuning $\delta\nu_{as}$ is trivial to measure but hard to control precisely, due to mode frequency drifts and backlash (Sec.~\ref{sub:cav_tuning}). In practice we acquired the rescan data starting with the highest-frequency candidate and tuning down: for each candidate, we tuned the TM$_{010}$ mode $\approx 100 - 200$~kHz above $\nu_{\ell(s)}$ and waited 20 minutes for the mode frequency to settle before starting the measurement. We proceeded with the measurement only if $|\delta\nu_{as}|<150$~kHz after this interval. We set $\delta\nu_{as}=0$ for each candidate in Eq.~\eqref{eq:tau_*} for simplicity; since the cavity was overcoupled and $N_\text{cav}$ (hence $N_\text{sys}$) also decreases for $\delta\nu_{as}\neq0$, $\tau^*_s$ is not too sensitive to small detunings. 

Another potentially more serious consequence of mode frequency drift is that for any given rescan iteration $s$, some or all of the processed spectrum bins contributing to the grand spectrum bin $\ell(s)$ may happen to coincide with a region of the analysis band contaminated by IF interference. We saw in Sec.~\ref{sub:badbins} that 11\% of analysis band bins were contaminated in this way -- thus there is a non-negligible chance that $\hat{R}^*_{\ell(s)}$ will be substantially smaller than the target value $R^*_T$ due to missing bins. 

We mitigate this effect by splitting the total integration time $\tau^*_s$ required for each candidate across 10 cavity noise measurements of duration $\tau^*_s/10$, and step $\nu_\text{LO}$ and $\nu_p$ together by 1~kHz (without tuning the cavity mode) between measurements. On average, we expect the candidate to fall in a contaminated part of the analysis band in about 1 of 10 such measurements: thus we set $\epsilon=0.1$ in Eq.~\eqref{eq:tau_*}.

Finally, unlike the experimental parameters discussed above, $\eta^*$ and $F_\text{ML}$ depend on fixed parameters of the rescan analysis procedure and cannot change from one rescan measurement to the next. We will see in Sec.~\ref{sub:rescan_analysis} that while  $F_\text{ML}$ will not change in the rescan analysis, $\eta^*$ will not in general be equal to $\eta$ and thus should be estimated in advance to avoid systematically biasing $\tau^*_s$. 

Applying Eq.~\eqref{eq:tau_*} to the 28 rescan candidates from the first HAYSTAC data run, we obtained rescan times $\tau^*_s$ ranging from 5.8 hours (for $s=28$ in Tab.~\ref{tab:rescans}) to 17.9 hours (for $s=11$). We had $|\nu_{\ell(s)}-\nu_{\ell(s+1)}|<200$~kHz for 3 of the 27 pairs of adjacent candidates ($s=7$, $13$, and $21$ in Tab.~\ref{tab:rescans}): thus there is a 100~kHz range for $\nu_{cs}$ in which the condition $|\delta\nu_{as}|<150$~kHz can be satisfied simultaneously for both candidates. In each of these cases, we acquired rescan data for both candidates together, taking the larger of the two calculated integration times. Thus we made 25 rescan measurements, for a total of 282 hours of rescan time (c.f.\ $N\tau=1692$~hours of initial scan time).\footnote{I will use $s$ to index quantities $a_s$ associated with each rescan measurement as well as quantities $b_s$ associated with each candidate, with the implicit understanding that in three cases, we will have $a_s=a_{s+1}$ but $b_s\neq b_{s+1}$}

At each iteration $s$, after tuning the cavity to the appropriate frequency $\nu_{cs}$ and setting $\beta_s\approx2$, we used a LabVIEW program to make 10 cavity noise measurements and acquire auxiliary data. Each cavity noise measurement was saved as an averaged power spectrum with frequency resolution $\Delta\nu_b$ as in the initial scan. The auxiliary data at each iteration comprised VNA measurements of the cavity mode in transmission and the JPA gain profile both before and after the set of cavity noise measurements, a VNA measurement of the cavity mode in reflection, and a $Y$-factor measurement. 

We use this auxiliary data to quantify $\hat{R}^*_{\ell(s)}$ as described in Sec.~\ref{sub:rescan_analysis}, and also to flag and cut anomalous iterations as in Sec.~\ref{sub:badscans}. Unlike in the initial scan analysis we must repeat any iterations we cut at this stage, to ensure that we have meaningful data around each rescan candidate. In the first HAYSTAC data run we had to repeat 6 of our 25 rescan measurements (before analyzing the data), in each case because of excessive mode frequency drift $|\nu_{c1}-\nu_{c2}|>130$~kHz.\footnote{Mode frequency drifts were generally larger than in the initial scan due to a combination of much larger tuning steps between iterations and much longer integration times. Four rescan measurements had drifts below 130 kHz but above the more conservative 60 kHz threshold used in the initial scan. The range over which the mode drifted was roughly centered on the candidate frequency $\nu_{\ell(s)}$ in these cases, so the systematic deviation from the correct ML weight for any processed spectrum bin contributing to the combined spectrum around $\nu_{\ell(s)}$ will be quite small. To bound this error we can consider the more extreme case where the mode frequency initially coincides with the candidate frequency and then drifts away slowly over the 10 subsequent measurements: with the maximum allowed drift and the minimum cavity bandwidth $\Delta\nu_{cs}$, the RMS fractional deviation from the true combined spectrum ML weights is 13\%. As noted in Sec.~\ref{sub:correlations}, the systematic effect on the combined spectrum bin values $\delta^{\text{c}*}_k$ and the SNR $R^{\text{\,c}*}_k$ will be much smaller.}

\subsection{The rescan analysis}\label{sub:rescan_analysis}
Once we have acquired a complete rescan dataset, the next step is to process and combine all rescan power spectra to produce a single rescan grand spectrum. We begin by truncating each of our 250 rescan spectra as in Sec.~\ref{sec:baseline}, normalizing each spectrum to the average baseline from the initial scan analysis, and using the list defined in Sec.~\ref{sub:badbins} to cut bins contaminated by IF interference from each spectrum.

Next we must use an SG filter to remove the residual baseline from each spectrum. At this stage it becomes important that $\tau^*_s/10>\tau$ even for the smallest value of $\tau^*_s$ obtained from Eq.~\eqref{eq:tau_*}; moreover the residual baselines for the 10 spectra from each iteration will be very similar, since we do not tune the cavity or rebias the JPA between power spectrum measurements. Thus, although the total averaging at each candidate frequency in the rescan data is comparable to the total averaging at that frequency in the initial scan, we should expect the amplitude (relative to $\sigma^\text{p}$) of any small-scale systematic structure in the rescan processed spectra to be enhanced by a factor $\sim\sqrt{\tau^*_s/\tau}$ if we use the same SG filter parameters as in the initial scan (see also discussion in appendix~\ref{app:sg_params}).

We have seen in this chapter that the statistics of the initial scan spectra are Gaussian at each stage of the processing, and in particular that the narrowing of the histogram of normalized grand spectrum bins $\delta^\text{g}_\ell/\sigma^\text{g}_\ell$ is completely explained by the stopband properties of the SG filter with parameters $d=4$ and $W=500$. This good agreement between the observed and expected statistics indicates that the amplitude of any small-scale systematic structure in the initial scan processed spectra must be $\ll\sigma^\text{p}$. The observation that baseline systematics will grow coherently over at least the single-spectrum integration time (and likely over the full rescan integration time) indicates that we cannot necessarily assume systematic structure will remain negligibly small in the rescan processed spectra. Studies of the effects of SG filters on simulated Gaussian white noise indicate that the parameters $d$ and $W$ used in the initial scan would produce unacceptable deviations from Gaussianity if applied to the rescan analysis. Thus we used an SG filter with $d^*=6$ and $W^*=300$ for the rescan analysis instead; Fig.~\ref{fig:filter} suggests that with these parameters we should expect $\xi^*<\xi$ and $\eta^*<\eta$, and we will see below that this is indeed the case. 

After applying the SG filter with parameters $d^*$ and $W^*$ to each rescan spectrum, we verify that the bins in each of the 10 processed spectra at iteration $s$ have the expected Gaussian distribution with mean 0 and standard deviation $\sigma^{\text{p}*}_s = 1/\sqrt{\Delta\nu_b\tau^*_s/10}$. We then rescale the spectra to obtain a mean power excess of 1 in any rescaled spectrum bin in which a KSVZ axion deposits a fraction $1/(CK)$ of its total conversion power. Formally, the required rescaling is given by Eqs.~\eqref{eq:delta_s} and \eqref{eq:sigma_s}, with the additional factor of $1/(CK)$ discussed at the beginning of Sec.~\ref{sub:rebinned_spectrum} absorbed into the definition of the signal power. Values for the factors in Eq.~\eqref{eq:power_ij} and Eq.~\eqref{eq:noise_ij} are obtained from the auxiliary data at each rescan measurement via the procedure described in Sec.~\ref{sub:rescale}; unlike in the initial scan analysis, no interpolation is required for $N^*_{sj}$ because we made a $Y$-factor measurement at each rescan iteration.\footnote{The $j$-dependent quantities in the rescaling factor should more properly be written with an additional index $t=1,\dots,10$ to account for the fact that the LO frequency varies across the 10 spectra at each iteration $s$. Apart from this small frequency offset, the rescaling factor is the same for all the spectra at a given iteration $s$.}

We then follow the procedure of Sec.~\ref{sub:combine} to construct a single ML-weighted combined spectrum from the set of 250 rescaled spectra. The frequency axis for the rescan combined spectrum extends from the smallest candidate frequency minus 651 kHz (i.e., half the analysis band) to the largest candidate frequency plus 651 kHz: there are thus formally a total of $1.02\times10^6$ combined spectrum bins, though about 70\% of these bins are empty because we only took data around candidate frequencies. The typical spacing between candidate frequencies is such that most (non-empty) combined spectrum bins $k$ are obtained by averaging only the $m_k=10$ spectra from a single rescan measurement. But the formal procedure of Sec.~\ref{sub:combine} also correctly treats the cases where adjacent candidates are sufficiently close that spectra from different iterations overlap, and thus $m_k>10$. As expected, the distribution of combined spectrum bins $\delta^{\text{c}*}_k/\sigma^{\text{c}*}_k$ is Gaussian with mean 0 and standard deviation 1.

Finally, we follow the procedure of Sec.~\ref{sub:rebinned_spectrum} and Sec.~\ref{sub:grand_spectrum} to obtain the rescan grand spectrum. Since we want to reproduce the initial scan sensitivity without changing any assumptions about the axion signal, we should use the same values of $C$, $K$, and $\bar{L}_q$ in Eqs.~\eqref{eq:d_r}, \eqref{eq:snr_r}, \eqref{eq:snr_g}, and \eqref{eq:delta_sigma_g}. However, we should expect $\xi^{\text{r}*}\neq\xi^\text{r}$ and $\xi^{\text{g}*}\neq\xi^\text{g}$ because we have used a different SG filter. Empirically, the distribution of rebinned spectrum bins $\delta^{\text{r}*}_\ell/\sigma^{\text{r}*}_\ell$ is Gaussian with mean 0 and standard deviation $\xi^{\text{r}*}=0.96$, and the distribution of grand spectrum bins $\delta^{\text{g}*}_\ell/\sigma^{\text{g}*}_\ell$ is Gaussian with mean 0 and standard deviation $\xi^*=0.83$, implying $\xi^{\text{g}*}=\xi^*/\xi^{\text{r}*}=0.86$.

As in the initial scan analysis, we are ultimately interested in the quantities $\delta^{\text{g}*}_\ell/\tilde{\sigma}^{\text{g}*}_\ell=\delta^{\text{g}*}_\ell/\big(\xi^*\sigma^{\text{g}*}_\ell\big)$ and $\tilde{R}^{\text{\,g}*}_\ell=\eta^*R^{\text{\,g}*}_\ell$ that have been corrected for filter effects. As before, we obtain the value of $\xi^*$ directly from the data; the value of $\eta^*$ can only be obtained from simulation, but the common origin of $\xi^*$ and $\eta^*$ and good agreement between the observed and simulated values of $\xi^*$ gives us confidence that we have applied the appropriate correction factor.

We validate the observed value of $\xi^*$ and measure $\eta^*$ using simulations very similar to the ones described in Sec.~\ref{sub:correlations} and Sec.~\ref{sub:axion_atten}, respectively. Formally, the rescan simulations only differ in two respects: we multiply each simulated white noise spectrum by the same sample baseline instead of a random sample baseline, and we assign the same mode frequency to each spectrum in the simulation to quantify $\eta^*$.\footnote{We did not assign frequency offsets to the spectra in the simulation used to measure $\xi$ in the initial scan analysis (see Sec.~\ref{sub:correlations}). The fact that we nonetheless obtained the same value of $\xi$ as in real data indicates that small changes in the baseline shape (associated with tuning the cavity or rebiasing the JPA) can suppress the growth of small-scale systematics substantially, even without frequency offsets between spectra.} We reproduced the observed values of $\xi^{\text{r}*}$ and $\xi^*$ and obtained $\eta^*=0.76$ from these simulations; we verified that the results in each case were independent of the number of averages $m$ and integration time $\tau$, at least for $m\tau\leq20$~hours, and thus independent of frequency (see discussion in Sec.~\ref{sub:correlations}).

At this point, we have obtained an explicit expression for the SNR $\tilde{R}^{\text{\,g}*}_{\ell}$ for a KSVZ axion signal in each bin $\ell$ of the rescan grand spectrum, whereas we care about the SNR for an axion signal with the threshold coupling $|g^\text{min}_\gamma|_\ell$. Naively we only care about evaluating the SNR in the $S$ bins $\ell(s)$ that passed the threshold in the initial scan, as the hypothesis that an axion signal with coupling $|g^\text{min}_\gamma|_\ell$ is present in any other bin has already been excluded with confidence $c_1$. 

The presence of grand spectrum correlations complicates this picture slightly. If several adjacent bins pass the threshold together, we associate the candidate with the bin whose power excess was largest, but in the presence of fluctuations the bin with larger power excess does not necessarily have the largest SNR. Thus it is possible in principle that the rescan candidate we have associated with bin $\ell(s)$ actually corresponds to an axion signal in any of the $2K-1$ grand spectrum bins $\ell'(s)$ correlated with $\ell(s)$. To be conservative we require \textit{each} such hypothesis be rejected with confidence $c_2$ before we can reject the candidate. The above discussion implies that we should define
\begin{equation}
\hat{R}^*_{\ell'(s)} = G^2_{\ell'(s)}\tilde{R}^{\text{\,g}*}_{\ell'(s)}\label{eq:hat_snr_*}
\end{equation}
with $\ell'(s)$ defined in the range $[\ell(s)-(K-1),\ell(s)+(K-1)]$. Values of $\hat{R}^*_{\ell'(s)}$ in the first HAYSTAC data run ranged from 4.26 to 7.19, with an average of 5.19.\footnote{The SNR was consistently above 6.4 for all the bins associated with two adjacent candidates that were separated in frequency by only 270~kHz: this was above our threshold for acquiring data for both candidates together, but still close enough that the integration at each candidate contributed significantly to the SNR for the other. The average SNR among all other candidates was 5.09, close to our target value $R^*_T=5.03$. The RMS variation in $\hat{R}^*_{\ell'(s)}$ among the bins $\ell'(s)$ associated with each candidate $s$ was typically less than 1\%, but was $\sim5\%$ in a few cases where the candidate frequency was close to a region of the grand spectrum with reduced exposure due to missing bins.} The effects of uncertainty in the factors used to calculate the rescan SNR are discussed in appendix~\ref{app:error}.

The appropriate coincidence threshold $\Theta^*_{\ell'(s)}$ for each bin correlated with each candidate is then obtained by using Eq.~\eqref{eq:hat_snr_*} in Eq.~\eqref{eq:threshold_*} with the substitution $\ell(s)\rightarrow\ell'(s)$. In the first HAYSTAC data run, $\delta^{\text{g}*}_{\ell'(s)}/\tilde{\sigma}^{\text{g}*}_{\ell'(s)}$ did not exceed $\Theta^*_{\ell'(s)}$ for any of the bins $\ell'(s)$ associated with any of our $S=28$ rescan candidates. The final result of the first HAYSTAC data run is thus a limit on the axion-photon coupling $|g_\gamma|$.

\section{Results}\label{sec:results}
\begin{figure}[h]
\centering\includegraphics[width=1.0\textwidth]{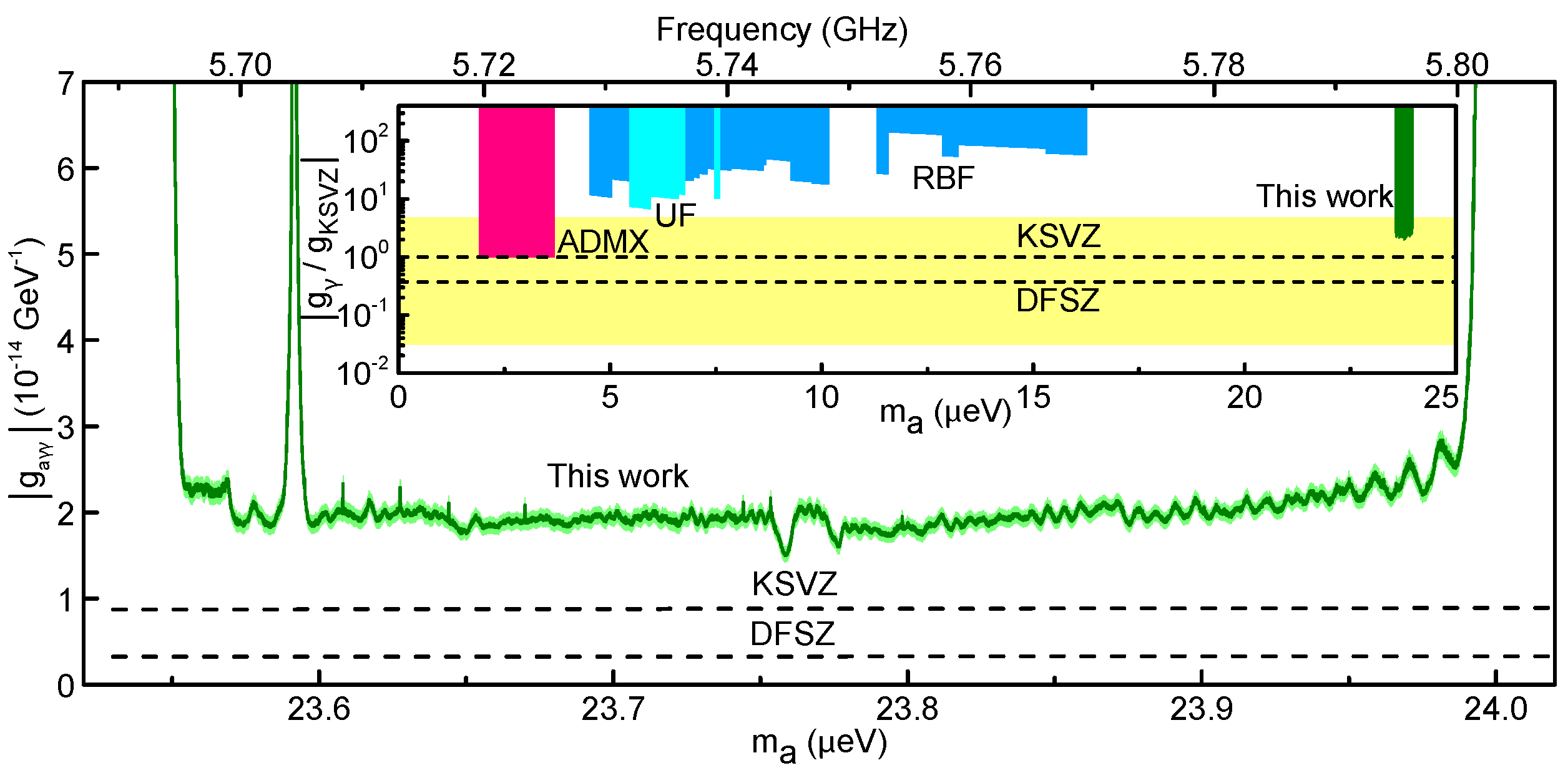}
\caption[Exclusion limit from the first HAYSTAC data run]{\label{fig:exclusion} The exclusion limit from the first HAYSTAC data run at 95\% confidence (see discussion in text). The light green shaded region is a rough estimate of our uncertainty, discussed in appendix~\ref{app:error}. The large notch near 5.704 GHz is the result of cutting spectra around the intruder mode discussed in Sec.~\ref{sub:badscans}. The narrow notches correspond to frequencies at which synthetic axion signals were injected during the winter scan (see appendix~\ref{app:fake_axions}). The inset is Fig.~\ref{fig:paramspace_lin} updated to include our limit (shown in green). The other colored regions are the axion model band (yellow, Ref.~\citep{cheng1995}) and exclusion limits from the ADMX (magenta, Refs.~\citep{ADMX1998,ADMX2002,ADMX2004,ADMX2010,ADMX2016}, $\text{C.L.} \geq 90\%$), RBF (blue, Refs.~\citep{RBF1987,RBF1989}, $\text{C.L.} = 95\%$) and UF (cyan, Ref.~\citep{UF1990,hagmann1990}, $\text{C.L.} = 95\%$) experiments.}
\end{figure}

The absence of any persistent candidates in the first HAYSTAC data run implies that $|g^\text{min}_\gamma|_\ell$ given by Eq.~\eqref{eq:g_min} may be interpreted as an exclusion limit on the dimensionless coupling $|g_\gamma|$ in each bin $\ell$ in our initial scan range. The corresponding limit on the physical coupling $|g_{a\gamma\gamma}|$ [related to $g_\gamma$ by Eq.~\eqref{eq:agammagamma}] that appears in the Lagrangian is plotted in Fig.~\ref{fig:exclusion}. Assuming an axion signal lineshape described by Eq.~\eqref{eq:f_dist}, we excluded $|g_{\gamma}|\geq2.3\times|g^\text{KSVZ}_\gamma|$ on average over the mass range $23.55 < m_a < 24.0~\mu$eV. 

What confidence should we ascribe to the exclusion of axions with the threshold coupling $|g^\text{min}_\gamma|_\ell$? Following Ref.~\citep{daw1998}, we initially chose $c_1=c_2=0.95$ to ensure the product $c_1c_2\geq0.9$, and interpreted this product as the net confidence level. But this interpretation is overly conservative, because we only acquired and analyzed rescan data at frequencies that exceeded the initial scan threshold. The hypothesis of an axion signal with the threshold coupling in any given bin is excluded with confidence $c_1$ if that bin did not exceed the initial scan threshold. In the bins correlated with each candidate, the appropriate confidence level is the conditional probability that a true axion signal would fail to exceed the coincidence threshold, having already exceeded the initial scan threshold; since the two scans are independent, this probability is just $c_2$. Thus, our result $|g^\text{min}_\gamma|_\ell$ should be interpreted as an exclusion limit at 95\% confidence.\footnote{We can equivalently interpret this result as a marginally more sensitive exclusion limit at lower confidence. Our threshold coupling at 90\% confidence would be smaller by a factor of $\big[\big(R_T-\Phi^{-1}(0.95)+\Phi^{-1}(0.9)\big)/R_T\big]^{1/2}\approx0.964$. Very recently as of this writing, discussions within the HAYSTAC collaboration have prompted closer scrutiny of the statistical underpinnings of hypothesis testing in the haloscope search. A Bayesian perspective is being investigated that might offer an alternative approach to establishing confidence levels~\citep{palken}.} 

In this chapter, I have described in detail the analysis procedure used to derive the first limits on cosmic axions from the HAYSTAC experiment. I have cited specific examples from the analysis of our first data run, but our formal procedure may easily be adapted to the analysis of data from other haloscope detectors. Throughout the preceding sections I have specifically emphasized our use of Savitzky-Golay filters to remove individual spectral baselines, our quantitative understanding of how filtering affects the statistics of the spectra, and our consistent application of maximum-likelihood weights to both the ``vertical'' sum of overlapping spectra and the ``horizontal'' sum of adjacent bins in the combined spectrum. All of these were innovations of the HAYSTAC analysis; taken together, they enable us to calculate our search sensitivity with minimal input from simulation, and obtain the relationship between sensitivity and confidence level directly from statistics. 

\chapter{Conclusion}\label{chap:conclusion}
\setlength\epigraphwidth{0.64\textwidth}\epigraph{\itshape Of course I didn't find a cube balancing on a tip under the couch. I didn't find the cube until I stepped on it the next morning.\\I did however quite literally find a missing puzzle piece --\\ and that's as much as a theoretical physicist can ask for.}{Sabine Hossenfelder}

\noindent In this dissertation I have described a laboratory search for cold dark matter in the form of axions. In chapter~\ref{chap:theory} I explained how the axion arises as a necessary consequence of the Peccei-Quinn mechanism, which remains the favored solution to the strong $CP$ problem, and the only solution capable of ensuring $\bar{\theta}=0$ to all orders in perturbation theory. The implementation of the Peccei-Quinn mechanism proved more complicated than theorists initially anticipated, but the associated axion turned out to have all the right properties to constitute dark matter. In chapter~\ref{chap:cosmo} we saw that axion cosmology depends on several quantities which are difficult to calculate and permits a number of loopholes, so a definitive prediction for the axion mass seems unlikely in the near future. Even so, given the remaining viable parameter space, it is likely that axions contribute significantly to the cosmic dark matter density $\Omega_\text{DM}$ if they exist at all. 

In chapter~\ref{chap:search} I introduced the conceptual design for a CDM axion detector called the axion haloscope. Thirty years after haloscope detection of dark matter axions was first proposed, it remains the only technique with proven sensitivity to cosmologically relevant couplings. Prior to the work described in this dissertation, haloscope experiments had only achieved this sensitivity in the few $\mu\text{eV}$ mass range, primarily because the effective detector volume $VC_{mn\ell}$ falls off rapidly with increasing frequency. In chapters~\ref{chap:detector} and \ref{chap:data} I described the design and operation of HAYSTAC, the first haloscope detector to incorporate a dilution refrigerator and a Josephson parametric amplifier and thereby demonstrate total noise $N_\text{sys}$ within a factor of three of the standard quantum limit for coherent detection. In chapter~\ref{chap:analysis} I described the analysis procedure used to derive limits on axion parameter space from data acquired in the first operation of HAYSTAC. This initial data run demonstrated that despite the challenges facing high-frequency haloscopes, a sufficiently low-noise detector can reach the model band above $20~\mu\text{eV}$. Specifically, the first HAYSTAC data run allowed us to set a limit $\left|g_\gamma\right| \gtrsim 2.3\times \left|g_\gamma^{\,\text{KSVZ}}\right|$ over the 100~MHz frequency range show in Fig.~\ref{fig:exclusion}. 

In the remainder of this brief concluding chapter, I will discuss near-term upgrades to HAYSTAC and R\&D proposals being pursued within the collaboration (Sec.~\ref{sec:haystac_future}), efforts in axion detection more broadly (Sec.~\ref{sec:other_exp}), and finally the long-term outlook for the field (Sec.~\ref{sec:outlook}).

\section{HAYSTAC upgrades and R\&D}\label{sec:haystac_future}
After the completion of the first HAYSTAC data run, the detector was taken offline to replace the rotary tuning system with a cryogenic piezoelectric actuator and improve the thermal link to the tuning rod. The results of these upgrades were briefly noted in Sec.~\ref{sub:cav_tuning} and Sec.~\ref{sub:hotrod} respectively; see also Ref.~\citep{zhong2017}. A second data run in the $5.6-5.7$~GHz range was conducted in spring/summer 2017, with analysis in progress as of this writing. In fall 2017, the operational elements of HAYSTAC will be transferred to a new BlueFors BF-LD250 dilution refrigerator with improved operational stability. The new DR also features improved vibration isolation, which should enable us to reduce the operating temperature $T_\text{mc}$ and realize a modest improvement in noise performance (see Sec.~\ref{sub:noise_offres}).

As noted in Sec.~\ref{sub:haystac}, HAYSTAC was initially conceived in part as an R\&D testbed in service of future haloscope detectors. Cavity R\&D within HAYSTAC has resulted in the development of widely tunable photonic bandgap (PBG) resonators, which do not support TE modes, and thus almost completely eliminate the mode crossings which have been the bane of efficient continuous parameter space coverage in the haloscope search. A prototype PBG resonator has been designed with a $\text{TM}_{010}$ tuning range of $7.5-9.5$~GHz, and values of $C_{010}$ and $Q_0$ comparable to more traditional cavity designs~\citep{lewis}. 

Other cavity R\&D projects within HAYSTAC focus on improving $Q_0$. One approach is to adapt the design principles of distributed Bragg reflectors, which use high-$\varepsilon$, low-loss dielectric shells to concentrate the field away from the lossy cavity walls. The prime challenge for this design is developing a tuning scheme which respects the more complex geometry. Another idea is to apply Type~II thin film superconducting coatings to the inner surface of the cavity barrel. For this to work the superconducting layer must be thin enough to remain lossless in a static 9~T field, but still exhibit good microwave reflectivity; simultaneously satisfying both of these constraints will likely require deposition of additional insulating layers~\citep{tanner}.

The first R\&D project which will be incorporated into HAYSTAC as soon as it is transfered to the new DR is a squeezed-state receiver based on the proposal in Ref.~\citep{zheng2016}. The essential idea behind the squeezed-state receiver is that a pair of JPAs driven by the same pump tone with a $90^\circ$ relative phase shift may be used to measure a single quadrature of the microwave field with sub-quantum-limited precision~\citep{mallet2011}: the first JPA squeezes the quantum noise at its input, and the second noiselessly amplifies the squeezed quadrature of the first. In practice the noise reduction achievable with such a system is limited by transmission inefficiency $\eta$ between the JPAs, which replaces a fraction $1-\eta$ of the squeezed state with unsqueezed vacuum or thermal noise.

It may not be \textit{a priori} obvious that a squeezed state receiver would improve the sensitivity of the haloscope search, as there is no way to squeeze the Johnson noise which originates in the cavity without also squeezing the axion signal to exactly the same degree. However, we have already seen that although the haloscope signal power Eq.~\eqref{eq:signal_power} is maximized on resonance with the cavity critically coupled to the receiver ($\delta\nu_a=0$, $\beta=1$), tuning steps in which the cavity is at finite detuning $\delta\nu_a\lesssim\Delta\nu_c$ from the putative signal frequency $\nu_a$ still contribute to the SNR, and indeed in thermal equilibrium the haloscope scan rate is maximized at $\beta=2$. We saw in Sec.~\ref{sub:yfactor} that with $\delta\nu_a\neq0$ and/or $\beta\neq1$, the total noise $N_\text{sys}$ comes partially from the Johnson noise of a termination at temperature $T_\text{mc}$, and this noise source can be replaced with a squeezed state without affecting the axion signal.\footnote{Recall that the axion is a noise signal, and thus rotates relative to the quadrature axes on measurement timescales $\tau\gg \tau_a$. With a single-quadrature measurement we are thus effectively measuring half of the axion signal (along with half of the input-referred noise) on average.}

We saw in a somewhat different context in Sec.~\ref{sub:hotrod} that when the off-resonant noise is colder that the noise that originates in the cavity, the scan rate is optimized at $\beta>2$. It is easy to show~\citep{zheng2016} that for any given level of squeezing (limited in practice by the transmission inefficiency $\eta$), by operating sufficiently far overcoupled we achieve sensitivity which is essentially unchanged relative to the simpler case with $\beta=2$ and no squeezing, but crucially this sensitivity is now achieved over a wider bandwidth, which implies an improved scan rate $\frac{\mathrm{d}\nu}{\mathrm{d}t}$. The goal of the prototype system being designed for HAYSTAC is to achieve 90\% transmission efficiency between the squeezer JPA and the cavity, and 90\% transmission efficiency between the cavity and the amplifier JPA ($\eta\approx0.81$ overall). If these loss specifications are realized, 8~dB squeezing and $\beta\approx8$ would result in a factor of 3.5 improvement in the scan rate at constant coupling, which is quite nontrivial given the difficulty of improving haloscope operating parameters at high frequencies (Sec.~\ref{sub:highfreq}).\footnote{Realizing this improved sensitivity will also require reconfiguring the receiver in a number of ways to optimize it for single-quadrature readout; such operational details are beyond the scope of my discussion here.}

\section{Other experiments}\label{sec:other_exp}
When I began to work on HAYSTAC in 2012, the known schemes for invisible axion and ALP detection divided themselves neatly into three classes: haloscopes which probe axion CDM, \textbf{helioscopes} attempting to detect axions produced by the sun, and experiments attempting to both produce and detect axions within the lab (with the prototypical examples being \textbf{light shining through walls (LSW)} experiments). As Fig.~\ref{fig:paramspace_big} indicates, helioscope and LSW experiments are more properly regarded as searches for generic ALPs (Sec.~\ref{sub:wisps}) than searches for axions per se. See Ref.~\citep{annurev2015} for a recent review of the present status of such searches and prospects for improving sensitivity with next-generation detectors.

In recent years, there has been a resurgence of interest in experimental probes of axions, with several new experiments proposed or in various stages of design. A prominent new player is the Center for Axion and Precision Physics (CAPP) in Daejeon, South Korea. The flagship project at CAPP is a haloscope called \textbf{CULTASK}~\citep{CAPP2016} which seeks to probe the same region of parameter space as HAYSTAC and is presently under construction. However, most new proposals have focused on parameter space inaccessible to the haloscope technique, both at low masses ($m_a \lesssim 100$~neV) and high masses ($m_a \gtrsim 50~\mu$eV).

Interest in low-mass axions was revitalized by the suggestion~\citep{graham2013} that the techniques of nuclear magnetic resonance (NMR) could be used to search for an oscillating neutron EDM induced by oscillations of the CDM axion field: this effect is a simple consequence of the fact that the axion behaves exactly like a dynamical version of $\bar\theta$ (see Sec.~\ref{sub:nedm} and Sec.~\ref{sub:pq_mechanism}). It is notable that this technique probes the anomaly-mediated coupling of the axion to QCD rather than the axion-photon coupling $g_{a\gamma\gamma}$. This suggestion matured into a proposal for the \textbf{CASPEr} experiment~\citep{casper2014}, which is presently under construction at Boston University and will probe very low masses $m_a\lesssim1~\text{neV}$, well into the anthropic regime (Sec.~\ref{sub:anthropic}). In the mass range between CASPEr and the practical $\sim\mu\text{eV}$ lower limit below which haloscopes become too large, several groups~\citep{sikivie2014,abracadabra2016} have proposed detectors based on the magnetoquasistatic limit of axion electrodynamics, in which the axion field sources an effective current oriented \textit{parallel} to an applied magnetic field, and the flux generated by this current can be measured by a pickup loop coupled to a sensitive magnetometer. A similar technique has been proposed for the detection of hidden photon dark matter in the same mass range~\citep{dmradio2016}.

Proposed techniques for the high-mass region include ``dielectric haloscopes'' which seek to achieve larger sensitive volumes at high frequencies than conventional haloscopes by exploiting discontinuities in the axion-sourced $E$-field at the interface between vacuum and high-$\varepsilon$ dielectrics~\citep{MADMAX2017}. This technique (which is being pursued by the \textbf{MADMAX} working group) bears some resemblance to an R\&D project within ADMX exploring the use of dielectrics within a cavity to deform the $E$-field profile of a higher-order mode and thus improve its form factor; the latter in turn grew out of a proposal to render higher-order modes sensitive to axions by shaping the spatial profile of the applied \textit{magnetic} field~\citep{rybka2015}. Finally, a recent paper~\citep{ariadne2014} has suggested that by combining NMR techniques with methods used to look for short-range modifications to gravity, it might be possible to detect forces \textit{mediated} by axions with $m_a\gtrsim100~\mu\text{eV}$; a prototype called \textbf{ARIADNE} has been conceived to explore this idea. Like CASPEr, ARIADNE exploits a coupling other than $g_{a\gamma\gamma}$ for axion detection, though in ARIADNE this coupling is not fixed by the QCD anomaly so it is somewhat more model-dependent. However, it should be emphasized that if the projected sensitivity is realized, ARIADNE would be sensitive to an interesting part of axion parameter space without any assumptions about whether axions constitute dark matter!

The point of this brief survey of recent proposals in axion detection is not to be comprehensive but rather to convey a sense of the recent influx of new ideas into this field. Moving beyond the domain of direct detection in the laboratory, next-generation CMB observatories may also be able to significantly constrain the allowed axion parameter space if bounds on \textit{thermal} axions continue to improve (see Sec.~\ref{sub:overclosure} and Ref.~\citep{baumann2016}).

\section{Outlook}\label{sec:outlook}
Throughout this dissertation I have taken the mystery of dark matter and the strong $CP$ problem as sufficient motivation for the haloscope search, but I would be remiss not to mention the immediate impact that axion detection would have on astronomy. If an axion signal were observed (and confirmed) in the mass range accessible to the haloscope technique, it would be relatively easy for multiple groups to quickly build detectors optimized for detection at the known value of $\nu_a$, and then the haloscope could truly live up to the last syllable in its name. 

Deviations of the spectral shape of a haloscope signal from the featureless boosted Maxwellian associated with an isothermal sphere would encode a wealth of information about the Milky Way's formation history; studying temporally resolved long integrations and spatial correlations between nearby detectors would also enable us to reconstruct a map of the local dark matter phase space. All of these topics are outside the scope of this dissertation (see Ref.~\citep{ohare2017} for further discussion), but serve to underscore how quickly elusive signals predicted by fundamental physics can find application following a first detection.\footnote{We are presently witnessing this phenomenon with the birth of ``gravitational wave astronomy'' in the wake of the 2016 LIGO detection.} 

Axion parameter space is very large, and in all likelihood it will be decades before it is fully explored.\footnote{Of course, a detection could happen at any time!} Nonetheless, the recent resurgence of interest in axion detection noted in Sec.~\ref{sec:other_exp}, along with new ideas about the role axions and the PQ mechanism might play in theoretical physics~\citep{graham2015,ballesteros2017}, provide some reason for optimism. This dissertation demonstrated the first operation of an axion haloscope integrated with a Josephson parametric amplifier: while the JPA is likely to be one of the key enabling technologies for the haloscope search in the $10 \lesssim m_a \lesssim50~\mu\text{eV}$ mass range, the impact of quantum measurement technology on axion detection is likely to be broader and more transformative still in the long run. A hypothetical future haloscope might operate with all base-temperature receiver components (i.e., switches and circulators\footnote{See Ref.~\citep{kerckhoff2015} for discussion of how such devices might be implemented using SQUID arrays as tunable linear inductors.} as well as the preamplifier) in the form of active superconducting circuits printed on a single chip, with off-chip launching only for coupling to the cavity and subsequent amplifiers. As noted in Sec.~\ref{sub:haystac}, superconducting qubits themselves may revolutionize the haloscope search by enabling the detection of high-efficiency single-photon detection at microwave frequencies. The invisible axion has tried its best to avoid detection, but it may not be able to hide forever!

\appendix 

\chapter{Axion electrodynamics}\label{app:maxwell}
\setlength\epigraphwidth{0.54\textwidth}\epigraph{\itshape The Graduate School requires that each dissertation\\be read by at least three persons but not more than five.}{Yale University}

\noindent The Lagrangian for the coupled axion and photon fields is
\begin{equation}
\lagr = \frac{1}{2}\left(\partial_\mu a\right)^2 - \frac{1}{2}m_a^2a^2 - \frac{1}{4}F_{\mu\nu}F^{\mu\nu} + \frac{1}{4}g_{a\gamma\gamma}aF_{\mu\nu}\widetilde{F}^{\mu\nu},
\label{eq:a_em_lagr}
\end{equation}
where
\[
F_{\mu\nu} = \partial_\mu A_\nu - \partial_\nu A_\mu 
\]
is the electromagnetic field strength tensor, and 
\[
\widetilde{F}^{\mu\nu} = \frac{1}{2}\epsilon^{\mu\nu\alpha\beta}F_{\alpha\beta} 
\]
is its dual. I use the sign convention $\epsilon^{0123} = +1$. Writing the Lagrangian density explicitly in terms of the photon field $A$ and re-indexing redundant terms, we obtain
\[
\lagr =  \frac{1}{2}\left(\partial_\mu a\right)^2 - \frac{1}{2}m_a^2a^2 -\frac{1}{2}\eta^{\,\mu\alpha}\eta^{\nu\beta}\left[\partial_\mu A_\nu\partial_\alpha A_\beta  - \partial_\mu A_\nu\partial_\beta A_\alpha  \right]  + \frac{1}{2}g_{a\gamma\gamma}\epsilon^{\mu\nu\alpha\beta}a\partial_\mu A_\nu\partial_\alpha A_\beta,
\]
where $\eta^{\,\mu\nu}$ is the Minkowski metric and I have used the antisymmetry of the Levi-Civita symbol. The Euler-Lagrange equations are
\begin{align*}
\pd{\lagr}{a} &- \partial_\lambda\left[\pd{\lagr}{\left(\partial_\lambda a\right)}\right] = 0, \\
\pd{\lagr}{A_\kappa} &- \partial_\lambda\left[\pd{\lagr}{\left(\partial_\lambda A_\kappa\right)}\right] = 0.
\end{align*}
The first line can be used to derive the equation of motion for the axion field, which is not usually of interest, because the coupling to electromagnetism does not appreciably perturb the free-field dynamics of halo axions. Thus I will just quote the result for reference:
\begin{equation}
\left(\partial^2 + m_a^2\right)a =  \frac{1}{8}g_{a\gamma\gamma}\epsilon^{\mu\nu\alpha\beta}F_{\mu\nu}F_{\alpha\beta}.
\label{eq:a_eom}
\end{equation}
The second Euler-Lagrange equation corresponds to modified Maxwell's equations with the axion field as a source term. The first term on the LHS contributes nothing; the second is
\begin{align*}
\pd{\lagr}{(\partial_\lambda A_\kappa)} &= -\frac{1}{2}\eta^{\,\mu\alpha}\eta^{\nu\beta}\left[\delta_\mu^\lambda\delta_\nu^\kappa\left(\partial_\alpha A_\beta  - \partial_\beta A_\alpha  \right) + \left(\delta_\alpha^\lambda\delta_\beta^\kappa - \delta_\beta^\lambda\delta_\alpha^\kappa\right)\partial_\mu A_\nu\right]  + \cdots \\
&\qquad\qquad \cdots + \frac{1}{2}g_{a\gamma\gamma}\epsilon^{\mu\nu\alpha\beta}\left[\delta_\mu^\lambda\delta_\nu^\kappa\partial_\alpha A_\beta + \delta_\alpha^\lambda\delta_\beta^\kappa\partial_\mu A_\nu \right]a \\
&= -\big[\partial^{\lambda} A^\kappa  - \partial^{\kappa} A^\lambda  \big] + \frac{1}{2}g_{a\gamma\gamma}\epsilon^{\mu\nu\lambda\kappa}\big[ \partial_\mu A_\nu -  \partial_\nu A_\mu\big]a \\
&= -F^{\lambda\kappa} + \frac{1}{2}g_{a\gamma\gamma}\epsilon^{\lambda\kappa\mu\nu}F_{\mu\nu}a.
\end{align*}
The equations of motion for the electromagnetic field are thus
\[
\partial_\alpha F^{\alpha\beta} = \frac{1}{2}g_{a\gamma\gamma}\epsilon^{\alpha\beta\mu\nu}\partial_\alpha(F_{\mu\nu}a).
\]
We can simplify these equations further by invoking Maxwell's constraint equations, which arise from the commutation of partial derivatives and are thus unaffected by the coupling to the axion field. The constraint equations have the covariant form
\begin{equation}
\partial_\alpha F_{\mu\nu} + \partial_\mu F_{\nu\alpha} + \partial_\nu F_{\alpha\mu} = 0.
\label{eq:constraint}
\end{equation}
The term in the equations of motion in which the derivative acts on $F_{\mu\nu}$ is proportional to
\begin{align*}
\epsilon^{\alpha\beta\mu\nu}\partial_\alpha F_{\mu\nu} &= \frac{1}{3}\left[\epsilon^{\alpha\beta\mu\nu}\partial_\alpha F_{\mu\nu} + \epsilon^{\mu\beta\nu\alpha}\partial_\mu F_{\nu\alpha} + \epsilon^{\nu\beta\alpha\mu}\partial_\nu F_{\alpha\mu}\right] \\
&= \frac{1}{3}\epsilon^{\alpha\beta\mu\nu}\left[\partial_\alpha F_{\mu\nu} + \partial_\mu F_{\nu\alpha} + \partial_\nu F_{\alpha\mu}\right] = 0,
\end{align*}
where I have permuted the Levi-Civita symbols in the first line. Maxwell's source equations thus become
\begin{equation}
\partial_\alpha F^{\alpha\beta} = \frac{1}{2}g_{a\gamma\gamma}\epsilon^{\alpha\beta\mu\nu}F_{\mu\nu}\partial_\alpha a.
\label{eq:em_eom}
\end{equation}
We can now rewrite our results in terms of the $\mathbf{E}$ and $\mathbf{B}$ fields using
\begin{align*}
F^{0i} = -E_i &\Rightarrow F_{0i} = E_i,\\
F^{ij} = -\epsilon_{ijk}B_k  &\Rightarrow F_{ij} = -\epsilon_{ijk}B_k.
\end{align*}
First consider the quantity
\begin{align*}
\epsilon^{\mu\nu\alpha\beta}F_{\mu\nu}F_{\alpha\beta} &= \big[\epsilon^{0123}F_{01}F_{23} + \epsilon^{1023}F_{10}F_{23} + \epsilon^{1032}F_{10}F_{32} + \epsilon^{0132}F_{01}F_{32} + \cdots \\ 
&\qquad \cdots +\epsilon^{2301}F_{23}F_{01} + \epsilon^{2310}F_{23}F_{10} + \epsilon^{3210}F_{32}F_{10} + \epsilon^{3201}F_{32}F_{01}+ \cdots\big],
\end{align*}
where I have omitted eight terms in which the indices 0 and 2 appear together in the field strength tensor and another eight terms in which the indices 0 and 3 appear together. Using the total antisymmetry of $\epsilon$ and $F$ and combining like terms, we see that
\begin{align*}
\epsilon^{\mu\nu\alpha\beta}F_{\mu\nu}F_{\alpha\beta} &= 2\left[F_{01}F_{23} - F_{10}F_{23} + F_{10}F_{32} - F_{01}F_{32} + \cdots \right] \\
&= - 8E_1B_1 + \cdots
\end{align*}
where we can invoke symmetry to write down the remaining terms. So the interaction term in the Lagrangian becomes
\begin{equation}\label{eq:a_em_lagr_EB}
+\frac{1}{4}g_{a\gamma\gamma}aF_{\mu\nu}\widetilde{F}^{\mu\nu} = -g_{a\gamma\gamma}a\mathbf{E}\cdot\mathbf{B},
\end{equation}
which is the second equality in Eq.~\eqref{eq:agammagamma}; the same result can be used to rewrite the equation of motion Eq.~\eqref{eq:a_eom}. Next we consider Eq.~\eqref{eq:em_eom} with the index $\beta = 0$. Using the antisymmetry of $F$ and $\epsilon$ to discard terms with repeated indices, we obtain
\begin{align*}
\partial_i F^{i0} &= \frac{1}{2}g_{a\gamma\gamma}\epsilon^{i0\mu\nu}F_{\mu\nu}\partial_i a \\
&= -g_{a\gamma\gamma}\left[F_{23}\partial_1a+ F_{31}\partial_2a + F_{12}\partial_3a\right] \\
&= g_{a\gamma\gamma}\left[\epsilon_{231}B_1\partial_1a+ \epsilon_{312}B_2\partial_2a + \epsilon_{123}B_3\partial_3a\right] \\
\Rightarrow\partial_iE_i &= g_{a\gamma\gamma}B_i\partial_ia.
\end{align*}
Taking $\beta = 1$ in Eq.~\eqref{eq:em_eom}, we get
\begin{align*}
\partial_0 F^{01} + \partial_2 F^{21} + \partial_3 F^{31} &= \frac{1}{2}g_{a\gamma\gamma}\left[\epsilon^{01\mu\nu}F_{\mu\nu}\partial_0 a + \epsilon^{21\mu\nu}F_{\mu\nu}\partial_2 a + \epsilon^{31\mu\nu}F_{\mu\nu}\partial_3 a\right] \\
&= g_{a\gamma\gamma}\left[F_{23}\partial_0 a + F_{30}\partial_2 a + F_{02}\partial_3 a\right] \\
\Rightarrow -\partial_0E_1 - \epsilon_{213}\partial_2B_3 -\epsilon_{312}\partial_3B_2 &= g_{a\gamma\gamma}\left[-\epsilon_{231}B_1\partial_0 a - E_3\partial_2 a + E_2\partial_3 a\right] \\
\Rightarrow -\partial_0E_1  +\left(\partial_2B_3 - \partial_3B_2\right) &= g_{a\gamma\gamma}\left[-B_1\partial_0 a + \left(E_2\partial_3 a - E_3\partial_2 a\right)\right].
\end{align*}
The $\beta=2$ and $\beta=3$ terms are obtained by cyclic permutation of spatial indices. Putting this all together, we obtain Maxwell's equations in the presence of the axion field
\begin{align}
\gv{\del}\cdot\mathbf{E} &= g_{a\gamma\gamma}\mathbf{B}\cdot\gv{\del}a \label{eq:div_E} \\
\gv{\del}\times\mathbf{B} - \partial_t\mathbf{E} &= g_{a\gamma\gamma}\left(\mathbf{E}\times\gv{\del}a - \mathbf{B}\partial_ta\right) \label{eq:curl_B} \\
\gv{\del}\times\mathbf{E} + \partial_t\mathbf{B} &= 0 \label{eq:curl_E} \\
\gv{\del}\cdot\mathbf{B} &= 0 \label{eq:div_B}
\end{align}
I have assumed the absence of ordinary electromagnetic sources and that the vacuum permittivity and permeability are good approximations to any media we will consider. See Ref.~\citep{millar2017} for a thorough discussion of axion electrodynamics in dielectrics.

\chapter{Magnet quench}\label{app:quench}
I noted the advantages of ``dry'' cryogenic systems for the operation of a haloscope detector in Sec.~\ref{sub:cryo}. Of course, such systems rely on uninterrupted electrical power to cool their 4~K and 70~K stages. Nothing particularly dramatic would happen if HAYSTAC were to experience a power outage with the magnet off: the DR's $^3$He/$^4$He mixture would stop circulating, and would begin to evaporate out of the mixing chamber as the low-temperature stages of the DR slowly warmed up, increasing the pressure in the circulation path. Eventually, the increased pressure would open an overpressure valve which would route the mixture back to its tank.

With 72~A circulating in the main magnet (at $B_0=9$~T; see Sec.~\ref{sub:magnet}), the effects of a power outage are much less benign. With no cooling power to balance the ambient heat load, $T_\text{mag}$ will slowly increase, and as soon as it exceeds 4.15~K, this enormous current will suddenly be flowing through a very lossy normal conductor: the resulting rapid dissipation of the magnet's stored energy is called a ``quench.'' Part of good magnet design is ensuring that the energy dissipated in a quench is efficiently distributed throughout the magnet's support structure, so it doesn't lead to localized melting or other permanent damage to the magnet itself.

To appreciate the gravity of the situation, note that the total stored energy of the fully charged HAYSTAC magnet is $U_B = \frac{1}{2}LI^2=500$~kJ, or approximately the kinetic energy of a small car ($\approx3000$~pounds) traveling at 45~miles per hour. All of this energy is dissipated within seconds in the event of a quench; the time for the magnet coils to warm up from $3.6$ to $4.15$~K without active cooling is about 4 minutes.

In principle, Yale has fast acting emergency backup power, but due to system upgrades and other issues, this emergency power was not available in early March 2016, when an unscheduled power outage resulted in a magnet quench shortly after the first complete pass through the tuning range. The magnet's built-in quench protection did its job, and there was no direct damage to the experiment from the heat of the quench itself. However, quenching also causes the flux through the magnet coils to drop to zero extremely rapidly, and by Faraday's law this generates large eddy currents in the components of the cryostat inserted into the magnet bore. The dipole interaction between these induced eddy currents and the decaying current in the main coil produces a net downward force on the cryostat. These forces resulted in significant structural damage to the DR during the March 2016 quench.

\begin{figure}[h]
\centering\includegraphics[width=0.8\textwidth]{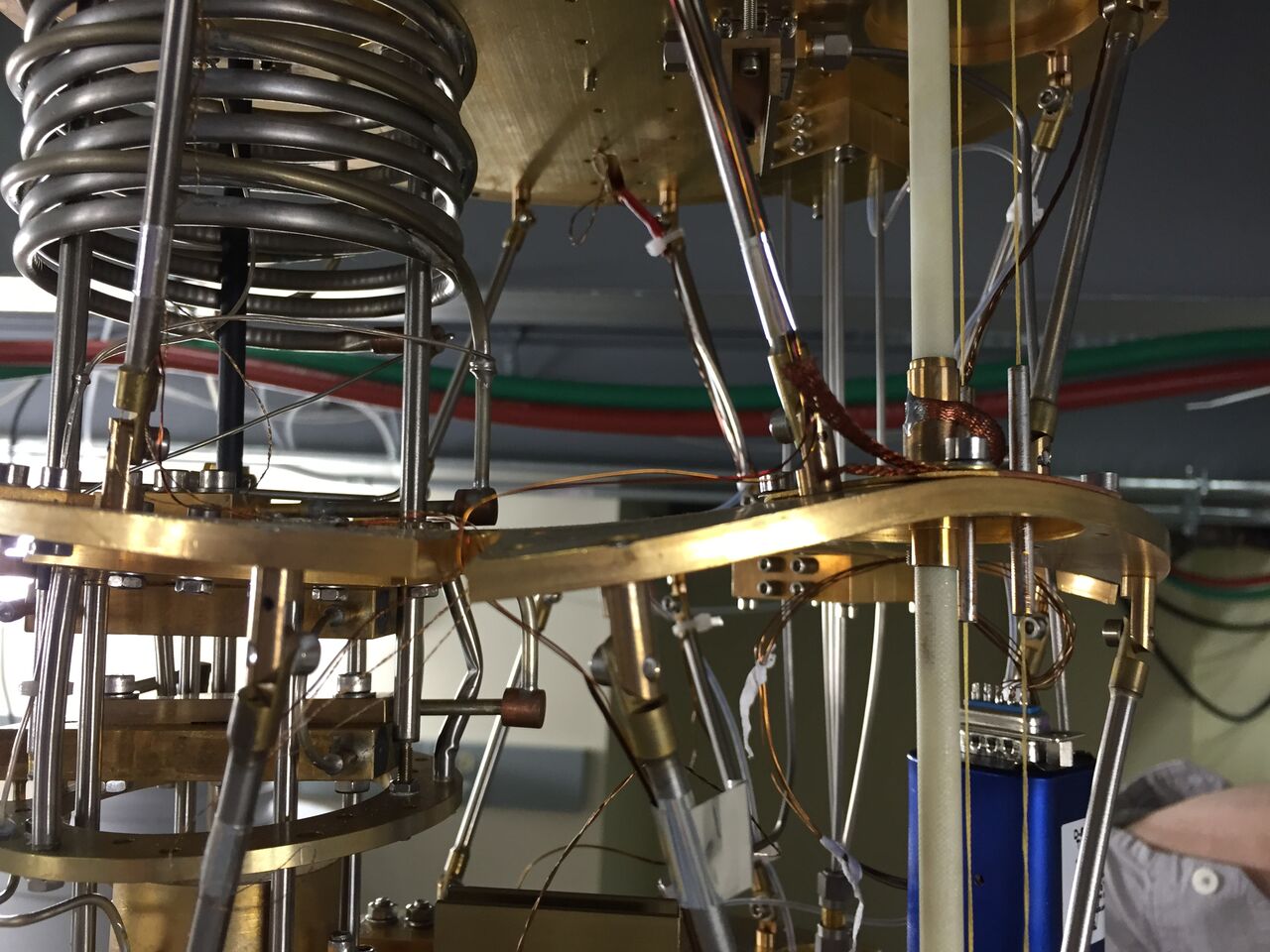}
\caption[Damage from the magnet quench]{\label{fig:quench} Warping of the DR plate between the still and mixing chamber stages due to eddy current forces during the March 2016 magnet quench. The elbow of a distressed experimentalist is visible in the lower right corner.}
\end{figure}

Because Faraday's law sets the magnitude of the voltage, larger eddy currents are generated in materials with high conductivity. In HAYSTAC the biggest offenders were the copper still shield extensions and the large copper rings in the lower part of the gantry. The still shield extensions themselves were warped in the quench, and were subsequently rebuilt using copper-plated stainless steel as noted in Sec.~\ref{sub:cryo}.\footnote{In a future design, most of the gantry components will likewise be constructed from copper-plated stainless steel, which should lead to more than a 100-fold reduction in forces.} The quench also led to serious warping of the DR plate between the mixing chamber and still plates (see Fig.~\ref{fig:quench}), because the net downward force on the DR plates was localized at a few points around the circumference of each stage. The mixing chamber plate itself fared better probably because of the extra structural support provided by the circulator field cancellation coil.

\begin{figure}[h]
\centering\includegraphics[width=0.8\textwidth]{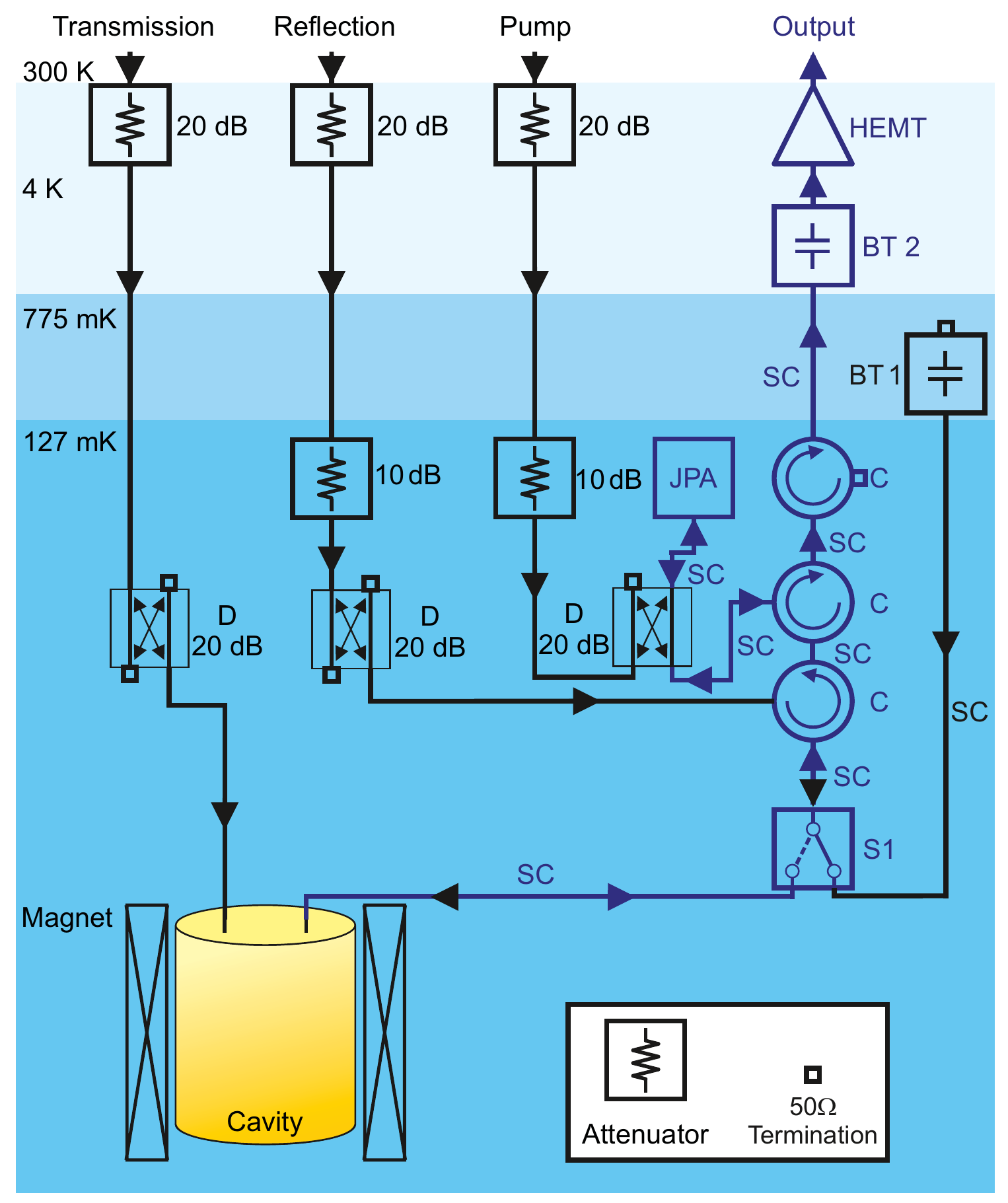}
\caption[The cryogenic microwave layout]{\label{fig:cryo_setup} The cryogenic microwave layout. Blue arrows indicate the receiver signal path from the cavity to room temperature; black arrows indicate other paths used for network analysis, noise calibration, and JPA biasing. Component part numbers and manufacturers are listed in Table~\ref{tab:parts}.}
\end{figure}

The structural supports between the 70~K and 4~K stages of the DR are fiberglass rather than stainless steel, whose thermal conductivity would be too high at these temperatures. During the quench two of these structural supports were pulled free of the 70~K plate. At this point the microwave coaxial lines (which are rigidly clamped at each DR stage; see Sec.~\ref{sub:receiver_cryo}) bore the weight of the DR, causing the connectors to break off at the interface just above the 4~K stage (Fig.~\ref{fig:cryo_setup}). When we opened the DR in the wake of the quench, we found it hanging off-axis in the magnet bore, with the angular misalignment constrained primarily by the radiation shields.

Fortunately, none of the gas flow lines were compromised during the quench, so none of the DR's $^3$He was lost. Likewise, the cavity, the JPA, and discrete receiver components were not damaged. After repairing the structural damage to the DR and replacing the ruined coaxial lines, the HAYSTAC detector worked just as well as it did before the quench.

\chapter{Receiver layout diagrams}\label{app:diagrams}
\begin{table}[h]
\centering
\begin{adjustbox}{max width=\textwidth}
\begin{tabular}{llll}
\hline
\hline
   & Label & Type & Supplier: Part \#\\
\hline
\hline
 Cryogenic & & & \\ 
  	& BT1 & Bias tee & Mini-Circuits: ZX85-12G+ (ferrite removed)\\
	& BT2 & Bias tee& Anritsu: K250\\
	& C & Circulator& Pamtech Inc.: CTH1184K18 \\
	& D & Directional coupler& Pasternack: PE2211-20\\
	& HEMT & HEMT amplifier& Low Noise Factory: LNF-LNC4\textunderscore8A\\
  	& S1& Switch & Radiall: R577443005\\
	& SC& NbTi/NbTi coax & Keycom: UPJ07\\
\hline
 Room-temperature & & & \\ 
  	& A1 & RF amplifier& Miteq: AMF-4F-04001200-15-10P\\
	& A2 & IF amplifier& Homemade: based on Fig.~2 in \citep{ca3018}\\
	& A3 & IF amplifier& Mini-Circuits: ZFL-500LN-BNC+ \\
	& A4 & IF amplifier& Stanford Research: SR445A\\
	& A5 & IF amplifier& Stanford Research: SR560\\
  	& AT1 & Step attenuator& Agilent Technologies: 8496H\\
	& AT2 & Step attenuator& Agilent Technologies: 8494H\\
	& B & Balun & North Hills: 0017CC \\
	& DC & DC block& Inmet: 8039\\
  	& F1 & Low-pass filter& Mini-Circuits: VLFX-80\\
	& F2 & Low-pass filter& Mini-Circuits: SLP-1.9+\\
	& F3 & Low-pass filter& Mini-Circuits: BLP-2.5+\\
	& F4 & Low-pass filter& Homemade: 3.39~kHz single pole RC filter\\
	& I1 & Isolator& Ditom Microwave: D314080\\
	& I2 & Double isolator& Ditom Microwave: D414080\\
	& M1 & Mixer& Marki Microwave: M1-0408\\
	& M2 & IQ mixer& Marki Microwave: IQ0307LXP\\
	& M3 & Mixer& Mini-Circuits: ZAD-8+\\
	& PS1 & Power splitter/combiner& Mini-Circuits: ZX10-2-71-s+\\
	& PS2 & Power splitter/combiner& Mini-Circuits: ZFRSC-2050+\\
	& S2 & Switch& Mini-Circuits: ZFSWA2-63DR+\\
\hline
\hline
\end{tabular}
\end{adjustbox}
\caption{\label{tab:parts} Receiver component part numbers.}
\end{table}

I have included full diagrams of the cryogenic and room-temperature signal paths through the HAYSTAC detector (referenced in Secs.~\ref{sec:receiver}, \ref{sub:feedback}, \ref{sub:power_spectra}, and \ref{sec:noise})\footnote{If you are reading this in PDF form, these section references are hyperlinked to take you back to whichever section you came from.} in this appendix for ease of access. DC lines, power lines, 10~MHz reference signals from the FS725 frequency standard, and GPIB connections to the DAQ PC are not shown.

\begin{figure}[h]
\centering\includegraphics[width=0.74\textwidth]{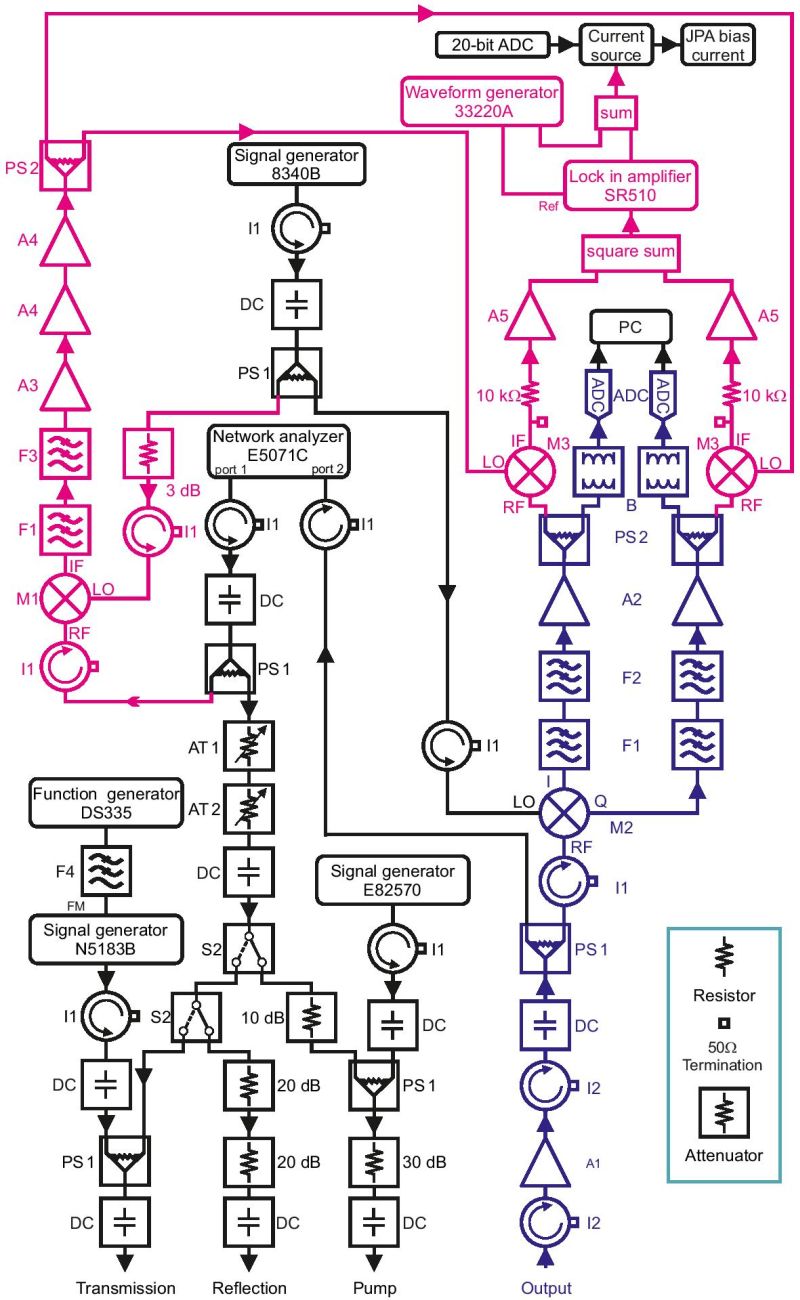}
\caption[The room-temperature microwave/IF layout]{\label{fig:RT_setup} The room-temperature microwave/IF layout. Blue arrows indicate the receiver signal path from the top of the DR to the ADCs; black arrows indicate other paths used for network analysis, JPA biasing, providing LO power, and synthetic axion signal injection. Parts of the chain used exclusively by the JPA flux feedback system are indicated in pink. Component part numbers and manufacturers are listed in table~\ref{tab:parts}; those shown on the diagram are from Keysight or Stanford Research Systems.}
\end{figure}

\chapter{Synthetic signal injections}\label{app:fake_axions}
As noted in Sec.~\ref{sub:daq_procedure}, we injected synthetic axion signals (narrowband noise signals with $\Delta\nu \approx 5$~kHz) through the cavity transmission line at ten random frequencies during the initial data acquisition period in winter 2016. We generated these signals by injecting band-limited white noise into the FM port of the N5183B microwave signal generator shown in Fig.~\ref{fig:RT_setup}; the linewidth is then controlled by the modulation depth. 

Our goal in injecting these signals into the detector was not to demonstrate an alternative approach to calibrating the search sensitivity, as obtaining sufficiently good statistics would entail polluting our spectrum with a large number of synthetic axions; moreover the precision of such a calibration would be limited by the $\pm3$~dB uncertainty in the synthetic axion power delivered to the cavity, due to unknown cryogenic insertion losses of individual components. Instead, we used synthetic signal injections as a simple fail-safe check on our data acquisition and analysis procedures, to verify that faint narrowband signals injected into the cavity did indeed result in large excess power in the expected grand spectrum bins.

We decided on a nominal signal power of $10^{-22}$~W, roughly equal to the expected conversion power for an axion with $|g_\gamma|=4|g^\text{KSVZ}_\gamma|$ and sufficiently far above our target sensitivity to allow us to immediately establish the presence or absence of excess power with only a single pass over the tuning range. Due to a miscalculation, we set the power lower than this by a factor of 2.5 for the three highest-frequency signals, and moreover the exposure was lowest at these frequencies in the winter run: thus the expected SNR for these three signals was $\approx1.5$. We observed excess power consistent with this estimate (though of course also consistent with the absence of a signal) at these three frequencies. After correcting the signal power, we observed $\delta^\text{g}_\ell/\tilde{\sigma}^\text{g}_\ell > 5$ in all bins corresponding to the remaining injected signals (see e.g., Fig.~\ref{fig:fakeaxion}). Having demonstrated to our satisfaction that our analysis procedure can detect real axion-like signals, we opted not to inject signals during the summer run. 

Before constructing the combined spectrum used in the final analysis, we cut RF bins around each injected signal in which we expect more than 1\% of the peak power given the measured signal lineshape. Thus, synthetic signal injections manifest as small notches in our coverage in which $|g^\text{min}_\gamma|_\ell$ increases sharply over a very narrow range. Seven such notches are visible in Fig.~\ref{fig:exclusion}. In the two lowest-frequency notches, $|g^\text{min}_\gamma|_\ell$ increases by about a factor of $2^{1/4}$ because roughly half the data contributing to the SNR at these frequencies was acquired during the winter run. At higher frequencies, a larger fraction of the data came from the summer run, and thus the depth of the notches gets progressively smaller. In particular, the effects of cutting data from the winter run around two injected signals above 5.76~GHz are not visible at the resolution of Fig.~\ref{fig:exclusion}. The last injected signal happened to fall in the range where we cut spectra around an intruder mode (see Sec.~\ref{sub:badscans}), so it is also not visible in Fig.~\ref{fig:exclusion}.

\chapter{Maximum likelihood estimation}\label{app:mle}
Taking the discussion at the beginning of Sec.~\ref{sec:rescale_combine} as motivation, I will assume we have $m$ independent Gaussian random variables $y_k$ drawn from distributions with the same mean $\mu$ but different variances $\sigma_k^2$. We are interested in finding an estimate of $\mu$ that maximizes the likelihood function, which is just the joint probability distribution of the observations $y_k$ considered as a function of $\mu$:
\begin{equation}
\mathcal{L}(\mu)=\exp\left(-\frac{1}{2}\sum_k\left(\frac{y_k-\mu}{\sigma_k}\right)^2\right).\label{eq:likelihood_ind}
\end{equation}
We can equivalently maximize $\log\mathcal{L}$, since the logarithm is monotonically increasing. So we take
\begin{equation*}
\frac{\mathrm{d}}{\mathrm{d}\mu}\log\mathcal{L}=\sum_k\left(\frac{y_k-\mu}{\sigma_k^2}\right)=0.
\end{equation*}
Solving for $\mu$ yields
\begin{equation}
\mu=\frac{\sum_k y_k/\sigma_k^2}{\sum_k 1/\sigma_k^2},
\label{eq:ml_ind}
\end{equation}
which may be compared to Eq.~\eqref{eq:delta_c}.

If our observations are not independent but rather correlated, Eq.~\eqref{eq:likelihood_ind} should be replaced with
\begin{equation}
\mathcal{L}(\mu)=\exp\left(-\frac{1}{2}\left(\mathbf{y}-\mu\mathbf{i}\right)^\intercal\mathbf{\Sigma}^{-1}\left(\mathbf{y}-\mu\mathbf{i}\right)\right)\label{eq:likelihood_cor},
\end{equation}
where $\mathbf{i}$ is the $m$-vector $(1,1,\dots,1)$, and $\mathbf{\Sigma}$ is the covariance matrix whose diagonal elements are $\sigma_k^2$. Maximizing with respect to $\mu$ we obtain
\begin{equation}
\mu=\frac{\mathbf{y}^\intercal\mathbf{\Sigma}^{-1}\mathbf{i}}{\mathbf{i}^\intercal\mathbf{\Sigma}^{-1}\mathbf{i}}.
\label{eq:ml_cor}
\end{equation}
We see that the (unnormalized) ML weight for each $y_k$ is a sum over the $k$th row of $\mathbf{\Sigma}^{-1}$. A useful approximation to this sum for sufficiently small correlations is
\begin{equation}
\sum_{k}\big(\mathbf{\Sigma}^{-1}\big)_{kk'} \approx \frac{1}{\sigma^2_{k'}}\left[1 - \sum_{k\neq k'}\frac{\Sigma_{kk'}}{\sigma^2_{k}}\right],
\label{eq:weights_cor}
\end{equation}
where I have neglected all terms that are higher than first order in the ratio of any off-diagonal element to any diagonal element of $\mathbf{\Sigma}$; to first order the normalization is then just the sum of Eq.~\eqref{eq:weights_cor} over $k'$. In Sec.~\ref{sec:rebin} I discuss ML weighting in the presence of small correlations: I continue to use Eq.~\eqref{eq:ml_ind} rather than Eq.~\eqref{eq:ml_cor}, and argue in Sec.~\ref{sub:correlations} that deviations from the true optimal weights are acceptably small.

The ML estimate of the mean of a multivariate Gaussian distribution with arbitrary covariance matrix $\mathbf{\Sigma}$ can also be obtained from a least-squares perspective. To see this, consider a linear regression model $\mathbf{y} = \mu\mathbf{x} + \boldsymbol{\epsilon}$, where we would like to estimate the slope $\mu$ in the presence of noise $\boldsymbol{\epsilon}$, assumed to be drawn from a Gaussian distribution with zero mean and covariance matrix $\mathbf{\Sigma}$. The generalized least squares (GLS) estimate of $\mu$ is the value that minimizes the mean squared error
\begin{equation}
\chi^2(\mu) = \frac{1}{m}\left(\mathbf{y}-\mu\mathbf{x}\right)^\intercal\mathbf{\Sigma}^{-1}\left(\mathbf{y}-\mu\mathbf{x}\right).
\end{equation}
For $\mathbf{x}=\mathbf{i}$, $\chi^2(\mu)\propto\log\mathcal{L}$, so the estimate that extremizes either criterion will also extremize the other. This equivalence between the ML and GLS methods requires only that the statistics of the underlying noise distribution be Gaussian, and this condition will always be satisfied in our haloscope analysis. It can be proved that the variance of the GLS estimator is smaller than the variance of any other unbiased linear estimator~\citep{GLS1935}.

Finally, note that if we allow the elements of $\mathbf{x}$ to vary, and take $\mathbf{\Sigma}$ to be diagonal for simplicity, the least squares estimate of $\mu$ becomes
\begin{equation}
\mu=\frac{\sum_k x_ky_k/\sigma_k^2}{\sum_k (x_k/\sigma_k)^2},
\label{eq:ls_x}
\end{equation}
The elements of $x_k$ here play the role of the rescaling factor discussed in Sec.~\ref{sub:rescale}; thus from a least-squares perspective the rescaling of the spectra need not be regarded as a distinct step of the analysis procedure. I stick to the ML perspective in the text to emphasize the value of using units in which the expected axion conversion power is 1, and thus the $R=\sigma^{-1}$ correspondence has an intuitive interpretation.

\chapter{Optimizing SG filter parameters}\label{app:sg_params}
I discussed the optimization of the SG filter parameters $d$ and $W$ briefly at the end of Sec.~\ref{sub:sg_filter}, but it is instructive to revisit this question after having observed the filter-induced narrowing $\xi$ of the distribution of grand spectrum bins (Sec.~\ref{sub:correlations}) and the filter-induced attenuation of the SNR (Sec.~\ref{sub:axion_atten}). Fig.~\ref{fig:filter} indicates that reducing $d/W$ moves the 3~dB point of the SG filter down towards larger spectral scales and increases the stopband attenuation on the small spectral scales of interest ($\leq CK$ bins). Thus we should expect $\xi,\eta\rightarrow1$ as we reduce $d/W$. 

However, as noted in Sec.~\ref{sub:sg_filter}, reducing the 3~dB point of the SG filter invariably moves progressively larger-amplitude components of the baseline from the filter's passband into its stopband. This claim implicitly assumes that the power spectrum \textit{of the residual baseline} falls off monotonically towards smaller spectral scales, and we can confirm this empirically: on small spectral scales the residual baseline power spectrum follows a power law distribution with spectral index $\alpha\approx-2$.

The largest-amplitude baseline component that is not removed by the SG filter (and thus remains in the processed spectra) will coincide with the first zero of the filter's transfer function; let us call the corresponding bin separation $\kappa$. As we reduce $d/W$ at fixed integration time $\tau$ (or increase $\tau$ for a given filter), the baseline amplitude $A(\kappa)$ will grow relative to the statistical fluctuations $\sigma^\text{p}=1/\sqrt{\Delta\nu_b\tau}$. For $A(\kappa)/\sigma^\text{p}$ sufficiently large, the distribution of processed spectrum bins $\delta^\text{p}_{ij}$ will appear non-Gaussian. Of course, each bin in each processed spectrum is still a Gaussian random variable with standard deviation $\sigma^\text{p}$; the apparent breakdown of Gaussianity just indicates that $\mu^\text{p}_{ij}=0$ for each bin $j$ has become a poor approximation given our failure to completely remove the spectral baseline.

Even if the distribution of $\delta^\text{p}_{ij}$ exhibits no signs of non-Gaussianity, $A(\kappa)\neq0$ implies \textit{positive} correlations in the processed spectra on scales $\leq\kappa/2$; since $\kappa>CK$, this effect tends to counteract the negative correlations due to the SG filter stopband alone (i.e., independent of the spectrum of the baseline). In other words, systematic effects due to the shape of the baseline grow coherently in the horizontal sum over adjacent bins. They can also grow coherently in the vertical sum if the $m_k$ contributing spectra have small detunings and similar baselines, as in the rescan data set (Sec~\ref{sub:rescan_analysis}).

Thus, we find that unless $A(\kappa) \ll 1/\sqrt{\Delta\nu_b\tau}$, $\xi$ and $\eta$ will depend on the integration time $\tau$ (and possibly also on $m_k$). The simulations discussed in Secs.~\ref{sub:correlations}, \ref{sub:axion_atten}, and \ref{sub:rescan_analysis} demonstrate that we are safely in the $A(\kappa) \ll 1/\sqrt{\Delta\nu_b\tau}$ regime with filter parameters $d,W$ ($d^*,W^*$) for the initial scan (rescan) analysis.

\chapter{Parameter uncertainties}\label{app:error}
The corrected grand spectrum SNR $\tilde{R}^\text{\,g}_\ell$ depends on many measured parameters whose uncertainties I have thus far ignored. Here I will quantify the effects of these uncertainties on the analysis, but first I should note that there is potential for terminological confusion because ``confidence level'' is a generic statistical term often used to quantify uncertainty. The axion search confidence level $c_1$ defined in Sec.~\ref{sec:candidates} is the probability that an axion with SNR $R_T$ in any given grand spectrum bin will exceed the threshold -- since the value of $R_T$ is not fixed by measurement, $c_1$ is completely independent of parameter uncertainties. Rather, uncertainty in $\tilde{R}^\text{\,g}_\ell$ translates [via Eqs.~\eqref{eq:g_ell} and \eqref{eq:g_min}] into uncertainty in the threshold coupling $|g^\text{min}_\gamma|_\ell$ for which we obtain SNR $R_T$ in each bin $\ell$.

We can estimate the size of the fractional uncertainty $\delta|g^\text{min}_\gamma|/|g^\text{min}_\gamma|$ in a typical grand spectrum bin by first noting that
\begin{equation}
|g^\text{min}_\gamma| \propto \left(\frac{N_\text{eff}}{\eta\phi(\delta\nu_r)\eta_0C_{010}}\right)^{1/2},\label{eq:g_error}
\end{equation}
where I have elided factors without uncertainty and quantities like $Q_L$ and $B_0$ that are easily measured with fractional uncertainty $\leq1\%$, and introduced an effective number of noise quanta $N_\text{eff}$ and a function $\phi(\delta\nu_r)$ to quantify the effects of misalignment. It is easy to estimate the error in the factor $\eta_0$ introduced in Sec.~\ref{sub:yfactor} to quantify loss between the cavity and JPA. We estimated this loss to be $-0.60 \pm 0.15$~dB, which implies $\delta\eta_0/\eta_0\approx3.5\%$.

The filter-induced attenuation $\eta$ and cavity mode form factor $C_{010}$ are both obtained from simulation, and thus estimating the uncertainty in these parameters is not necessarily straightforward. Nonetheless, our result for $\eta$ is very robust against changes in the parameters of the simulation (see discussion in Sec.~\ref{sub:axion_atten}), and this implies a fractional uncertainty of $\delta\eta/\eta\lesssim1\%$ which we can safely neglect. I have not included uncertainty in $C_{010}$ in this error budget because we do not yet have a reliable way to quantify it. Preliminary field profiling measurements suggest that the simulated form factors are reliable to better than $10\%$, so a careful treatment of the form factor uncertainty would likely change our final result $\delta|g^\text{min}_\gamma|/|g^\text{min}_\gamma|\approx4\%$ by at most a factor of 2 and possibly much less.

In the denominator of Eq.~\eqref{eq:g_error} I have defined
\begin{equation}
\phi(\delta\nu_r)=\sqrt{\sum_q\bar{L}_qL_q(\delta\nu_r)/K^2}\label{eq:phi_align}
\end{equation} 
to encode the dependence of the SNR on the misalignment $\delta\nu_r$ of the axion mass relative to the lower edge of the grand spectrum bin in which the SNR is maximized. $\phi$ is related to the misalignment attenuation factor $\eta_m$ introduced in Sec.~\ref{sub:lineshape} by $\eta_m\approx\bar{\phi}/\phi(0)$, where $\bar{\phi}$ is the mean value of $\phi(\delta\nu_r)$ over the range of possible misalignments; note also the formal similarity of Eq.~\eqref{eq:phi_align} to Eq.~\eqref{eq:fom_ml}.\footnote{Technically, $\eta_m$ as defined in Sec.~\ref{sub:lineshape} is obtained by replacing each $L_q(\delta\nu_r)$ by its average value $\bar{L}_q$ and then normalizing to $\phi(0)$, which is not quite the same as averaging $\phi$ because $\phi$ is not linear in $L_q$. In practice, the difference is negligible.} With the misalignment error $\delta\phi$ defined as the standard deviation of $\phi(\delta\nu_r)$ over this same range, we obtain $\delta\phi/\bar{\phi}\approx2\%$.

Finally, in any given grand spectrum bin, the effective system noise $N_\text{eff}$ is formally given by a ML-weighted average of $N_{ij}$ across all contributing processed spectrum bins. Since we are only interested in estimating the typical fractional uncertainty in the noise temperature, we can just average $N_{ij}$ over all spectra and evaluate it in the IF bin $j$ corresponding to the middle of the analysis band, where the ML weight is largest.\footnote{This same approximation was used to set $N_s$ in the calculation of the rescan time in Sec.~\ref{sub:rescan_daq}.} We can then write $N_\text{eff}=N_\text{mc} + \Delta N_\text{cav} + N_A$. As noted in Sec.~\ref{sec:noise}, we obtain $N_A=1.35\pm0.05$ quanta from the average of off-resonance $Y$-factor measurements and $\Delta N_\text{cav}=1.00\pm0.17$ quanta from the average of $Y$-factor measurements during the data run. Even allowing for a $\pm20$~mK uncertainty in the calibration of the mixing chamber thermometer, the uncertainty in $N_\text{mc}=0.63$ remains negligibly small, in part because the nominal HAYSTAC operating temperature $T_\text{mc}=127$~mK is sufficiently far into the Wien limit that $N_\text{mc}$ depends only weakly on the physical temperature, and in part because errors in different contributions to the total noise $N_\text{eff}$ are somewhat anti-correlated. Negative correlations arise because increasing any of the additive terms in $N_\text{eff}$ while holding the others constant would reduce the measured value of the hot/cold noise power ratio $Y$.

Adding the uncertainties cited in the above paragraph in quadrature and using $N_\text{eff}\approx3$ we obtain $\delta N_\text{eff}/N_\text{eff}\approx 6\%$. This estimate (dominated by the variation in measurements of $\Delta N_\text{cav}$) is conservative in that I have neglected the anticorrelation between $\delta N_A$ and $\delta(\Delta N_\text{cav})$ are negatively correlated, and included the RMS systematic variation of $\Delta N_\text{cav}$ across the tuning range in the ``uncertainty'' $\delta(\Delta N_\text{cav})$. Miscalibration of the still thermometer would need to be larger than $\pm20$ mK to affect this estimate of $\delta N_\text{eff}$.

Combining the results of the preceding paragraphs, we obtain
\begin{equation*}
\frac{\delta |g^\text{min}_\gamma|}{|g^\text{min}_\gamma|} \approx \sqrt{\Bigg(\frac{1}{2}\frac{\delta N_\text{eff}}{N_\text{eff}}\Bigg)^2 + \Bigg(\frac{1}{2}\frac{\delta\phi}{\bar\phi}\Bigg)^2 + \Bigg(\frac{1}{2}\frac{\delta\eta_0}{\eta_0}\Bigg)^2} \approx4\%.
\end{equation*}
This result (represented by the light green shaded region in Fig.~\ref{fig:exclusion}) should be interpreted as a rough estimate of the uncertainty in our exclusion limit, not a formal $1\sigma$ error bar on the threshold coupling $|g^\text{min}_\gamma|_\ell$ in each bin.

We should also consider the effects of miscalibrating the SNR in the rescan analysis. We can distinguish between ``global'' effects (e.g., overall miscalibration of $N_\text{sys}$ or uncertainty in $\eta_0$) and effects confined to the rescan analysis (e.g., miscalibration of $\eta^*$ or mode frequency drifts in particular rescan measurements). The former affect $\tilde{R}^{\text{\,g}*}_\ell$ and $\tilde{R}^\text{\,g}_\ell$ in the same way: thus they do not change the candidate SNR $\hat{R}^*_{\ell'(s)}$ obtained from Eq.~\eqref{eq:hat_snr_*}, and cannot change the results of the rescan analysis.

Conversely, miscalibration of $\tilde{R}^{\text{\,g}*}_{\ell}$ relative to $\tilde{R}^\text{\,g}_{\ell}$ around any given candidate $s$ implies that we have either underestimated or overestimated $\hat{R}^*_{\ell'(s)}$, which in turn implies that the coincidence thresholds $\Theta^*_{\ell'(s)}$ we imposed on the bins correlated with $\ell(s)$ were either unnecessarily low or too high. Clearly, the latter possibility is the one that should concern us: it implies that relative miscalibration of the rescan SNR can cause the probability that we miss a real persistent signal to exceed $1-c_2$.

Empirically, in the first HAYSTAC data run, we could reduce each $\hat{R}^*_{\ell'(s)}$ by 17\% before any of the $(2K-1)S$ bins we examined exceeded the corresponding threshold.\footnote{The first bin to do so (associated with the rescan candidate $s=26$) had $\delta^{\text{g}*}_{\ell}/\tilde{\sigma}^{\text{g}*}_{\ell}=2.7$. Among $S\times n_K$ independent bins, we expect 0.5 bins with power excess this large, so the observation of one should not surprise us.} All of the parameter uncertainties whose contributions to $\delta |g^\text{min}_\gamma|/|g^\text{min}_\gamma|$ we have considered in this section are global effects to which the coincidence thresholds are insensitive. We conclude that miscalibration of $\tilde{R}^{\text{\,g}*}_{\ell}$ relative to $\tilde{R}^\text{\,g}_{\ell}$ by more than 17\% is extremely unlikely. A more formal way to account for the possibility of relative miscalibration is to require a rescan confidence level $c_2>c_1$; future HAYSTAC analyses will adopt this approach.

\chapter{Effects of a wider lineshape}\label{app:axion_width}
As noted in Sec.~\ref{sub:lineshape}, the analysis published in Ref.~\citep{PRL2017} and described in this dissertation assumed the spectral distribution of axion conversion power is given by Eq.~\eqref{eq:f_dist} instead of Eq.~\eqref{eq:f_dist_2}, but we should actually expect the latter distribution in a terrestrial experiment if the halo axions are fully virialized with a pseudo-isothermal density profile and RMS velocity $\sqrt{\left<v^2\right>}=270$~km/s.

\begin{figure}[h]
\centering\includegraphics[width=0.7\textwidth]{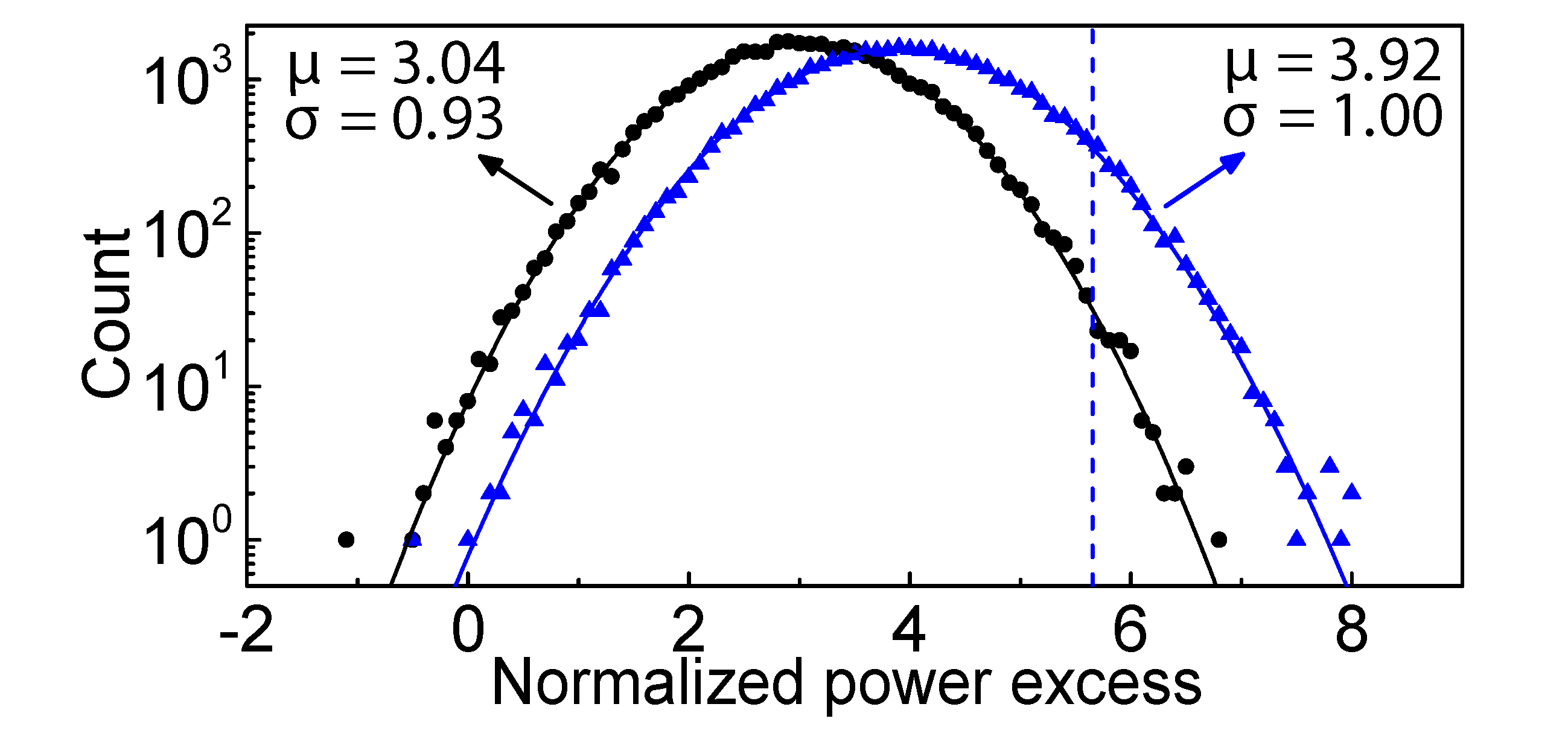}
\caption[Filter-induced signal attenuation with a wider lineshape]{\label{fig:wide_axion} The results of a simulation to quantify the reduction in SNR for an axion signal with the wider lineshape of Eq.~\eqref{eq:f_dist_2}. As in Fig.~\ref{fig:simu_hist}, the distribution of excess power $\delta^\text{g}_{\ell'}/\sigma^\text{g}_{\ell'}$ in a grand spectrum bin $\ell'$ containing an axion signal is histogrammed over iterations of the simulation; the two histograms correspond to two different analysis pipelines. Parameters obtained from Gaussian fits to the both histograms are displayed on the plot. The SNR $R^\text{\,g}_{\ell'}=5.66$ calculated assuming the narrower lineshape of Eq.~\eqref{eq:f_dist} is indicated by the dashed vertical line. With an analysis that neglects SG filter effects (blue triangles), the distribution is Gaussian with standard deviation 1 and mean smaller than $R^\text{\,g}_{\ell'}$ by a factor $\zeta_0=0.69$. From the analysis that takes into account effects of the SG filter (black circles), we obtain an additional SNR attenuation factor $\eta_\text{lab}=0.84$. The net reduction of the corrected grand spectrum SNR $\tilde{R}^\text{\,g}_{\ell'}$ is thus $\zeta=(\eta_\text{lab}/\eta)\zeta_0=0.64$.}
\end{figure}

To quantify the degradation of our exclusion limit $|g^\text{min}_\gamma|_\ell$ for an axion signal with the lab frame spectral distribution $f'(\nu)$, we repeated the simulation of Sec.~\ref{sub:axion_atten}, using Eq.~\eqref{eq:f_dist_2} instead of Eq.~\eqref{eq:f_dist} for the simulated axion signal but leaving the lineshape $\bar{L}_q$ used in both the ``standard'' and ``ideal'' analysis pipelines unchanged. As in Sec.~\ref{sub:axion_atten}, the main results of the simulation are two histograms (corresponding to the two analysis pipelines) representing the excess power distribution in the grand spectrum bin $\ell'$ best aligned with the simulated axion signal. These histograms are plotted in Fig.~\ref{fig:wide_axion}.

We see that the mean value of the ideal analysis histogram $E[\delta^\text{g}_{\ell'}/\sigma^\text{g}_{\ell'}]_\text{i}$ is no longer equal to the calculated SNR $R^\text{\,g}_{\ell'}$ represented by the dashed vertical line. This is unsurprising, as $R^\text{\,g}_{\ell'}$ is still calculated using the lineshape $\bar{L}_q$ obtained by integrating Eq.~\eqref{eq:f_dist}. Thus, neglecting SG filter effects, the ratio 
\begin{equation}
\zeta_0=E[\delta^\text{g}_{\ell'}/\sigma^\text{g}_{\ell'}]_\text{i}/R^\text{\,g}_{\ell'} = 0.69
\end{equation}
quantifies the reduction in SNR we should expect when we use an analysis optimized for signals with spectral distribution $f(\nu)$ to search for signals governed by the wider lab frame distribution $f'(\nu)$.

Next we can consider how $\zeta_0$ is modified by the imperfect SG filter stopband. From the width of the histogram obtained from the standard analysis, we obtain $\xi=0.93$, as we should expect given that we have not changed the parameters of the horizontal sum. Comparing the two histograms in Fig.~\ref{fig:wide_axion}, we obtain $\eta_\text{lab}=E[\delta^\text{g}_{\ell'}/\sigma^\text{g}_{\ell'}]_\text{s}/\big(\xi E[\delta^\text{g}_{\ell'}/\sigma^\text{g}_{\ell'}]_\text{i}\big) = 0.83$ [c.f.\ $\eta=0.90$ obtained in Sec.~\ref{sub:axion_atten} assuming the narrower distribution $f(\nu)$]. The result $\eta_\text{lab}<\eta$ is also expected, as the SG filter stopband attenuation gets worse towards larger spectral scales (See Fig.~\ref{fig:filter}). The net reduction of the corrected KSVZ SNR $\tilde{R}^\text{\,g}_{\ell'}$ is thus
\begin{equation}
\zeta=(\eta_\text{lab}/\eta)\zeta_0=0.64.
\end{equation}
Equivalently, at fixed $R_T$, $|g_\gamma|$ is increased by a factor $1/\sqrt{\zeta}=1.25$. Since we cannot change the threshold in a reanalysis of a completed run without acquiring more rescan data, we conclude that our published exclusion limit $|g^\text{min}_\gamma|_\ell$ is degraded by 25\% for axion signals with spectrum given by Eq.~\eqref{eq:f_dist_2}. The modified limits still fall within the axion model band~\citep{cheng1995}; thus the qualitative conclusions of Ref.~\citep{PRL2017} remain unchanged.

It should be emphasized that the value of $\zeta_0$ derived from simulation above arises from the combination of two conceptually distinct effects. First, $f'(\nu)$ is wider than $f(\nu)$, and thus any analysis assuming the former will be less sensitive for any given value of the noise per unit bandwidth $N_\text{sys}$. Second, our analysis used values of $K$ and $\bar{L}_q$ appropriate for the distribution $f(\nu)$, so the horizontal sum is not optimally weighted if the true signal spectrum is $f'(\nu)$. With $C=10$, $K=7$, and $f(\nu)\rightarrow f'(\nu)$ in Eq.~\eqref{eq:int_lineshape}, we can obtain $\zeta_0=0.78$ analytically using Eq.~\eqref{eq:fom_ml}; simulation confirms this value and indicates that $\eta_\text{lab}$ is unchanged. Thus we should expect $\zeta=0.72$ for an analysis optimized for the wider signal distribution, or equivalently $|g_\gamma|$ larger than our present limit by 18\%, up to changes in other factors affecting the SNR. 

\chapter{Scaling with integration time}\label{app:avar}
Eq.~\eqref{eq:dicke} indicates that the haloscope search SNR scales as $\sqrt{\tau}$ as a result of the $\tau^{-1/2}$ scaling of the RMS noise power in each bin expected from Gaussian statistics [Eq.~\eqref{eq:delta_P}]. In a real detector, we should not expect this scaling to hold out to arbitrarily large $\tau$, and it is important to confirm that we are still operating in a regime where the radiometer equation is valid. In HAYSTAC, the observed standard normal distribution of the combined spectrum power excess $\delta^\text{c}_k/\sigma^\text{c}_k$ in both the initial scan and rescan analyses implicitly indicates that the RMS noise has the correct scaling. We also demonstrated more directly that this $\tau^{-1/2}$ scaling holds for real data out to $\tau>\text{max}(\tau^*_s)$ with a dedicated measurement described below.

\begin{figure}[h]
\centering\includegraphics[width=0.7\textwidth]{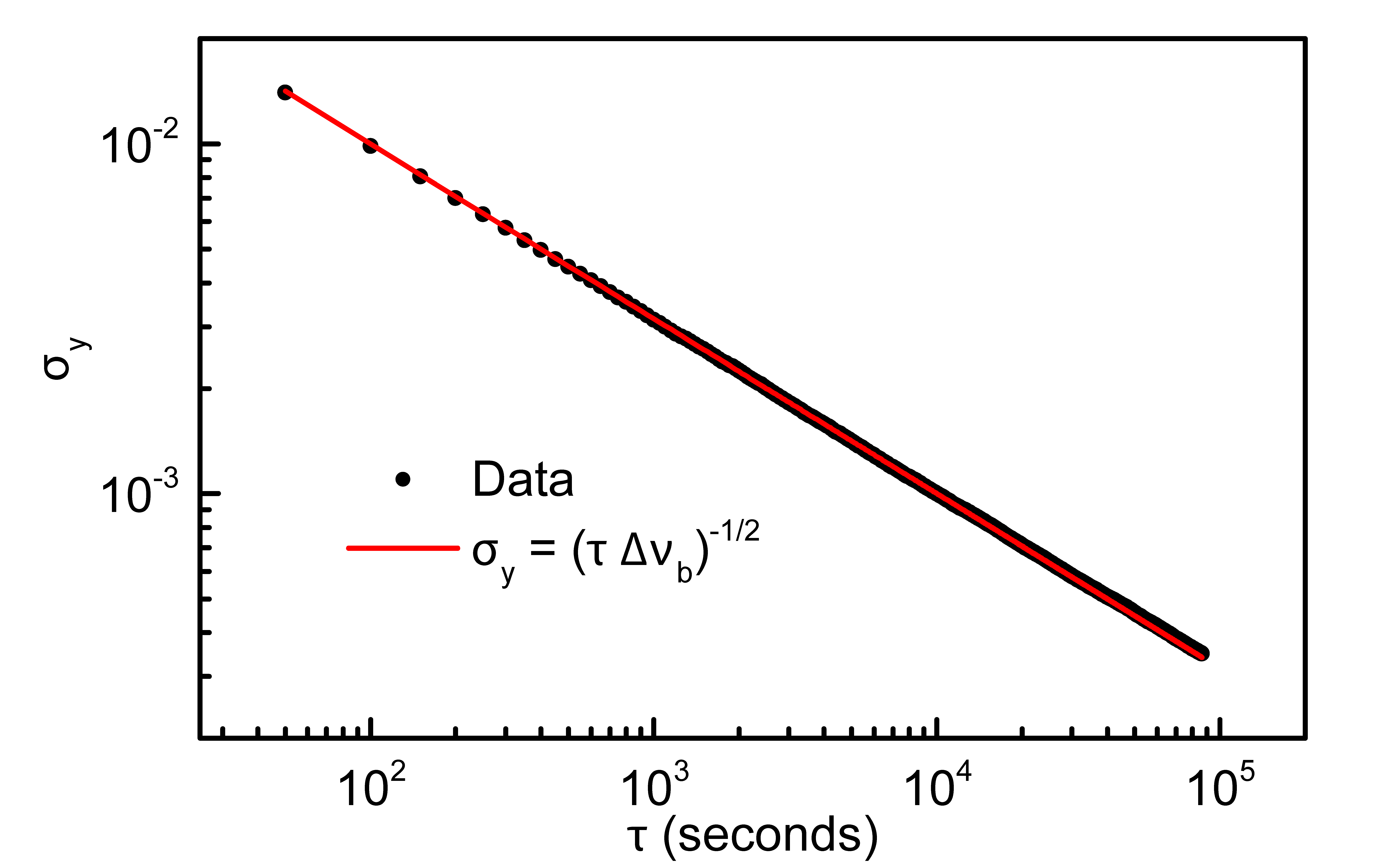}
\caption[Testing the limits of HAYSTAC statistical sensitivity]{\label{fig:avar} Results of a direct measurement of the scaling of RMS noise $\sigma_y$ with integration time $\tau$ in HAYSTAC, demonstrating that $\sigma_y\propto\tau^{-1/2}$ as expected out to at least 24 hours.}
\end{figure}

For this measurement, we acquired 24 hours of noise data at a single frequency with the JPA gain maintained by feedback as in the data run. The data was saved to disk as a set of $N=17280$ raw spectra obtained from $\tau_0=5$ s of averaging each. In offline analysis we removed bins contaminated by known IF interference, divided by the average baseline as in Sec.~\ref{sec:baseline}, and averaged every $m=10$ adjacent spectra. We used this set of $N/m$ spectra to probe the behavior of the RMS noise $\sigma_y$ as a function of the integration time $\tau_k=k\,m\,\tau_0$, for $k=1,\dots,N/m$.

To measure $\sigma_y(\tau)$, we apply a Savitzky-Golay filter with parameters $d^*$ and $W^*$ (Sec.~\ref{sub:rescan_analysis}) to each of the $N/m$ averages. Then for each $k=1,\dots,N/m$ we average $k$ filtered spectra and take $\sigma_y(\tau_k)$ to be the sample standard deviation of all bins in this $k$-spectrum average. We expect $\sigma_y(\tau)=1/\sqrt{\Delta\nu_b\tau}$.\footnote{Formally, $\sigma_y=\sigma^\text{p}$ considered as a function of the integration time $\tau$; we call this quantity $\sigma_y$ in analogy to the Allan deviation, a time-domain measure of the dependence of the RMS noise on $\tau$.} The measured values of $\sigma_y(\tau_k)$ (plotted in Fig.~\ref{fig:avar}) exhibit this expected behavior out to at least $\tau=24$~hours. 

\backmatter


\begin{thebibliography}{191}%
\makeatletter
\providecommand \@ifxundefined [1]{%
 \@ifx{#1\undefined}
}%
\providecommand \@ifnum [1]{%
 \ifnum #1\expandafter \@firstoftwo
 \else \expandafter \@secondoftwo
 \fi
}%
\providecommand \@ifx [1]{%
 \ifx #1\expandafter \@firstoftwo
 \else \expandafter \@secondoftwo
 \fi
}%
\providecommand \natexlab [1]{#1}%
\providecommand \enquote  [1]{``#1''}%
\providecommand \bibnamefont  [1]{#1}%
\providecommand \bibfnamefont [1]{#1}%
\providecommand \citenamefont [1]{#1}%
\providecommand \href@noop [0]{\@secondoftwo}%
\providecommand \href [0]{\begingroup \@sanitize@url \@href}%
\providecommand \@href[1]{\@@startlink{#1}\@@href}%
\providecommand \@@href[1]{\endgroup#1\@@endlink}%
\providecommand \@sanitize@url [0]{\catcode `\\12\catcode `\$12\catcode
  `\&12\catcode `\#12\catcode `\^12\catcode `\_12\catcode `\%12\relax}%
\providecommand \@@startlink[1]{}%
\providecommand \@@endlink[0]{}%
\providecommand \url  [0]{\begingroup\@sanitize@url \@url }%
\providecommand \@url [1]{\endgroup\@href {#1}{\urlprefix }}%
\providecommand \urlprefix  [0]{URL }%
\providecommand \Eprint [0]{\href }%
\providecommand \doibase [0]{http://dx.doi.org/}%
\providecommand \selectlanguage [0]{\@gobble}%
\providecommand \bibinfo  [0]{\@secondoftwo}%
\providecommand \bibfield  [0]{\@secondoftwo}%
\providecommand \translation [1]{[#1]}%
\providecommand \BibitemOpen [0]{}%
\providecommand \bibitemStop [0]{}%
\providecommand \bibitemNoStop [0]{.\EOS\space}%
\providecommand \EOS [0]{\spacefactor3000\relax}%
\providecommand \BibitemShut  [1]{\csname bibitem#1\endcsname}%
\let\auto@bib@innerbib\@empty
\bibitem [{\citenamefont {Lubej}()}]{SMfig}%
  \BibitemOpen
  \bibfield  {author} {\bibinfo {author} {\bibfnamefont {M.}~\bibnamefont
  {Lubej}},\ }\href@noop {} {}\bibinfo {howpublished}
  {\url{http://www-f9.ijs.si/\~lubej/}},\ \bibinfo {note} {accessed
  5/10/2017}\BibitemShut {NoStop}%
\bibitem [{\citenamefont {{European Space Agency / Planck
  Sattelite}}(2013)}]{LCDMfig}%
  \BibitemOpen
  \bibfield  {author} {\bibinfo {author} {\bibnamefont {{European Space Agency
  / Planck Sattelite}}},\ }\href@noop {} {}\bibinfo {howpublished}
  {\url{http://planck.cf.ac.uk/results/cosmic-microwave-background}} (\bibinfo
  {year} {2013}),\ \bibinfo {note} {accessed 5/10/2017}\BibitemShut {NoStop}%
\bibitem [{\citenamefont {Wilczek}(2016)}]{soap}%
  \BibitemOpen
  \bibfield  {author} {\bibinfo {author} {\bibfnamefont {F.}~\bibnamefont
  {Wilczek}},\ }\href@noop {} {\enquote {\bibinfo {title} {{Time's (Almost)
  Reversible Arrow}},}\ }\bibinfo {howpublished} {Quanta Magazine} (\bibinfo
  {year} {2016})\BibitemShut {NoStop}%
\bibitem [{\citenamefont {Dine}\ and\ \citenamefont {Draper}(2015)}]{dd2015}%
  \BibitemOpen
  \bibfield  {author} {\bibinfo {author} {\bibfnamefont {M.}~\bibnamefont
  {Dine}}\ and\ \bibinfo {author} {\bibfnamefont {P.}~\bibnamefont {Draper}},\
  }\href@noop {} {\bibfield  {journal} {\bibinfo  {journal} {J. High Energ.
  Phys.}\ }\textbf {\bibinfo {volume} {2015}},\ \bibinfo {pages} {132}
  (\bibinfo {year} {2015})},\ \Eprint
  {http://arxiv.org/abs/1506.05433}{arXiv:1506.05433}\BibitemShut {NoStop}%
\bibitem [{\citenamefont {Brubaker}\ \emph {et~al.}(2017)\citenamefont
  {Brubaker} \emph {et~al.}}]{PRL2017}%
  \BibitemOpen
  \bibfield  {author} {\bibinfo {author} {\bibfnamefont {B.~M.}\ \bibnamefont
  {Brubaker}} \emph {et~al.},\ }\href@noop {} {\bibfield  {journal} {\bibinfo
  {journal} {Phys. Rev. Lett.}\ }\textbf {\bibinfo {volume} {118}},\ \bibinfo
  {pages} {061302} (\bibinfo {year} {2017})},\ \Eprint
  {http://arxiv.org/abs/1610.02580}{arXiv:1610.02580}\BibitemShut {NoStop}%
\bibitem [{\citenamefont {Al~Kenany}\ \emph {et~al.}(2017)\citenamefont
  {Al~Kenany} \emph {et~al.}}]{NIM2017}%
  \BibitemOpen
  \bibfield  {author} {\bibinfo {author} {\bibfnamefont {S.}~\bibnamefont
  {Al~Kenany}} \emph {et~al.},\ }\href@noop {} {\bibfield  {journal} {\bibinfo
  {journal} {Nucl. Instrum. Meth. A}\ }\textbf {\bibinfo {volume} {854}},\
  \bibinfo {pages} {11} (\bibinfo {year} {2017})},\ \Eprint
  {http://arxiv.org/abs/1611.07123}{arXiv:1611.07123}\BibitemShut {NoStop}%
\bibitem [{\citenamefont {Brubaker}\ \emph {et~al.}()\citenamefont {Brubaker},
  \citenamefont {Zhong}, \citenamefont {Lamoreaux}, \citenamefont {Lehnert},\
  and\ \citenamefont {van Bibber}}]{PRD2017}%
  \BibitemOpen
  \bibfield  {author} {\bibinfo {author} {\bibfnamefont {B.~M.}\ \bibnamefont
  {Brubaker}}, \bibinfo {author} {\bibfnamefont {L.}~\bibnamefont {Zhong}},
  \bibinfo {author} {\bibfnamefont {S.~K.}\ \bibnamefont {Lamoreaux}}, \bibinfo
  {author} {\bibfnamefont {K.~W.}\ \bibnamefont {Lehnert}}, \ and\ \bibinfo
  {author} {\bibfnamefont {K.~A.}\ \bibnamefont {van Bibber}},\ }\href@noop {}
  {\ }\Eprint {http://arxiv.org/abs/1706.08388}{arXiv:1706.08388}\BibitemShut
  {NoStop}%
\bibitem [{\citenamefont {Griffiths}(2008)}]{griffiths2008}%
  \BibitemOpen
  \bibfield  {author} {\bibinfo {author} {\bibfnamefont {D.~J.}\ \bibnamefont
  {Griffiths}},\ }\href@noop {} {\emph {\bibinfo {title} {Introduction to
  {Elementary} {Particles}}}},\ \bibinfo {edition} {2nd}\ ed.\ (\bibinfo
  {publisher} {Wiley},\ \bibinfo {year} {2008})\BibitemShut {NoStop}%
\bibitem [{\citenamefont {Kolb}\ and\ \citenamefont {Turner}(1994)}]{KT1994}%
  \BibitemOpen
  \bibfield  {author} {\bibinfo {author} {\bibfnamefont {E.}~\bibnamefont
  {Kolb}}\ and\ \bibinfo {author} {\bibfnamefont {M.}~\bibnamefont {Turner}},\
  }\href@noop {} {\emph {\bibinfo {title} {The {Early} {Universe}}}}\ (\bibinfo
   {publisher} {Westview Press},\ \bibinfo {year} {1994})\BibitemShut {NoStop}%
\bibitem [{\citenamefont {Peskin}\ and\ \citenamefont
  {Schroeder}(1995)}]{PS1995}%
  \BibitemOpen
  \bibfield  {author} {\bibinfo {author} {\bibfnamefont {M.~E.}\ \bibnamefont
  {Peskin}}\ and\ \bibinfo {author} {\bibfnamefont {D.~V.}\ \bibnamefont
  {Schroeder}},\ }\href@noop {} {\emph {\bibinfo {title} {An {Introduction} to
  {Quantum} {Field} {Theory}}}}\ (\bibinfo  {publisher} {Westview Press},\
  \bibinfo {year} {1995})\BibitemShut {NoStop}%
\bibitem [{\citenamefont {Schwartz}(2014)}]{schwartz2014}%
  \BibitemOpen
  \bibfield  {author} {\bibinfo {author} {\bibfnamefont {M.~D.}\ \bibnamefont
  {Schwartz}},\ }\href@noop {} {\emph {\bibinfo {title} {Quantum {Field}
  {Theory} and the {Standard} {Model}}}}\ (\bibinfo  {publisher} {Cambridge
  University Press},\ \bibinfo {year} {2014})\BibitemShut {NoStop}%
\bibitem [{\citenamefont {Srednicki}(2007)}]{srednicki2007}%
  \BibitemOpen
  \bibfield  {author} {\bibinfo {author} {\bibfnamefont {M.}~\bibnamefont
  {Srednicki}},\ }\href@noop {} {\emph {\bibinfo {title} {Quantum {Field}
  {Theory}}}}\ (\bibinfo  {publisher} {Cambridge University Press},\ \bibinfo
  {year} {2007})\BibitemShut {NoStop}%
\bibitem [{\citenamefont {Cheng}(1988)}]{cheng1988}%
  \BibitemOpen
  \bibfield  {author} {\bibinfo {author} {\bibfnamefont {H.-Y.}\ \bibnamefont
  {Cheng}},\ }\href@noop {} {\bibfield  {journal} {\bibinfo  {journal} {Phys.
  Rep.}\ }\textbf {\bibinfo {volume} {158}},\ \bibinfo {pages} {1} (\bibinfo
  {year} {1988})}\BibitemShut {NoStop}%
\bibitem [{\citenamefont {Peccei}(1996)}]{peccei1996}%
  \BibitemOpen
  \bibfield  {author} {\bibinfo {author} {\bibfnamefont {R.~D.}\ \bibnamefont
  {Peccei}},\ }\href@noop {} {\bibfield  {journal} {\bibinfo  {journal} {J.
  Korean Phys. Soc.}\ }\textbf {\bibinfo {volume} {29}},\ \bibinfo {pages}
  {S199} (\bibinfo {year} {1996})},\ \Eprint
  {http://arxiv.org/abs/hep-ph/9606475}{arXiv:hep-ph/9606475}\BibitemShut
  {NoStop}%
\bibitem [{\citenamefont {Peccei}(2008)}]{peccei2008}%
  \BibitemOpen
  \bibfield  {author} {\bibinfo {author} {\bibfnamefont {R.~D.}\ \bibnamefont
  {Peccei}},\ }in\ \href@noop {} {\emph {\bibinfo {booktitle} {Axions:
  {Theory}, {Cosmology}, and {Experimental} {Searches}}}},\ \bibinfo {series}
  {Lecture {Notes} in {Physics}}, Vol.\ \bibinfo {volume} {741}\ (\bibinfo
  {publisher} {Springer},\ \bibinfo {year} {2008})\ pp.\ \bibinfo {pages}
  {3--17},\ \Eprint
  {http://arxiv.org/abs/hep-ph/0607268}{arXiv:hep-ph/0607268}\BibitemShut
  {NoStop}%
\bibitem [{\citenamefont {Aitchinson}(1984)}]{aitchinson1984}%
  \BibitemOpen
  \bibfield  {author} {\bibinfo {author} {\bibfnamefont {I.~J.~R.}\
  \bibnamefont {Aitchinson}},\ }\href@noop {} {\emph {\bibinfo {title} {An
  {Informal} {Introduction} to {Gauge} {Theories}}}}\ (\bibinfo  {publisher}
  {Cambridge University Press},\ \bibinfo {year} {1984})\BibitemShut {NoStop}%
\bibitem [{\citenamefont {Wikipedia}()}]{ssbfig}%
  \BibitemOpen
  \bibfield  {author} {\bibinfo {author} {\bibnamefont {Wikipedia}},\
  }\href@noop {} {}\bibinfo {howpublished}
  {\url{https://en.wikipedia.org/wiki/Spontaneous\_symmetry\_breaking}},\
  \bibinfo {note} {accessed 5/19/2017}\BibitemShut {NoStop}%
\bibitem [{\citenamefont {Kirzhnits}(1972)}]{kirzhnits1972}%
  \BibitemOpen
  \bibfield  {author} {\bibinfo {author} {\bibfnamefont {D.~A.}\ \bibnamefont
  {Kirzhnits}},\ }\href@noop {} {\bibfield  {journal} {\bibinfo  {journal} {J.
  Exp. Theor. Phys. Lett.}\ }\textbf {\bibinfo {volume} {15}},\ \bibinfo
  {pages} {745} (\bibinfo {year} {1972})}\BibitemShut {NoStop}%
\bibitem [{\citenamefont {Weinberg}(1974)}]{weinberg1974}%
  \BibitemOpen
  \bibfield  {author} {\bibinfo {author} {\bibfnamefont {S.}~\bibnamefont
  {Weinberg}},\ }\href@noop {} {\bibfield  {journal} {\bibinfo  {journal}
  {Phys. Rev. D}\ }\textbf {\bibinfo {volume} {9}},\ \bibinfo {pages} {3357}
  (\bibinfo {year} {1974})}\BibitemShut {NoStop}%
\bibitem [{\citenamefont {\'Alvarez-Gaum\'e}\ and\ \citenamefont
  {Ellis}(2011)}]{goldstonefig}%
  \BibitemOpen
  \bibfield  {author} {\bibinfo {author} {\bibfnamefont {L.}~\bibnamefont
  {\'Alvarez-Gaum\'e}}\ and\ \bibinfo {author} {\bibfnamefont {J.}~\bibnamefont
  {Ellis}},\ }\href@noop {} {\bibfield  {journal} {\bibinfo  {journal} {Nature
  Phys.}\ }\textbf {\bibinfo {volume} {7}},\ \bibinfo {pages} {2} (\bibinfo
  {year} {2011})}\BibitemShut {NoStop}%
\bibitem [{\citenamefont {Khriplovich}\ and\ \citenamefont
  {Lamoreaux}(1997)}]{kl1997}%
  \BibitemOpen
  \bibfield  {author} {\bibinfo {author} {\bibfnamefont {I.~B.}\ \bibnamefont
  {Khriplovich}}\ and\ \bibinfo {author} {\bibfnamefont {S.~K.}\ \bibnamefont
  {Lamoreaux}},\ }\href@noop {} {\emph {\bibinfo {title} {{$CP$} {Violation}
  {Without} {Strangeness}: {Electric} {Dipole} {Moments} of {Particles},
  {Atoms}, and {Molecules}}}}\ (\bibinfo  {publisher} {Springer},\ \bibinfo
  {year} {1997})\BibitemShut {NoStop}%
\bibitem [{\citenamefont {Kobayashi}\ and\ \citenamefont
  {Maskawa}(1973)}]{KM1973}%
  \BibitemOpen
  \bibfield  {author} {\bibinfo {author} {\bibfnamefont {M.}~\bibnamefont
  {Kobayashi}}\ and\ \bibinfo {author} {\bibfnamefont {T.}~\bibnamefont
  {Maskawa}},\ }\href@noop {} {\bibfield  {journal} {\bibinfo  {journal} {Prog.
  Theor. Phys.}\ }\textbf {\bibinfo {volume} {49}},\ \bibinfo {pages} {652}
  (\bibinfo {year} {1973})}\BibitemShut {NoStop}%
\bibitem [{\citenamefont {Patrignani}\ \emph {et~al.}(2016)\citenamefont
  {Patrignani} \emph {et~al.}}]{pdg2016}%
  \BibitemOpen
  \bibfield  {author} {\bibinfo {author} {\bibfnamefont {C.}~\bibnamefont
  {Patrignani}} \emph {et~al.} (\bibinfo {collaboration} {Particle Data
  Group}),\ }\href@noop {} {\bibfield  {journal} {\bibinfo  {journal} {Chinese
  Phys. C}\ }\textbf {\bibinfo {volume} {40}},\ \bibinfo {pages} {100001}
  (\bibinfo {year} {2016})}\BibitemShut {NoStop}%
\bibitem [{\citenamefont {Weinberg}(1975)}]{weinberg1975}%
  \BibitemOpen
  \bibfield  {author} {\bibinfo {author} {\bibfnamefont {S.}~\bibnamefont
  {Weinberg}},\ }\href@noop {} {\bibfield  {journal} {\bibinfo  {journal}
  {Phys. Rev. D}\ }\textbf {\bibinfo {volume} {11}},\ \bibinfo {pages} {3583}
  (\bibinfo {year} {1975})}\BibitemShut {NoStop}%
\bibitem [{\citenamefont {Adler}(1969)}]{adler1969}%
  \BibitemOpen
  \bibfield  {author} {\bibinfo {author} {\bibfnamefont {S.~L.}\ \bibnamefont
  {Adler}},\ }\href@noop {} {\bibfield  {journal} {\bibinfo  {journal} {Phys.
  Rev.}\ }\textbf {\bibinfo {volume} {177}},\ \bibinfo {pages} {2426} (\bibinfo
  {year} {1969})}\BibitemShut {NoStop}%
\bibitem [{\citenamefont {Bell}\ and\ \citenamefont {Jackiw}(1969)}]{bell1969}%
  \BibitemOpen
  \bibfield  {author} {\bibinfo {author} {\bibfnamefont {J.~S.}\ \bibnamefont
  {Bell}}\ and\ \bibinfo {author} {\bibfnamefont {R.}~\bibnamefont {Jackiw}},\
  }\href@noop {} {\bibfield  {journal} {\bibinfo  {journal} {Nuovo Cimento A}\
  }\textbf {\bibinfo {volume} {60}},\ \bibinfo {pages} {47} (\bibinfo {year}
  {1969})}\BibitemShut {NoStop}%
\bibitem [{\citenamefont {Primakoff}(1951)}]{primakoff1951}%
  \BibitemOpen
  \bibfield  {author} {\bibinfo {author} {\bibfnamefont {H.}~\bibnamefont
  {Primakoff}},\ }\href@noop {} {\bibfield  {journal} {\bibinfo  {journal}
  {Phys. Rev.}\ }\textbf {\bibinfo {volume} {81}},\ \bibinfo {pages} {899}
  (\bibinfo {year} {1951})}\BibitemShut {NoStop}%
\bibitem [{\citenamefont {Belavin}\ \emph {et~al.}(1975)\citenamefont
  {Belavin}, \citenamefont {Polyakov}, \citenamefont {Schwartz},\ and\
  \citenamefont {Tyupkin}}]{belavin1975}%
  \BibitemOpen
  \bibfield  {author} {\bibinfo {author} {\bibfnamefont {A.~A.}\ \bibnamefont
  {Belavin}}, \bibinfo {author} {\bibfnamefont {A.~M.}\ \bibnamefont
  {Polyakov}}, \bibinfo {author} {\bibfnamefont {A.~S.}\ \bibnamefont
  {Schwartz}}, \ and\ \bibinfo {author} {\bibfnamefont {Y.~S.}\ \bibnamefont
  {Tyupkin}},\ }\href@noop {} {\bibfield  {journal} {\bibinfo  {journal} {Phys.
  Lett. B}\ }\textbf {\bibinfo {volume} {59}},\ \bibinfo {pages} {85} (\bibinfo
  {year} {1975})}\BibitemShut {NoStop}%
\bibitem [{\citenamefont {'t~Hooft}(1976)}]{thooft1976}%
  \BibitemOpen
  \bibfield  {author} {\bibinfo {author} {\bibfnamefont {G.}~\bibnamefont
  {'t~Hooft}},\ }\href@noop {} {\bibfield  {journal} {\bibinfo  {journal}
  {Phys. Rev. Lett.}\ }\textbf {\bibinfo {volume} {37}},\ \bibinfo {pages} {8}
  (\bibinfo {year} {1976})}\BibitemShut {NoStop}%
\bibitem [{\citenamefont {Callan}\ \emph {et~al.}(1976)\citenamefont {Callan},
  \citenamefont {Dashen},\ and\ \citenamefont {Gross}}]{callan1976}%
  \BibitemOpen
  \bibfield  {author} {\bibinfo {author} {\bibfnamefont {C.~G.}\ \bibnamefont
  {Callan}}, \bibinfo {author} {\bibfnamefont {R.~F.}\ \bibnamefont {Dashen}},
  \ and\ \bibinfo {author} {\bibfnamefont {D.~J.}\ \bibnamefont {Gross}},\
  }\href@noop {} {\bibfield  {journal} {\bibinfo  {journal} {Phys. Lett. B}\
  }\textbf {\bibinfo {volume} {63}},\ \bibinfo {pages} {334} (\bibinfo {year}
  {1976})}\BibitemShut {NoStop}%
\bibitem [{\citenamefont {Jackiw}\ and\ \citenamefont
  {Rebbi}(1976)}]{jackiw1976}%
  \BibitemOpen
  \bibfield  {author} {\bibinfo {author} {\bibfnamefont {R.}~\bibnamefont
  {Jackiw}}\ and\ \bibinfo {author} {\bibfnamefont {C.}~\bibnamefont {Rebbi}},\
  }\href@noop {} {\bibfield  {journal} {\bibinfo  {journal} {Phys. Rev. Lett.}\
  }\textbf {\bibinfo {volume} {37}},\ \bibinfo {pages} {172} (\bibinfo {year}
  {1976})}\BibitemShut {NoStop}%
\bibitem [{\citenamefont {Gross}\ \emph {et~al.}(1981)\citenamefont {Gross},
  \citenamefont {Pisarski},\ and\ \citenamefont {Yaffe}}]{gross1981}%
  \BibitemOpen
  \bibfield  {author} {\bibinfo {author} {\bibfnamefont {D.~J.}\ \bibnamefont
  {Gross}}, \bibinfo {author} {\bibfnamefont {R.~D.}\ \bibnamefont {Pisarski}},
  \ and\ \bibinfo {author} {\bibfnamefont {L.~G.}\ \bibnamefont {Yaffe}},\
  }\href@noop {} {\bibfield  {journal} {\bibinfo  {journal} {Rev. Mod. Phys.}\
  }\textbf {\bibinfo {volume} {53}},\ \bibinfo {pages} {43} (\bibinfo {year}
  {1981})}\BibitemShut {NoStop}%
\bibitem [{\citenamefont {Fortson}\ \emph {et~al.}(2003)\citenamefont
  {Fortson}, \citenamefont {Sandars},\ and\ \citenamefont {Barr}}]{edmfig}%
  \BibitemOpen
  \bibfield  {author} {\bibinfo {author} {\bibfnamefont {N.}~\bibnamefont
  {Fortson}}, \bibinfo {author} {\bibfnamefont {P.}~\bibnamefont {Sandars}}, \
  and\ \bibinfo {author} {\bibfnamefont {S.}~\bibnamefont {Barr}},\ }\href@noop
  {} {\bibfield  {journal} {\bibinfo  {journal} {Phys. Today}\ }\textbf
  {\bibinfo {volume} {56}},\ \bibinfo {pages} {33} (\bibinfo {year}
  {2003})}\BibitemShut {NoStop}%
\bibitem [{\citenamefont {Baluni}(1979)}]{baluni1979}%
  \BibitemOpen
  \bibfield  {author} {\bibinfo {author} {\bibfnamefont {V.}~\bibnamefont
  {Baluni}},\ }\href@noop {} {\bibfield  {journal} {\bibinfo  {journal} {Phys.
  Rev. D}\ }\textbf {\bibinfo {volume} {19}},\ \bibinfo {pages} {2227}
  (\bibinfo {year} {1979})}\BibitemShut {NoStop}%
\bibitem [{\citenamefont {Crewther}\ \emph {et~al.}(1979)\citenamefont
  {Crewther}, \citenamefont {Di~Vecchia}, \citenamefont {Veneziano},\ and\
  \citenamefont {Witten}}]{crewther1979}%
  \BibitemOpen
  \bibfield  {author} {\bibinfo {author} {\bibfnamefont {R.~J.}\ \bibnamefont
  {Crewther}}, \bibinfo {author} {\bibfnamefont {P.}~\bibnamefont
  {Di~Vecchia}}, \bibinfo {author} {\bibfnamefont {G.}~\bibnamefont
  {Veneziano}}, \ and\ \bibinfo {author} {\bibfnamefont {E.}~\bibnamefont
  {Witten}},\ }\href@noop {} {\bibfield  {journal} {\bibinfo  {journal} {Phys.
  Lett. B}\ }\textbf {\bibinfo {volume} {88}},\ \bibinfo {pages} {123}
  (\bibinfo {year} {1979})}\BibitemShut {NoStop}%
\bibitem [{\citenamefont {Dress}\ \emph {et~al.}(1977)\citenamefont {Dress},
  \citenamefont {Miller}, \citenamefont {Pendlebury}, \citenamefont {Perrin},\
  and\ \citenamefont {Ramsey}}]{dress1977}%
  \BibitemOpen
  \bibfield  {author} {\bibinfo {author} {\bibfnamefont {W.~B.}\ \bibnamefont
  {Dress}}, \bibinfo {author} {\bibfnamefont {P.~D.}\ \bibnamefont {Miller}},
  \bibinfo {author} {\bibfnamefont {J.~M.}\ \bibnamefont {Pendlebury}},
  \bibinfo {author} {\bibfnamefont {P.}~\bibnamefont {Perrin}}, \ and\ \bibinfo
  {author} {\bibfnamefont {N.~F.}\ \bibnamefont {Ramsey}},\ }\href@noop {}
  {\bibfield  {journal} {\bibinfo  {journal} {Phys. Rev. D}\ }\textbf {\bibinfo
  {volume} {15}},\ \bibinfo {pages} {9} (\bibinfo {year} {1977})}\BibitemShut
  {NoStop}%
\bibitem [{\citenamefont {Pendlebury}\ \emph {et~al.}(2015)\citenamefont
  {Pendlebury} \emph {et~al.}}]{pendlebury2015}%
  \BibitemOpen
  \bibfield  {author} {\bibinfo {author} {\bibfnamefont {J.~M.}\ \bibnamefont
  {Pendlebury}} \emph {et~al.},\ }\href@noop {} {\bibfield  {journal} {\bibinfo
   {journal} {Phys. Rev. D}\ }\textbf {\bibinfo {volume} {92}},\ \bibinfo
  {pages} {092003} (\bibinfo {year} {2015})},\ \Eprint
  {http://arxiv.org/abs/1509.04411}{arXiv:1509.04411}\BibitemShut {NoStop}%
\bibitem [{\citenamefont {Ellis}\ and\ \citenamefont
  {Gaillard}(1979)}]{ellis1979}%
  \BibitemOpen
  \bibfield  {author} {\bibinfo {author} {\bibfnamefont {J.}~\bibnamefont
  {Ellis}}\ and\ \bibinfo {author} {\bibfnamefont {M.~K.}\ \bibnamefont
  {Gaillard}},\ }\href@noop {} {\bibfield  {journal} {\bibinfo  {journal}
  {Nucl. Phys. B}\ }\textbf {\bibinfo {volume} {150}},\ \bibinfo {pages} {141}
  (\bibinfo {year} {1979})}\BibitemShut {NoStop}%
\bibitem [{\citenamefont {Khriplovich}\ and\ \citenamefont
  {Vainshtein}(1994)}]{khriplovich1994}%
  \BibitemOpen
  \bibfield  {author} {\bibinfo {author} {\bibfnamefont {I.~B.}\ \bibnamefont
  {Khriplovich}}\ and\ \bibinfo {author} {\bibfnamefont {A.~I.}\ \bibnamefont
  {Vainshtein}},\ }\href@noop {} {\bibfield  {journal} {\bibinfo  {journal}
  {Nucl. Phys. B}\ }\textbf {\bibinfo {volume} {414}},\ \bibinfo {pages} {27}
  (\bibinfo {year} {1994})},\ \Eprint
  {http://arxiv.org/abs/hep-ph/9308334}{arXiv:hep-ph/9308334}\BibitemShut
  {NoStop}%
\bibitem [{\citenamefont {'t~Hooft}(1980)}]{thooft1980}%
  \BibitemOpen
  \bibfield  {author} {\bibinfo {author} {\bibfnamefont {G.}~\bibnamefont
  {'t~Hooft}},\ }in\ \href@noop {} {\emph {\bibinfo {booktitle} {Recent
  {Developments} in {Gauge} {Theories}}}}\ (\bibinfo  {publisher} {Springer},\
  \bibinfo {year} {1980})\ pp.\ \bibinfo {pages} {135--157}\BibitemShut
  {NoStop}%
\bibitem [{\citenamefont {Dine}(2015)}]{dine2015}%
  \BibitemOpen
  \bibfield  {author} {\bibinfo {author} {\bibfnamefont {M.}~\bibnamefont
  {Dine}},\ }\href@noop {} {\bibfield  {journal} {\bibinfo  {journal} {Annu.
  Rev. Nucl. Part. Sci.}\ }\textbf {\bibinfo {volume} {65}},\ \bibinfo {pages}
  {43} (\bibinfo {year} {2015})},\ \Eprint
  {http://arxiv.org/abs/1501.01035}{arXiv:1501.01035}\BibitemShut {NoStop}%
\bibitem [{\citenamefont {Nelson}(1984)}]{nelson1984}%
  \BibitemOpen
  \bibfield  {author} {\bibinfo {author} {\bibfnamefont {A.}~\bibnamefont
  {Nelson}},\ }\href@noop {} {\bibfield  {journal} {\bibinfo  {journal} {Phys.
  Lett. B}\ }\textbf {\bibinfo {volume} {136}},\ \bibinfo {pages} {387}
  (\bibinfo {year} {1984})}\BibitemShut {NoStop}%
\bibitem [{\citenamefont {Barr}(1984{\natexlab{a}})}]{barr1984a}%
  \BibitemOpen
  \bibfield  {author} {\bibinfo {author} {\bibfnamefont {S.~M.}\ \bibnamefont
  {Barr}},\ }\href@noop {} {\bibfield  {journal} {\bibinfo  {journal} {Phys.
  Rev. Lett.}\ }\textbf {\bibinfo {volume} {53}},\ \bibinfo {pages} {329}
  (\bibinfo {year} {1984}{\natexlab{a}})}\BibitemShut {NoStop}%
\bibitem [{\citenamefont {Barr}(1984{\natexlab{b}})}]{barr1984b}%
  \BibitemOpen
  \bibfield  {author} {\bibinfo {author} {\bibfnamefont {S.~M.}\ \bibnamefont
  {Barr}},\ }\href@noop {} {\bibfield  {journal} {\bibinfo  {journal} {Phys.
  Rev. D}\ }\textbf {\bibinfo {volume} {30}},\ \bibinfo {pages} {1805}
  (\bibinfo {year} {1984}{\natexlab{b}})}\BibitemShut {NoStop}%
\bibitem [{\citenamefont {Peccei}\ and\ \citenamefont
  {Quinn}(1977{\natexlab{a}})}]{pq1977a}%
  \BibitemOpen
  \bibfield  {author} {\bibinfo {author} {\bibfnamefont {R.~D.}\ \bibnamefont
  {Peccei}}\ and\ \bibinfo {author} {\bibfnamefont {H.~R.}\ \bibnamefont
  {Quinn}},\ }\href@noop {} {\bibfield  {journal} {\bibinfo  {journal} {Phys.
  Rev. Lett.}\ }\textbf {\bibinfo {volume} {38}},\ \bibinfo {pages} {1440}
  (\bibinfo {year} {1977}{\natexlab{a}})}\BibitemShut {NoStop}%
\bibitem [{\citenamefont {Peccei}\ and\ \citenamefont
  {Quinn}(1977{\natexlab{b}})}]{pq1977b}%
  \BibitemOpen
  \bibfield  {author} {\bibinfo {author} {\bibfnamefont {R.~D.}\ \bibnamefont
  {Peccei}}\ and\ \bibinfo {author} {\bibfnamefont {H.~R.}\ \bibnamefont
  {Quinn}},\ }\href@noop {} {\bibfield  {journal} {\bibinfo  {journal} {Phys.
  Rev. D}\ }\textbf {\bibinfo {volume} {16}},\ \bibinfo {pages} {1791}
  (\bibinfo {year} {1977}{\natexlab{b}})}\BibitemShut {NoStop}%
\bibitem [{\citenamefont {Weinberg}(1978)}]{weinberg1978}%
  \BibitemOpen
  \bibfield  {author} {\bibinfo {author} {\bibfnamefont {S.}~\bibnamefont
  {Weinberg}},\ }\href@noop {} {\bibfield  {journal} {\bibinfo  {journal}
  {Phys. Rev. Lett.}\ }\textbf {\bibinfo {volume} {40}},\ \bibinfo {pages}
  {223} (\bibinfo {year} {1978})}\BibitemShut {NoStop}%
\bibitem [{\citenamefont {Wilczek}(1978)}]{wilczek1978}%
  \BibitemOpen
  \bibfield  {author} {\bibinfo {author} {\bibfnamefont {F.}~\bibnamefont
  {Wilczek}},\ }\href@noop {} {\bibfield  {journal} {\bibinfo  {journal} {Phys.
  Rev. Lett.}\ }\textbf {\bibinfo {volume} {40}},\ \bibinfo {pages} {279}
  (\bibinfo {year} {1978})}\BibitemShut {NoStop}%
\bibitem [{\citenamefont {Vafa}\ and\ \citenamefont {Witten}(1984)}]{vafa1984}%
  \BibitemOpen
  \bibfield  {author} {\bibinfo {author} {\bibfnamefont {C.}~\bibnamefont
  {Vafa}}\ and\ \bibinfo {author} {\bibfnamefont {E.}~\bibnamefont {Witten}},\
  }\href@noop {} {\bibfield  {journal} {\bibinfo  {journal} {Phys. Rev. Lett.}\
  }\textbf {\bibinfo {volume} {53}},\ \bibinfo {pages} {535} (\bibinfo {year}
  {1984})}\BibitemShut {NoStop}%
\bibitem [{\citenamefont {Bardeen}\ and\ \citenamefont
  {Tye}(1978)}]{bardeen1978}%
  \BibitemOpen
  \bibfield  {author} {\bibinfo {author} {\bibfnamefont {W.~A.}\ \bibnamefont
  {Bardeen}}\ and\ \bibinfo {author} {\bibfnamefont {S.-H.~H.}\ \bibnamefont
  {Tye}},\ }\href@noop {} {\bibfield  {journal} {\bibinfo  {journal} {Phys.
  Lett. B}\ }\textbf {\bibinfo {volume} {74}},\ \bibinfo {pages} {229}
  (\bibinfo {year} {1978})}\BibitemShut {NoStop}%
\bibitem [{\citenamefont {Edwards}\ \emph {et~al.}(1982)\citenamefont {Edwards}
  \emph {et~al.}}]{edwards1982}%
  \BibitemOpen
  \bibfield  {author} {\bibinfo {author} {\bibfnamefont {C.}~\bibnamefont
  {Edwards}} \emph {et~al.} (\bibinfo {collaboration} {Crystal Ball
  Collaboration}),\ }\href@noop {} {\bibfield  {journal} {\bibinfo  {journal}
  {Phys. Rev. Lett.}\ }\textbf {\bibinfo {volume} {48}},\ \bibinfo {pages}
  {903} (\bibinfo {year} {1982})}\BibitemShut {NoStop}%
\bibitem [{\citenamefont {Sivertz}\ \emph {et~al.}(1982)\citenamefont {Sivertz}
  \emph {et~al.}}]{sivertz1982}%
  \BibitemOpen
  \bibfield  {author} {\bibinfo {author} {\bibfnamefont {M.}~\bibnamefont
  {Sivertz}} \emph {et~al.},\ }\href@noop {} {\bibfield  {journal} {\bibinfo
  {journal} {Phys. Rev. D}\ }\textbf {\bibinfo {volume} {26}},\ \bibinfo
  {pages} {717} (\bibinfo {year} {1982})}\BibitemShut {NoStop}%
\bibitem [{\citenamefont {Alam}\ \emph {et~al.}(1983)\citenamefont {Alam} \emph
  {et~al.}}]{alam1983}%
  \BibitemOpen
  \bibfield  {author} {\bibinfo {author} {\bibfnamefont {M.~S.}\ \bibnamefont
  {Alam}} \emph {et~al.},\ }\href@noop {} {\bibfield  {journal} {\bibinfo
  {journal} {Phys. Rev. D}\ }\textbf {\bibinfo {volume} {27}},\ \bibinfo
  {pages} {1665} (\bibinfo {year} {1983})}\BibitemShut {NoStop}%
\bibitem [{\citenamefont {Kim}(1979)}]{kim1979}%
  \BibitemOpen
  \bibfield  {author} {\bibinfo {author} {\bibfnamefont {J.~E.}\ \bibnamefont
  {Kim}},\ }\href@noop {} {\bibfield  {journal} {\bibinfo  {journal} {Phys.
  Rev. Lett.}\ }\textbf {\bibinfo {volume} {43}},\ \bibinfo {pages} {103}
  (\bibinfo {year} {1979})}\BibitemShut {NoStop}%
\bibitem [{\citenamefont {Shifman}\ \emph {et~al.}(1980)\citenamefont
  {Shifman}, \citenamefont {Vainshtein},\ and\ \citenamefont
  {Zakharov}}]{SVZ1980}%
  \BibitemOpen
  \bibfield  {author} {\bibinfo {author} {\bibfnamefont {M.~A.}\ \bibnamefont
  {Shifman}}, \bibinfo {author} {\bibfnamefont {A.~I.}\ \bibnamefont
  {Vainshtein}}, \ and\ \bibinfo {author} {\bibfnamefont {V.~I.}\ \bibnamefont
  {Zakharov}},\ }\href@noop {} {\bibfield  {journal} {\bibinfo  {journal}
  {Nucl. Phys. B}\ }\textbf {\bibinfo {volume} {166}},\ \bibinfo {pages} {493}
  (\bibinfo {year} {1980})}\BibitemShut {NoStop}%
\bibitem [{\citenamefont {Zhitnitsky}(1980)}]{zhitnitsky1980}%
  \BibitemOpen
  \bibfield  {author} {\bibinfo {author} {\bibfnamefont {A.~R.}\ \bibnamefont
  {Zhitnitsky}},\ }\href@noop {} {\bibfield  {journal} {\bibinfo  {journal}
  {Sov. J. Nucl. Phys.}\ }\textbf {\bibinfo {volume} {31}},\ \bibinfo {pages}
  {260} (\bibinfo {year} {1980})}\BibitemShut {NoStop}%
\bibitem [{\citenamefont {Dine}\ \emph {et~al.}(1981)\citenamefont {Dine},
  \citenamefont {Fischler},\ and\ \citenamefont {Srednicki}}]{DFS1981}%
  \BibitemOpen
  \bibfield  {author} {\bibinfo {author} {\bibfnamefont {M.}~\bibnamefont
  {Dine}}, \bibinfo {author} {\bibfnamefont {W.}~\bibnamefont {Fischler}}, \
  and\ \bibinfo {author} {\bibfnamefont {M.}~\bibnamefont {Srednicki}},\
  }\href@noop {} {\bibfield  {journal} {\bibinfo  {journal} {Phys. Lett. B}\
  }\textbf {\bibinfo {volume} {104}},\ \bibinfo {pages} {199} (\bibinfo {year}
  {1981})}\BibitemShut {NoStop}%
\bibitem [{\citenamefont {Wise}\ \emph {et~al.}(1981)\citenamefont {Wise},
  \citenamefont {Georgi},\ and\ \citenamefont {Glashow}}]{wise1981}%
  \BibitemOpen
  \bibfield  {author} {\bibinfo {author} {\bibfnamefont {M.~B.}\ \bibnamefont
  {Wise}}, \bibinfo {author} {\bibfnamefont {H.}~\bibnamefont {Georgi}}, \ and\
  \bibinfo {author} {\bibfnamefont {S.~L.}\ \bibnamefont {Glashow}},\
  }\href@noop {} {\bibfield  {journal} {\bibinfo  {journal} {Phys. Rev. Lett.}\
  }\textbf {\bibinfo {volume} {47}},\ \bibinfo {pages} {402} (\bibinfo {year}
  {1981})}\BibitemShut {NoStop}%
\bibitem [{\citenamefont {Bartelmann}(2010)}]{bartelmann2010}%
  \BibitemOpen
  \bibfield  {author} {\bibinfo {author} {\bibfnamefont {M.}~\bibnamefont
  {Bartelmann}},\ }\href@noop {} {\bibfield  {journal} {\bibinfo  {journal}
  {Rev. Mod. Phys.}\ }\textbf {\bibinfo {volume} {82}},\ \bibinfo {pages} {331}
  (\bibinfo {year} {2010})},\ \Eprint
  {http://arxiv.org/abs/0906.5036}{arXiv:0906.5036}\BibitemShut {NoStop}%
\bibitem [{\citenamefont {Bertone}\ and\ \citenamefont
  {Hooper}()}]{bertone2016}%
  \BibitemOpen
  \bibfield  {author} {\bibinfo {author} {\bibfnamefont {G.}~\bibnamefont
  {Bertone}}\ and\ \bibinfo {author} {\bibfnamefont {D.}~\bibnamefont
  {Hooper}},\ }\href@noop {} {\ }\Eprint
  {http://arxiv.org/abs/1605.04909}{arXiv:1605.04909}\BibitemShut {NoStop}%
\bibitem [{\citenamefont {Turner}(1990{\natexlab{a}})}]{windows1990}%
  \BibitemOpen
  \bibfield  {author} {\bibinfo {author} {\bibfnamefont {M.~S.}\ \bibnamefont
  {Turner}},\ }\href@noop {} {\bibfield  {journal} {\bibinfo  {journal} {Phys.
  Rep.}\ }\textbf {\bibinfo {volume} {197}},\ \bibinfo {pages} {67} (\bibinfo
  {year} {1990}{\natexlab{a}})}\BibitemShut {NoStop}%
\bibitem [{\citenamefont {Sikivie}(2008)}]{sikivie2008}%
  \BibitemOpen
  \bibfield  {author} {\bibinfo {author} {\bibfnamefont {P.}~\bibnamefont
  {Sikivie}},\ }in\ \href@noop {} {\emph {\bibinfo {booktitle} {Axions:
  {Theory}, {Cosmology}, and {Experimental} {Searches}}}},\ \bibinfo {series}
  {Lecture {Notes} in {Physics}}, Vol.\ \bibinfo {volume} {741}\ (\bibinfo
  {publisher} {Springer},\ \bibinfo {year} {2008})\ pp.\ \bibinfo {pages}
  {19--50}\BibitemShut {NoStop}%
\bibitem [{\citenamefont {Wantz}\ and\ \citenamefont
  {Shellard}(2010)}]{wantz2010}%
  \BibitemOpen
  \bibfield  {author} {\bibinfo {author} {\bibfnamefont {O.}~\bibnamefont
  {Wantz}}\ and\ \bibinfo {author} {\bibfnamefont {E.~P.~S.}\ \bibnamefont
  {Shellard}},\ }\href@noop {} {\bibfield  {journal} {\bibinfo  {journal}
  {Phys. Rev. D}\ }\textbf {\bibinfo {volume} {82}},\ \bibinfo {pages} {123508}
  (\bibinfo {year} {2010})},\ \Eprint
  {http://arxiv.org/abs/0910.1066}{arXiv:0910.1066}\BibitemShut {NoStop}%
\bibitem [{\citenamefont {Riess}\ \emph {et~al.}(2016)\citenamefont {Riess}
  \emph {et~al.}}]{riess2016}%
  \BibitemOpen
  \bibfield  {author} {\bibinfo {author} {\bibfnamefont {A.~G.}\ \bibnamefont
  {Riess}} \emph {et~al.},\ }\href@noop {} {\bibfield  {journal} {\bibinfo
  {journal} {Astrophys. J.}\ }\textbf {\bibinfo {volume} {826}},\ \bibinfo
  {pages} {56} (\bibinfo {year} {2016})},\ \Eprint
  {http://arxiv.org/abs/1604.01424}{arXiv:1604.01424}\BibitemShut {NoStop}%
\bibitem [{\citenamefont {Ade}\ \emph {et~al.}(2016)\citenamefont {Ade} \emph
  {et~al.}}]{planck2016}%
  \BibitemOpen
  \bibfield  {author} {\bibinfo {author} {\bibfnamefont {P.~A.~R.}\
  \bibnamefont {Ade}} \emph {et~al.} (\bibinfo {collaboration} {Planck
  collaboration}),\ }\href@noop {} {\bibfield  {journal} {\bibinfo  {journal}
  {Astron. Astrophys.}\ }\textbf {\bibinfo {volume} {594}},\ \bibinfo {pages}
  {A13} (\bibinfo {year} {2016})},\ \Eprint
  {http://arxiv.org/abs/1502.01589}{arXiv:1502.01589}\BibitemShut {NoStop}%
\bibitem [{\citenamefont {{Fermilab}}()}]{timeline}%
  \BibitemOpen
  \bibfield  {author} {\bibinfo {author} {\bibnamefont {{Fermilab}}},\
  }\href@noop {} {}\bibinfo {howpublished}
  {\url{http://vms.fnal.gov/asset/detail?recid=1742861}},\ \bibinfo {note}
  {image No.\ 85-0138, Accessed 6/12/2017}\BibitemShut {NoStop}%
\bibitem [{\citenamefont {Bertone}\ \emph {et~al.}(2005)\citenamefont
  {Bertone}, \citenamefont {Hooper},\ and\ \citenamefont {Silk}}]{bertone2005}%
  \BibitemOpen
  \bibfield  {author} {\bibinfo {author} {\bibfnamefont {G.}~\bibnamefont
  {Bertone}}, \bibinfo {author} {\bibfnamefont {D.}~\bibnamefont {Hooper}}, \
  and\ \bibinfo {author} {\bibfnamefont {J.}~\bibnamefont {Silk}},\ }\href@noop
  {} {\bibfield  {journal} {\bibinfo  {journal} {Phys. Rep.}\ }\textbf
  {\bibinfo {volume} {405}},\ \bibinfo {pages} {279} (\bibinfo {year}
  {2005})},\ \Eprint
  {http://arxiv.org/abs/hep-ph/0404175}{arXiv:hep-ph/0404175}\BibitemShut
  {NoStop}%
\bibitem [{\citenamefont {van Dokkum}\ \emph {et~al.}(2016)\citenamefont {van
  Dokkum}, \citenamefont {Abraham}, \citenamefont {Brodie}, \citenamefont
  {Conroy}, \citenamefont {Danieli}, \citenamefont {Merritt}, \citenamefont
  {Mowla}, \citenamefont {Romanowsky},\ and\ \citenamefont
  {Zhang}}]{vandokkum2016}%
  \BibitemOpen
  \bibfield  {author} {\bibinfo {author} {\bibfnamefont {P.}~\bibnamefont {van
  Dokkum}}, \bibinfo {author} {\bibfnamefont {R.}~\bibnamefont {Abraham}},
  \bibinfo {author} {\bibfnamefont {J.}~\bibnamefont {Brodie}}, \bibinfo
  {author} {\bibfnamefont {C.}~\bibnamefont {Conroy}}, \bibinfo {author}
  {\bibfnamefont {S.}~\bibnamefont {Danieli}}, \bibinfo {author} {\bibfnamefont
  {A.}~\bibnamefont {Merritt}}, \bibinfo {author} {\bibfnamefont
  {L.}~\bibnamefont {Mowla}}, \bibinfo {author} {\bibfnamefont
  {A.}~\bibnamefont {Romanowsky}}, \ and\ \bibinfo {author} {\bibfnamefont
  {J.}~\bibnamefont {Zhang}},\ }\href@noop {} {\bibfield  {journal} {\bibinfo
  {journal} {Astrophys. J. Lett.}\ }\textbf {\bibinfo {volume} {828}},\
  \bibinfo {pages} {L6} (\bibinfo {year} {2016})},\ \Eprint
  {http://arxiv.org/abs/1606.06291}{arXiv:1606.06291}\BibitemShut {NoStop}%
\bibitem [{\citenamefont {Peebles}(1982)}]{peebles1982}%
  \BibitemOpen
  \bibfield  {author} {\bibinfo {author} {\bibfnamefont {P.~J.~E.}\
  \bibnamefont {Peebles}},\ }\href@noop {} {\bibfield  {journal} {\bibinfo
  {journal} {Astrophys. J. Lett.}\ }\textbf {\bibinfo {volume} {263}},\
  \bibinfo {pages} {L1} (\bibinfo {year} {1982})}\BibitemShut {NoStop}%
\bibitem [{\citenamefont {Hu}()}]{waynehu}%
  \BibitemOpen
  \bibfield  {author} {\bibinfo {author} {\bibfnamefont {W.}~\bibnamefont
  {Hu}},\ }\href@noop {} {\enquote {\bibinfo {title} {{Cosmological}
  {Parameter} {Animations}},}\ }\bibinfo {howpublished}
  {\url{http://background.uchicago.edu/~whu/metaanim.html}},\ \bibinfo {note}
  {accessed 6/12/2017}\BibitemShut {NoStop}%
\bibitem [{\citenamefont {Samtleben}\ \emph {et~al.}(2007)\citenamefont
  {Samtleben}, \citenamefont {Staggs},\ and\ \citenamefont
  {Winstein}}]{samtleben2007}%
  \BibitemOpen
  \bibfield  {author} {\bibinfo {author} {\bibfnamefont {D.}~\bibnamefont
  {Samtleben}}, \bibinfo {author} {\bibfnamefont {S.}~\bibnamefont {Staggs}}, \
  and\ \bibinfo {author} {\bibfnamefont {B.}~\bibnamefont {Winstein}},\
  }\href@noop {} {\bibfield  {journal} {\bibinfo  {journal} {Annu. Rev. Nucl.
  Part. Sci.}\ }\textbf {\bibinfo {volume} {57}},\ \bibinfo {pages} {245}
  (\bibinfo {year} {2007})},\ \Eprint
  {http://arxiv.org/abs/0803.0834}{arXiv:0803.0834}\BibitemShut {NoStop}%
\bibitem [{\citenamefont {Blumenthal}\ \emph {et~al.}(1984)\citenamefont
  {Blumenthal}, \citenamefont {Faber}, \citenamefont {Primack},\ and\
  \citenamefont {Rees}}]{blumenthal1984}%
  \BibitemOpen
  \bibfield  {author} {\bibinfo {author} {\bibfnamefont {G.~R.}\ \bibnamefont
  {Blumenthal}}, \bibinfo {author} {\bibfnamefont {S.~M.}\ \bibnamefont
  {Faber}}, \bibinfo {author} {\bibfnamefont {J.~R.}\ \bibnamefont {Primack}},
  \ and\ \bibinfo {author} {\bibfnamefont {M.~J.}\ \bibnamefont {Rees}},\
  }\href@noop {} {\bibfield  {journal} {\bibinfo  {journal} {Nature}\ }\textbf
  {\bibinfo {volume} {311}},\ \bibinfo {pages} {517} (\bibinfo {year}
  {1984})}\BibitemShut {NoStop}%
\bibitem [{\citenamefont {Falkowski}(2016)}]{resonaances}%
  \BibitemOpen
  \bibfield  {author} {\bibinfo {author} {\bibfnamefont {A.}~\bibnamefont
  {Falkowski}},\ }\href@noop {} {}\bibinfo {howpublished}
  {\url{http://resonaances.blogspot.com/2016/09/weekend-plot-update-on-wimps.html}}
  (\bibinfo {year} {2016}),\ \bibinfo {note} {accessed 6/12/2017}\BibitemShut
  {NoStop}%
\bibitem [{\citenamefont {Bertone}(2010)}]{bertone2010}%
  \BibitemOpen
  \bibfield  {author} {\bibinfo {author} {\bibfnamefont {G.}~\bibnamefont
  {Bertone}},\ }\href@noop {} {\bibfield  {journal} {\bibinfo  {journal}
  {Nature}\ }\textbf {\bibinfo {volume} {468}},\ \bibinfo {pages} {389}
  (\bibinfo {year} {2010})},\ \Eprint
  {http://arxiv.org/abs/1011.3532}{arXiv:1011.3532}\BibitemShut {NoStop}%
\bibitem [{\citenamefont {Moroi}\ and\ \citenamefont
  {Murayama}(1998)}]{moroi1998}%
  \BibitemOpen
  \bibfield  {author} {\bibinfo {author} {\bibfnamefont {T.}~\bibnamefont
  {Moroi}}\ and\ \bibinfo {author} {\bibfnamefont {H.}~\bibnamefont
  {Murayama}},\ }\href@noop {} {\bibfield  {journal} {\bibinfo  {journal}
  {Phys. Lett. B}\ }\textbf {\bibinfo {volume} {440}},\ \bibinfo {pages} {69}
  (\bibinfo {year} {1998})},\ \Eprint
  {http://arxiv.org/abs/hep-ph/9804291}{arXiv:hep-ph/9804291}\BibitemShut
  {NoStop}%
\bibitem [{\citenamefont {Nelson}\ and\ \citenamefont
  {Scholtz}(2011)}]{nelson2011}%
  \BibitemOpen
  \bibfield  {author} {\bibinfo {author} {\bibfnamefont {A.~E.}\ \bibnamefont
  {Nelson}}\ and\ \bibinfo {author} {\bibfnamefont {J.}~\bibnamefont
  {Scholtz}},\ }\href@noop {} {\bibfield  {journal} {\bibinfo  {journal} {Phys.
  Rev. D}\ }\textbf {\bibinfo {volume} {84}},\ \bibinfo {pages} {103501}
  (\bibinfo {year} {2011})},\ \Eprint
  {http://arxiv.org/abs/1105.2812}{arXiv:1105.2812}\BibitemShut {NoStop}%
\bibitem [{\citenamefont {Arias}\ \emph {et~al.}(2012)\citenamefont {Arias},
  \citenamefont {Cadamuro}, \citenamefont {Goodsell}, \citenamefont {Jaeckel},
  \citenamefont {Redondo},\ and\ \citenamefont {Ringwald}}]{arias2012}%
  \BibitemOpen
  \bibfield  {author} {\bibinfo {author} {\bibfnamefont {P.}~\bibnamefont
  {Arias}}, \bibinfo {author} {\bibfnamefont {D.}~\bibnamefont {Cadamuro}},
  \bibinfo {author} {\bibfnamefont {M.}~\bibnamefont {Goodsell}}, \bibinfo
  {author} {\bibfnamefont {J.}~\bibnamefont {Jaeckel}}, \bibinfo {author}
  {\bibfnamefont {J.}~\bibnamefont {Redondo}}, \ and\ \bibinfo {author}
  {\bibfnamefont {A.}~\bibnamefont {Ringwald}},\ }\href@noop {} {\bibfield
  {journal} {\bibinfo  {journal} {J. Cosmol. Astropart. Phys.}\ }\textbf
  {\bibinfo {volume} {2012}},\ \bibinfo {pages} {013} (\bibinfo {year}
  {2012})},\ \Eprint
  {http://arxiv.org/abs/1201.5902}{arXiv:1201.5902}\BibitemShut {NoStop}%
\bibitem [{\citenamefont {Hui}\ \emph {et~al.}(2017)\citenamefont {Hui},
  \citenamefont {Ostriker}, \citenamefont {Tremaine},\ and\ \citenamefont
  {Witten}}]{hui2017}%
  \BibitemOpen
  \bibfield  {author} {\bibinfo {author} {\bibfnamefont {L.}~\bibnamefont
  {Hui}}, \bibinfo {author} {\bibfnamefont {J.~P.}\ \bibnamefont {Ostriker}},
  \bibinfo {author} {\bibfnamefont {S.}~\bibnamefont {Tremaine}}, \ and\
  \bibinfo {author} {\bibfnamefont {E.}~\bibnamefont {Witten}},\ }\href@noop {}
  {\bibfield  {journal} {\bibinfo  {journal} {Phys. Rev. D}\ }\textbf {\bibinfo
  {volume} {95}},\ \bibinfo {pages} {043541} (\bibinfo {year} {2017})},\
  \Eprint {http://arxiv.org/abs/1610.08297}{arXiv:1610.08297}\BibitemShut
  {NoStop}%
\bibitem [{\citenamefont {Preskill}\ \emph {et~al.}(1983)\citenamefont
  {Preskill}, \citenamefont {Wise},\ and\ \citenamefont {Wilczek}}]{pww1983}%
  \BibitemOpen
  \bibfield  {author} {\bibinfo {author} {\bibfnamefont {J.}~\bibnamefont
  {Preskill}}, \bibinfo {author} {\bibfnamefont {M.~B.}\ \bibnamefont {Wise}},
  \ and\ \bibinfo {author} {\bibfnamefont {F.}~\bibnamefont {Wilczek}},\
  }\href@noop {} {\bibfield  {journal} {\bibinfo  {journal} {Phys. Lett. B}\
  }\textbf {\bibinfo {volume} {120}},\ \bibinfo {pages} {127} (\bibinfo {year}
  {1983})}\BibitemShut {NoStop}%
\bibitem [{\citenamefont {Abbott}\ and\ \citenamefont
  {Sikivie}(1983)}]{as1983}%
  \BibitemOpen
  \bibfield  {author} {\bibinfo {author} {\bibfnamefont {L.~F.}\ \bibnamefont
  {Abbott}}\ and\ \bibinfo {author} {\bibfnamefont {P.}~\bibnamefont
  {Sikivie}},\ }\href@noop {} {\bibfield  {journal} {\bibinfo  {journal} {Phys.
  Lett. B}\ }\textbf {\bibinfo {volume} {120}},\ \bibinfo {pages} {133}
  (\bibinfo {year} {1983})}\BibitemShut {NoStop}%
\bibitem [{\citenamefont {Dine}\ and\ \citenamefont {Fischler}(1983)}]{df1983}%
  \BibitemOpen
  \bibfield  {author} {\bibinfo {author} {\bibfnamefont {M.}~\bibnamefont
  {Dine}}\ and\ \bibinfo {author} {\bibfnamefont {W.}~\bibnamefont
  {Fischler}},\ }\href@noop {} {\bibfield  {journal} {\bibinfo  {journal}
  {Phys. Lett. B}\ }\textbf {\bibinfo {volume} {120}},\ \bibinfo {pages} {137}
  (\bibinfo {year} {1983})}\BibitemShut {NoStop}%
\bibitem [{\citenamefont {Turner}(1983)}]{turner1983}%
  \BibitemOpen
  \bibfield  {author} {\bibinfo {author} {\bibfnamefont {M.~S.}\ \bibnamefont
  {Turner}},\ }\href@noop {} {\bibfield  {journal} {\bibinfo  {journal} {Phys.
  Rev. D}\ }\textbf {\bibinfo {volume} {28}},\ \bibinfo {pages} {1243}
  (\bibinfo {year} {1983})}\BibitemShut {NoStop}%
\bibitem [{\citenamefont {Davis}(1986)}]{davis1986}%
  \BibitemOpen
  \bibfield  {author} {\bibinfo {author} {\bibfnamefont {R.~L.}\ \bibnamefont
  {Davis}},\ }\href@noop {} {\bibfield  {journal} {\bibinfo  {journal} {Phys.
  Lett. B}\ }\textbf {\bibinfo {volume} {180}},\ \bibinfo {pages} {225}
  (\bibinfo {year} {1986})}\BibitemShut {NoStop}%
\bibitem [{\citenamefont {Di~Valentino}\ \emph {et~al.}(2016)\citenamefont
  {Di~Valentino}, \citenamefont {Giusarma}, \citenamefont {Lattanzi},
  \citenamefont {Mena}, \citenamefont {Melchiorri},\ and\ \citenamefont
  {Silk}}]{divalentino2016}%
  \BibitemOpen
  \bibfield  {author} {\bibinfo {author} {\bibfnamefont {E.}~\bibnamefont
  {Di~Valentino}}, \bibinfo {author} {\bibfnamefont {E.}~\bibnamefont
  {Giusarma}}, \bibinfo {author} {\bibfnamefont {M.}~\bibnamefont {Lattanzi}},
  \bibinfo {author} {\bibfnamefont {O.}~\bibnamefont {Mena}}, \bibinfo {author}
  {\bibfnamefont {A.}~\bibnamefont {Melchiorri}}, \ and\ \bibinfo {author}
  {\bibfnamefont {J.}~\bibnamefont {Silk}},\ }\href@noop {} {\bibfield
  {journal} {\bibinfo  {journal} {Phys. Lett. B}\ }\textbf {\bibinfo {volume}
  {752}},\ \bibinfo {pages} {182} (\bibinfo {year} {2016})},\ \Eprint
  {http://arxiv.org/abs/1507.08665}{arXiv:1507.08665}\BibitemShut {NoStop}%
\bibitem [{\citenamefont {Turner}(1986)}]{turner1986}%
  \BibitemOpen
  \bibfield  {author} {\bibinfo {author} {\bibfnamefont {M.~S.}\ \bibnamefont
  {Turner}},\ }\href@noop {} {\bibfield  {journal} {\bibinfo  {journal} {Phys.
  Rev. D}\ }\textbf {\bibinfo {volume} {33}},\ \bibinfo {pages} {889} (\bibinfo
  {year} {1986})}\BibitemShut {NoStop}%
\bibitem [{\citenamefont {Borsanyi}\ \emph {et~al.}(2016)\citenamefont
  {Borsanyi} \emph {et~al.}}]{borsanyi2016}%
  \BibitemOpen
  \bibfield  {author} {\bibinfo {author} {\bibfnamefont {S.}~\bibnamefont
  {Borsanyi}} \emph {et~al.},\ }\href@noop {} {\bibfield  {journal} {\bibinfo
  {journal} {Nature (London)}\ }\textbf {\bibinfo {volume} {539}},\ \bibinfo
  {pages} {69} (\bibinfo {year} {2016})}\BibitemShut {NoStop}%
\bibitem [{\citenamefont {Bonati}\ \emph {et~al.}(2016)\citenamefont {Bonati},
  \citenamefont {D'Elia}, \citenamefont {Mariti}, \citenamefont {Martinelli},
  \citenamefont {Mesiti}, \citenamefont {Negro}, \citenamefont {Sanfilippo},\
  and\ \citenamefont {Villadoro}}]{bonati2016}%
  \BibitemOpen
  \bibfield  {author} {\bibinfo {author} {\bibfnamefont {C.}~\bibnamefont
  {Bonati}}, \bibinfo {author} {\bibfnamefont {M.}~\bibnamefont {D'Elia}},
  \bibinfo {author} {\bibfnamefont {M.}~\bibnamefont {Mariti}}, \bibinfo
  {author} {\bibfnamefont {G.}~\bibnamefont {Martinelli}}, \bibinfo {author}
  {\bibfnamefont {M.}~\bibnamefont {Mesiti}}, \bibinfo {author} {\bibfnamefont
  {F.}~\bibnamefont {Negro}}, \bibinfo {author} {\bibfnamefont
  {F.}~\bibnamefont {Sanfilippo}}, \ and\ \bibinfo {author} {\bibfnamefont
  {G.}~\bibnamefont {Villadoro}},\ }\href@noop {} {\bibfield  {journal}
  {\bibinfo  {journal} {J. High Energ. Phys.}\ }\textbf {\bibinfo {volume}
  {2016}},\ \bibinfo {pages} {155} (\bibinfo {year} {2016})},\ \Eprint
  {http://arxiv.org/abs/1512.06746}{arXiv:1512.06746}\BibitemShut {NoStop}%
\bibitem [{\citenamefont {Steinhardt}\ and\ \citenamefont
  {Turner}(1983)}]{steinhardt1983}%
  \BibitemOpen
  \bibfield  {author} {\bibinfo {author} {\bibfnamefont {P.~J.}\ \bibnamefont
  {Steinhardt}}\ and\ \bibinfo {author} {\bibfnamefont {M.~S.}\ \bibnamefont
  {Turner}},\ }\href@noop {} {\bibfield  {journal} {\bibinfo  {journal} {Phys.
  Lett. B}\ }\textbf {\bibinfo {volume} {129}},\ \bibinfo {pages} {51}
  (\bibinfo {year} {1983})}\BibitemShut {NoStop}%
\bibitem [{\citenamefont {Vilenkin}(1985)}]{vilenkin1985}%
  \BibitemOpen
  \bibfield  {author} {\bibinfo {author} {\bibfnamefont {A.}~\bibnamefont
  {Vilenkin}},\ }\href@noop {} {\bibfield  {journal} {\bibinfo  {journal}
  {Phys. Rep.}\ }\textbf {\bibinfo {volume} {121}},\ \bibinfo {pages} {263}
  (\bibinfo {year} {1985})}\BibitemShut {NoStop}%
\bibitem [{\citenamefont {Fleury}\ and\ \citenamefont
  {Moore}(2016)}]{fleury2016}%
  \BibitemOpen
  \bibfield  {author} {\bibinfo {author} {\bibfnamefont {L.}~\bibnamefont
  {Fleury}}\ and\ \bibinfo {author} {\bibfnamefont {G.~D.}\ \bibnamefont
  {Moore}},\ }\href@noop {} {\bibfield  {journal} {\bibinfo  {journal} {J.
  Cosmol. Astropart. Phys.}\ }\textbf {\bibinfo {volume} {2016}},\ \bibinfo
  {pages} {004} (\bibinfo {year} {2016})},\ \Eprint
  {http://arxiv.org/abs/1509.00026}{arXiv:1509.00026}\BibitemShut {NoStop}%
\bibitem [{\citenamefont {Sikivie}(1982)}]{sikivie1982}%
  \BibitemOpen
  \bibfield  {author} {\bibinfo {author} {\bibfnamefont {P.}~\bibnamefont
  {Sikivie}},\ }\href@noop {} {\bibfield  {journal} {\bibinfo  {journal} {Phys.
  Rev. Lett.}\ }\textbf {\bibinfo {volume} {48}},\ \bibinfo {pages} {1156}
  (\bibinfo {year} {1982})}\BibitemShut {NoStop}%
\bibitem [{\citenamefont {Turner}\ \emph {et~al.}(1983)\citenamefont {Turner},
  \citenamefont {Wilczek},\ and\ \citenamefont {Zee}}]{twz1983}%
  \BibitemOpen
  \bibfield  {author} {\bibinfo {author} {\bibfnamefont {M.~S.}\ \bibnamefont
  {Turner}}, \bibinfo {author} {\bibfnamefont {F.}~\bibnamefont {Wilczek}}, \
  and\ \bibinfo {author} {\bibfnamefont {A.}~\bibnamefont {Zee}},\ }\href@noop
  {} {\bibfield  {journal} {\bibinfo  {journal} {Phys. Lett. B}\ }\textbf
  {\bibinfo {volume} {125}},\ \bibinfo {pages} {35} (\bibinfo {year}
  {1983})}\BibitemShut {NoStop}%
\bibitem [{\citenamefont {Axenides}\ \emph {et~al.}(1983)\citenamefont
  {Axenides}, \citenamefont {Brandenberger},\ and\ \citenamefont
  {Turner}}]{axenides1983}%
  \BibitemOpen
  \bibfield  {author} {\bibinfo {author} {\bibfnamefont {M.}~\bibnamefont
  {Axenides}}, \bibinfo {author} {\bibfnamefont {R.}~\bibnamefont
  {Brandenberger}}, \ and\ \bibinfo {author} {\bibfnamefont {M.}~\bibnamefont
  {Turner}},\ }\href@noop {} {\bibfield  {journal} {\bibinfo  {journal} {Phys.
  Lett. B}\ }\textbf {\bibinfo {volume} {126}},\ \bibinfo {pages} {178}
  (\bibinfo {year} {1983})}\BibitemShut {NoStop}%
\bibitem [{\citenamefont {Ipser}\ and\ \citenamefont
  {Sikivie}(1983)}]{ipser1983}%
  \BibitemOpen
  \bibfield  {author} {\bibinfo {author} {\bibfnamefont {J.}~\bibnamefont
  {Ipser}}\ and\ \bibinfo {author} {\bibfnamefont {P.}~\bibnamefont
  {Sikivie}},\ }\href@noop {} {\bibfield  {journal} {\bibinfo  {journal} {Phys.
  Rev. Lett.}\ }\textbf {\bibinfo {volume} {50}},\ \bibinfo {pages} {925}
  (\bibinfo {year} {1983})}\BibitemShut {NoStop}%
\bibitem [{\citenamefont {Stecker}\ and\ \citenamefont
  {Shafi}(1983)}]{stecker1983}%
  \BibitemOpen
  \bibfield  {author} {\bibinfo {author} {\bibfnamefont {F.~W.}\ \bibnamefont
  {Stecker}}\ and\ \bibinfo {author} {\bibfnamefont {Q.}~\bibnamefont
  {Shafi}},\ }\href@noop {} {\bibfield  {journal} {\bibinfo  {journal} {Phys.
  Rev. Lett.}\ }\textbf {\bibinfo {volume} {50}},\ \bibinfo {pages} {928}
  (\bibinfo {year} {1983})}\BibitemShut {NoStop}%
\bibitem [{\citenamefont {Sikivie}\ and\ \citenamefont
  {Yang}(2009)}]{sikivie2009}%
  \BibitemOpen
  \bibfield  {author} {\bibinfo {author} {\bibfnamefont {P.}~\bibnamefont
  {Sikivie}}\ and\ \bibinfo {author} {\bibfnamefont {Q.}~\bibnamefont {Yang}},\
  }\href@noop {} {\bibfield  {journal} {\bibinfo  {journal} {Phys. Rev. Lett.}\
  }\textbf {\bibinfo {volume} {103}},\ \bibinfo {pages} {111301} (\bibinfo
  {year} {2009})},\ \Eprint
  {http://arxiv.org/abs/0901.1106}{arXiv:0901.1106}\BibitemShut {NoStop}%
\bibitem [{\citenamefont {Guth}\ \emph {et~al.}(2015)\citenamefont {Guth},
  \citenamefont {Hertzberg},\ and\ \citenamefont
  {Prescod-Weinstein}}]{guth2015}%
  \BibitemOpen
  \bibfield  {author} {\bibinfo {author} {\bibfnamefont {A.~H.}\ \bibnamefont
  {Guth}}, \bibinfo {author} {\bibfnamefont {M.~P.}\ \bibnamefont {Hertzberg}},
  \ and\ \bibinfo {author} {\bibfnamefont {C.}~\bibnamefont
  {Prescod-Weinstein}},\ }\href@noop {} {\bibfield  {journal} {\bibinfo
  {journal} {Phys. Rev. D}\ }\textbf {\bibinfo {volume} {92}},\ \bibinfo
  {pages} {103513} (\bibinfo {year} {2015})},\ \Eprint
  {http://arxiv.org/abs/1412.5930}{arXiv:1412.5930}\BibitemShut {NoStop}%
\bibitem [{\citenamefont {Visinelli}\ and\ \citenamefont
  {Gondolo}(2014)}]{visinelli2014}%
  \BibitemOpen
  \bibfield  {author} {\bibinfo {author} {\bibfnamefont {L.}~\bibnamefont
  {Visinelli}}\ and\ \bibinfo {author} {\bibfnamefont {P.}~\bibnamefont
  {Gondolo}},\ }\href@noop {} {\bibfield  {journal} {\bibinfo  {journal} {Phys.
  Rev. Lett.}\ }\textbf {\bibinfo {volume} {113}},\ \bibinfo {pages} {011802}
  (\bibinfo {year} {2014})},\ \Eprint
  {http://arxiv.org/abs/1403.4594}{arXiv:1403.4594}\BibitemShut {NoStop}%
\bibitem [{\citenamefont {Di~Valentino}\ \emph {et~al.}(2014)\citenamefont
  {Di~Valentino}, \citenamefont {Giusarma}, \citenamefont {Lattanzi},
  \citenamefont {Melchiorri},\ and\ \citenamefont {Mena}}]{divalentino2014}%
  \BibitemOpen
  \bibfield  {author} {\bibinfo {author} {\bibfnamefont {E.}~\bibnamefont
  {Di~Valentino}}, \bibinfo {author} {\bibfnamefont {E.}~\bibnamefont
  {Giusarma}}, \bibinfo {author} {\bibfnamefont {M.}~\bibnamefont {Lattanzi}},
  \bibinfo {author} {\bibfnamefont {A.}~\bibnamefont {Melchiorri}}, \ and\
  \bibinfo {author} {\bibfnamefont {O.}~\bibnamefont {Mena}},\ }\href@noop {}
  {\bibfield  {journal} {\bibinfo  {journal} {Phys. Rev. D}\ }\textbf {\bibinfo
  {volume} {90}},\ \bibinfo {pages} {043534} (\bibinfo {year} {2014})},\
  \Eprint {http://arxiv.org/abs/1405.1860}{arXiv:1405.1860}\BibitemShut
  {NoStop}%
\bibitem [{\citenamefont {Olive}\ \emph {et~al.}(2014)\citenamefont {Olive}
  \emph {et~al.}}]{pdg2014}%
  \BibitemOpen
  \bibfield  {author} {\bibinfo {author} {\bibfnamefont {K.~A.}\ \bibnamefont
  {Olive}} \emph {et~al.} (\bibinfo {collaboration} {Particle Data Group}),\
  }\href@noop {} {\bibfield  {journal} {\bibinfo  {journal} {Chinese Phys. C}\
  }\textbf {\bibinfo {volume} {38}},\ \bibinfo {pages} {090001} (\bibinfo
  {year} {2014})}\BibitemShut {NoStop}%
\bibitem [{\citenamefont {Graham}\ \emph
  {et~al.}(2015{\natexlab{a}})\citenamefont {Graham}, \citenamefont
  {Irastorza}, \citenamefont {Lamoreaux}, \citenamefont {Lindner},\ and\
  \citenamefont {van Bibber}}]{annurev2015}%
  \BibitemOpen
  \bibfield  {author} {\bibinfo {author} {\bibfnamefont {P.~W.}\ \bibnamefont
  {Graham}}, \bibinfo {author} {\bibfnamefont {I.~G.}\ \bibnamefont
  {Irastorza}}, \bibinfo {author} {\bibfnamefont {S.~K.}\ \bibnamefont
  {Lamoreaux}}, \bibinfo {author} {\bibfnamefont {A.}~\bibnamefont {Lindner}},
  \ and\ \bibinfo {author} {\bibfnamefont {K.~A.}\ \bibnamefont {van Bibber}},\
  }\href@noop {} {\bibfield  {journal} {\bibinfo  {journal} {Annu. Rev. Nucl.
  Part. Sci.}\ }\textbf {\bibinfo {volume} {65}},\ \bibinfo {pages} {485}
  (\bibinfo {year} {2015}{\natexlab{a}})},\ \Eprint
  {http://arxiv.org/abs/1602.00039}{arXiv:1602.00039}\BibitemShut {NoStop}%
\bibitem [{\citenamefont {Cheng}\ \emph {et~al.}(1995)\citenamefont {Cheng},
  \citenamefont {Geng},\ and\ \citenamefont {Ni}}]{cheng1995}%
  \BibitemOpen
  \bibfield  {author} {\bibinfo {author} {\bibfnamefont {S.~L.}\ \bibnamefont
  {Cheng}}, \bibinfo {author} {\bibfnamefont {C.~Q.}\ \bibnamefont {Geng}}, \
  and\ \bibinfo {author} {\bibfnamefont {W.-T.}\ \bibnamefont {Ni}},\
  }\href@noop {} {\bibfield  {journal} {\bibinfo  {journal} {Phys. Rev. D}\
  }\textbf {\bibinfo {volume} {52}},\ \bibinfo {pages} {3132} (\bibinfo {year}
  {1995})},\ \Eprint
  {http://arxiv.org/abs/hep-ph/9506295}{arXiv:hep-ph/9506295}\BibitemShut
  {NoStop}%
\bibitem [{\citenamefont {Kaplan}(1985)}]{kaplan1985}%
  \BibitemOpen
  \bibfield  {author} {\bibinfo {author} {\bibfnamefont {D.~B.}\ \bibnamefont
  {Kaplan}},\ }\href@noop {} {\bibfield  {journal} {\bibinfo  {journal} {Nucl.
  Phys. B}\ }\textbf {\bibinfo {volume} {260}},\ \bibinfo {pages} {215}
  (\bibinfo {year} {1985})}\BibitemShut {NoStop}%
\bibitem [{\citenamefont {Kim}(1998)}]{kim1998}%
  \BibitemOpen
  \bibfield  {author} {\bibinfo {author} {\bibfnamefont {J.~E.}\ \bibnamefont
  {Kim}},\ }\href@noop {} {\bibfield  {journal} {\bibinfo  {journal} {Phys.
  Rev. D}\ }\textbf {\bibinfo {volume} {58}},\ \bibinfo {pages} {055006}
  (\bibinfo {year} {1998})},\ \Eprint
  {http://arxiv.org/abs/hep-ph/9802220}{arXiv:hep-ph/9802220}\BibitemShut
  {NoStop}%
\bibitem [{\citenamefont {Espriu}\ \emph {et~al.}(2015)\citenamefont {Espriu},
  \citenamefont {Mescia},\ and\ \citenamefont {Renau}}]{espriu2015}%
  \BibitemOpen
  \bibfield  {author} {\bibinfo {author} {\bibfnamefont {D.}~\bibnamefont
  {Espriu}}, \bibinfo {author} {\bibfnamefont {F.}~\bibnamefont {Mescia}}, \
  and\ \bibinfo {author} {\bibfnamefont {A.}~\bibnamefont {Renau}},\
  }\href@noop {} {\bibfield  {journal} {\bibinfo  {journal} {Phys. Rev. D}\
  }\textbf {\bibinfo {volume} {92}},\ \bibinfo {pages} {095013} (\bibinfo
  {year} {2015})},\ \Eprint
  {http://arxiv.org/abs/1503.02953}{arXiv:1503.02953}\BibitemShut {NoStop}%
\bibitem [{\citenamefont {Di~Luzio}\ \emph {et~al.}(2017)\citenamefont
  {Di~Luzio}, \citenamefont {Mescia},\ and\ \citenamefont
  {Nardi}}]{diluzio2017}%
  \BibitemOpen
  \bibfield  {author} {\bibinfo {author} {\bibfnamefont {L.}~\bibnamefont
  {Di~Luzio}}, \bibinfo {author} {\bibfnamefont {F.}~\bibnamefont {Mescia}}, \
  and\ \bibinfo {author} {\bibfnamefont {E.}~\bibnamefont {Nardi}},\
  }\href@noop {} {\bibfield  {journal} {\bibinfo  {journal} {Phys. Rev. Lett.}\
  }\textbf {\bibinfo {volume} {118}},\ \bibinfo {pages} {031801} (\bibinfo
  {year} {2017})},\ \Eprint
  {http://arxiv.org/abs/1610.07593}{arXiv:1610.07593}\BibitemShut {NoStop}%
\bibitem [{\citenamefont {Raffelt}(2008)}]{raffelt2008}%
  \BibitemOpen
  \bibfield  {author} {\bibinfo {author} {\bibfnamefont {G.~G.}\ \bibnamefont
  {Raffelt}},\ }in\ \href@noop {} {\emph {\bibinfo {booktitle} {Axions:
  {Theory}, {Cosmology}, and {Experimental} {Searches}}}},\ \bibinfo {series}
  {Lecture {Notes} in {Physics}}, Vol.\ \bibinfo {volume} {741}\ (\bibinfo
  {publisher} {Springer},\ \bibinfo {year} {2008})\ pp.\ \bibinfo {pages}
  {51--71},\ \Eprint
  {http://arxiv.org/abs/hep-ph/0611350}{arXiv:hep-ph/0611350}\BibitemShut
  {NoStop}%
\bibitem [{\citenamefont {Kawasaki}\ \emph {et~al.}(2015)\citenamefont
  {Kawasaki}, \citenamefont {Saikawa},\ and\ \citenamefont
  {Sekiguchi}}]{kawasaki2015}%
  \BibitemOpen
  \bibfield  {author} {\bibinfo {author} {\bibfnamefont {M.}~\bibnamefont
  {Kawasaki}}, \bibinfo {author} {\bibfnamefont {K.}~\bibnamefont {Saikawa}}, \
  and\ \bibinfo {author} {\bibfnamefont {T.}~\bibnamefont {Sekiguchi}},\
  }\href@noop {} {\bibfield  {journal} {\bibinfo  {journal} {Phys. Rev. D}\
  }\textbf {\bibinfo {volume} {91}},\ \bibinfo {pages} {065014} (\bibinfo
  {year} {2015})},\ \Eprint
  {http://arxiv.org/abs/1412.0789}{arXiv:1412.0789}\BibitemShut {NoStop}%
\bibitem [{\citenamefont {Jimenez}\ \emph {et~al.}(2003)\citenamefont
  {Jimenez}, \citenamefont {Verde},\ and\ \citenamefont {Oh}}]{jimenez2003}%
  \BibitemOpen
  \bibfield  {author} {\bibinfo {author} {\bibfnamefont {R.}~\bibnamefont
  {Jimenez}}, \bibinfo {author} {\bibfnamefont {L.}~\bibnamefont {Verde}}, \
  and\ \bibinfo {author} {\bibfnamefont {S.~P.}\ \bibnamefont {Oh}},\
  }\href@noop {} {\bibfield  {journal} {\bibinfo  {journal} {Mon. Not. R.
  Astron. Soc.}\ }\textbf {\bibinfo {volume} {339}},\ \bibinfo {pages} {243}
  (\bibinfo {year} {2003})},\ \Eprint
  {http://arxiv.org/abs/astro-ph/0201352}{arXiv:astro-ph/0201352}\BibitemShut
  {NoStop}%
\bibitem [{\citenamefont {Read}(2014)}]{read2014}%
  \BibitemOpen
  \bibfield  {author} {\bibinfo {author} {\bibfnamefont {J.~I.}\ \bibnamefont
  {Read}},\ }\href@noop {} {\bibfield  {journal} {\bibinfo  {journal} {J. Phys.
  G}\ }\textbf {\bibinfo {volume} {41}},\ \bibinfo {pages} {063101} (\bibinfo
  {year} {2014})},\ \Eprint
  {http://arxiv.org/abs/1404.1938}{arXiv:1404.1938}\BibitemShut {NoStop}%
\bibitem [{\citenamefont {Iocco}\ \emph {et~al.}(2015)\citenamefont {Iocco},
  \citenamefont {Pato},\ and\ \citenamefont {Bertone}}]{iocco2015}%
  \BibitemOpen
  \bibfield  {author} {\bibinfo {author} {\bibfnamefont {F.}~\bibnamefont
  {Iocco}}, \bibinfo {author} {\bibfnamefont {M.}~\bibnamefont {Pato}}, \ and\
  \bibinfo {author} {\bibfnamefont {G.}~\bibnamefont {Bertone}},\ }\href@noop
  {} {\bibfield  {journal} {\bibinfo  {journal} {Nature Phys.}\ }\textbf
  {\bibinfo {volume} {11}},\ \bibinfo {pages} {245} (\bibinfo {year} {2015})},\
  \Eprint {http://arxiv.org/abs/1502.03821}{arXiv:1502.03821}\BibitemShut
  {NoStop}%
\bibitem [{\citenamefont {Turner}(1990{\natexlab{b}})}]{turner1990}%
  \BibitemOpen
  \bibfield  {author} {\bibinfo {author} {\bibfnamefont {M.~S.}\ \bibnamefont
  {Turner}},\ }\href@noop {} {\bibfield  {journal} {\bibinfo  {journal} {Phys.
  Rev. D}\ }\textbf {\bibinfo {volume} {42}},\ \bibinfo {pages} {3572}
  (\bibinfo {year} {1990}{\natexlab{b}})}\BibitemShut {NoStop}%
\bibitem [{\citenamefont {Krauss}\ \emph {et~al.}(1985)\citenamefont {Krauss},
  \citenamefont {Moody}, \citenamefont {Wilczek},\ and\ \citenamefont
  {Morris}}]{krauss1985}%
  \BibitemOpen
  \bibfield  {author} {\bibinfo {author} {\bibfnamefont {L.}~\bibnamefont
  {Krauss}}, \bibinfo {author} {\bibfnamefont {J.}~\bibnamefont {Moody}},
  \bibinfo {author} {\bibfnamefont {F.}~\bibnamefont {Wilczek}}, \ and\
  \bibinfo {author} {\bibfnamefont {D.~E.}\ \bibnamefont {Morris}},\
  }\href@noop {} {\bibfield  {journal} {\bibinfo  {journal} {Phys. Rev. Lett.}\
  }\textbf {\bibinfo {volume} {55}},\ \bibinfo {pages} {1797} (\bibinfo {year}
  {1985})}\BibitemShut {NoStop}%
\bibitem [{\citenamefont {Sikivie}(1983)}]{sikivie1983}%
  \BibitemOpen
  \bibfield  {author} {\bibinfo {author} {\bibfnamefont {P.}~\bibnamefont
  {Sikivie}},\ }\href@noop {} {\bibfield  {journal} {\bibinfo  {journal} {Phys.
  Rev. Lett.}\ }\textbf {\bibinfo {volume} {51}},\ \bibinfo {pages} {1415}
  (\bibinfo {year} {1983})}\BibitemShut {NoStop}%
\bibitem [{\citenamefont {Sikivie}(1985)}]{sikivie1985}%
  \BibitemOpen
  \bibfield  {author} {\bibinfo {author} {\bibfnamefont {P.}~\bibnamefont
  {Sikivie}},\ }\href@noop {} {\bibfield  {journal} {\bibinfo  {journal} {Phys.
  Rev. D}\ }\textbf {\bibinfo {volume} {32}},\ \bibinfo {pages} {2988}
  (\bibinfo {year} {1985})}\BibitemShut {NoStop}%
\bibitem [{\citenamefont {Cadamuro}(2012)}]{cadamuro2012}%
  \BibitemOpen
  \bibfield  {author} {\bibinfo {author} {\bibfnamefont {D.}~\bibnamefont
  {Cadamuro}},\ }\emph {\bibinfo {title} {Cosmological limits on axions and
  axion-like particles}},\ \href@noop {} {Ph.D. thesis},\ \bibinfo  {school}
  {Ludwig Maximilian University of Munich} (\bibinfo {year} {2012}),\ \Eprint
  {http://arxiv.org/abs/1210.3196}{arXiv:1210.3196}\BibitemShut {NoStop}%
\bibitem [{\citenamefont {Blout}\ \emph {et~al.}(2001)\citenamefont {Blout},
  \citenamefont {Daw}, \citenamefont {Decowski}, \citenamefont {Ho},
  \citenamefont {Rosenberg},\ and\ \citenamefont {Yu}}]{blout2001}%
  \BibitemOpen
  \bibfield  {author} {\bibinfo {author} {\bibfnamefont {B.~D.}\ \bibnamefont
  {Blout}}, \bibinfo {author} {\bibfnamefont {E.~J.}\ \bibnamefont {Daw}},
  \bibinfo {author} {\bibfnamefont {M.~P.}\ \bibnamefont {Decowski}}, \bibinfo
  {author} {\bibfnamefont {P.~T.~P.}\ \bibnamefont {Ho}}, \bibinfo {author}
  {\bibfnamefont {L.~J.}\ \bibnamefont {Rosenberg}}, \ and\ \bibinfo {author}
  {\bibfnamefont {D.~B.}\ \bibnamefont {Yu}},\ }\href@noop {} {\bibfield
  {journal} {\bibinfo  {journal} {Astrophys. J.}\ }\textbf {\bibinfo {volume}
  {546}},\ \bibinfo {pages} {825} (\bibinfo {year} {2001})},\ \Eprint
  {http://arxiv.org/abs/astro-ph/0006310}{arXiv:astro-ph/0006310}\BibitemShut
  {NoStop}%
\bibitem [{\citenamefont {Chou}()}]{aaron}%
  \BibitemOpen
  \bibfield  {author} {\bibinfo {author} {\bibfnamefont {A.~S.}\ \bibnamefont
  {Chou}},\ }\href@noop {} {}\bibinfo {howpublished}
  {\url{http://home.fnal.gov/\~achou/}},\ \bibinfo {note} {accessed
  6/28/2017}\BibitemShut {NoStop}%
\bibitem [{\citenamefont {Zheng}\ \emph {et~al.}()\citenamefont {Zheng},
  \citenamefont {Silveri}, \citenamefont {Brierley}, \citenamefont {Girvin},\
  and\ \citenamefont {Lehnert}}]{zheng2016}%
  \BibitemOpen
  \bibfield  {author} {\bibinfo {author} {\bibfnamefont {H.}~\bibnamefont
  {Zheng}}, \bibinfo {author} {\bibfnamefont {M.}~\bibnamefont {Silveri}},
  \bibinfo {author} {\bibfnamefont {R.~T.}\ \bibnamefont {Brierley}}, \bibinfo
  {author} {\bibfnamefont {S.~M.}\ \bibnamefont {Girvin}}, \ and\ \bibinfo
  {author} {\bibfnamefont {K.~W.}\ \bibnamefont {Lehnert}},\ }\href@noop {} {\
  }\Eprint {http://arxiv.org/abs/1607.02529}{arXiv:1607.02529}\BibitemShut
  {NoStop}%
\bibitem [{\citenamefont {Nyquist}(1928)}]{nyquist1928}%
  \BibitemOpen
  \bibfield  {author} {\bibinfo {author} {\bibfnamefont {H.}~\bibnamefont
  {Nyquist}},\ }\href@noop {} {\bibfield  {journal} {\bibinfo  {journal} {Phys.
  Rev.}\ }\textbf {\bibinfo {volume} {32}},\ \bibinfo {pages} {110} (\bibinfo
  {year} {1928})}\BibitemShut {NoStop}%
\bibitem [{\citenamefont {Callen}\ and\ \citenamefont {Welton}(1951)}]{cw1951}%
  \BibitemOpen
  \bibfield  {author} {\bibinfo {author} {\bibfnamefont {H.~B.}\ \bibnamefont
  {Callen}}\ and\ \bibinfo {author} {\bibfnamefont {T.~A.}\ \bibnamefont
  {Welton}},\ }\href@noop {} {\bibfield  {journal} {\bibinfo  {journal} {Phys.
  Rev.}\ }\textbf {\bibinfo {volume} {83}},\ \bibinfo {pages} {34} (\bibinfo
  {year} {1951})}\BibitemShut {NoStop}%
\bibitem [{\citenamefont {Dicke}(1946)}]{dicke1946}%
  \BibitemOpen
  \bibfield  {author} {\bibinfo {author} {\bibfnamefont {R.~H.}\ \bibnamefont
  {Dicke}},\ }\href@noop {} {\bibfield  {journal} {\bibinfo  {journal} {Rev.
  Sci. Instrum.}\ }\textbf {\bibinfo {volume} {17}},\ \bibinfo {pages} {268}
  (\bibinfo {year} {1946})}\BibitemShut {NoStop}%
\bibitem [{\citenamefont {Zmuidzinas}(2003)}]{zmuidzinas2003}%
  \BibitemOpen
  \bibfield  {author} {\bibinfo {author} {\bibfnamefont {J.}~\bibnamefont
  {Zmuidzinas}},\ }\href@noop {} {\bibfield  {journal} {\bibinfo  {journal}
  {Appl. Opt.}\ }\textbf {\bibinfo {volume} {42}},\ \bibinfo {pages} {4989}
  (\bibinfo {year} {2003})}\BibitemShut {NoStop}%
\bibitem [{\citenamefont {Pozar}(2012)}]{pozar2012}%
  \BibitemOpen
  \bibfield  {author} {\bibinfo {author} {\bibfnamefont {D.~M.}\ \bibnamefont
  {Pozar}},\ }\href@noop {} {\emph {\bibinfo {title} {Microwave
  {Engineering}}}},\ \bibinfo {edition} {4th}\ ed.\ (\bibinfo  {publisher}
  {Wiley},\ \bibinfo {year} {2012})\BibitemShut {NoStop}%
\bibitem [{\citenamefont {Haus}\ and\ \citenamefont {Mullen}(1962)}]{haus1962}%
  \BibitemOpen
  \bibfield  {author} {\bibinfo {author} {\bibfnamefont {H.~A.}\ \bibnamefont
  {Haus}}\ and\ \bibinfo {author} {\bibfnamefont {J.~A.}\ \bibnamefont
  {Mullen}},\ }\href@noop {} {\bibfield  {journal} {\bibinfo  {journal} {Phys.
  Rev.}\ }\textbf {\bibinfo {volume} {128}},\ \bibinfo {pages} {2407} (\bibinfo
  {year} {1962})}\BibitemShut {NoStop}%
\bibitem [{\citenamefont {Caves}(1982)}]{caves1982}%
  \BibitemOpen
  \bibfield  {author} {\bibinfo {author} {\bibfnamefont {C.~M.}\ \bibnamefont
  {Caves}},\ }\href@noop {} {\bibfield  {journal} {\bibinfo  {journal} {Phys.
  Rev. D}\ }\textbf {\bibinfo {volume} {26}},\ \bibinfo {pages} {1817}
  (\bibinfo {year} {1982})}\BibitemShut {NoStop}%
\bibitem [{\citenamefont {Lamoreaux}\ \emph {et~al.}(2013)\citenamefont
  {Lamoreaux}, \citenamefont {van Bibber}, \citenamefont {Lehnert},\ and\
  \citenamefont {Carosi}}]{lamoreaux2013}%
  \BibitemOpen
  \bibfield  {author} {\bibinfo {author} {\bibfnamefont {S.~K.}\ \bibnamefont
  {Lamoreaux}}, \bibinfo {author} {\bibfnamefont {K.~A.}\ \bibnamefont {van
  Bibber}}, \bibinfo {author} {\bibfnamefont {K.~W.}\ \bibnamefont {Lehnert}},
  \ and\ \bibinfo {author} {\bibfnamefont {G.}~\bibnamefont {Carosi}},\
  }\href@noop {} {\bibfield  {journal} {\bibinfo  {journal} {Phys. Rev. D}\
  }\textbf {\bibinfo {volume} {88}},\ \bibinfo {pages} {035020} (\bibinfo
  {year} {2013})},\ \Eprint
  {http://arxiv.org/abs/1306.3591}{arXiv:1306.3591}\BibitemShut {NoStop}%
\bibitem [{\citenamefont {DePanfilis}\ \emph {et~al.}(1987)\citenamefont
  {DePanfilis}, \citenamefont {Melissinos}, \citenamefont {Moskowitz},
  \citenamefont {Rogers}, \citenamefont {Semertzidis}, \citenamefont {Wuensch},
  \citenamefont {Halama}, \citenamefont {Prodell}, \citenamefont {Fowler},\
  and\ \citenamefont {Nezrick}}]{RBF1987}%
  \BibitemOpen
  \bibfield  {author} {\bibinfo {author} {\bibfnamefont {S.}~\bibnamefont
  {DePanfilis}}, \bibinfo {author} {\bibfnamefont {A.~C.}\ \bibnamefont
  {Melissinos}}, \bibinfo {author} {\bibfnamefont {B.~E.}\ \bibnamefont
  {Moskowitz}}, \bibinfo {author} {\bibfnamefont {J.~T.}\ \bibnamefont
  {Rogers}}, \bibinfo {author} {\bibfnamefont {Y.~K.}\ \bibnamefont
  {Semertzidis}}, \bibinfo {author} {\bibfnamefont {W.~U.}\ \bibnamefont
  {Wuensch}}, \bibinfo {author} {\bibfnamefont {H.~J.}\ \bibnamefont {Halama}},
  \bibinfo {author} {\bibfnamefont {A.~G.}\ \bibnamefont {Prodell}}, \bibinfo
  {author} {\bibfnamefont {W.~B.}\ \bibnamefont {Fowler}}, \ and\ \bibinfo
  {author} {\bibfnamefont {F.~A.}\ \bibnamefont {Nezrick}},\ }\href@noop {}
  {\bibfield  {journal} {\bibinfo  {journal} {Phys. Rev. Lett.}\ }\textbf
  {\bibinfo {volume} {59}},\ \bibinfo {pages} {839} (\bibinfo {year}
  {1987})}\BibitemShut {NoStop}%
\bibitem [{\citenamefont {Hagmann}\ \emph
  {et~al.}(1990{\natexlab{a}})\citenamefont {Hagmann}, \citenamefont {Sikivie},
  \citenamefont {Sullivan},\ and\ \citenamefont {Tanner}}]{UF1990}%
  \BibitemOpen
  \bibfield  {author} {\bibinfo {author} {\bibfnamefont {C.}~\bibnamefont
  {Hagmann}}, \bibinfo {author} {\bibfnamefont {P.}~\bibnamefont {Sikivie}},
  \bibinfo {author} {\bibfnamefont {N.~S.}\ \bibnamefont {Sullivan}}, \ and\
  \bibinfo {author} {\bibfnamefont {D.~B.}\ \bibnamefont {Tanner}},\
  }\href@noop {} {\bibfield  {journal} {\bibinfo  {journal} {Phys. Rev. D}\
  }\textbf {\bibinfo {volume} {42}},\ \bibinfo {pages} {1297} (\bibinfo {year}
  {1990}{\natexlab{a}})}\BibitemShut {NoStop}%
\bibitem [{\citenamefont {Wuensch}\ \emph {et~al.}(1989)\citenamefont
  {Wuensch}, \citenamefont {De~Panfilis-Wuensch}, \citenamefont {Semertzidis},
  \citenamefont {Rogers}, \citenamefont {Melissinos}, \citenamefont {Halama},
  \citenamefont {Moskowitz}, \citenamefont {Prodell}, \citenamefont {Fowler},\
  and\ \citenamefont {Nezrick}}]{RBF1989}%
  \BibitemOpen
  \bibfield  {author} {\bibinfo {author} {\bibfnamefont {W.~U.}\ \bibnamefont
  {Wuensch}}, \bibinfo {author} {\bibfnamefont {S.}~\bibnamefont
  {De~Panfilis-Wuensch}}, \bibinfo {author} {\bibfnamefont {Y.~K.}\
  \bibnamefont {Semertzidis}}, \bibinfo {author} {\bibfnamefont {J.~T.}\
  \bibnamefont {Rogers}}, \bibinfo {author} {\bibfnamefont {A.~C.}\
  \bibnamefont {Melissinos}}, \bibinfo {author} {\bibfnamefont {H.~J.}\
  \bibnamefont {Halama}}, \bibinfo {author} {\bibfnamefont {B.~E.}\
  \bibnamefont {Moskowitz}}, \bibinfo {author} {\bibfnamefont {A.~G.}\
  \bibnamefont {Prodell}}, \bibinfo {author} {\bibfnamefont {W.~B.}\
  \bibnamefont {Fowler}}, \ and\ \bibinfo {author} {\bibfnamefont {F.~A.}\
  \bibnamefont {Nezrick}},\ }\href@noop {} {\bibfield  {journal} {\bibinfo
  {journal} {Phys. Rev. D}\ }\textbf {\bibinfo {volume} {40}},\ \bibinfo
  {pages} {3153} (\bibinfo {year} {1989})}\BibitemShut {NoStop}%
\bibitem [{\citenamefont {Hagmann}\ \emph {et~al.}(1998)\citenamefont {Hagmann}
  \emph {et~al.}}]{ADMX1998}%
  \BibitemOpen
  \bibfield  {author} {\bibinfo {author} {\bibfnamefont {C.}~\bibnamefont
  {Hagmann}} \emph {et~al.},\ }\href@noop {} {\bibfield  {journal} {\bibinfo
  {journal} {Phys. Rev. Lett.}\ }\textbf {\bibinfo {volume} {80}},\ \bibinfo
  {pages} {2043} (\bibinfo {year} {1998})},\ \Eprint
  {http://arxiv.org/abs/astro-ph/9801286}{arXiv:astro-ph/9801286}\BibitemShut
  {NoStop}%
\bibitem [{\citenamefont {Peng}\ \emph {et~al.}(2000)\citenamefont {Peng} \emph
  {et~al.}}]{ADMX2000}%
  \BibitemOpen
  \bibfield  {author} {\bibinfo {author} {\bibfnamefont {H.}~\bibnamefont
  {Peng}} \emph {et~al.},\ }\href@noop {} {\bibfield  {journal} {\bibinfo
  {journal} {Nucl. Instrum. Meth. A}\ }\textbf {\bibinfo {volume} {444}},\
  \bibinfo {pages} {569} (\bibinfo {year} {2000})}\BibitemShut {NoStop}%
\bibitem [{\citenamefont {Asztalos}\ \emph {et~al.}(2002)\citenamefont
  {Asztalos} \emph {et~al.}}]{ADMX2002}%
  \BibitemOpen
  \bibfield  {author} {\bibinfo {author} {\bibfnamefont {S.~J.}\ \bibnamefont
  {Asztalos}} \emph {et~al.},\ }\href@noop {} {\bibfield  {journal} {\bibinfo
  {journal} {Astrophys. J. Lett.}\ }\textbf {\bibinfo {volume} {571}},\
  \bibinfo {pages} {L27} (\bibinfo {year} {2002})},\ \Eprint
  {http://arxiv.org/abs/astro-ph/0104200}{arXiv:astro-ph/0104200}\BibitemShut
  {NoStop}%
\bibitem [{\citenamefont {Asztalos}\ \emph {et~al.}(2004)\citenamefont
  {Asztalos} \emph {et~al.}}]{ADMX2004}%
  \BibitemOpen
  \bibfield  {author} {\bibinfo {author} {\bibfnamefont {S.~J.}\ \bibnamefont
  {Asztalos}} \emph {et~al.},\ }\href@noop {} {\bibfield  {journal} {\bibinfo
  {journal} {Phys. Rev. D}\ }\textbf {\bibinfo {volume} {69}},\ \bibinfo
  {pages} {011101} (\bibinfo {year} {2004})},\ \Eprint
  {http://arxiv.org/abs/astro-ph/0310042}{arXiv:astro-ph/0310042}\BibitemShut
  {NoStop}%
\bibitem [{\citenamefont {Matsuki}\ \emph {et~al.}(1996)\citenamefont
  {Matsuki}, \citenamefont {Ogawa}, \citenamefont {Nakamura}, \citenamefont
  {Tada}, \citenamefont {Yamamoto},\ and\ \citenamefont
  {Masaike}}]{matsuki1996}%
  \BibitemOpen
  \bibfield  {author} {\bibinfo {author} {\bibfnamefont {S.}~\bibnamefont
  {Matsuki}}, \bibinfo {author} {\bibfnamefont {I.}~\bibnamefont {Ogawa}},
  \bibinfo {author} {\bibfnamefont {S.}~\bibnamefont {Nakamura}}, \bibinfo
  {author} {\bibfnamefont {M.}~\bibnamefont {Tada}}, \bibinfo {author}
  {\bibfnamefont {K.}~\bibnamefont {Yamamoto}}, \ and\ \bibinfo {author}
  {\bibfnamefont {A.}~\bibnamefont {Masaike}},\ }\href@noop {} {\bibfield
  {journal} {\bibinfo  {journal} {Nucl. Phys. B (Proc. Suppl.)}\ }\textbf
  {\bibinfo {volume} {51}},\ \bibinfo {pages} {213} (\bibinfo {year}
  {1996})}\BibitemShut {NoStop}%
\bibitem [{\citenamefont {Tada}\ \emph {et~al.}(1999)\citenamefont {Tada},
  \citenamefont {Kishimoto}, \citenamefont {Kominato}, \citenamefont {Shibata},
  \citenamefont {Funahashi}, \citenamefont {Yamamoto}, \citenamefont
  {Masaike},\ and\ \citenamefont {Matsuki}}]{tada1999}%
  \BibitemOpen
  \bibfield  {author} {\bibinfo {author} {\bibfnamefont {M.}~\bibnamefont
  {Tada}}, \bibinfo {author} {\bibfnamefont {Y.}~\bibnamefont {Kishimoto}},
  \bibinfo {author} {\bibfnamefont {K.}~\bibnamefont {Kominato}}, \bibinfo
  {author} {\bibfnamefont {M.}~\bibnamefont {Shibata}}, \bibinfo {author}
  {\bibfnamefont {H.}~\bibnamefont {Funahashi}}, \bibinfo {author}
  {\bibfnamefont {K.}~\bibnamefont {Yamamoto}}, \bibinfo {author}
  {\bibfnamefont {A.}~\bibnamefont {Masaike}}, \ and\ \bibinfo {author}
  {\bibfnamefont {S.}~\bibnamefont {Matsuki}},\ }\href@noop {} {\bibfield
  {journal} {\bibinfo  {journal} {Nucl. Phys. B (Proc. Suppl.)}\ }\textbf
  {\bibinfo {volume} {72}},\ \bibinfo {pages} {164} (\bibinfo {year}
  {1999})}\BibitemShut {NoStop}%
\bibitem [{\citenamefont {M\"{u}ck}\ \emph {et~al.}(1998)\citenamefont
  {M\"{u}ck}, \citenamefont {Andr\'{e}}, \citenamefont {Clarke}, \citenamefont
  {Gail},\ and\ \citenamefont {Heiden}}]{msa1998}%
  \BibitemOpen
  \bibfield  {author} {\bibinfo {author} {\bibfnamefont {M.}~\bibnamefont
  {M\"{u}ck}}, \bibinfo {author} {\bibfnamefont {M.-O.}\ \bibnamefont
  {Andr\'{e}}}, \bibinfo {author} {\bibfnamefont {J.}~\bibnamefont {Clarke}},
  \bibinfo {author} {\bibfnamefont {J.}~\bibnamefont {Gail}}, \ and\ \bibinfo
  {author} {\bibfnamefont {C.}~\bibnamefont {Heiden}},\ }\href@noop {}
  {\bibfield  {journal} {\bibinfo  {journal} {Appl. Phys. Lett.}\ }\textbf
  {\bibinfo {volume} {72}},\ \bibinfo {pages} {2885} (\bibinfo {year}
  {1998})}\BibitemShut {NoStop}%
\bibitem [{\citenamefont {M\"{u}ck}\ \emph {et~al.}(1999)\citenamefont
  {M\"{u}ck}, \citenamefont {Andr\'{e}}, \citenamefont {Clarke}, \citenamefont
  {Gail},\ and\ \citenamefont {Heiden}}]{msa1999}%
  \BibitemOpen
  \bibfield  {author} {\bibinfo {author} {\bibfnamefont {M.}~\bibnamefont
  {M\"{u}ck}}, \bibinfo {author} {\bibfnamefont {M.-O.}\ \bibnamefont
  {Andr\'{e}}}, \bibinfo {author} {\bibfnamefont {J.}~\bibnamefont {Clarke}},
  \bibinfo {author} {\bibfnamefont {J.}~\bibnamefont {Gail}}, \ and\ \bibinfo
  {author} {\bibfnamefont {C.}~\bibnamefont {Heiden}},\ }\href@noop {}
  {\bibfield  {journal} {\bibinfo  {journal} {Appl. Phys. Lett.}\ }\textbf
  {\bibinfo {volume} {75}},\ \bibinfo {pages} {3545} (\bibinfo {year}
  {1999})}\BibitemShut {NoStop}%
\bibitem [{\citenamefont {Asztalos}\ \emph {et~al.}(2011)\citenamefont
  {Asztalos} \emph {et~al.}}]{ADMX2011}%
  \BibitemOpen
  \bibfield  {author} {\bibinfo {author} {\bibfnamefont {S.~J.}\ \bibnamefont
  {Asztalos}} \emph {et~al.},\ }\href@noop {} {\bibfield  {journal} {\bibinfo
  {journal} {Nucl. Instrum. Meth. A}\ }\textbf {\bibinfo {volume} {656}},\
  \bibinfo {pages} {39} (\bibinfo {year} {2011})}\BibitemShut {NoStop}%
\bibitem [{\citenamefont {Asztalos}\ \emph {et~al.}(2010)\citenamefont
  {Asztalos} \emph {et~al.}}]{ADMX2010}%
  \BibitemOpen
  \bibfield  {author} {\bibinfo {author} {\bibfnamefont {S.~J.}\ \bibnamefont
  {Asztalos}} \emph {et~al.},\ }\href@noop {} {\bibfield  {journal} {\bibinfo
  {journal} {Phys. Rev. Lett.}\ }\textbf {\bibinfo {volume} {104}},\ \bibinfo
  {pages} {041301} (\bibinfo {year} {2010})},\ \Eprint
  {http://arxiv.org/abs/0910.5914}{arXiv:0910.5914}\BibitemShut {NoStop}%
\bibitem [{\citenamefont {Sloan}\ \emph {et~al.}(2016)\citenamefont {Sloan}
  \emph {et~al.}}]{ADMX2016}%
  \BibitemOpen
  \bibfield  {author} {\bibinfo {author} {\bibfnamefont {J.~V.}\ \bibnamefont
  {Sloan}} \emph {et~al.},\ }\href@noop {} {\bibfield  {journal} {\bibinfo
  {journal} {Phys. Dark Univ.}\ }\textbf {\bibinfo {volume} {14}},\ \bibinfo
  {pages} {95} (\bibinfo {year} {2016})}\BibitemShut {NoStop}%
\bibitem [{\citenamefont {Duffy}\ \emph {et~al.}(2006)\citenamefont {Duffy},
  \citenamefont {Sikivie}, \citenamefont {Tanner}, \citenamefont {Asztalos},
  \citenamefont {Hagmann}, \citenamefont {Kinion}, \citenamefont {Rosenberg},
  \citenamefont {van Bibber}, \citenamefont {Yu},\ and\ \citenamefont
  {Bradley}}]{duffy2006}%
  \BibitemOpen
  \bibfield  {author} {\bibinfo {author} {\bibfnamefont {L.~D.}\ \bibnamefont
  {Duffy}}, \bibinfo {author} {\bibfnamefont {P.}~\bibnamefont {Sikivie}},
  \bibinfo {author} {\bibfnamefont {D.~B.}\ \bibnamefont {Tanner}}, \bibinfo
  {author} {\bibfnamefont {S.~J.}\ \bibnamefont {Asztalos}}, \bibinfo {author}
  {\bibfnamefont {C.}~\bibnamefont {Hagmann}}, \bibinfo {author} {\bibfnamefont
  {D.}~\bibnamefont {Kinion}}, \bibinfo {author} {\bibfnamefont {L.~J.}\
  \bibnamefont {Rosenberg}}, \bibinfo {author} {\bibfnamefont {K.}~\bibnamefont
  {van Bibber}}, \bibinfo {author} {\bibfnamefont {D.~B.}\ \bibnamefont {Yu}},
  \ and\ \bibinfo {author} {\bibfnamefont {R.~F.}\ \bibnamefont {Bradley}},\
  }\href@noop {} {\bibfield  {journal} {\bibinfo  {journal} {Phys. Rev. D}\
  }\textbf {\bibinfo {volume} {74}},\ \bibinfo {pages} {012006} (\bibinfo
  {year} {2006})},\ \Eprint
  {http://arxiv.org/abs/astro-ph/0603108}{arXiv:astro-ph/0603108}\BibitemShut
  {NoStop}%
\bibitem [{\citenamefont {Hoskins}\ \emph {et~al.}(2011)\citenamefont {Hoskins}
  \emph {et~al.}}]{hoskins2011}%
  \BibitemOpen
  \bibfield  {author} {\bibinfo {author} {\bibfnamefont {J.}~\bibnamefont
  {Hoskins}} \emph {et~al.},\ }\href@noop {} {\bibfield  {journal} {\bibinfo
  {journal} {Phys. Rev. D}\ }\textbf {\bibinfo {volume} {84}},\ \bibinfo
  {pages} {121302} (\bibinfo {year} {2011})},\ \Eprint
  {http://arxiv.org/abs/1109.4128}{arXiv:1109.4128}\BibitemShut {NoStop}%
\bibitem [{\citenamefont {Hagmann}(1990)}]{hagmann1990}%
  \BibitemOpen
  \bibfield  {author} {\bibinfo {author} {\bibfnamefont {C.}~\bibnamefont
  {Hagmann}},\ }\emph {\bibinfo {title} {A {Search} for {Cosmic} {Axions}}},\
  \href@noop {} {Ph.D. thesis},\ \bibinfo  {school} {University of Florida}
  (\bibinfo {year} {1990})\BibitemShut {NoStop}%
\bibitem [{\citenamefont {Kinion}(2001)}]{kinion2001}%
  \BibitemOpen
  \bibfield  {author} {\bibinfo {author} {\bibfnamefont {D.}~\bibnamefont
  {Kinion}},\ }\emph {\bibinfo {title} {First {Results} from a
  {Multiple}-{Microwave}-{Cavity} {Search} for {Dark} {Matter} {Axions}}},\
  \href@noop {} {Ph.D. thesis},\ \bibinfo  {school} {University of California
  Davis} (\bibinfo {year} {2001})\BibitemShut {NoStop}%
\bibitem [{\citenamefont {Hagmann}\ \emph
  {et~al.}(1990{\natexlab{b}})\citenamefont {Hagmann}, \citenamefont {Sikivie},
  \citenamefont {Sullivan}, \citenamefont {Tanner},\ and\ \citenamefont
  {Cho}}]{cavity1990}%
  \BibitemOpen
  \bibfield  {author} {\bibinfo {author} {\bibfnamefont {C.}~\bibnamefont
  {Hagmann}}, \bibinfo {author} {\bibfnamefont {P.}~\bibnamefont {Sikivie}},
  \bibinfo {author} {\bibfnamefont {N.}~\bibnamefont {Sullivan}}, \bibinfo
  {author} {\bibfnamefont {D.~B.}\ \bibnamefont {Tanner}}, \ and\ \bibinfo
  {author} {\bibfnamefont {S.-I.}\ \bibnamefont {Cho}},\ }\href@noop {}
  {\bibfield  {journal} {\bibinfo  {journal} {Rev. Sci. Instrum.}\ }\textbf
  {\bibinfo {volume} {61}},\ \bibinfo {pages} {1076} (\bibinfo {year}
  {1990}{\natexlab{b}})}\BibitemShut {NoStop}%
\bibitem [{\citenamefont {Pippard}(1947)}]{pippard1947}%
  \BibitemOpen
  \bibfield  {author} {\bibinfo {author} {\bibfnamefont {A.~B.}\ \bibnamefont
  {Pippard}},\ }\href@noop {} {\bibfield  {journal} {\bibinfo  {journal} {Proc.
  R. Soc. A}\ }\textbf {\bibinfo {volume} {191}},\ \bibinfo {pages} {385}
  (\bibinfo {year} {1947})}\BibitemShut {NoStop}%
\bibitem [{\citenamefont {Carosi}()}]{gp}%
  \BibitemOpen
  \bibfield  {author} {\bibinfo {author} {\bibfnamefont {G.}~\bibnamefont
  {Carosi}},\ }\href@noop {} {}\bibinfo {howpublished} {private
  communication}\BibitemShut {NoStop}%
\bibitem [{\citenamefont {Finger}\ and\ \citenamefont
  {Kerr}(2008)}]{finger2008}%
  \BibitemOpen
  \bibfield  {author} {\bibinfo {author} {\bibfnamefont {R.}~\bibnamefont
  {Finger}}\ and\ \bibinfo {author} {\bibfnamefont {A.~R.}\ \bibnamefont
  {Kerr}},\ }\href@noop {} {\bibfield  {journal} {\bibinfo  {journal} {Int. J.
  Infrared Milli. Waves}\ }\textbf {\bibinfo {volume} {29}},\ \bibinfo {pages}
  {924} (\bibinfo {year} {2008})},\ \Eprint
  {http://arxiv.org/abs/1509.05273}{arXiv:1509.05273}\BibitemShut {NoStop}%
\bibitem [{\citenamefont {Lehnert}()}]{konrad}%
  \BibitemOpen
  \bibfield  {author} {\bibinfo {author} {\bibfnamefont {K.~W.}\ \bibnamefont
  {Lehnert}},\ }\href@noop {} {}\bibinfo {howpublished} {private
  communication}\BibitemShut {NoStop}%
\bibitem [{\citenamefont {Slocum}\ \emph {et~al.}(2015)\citenamefont {Slocum},
  \citenamefont {Baker}, \citenamefont {Hirshfield}, \citenamefont {Jiang},
  \citenamefont {Malagon}, \citenamefont {Martin}, \citenamefont
  {Shchelkunov},\ and\ \citenamefont {Szymkowiak}}]{YMCE2015}%
  \BibitemOpen
  \bibfield  {author} {\bibinfo {author} {\bibfnamefont {P.~L.}\ \bibnamefont
  {Slocum}}, \bibinfo {author} {\bibfnamefont {O.~K.}\ \bibnamefont {Baker}},
  \bibinfo {author} {\bibfnamefont {J.~L.}\ \bibnamefont {Hirshfield}},
  \bibinfo {author} {\bibfnamefont {Y.}~\bibnamefont {Jiang}}, \bibinfo
  {author} {\bibfnamefont {A.~T.}\ \bibnamefont {Malagon}}, \bibinfo {author}
  {\bibfnamefont {A.~J.}\ \bibnamefont {Martin}}, \bibinfo {author}
  {\bibfnamefont {S.}~\bibnamefont {Shchelkunov}}, \ and\ \bibinfo {author}
  {\bibfnamefont {A.}~\bibnamefont {Szymkowiak}},\ }\href@noop {} {\bibfield
  {journal} {\bibinfo  {journal} {Nucl. Instrum. Meth. A}\ }\textbf {\bibinfo
  {volume} {770}},\ \bibinfo {pages} {76} (\bibinfo {year} {2015})},\ \Eprint
  {http://arxiv.org/abs/1410.1807}{arXiv:1410.1807}\BibitemShut {NoStop}%
\bibitem [{\citenamefont {Malagon}(2014)}]{malagon2014}%
  \BibitemOpen
  \bibfield  {author} {\bibinfo {author} {\bibfnamefont {A.~T.}\ \bibnamefont
  {Malagon}},\ }\emph {\bibinfo {title} {Search for 140 {microeV}
  {Pseudoscalar} and {Vector} {Dark} {Matter} {Using} {Microwave}
  {Cavities}}},\ \href@noop {} {Ph.D. thesis},\ \bibinfo  {school} {Yale
  University} (\bibinfo {year} {2014})\BibitemShut {NoStop}%
\bibitem [{\citenamefont {Castellanos-Beltran}\ and\ \citenamefont
  {Lehnert}(2007)}]{castellanos2007}%
  \BibitemOpen
  \bibfield  {author} {\bibinfo {author} {\bibfnamefont {M.~A.}\ \bibnamefont
  {Castellanos-Beltran}}\ and\ \bibinfo {author} {\bibfnamefont {K.~W.}\
  \bibnamefont {Lehnert}},\ }\href@noop {} {\bibfield  {journal} {\bibinfo
  {journal} {Appl. Phys. Lett.}\ }\textbf {\bibinfo {volume} {91}},\ \bibinfo
  {pages} {083509} (\bibinfo {year} {2007})},\ \Eprint
  {http://arxiv.org/abs/0706.2373}{arXiv:0706.2373}\BibitemShut {NoStop}%
\bibitem [{\citenamefont {Castellanos-Beltran}\ \emph
  {et~al.}(2008)\citenamefont {Castellanos-Beltran}, \citenamefont {Irwin},
  \citenamefont {Hilton}, \citenamefont {Vale},\ and\ \citenamefont
  {Lehnert}}]{castellanos2008}%
  \BibitemOpen
  \bibfield  {author} {\bibinfo {author} {\bibfnamefont {M.~A.}\ \bibnamefont
  {Castellanos-Beltran}}, \bibinfo {author} {\bibfnamefont {K.~D.}\
  \bibnamefont {Irwin}}, \bibinfo {author} {\bibfnamefont {G.~C.}\ \bibnamefont
  {Hilton}}, \bibinfo {author} {\bibfnamefont {L.~R.}\ \bibnamefont {Vale}}, \
  and\ \bibinfo {author} {\bibfnamefont {K.~W.}\ \bibnamefont {Lehnert}},\
  }\href@noop {} {\bibfield  {journal} {\bibinfo  {journal} {Nature Phys.}\
  }\textbf {\bibinfo {volume} {4}},\ \bibinfo {pages} {929} (\bibinfo {year}
  {2008})},\ \Eprint
  {http://arxiv.org/abs/0806.0659}{arXiv:0806.0659}\BibitemShut {NoStop}%
\bibitem [{\citenamefont {Mallet}\ \emph {et~al.}(2011)\citenamefont {Mallet},
  \citenamefont {Castellanos-Beltran}, \citenamefont {Ku}, \citenamefont
  {Glancy}, \citenamefont {Knill}, \citenamefont {Irwin}, \citenamefont
  {Hilton}, \citenamefont {Vale},\ and\ \citenamefont {Lehnert}}]{mallet2011}%
  \BibitemOpen
  \bibfield  {author} {\bibinfo {author} {\bibfnamefont {F.}~\bibnamefont
  {Mallet}}, \bibinfo {author} {\bibfnamefont {M.~A.}\ \bibnamefont
  {Castellanos-Beltran}}, \bibinfo {author} {\bibfnamefont {H.~S.}\
  \bibnamefont {Ku}}, \bibinfo {author} {\bibfnamefont {S.}~\bibnamefont
  {Glancy}}, \bibinfo {author} {\bibfnamefont {E.}~\bibnamefont {Knill}},
  \bibinfo {author} {\bibfnamefont {K.~D.}\ \bibnamefont {Irwin}}, \bibinfo
  {author} {\bibfnamefont {G.~C.}\ \bibnamefont {Hilton}}, \bibinfo {author}
  {\bibfnamefont {L.~R.}\ \bibnamefont {Vale}}, \ and\ \bibinfo {author}
  {\bibfnamefont {K.~W.}\ \bibnamefont {Lehnert}},\ }\href@noop {} {\bibfield
  {journal} {\bibinfo  {journal} {Phys. Rev. Lett.}\ }\textbf {\bibinfo
  {volume} {106}},\ \bibinfo {pages} {220502} (\bibinfo {year} {2011})},\
  \Eprint {http://arxiv.org/abs/1012.0007}{arXiv:1012.0007}\BibitemShut
  {NoStop}%
\bibitem [{\citenamefont {Pobell}(1996)}]{pobell1996}%
  \BibitemOpen
  \bibfield  {author} {\bibinfo {author} {\bibfnamefont {F.}~\bibnamefont
  {Pobell}},\ }\href@noop {} {\emph {\bibinfo {title} {Matter and {Methods} at
  {Low} {Temperatures}}}}\ (\bibinfo  {publisher} {Springer},\ \bibinfo {year}
  {1996})\BibitemShut {NoStop}%
\bibitem [{\citenamefont {Brigham}(1988)}]{brigham1988}%
  \BibitemOpen
  \bibfield  {author} {\bibinfo {author} {\bibfnamefont {E.~O.}\ \bibnamefont
  {Brigham}},\ }\href@noop {} {\emph {\bibinfo {title} {The {Fast} {Fourier}
  {Transform} and its {Applications}}}}\ (\bibinfo  {publisher} {Prentice
  Hall},\ \bibinfo {year} {1988})\BibitemShut {NoStop}%
\bibitem [{\citenamefont {Zhong}\ \emph {et~al.}()\citenamefont {Zhong},
  \citenamefont {Brubaker}, \citenamefont {Cahn},\ and\ \citenamefont
  {Lamoreaux}}]{zhong2017}%
  \BibitemOpen
  \bibfield  {author} {\bibinfo {author} {\bibfnamefont {L.}~\bibnamefont
  {Zhong}}, \bibinfo {author} {\bibfnamefont {B.~M.}\ \bibnamefont {Brubaker}},
  \bibinfo {author} {\bibfnamefont {S.~B.}\ \bibnamefont {Cahn}}, \ and\
  \bibinfo {author} {\bibfnamefont {S.~K.}\ \bibnamefont {Lamoreaux}},\
  }\href@noop {} {\ }\Eprint
  {http://arxiv.org/abs/1706.03676}{arXiv:1706.03676}\BibitemShut {NoStop}%
\bibitem [{\citenamefont {Stern}\ \emph {et~al.}(2015)\citenamefont {Stern},
  \citenamefont {Chisholm}, \citenamefont {Hoskins}, \citenamefont {Sikivie},
  \citenamefont {Sullivan}, \citenamefont {Tanner}, \citenamefont {Carosi},\
  and\ \citenamefont {van Bibber}}]{stern2015}%
  \BibitemOpen
  \bibfield  {author} {\bibinfo {author} {\bibfnamefont {I.}~\bibnamefont
  {Stern}}, \bibinfo {author} {\bibfnamefont {A.~A.}\ \bibnamefont {Chisholm}},
  \bibinfo {author} {\bibfnamefont {J.}~\bibnamefont {Hoskins}}, \bibinfo
  {author} {\bibfnamefont {P.}~\bibnamefont {Sikivie}}, \bibinfo {author}
  {\bibfnamefont {N.~S.}\ \bibnamefont {Sullivan}}, \bibinfo {author}
  {\bibfnamefont {D.~B.}\ \bibnamefont {Tanner}}, \bibinfo {author}
  {\bibfnamefont {G.}~\bibnamefont {Carosi}}, \ and\ \bibinfo {author}
  {\bibfnamefont {K.}~\bibnamefont {van Bibber}},\ }\href@noop {} {\bibfield
  {journal} {\bibinfo  {journal} {Rev. Sci. Instrum.}\ }\textbf {\bibinfo
  {volume} {86}},\ \bibinfo {pages} {123305} (\bibinfo {year} {2015})},\
  \Eprint {http://arxiv.org/abs/1603.06990}{arXiv:1603.06990}\BibitemShut
  {NoStop}%
\bibitem [{\citenamefont {Nation}\ \emph {et~al.}(2012)\citenamefont {Nation},
  \citenamefont {Johansson}, \citenamefont {Blencowe},\ and\ \citenamefont
  {Nori}}]{swing}%
  \BibitemOpen
  \bibfield  {author} {\bibinfo {author} {\bibfnamefont {P.~D.}\ \bibnamefont
  {Nation}}, \bibinfo {author} {\bibfnamefont {J.~R.}\ \bibnamefont
  {Johansson}}, \bibinfo {author} {\bibfnamefont {M.~P.}\ \bibnamefont
  {Blencowe}}, \ and\ \bibinfo {author} {\bibfnamefont {F.}~\bibnamefont
  {Nori}},\ }\href@noop {} {\bibfield  {journal} {\bibinfo  {journal} {Rev.
  Mod. Phys.}\ }\textbf {\bibinfo {volume} {84}},\ \bibinfo {pages} {1}
  (\bibinfo {year} {2012})},\ \Eprint
  {http://arxiv.org/abs/1103.0835}{arXiv:1103.0835}\BibitemShut {NoStop}%
\bibitem [{\citenamefont {Yurke}\ and\ \citenamefont {Buks}(2006)}]{yurke2006}%
  \BibitemOpen
  \bibfield  {author} {\bibinfo {author} {\bibfnamefont {B.}~\bibnamefont
  {Yurke}}\ and\ \bibinfo {author} {\bibfnamefont {E.}~\bibnamefont {Buks}},\
  }\href@noop {} {\bibfield  {journal} {\bibinfo  {journal} {J. Lightwave
  Technol.}\ }\textbf {\bibinfo {volume} {24}},\ \bibinfo {pages} {5054}
  (\bibinfo {year} {2006})},\ \Eprint
  {http://arxiv.org/abs/quant-ph/0505018}{arXiv:quant-ph/0505018}\BibitemShut
  {NoStop}%
\bibitem [{\citenamefont {Castellanos-Beltran}(2010)}]{castellanos2010}%
  \BibitemOpen
  \bibfield  {author} {\bibinfo {author} {\bibfnamefont {M.~A.}\ \bibnamefont
  {Castellanos-Beltran}},\ }\emph {\bibinfo {title} {Development of a
  {Josephson} parametric amplifier for the preparation and detection of
  nonclassical states of microwave fields}},\ \href@noop {} {Ph.D. thesis},\
  \bibinfo  {school} {University of Colorado} (\bibinfo {year}
  {2010})\BibitemShut {NoStop}%
\bibitem [{\citenamefont {Krantz}\ \emph {et~al.}(2013)\citenamefont {Krantz},
  \citenamefont {Reshitnyk}, \citenamefont {Wustmann}, \citenamefont
  {Bylander}, \citenamefont {Gustavsson}, \citenamefont {Oliver}, \citenamefont
  {Duty}, \citenamefont {Shumeiko},\ and\ \citenamefont
  {Delsing}}]{krantz2013}%
  \BibitemOpen
  \bibfield  {author} {\bibinfo {author} {\bibfnamefont {P.}~\bibnamefont
  {Krantz}}, \bibinfo {author} {\bibfnamefont {Y.}~\bibnamefont {Reshitnyk}},
  \bibinfo {author} {\bibfnamefont {W.}~\bibnamefont {Wustmann}}, \bibinfo
  {author} {\bibfnamefont {J.}~\bibnamefont {Bylander}}, \bibinfo {author}
  {\bibfnamefont {S.}~\bibnamefont {Gustavsson}}, \bibinfo {author}
  {\bibfnamefont {W.~D.}\ \bibnamefont {Oliver}}, \bibinfo {author}
  {\bibfnamefont {T.}~\bibnamefont {Duty}}, \bibinfo {author} {\bibfnamefont
  {V.}~\bibnamefont {Shumeiko}}, \ and\ \bibinfo {author} {\bibfnamefont
  {P.}~\bibnamefont {Delsing}},\ }\href@noop {} {\bibfield  {journal} {\bibinfo
   {journal} {New J. Phys.}\ }\textbf {\bibinfo {volume} {15}},\ \bibinfo
  {pages} {105002} (\bibinfo {year} {2013})},\ \Eprint
  {http://arxiv.org/abs/1310.3966}{arXiv:1310.3966}\BibitemShut {NoStop}%
\bibitem [{\citenamefont {Castellanos-Beltran}\ \emph
  {et~al.}(2009)\citenamefont {Castellanos-Beltran}, \citenamefont {Irwin},
  \citenamefont {Vale}, \citenamefont {Hilton},\ and\ \citenamefont
  {Lehnert}}]{castellanos2009}%
  \BibitemOpen
  \bibfield  {author} {\bibinfo {author} {\bibfnamefont {M.~A.}\ \bibnamefont
  {Castellanos-Beltran}}, \bibinfo {author} {\bibfnamefont {K.~D.}\
  \bibnamefont {Irwin}}, \bibinfo {author} {\bibfnamefont {L.~R.}\ \bibnamefont
  {Vale}}, \bibinfo {author} {\bibfnamefont {G.~C.}\ \bibnamefont {Hilton}}, \
  and\ \bibinfo {author} {\bibfnamefont {K.~W.}\ \bibnamefont {Lehnert}},\
  }\href@noop {} {\bibfield  {journal} {\bibinfo  {journal} {IEEE Trans. Appl.
  Supercond.}\ }\textbf {\bibinfo {volume} {19}},\ \bibinfo {pages} {944}
  (\bibinfo {year} {2009})},\ \Eprint
  {http://arxiv.org/abs/0903.1243}{arXiv:0903.1243}\BibitemShut {NoStop}%
\bibitem [{\citenamefont {Asztalos}\ \emph {et~al.}(2001)\citenamefont
  {Asztalos} \emph {et~al.}}]{ADMX2001}%
  \BibitemOpen
  \bibfield  {author} {\bibinfo {author} {\bibfnamefont {S.~J.}\ \bibnamefont
  {Asztalos}} \emph {et~al.},\ }\href@noop {} {\bibfield  {journal} {\bibinfo
  {journal} {Phys. Rev. D}\ }\textbf {\bibinfo {volume} {64}},\ \bibinfo
  {pages} {092003} (\bibinfo {year} {2001})}\BibitemShut {NoStop}%
\bibitem [{\citenamefont {Savitzky}\ and\ \citenamefont
  {Golay}(1964)}]{sg1964}%
  \BibitemOpen
  \bibfield  {author} {\bibinfo {author} {\bibfnamefont {A.}~\bibnamefont
  {Savitzky}}\ and\ \bibinfo {author} {\bibfnamefont {M.~J.~E.}\ \bibnamefont
  {Golay}},\ }\href@noop {} {\bibfield  {journal} {\bibinfo  {journal} {Anal.
  Chem.}\ }\textbf {\bibinfo {volume} {36}},\ \bibinfo {pages} {1627} (\bibinfo
  {year} {1964})}\BibitemShut {NoStop}%
\bibitem [{\citenamefont {Schafer}(2011)}]{schafer2011}%
  \BibitemOpen
  \bibfield  {author} {\bibinfo {author} {\bibfnamefont {R.~W.}\ \bibnamefont
  {Schafer}},\ }\href@noop {} {\emph {\bibinfo {title} {On the frequency-domain
  properties of {Savitzky}-{Golay} filters}}},\ \bibinfo {type} {Tech. Rep.}\
  \bibinfo {number} {HPL-2010-109}\ (\bibinfo  {institution} {HP
  Laboratories},\ \bibinfo {year} {2011})\BibitemShut {NoStop}%
\bibitem [{\citenamefont {Daw}(1998)}]{daw1998}%
  \BibitemOpen
  \bibfield  {author} {\bibinfo {author} {\bibfnamefont {E.~J.}\ \bibnamefont
  {Daw}},\ }\emph {\bibinfo {title} {A {Search} for {Halo} {Axions}}},\
  \href@noop {} {Ph.D. thesis},\ \bibinfo  {school} {Massachusetts Institute of
  Technology} (\bibinfo {year} {1998})\BibitemShut {NoStop}%
\bibitem [{\citenamefont {Yu}(2004)}]{yu2004}%
  \BibitemOpen
  \bibfield  {author} {\bibinfo {author} {\bibfnamefont {D.~B.}\ \bibnamefont
  {Yu}},\ }\emph {\bibinfo {title} {An {Improved} {RF} {Cavity} {Search} for
  {Halo} {Axions}}},\ \href@noop {} {Ph.D. thesis},\ \bibinfo  {school}
  {Massachusetts Institute of Technology} (\bibinfo {year} {2004})\BibitemShut
  {NoStop}%
\bibitem [{\citenamefont {Hotz}(2013)}]{hotz2013}%
  \BibitemOpen
  \bibfield  {author} {\bibinfo {author} {\bibfnamefont {M.~T.}\ \bibnamefont
  {Hotz}},\ }\emph {\bibinfo {title} {A {SQUID}-{Based} {RF} {Cavity} {Search}
  for {Dark} {Matter} {Axions}}},\ \href@noop {} {Ph.D. thesis},\ \bibinfo
  {school} {University of Washington} (\bibinfo {year} {2013})\BibitemShut
  {NoStop}%
\bibitem [{\citenamefont {Lyapustin}(2015)}]{lyapustin2015}%
  \BibitemOpen
  \bibfield  {author} {\bibinfo {author} {\bibfnamefont {D.}~\bibnamefont
  {Lyapustin}},\ }\emph {\bibinfo {title} {An improved low-temperature
  {RF}-cavity search for dark-matter axions}},\ \href@noop {} {Ph.D. thesis},\
  \bibinfo  {school} {University of Washington} (\bibinfo {year}
  {2015})\BibitemShut {NoStop}%
\bibitem [{\citenamefont {Palken}()}]{palken}%
  \BibitemOpen
  \bibfield  {author} {\bibinfo {author} {\bibfnamefont {D.~A.}\ \bibnamefont
  {Palken}},\ }\href@noop {} {}\bibinfo {howpublished} {private
  communication}\BibitemShut {NoStop}%
\bibitem [{\citenamefont {Lewis}()}]{lewis}%
  \BibitemOpen
  \bibfield  {author} {\bibinfo {author} {\bibfnamefont {S.~M.}\ \bibnamefont
  {Lewis}},\ }\href@noop {} {}\bibinfo {howpublished} {private
  communication}\BibitemShut {NoStop}%
\bibitem [{\citenamefont {Tanner}()}]{tanner}%
  \BibitemOpen
  \bibfield  {author} {\bibinfo {author} {\bibfnamefont {D.~B.}\ \bibnamefont
  {Tanner}},\ }\href@noop {} {}\bibinfo {howpublished} {private
  communication}\BibitemShut {NoStop}%
\bibitem [{\citenamefont {Chung}(2016)}]{CAPP2016}%
  \BibitemOpen
  \bibfield  {author} {\bibinfo {author} {\bibfnamefont {W.}~\bibnamefont
  {Chung}},\ }\href@noop {} {\bibfield  {journal} {\bibinfo  {journal} {PoS}\
  }\textbf {\bibinfo {volume} {CORFU2015}},\ \bibinfo {pages} {047} (\bibinfo
  {year} {2016})}\BibitemShut {NoStop}%
\bibitem [{\citenamefont {Graham}\ and\ \citenamefont
  {Rajendran}(2013)}]{graham2013}%
  \BibitemOpen
  \bibfield  {author} {\bibinfo {author} {\bibfnamefont {P.~W.}\ \bibnamefont
  {Graham}}\ and\ \bibinfo {author} {\bibfnamefont {S.}~\bibnamefont
  {Rajendran}},\ }\href@noop {} {\bibfield  {journal} {\bibinfo  {journal}
  {Phys. Rev. D}\ }\textbf {\bibinfo {volume} {88}},\ \bibinfo {pages} {035023}
  (\bibinfo {year} {2013})},\ \Eprint
  {http://arxiv.org/abs/1306.6088}{arXiv:1306.6088}\BibitemShut {NoStop}%
\bibitem [{\citenamefont {Budker}\ \emph {et~al.}(2014)\citenamefont {Budker},
  \citenamefont {Graham}, \citenamefont {Ledbetter}, \citenamefont
  {Rajendran},\ and\ \citenamefont {Sushkov}}]{casper2014}%
  \BibitemOpen
  \bibfield  {author} {\bibinfo {author} {\bibfnamefont {D.}~\bibnamefont
  {Budker}}, \bibinfo {author} {\bibfnamefont {P.~W.}\ \bibnamefont {Graham}},
  \bibinfo {author} {\bibfnamefont {M.}~\bibnamefont {Ledbetter}}, \bibinfo
  {author} {\bibfnamefont {S.}~\bibnamefont {Rajendran}}, \ and\ \bibinfo
  {author} {\bibfnamefont {A.~O.}\ \bibnamefont {Sushkov}},\ }\href@noop {}
  {\bibfield  {journal} {\bibinfo  {journal} {Phys. Rev. X}\ }\textbf {\bibinfo
  {volume} {4}},\ \bibinfo {pages} {021030} (\bibinfo {year} {2014})},\ \Eprint
  {http://arxiv.org/abs/1306.6089}{arXiv:1306.6089}\BibitemShut {NoStop}%
\bibitem [{\citenamefont {Sikivie}\ \emph {et~al.}(2014)\citenamefont
  {Sikivie}, \citenamefont {Sullivan},\ and\ \citenamefont
  {Tanner}}]{sikivie2014}%
  \BibitemOpen
  \bibfield  {author} {\bibinfo {author} {\bibfnamefont {P.}~\bibnamefont
  {Sikivie}}, \bibinfo {author} {\bibfnamefont {N.}~\bibnamefont {Sullivan}}, \
  and\ \bibinfo {author} {\bibfnamefont {D.~B.}\ \bibnamefont {Tanner}},\
  }\href@noop {} {\bibfield  {journal} {\bibinfo  {journal} {Phys. Rev. Lett.}\
  }\textbf {\bibinfo {volume} {112}},\ \bibinfo {pages} {131301} (\bibinfo
  {year} {2014})},\ \Eprint
  {http://arxiv.org/abs/1310.8545}{arXiv:1310.8545}\BibitemShut {NoStop}%
\bibitem [{\citenamefont {Kahn}\ \emph {et~al.}(2016)\citenamefont {Kahn},
  \citenamefont {Safdi},\ and\ \citenamefont {Thaler}}]{abracadabra2016}%
  \BibitemOpen
  \bibfield  {author} {\bibinfo {author} {\bibfnamefont {Y.}~\bibnamefont
  {Kahn}}, \bibinfo {author} {\bibfnamefont {B.~R.}\ \bibnamefont {Safdi}}, \
  and\ \bibinfo {author} {\bibfnamefont {J.}~\bibnamefont {Thaler}},\
  }\href@noop {} {\bibfield  {journal} {\bibinfo  {journal} {Phys. Rev. Lett.}\
  }\textbf {\bibinfo {volume} {117}},\ \bibinfo {pages} {141801} (\bibinfo
  {year} {2016})},\ \Eprint
  {http://arxiv.org/abs/1602.01086}{arXiv:1602.01086}\BibitemShut {NoStop}%
\bibitem [{\citenamefont {Silva-Feaver}\ \emph {et~al.}()\citenamefont
  {Silva-Feaver} \emph {et~al.}}]{dmradio2016}%
  \BibitemOpen
  \bibfield  {author} {\bibinfo {author} {\bibfnamefont {M.}~\bibnamefont
  {Silva-Feaver}} \emph {et~al.},\ }\href@noop {} {\ }\Eprint
  {http://arxiv.org/abs/1610.09344}{arXiv:1610.09344}\BibitemShut {NoStop}%
\bibitem [{\citenamefont {Caldwell}\ \emph {et~al.}(2017)\citenamefont
  {Caldwell}, \citenamefont {Dvali}, \citenamefont {Majorovits}, \citenamefont
  {Millar}, \citenamefont {Raffelt}, \citenamefont {Redondo}, \citenamefont
  {Reimann}, \citenamefont {Simon},\ and\ \citenamefont
  {Steffen}}]{MADMAX2017}%
  \BibitemOpen
  \bibfield  {author} {\bibinfo {author} {\bibfnamefont {A.}~\bibnamefont
  {Caldwell}}, \bibinfo {author} {\bibfnamefont {G.}~\bibnamefont {Dvali}},
  \bibinfo {author} {\bibfnamefont {B.}~\bibnamefont {Majorovits}}, \bibinfo
  {author} {\bibfnamefont {A.}~\bibnamefont {Millar}}, \bibinfo {author}
  {\bibfnamefont {G.}~\bibnamefont {Raffelt}}, \bibinfo {author} {\bibfnamefont
  {J.}~\bibnamefont {Redondo}}, \bibinfo {author} {\bibfnamefont
  {O.}~\bibnamefont {Reimann}}, \bibinfo {author} {\bibfnamefont
  {F.}~\bibnamefont {Simon}}, \ and\ \bibinfo {author} {\bibfnamefont
  {F.}~\bibnamefont {Steffen}} (\bibinfo {collaboration} {MADMAX Working
  Group}),\ }\href@noop {} {\bibfield  {journal} {\bibinfo  {journal} {Phys.
  Rev. Lett.}\ }\textbf {\bibinfo {volume} {118}},\ \bibinfo {pages} {091801}
  (\bibinfo {year} {2017})},\ \Eprint
  {http://arxiv.org/abs/1611.05865}{arXiv:1611.05865}\BibitemShut {NoStop}%
\bibitem [{\citenamefont {Rybka}\ \emph {et~al.}(2015)\citenamefont {Rybka},
  \citenamefont {Wagner}, \citenamefont {Brill}, \citenamefont {Patel},
  \citenamefont {Percival},\ and\ \citenamefont {Ramos}}]{rybka2015}%
  \BibitemOpen
  \bibfield  {author} {\bibinfo {author} {\bibfnamefont {G.}~\bibnamefont
  {Rybka}}, \bibinfo {author} {\bibfnamefont {A.}~\bibnamefont {Wagner}},
  \bibinfo {author} {\bibfnamefont {A.}~\bibnamefont {Brill}}, \bibinfo
  {author} {\bibfnamefont {K.}~\bibnamefont {Patel}}, \bibinfo {author}
  {\bibfnamefont {R.}~\bibnamefont {Percival}}, \ and\ \bibinfo {author}
  {\bibfnamefont {K.}~\bibnamefont {Ramos}},\ }\href@noop {} {\bibfield
  {journal} {\bibinfo  {journal} {Phys. Rev. D}\ }\textbf {\bibinfo {volume}
  {91}},\ \bibinfo {pages} {011701} (\bibinfo {year} {2015})},\ \Eprint
  {http://arxiv.org/abs/1403.3121}{arXiv:1403.3121}\BibitemShut {NoStop}%
\bibitem [{\citenamefont {Arvanitaki}\ and\ \citenamefont
  {Geraci}(2014)}]{ariadne2014}%
  \BibitemOpen
  \bibfield  {author} {\bibinfo {author} {\bibfnamefont {A.}~\bibnamefont
  {Arvanitaki}}\ and\ \bibinfo {author} {\bibfnamefont {A.~A.}\ \bibnamefont
  {Geraci}},\ }\href@noop {} {\bibfield  {journal} {\bibinfo  {journal} {Phys.
  Rev. Lett.}\ }\textbf {\bibinfo {volume} {113}},\ \bibinfo {pages} {161801}
  (\bibinfo {year} {2014})},\ \Eprint
  {http://arxiv.org/abs/1403.1290}{arXiv:1403.1290}\BibitemShut {NoStop}%
\bibitem [{\citenamefont {Baumann}\ \emph {et~al.}(2016)\citenamefont
  {Baumann}, \citenamefont {Green},\ and\ \citenamefont
  {Wallisch}}]{baumann2016}%
  \BibitemOpen
  \bibfield  {author} {\bibinfo {author} {\bibfnamefont {D.}~\bibnamefont
  {Baumann}}, \bibinfo {author} {\bibfnamefont {D.}~\bibnamefont {Green}}, \
  and\ \bibinfo {author} {\bibfnamefont {B.}~\bibnamefont {Wallisch}},\
  }\href@noop {} {\bibfield  {journal} {\bibinfo  {journal} {Phys. Rev. Lett.}\
  }\textbf {\bibinfo {volume} {117}},\ \bibinfo {pages} {171301} (\bibinfo
  {year} {2016})},\ \Eprint
  {http://arxiv.org/abs/1604.08614}{arXiv:1604.08614}\BibitemShut {NoStop}%
\bibitem [{\citenamefont {O'Hare}\ and\ \citenamefont
  {Green}(2017)}]{ohare2017}%
  \BibitemOpen
  \bibfield  {author} {\bibinfo {author} {\bibfnamefont {C.~A.~J.}\
  \bibnamefont {O'Hare}}\ and\ \bibinfo {author} {\bibfnamefont {A.~M.}\
  \bibnamefont {Green}},\ }\href@noop {} {\bibfield  {journal} {\bibinfo
  {journal} {Phys. Rev. D}\ }\textbf {\bibinfo {volume} {95}},\ \bibinfo
  {pages} {063017} (\bibinfo {year} {2017})},\ \Eprint
  {http://arxiv.org/abs/1701.03118}{arXiv:1701.03118}\BibitemShut {NoStop}%
\bibitem [{\citenamefont {Graham}\ \emph
  {et~al.}(2015{\natexlab{b}})\citenamefont {Graham}, \citenamefont {Kaplan},\
  and\ \citenamefont {Rajendran}}]{graham2015}%
  \BibitemOpen
  \bibfield  {author} {\bibinfo {author} {\bibfnamefont {P.~W.}\ \bibnamefont
  {Graham}}, \bibinfo {author} {\bibfnamefont {D.~E.}\ \bibnamefont {Kaplan}},
  \ and\ \bibinfo {author} {\bibfnamefont {S.}~\bibnamefont {Rajendran}},\
  }\href@noop {} {\bibfield  {journal} {\bibinfo  {journal} {Phys. Rev. Lett.}\
  }\textbf {\bibinfo {volume} {115}},\ \bibinfo {pages} {221801} (\bibinfo
  {year} {2015}{\natexlab{b}})},\ \Eprint
  {http://arxiv.org/abs/1504.07551}{arXiv:1504.07551}\BibitemShut {NoStop}%
\bibitem [{\citenamefont {Ballesteros}\ \emph {et~al.}(2017)\citenamefont
  {Ballesteros}, \citenamefont {Redondo}, \citenamefont {Ringwald},\ and\
  \citenamefont {Tamarit}}]{ballesteros2017}%
  \BibitemOpen
  \bibfield  {author} {\bibinfo {author} {\bibfnamefont {G.}~\bibnamefont
  {Ballesteros}}, \bibinfo {author} {\bibfnamefont {J.}~\bibnamefont
  {Redondo}}, \bibinfo {author} {\bibfnamefont {A.}~\bibnamefont {Ringwald}}, \
  and\ \bibinfo {author} {\bibfnamefont {C.}~\bibnamefont {Tamarit}},\
  }\href@noop {} {\bibfield  {journal} {\bibinfo  {journal} {Phys. Rev. Lett.}\
  }\textbf {\bibinfo {volume} {118}},\ \bibinfo {pages} {071802} (\bibinfo
  {year} {2017})},\ \Eprint
  {http://arxiv.org/abs/1608.05414}{arXiv:1608.05414}\BibitemShut {NoStop}%
\bibitem [{\citenamefont {Kerckhoff}\ \emph {et~al.}(2015)\citenamefont
  {Kerckhoff}, \citenamefont {LalumiÃ¨re}, \citenamefont {Chapman},
  \citenamefont {Blais},\ and\ \citenamefont {Lehnert}}]{kerckhoff2015}%
  \BibitemOpen
  \bibfield  {author} {\bibinfo {author} {\bibfnamefont {J.}~\bibnamefont
  {Kerckhoff}}, \bibinfo {author} {\bibfnamefont {K.}~\bibnamefont
  {LalumiÃ¨re}}, \bibinfo {author} {\bibfnamefont {B.~J.}\ \bibnamefont
  {Chapman}}, \bibinfo {author} {\bibfnamefont {A.}~\bibnamefont {Blais}}, \
  and\ \bibinfo {author} {\bibfnamefont {K.~W.}\ \bibnamefont {Lehnert}},\
  }\href@noop {} {\bibfield  {journal} {\bibinfo  {journal} {Phys. Rev.
  Applied}\ }\textbf {\bibinfo {volume} {4}},\ \bibinfo {pages} {034002}
  (\bibinfo {year} {2015})},\ \Eprint
  {http://arxiv.org/abs/1502.06041}{arXiv:1502.06041}\BibitemShut {NoStop}%
\bibitem [{\citenamefont {Millar}\ \emph {et~al.}(2017)\citenamefont {Millar},
  \citenamefont {Raffelt}, \citenamefont {Redondo},\ and\ \citenamefont
  {Steffen}}]{millar2017}%
  \BibitemOpen
  \bibfield  {author} {\bibinfo {author} {\bibfnamefont {A.~J.}\ \bibnamefont
  {Millar}}, \bibinfo {author} {\bibfnamefont {G.~G.}\ \bibnamefont {Raffelt}},
  \bibinfo {author} {\bibfnamefont {J.}~\bibnamefont {Redondo}}, \ and\
  \bibinfo {author} {\bibfnamefont {F.~D.}\ \bibnamefont {Steffen}},\
  }\href@noop {} {\bibfield  {journal} {\bibinfo  {journal} {J. Cosmol.
  Astropart. Phys.}\ }\textbf {\bibinfo {volume} {2017}},\ \bibinfo {pages}
  {061} (\bibinfo {year} {2017})},\ \Eprint
  {http://arxiv.org/abs/1612.07057}{arXiv:1612.07057}\BibitemShut {NoStop}%
\bibitem [{\citenamefont {Theriault}\ \emph {et~al.}(2012)\citenamefont
  {Theriault}, \citenamefont {Leidich},\ and\ \citenamefont
  {Campbell}}]{ca3018}%
  \BibitemOpen
  \bibfield  {author} {\bibinfo {author} {\bibfnamefont {G.~E.}\ \bibnamefont
  {Theriault}}, \bibinfo {author} {\bibfnamefont {A.~J.}\ \bibnamefont
  {Leidich}}, \ and\ \bibinfo {author} {\bibfnamefont {T.~H.}\ \bibnamefont
  {Campbell}},\ }\href@noop {} {\emph {\bibinfo {title} {Application of the
  {CA3018} {Integrated-Circuit} {Transistor} {Array}}}},\ \bibinfo {type}
  {Tech. Rep.}\ \bibinfo {number} {AN5296.0}\ (\bibinfo  {institution}
  {Intersil},\ \bibinfo {year} {2012})\BibitemShut {NoStop}%
\bibitem [{\citenamefont {Aitken}(1935)}]{GLS1935}%
  \BibitemOpen
  \bibfield  {author} {\bibinfo {author} {\bibfnamefont {A.~C.}\ \bibnamefont
  {Aitken}},\ }\href@noop {} {\bibfield  {journal} {\bibinfo  {journal} {Proc.
  R. Soc. Edinb.}\ }\textbf {\bibinfo {volume} {55}},\ \bibinfo {pages} {42}
  (\bibinfo {year} {1935})}\BibitemShut {NoStop}%
\end{thebibliography}
\end{document}